\documentclass[11pt,letterpaper]{report} 	

\usepackage{amsmath}
\usepackage{amssymb}
\usepackage{amsthm}
\usepackage{geometry}
\usepackage{setspace}               
\usepackage{graphicx}
\usepackage{rotating}
\usepackage{lscape}
\usepackage{caption}
\usepackage{subfigure}
\usepackage{appendix}
\usepackage[english]{babel}         
\usepackage{blindtext}              
\usepackage{unh_thesis}
\usepackage{slashed}
\newcommand{\pvec}[1]{\vec{#1}\mkern2mu\vphantom{#1}}
\usepackage{braket}
\usepackage{multirow}
\usepackage{notoccite}

\usepackage{color}
\usepackage[colorlinks=true,hypertexnames=false]{hyperref}
\usepackage[all]{hypcap}
\usepackage{bookmark}

\begin{document}

\title{\sc The $g_2^p$ Experiment: A Measurement of the Proton's Spin Structure Functions}           	 
\author{RYAN ZIELINSKI}                     		 
\AbstractAuthor{Ryan Zielinski}                   
\prevdegrees{B.S. in Physics, College of William and Mary, 2010 \\
             }	
\major{Physics}                               						
\degree{Doctor of Philosophy}                          				
\degreemonth{September}                                   			
\degreeyear{2017}                                      							 %
\approvaldate{June 9, 2017}
\thesisdate{June 9, 2017}
\DOCUMENTtype{DISSERTATION}
\Documenttype{Dissertation}
\documenttype{dissertation}
\maketitle

\copyrightyear{2017}                                   	    
\makecopyright                                         	 		

\supervisor{Karl Slifer}{Associate Professor of Phyiscs}
\committee{Per Berglund}{Professor of Physics}
\committee{James Connell}{Associate Professor of Physics}
\committee{Maurik Holtrop}{Professor of Physics}
\committee{Elena Long}{Assistant Professor of Physics}
\makeapproval 


\begin{dedication}
To my parents, Ron and Judy Zielinski, who made all of this possible. And no, you don't have to read any further. There isn't going to be a test at the end.
\end{dedication}

\acknowledgments
\indent
\singlespacing
I apologize in advance for this section being as brief as it is. It isn't possible to list and thank everyone who helped complete this thesis, but I won't forget your contributions. Even if you are not mentioned explicitly, I was still thinking about you as I wrote this.
\newline

I didn't take the decision to graduate after (only) seven years lightly. The option of potentially continuing on for an eighth year, knowing that I would still be working with my advisor Dr. Karl Slifer was very enticing. Karl is the epitome of a good PhD advisor. His support and guidance were very much appreciated during this entire process. I will always be thankful for the time I spent in his lab and research group.
\newline

I was very fortunate that the E08-027 collaboration was filled with talented physicists. I learned a lot from the spokespeople (Karl was also a spokesperson), Alexandre Camsonne, Don Crabb, and J.-P. Chen. This thesis would not haven been possible if not for their steady hands as they navigated the experiment through its fair-share of problems. Thank you to the Hall A collaboration and the JLab target group who also greatly contributed to the success of the experiment. I would also be remiss if I didn't mention my fellow graduate students who contributed in many ways to the work presented in this thesis: Toby Badman, Melissa Cummings, Chao Gu, Min Huang, Jie Liu, and Pengjia Zhu. The experiment had several post-docs whose hard work was crucial to its success; many thanks to Kalyan Allada, Ellie Long, James Maxwell, Vince Sulkosky, and Jixie Zhang. Special thanks to Vince for his incalculable amount of help from the start of the experiment to my final analysis. And also for not rubbing it in when the Steelers beat the Jets.
\newline

Thank you to the many friends I've met at UNH: Kris, Narges, Toby, Max, Wei, Alex, Luna, Jon, Dan, Muji, Amanda, Ian. Special thanks to my fellow UNH physics graduate students who helped me get through the long slog of classes and the longer slog of the PhD candidate life. You all will be missed.
\newline  

Thank you to Carolyn for all that you've done. Words aren't enough to express my gratitude for your help and encouragement as I finished this long journey through graduate school.
\newline 

Additional thank yous are in order for the rest of the UNH polarized target group. I couldn't think of a better way to spend twelve plus hours than on a cool down with you all (indium seals notwithstanding). It was a very welcome distraction from g2p analysis.  Also, I believe there is a free dime on the floor of the DeMeritt 103 lab. 
\newline

To Kyle, Mike, Lee and Damian: I think it's finally time to get the band back together!


\endacknowledgments
\doublespacing
\setcounter{secnumdepth}{2} 
\setcounter{tocdepth}{3}  
\hypersetup{linkcolor=black}
\tableofcontents
\listoftables
\listoffigures

\begin{abstractpage}
\indent
\singlespacing

The E08-027 (g2p) experiment measured the spin structure functions of the proton at Jefferson Laboratory in Newport News, Va.  Longitudinally polarized electrons were scattered from a transversely and longitudinally polarized solid ammonia target in Jefferson Lab's Hall A, with the polarized NH$_3$ acting as an effective proton target. Focusing on small scattering angle events at the electron energies available at Jefferson Lab, the experiment covered a kinematic phase space of 0.02 GeV$^2$ $<$ $Q^2$ $<$ 0.20 GeV$^2$ in the proton's resonance region. The spin structure functions, $g_{1}^p(x,Q^2)$ and $g_{2}^p(x,Q^2)$ , are extracted from an inclusive polarized cross section measurement of the electron-proton interaction.
 
Low momentum transfer measurements, such as this experiment, are critical to enhance the understanding of the proton because of its complex internal structure and finite size. These internal interactions influence the proton's global properties and even the energy levels in atomic hydrogen.  The non-pertubative nature of the theory governing the interactions of the internal quarks and gluons, Quantum Chromodynamics (QCD), makes it difficult to calculate the effect from the internal interactions in the underlying theory. While not able to calculate the structure functions directly, QCD can make predictions of the spin structure functions integrated over the kinematic phase space. 

This thesis will present results for the proton spin structure functions $g_1(x,Q^2)$ and $g_2(x,Q^2)$ from the E08-027 experimental data. Integrated moments of $g_1(x,Q^2)$ are calculated and compared to theoretical predictions made by Chiral Perturbation Theory. The $g_1(x,Q^2)$ results are in agreement with previous measurements, but include a significant increase in statistical precision. The spin structure function contributions to the hyperfine energy levels in the hydrogen atom are also investigated. The $g_2(x,Q^2)$ measured contribution to the hyperfine splitting is the first ever experimental determination of this quantity. The results of this thesis suggest a disagreement of over 100\% with previously published model results.

\doublespacing
\end{abstractpage}

\pagenumbering{arabic}
\chapter{\sc Introduction}
\label{ch:Introduction}
Throughout recorded history, humans have attempted to describe the world around them. The descriptions have evolved over time, becoming smaller in scale; starting with the four classical elements of earth, water, air, and fire and continuing on to periodic table of the elements and the atom. In 1897, J.J. Thomson discovered the electron and, with it, the first elementary particle. Developed through the following decades, quantum mechanics provided a theoretical description of such particles. A key tenant of quantum theory states that some physical observables exist in discrete levels. In 1922, Otto Stern and Walther Gerlach~\cite{Stern} demonstrated that $spin$ was one such observable. By sending a beam of silver atoms through an inhomogeneous magnetic field, they were able to observe deflected electrons that localized to discrete points as opposed to the continuum predicted by classical theory.

But what is spin? In a classical theory, an object's spin is related to its rotation about its own axis and is a form of angular momentum. The day-night cycle of the Earth is a direct result of classical spin. If the classical object carries a charge, then a magnetic moment is associated with this spin. In a quantum theory, an object's spin cannot be associated with any physical rotation; a point-like, elementary particle would need to spin infinitely fast, but quantum objects that possess spin and charge also possess a magnetic moment. Quantum spin is therefore best described as a fundamental property of an object.

In 1928, Paul Dirac derived a quantum mechanical theory that described massive spin-$\frac{1}{2}$ point-like particles~\cite{Dirac610}. A year prior, in 1927, David Dennison showed that like an electron, a proton (discovered by Ernest Rutherford in 1917~\cite{Rutherford374,soddy1920name}) has spin-$\frac{1}{2}$. Unlike an electron though, the proton is not a point-like particle.  In 1933,  Immanuel Estermann and Otto Stern found that the magnetic moment of the proton was roughly two times larger than predicted in Dirac's theory for a structureless spin-$\frac{1}{2}$ particle~\cite{MagMom}. Years later in the late 1960s, high energy scattering experiments at the Stanford Linear Accelerator (SLAC) revealed that the inner structure of the proton is comprised of two kinds of constituent particles: quarks and gluons.

Around the same time as the SLAC experiments, Sin-Itiro Tomonaga, Julian Schwinger and Richard Feynman extended Dirac's theory to include the massless photon and created Quantum Electrodynamics (QED). The theory governs the interactions of all electromagnetically charged particles and has produced highly accurate results, where measurements of the electron's anomalous magnetic moment agree with QED beyond 10 significant figures~\cite{Feynman}. Quantum Chromodynamics (QCD) is the attempt to extend the rules of QED to the theory of the strong interaction between quarks and gluons. In QCD, the force carrying particle is the gluon instead of the photon and electromagnetic charge is replaced by color charge. 

There are two distinct differences between QCD and QED: confinement and asymptotic freedom. Confinement states that color charged particles do not exist singularly but only as a group and, consequentially, individual quarks cannot be directly observed. Asymptotic freedom means that in higher energy interactions the quarks and gluons interact $less$. Both confinement and asymptotic freedom are related to the idea that the gluon carries color charge (in QED the photon is electromagnetically neutral)~\cite{Thomas,Greiner}. 

Asymptotic freedom makes QCD perturbative only at high energies\footnote{High is generally considered greater than a few GeV$^2$, with respect to the squared four-momentum transfer ($Q^2$) in the relevant process.}, where the quarks and gluons interact very weakly. How then to describe the role quarks and gluons play at low energies where their complex many-body interactions give rise to the global properties of the proton? Effective fields theories, where the degrees of freedom in the theory are adjusted to the relevant energy scale, are one possible solution. Chiral Perturbation Theory ($\chi$PT) is an example and it replaces the quark and gluon degrees of freedom of QCD with hadron degrees of freedom. This is a consequence of confinement. 

Another way to study the low-energy non-perturbative aspect of QCD is through experiment. Before the 1980s, it was assumed that the quarks carried all of the proton's spin. In 1988, data from the European Muon Collaboration at CERN suggested that the intrinsic quark spin only contributes 30\% of the total proton spin~\cite{SpinCrisis}. This ``$spin$ $crisis$" inspired many new experiments, all trying to describe the spin structure of the nucleon. While these experiments and similar ones that followed~\cite{SpinCrisis2} collected data in the pertrubative QCD regime, the same general idea applies to probes of low-energy QCD: use experimentally collected data to verify the governing theory.

This thesis will focus on the analysis of one such spin structure experiment, E08-027 (g2p), which ran during the spring of 2012 at the Thomas Jefferson National Accelerator Facility's (Jefferson Lab) Hall A. The experiment collected data in the non-pertubative region of QCD, where $\chi$PT is a more theoretically tractable description of the underlying interactions. The data is a direct test of this effective field theory. The nature of the low-energy\footnote{Low energy here refers to $Q^2$ $\lessapprox$ 0.5 GeV$^2$.} QCD dynamics also make this data a measurement of how the internal interactions of the nucleons manifest themselves in the nucleon's global properties. A link between this behavior, via the nucleon structure functions, and the energy levels in the hydrogen atom is investigated in this thesis.

The theoretical background of the experiment is discussed in Chapter~\ref{ch:epScatter} and Chapter~\ref{ch:Tools}. Existing spin structure function data, with a focus on the applications of the low momentum transfer data is the focus of Chapter~\ref{ch:ExistingData}. Chapter~\ref{ch:Experiment} discusses the experimental apparatus in Jefferson Lab's Hall A, and Chapters~\ref{ch:Analysis} and~\ref{ch:RC} provide the details of the experimental data analysis and radiative corrections to the data, respectively. Results are presented in Chapter~\ref{ch:Results} and the thesis ends with the conclusion in Chapter~\ref{ch:Conclusion}.

\chapter{\sc Inclusive Electron Scattering}
\label{ch:epScatter}
Experimental attempts to the study the proton's structure rely on scattering techniques first pioneered by Ernest Rutherford around a century ago. By scattering alpha particles through a thin gold foil, Rutherford and his colleagues determined that the atom consisted of a compact and positively charged nucleus~\cite{Rutherford:1911}. Present-day scattering experiments probe structure with a variety of particles, but the concept is the same: use the scattering interaction to determine the properties of the target. In terms of measurable quantities, the scattering cross section quantifies the scattering interaction as a function of the interaction rate, and energy and angular distribution of the colliding particles.

In a typical proton structure experiment, a beam of incoming electrons is scattered from a fixed proton target. For inclusive scattering experiments, only the resultant scattered electrons are detected. The length scale that the electrons are able to probe is inversely proportional to their momentum, through the (Louis) de Broglie relations~\cite{debroglie}. By controlling the energy of the electron beam, experimentalists control the resolution of their probe. For example, when the electron wavelength is greater than the size of the proton, the electron lacks the momentum to penetrate inside the proton and the scattering interaction describes the collective behavior of the quarks and gluons ($i.e.$ the bulk properties of the proton).


\section{Kinematic Variables}
\begin{figure}[htp]
\begin{center}
\includegraphics[scale=.4]{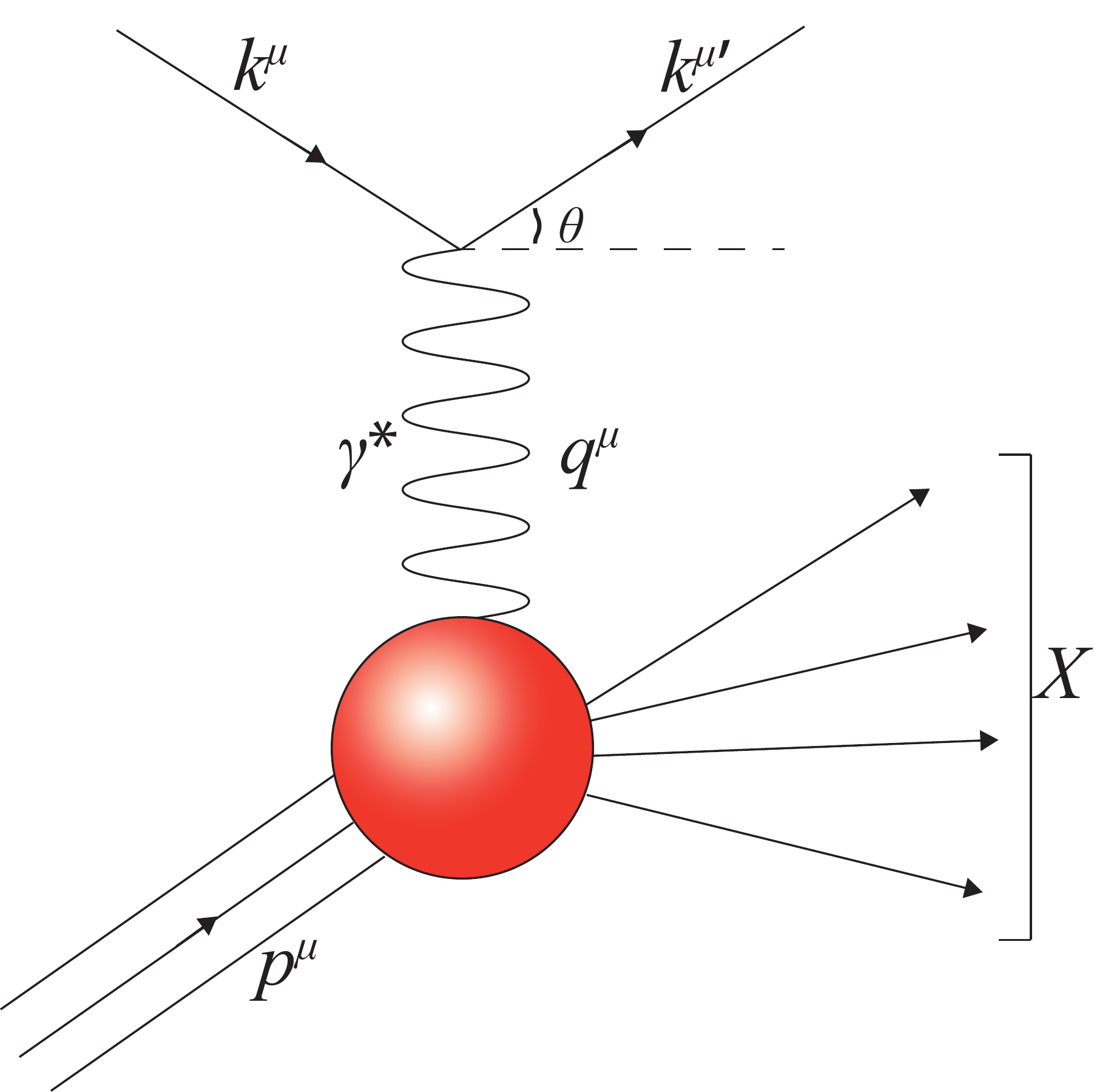}
\caption{The leading order  $ep \rightarrow e'X$ scattering interaction is mediated by the virtual photon, $\gamma*$. The outgoing electron, $k'^\mu$, is scattered at angle $\theta$ into the detectors. }
\label{fig:Born_Feynman}
\end{center}
\end{figure}

In the leading order electron-proton scattering process shown in Figure~\ref{fig:Born_Feynman}, an electron with four-momentum $k^\mu = (E,\bf{k})$ interacts with a proton of four-momentum  $p^\mu = (\epsilon,{\bf P})$ through the exchange of a virtual photon. The electron is scattered at an angle $\theta$ and with four-momentum $k'^\mu = (E',\bf{k'})$. The space-like virtual photon  ($q^2 < 0$) has four-momentum $q^\mu = (k - k')^\mu = (\nu,{\bf q})$, where $\nu$ represents the energy loss of the electron (or energy transferred to the target). The invariant mass of the undetected hadronic system, $X$, is $W = \sqrt{(p^{\mu}+q^{\mu})^2}$. In the preceding paragraph, bold quantities are three-vectors. 

In the laboratory frame of reference the proton is at rest, which leads to the following kinematic definitions:
\begin{align}
\nu &= E - E'\,, \\
Q^2 &= -q^2 \simeq 4EE'\mathrm{sin}^2\frac{\theta}{2}\,, \\ 
W^2    &= M^2 + 2M\nu - Q^2 \,,
\end{align}
where $M$ is the mass of the proton and $E$ and $E'$ are assumed to be much greater than $m_e$, so the electron mass is safely ignored.

Two additional invariants complete the list of variables typically used in inclusive electron scattering. The scalar quantity $x$, first introduced by James Bjorken~\cite{Bjorken:1968dy}, refers to the momentum fraction carried by the particle struck in the interaction, and $y$ refers to the fractional energy loss (sometimes called the $inelasticity$) of the electron: 
\begin{align}
x &= \frac{Q^2}{2M\nu}\,,\\
y&= \frac{\nu}{E}\,.
\end{align}
\section{Scattering Processes}
The scattering process in Figure~\ref{fig:Born_Feynman} is divided into multiple kinematic regions: elastic, inelastic, and deep inelastic scattering.  In elastic scattering, the proton remains in its ground state and the energy and momentum transfer are absorbed by the recoil proton. The invariant mass, $W$, is equal to the mass of the proton and the momentum fraction, $x$, is equal to one. Conservation of energy and momentum dictate that in elastic scattering the energy of the scattered electron is
\begin{equation}
E'_{\mathrm{elas}} = \frac{E}{1 + \frac{2E}{M}\mathrm{sin}^2\frac{\theta}{2}}\,.
\end{equation} 

Increases in momentum transfer, create inelastic scattering and the resonant excitation of the proton. The resonance region begins at the pion production threshold of $W = M_p + M_{\pi} = 1072$ MeV and continues until $W$ $\sim$ 2000 MeV. The first of the inelastic resonant states is the $\Delta$-baryon at $W$= 1232 MeV. The $\Delta$(1232) has little overlap with other resonances and is the most prominent. After the $\Delta$ is the $N^*$(1440) at $W$ = 1440 MeV. Following that are the other $N^*$ resonances, which include an observed excitation  at $W \approx$ 1500 MeV. This $N_1^*$ resonance is the combination of two resonant states: $N^*$(1520) and $N^*$(1535). The final observed resonance, $N_2^*$, is at  $W \approx$  1700 MeV and consists of many resonant states, but $N^*$(1680) is the strongest at low $Q^2$.


Past the resonance region, the energy of the electron probe is sufficient to scatter off quarks inside the proton. This kinematic region is referred to as deep inelastic scattering (DIS). There are no more resonance peaks in inclusive DIS. For sufficiently large energy, asymptotic freedom requires that the proton's constituents are non-interacting. This means that the DIS scattering process occurs from an incoherent sum over the individual quarks and gluons inside a proton.

\begin{figure}[htp]
\begin{center}
\includegraphics[scale=1.1]{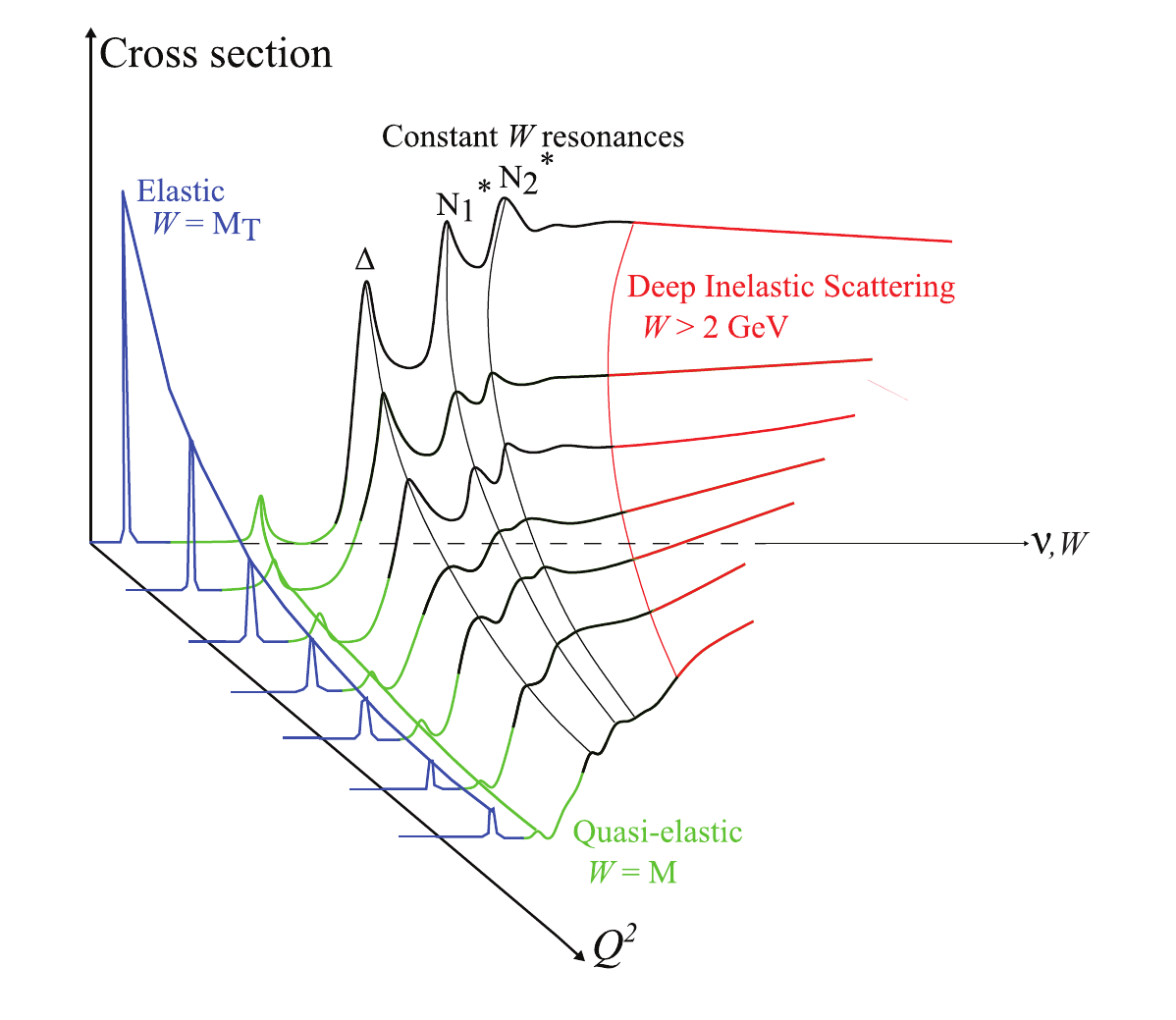}
\caption{The cross section (arbitrary units) as a function of $Q^2$ and $\nu$, $W$ from inclusive electron scattering of an arbitrary nuclear target of mass $M_T$. The mass of the nucleon is $M$. Based on a figure from Ref~\cite{VinceT}.}
\label{XS_Res}
\end{center}
\end{figure}
The idea is the same for  nuclear targets, except that the elastic peak is located at a line of constant $W$ corresponding to the mass of the nucleus and nuclear excitations are also visible. There is an additional, quasi-elastic, peak  corresponding to elastic scattering from nucleons inside the nucleus.  The location of the peak is slightly shifted away from $W = M$ due to the nuclear binding energy and is broadened due to the Fermi motion of the nucleons inside the nucleus~\cite{Povh}. The cross section spectrum for an arbitrary nuclear target is shown in Figure~\ref{XS_Res}.
\section{Scattering Cross Sections}
As an experimental quantity, the total cross section is defined as
\begin{equation}
\sigma_{\mathrm{tot}} = \frac{\#\,\mathrm{reactions\, per\, unit\, time}}{\#\,\mathrm{beam\, particles\, per\, unit\, time}}\cdot\frac{1}{\mathrm{scattering\, centers\, per\, unit\, area}}\,.
\end{equation}
A glance at the units reveals that $\sigma_{\mathrm{tot}}$ is a function of area. It should be noted that this area is less a geometrical construct, but related more to the structure of the interaction potential that led to the scattering. In general, only a small sample of the reaction products are experimentally measured. This is accounted for by normalizing the total cross section to the solid angle acceptance, $\Omega$,  of the detector area and the energy of the scattered particles, $E'$, to produce a doubly differential cross section $d^2\sigma / d\Omega dE'$.

From a theoretical viewpoint, the reactions per unit time in the scattering process A + B $\rightarrow$ C + D is characterized, using Fermi's golden rule, as the quantity
\begin{equation}
\label{TransProb}
\mathcal{M}_{fi} = \frac{(2\pi)^4}{V^2}|\mathcal{T}|^2\frac{d^3p_C}{(2\pi)^32E_C}\frac{d^3p_D}{(2\pi)^32E_D}\delta^4(p_C + p_D - p_A - p_B)\,,
\end{equation}
where the four momentums of the interacting particles are given by $p_{A,B,C,D}$ and their energies are $E_{A,B,C,D}$. The number of reactions per unit time is a function of the probability amplitude of the interaction, $|\mathcal{T}|^2$, the volume, $V$, scattered into and the density of available final states, $d^3p_C$ and $d^3p_D$~\cite{Povh}. The $\delta$-function ensures the conservation of four-momentum. The probability amplitude is a measure of the coupling between the initial and final states and is worked out for the specific case of $ep$ scattering in the remainder of the chapter. The number of beam particles per unit time is $|v_A|$$2E_A/V$ ($v_A$ = ${\bf p_A}/E_A$) and the number of scattering centers per unit area is $2E_B/V$~\cite{Quarks}. The combination of the former two quantities ($|v_A|$$2E_A/V$, $2E_B/V$) is sometimes referred to as the initial flux, $\Phi$.
\subsection{Tensor Formulation}
For the electron-proton scattering process, 
\begin{equation}
e(k) + p(p) \rightarrow e'(k')+X(p_X)\,,
\end{equation}
 the electron component of the theoretical probability amplitude is computed using the Feynman rules for spinor electrodynamics~\cite{Srednicki}. The proton part is trickier because its exact form is unknown, and is instead parameterized as an initial state $\Ket{P,S}$ with momentum $P$ and spin $S$, and a final (unobserved) state $\Bra{X}$. Combining the lepton and hadron components gives the probability amplitude as
\begin{align}
\label{eq:TMatrix}
|\mathcal{T}|^2 = &\frac{e^4}{Q^4} \bar{u}_{s'}({ k'})\gamma^{\mu}u_s({ k}) \bar{u}_{s}({ k})\gamma^{\nu}u_{s'}({ k'}) \times  \\
                               & \Bra{P,S}J^{\mu}(0)\Ket{X}\Bra{X}J^{\nu}(0)\Ket{P,S}\,,
\end{align}
where $u_s(k)$ ($u_{s'}(k')$) are the spinor states of the incoming (outgoing) electron with spin $s$ ($s'$) and four-momentum ${k}$ (${k'}$) and the virtual photon has four-momentum $Q^2$. The interaction current between the initial and final hadron states is $eJ^{\mu}(0)$ and the $\gamma^{\mu}$ are the gamma matrices.

In terms of the T-matrix element and initial flux, $\Phi$, the differential cross section is~\cite{Quarks}
\begin{equation}
\label{Trans}
d\sigma = \frac{|\mathcal{T}|^2}{\Phi}dQ\\,
\end{equation}
where $dQ$ is the Lorentz invariant phase space factor for a scattered electron and proton of energy $E'$ and $p_{x0}$, respectively, 
\begin{equation}
\label{dLips}
dQ = (2\pi)^4 \delta^{(4)}( { k'} + { p_x} - { p} - { k})\frac{d^3k'}{(2\pi)^32E'}\frac{d^3p_x}{(2\pi)^32p_{x0}}\,,
\end{equation}
and the flux is 
\begin{align}
\label{Flux}
\Phi &= 4 ( ({ k} \cdot { p})^2 - m_e^2 M^2)^{1/2}\,,\\
   &= 4 E M\,,
\end{align}
 where the electron mass, $m_e$, is neglected but its incident energy is $E$. The proton is initially at rest and has mass $M$.

Using this form of the probability amplitude, the differential scattering cross section is 
\begin{equation} 
d\sigma = \frac{\alpha^2}{Q^4} \int \frac{d^3 k'}{E'} \frac{1}{E}L^{\mu\nu}W_{\mu\nu}\,,
\end{equation}
where $\alpha$ is the fine structure constant and the lepton tensor, $L^{\mu\nu}$, is  
\begin{equation}
\label{LeppyT}
L^{\mu\nu} ({k},s; {k'},s') =  \bar{u}_{s'}({k'})\gamma^{\mu}u_s({k}) \bar{u}_{s}({k})\gamma^{\nu}u_{s'}({k'})\,,
\end{equation}
 and the proton (hadron) tensor, $W_{\mu\nu}$, is 
\begin{align}
\label{HadronT}
W_{\mu\nu} (P,S)= \frac{1}{2M}\sum_{N_x} \int \frac{d^3p_x}{(2\pi)^32E_x} &\Bra{P,S}J^{\mu}(0)\Ket{X}\\  \nonumber
                          &\Bra{X}J^{\nu}(0)\Ket{P,S}(2\pi)^3\delta^{(4)}(k' + p_X - p - k)\,,
\end{align}
where the sum runs over possible proton states, $N_x$, each of energy $E_x$ and three-momentum $p_x$.
Using the completeness relation and invoking translational invariance~\cite{Thomas}, the proton tensor is also written as
\begin{align}
\label{BetterHadronT}
W_{\mu\nu}(P,S) = \frac{1}{2M} \int \frac{d^4 x}{2\pi} e^{iq\cdot x}  \Bra{P,S}J^{\nu}(x)J^{\mu}(0)\Ket{P,S}\,.
\end{align}

 The integral with respect to the phase space of the detected electron is replaced with $d^3k' = k'^2 dk'd\Omega \simeq E'^2 dE' d\Omega$~\cite{Srednicki} to produce the doubly differential cross section 
\begin{equation} 
\label{XS}
\frac{d^2\sigma}{dE'd\Omega} = \frac{\alpha^2}{Q^4} \frac{E'}{E} L^{\mu\nu}W_{\mu\nu}\,,
\end{equation}
where $E'$ is the detected electron's energy and $\Omega$ is the solid angle into which the outgoing electron is scattered.
\subsection{Lepton Tensor}
The lepton tensor of equation~\eqref{LeppyT} has two general forms: one where the initial electron is polarized and another where it is not. For reasons that will become readily apparent, the unpolarized tensor is denoted with the subscript $S$ and the polarized tensor with the subscript $A$. In the case of an unpolarized electron beam, the tensor is averaged over the initial and final electron spins. This is equivalent to $\langle | \mathcal{T}|^2\rangle = \frac{1}{2} \Sigma_{\mathrm{spins}} |\mathcal{T}|^2$. The spin-averaged lepton tensor becomes
\begin{align}
\label{LeptonTrace}
L^{\mu\nu}_S ({k},s; {k'},s') &= \frac{1}{2} \mathrm{Tr}((-\slashed{k} + m)\gamma^{\mu}(-{\slashed k'} + m)\gamma^{\nu})\,,\\
&= \frac{1}{2}k_{\sigma}k'_{\rho} \mathrm{Tr}( \gamma^{\sigma} \gamma^{\mu} \gamma^{\rho} \gamma^{\nu}) + \frac{1}{2}m^2 \mathrm{Tr}(\gamma^{\mu}\gamma^{\nu})\,,\\
\label{LeptonTensor}
&=2(k'^{\mu}k^{\nu}+k'^{\nu} k^{\mu} - ({k'} \cdot {k} - m^2)g^{\mu\nu})\,,
\end{align}
which makes use of the normalization condition
\begin{equation}
\label{SpinSum}
\Sigma_{\mathrm{spins}}~u_s({k})\bar{u}_s({k}) = -\slashed k + m\,,
\end{equation}
and the Feynman slash notation:
\begin{equation}
\slashed A \equiv \gamma^{\mu} A_{\mu}\,.
\end{equation}
If the incident electron has an initial polarization, then the normalization condition of equation~\eqref{SpinSum} is no longer valid for the incoming electron. A better choice is
\begin{equation}
u_s({k})\bar{u}_s({k}) = \frac{1}{2}(1-s\gamma_5 \slashed z)(-\slashed k + m)\,,
\end{equation}
where $s = \pm$ gives the direction of the electron spin along the spin quantization axis $z$~\cite{Srednicki}. The polarization of the outgoing electron is not measured so equation~\eqref{SpinSum} is still correct for that portion of the tensor. The lepton $A$ tensor is 
\begin{align}
L^{\mu\nu}_A({ k},s; { k'},s')&=\frac{1}{2}\mathrm{Tr}{\bigg [}(\slashed k' + m)\gamma^{\nu}(\slashed k + m){\bigg (}\frac{1+\gamma_5\slashed s/m}{2}{\bigg)}\gamma^{\mu}{\bigg]}\,,\\
\label{LeptonAnti}
&=2(k'^{\mu}k^{\nu}+k'^{\nu} k^{\mu} - ({ k'} \cdot { k} - m^2)g^{\mu\nu} + i\epsilon^{\mu\nu\rho\sigma}q_{\rho}s_{\sigma})\,,
\end{align}
where $\slashed s = ms \slashed z$, and  $\epsilon^{\mu\nu\rho\sigma}$ is the totally antisymmetric tensor.

Comparing equation~\eqref{LeptonAnti} to equation~\eqref{LeptonTensor} shows that the unpolarized tensor is symmetric (hence the $S$ subscript) and the polarized lepton tensor is the unpolarized tensor with an additional spin-dependent, antisymmetric term (hence the $A$ subscript).

\subsection{Hadron Tensor: Elastic Scattering}
In elastic scattering, the proton stays in its ground state and the energy and momentum transfer are absorbed by the recoil proton. This simplifies the hadron tensor greatly because it no longer needs to take into account a multitude of final states and particles. The electron tensor of equation~\eqref{LeppyT} is a good starting part for writing down the  form of the elastic hadron tensor (reminder $u(P)$ are the proton spinors):
\begin{equation}
W_{\mathrm{el}}^{\mu\nu} = \bar{u}({ P'})\Gamma^{\mu}u({ P}) \bar{u}({ P})\Gamma^{\nu}u({ P'})\,.
\end{equation}
 The proton is not a point particle like the electron so the transition current, $\Gamma^{\mu}$, is not a simple $\gamma^{\mu}$ matrix~\cite{Peskin}. The most general form of $\Gamma^{\mu}$ that respects the necessary invariance laws and symmetries is 
\begin{align}
\label{ElasticVertex}
\Gamma^{\mu} &= \gamma^{\mu}F_1(Q^2) + \frac{i\sigma^{\mu\nu}q_{\nu}\kappa}{2M}F_2(Q^2)\,,\\
&= \gamma^{\mu}(F_1(Q^2) + \kappa F_2(Q^2)) - \frac{(P' + P)^{\mu}}{2M}\kappa F_2(Q^2)\,,
\end{align}
where $F_1(Q^2)$ and $F_2(Q^2)$\footnote{$F_1(Q^2)$ and $F_2(Q^2)$ are referred to as the Pauli and Dirac form factors, respectively.} are two independent form factors~\cite{Quarks} and $\kappa$ is the anomalous magnetic moment. The form factors parameterize the unknown structure of the target proton. With an appropriate form of $\Gamma^{\mu}$, the elastic proton tensor is calculated and contracted with the lepton tensor $L_S$ to produce the doubly differential cross section
\begin{align}
\frac{d^2\sigma}{dE'd\Omega} = \frac{(2\alpha)^2 E'^2 }{Q^4A}   &{\bigg [} {\bigg (} F_1^2(Q^2) - \frac{q^2\kappa^2}{4M^2}F_2^2(Q^2){\bigg )}\mathrm{cos}^2\frac{\theta}{2} \\ \nonumber
&- \frac{q^2}{2M^2}{\bigg(}F_1(Q^2) + \kappa F_2(Q^2){\bigg)}^2\mathrm{sin}\frac{\theta}{2}{\bigg ]}\delta(E' - E/A)\,,
\end{align}
where $ A = 1 + (2E/M)\mathrm{sin}^2\frac{\theta}{2}$, and is a result of the conservation of energy and momentum during an elastic collision.
The $\delta$-function collapses the $dE'$ integral, yielding the Rosenbluth formula~\cite{Rosenbluth} (named after Marshall Rosenbluth)
\begin{align}
\label{Rosenbluth}
\frac{d\sigma}{d\Omega} = \frac{\alpha^2}{4E^2\mathrm{sin}^4\frac{\theta}{2}}\frac{E'}{E}  {\bigg [} {\bigg (} F_1^2(Q^2) &- \frac{q^2\kappa^2}{4M^2}F_2(Q^2)^2{\bigg )}\mathrm{cos} ^2\frac{\theta}{2} \\ \nonumber
&-\frac{q^2}{2M^2}{\bigg (}F_1(Q^2) + \kappa F_2(Q^2){\bigg)}^2\mathrm{sin}^2\frac{\theta}{2}{\bigg ]}\,.
\end{align}
The Rosenbluth formula is recast into a more instructive form by factoring out the Mott cross section (named after Neville Mott), which is  defined as~\cite{MOTT}
\begin{equation}
{\bigg(}\frac{d\sigma}{d\Omega}{\bigg)}_{\mathrm{Mott}} = \frac{\alpha^2 \mathrm{cos}^2\frac{\theta}{2}}{4E^2\mathrm{sin}^4\frac{\theta}{2}} \frac{E'}{E}\,.
\end{equation}
The Mott cross section represents scattering of electrons from a point charge, so the redefined Rosenbluth formula,
\begin{equation}
\label{RosieQ}
\frac{d\sigma}{d\Omega} = {\bigg(}\frac{d\sigma}{d\Omega}{\bigg)}_{\mathrm{Mott}} {\bigg[}F_1^2(Q^2) + \tau [\kappa F_2(Q^2)]^2 + 2\tau(F_1 + \kappa F_2)^2\mathrm{tan}^2\frac{\theta}{2}{\bigg ]}\,,
\end{equation}
(where $\tau = Q^2 / 4M^2$) is just the structureless, point charge cross section modified to account for the internal structure of the proton as defined by the $F_1(Q^2)$ and $F_2(Q^2)$ terms.

The structure functions $F_1(Q^2)$ and $F_2(Q^2)$ themselves can be recast into the electric, $G_E(Q^2)$, and magnetic, $G_M(Q^2)$, Sachs form factors such that
\begin{align}
G_E(Q^2) &= F_1(Q^2) - \tau\kappa F_2(Q^2)\,,\\
G_M(Q^2) &= F_1(Q^2) + \kappa F_2(Q^2)\,,
\end{align}
where in the static limit of $Q^2$ = 0, they normalize to the charge and magnetic moment of the proton in units of the electron charge and of the nuclear magneton
$\mu_K$: $G_E(0) = 1$ and $G_M(0) = \mu_p$. These form factors carry information on the charge and current distributions inside the proton.



\subsection{Hadron Tensor: Inelastic Scattering}
In an inelastic scattering process, the proton is able to transition from its ground state $\Ket{P,S}$ to any excited state $\bra{X}$. This complicates the hadron tensor because the final state is no longer a single proton of the form  $\bar{u}\Gamma^{\mu}u$, as was the case for elastic scattering. Instead the inelastic hadron interaction is parametrized in the most general form, using the independent momentum $P$, $q$ and metric tensor, $g^{\mu\nu}$~\cite{Quarks}. As with the lepton tensor, the inelastic hadron tensor has two general forms: a symmetric unpolarized tensor and an antisymmetric polarized tensor.

The unpolarized hadron tensor must be symmetric to ensure a non-zero result when it is contracted with the unpolarized and symmetric lepton tensor. The result is
\begin{equation}
W_S^{\mu\nu}= W_1{\bigg(}-g^{\mu\nu} + \frac{q^{\mu}q^{\nu}}{q^2}{\bigg)} + W_2 \frac{1}{M^2}{\bigg(}P^{\mu} - \frac{P\cdot q}{q^2}q^{\mu}{\bigg)}{\bigg(}P^{\nu} - \frac{P\cdot q}{q^2}q^{\nu}{\bigg)}\,,
\end{equation}
where  $W_1(\nu,Q^2)$ and $W_2(\nu,Q^2)$ are the inelastic structure functions and describe the internal structure of the proton. Carrying out the contraction of the lepton and hadron tensors yields
\begin{equation}
L^{\mu\nu}W_{\mu\nu}= 4W_1(k' \cdot k) + 2 \frac{W_2}{m_k^2}[2(P \cdot k)(P \cdot k') - M^2(k \cdot k')]\,,
\end{equation}
having used the on-shell mass condition, $P^2 = -M^2$, and also neglecting terms involving $m^2$. In the lab reference frame (neglecting the electron mass), the relevant four vectors are:
\begin{align}
 k &= (E,0,0,E)\,, \\
 P &= (M,0,0,0)\,,\\
 k' &= (E', E'\mathrm{sin}\theta \mathrm{cos}\phi,E'\mathrm{sin}\theta \mathrm{sin}\phi, E'\mathrm{cos}\theta)\,,
 \end{align}
  which leads to the doubly differential cross section
\begin{equation}
\label{XSfinal}
\frac{d^2\sigma}{dE'd\Omega} = {\bigg(}\frac{d\sigma}{d\Omega}{\bigg)}_{\mathrm{Mott}} {\bigg[}2 W_1(\nu,Q^2) \mathrm{tan}^2\frac{\theta}{2} +W_2(\nu,Q^2){\bigg]}\,.
\end{equation}

The polarized lepton tensor contains an antisymmetric term, so the polarized hadron tensor also needs an antisymmetric term for a non-zero result. This implies that, to gain any insight into the spin-dependent properties of the proton, both the electron probe and proton target must be polarized. Taking into account the extra polarization degree of freedom, represented by the proton spin vector $s$, the antisymmetric hadron tensor is~\cite{Thomas} 
\begin{equation}
W^{\mu\nu}_A =  i\epsilon^{\mu\nu\rho\sigma}q_{\rho}{\bigg (}MG_1(\nu,Q^2)s_{\sigma} + \frac{G_2(\nu,Q^2)}{M}(s_{\sigma}P\cdot q - P_{\sigma}s\cdot q){\bigg)}\,,
\end{equation}
where $G_1(\nu,Q^2)$ and $G_2(\nu,Q^2)$ are the polarized structure functions and $\epsilon^{\mu\nu\rho\sigma}$ is the totally antisymmetric tensor.  After contracting the two tensors ($L = L_S + iL_A$ and $W = W_S + iW_A$), the polarized cross section is
\begin{align}
\label{PolXS}
\frac{d^2\sigma}{dE'd\Omega} = \frac{\alpha^2}{Q^2}\frac{E'}{E} {\bigg[}&2 W_1(\nu,Q^2) + W_2(\nu,Q^2)\frac{1}{\mathrm{tan}^2\frac{\theta}{2}}\\ \nonumber
&+ 4MG_1(\nu,Q^2)\{(s^l \cdot s^p)+\frac{1}{Q^2}(s^l \cdot q)(q\cdot s^p)\}\\ \nonumber
&+\frac{4G_2(\nu,Q^2)}{M}\{(P\cdot q)(s^l \cdot s^p) -(s^p \cdot q)(P \cdot s^l)\}{\bigg]}  \,,
\end{align}
where $s^l$ is the incoming electron spin vector and $s^p$ is the spin vector of the target proton.

\begin{figure}[htp]
\begin{center}
\subfigure[Angle definitions]{\label{fig:a}\includegraphics[width=.45\textwidth]{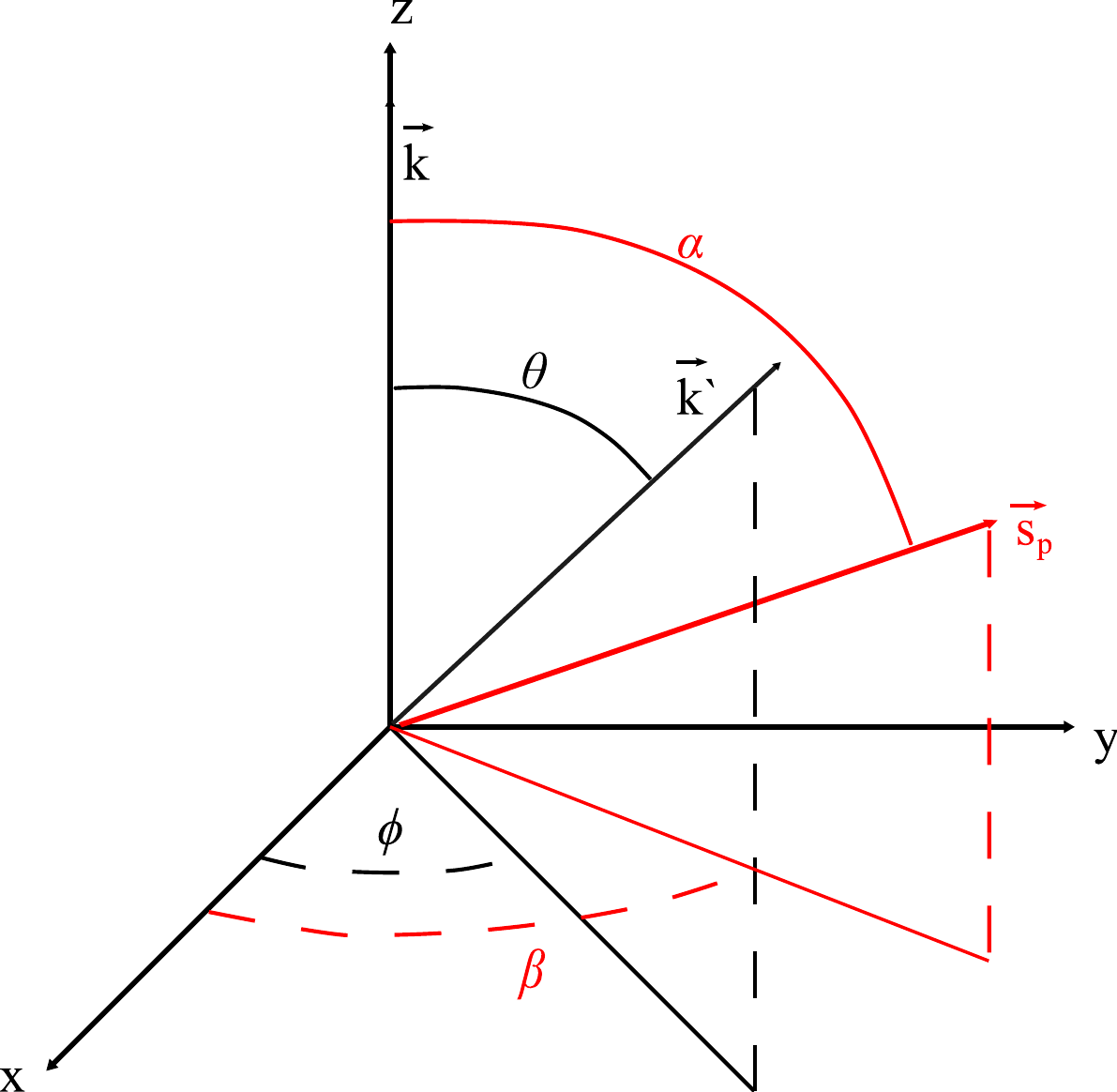}}
\qquad
\subfigure[Scattering and polarization planes]{\label{fig:b}\includegraphics[width=.45\textwidth]{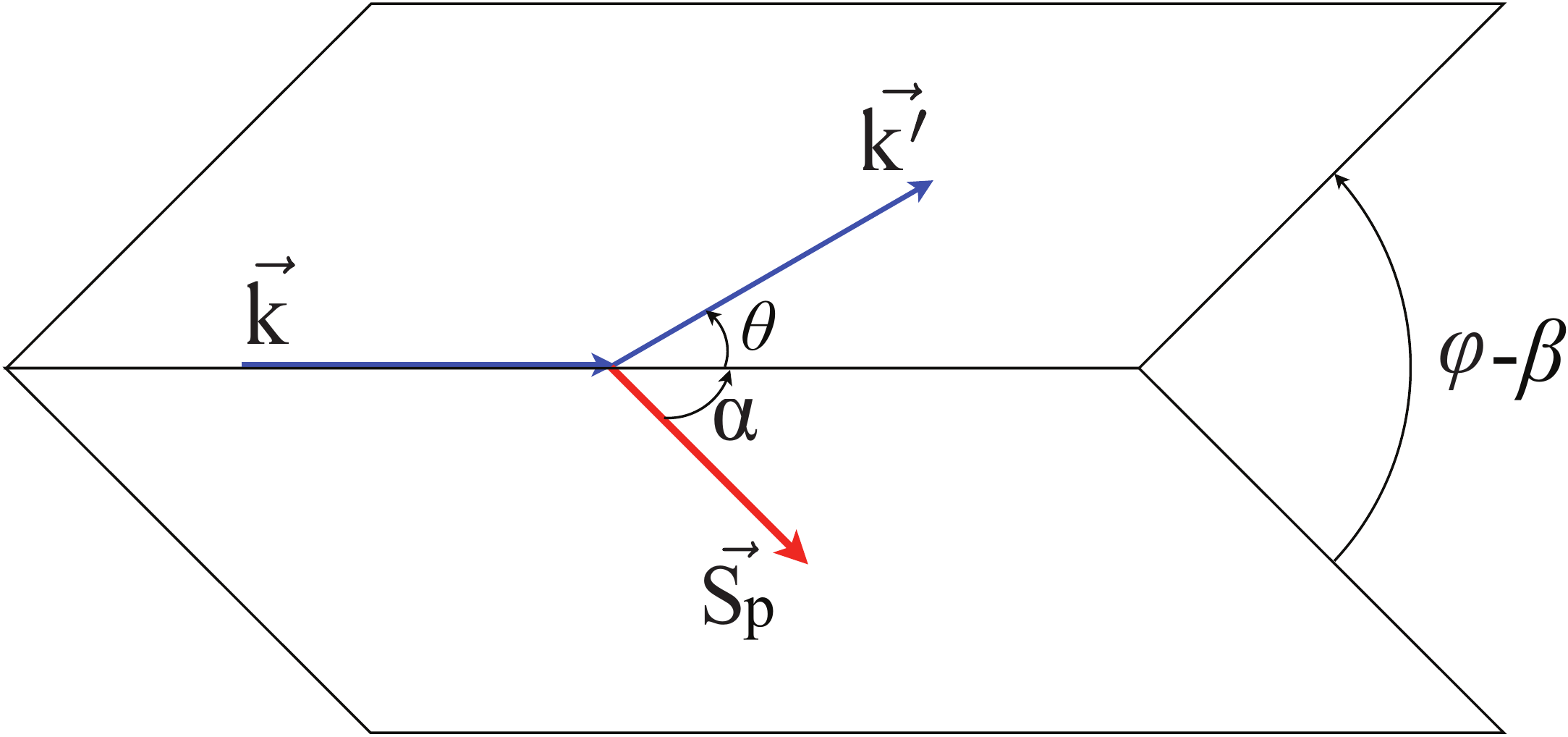}}
\caption{\label{CoordPol}Scattering coordinate system and angle definitions. Relevant for transversely polarized targets, the out-of-plane polarization angle, defined as $\phi$ - $\beta$, is the angle between the polarization and scattering planes. Reproduced from~\cite{Leader}.}
\end{center}
\end{figure}

The polarized structure functions are isolated from the polarized cross section by looking at the difference between the cross sections of two opposite electron spin (helicity) states:
\begin{align}
\label{DiffXS}
\frac{d^2\sigma^{\uparrow}}{dE'd\Omega} - \frac{d^2\sigma^{\downarrow}}{dE'd\Omega} = \frac{8\alpha^2}{Q^2}\frac{E'}{E}&{\bigg [}M G_1(\nu,Q^2)\{(s^l \cdot s^p)+\frac{1}{Q^2}(s^l \cdot q)(q\cdot s^p)\}\\ \nonumber
&+\frac{G_2(\nu,Q^2)}{M}\{(P\cdot q)(s^l \cdot s^p) -(s^p \cdot q)(P \cdot s^l)\} {\bigg ]}\,,
\end{align}
where $\uparrow$ and $\downarrow$ correspond to the two electron helicity states. The electrons are polarized with their spins along and opposite the direction of their motion. This is simplified further by fixing the target proton's spin vector relative to the incoming electron's spin vector ($i.e$ picking a polarization axis). With the aid of Figure~\ref{CoordPol}, the relevant four vectors are
\begin{align}
k' & = (E', E'\mathrm{sin}\theta\mathrm{cos}\phi,E'\mathrm{sin}\theta\mathrm{sin}\phi,E'\mathrm{cos}\theta)\,, \\
k  & = (E,0,0,E)\,,\\
s^l &=\frac{1}{m}(E,0,0,E)\,, \\
s^p&= (0,\mathrm{sin}\alpha\mathrm{cos}\beta,\mathrm{sin}\alpha\mathrm{sin}\beta,\mathrm{cos}\alpha)\,,\\
P &=(M,0,0,0)\,.
\end{align}

If the electron and proton spins are parallel, then $\alpha = 0,\pi$ and the cross section difference is
\begin{equation}
\label{DiffXSPara}
\frac{d^2\sigma^{\uparrow\Uparrow}}{dE'd\Omega} - \frac{d^2\sigma^{\downarrow\Uparrow}}{dE'd\Omega} = \frac{4\alpha^2}{Q^2}\frac{E'}{E}{\bigg [} MG_1(\nu,Q^2)\{E +E'\mathrm{cos}\theta\}  -Q^2G_2(\nu,Q^2){\bigg ]}\,,
\end{equation}
where $\Uparrow$ refers to the direction of the proton's spin. If the electron and proton spins are  perpendicular, then $\alpha = \frac{\pi}{2},\frac{3\pi}{2}$ and the cross section difference is
\begin{equation}
\label{DiffXSPerp}
\frac{d^2\sigma^{\uparrow\Rightarrow}}{dE'd\Omega} - \frac{d^2\sigma^{\downarrow\Rightarrow}}{dE'd\Omega} = \frac{4\alpha^2}{Q^2}\frac{E'^2}{E}\mathrm{sin}\theta\mathrm{cos}(\phi - \beta){\bigg [} MG_1(\nu,Q^2) + 2EG_2(\nu,Q^2){\bigg ]}\,,
\end{equation}
where $\phi-\beta$ is the azimuthal angle between the polarization and scattering planes. The polarization plane spans the incoming electron vector and proton polarization vector, while the scattering plane spans the incoming and outgoing electron vectors.

\section{Virtual Photon Proton Cross Section}
\label{VPPXS}
\begin{figure}[htp]
\begin{center}
\includegraphics[scale=0.7]{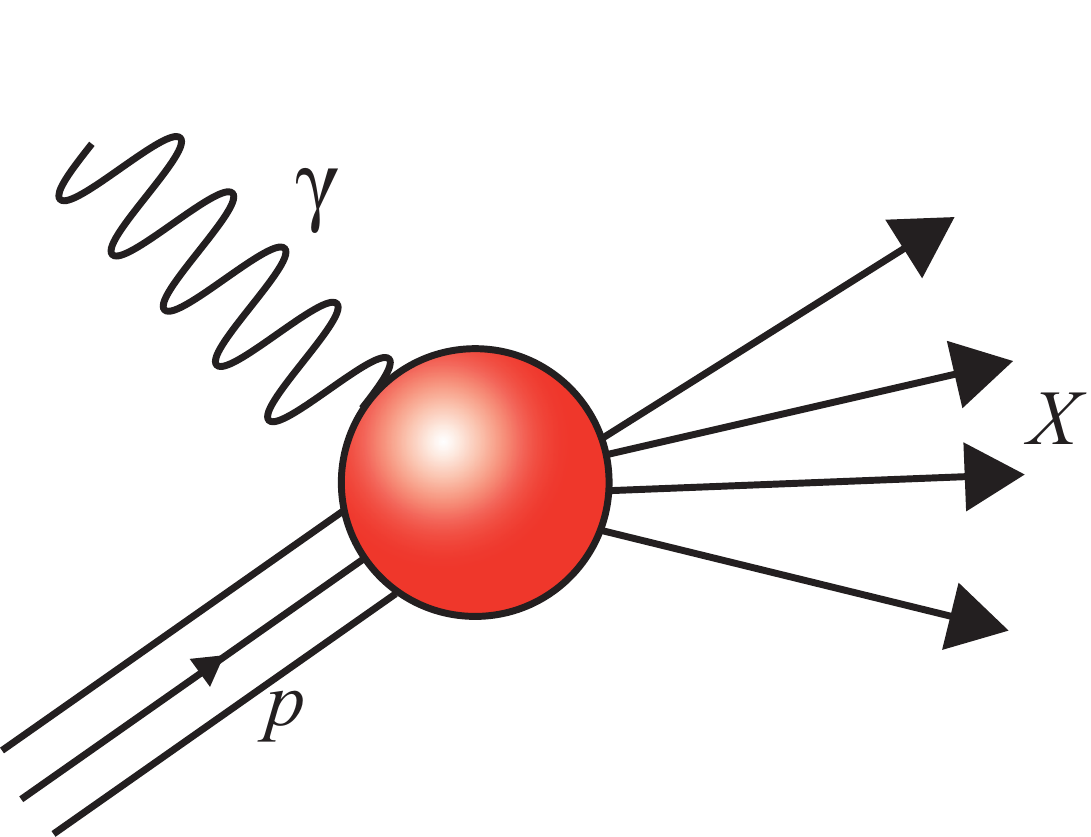}
\caption{\label{PhotonProton}Compton scattering process for a real photon, $\gamma$, at $Q^2$ = 0.}
\end{center}
\end{figure}

Referring back to Figure~\ref{fig:Born_Feynman}, the virtual photon is what actually probes the extent of the proton; the electron is used only to produce this virtual photon. It stands to reason that there is an equivalent formulation of the proton structure and its structure functions in terms of a photon-proton scattering cross section. It is easiest to do this, first, by assuming a real ($Q^2 = 0$) photon-proton scattering interaction, like that in Figure~\ref{PhotonProton},  and then making the necessary corrections to account for a virtual photon. 

 Using the Feynman rules, the total cross section for a real photon of energy $q^0 = \nu = K$ inelastically scattering off of an unpolarized proton is
\begin{align}
\sigma_{\mathrm{tot}}(\gamma p \rightarrow X) =  \frac{1}{2M 2K}\sum_{N_x} \int &\frac{d^3p_X}{(2\pi)^32E_{X0}'}e^2 \epsilon^{\mu*}_{\lambda}\epsilon^{\nu}_{\lambda}\Bra{P,S}J^{\mu}(0)\Ket{X}\\  \nonumber
                          &\Bra{X}J^{\nu}(0)\Ket{P,S}(2\pi)^3\delta^{(4)}(  p_x - p +q)\,,                         
\end{align}
 \begin{equation}
 \sigma_{\mathrm{tot}}(\gamma p \rightarrow X) = \frac{4\pi^2\alpha}{K}\epsilon^{\mu*}_{\lambda}\epsilon^{\nu}_{\lambda}W_{\mu\nu}\,,
 \end{equation}
  where $1/2M2K$ is the flux factor, $\epsilon_{\lambda}$ is the photon polarization vector and $W_{\mu\nu}$ is the same hadronic tensor defined in equation~\eqref{HadronT}. The photon helicity, $\lambda$, takes values of $\pm$ 1 for a real photon~\cite{Weinberg:1995mt}.
  The distinction between real and virtual photons arises in both the definition of the polarization vectors and the flux factor. In fact, the definition of $K$ when $Q^2 \neq 0$ is completely dependent on the choice of  convention. The (Louis) Hand convention~\cite{Hand},
 \begin{equation}
 K_H = \frac{W^2 - M^2}{2M} = \nu - \frac{Q^2}{2M}\,,
 \end{equation}
 is the most commonly used.  Others include
 \begin{align}
 K_A &= \nu\,,\\
 K_G &= \nu\sqrt{1 + Q/\nu}\,,
 \end{align}
where $K_G$ was first proposed by Fred Gilman~\cite{Gilman}. 

For a real photon, the relevant degrees of freedom create two independent and transverse polarization states, $\epsilon^T_+$ and $\epsilon^T_-$. The positive helicity vector,  $\epsilon^T_+$, corresponds to a $J_z = \frac{3}{2}$ total spin state\footnote{The photon has spin-1 and the proton has spin-$\frac{1}{2}$.} ($J_z = 1 + \frac{1}{2}$), while the negative helicity vector, $\epsilon^T_-$, is a $J_z = \frac{1}{2} $ total spin state ($J_z = 1 - \frac{1}{2}$).  A virtual photon is off-shell so a correlation is drawn to massive spin-1 particles. The result is that the virtual photon  gains a third, independent, polarization vector, $\epsilon_0$. This corresponds to a helicty state of $\lambda$ = 0. A suitable choice of the polarization vectors is~\cite{Quarks}:
 \begin{align}
 \epsilon^T_+ &=- (0,0,1,i,0)/\sqrt{2}\,,\\
 \epsilon^T_-  &=(0,0,1,-i,0)/\sqrt{2}\,,\\
 \epsilon_0 &=(\sqrt{\nu^2 +Q^2},0,0,\nu)/\sqrt{Q^2}\,.
 \end{align}
In the lab frame, $p = (M,0,0,0)$ and $q  =(\nu,0,0,\sqrt{\nu^2 +Q^2})$ which gives
\begin{align}
\sigma^T_+  = \sigma^T_- &=\frac{4\pi^2\alpha}{K}W_1(\nu,Q^2)\,, \\
 \sigma_0 = \sigma_L &=  \frac{4\pi^2\alpha}{K}{\bigg(} -W_1(\nu,Q^2) + {\bigg[}1 + \frac{\nu^2}{Q^2}{\bigg]}W_2(\nu,Q^2){\bigg)}\,,
 \end{align}
where $\sigma^T_{\pm}$ are the total cross sections for scattering of a photon with helicity $\lambda = \pm1$ from a proton at rest. The total cross section $\sigma_0$ is for scattering of a $\lambda = 0$ helicity photon and is analogous to a longitudinal polarization state.

Solving for the unpolarized structure functions in the above equations and then applying the result to the doubly differential electron-proton cross section in equation~\eqref{XSfinal} gives
\begin{align}
\frac{d\sigma}{dE'd\Omega} & = \Gamma (\sigma_T + \epsilon \sigma_L)\,,
\end{align}
with the following defintions of the kinematic factors and total photon cross sections
\begin{align}
 \sigma_T & = \frac{1}{2}(\sigma^T_+ + \sigma^T_-)\,,\\
\Gamma & = \frac{\alpha K}{2\pi^2Q^2}\frac{E'}{E}\frac{1}{1-\epsilon}\,,\\
\epsilon  & = {\bigg(}1 + 2(1+\nu^2/Q^2)\mathrm{tan}^2\frac{\theta}{2}{\bigg)}^{-1}\,.
\end{align}
For clarity, $\sigma_T$ ($\sigma_L$) represents the cross section related to transverse (longitudinal) polarization states of the photon. The positive and negative transverse cross sections are often denoted with $\sigma^T_{3/2} = \sigma^T_+$ and $\sigma^T_{1/2}=\sigma^T_-$, to make the spin component explicit. The longitudinal states only exist for virtual, off-shell photons and vanish as $Q^2 \rightarrow 0$~\cite{Weinberg:1995mt}. All three virtual photon flux conventions reduce to $\nu$ in the real photon limit at $Q^2 = 0$.

If both the electron and proton are polarized, then two interference terms are added to the cross section giving~\cite{PhotonPol}
\begin{align}
\frac{d\sigma}{dE'd\Omega} & = \Gamma (\sigma_T + \epsilon \sigma_L - hP_x\sqrt{2\epsilon(1-\epsilon)}\sigma_{LT} - hP_z\sqrt{1-\epsilon^2}\sigma_{TT})\,,
\end{align}
where $h = \pm 1$ is the helicity of the incoming longitudinally polarized electrons, $P_z$ ($P_x$) is the polarization component of the target parallel (perpendicular) to the lab momentum of the virtual photon. The photon cross sections, $\sigma_{LT}$ and $\sigma_{TT}$, are expressed in terms of the polarized structure functions such that
\begin{align}
\label{sigmaLT}
\sigma_{LT} &= \frac{4\pi^2\alpha QM}{K}{\bigg(}G_1(\nu,Q^2) + \frac{\nu}{M}G_2(\nu,Q^2){\bigg)}\,,\\
\label{sigmaTT}
\sigma_{TT} &= \frac{4\pi^2\alpha M}{K}{\bigg (}\nu G_1(\nu,Q^2) -\frac{Q^2}{M}G_2(\nu,Q^2){\bigg)}\,,\\
\sigma_{TT} & = \frac{1}{2}(\sigma^T_{3/2} - \sigma^T_{1/2})\,.
\end{align}
Normalizing equation~\eqref{sigmaLT} and equation~\eqref{sigmaTT} by the transverse virtual photon cross section gives two virtual photon asymmetries
\begin{align}
A_1 &= \frac{\sigma_{TT}}{\sigma_{T}} = \frac{\nu G_1(\nu,Q^2) - \frac{Q^2}{M}G_2(\nu,Q^2)}{W_1(\nu,Q^2)}\,,\\
A_2 &= \frac{\sigma_{LT}}{\sigma_{T}} = \frac{QG_1(\nu,Q^2) + \frac{Q\nu}{M}G_2(\nu,Q^2)}{W_1(\nu,Q^2)}\,,
\end{align}
which have the benefit of canceling out the ambiguous virtual photon flux. 

\section{The Structure Functions}
Some physical meaning is attached to the structure functions by studying them in the deep inelastic region. It is easiest to describe this theoretically in the Breit (named after Gregory Breit) reference frame and for a fast moving proton. 

\begin{figure}
\centering     
\subfigure[Striking a parton]{\label{fig:a}\includegraphics[width=.45\textwidth]{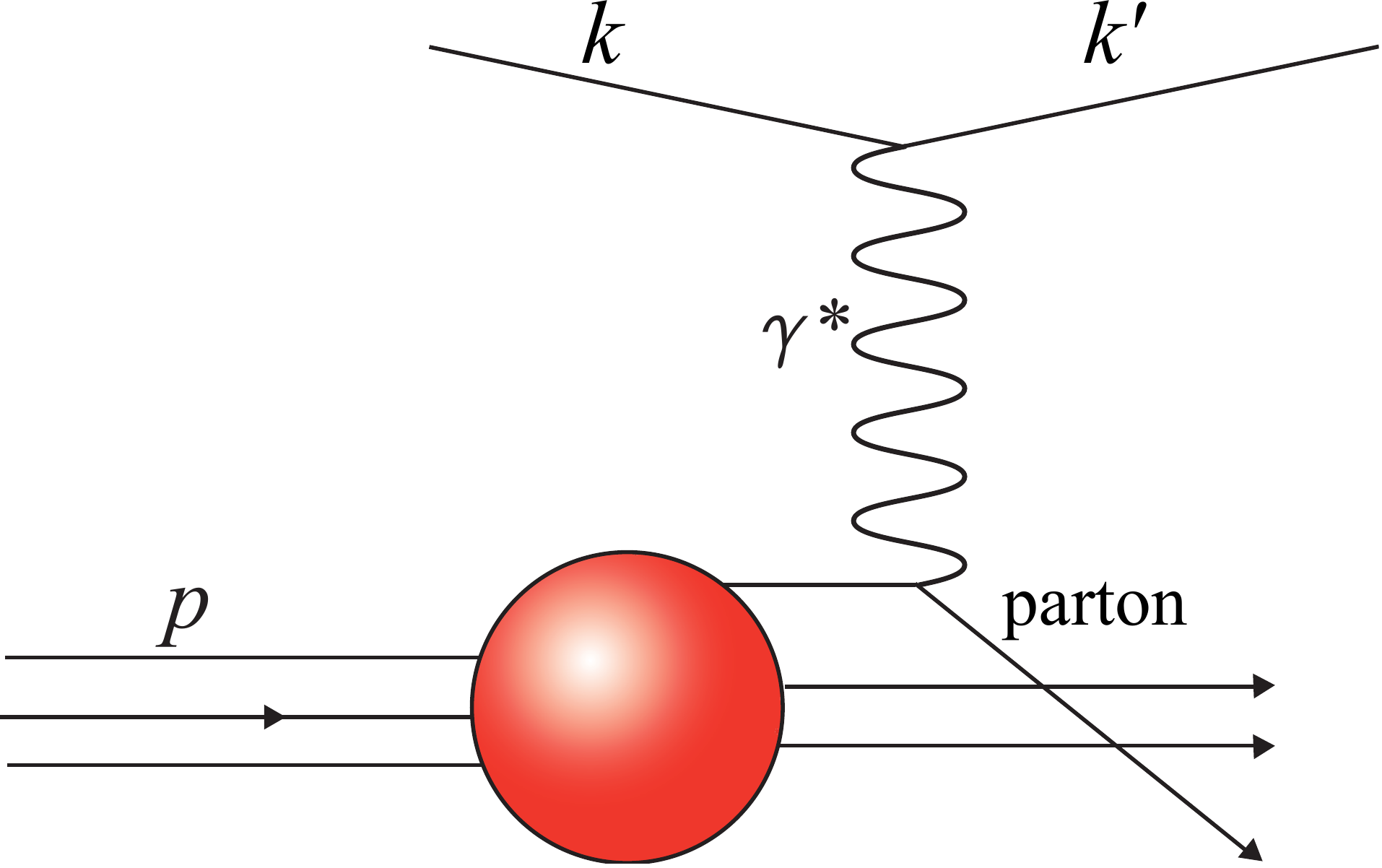}}
\qquad
\subfigure[Elastic recoil]{\label{fig:b}\includegraphics[width=.45\textwidth]{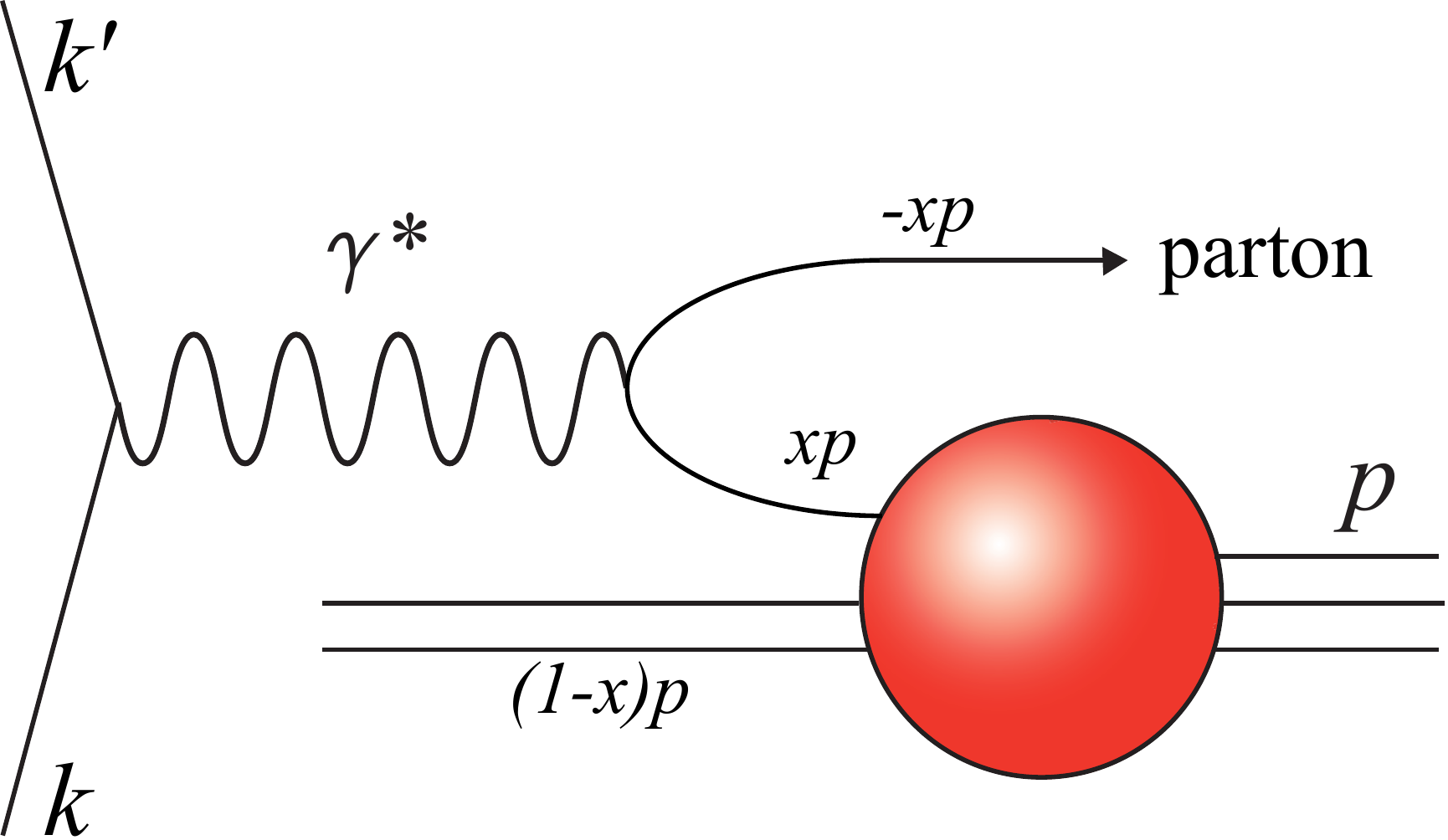}}
\caption{The Breit (infinite-momentum) reference frame is also referred to as the $brick-wall$ frame because the struck parton has its momentum reversed in the collision.}
\label{Parton}
\end{figure}

In the Breit frame (see Figure~\ref{Parton}),  the virtual photon does not transfer any energy to the struck particle and thus ${ |q|}$ = $\sqrt{Q^2}$~\cite{Quarks}. Applying the de Broglie relation, $\lambda = \hbar / { |q|}$, in the  Breit frame shows that the spatial resolution of the virtual photon is directly tied to its four-momentum transfer. 

If the proton is moving very fast then it is safe to assume that its constituents are also going to be moving very fast with it. Under this condition, the proton structure is approximately given by the longitudinal momenta of its constituents and transverse momenta and the proton rest mass can be neglected. Any scattering interaction in this frame must conserve helicity~\cite{Thomas}.

For large values of the momentum transfer, corresponding to the deep inelastic region, the spatial resolution of the virtual photon is sufficient to scatter from the proton's constituents. Richard Feynman collectively called these constituents $partons$~\cite{Feynman1988}, but they are now known as quarks and gluons. The asymptoticly free nature of the theory means that at small distance scales the partons are free moving (non-interacting) and the electron-proton interaction is described as the incoherent sum of the individual partons.

\subsection{Bjorken Scaling}
An obvious question to ask next is: what happens as the momentum transfer increases? Can the virtual photon probe inside these so-called partons? This is answered by experimentally looking at the behavior of the structure functions over a range of momentum transfer. It is convenient to do this by first making the following redefinitions of the structure functions:
\begin{align}
F_1(x,Q^2) &= MW_1(\nu,Q^2)\,,\\
F_2(x,Q^2) & = \nu W_2(\nu,Q^2)\,,\\ 
g_1(x,Q^2) &= M^2\nu G_1(\nu,Q^2)\,,\\
g_2(x,Q^2) &= M\nu^2G_2(\nu,Q^2)\,,
\end{align}
where $F_1(x,Q^2)$ and $F_2(x,Q^2)$ are inelastic form factors and should not be confused with the Pauli and Dirac form factors, $F_1(Q^2)$ and $F_2(Q^2)$, of elastic scattering.

\begin{figure}[htp]
\begin{center}
\includegraphics[scale=0.7]{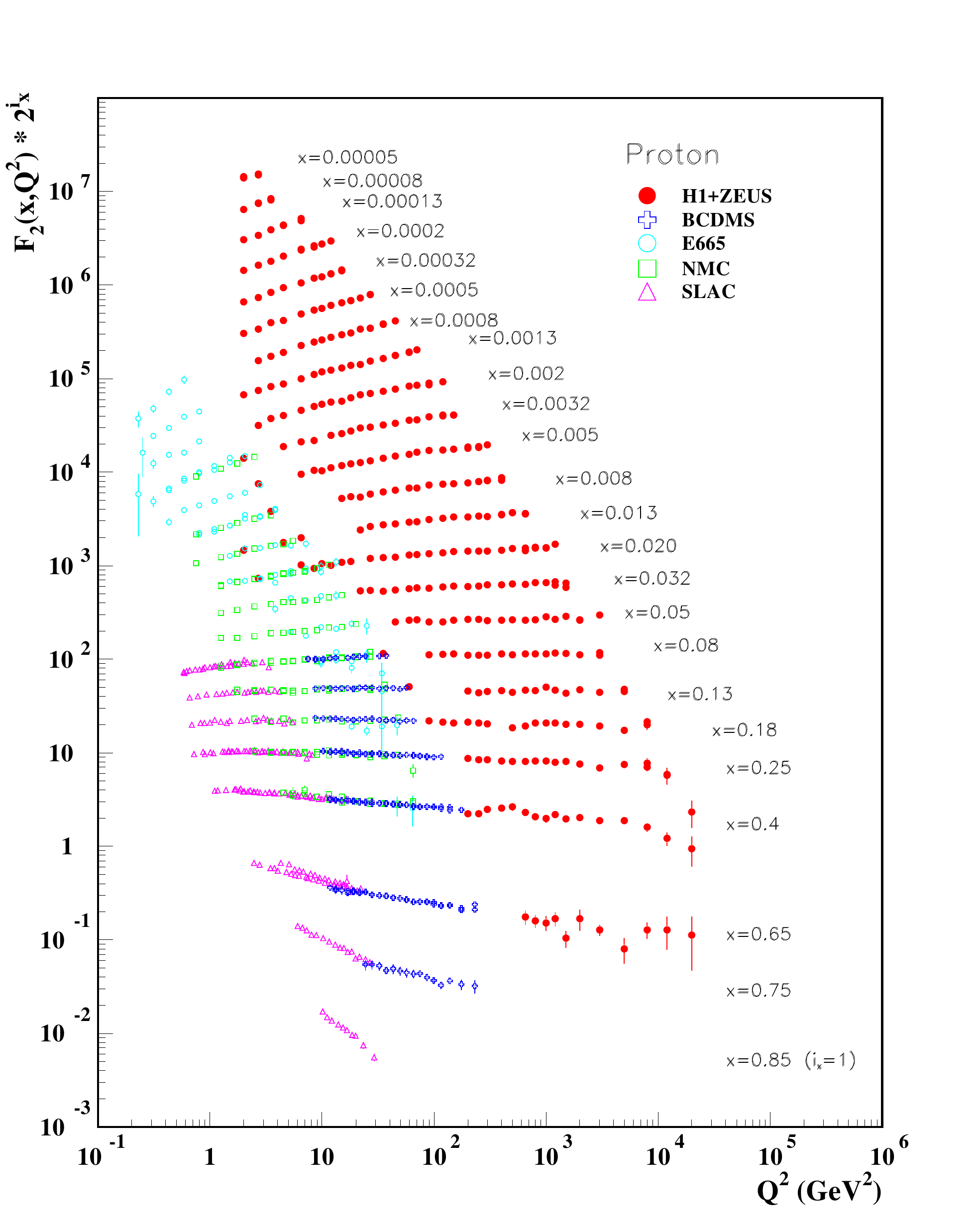}
\caption{\label{F2}World data for the proton structure function, $F_2^p$. The logarithmic dependence of $F_2(Q^2)$ is directly related the complex internal interactions of the proton. Reproduced from~\cite{PDG}.}
\end{center}
\end{figure}

At large $Q^2$, measurements of $F_1(x,Q^2)$ and $F_2(x,Q^2)$ show that for fixed values of $x$ they depend very weakly on momentum transfer. This effect is seen in Figure~\ref{F2} for data on electron-proton and positron-proton scattering. For plotting purposes, there is an additional scale factor\footnote{$i_x$ is the number of the $x$ bin ranging from $i_x$ = 1 to $i_x$ = 21.}, $2^{i_x}$, that produces an y-offset in the data. The observation that the structure functions are independent of $Q^2$ in the deep inelastic scattering region implies that the electrons are elastically scattering from point particles (quarks), and therefore the substructure of the proton is composed of point-like constituents~\cite{Povh}. This is referred to as Bjorken (named after James Bjorken) scaling. 

Bjorken scaling is only an approximation; quarks can radiate gluons before and after the scattering process and gluons can split into $q\bar{q}$ pairs or emit gluons themselves. At lower $Q^2$ these processes cannot be separated from electron scattering from a quark without gluon radiation. At higher $Q^2$ the higher energy resolution means that the process is more likely to isolate a single part of the complex quark-gluon radiation process. This causes the structure functions to develop a logarithmic dependence on $Q^2$ and weakly violate Bjorken scaling.
\subsection{Parton Model}

Feynman and Bjorken fold the point-like partons into the global properties of the proton through momentum distribution functions. Each parton has some probability $f(x)$ to carry an $x$-fraction of the proton's momentum. The sum over each individual $x$ returns the total momentum fraction of the proton
\begin{equation}
\sum_i \int dx\, x\, f_i(x) = 1\,,
\end{equation}
for $i$ types of partons. This sum includes both charged partons (quarks and antiquarks) and neutral partons (gluons). In electron-proton scattering, the electron predominately interacts with the charged quarks, so $f_i(x)$ is actually just $q_f(x) + \bar{q}_f(x)$, where $q_f(x)$ ($\bar{q}_f(x)$) is the probability for a quark (anti-quark) of flavor\footnote{The quark flavors are: up, down, strange, charm, top and bottom.} $f$ to have a momentum fraction $x$ inside the proton.  In the parton model, these quark momentum distribution functions are related to the structure functions via
\begin{align}
F_2(x) &= x \sum_f z^2_f{\big(}q_f(x)+\bar{q}_f(x){\big)}\,,\\
2xF_1 (x) &= F_2(x)\,,
\end{align}
where $z_f$ is the charge for a quark of flavor $f$. The equality between $F_1(x)$ and $F_2(x)$ is a consequence of the quarks having spin-$\frac{1}{2}$ and is known as the Callan-Gross relation~\cite{PhysRevLett.22.156}. Again, this statement is just the incoherent sum  of the non-interacting quarks and anti-quarks (partons) inside the proton.

 The above relations can also be described in terms of the spin of the quarks, electron and virtual photon probe. This discussion naturally leads to polarized structure functions within the parton model. In the scattering interaction, the incident electron imparts some of its helicity onto the virtual photon. Helicity conservation states that the virtual photon is only absorbed by a quark of the same helicity. Relating this back to the discussion in Chapter~\ref{VPPXS} gives
 
\begin{equation}
\sigma_+^T + \sigma_-^T \propto F_1 = \frac{1}{2} \sum_i z^2_i{\big(}f^+_i(x)+f^-_i(x){\big)}\,,
\end{equation}
 where, for simplicity, the sum is in the parton notation and includes both quarks and anti-quarks of positive and negative helicity. The polarized structure function $g_1(x)$ is 
 \begin{align}
\sigma_+^T - \sigma_-^T \propto\, g_1 &= \frac{1}{2} \sum_i z^2_i{\big(}f^+_i(x)-f^-_i(x){\big)}\,,\\
g_1(x)& = \frac{1}{2}\sum_f z^2_f{\big(}\Delta q_f(x)+\Delta\bar{q}_f(x){\big)}\,,
\end{align}
 where the polarized parton distribution $\Delta q_f(x)$ ($\Delta\bar{q}_f(x)$)  is defined as the difference between quarks (anti-quarks) with positive and negative helicity. 

What about the other spin structure function? Unfortunately, there is no simple interpretation of $g_2(x)$ in the parton model.  The relevant photon absorption cross section describing $g_1(x) + g_2(x) \propto \sigma_{LT}$ requires the absorption of a helicity zero, longitudinal photon. This can only happen if the quark undergoes a helicity flip~\cite{Ji}. Consequently, in QCD this requires that the quark has a mass, which is in contrast with the  zero-mass assumption of the parton model. 

Non-zero values of $g_2(x)$ may be obtained by adding transverse momentum to the parton model, but these formulations have an extreme sensitivity to the quark mass~\cite{Thomas}. Abandoning the parton model all together, the problematic quark spin flip is avoided if the quark is allowed to absorb a gluon. In any case, it is clear that a different approach is needed to describe this structure function.

\chapter{\sc Theoretical Tools and Phenomenological Models}
\label{ch:Tools} 

Driven by the {\it spin crisis}\footnote{Reminder: the European Muon Collaboration at CERN found that the intrinsic quark spin only contributes 30\% of the total spin of the nucleon from measuring $g_1(x,Q^2)$~\cite{Ashman}.}, experimental knowledge of the spin structure functions now covers a large $Q^2$ range. At low $Q^2$, the data are compared to predictions from Chiral Perturbation Theory, while at larger $Q^2$ the operator product expansion is typically used. These theoretical methods calculate scattering amplitudes for the virtual photon-proton interaction. The Compton amplitudes cannot be experimentally measured for space-like virtual photons ($Q^2 >$ 0); instead the data are compared to theory using a combination of dispersion relations and the optical theorem. The wealth of experimental data has also lead to the development of several phenomenological models for the nucleon structure functions. These models are empirical fits to the current world data and provide fairly accurate predictions in the kinematic region relevant to E08-027.

The theoretical methods in this chapter are presented to give an overview of the options available to relate experimental data to theoretical calculations and vice versa. There is a kinematic separation that occurs on the applicability of a given method. For example, the work of this thesis is best described theoretically by Chiral Perturbation Theory. The Operator Product Expansion (OPE), also described in this chapter, is an extension of perturbative QCD. The OPE is included for completeness and to elucidate the proton spin crisis. The common thread linking the two methods is the development of the concept of a moment of a spin structure function.  


\section{Moments of the Spin Structure Functions}
\label{Section:Moments}
\begin{figure}[htp]
\begin{center}
\includegraphics[scale=0.8]{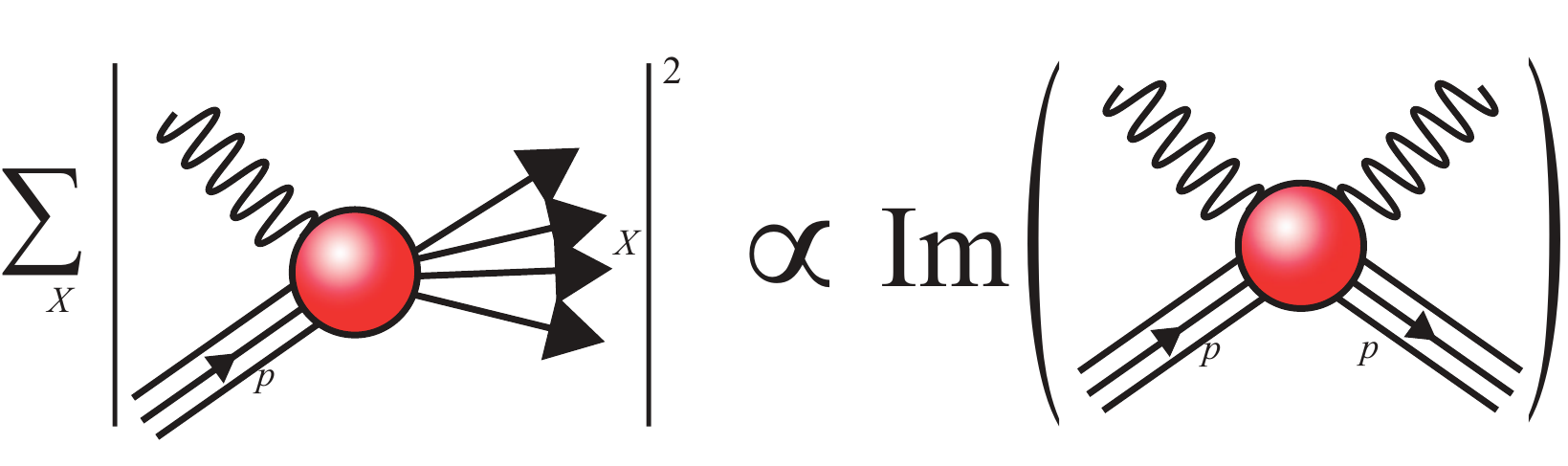}
\caption{\label{Optical}The optical theorem relates the total scattering cross section to the imaginary part of the forward scattering amplitude. Shown here is its application to (virtual) Compton scattering. Compton scattering is named after American physicist Arthur Holly Compton.}
\end{center}
\end{figure}

While straightforward to measure, the proton structure functions are difficult to calculate from first principles using QCD because the relevant Feynman diagrams contain sums over all final nucleon states. In terms of the photon-proton cross section, this is the same as summing over Figure~\ref{PhotonProton} for each possible final state, $X$. Theoretical methods, such as Chiral Perturbation Theory, are more suitable for (virtual) photon-proton Compton scattering. Fortunately, through the use of the optical theorem (see Figure~\ref{Optical} and Ref~\cite{jacksonT}) and dispersion relations, it is possible to relate a calculated photon-proton scattering amplitude to the experimentally measured structure functions. In this manner, the sum of all possible final states $X$ in virtual photon-proton scattering is proportional the imaginary part of the forward Compton scattering ($k = k'$) amplitude. The relevant experimental quantities are $x$-weighted integrals of the structure functions, known as moments or sum rules and can be thought of, physically, in terms of a polarizability. 

Polarizability is the ability for a composite system to be polarized, and is an elementary property of the system in the same manner as the charge or mass. Polarizabilities determine the response of a bound system to external fields and provide insight into the internal structure of the bound system. In the case of photon-proton Compton scattering, the external field is an electromagnetic dipole field provided by the outgoing photon and the proton's polarizability shapes its response to this field. The incoming photon also has an electromagnetic field. Interference between the two photon fields contributes to the complex amplitude of the outgoing wave, which suggests the existence of a Kramers-Kronig dispersion relation~\cite{Drechsel2}. 

Dispersion theory relates the imaginary and real parts of the Compton amplitudes via a photon energy-weighted integral, $i.e.$ a Cauchy integral. This relation is only valid if the total amplitude is analytic in the positive-imaginary plane, but causality automatically satisfies this condition. The theory also assumes that the amplitudes vanish as photon energies approach infinity. This no subtraction hypothesis eliminates fixed poles in the complex plane at infinity.  Finally, the dispersion relations are tied back to experimental measurements using the optical theorem; the imaginary Compton amplitudes are replaced by the experimental photon-proton cross sections.



\subsection{Forward Compton Scattering}
The first step in constructing the various sum rules and moments is to write down the forward Compton scattering amplitude:
\begin{align}
\label{ComptonA}
T^{\mu\nu} = i \int d^4x e^{q\cdot x} \Bra{P,S}T J^{\mu}(x)J^{\nu}(0)\Ket{P,S}\,,
\end{align} 
where $T$ represents the time-ordered product~\cite{Thomas}.  If the scattering involves two virtual photons\footnote{For real photons, $f_L (\nu,Q^2)= f_{LT} (\nu,Q^2)= 0$.} (VVCS) then the amplitude has the following decomposition
\begin{align}
\label{ComptonAA}
T= \pvec{\epsilon}'^{*} &\cdot \vec{\epsilon} f_T(\nu,Q^2) + f_L(\nu,Q^2) \nonumber \\ &+ i\vec{\sigma} \cdot ( \pvec{\epsilon}'^{*} \times \pvec{\epsilon} )f_{TT}(\nu,Q^2) - i\vec{\sigma} \cdot [ \pvec{\epsilon}'^{*} -\vec{\epsilon} \times \hat{q}]f_{LT}(\nu,Q^2)\,,
\end{align}
where $\pvec{\epsilon}$' and $\vec{\epsilon}$ are the transverse photon polarizations and $\hat{q}$ is the longitudinal polarization~\cite{Drechsel}. The proton spin vector is $\sigma$. The individual amplitudes $f_T,\, f_L,\, f_{TT}$, and $f_{LT}$ are related to the (virtual) photon-proton cross sections using the optical theorem:
\begin{align}
\mathrm{Im}\, f_T (\nu,Q^2) &= \frac{K}{4\pi} \sigma_T (\nu,Q^2)\,,\\
\mathrm{Im}\, f_L (\nu,Q^2) &= \frac{K}{4\pi} \sigma_L (\nu,Q^2)\,,\\
\mathrm{Im}\, f_{LT} (\nu,Q^2) &= \frac{K}{4\pi} \sigma_{LT} (\nu,Q^2)\,,\\
\mathrm{Im}\, f_{TT} (\nu,Q^2) &= \frac{K}{4\pi} \sigma_{TT} (\nu,Q^2)\,,
\end{align}
where $K$ is the same virtual photon flux factor discussed in Chapter~\ref{VPPXS}.

A more direct comparison to the structure functions is made by noting that the forward Compton amplitude in equation~\eqref{ComptonA} only differs from the hadron tensor in equation~\eqref{BetterHadronT} by a time-ordered product. This allows for the Compton amplitude to also be written as
\begin{align}
T^{\mu\nu}= &{\bigg (} -g^{\mu\nu} + \frac{q^{\mu}q^{\nu}}{q^2}{\bigg )}S_1(\nu,Q^2) \nonumber \\ 
&+ \frac{1}{M^2}{\bigg (P^{\mu} - \frac{M\nu}{q^2}q^{\mu}}{\bigg)}{\bigg (P^{\nu} - \frac{M\nu}{q^2}q^{\nu}}{\bigg)}S_2(\nu,Q^2) \nonumber\\
&-\frac{i}{M} \epsilon^{\mu\nu\rho\sigma} q_{\rho}s_{\sigma}A_1(\nu,Q^2) \nonumber \\
&-\frac{i}{M^3}\epsilon^{\mu\nu\rho\sigma}q_{\rho}[(M\nu)s_{\sigma} - (q\cdot s)P_{\sigma}]A_2(\nu,Q^2)\,,
\end{align}
where $S_{1,2} (\nu,Q^2)$ and $A_{1,2} (\nu,Q^2)$ are the covariant amplitudes. These amplitudes are related to the structure functions according to
\begin{align}
\mathrm{Im}\, A_{1,2} (\nu,Q^2) &= 2 \pi G_{1,2}(\nu,Q^2)\,, \\
\mathrm{Im}\, S_{1,2} (\nu,Q^2) &= 2 \pi W_{1,2}(\nu,Q^2) \,,
\end{align}
with
\begin{equation}
W_{\mu\nu} (\nu,Q^2) = \frac{1}{2\pi M}\mathrm{Im}T_{\mu\nu}(\nu,Q^2)\,,
\end{equation}
and are a direct result of the optical theorem.
\subsection{Evaluating Dispersion Relations}
Taking the above analysis one step further and applying a non-subtracted dispersion relation to the Compton amplitude $f_i (\nu,Q^2)$ directly relates a real scattering amplitude to a real measured structure function. After using the Cauchy integral theorem, the result is~\cite{Pantforder}
\begin{equation}
f_i (\nu,Q^2) = \frac{2}{\pi} \int_{\nu_0}^{\infty} \frac{d\nu' \nu'}{\nu'^2 - \nu^2} \mathrm{Im}\, f_i(\nu',Q^2)\,.
\end{equation}
The lower limit of the dispersion integral corresponds to the pion-production threshold, so as to avoid the nucleon elastic pole. This $\pi N$ state has invariant mass $(p+q)^2 \geq (M + m_{\pi})^2$ in the $s$-channel diagram and $(p-q)^2 \geq (M + m_{\pi})^2$ in the $u$-channel. This defines the  cut-off as
\begin{equation}
\nu_0 = m_{\pi} + \frac{m_{\pi}^2}{2M} + Q^2\,.
\end{equation}
Focusing on the two spin-flip amplitudes, $f_{TT}$ and $f_{LT}$, the corresponding dispersion relations are
\begin{align}
\mathrm{Re}[f_{TT} (\nu,Q^2) - f_{TT}^{pole}(\nu,Q^2)] &= \frac{\nu}{2\pi^2} \mathcal{P} \int_{\nu_0}^{\infty} \frac{K(\nu',Q^2)\sigma_{TT}(\nu',Q^2)}{\nu'^2 -\nu^2}d\nu'\,,\\
\mathrm{Re}[f_{LT} (\nu,Q^2) - f_{LT}^{pole}(\nu,Q^2)] &= \frac{1}{2\pi^2} \mathcal{P} \int_{\nu_0}^{\infty} \frac{\nu'K(\nu',Q^2)\sigma_{LT}(\nu',Q^2)}{\nu'^2 -\nu^2}d\nu'\,,
\end{align}
where $g_{TT}^{pole}/g_{LT}^{pole}$ are the nucleon pole (elastic) contributions and $\mathcal{P}$ denotes the Cauchy principal value integral. The elastic contributions are calculated using the elastic vertex function of equation~\eqref{ElasticVertex} and are~\cite{Drechsel2}
\begin{align}
\mathrm{Re}\, f_{TT}^{pole} &= - \frac{\alpha \nu}{2M^2}\frac{Q^2}{\nu^2 - \nu^2_B} G^2_M(Q^2)\,,\\
\mathrm{Re}\, f_{LT}^{pole} &= \frac{\alpha Q}{2M^2}\frac{Q^2}{\nu^2 - \nu^2_B}G_E(Q^2) G_M(Q^2)\,,
\end{align}
where $\nu_B = Q^2 / 2M$ is the elastic scattering condition.

The analyticity condition on the amplitudes implies that the dispersion integrals may be expanded into a Taylor series~\cite{Drechsel}. Furthermore, the Compton amplitude has a crossing symmetry, which means that equation~\eqref{ComptonAA} need be invariant under the transformations $\pvec{\epsilon}' \longleftrightarrow \vec{\epsilon}$ and $\nu \longleftrightarrow -\nu$. This implies that $f_{TT}$ ($f_{LT}$) is odd (even) so only odd (even) powers of $\nu$ should appear in the expansion. The result is 
\begin{align}
\label{MomentSum}
\mathrm{Re}\, f_{TT} &= \sum_{n=0} {\bigg(}\frac{1}{2\pi^2}\int_{\nu_0}^{\infty} \frac{K(\nu',Q^2)\sigma_{TT}(\nu',Q^2)}{\nu'^{2n + 2}}\nu^{2n +1}d\nu'{\bigg )}\,,\\
\mathrm{Re}\, f_{LT} &=  \sum_{n=0} {\bigg(}\frac{1}{2\pi^2}\int_{\nu_0}^{\infty} \frac{K(\nu',Q^2)\sigma_{LT}(\nu',Q^2)}{\nu'^{2n+1}}\nu^{2n}d\nu'{\bigg )}\,,
\end{align}
and, alternatively, a  low energy expansion for $f_{TT}$ and $g_{TT}$ in terms of $\nu$ gives~\cite{Chen}
\begin{align}
\label{MomentExpansion}
\mathrm{Re}[f_{TT} (\nu,Q^2) - f_{TT}^{pole}(\nu,Q^2)]  &= {\bigg (} \frac{2\alpha}{M^2}{\bigg )}I_{TT} (Q^2) \nu + \gamma_{TT} (Q^2) \nu^3 + O(\nu^5)\,,\\
\mathrm{Re}[f_{LT} (\nu,Q^2) - f_{LT}^{pole}(\nu,Q^2)]  &= {\bigg (} \frac{2\alpha}{M^2}Q{\bigg )}I_{LT} (Q^2)  + Q \delta_{LT} (Q^2) \nu^2 + O(\nu^4)\,.
\end{align}
Comparing terms order by order in the two separate expansions defines the sum rules and generalized\footnote{Generalized here implies that $Q^2$ is not zero.} polarizabilities. The leading term in the $f_{TT}$ expansion yields the generalized Gerasimov-Drell-Hearn (GDH) sum rule
\begin{align}
I_{TT} (Q^2) &= \frac{M^2}{4\pi^2 \alpha} \int_{\nu_0}^{\infty} \frac{K(\nu',Q^2)\sigma_{TT}}{\nu'^2}d\nu'\,, \\
 \label{GDHeq}                     &= \frac{2M^2}{Q^2}\int_{0}^{x_0} {\bigg (} g_1(x,Q^2) - \frac{4M^2}{Q^2} x^2 g_2(x,Q^2){\bigg )} dx\,.
\end{align}
At the real photon point of $Q^2$ = 0, $I(0)$ is related directly to the anomalous magnetic moment of the proton, further strengthening the relation between these integrated quantities and static properties of the nucleon. The original GDH sum rule is
\begin{equation}
\label{GenGDH}
I(0)= \int_{\nu_0}^{\infty} \frac{\sigma_{1/2}(\nu)-\sigma_{3/2}(\nu)}{\nu}d\nu = -\frac{2\pi^2\alpha \kappa^2}{M^2},\\
\end{equation}
where $\kappa$ is the anomalous magnetic moment~\cite{GDH}.
The second order term gives the generalized forward spin polarizability 
\begin{align}
\label{GAMMAEQ}
\gamma_{TT} (Q^2) &= \frac{1}{2\pi^2 \alpha} \int_{\nu_0}^{\infty} \frac{K(\nu',Q^2)\sigma_{TT}}{\nu'^4}d\nu'\,, \\
                      &= \frac{16\alpha M^2}{Q^6}\int_{0}^{x_0} x^2{\bigg (} g_1(x,Q^2) - \frac{4M^2}{Q^2} x^2 g_2(x,Q^2){\bigg )} dx\,,
\end{align}
where $x_0 = \frac{Q^2}{2M\nu_0}$. The process is the same for the $f_{LT}$ expansion. The first order term leads to a sum rule
\begin{align}
I_{LT} (Q^2) &= \frac{M^2}{4\pi^2 \alpha} \int_{\nu_0}^{\infty} \frac{K(\nu',Q^2)\sigma_{LT}}{\nu'Q}d\nu' \,,\\
                      &= \frac{2M^2}{Q^2}\int_{0}^{x_0} {\bigg (} g_1(x,Q^2) + g_2(x,Q^2){\bigg )} dx\,.
\end{align}
The next-to-leading order term gives the generalized longitudinal-transverse polarizability
\begin{align}
\label{DLTEQ}
\delta_{LT} (Q^2) &= \frac{1}{2\pi^2 \alpha} \int_{\nu_0}^{\infty} \frac{K(\nu',Q^2)\sigma_{TT}}{\nu'^3Q}d\nu'\,, \\
                      &= \frac{16\alpha M^2}{Q^6}\int_{0}^{x_0} x^2{\bigg (} g_1(x,Q^2) + g_2(x,Q^2){\bigg )} dx\,.
\end{align}

At low momentum transfer, the majority of the integral strength comes from the resonance region and the high-$x$, DIS contributions are minimal. The $x^2$ weighting of the higher moments, $\delta_{LT} (Q^2)$ and $I_{TT} (Q^2)$, causes them to converge even faster than the first order moments, further minimizing DIS contributions. Put simply, it is still possible to accurately evaluate the structure function moments, even if only a portion of the integrand is measured.  

\subsection{The Burkhardt-Cottingham Sum Rule}
Returning to the covariant formulation, a dispersion relation for the Compton amplitude $A_2(\nu,Q^2)$ reads
\begin{equation}
\label{Snonu}
\mathrm{Re}\, [A_2(\nu,Q^2) - A_2^{pole}(\nu,Q^2)] =\frac{2\nu}{\pi} \mathcal{P} \int_{\nu_0}^{\infty} \frac{G_2(\nu',Q^2)}{\nu'^2 - \nu^2}d\nu'\,,
\end{equation}
which is odd in $\nu$. This is expected because $A_2$ is odd, $i.e$ $A_2 (\nu,Q^2) = -A_2(-\nu,Q^2)$. Assuming a high-energy Regge behavior, as $\nu \rightarrow \infty$ the amplitude, $A_2(\nu,Q^2)$, is described solely as a function of $\nu$ with $A_2(\nu,Q^2) \rightarrow \nu^{\alpha_2}$ and $\alpha_2 < -1$. This allows for another dispersion relation for the amplitude $\nu A_2(\nu,Q^2)$,
\begin{equation}
\label{Snu}
\mathrm{Re}\, [\nu A_2(\nu,Q^2) - \nu A_2^{pole}(\nu,Q^2)] =\frac{2\nu}{\pi} \mathcal{P} \int_{\nu_0}^{\infty} \frac{ \nu'^2 G_2(\nu',Q^2)}{\nu'^2 - \nu^2}d\nu'\,.
\end{equation}
Subtracting equation~\eqref{Snu} from equation~\eqref{Snonu} multiplied by $\nu$, gives the ``super-convergence relation" valid for any value of $Q^2$~\cite{Drechsel2}:
\begin{align}
\label{BCSum}
\int_0^{1} g_2(x,Q^2) dx = 0\,.
\end{align}
The integration includes both the elastic (pole) and inelastic contributions. This relation is known as the Burkhardt-Cottingham (B.C.) Sum Rule. The validity of the sum rule requires that the integral converges, $i.e$ Regge behavior is valid at all values of $x$, and that the Compton amplitude is free of fixed poles at $x=0$, $i.e.$ $g_2(x,0)$ is not a $\delta$-function~\cite{BC}. 

\subsection{The First Moment of $g_1(x,Q^2)$}
\label{GDHSLOPECHAPTER}
The asymptotic limits of the generalized GDH sum are
\begin{align}
 \lim_{Q^2\to 0}I_{TT}(Q^2)& = \frac{2M^2}{Q^2} \int_0^{x_0} g_1(x,Q^2)dx\,,\\
  \lim_{Q^2\to \infty}I_{TT}(Q^2) &=\int_0^1 g_1(x,Q^2)dx\,,
\end{align}
which only depend on the first moment of $g_1(x,Q^2)$\,. The small momentum transfer limit is also given by the standard GDH sum. 
Equating the two puts a constraint on the first moment of $g_1(x,Q^2)$ and after making a change of variables from $\nu$ to $x$ in the integration gives~\cite{Bodo} 
\begin{equation}
\lim_{Q^2\to 0} \Gamma_1 (Q^2) = -\frac{1}{8}\kappa^2 \frac{Q^2}{M^2} = \frac{-0.456}{\,\,\,\mathrm{GeV}^2}Q^2\,,
\end{equation}
where $\Gamma_1(0)$ is expected to be zero. This is referred to as the GDH slope in the literature.

It is also possible to arrive at the low momentum transfer limit from a low energy expansion of the Compton amplitude $A_1(\nu,Q^2)$. Following the analysis of the preceding section gives:
\begin{align}
\mathrm{Re}\, [A_1(\nu,Q^2) - A_1^{pole}(\nu,Q^2)] &=\frac{2\nu}{\pi} \mathcal{P} \int_{\nu_0}^{\infty} \frac{G_1(\nu',Q^2)}{\nu'^2 - \nu^2}d\nu'\,,\\
\label{g1expand}
\mathrm{Re}[A_1(\nu,Q^2) - A_1^{pole}(\nu,Q^2)]  &=\frac{4\alpha}{M}I_{1} (Q^2)  + \gamma_{g_1}(Q^2) \nu^2 + O(\nu^4)\,,
\end{align}
with
\begin{equation}
I_{1}(Q^2)= \frac{2M^2}{Q^2} \int_0^{x_0} g_1(x,Q^2)dx\,,
\end{equation}
as the first term in the expansion.
\section{Operator Product Expansion}
\label{sec:OPE}
 The Operator Product Expansion is a way to evaluate the hadronic tensor in the deep inelastic scattering regime. Formulated by Ken Wilson~\cite{Wilson}, the OPE states that the product of two operators allows for the following expansion at small distances:
\begin{equation}
\label{OPE}
\lim_{x\rightarrow 0} \mathcal{O}_a(x)\mathcal{O}_b(0) = \sum_k C_{abk}(x)\mathcal{O}_k(0)\,,
\end{equation} 
where $C_{abk}$ are known as the Wilson coefficients and the $\mathcal{O}_k(0)$ are local operators. This expansion into a linear combination of operators is easier to calculate than the original product~\cite{Thomas}. In the language of QCD, the $C_{abk} (x)$ are perturbative in the limit $x\rightarrow0$ as a result of asymptotic freedom. The operators, $\mathcal{O}_k$, are non-perturbative and represent the local quark and gluon fields. Put more simply, the OPE performs a separation of scales, where the Wilson coefficients represent free quarks and gluons and the local operators parameterize the unknown sum over final proton states.

In order to apply the OPE, the differential cross section for e-p scattering needs to be written as the product of operators. This requirement is satisfied with the Compton amplitude:
\begin{align}
T^{\mu\nu} = i \int d^4x e^{iq\cdot x} \Bra{P,S}T J^{\mu}(x)J^{\nu}(0)\Ket{P,S}\,,
\end{align}
where the optical theorem relates the Compton amplitude back to the hadron tensor $2MW^{\mu\nu} = \frac{1}{\pi}\mathrm{Im}T^{\mu\nu}$. The two operators, $J^{\mu}(x)J^{\nu}(0)$, are the quark electromagnetic currents.  The Fourier transform\footnote{The limit is now $Q^2 \rightarrow \infty$ instead of $x \rightarrow$ 0.} is taken of equation~\eqref{OPE} to give 
\begin{equation}
\lim_{x\rightarrow 0}\int d^4x e^{iq\cdot x}\mathcal{O}_a(x)\mathcal{O}_b(0) = \sum_k C_{abk}(q)\mathcal{O}_k(0)\,,
\end{equation}
which allows for direct application of the OPE to the Compton amplitude. The calculation of the Wilson coefficients in DIS is carried out in Refs~\cite{Thomas,Greiner,Peskin}. They contribute to the differential cross section on the order of
\begin{equation}
x^{-n} {\bigg (}\frac{M}{Q}{\bigg)}^{\tau -2}\,,
\end{equation}
where $\tau = D-n$ is defined as the twist for an operator of dimension $D$ and spin $n$.

 The leading twist term is twist-2. It contributes the largest in the Bjorken limit. In terms of the parton model, an OPE analysis to leading twist is related to the amplitude for scattering of asymptotically free quarks and the higher twist terms arise from the quark-gluon interaction and the quark mass effects. At lower $Q^2$, the higher twist terms become more important. Eventually the twist expansion becomes meaningless at low enough $Q^2$ and Chiral Perturbation Theory supersedes the OPE. 

A dispersion relation relates the OPE coefficients and operators to the experimentally measured structure functions. The result is (ignoring anything beyond twist-3)~\cite{Jaffe}
\begin{align}
\label{g1izzle}
\int_0^1 x^{n-1}g_1 (x,Q^2) dx &= \frac{1}{2}C^a_{n-1}a_{n-1}; \,\,\,n=1,3,5\\
\label{g2izzle}
\int_0^1 x^{n-1}g_2 (x,Q^2) dx &= \frac{n-1}{2n} (C^d_{n-1}d_{n-1} - C^a_{n-1}a_{n-1}); \,\,\,n=3,5
\end{align}
where $a_{n-1}$ and $d_{n-1}$ are the matrix elements for the twist-2 and twist-3 local operators, respectively. The Wilson coefficients, $C^a_{n-1}$ and $C^d_{n-1}$, are assumed to be functions of $\alpha_s$ and $Q^2$. Note that the OPE relation for $g_2$ does not have a $n=1$ term; the analysis implicitly assumes that the B.C. sum rule holds.
\subsection{Wandzura-Wilczek Relation}
\label{G2WW}
The OPE allows for a twist description of $g_2(x,Q^2)$ that is non-zero, unlike in the parton model. The leading twist  component of $g_2(x,Q^2)$ is completely defined by $g_1(x,Q^2)$. This is shown by combining equations~\eqref{g1izzle} and~\eqref{g2izzle}, which causes the twist-2 terms to cancel and results in
\begin{equation}
\label{Wandzura}
\int_0^1 x^{n-1}dx {\bigg (} g_1(x,Q^2) + \frac{n}{n-1}g_2(x,Q^2){\bigg)} = \frac{d_{n-1}}{2}\,,
\end{equation}
where $n \geq 3$. Setting the $d_3$ (twist-3) term to zero and applying Mellin transforms gives the Wandzura-Wilczek relation~\cite{Wandzura}
\begin{equation}
g_2^{WW} (x, Q^2) + g_1(x,Q^2) = \int_x^1 \frac{dy}{y} g_1(y,Q^2)\,,
\end{equation}
where $g_2^{WW}(x,Q^2)$ denotes  the ignoring of higher twist contributions. Of course $g_2(x,Q^2)$ still contains twist-3 and higher terms, and at lower $Q^2$ they would be expected to be larger and more relevant. In terms of these higher twist components 
\begin{align}
g_2 (x,Q^2) &= g_2^{WW} (x,Q^2) + \bar{g_2} (x,Q^2)\,,\\
\bar{g_2} (x,Q^2) &= -\int_x^1 \frac{\partial}{\partial y}{\bigg (} \frac{m_q}{M} h_T(y,Q^2) + \xi(y,Q^2){\bigg)}\,,
\end{align}
where $\xi$ is the twist-3 term and $h_T$ is a twist-2 term related to the transverse polarization distribution of the quark. The twist-3 term describes quark-gluon interactions inside the proton, so for example a quark absorbing a gluon as the quark absorbs a longitudinal photon; a process that is forbidden in the parton model.
\subsection{Higher Twist $g_2(x,Q^2)$}
The deviation from leading twist behavior of $g_2(x,Q^2)$ is evident in equation~\eqref{Wandzura} by considering the $n$=3 case (higher powers are suppressed by 1/$Q^2$)
\begin{equation}
d_2(Q^2) = \int_0^1 x^{2}dx {\bigg (} 2g_1(x,Q^2) + 3g_2(x,Q^2){\bigg)}\,.
\end{equation}
Using the Wandzura-Wilczek relation and integration by parts the $d_2$ matrix element is written as
\begin{equation}
\label{d2ducks}
d_2(Q^2) = 3\int_0^1 x^{2}dx {\bigg (} -g_2^{WW}(x,Q^2) + g_2(x,Q^2){\bigg)}\,.
\end{equation}
At large momentum transfers, $d_2(Q^2)$ describes how the color electric and magnetic fields interact with the nucleon spin~\cite{Proposal}. At low momentum transfer a non-zero $d_2(Q^2)$ indicates higher twist effects. Interpolating between the regimes helps map out the change between a hadronic description of the nucleon at low $Q^2$ and a partonic description, based on the OPE at large $Q^2$.

\subsection{Higher Twist $g_1(x,Q^2)$ and the Spin Crisis}
\label{g1ww}
The generalization of equation~\eqref{g1izzle} to all orders of twist gives
\begin{equation}
\Gamma_1 (Q^2)= \int_0^1 g_1(x,Q^2)dx = \sum_{\tau = 2,4, ...} \frac{\mu_{\tau}(Q^2)}{Q^{\tau-2}}\,,
\end{equation}
where the $\mu_{\tau}(Q^2)$ are the coefficients of the twist matrix operators. At leading twist the $\mu_2(Q^2)$ term gives~\cite{G1Twist}
\begin{align}
\mu_2(Q^2) &= {\bigg(}\frac{1}{12}g_A + \frac{1}{36}a_8{\bigg)} + \frac{1}{9}\Delta\Sigma + \mathcal{O}(\alpha_s)\,,\\
a_8 &= \Delta u + \Delta d - 2\Delta s\,,\\
\Delta\Sigma & =  \Delta u + \Delta d +\Delta s\,,
\end{align}
where $g_A$ is the ratio of the axial and vector coupling constants and is determined from neutron beta decay. Higher order QCD corrections of order $\mathcal{O}(\alpha_s)$ are determined in perturbative QCD. The $a_8$ and $\Delta \Sigma$ terms are directly related to the proton spin carried by the intrinsic quarks in the parton model. The $a_8$ contribution is determined from hyperon beta decay, leaving just $\Delta\Sigma$~\cite{Thomas}. At large momentum transfer, where leading twist terms dominate, a measurement of the first moment of $g_1(x,Q^2)$ is a determination of the spin content of the proton.

Assuming that the up and down quarks carry all of the spin leads to the Ellis-Jaffe sum rule and a prediction for $\Delta\Sigma$~\cite{JaffeSum}. Initial attempts to verify the Ellis-Jaffe sum rule by the EMC experiment revealed a quantity considerably smaller than expected~\cite{Ashman}. The violation of the sum rule was dubbed the ``$spin$ $crisis$" and indicated that not all of the spin is carried by the up and down quarks.

The first term beyond leading order in the twist expansion is~\cite{Chen}
\begin{equation}
\mu_4 (Q^2) = \frac{M^2}{9}{\bigg(}a_2(Q^2) + 4d_2(Q^2) + 4f_2(Q^2){\bigg)}\,,
\end{equation}
where $a_2(Q^2)$ is the second moment of $g_1(x,Q^2)$ and is equivalent to $\gamma_{g_1}(Q^2)$ in equation~\eqref{g1expand} for low $Q^2$. The $d_2(Q^2)$ is the same as equation~\eqref{d2ducks}. The twist four term is $f_2(Q^2)$ and, in conjunction with $d_2(Q^2)$, they describe the color electric and magnetic polarizabilities of the proton: the response of the proton's spin to the color electric and magnetic fields.
\section{Chiral Perturbation Theory}
\label{sec:ChiPT}
An effective field theory is a subset of a larger theory that describes aspects of the full theory at a specific energy scale. This allows for a separation of the small and large (with respect to the energy scale) parameters of the theory. The finite effects of the small parameters are included as perturbations to the effective Lagrangian. In general, the presence of the large parameters causes the effective theory to be non-renormalizable~\cite{Weinberg}. This limits the scope of the new theory to a small region around the specified energy scale, but rigorous predictions are still possible in this region~\cite{Thomas}.

The asymptotically free nature of QCD makes understanding low-energy properties of the strong interaction theoretically difficult. At the same time, in the low-energy regime the consequences of asymptotic freedom, $i.e.$ confinement, means that theory no longer describes individual quarks and gluons but instead hadrons comprised of quarks and gluons. This allows for the construction of an effective field theory of QCD, that specifically describes QCD's low-energy properties. The effective theory, Chiral Perturbation\footnote{As with all  effective field theories, $\chi PT$ allows for a perturbative treatment in terms of momentum, instead of the coupling constant.}  Theory ($\chi PT$), respects all the symmetry patterns of QCD but its Lagrangian is constructed  from hadron degrees of freedom. 

\begin{figure}[htp]
\begin{center}
\includegraphics[scale=0.6]{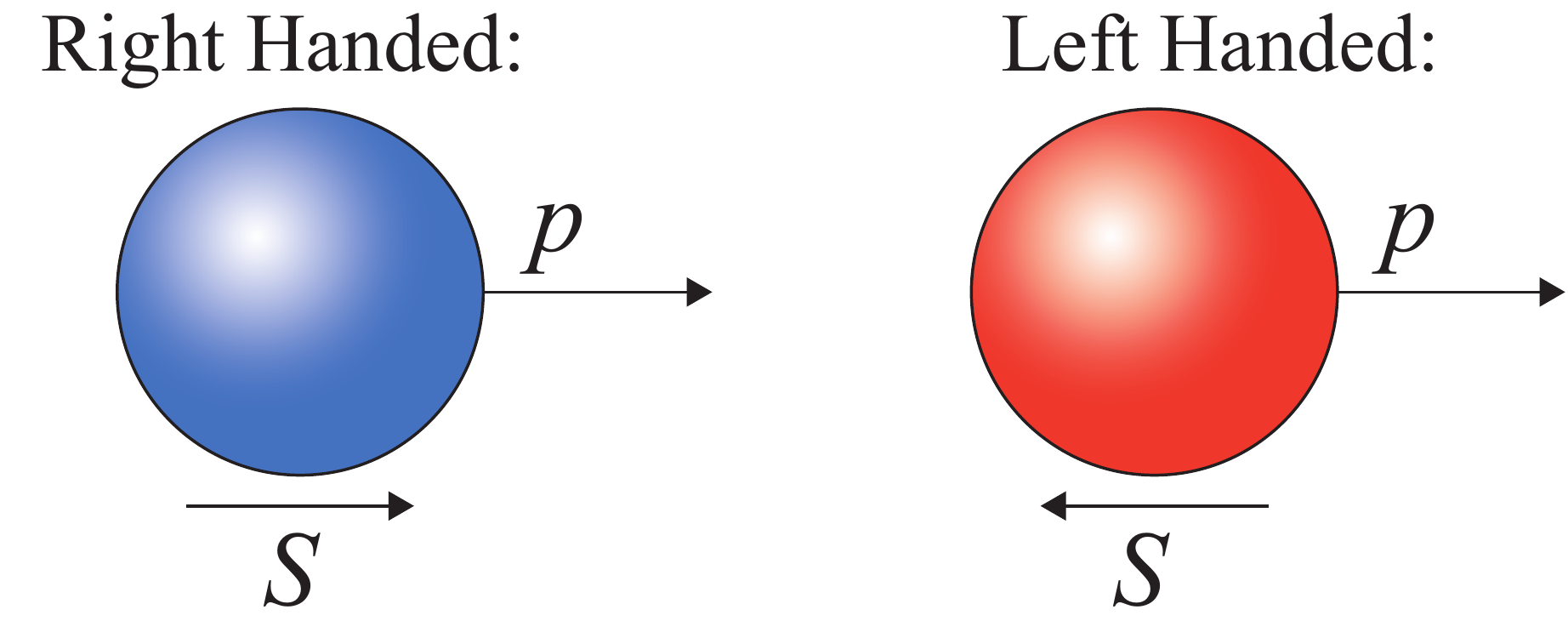}
\caption{\label{Chiral} For massless particles, chirality is the same as the helicity. Right (left) handed chiral quarks have their spin and momentum (anti) aligned.}
\end{center}
\end{figure}

The complete QCD Lagrangian is~\cite{Scherer}
\begin{equation}
\mathcal{L}_{\mathrm{QCD}} = \sum_f\bar{q}_f(i\slashed{D} - m_f)q_f - \frac{1}{4g^2}G_{\mu\nu,a}G_a^{\mu\nu}\,,
\end{equation}
where the sum $f$ is over the six flavors of quarks which correspond to quark fields, $q_f$ of mass $m_f$; the gluon field strength tensor is $G$. The lightest quarks are the up and down quarks with masses less than 10 MeV~\cite{Scherer}. At the baryonic  scale, $e.g.$ $\sim$ 1 GeV, the light quarks are essentially massless and the heavier quarks can be ignored because their masses are above 1 GeV. The quark fields are separated into left and right handed projections  (see Figure~\ref{Chiral}) such that
\begin{align}
P^{L,R}   &= \frac{1}{2}(1 \mp \gamma_5)\,,\\
q^{L,R}_f &= P^{L,R}q_f\,,\\
\bar{q}_f^{L,R} & = \bar{q}_fP^{R,L}\,.
\end{align}
In the chiral limit of massless quarks ($\mathrm{m_f\rightarrow 0}$), the left and right handed quarks do not interact and the Lagrangian becomes
\begin{equation}
\mathcal{L}^0_{\mathrm{QCD}} = \sum_f\bar{q}^R_fi\slashed{D} q^R_f + \bar{q}^L_fi\slashed{D} q^L_f - \frac{1}{4g^2}G_{\mu\nu,a}G_a^{\mu\nu}\,,
\end{equation}
using that $P^R + P^L = 1$ and $P^LP^R = P^RP^L = 0$.
This new Lagrangian has a $SU(2)_L \times SU(2)_R \times U(1)_V$ symmetry~\cite{Srednicki}.

According to (Emmy) Noether's Theorem~\cite{Noether1918} for every symmetry there is a corresponding conversation law.  The $U(1)_V$ symmetry corresponds to conservation of baryon number. The $SU(2)_L \times SU(2)_R$ symmetry is trickier because it says that there should be two of every hadron (one for each parity). A quick glance at the particle spectrum shows that this is not the case. Therefore the $SU(2)_L \times SU(2)_R$ symmetry is said to be spontaneously broken  down to a single $SU(2)_V$ symmetry, which conserves isospin~\cite{Srednicki}. The spontaneously broken symmetry leads to three Goldstone bosons~\cite{Goldstone} and they are the three pions\footnote{This entire argument also holds for the strange quark ($m_s \simeq$ 100 MeV), except that the symmetry is  $SU(3)_L \times SU(3)_R$, and the kaons ($K^{\pm},\bar{K}^0,K^0$) and the eta ($\eta$) are added to the list of Goldstone bosons.}: $\pi^{\pm},\pi^0$.

The $SU(2)_L \times SU(2)_R$ symmetry is explicitly broken as well because the quarks are not exactly massless. This is seen by adding the mass term back into the Lagrangian
\begin{equation}
\mathcal{L}^m_{\mathrm{QCD}} = -\sum_f\bar{q}_fm_fq_f = -\sum_f(\bar{q}_f^Rm_fq_f^L + \bar{q}_f^L m_f q_f^R)\,,
\end{equation}
and observing the mixing of left and right handed quark fields. The interaction between quarks and, ultimately, the Goldstone bosons at low energies is weak, which allows for a perturbative treatment of the explicit chiral symmetry breaking~\cite{Bernard}. This is the basis for Chiral Perturbation Theory; the formulation of the Lagrangian in terms of pion fields is done nicely in Ref~\cite{Srednicki}.

Combining the chiral and symmetry breaking Lagrangians gives the low-energy effective QCD Lagrangian
\begin{equation}
\mathcal{L}^{\mathrm{eff}}_{\mathrm{QCD}} = \mathcal{L}^0_{\mathrm{QCD}} + \mathcal{L}^m_{\mathrm{QCD}}\,,
\end{equation} 
where the symmetry breaking term is perturbative in powers of momenta. The low energy expansion is in terms of pion loops, instead of the quark and gluon loops of the full QCD Lagrangian. It is constructed by ordering all the possible interactions of particles based on the number of momentum and mass powers: lowest order corrections are proportional to $m_{\pi}^2$ and one loop corrections are proportional to $m_{\pi}^4$. All matrix elements and scattering amplitudes derived from the effective Lagrangian are organized in this ``power-counting" manner.

Applying the previously described formalism to the next simplest bound-quark system is a problem because the lightest baryon masses are approximately 1 GeV. They are not strictly pertubative with respect to the chiral length scale of $\Lambda_{\chi}\sim$ 1GeV; there is no guarantee that the power series converges~\cite{ChiPT_1}. Described below are two approaches theorists take to deal with this issue: Relativistic Baryon $\chi$PT~\cite{RelChiPT_1} and Heavy Baryon $\chi$PT~\cite{HeavyChiPT_1,HeavyChiPT_2}. Both rely on the formalism of a $\pi$-nucleon theory described in Ref~\cite{BaryonChiPT}, where divergent QCD effects are contained within a few phenomenological constants determined from experiment~\cite{Chi_Con}.

{\bf\noindent Heavy Baryon $\chi$PT :}
\\
\hangindent=0.6cm \hangafter=1
Heavy Baryon (HB) $\chi$PT treats the baryon as a static, very heavy particle. In this scheme, the power series converges, albiet slowly~\cite{ChiPT_1}, because baryon corrections are suppressed by powers of the baryon mass. One difficulty with HB-$\chi$PT is the large number of phenomenological constants appearing in the effective Lagrangian.

{\bf\noindent Relativistic Baryon $\chi$PT : }
\\
\hangindent=0.6cm \hangafter=1
In relativistic baryon $\chi$PT, large and small momentum effects are separated such that that the former are absorbed into the low-energy constants, while the latter are evaluated~\cite{RelChiPT_1}. Baryons are responsible for the large momentum effects and are included through a separate renormalization term. Small momentum results are obtained from the perturbative chiral expansion.

\noindent The two approaches produce comparable results on the Compton amplitudes for the polarized and unpolarized structure functions in the low momentum region~\cite{Ji,Ji2,Bernard2,Bernard3,Kao}. It should also be noted that resonance contributions are a further complication to the $\pi$-nucleon scattering picture. Ideally these resonances would be included as extra degrees of freedom in the effective Lagrangian, but currently this theory does not exist. Instead the resonances are included systematically through additional low energy constants. The theoretical predictions typically have a band of values due to the uncertainties in these parameters.

\section{Lattice QCD}
Lattice QCD is a non-perturbative approach to evaluating the QCD Lagrangian where calculations are made on a discrete grid of space-time points~\cite{Thomas,PDG}. In contrast to both $\chi$PT and the OPE, lattice QCD calculations are not an approximation; in the limit of small lattice spacing, $a$ and large volume lattice QCD returns the continuum QCD Lagrangian. In principle, lattice QCD should then be able to compute any observable governed by QCD ($e.g.$ the structure functions ) at any energy scale and return the exact result. In practice, the calculations are computationally costly and verified analytic solutions for the structure functions do not currently exist; however lattice QCD is a rapidly developing field and progress is made continually.

\section{Phenomological Models}
\label{PhenoM}
There are several different empirical fits to the existing world data that can make predictions at the kinematics of E08-027. These empirical models are divided into two broad categories, one corresponding to polarized structure functions and the other to unpolarized structure functions.  At the core of each model is a relativistic Breit-Wigner (named after Gregory Breit and Eugene Wigner) distribution fit to the nucleon resonances~\cite{BreitWigner}. The probability of producing a resonance at energy $W$ is given as 
\begin{equation}
f(W) = \frac{k}{(W^2 -M^2)^2+M^2\Gamma^2}\,,
\end{equation}
where $M$ is the mass of the resonant state, $W$ is the center of mass energy, $\Gamma$ is the width of the resonance, and $k$ is a constant of proportionality equal to
\begin{align}
k &= \frac{2\sqrt{2}M\Gamma\gamma}{\pi\sqrt{M^2+\gamma}}\,,\\\,
\gamma &=\sqrt{M^2(M^2+\Gamma^2)}\,.
\end{align}

The resonance masses and widths are cataloged in Ref~\cite{PDG}. Additional fit parameters account for additional physics inherent in the data, such as quasi-elastic scattering, deep inelastic contributions and the so-called dip-region around the pion-production threshold.

\subsection{Polarized Model: MAID 2007}
The unitary isobar model MAID~\cite{MAID2007} is a fit to the world data of pion photo- and electroproduction from the single pion-production threshold to the onset of DIS at $W$ = 2 GeV. The model includes Breit-Wigner forms for 13 resonance channels\footnote{All the 4-star resonances below $W$ = 2 GeV from Ref~\cite{PDG} are included.}. The contribution from the production of non-resonant background is also included. For the purpose of this thesis MAID is considered a polarized model but it also can reproduce unpolarized cross sections. Its predictions are in terms of the virtual photon-proton cross sections for four production channels: $\pi^0p$, $\pi^0n$, $\pi^+n$, $\pi^-p$.

\subsection{Polarized Model: CLAS EG1B}
\label{CLASHALLBMOD}
The CLAS EG1B model~\cite{HallB} (also referred to as the Hall B model) is a fit to the virtual photon asymmetries $A_1$ and $A_2$, which are related to the spin structure functions via
\begin{align}
g_1(x,Q^2) &= \frac{F_1(x,Q^2)}{1+\gamma^2}[A_1(x,Q^2) + \gamma A_2(x,Q^2)]\,,\\
g_2(x,Q^2) &= \frac{F_1(x,Q^2)}{1+\gamma^2}[A_2(x,Q^2)/\gamma -A_1(x,Q^2)]\,,
\end{align}
for $\gamma^2 = 4x^2M^2/Q^2$. Photo-production data constrain the fit as $Q^2 \rightarrow 0$ and the model parameterizes data in both the resonance and DIS regimes. For DIS predictions, the Wandzura-Wilczek relation is applied to existing $g_1(x,Q^2)$ data for $g_2(x,Q^2)$ contributions and the Burkhardt-Cottingham Sum rule further constrains the mostly unmeasured $g_2(x,Q^2)$ component.
\subsection{Unpolarized Model: Bosted-Christy-Mamyan}
\label{BostedModSec}
The Bosted-Christy-Mamyan model is an empirical fit to inclusive inelastic e-p~\cite{Bosted3}, e-d and e-n,~\cite{Bosted1} and e-N~\cite{Bosted2} scattering. The output is the unpolarized structure functions $F_1(x,Q^2)$ and $F_2(x,Q^2)$. The kinematic coverage of the three separate fits is shown in Table~\ref{BostedKin} and is largely driven by the coverage of the available data. To that end, the quality of the e-N fit is largely nucleus dependent and is directly correlated with the amount of data available to fit.

\begin{table}[htp]
\centering
\begin{tabular}{ l  c  r } \hline
   & $Q^2$ (GeV) & W (GeV) \\ \hline
  $e-p$ & 0.0 - 8.0 & 1.1 - 3.1 \\
  $e-d/n$ & 0.0 - 10.0 & 1.1 - 3.2 \\
  $e-N$ & 0.2 - 5.0 & 0.0 - 3.2 \\
  \hline  
\end{tabular}
\caption{\label{BostedKin} Kinematic applicability of the Bosted-Christy-Mamyan fit.} 
\end{table}

For $A >$ 1 the bulk of the fit is the free nucleon Breit-Wigner result with the addition of a Plane Wave Impulse Approximation (PWIA) analysis to account for the Fermi motion of the nucleons. The additional quasi-elastic peak is built up from the free nucleon form factors with the Fermi and binding energy of the nucleons from Ref~\cite{Superscale}.  The dip region contribution at the pion-production threshold is a purely empirical fit. 

\subsection{Unpolarized Model: Quasi-Free Scattering}
The Quasi-Free Scattering (QFS) model~\cite{QFS} parameterizes electron nucleus scattering using five reaction channels for electron beam energies between 0.5-5.0 GeV:
\begin{itemize}
\item quasi-elastic scattering
\item two nucleon processes in the dip region
\item $\Delta$(1232) resonance production
\item $N^*$(1500/1700) resonance production
\item deep inelastic scattering
\end{itemize}
The quasi-elastic peak is a Gaussian fit where the user controls the width by adjusting the Fermi momentum. The user can also adjust the location of the quasi-elastic peak using the nucleon separation energy parameter. There is also a similar term for the $\Delta$(1232) resonance. The accumulation of newer data since the QFS model was introduced means that its accuracy is surpassed by the newer Bosted-Christy-Mamyan fit. The analysis contained within this thesis does not use the QFS model, and it is instead presented for completeness.




\chapter{\sc Motivating the Measurement}
\label{ch:ExistingData}

While there has been considerable advancement in the knowledge of the spin structure functions over the past 40 years, there is still a largely unmeasured low $Q^2$ region for $g_2(x,Q^2)$ of the proton. This is in part due to the technical difficulty in operating a transversely polarized proton target for forward angle, low $Q^2$ scattering. The low momentum transfer region, 0.02  GeV$^2$ $<$ $Q^2$ $<$ 0.20 GeV$^2$, accessed by E08-027 is useful in testing calculations of Chiral Perturbation Theory and the Burkhardt-Cottingham Sum Rule.  At low $Q^2$, the scattering interaction is sensitive to finite size effects of the proton and the collective response of its internal structure to the electromagnetic probe. In that regard, the low $Q^2$ data will help improve accuracy of the theoretical calculations of the hydrogen hyperfine splitting.

Current measurements on a longitudinally polarized proton target exist all the way down to a momentum transfer of $Q^2$ = 0.05 GeV$^2$ in the resonance region. The E08-027 data\footnote{E08-027 took data at one kinematic setting on a longitudinally polarized target. The details of the collected data are discussed in Chapter~\ref{ch:Experiment}.} will be a useful cross-check on these existing measurements and provide a calibration point for the moment extraction procedure. For example, the the first moment of $g_1(x,Q^2)$ can be checked for agreement with the GDH slope, which also has implications for the hydrogen hyperfine splitting calculations. Additionally, the data will be an independent test on the accuracy of the CLAS EG1b and MAID models discussed in Chapter~\ref{PhenoM}.	

\section{Existing $g_2(x,Q^2)$ Measurements}
The first proton $g_2(x,Q^2)$ measurements were in the DIS region and were performed at the Stanford Linear Accelerator~\cite{E143,E155} (SLAC) and by the Spin Muon Collaboration~\cite{SMC} (SMC) group at CERN in the early 1990s\footnote{The SMC (SLAC) experiments used deep inelastic muon (electron) scattering.}. These experiments measured the virtual photon asymmetries $A_1(x,Q^2)$ and $A_2(x,Q^2)$ for a transversely and longitudinally polarized proton target (see Chapter~\ref{VPPXS}). The experiments relied on existing data for $F_1(x,Q^2)$ to extract $g_2(x,Q^2)$. The most precise DIS measurement of proton $g_2(x,Q^2)$ was carried out by SLAC E155x~\cite{E155x}, and their results show good agreement with the leading twist prediction based upon measured $g_1(x,Q^2)$ data. The SMC experiments also found consistency with leading twist effects. SLAC produced some limited resonance region data with E143~\cite{E143} but the statistical error bars are not sufficient to notice any deviation from leading twist behavior.
 \begin{figure}[htp]
\begin{center}
\includegraphics[scale=.60]{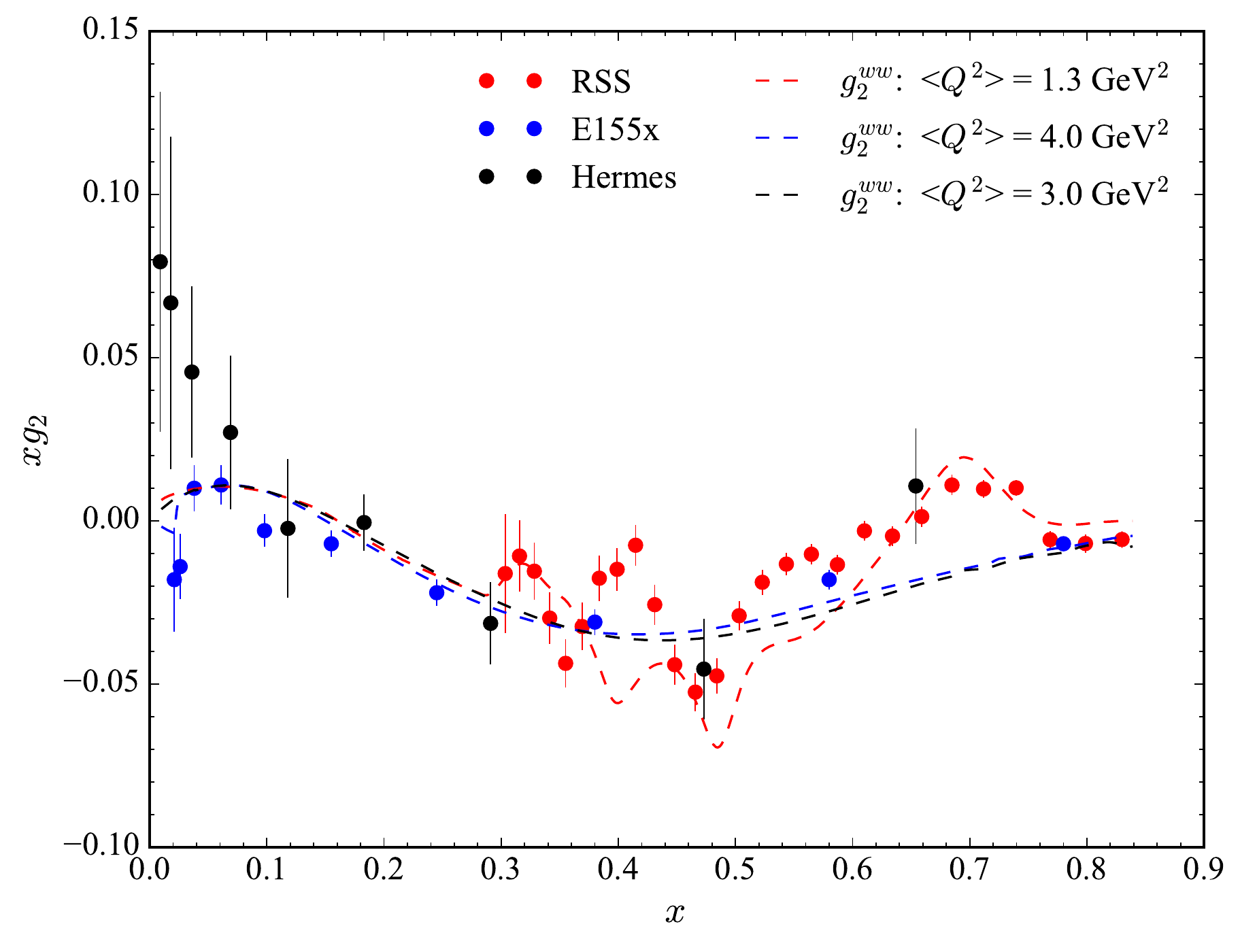}
\includegraphics[scale=.60]{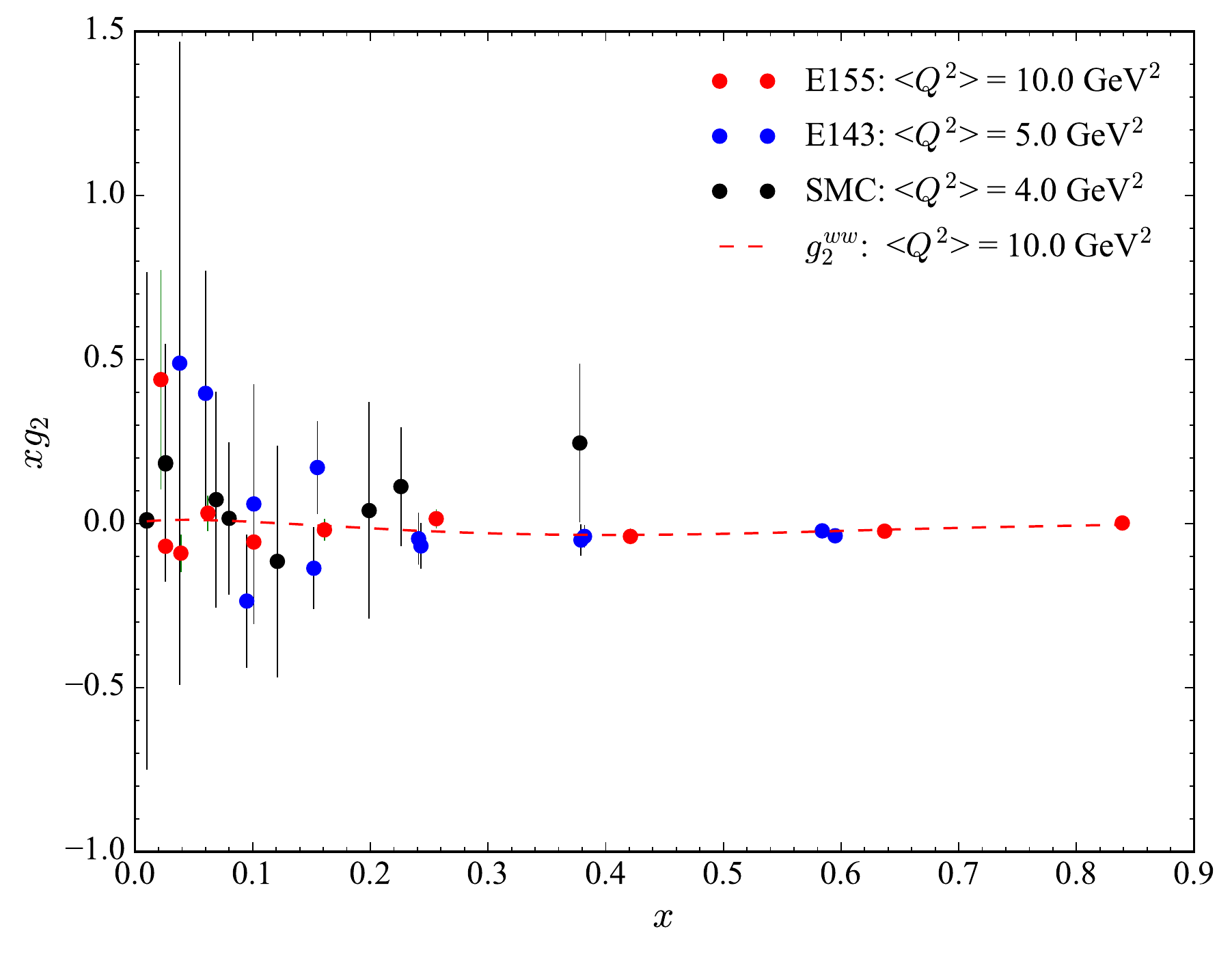}
\caption{\label{g2data}The existing $g_2^p(x,Q^2)$ data and leading twist prediction from the EG1b model. The data is roughly separated into two figures based upon $Q^2$, in order to highlight the applicability of the $g_2^{ww}(x,Q^2)$ prediction.}  
\end{center}
\end{figure}

\begin{table}[htp]
\centering
\begin{tabular}{ l  c  r } \hline
   Exp. & $Q^2$ (GeV) & $x$ range \\ \hline
  SMC & 1.0 - 30.0 & 0.003 - 0.7 \\
  HERMES & 0.2 - 20.0 & 0.004 - 0.9 \\
  E155x & 0.7 - 20.0 & 0.02 - 0.8 \\
  E155  & 1.0 - 30.0 & 0.02 - 0.8 \\
  E143 & 1.3 - 10.0 & 0.03 - 0.8 \\

  RSS &  1.3 & 0.03 - 0.8 \\
  SANE$^{\dagger}$ &  2.5 - 6.5 & 0.3 - 0.8 \\
  \hline  
\end{tabular}
\caption{\label{g2Kin} Kinematics of proton $g_2(x,Q^2)$ measurements. $^{\dagger}$Data still under analysis.} 
\end{table}
More recent measurements of proton $g_2(x,Q^2)$ via polarized electron scattering include the Resonance Spin Structure (RSS)~\cite{RSS} experiment at Jefferson Laboratory's Hall C and an experiment by the HERMES~\cite{Hermes} collaboration at DESY. RSS made its measurement at $Q^2$ = 1.3 GeV$^2$ and it is currently the lowest $Q^2$ measurement of $g_2^p(x,Q^2)$. They see clear deviations from the leading twist behavior in their data, which indicates a stronger quark-gluon interaction in the resonance region. The HERMES data includes polarized positron scattering, in combination with the electron data, and primarily focused on the DIS region. Their data lacks the statistics to  detect a deviation of $g_2(x,Q^2)$ from the Wandzura-Wilczek relation. The existing data with leading twist predictions provided by the CLAS EG1b model is shown in Figure~\ref{g2data}, and the kinematic footprint of these experiments is shown in Table~\ref{g2Kin}. For the DIS experiments, the leading twist prediction represents an average $Q^2$ of the experiments. The similarity between the DIS curves is a direct result of Bjorken scaling. The Spin Asymmetries of the Nucleon~\cite{SANE} (SANE) experiment took data at Jefferson Laboratory's Hall C in both DIS and the resonance regions. The data is currently in the analysis phase.

\section{Existing $g_1(x,Q^2)$ Measurements}
\begin{figure}[htp]
\begin{center}
\includegraphics[scale=.80]{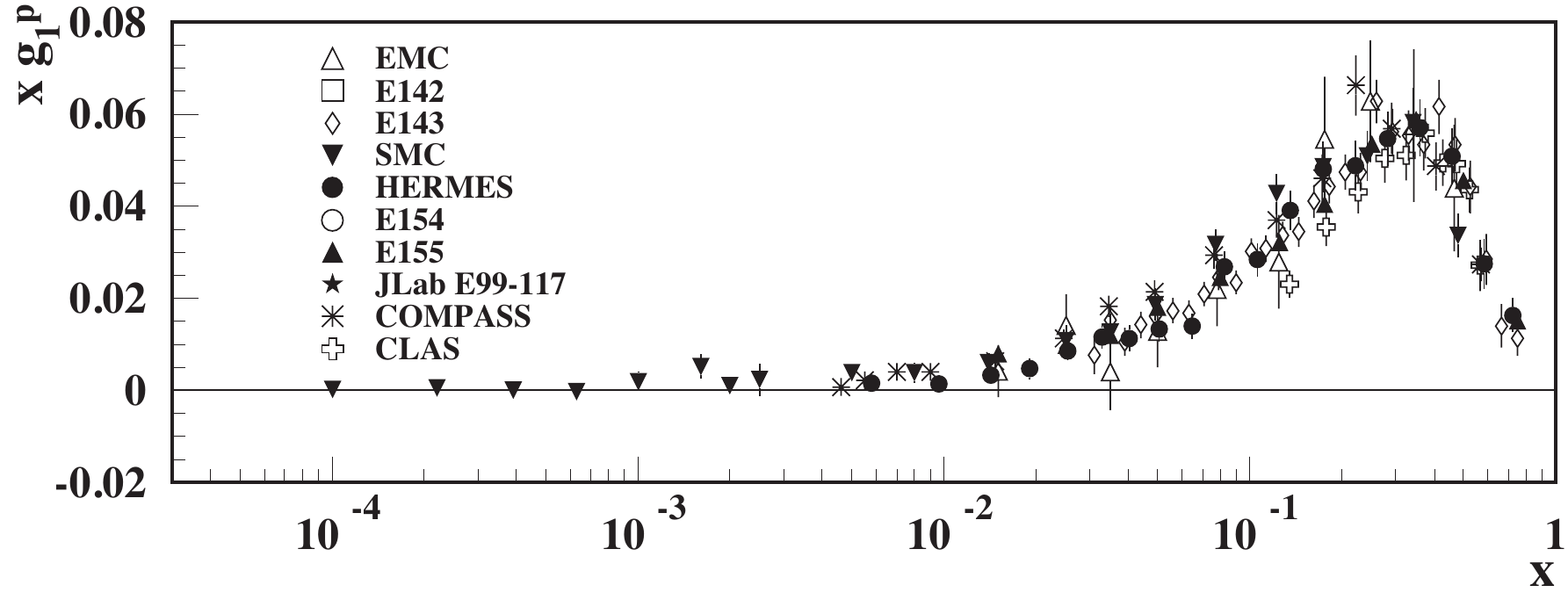}
\caption{\label{g1data}Existing $g_1^p(x,Q^2)$ DIS data. Figure reproduced from Ref~\cite{PDG}.}  
\end{center}
\end{figure}

The first proton $A_1(x,Q^2)$ and $g_1(x,Q^2)$ measurement was made by the European Muon Collaboration (EMC) in the late 1980s~\cite{Ashman2}. The results of their measurement of polarized muon scattering in the deep inelastic region disagreed significantly with the Ellis-Jaffe sum rule, discussed in Chapter~\ref{g1ww}, and showed that the total quark spin is a small fraction of the proton's spin. This result created the $spin$ $crisis$. Many measurements followed, including additional polarized muon scattering data from SMC~\cite{SMC2,SMC3} and COMPASS~\cite{COMPASS}, and polarized electron data from HERMES~\cite{Hermesg1}, SLAC~\cite{E143_2,E155g1} and Jefferson Lab~\cite{CLASg1}. The results of these  DIS measurements are shown in Figure~\ref{g1data}.

Less is known about  $A_1(x,Q^2)$ and $g_1(x,Q^2)$ in the resonance region, with the measured data coming from two experimental collaborations: CLAS EG1~\cite{CLASg1,EG1b} (also referred to as EG1a and EG1b) and RSS~\cite{RSS}. This data in the low momentum transfer  and resonance regions ($Q^2$ $<$ 2 GeV$^2$) is a good test of Chiral Perturbation Theory and the current phenomenological models that provide a bulk of the description of the physics in this kinematic region. The CLAS data covers a momentum transfer range of approximately 0.05 $< Q^2 <$ 5.0 GeV$^2$, taken at 27 individual $Q^2$ points. The RSS data mirrors the $g_2(x,Q^2)$ point and is at $Q^2$ = 1.3 GeV$^2$. Additional, lower momentum transfer data, covering 0.01 $< Q^2 <$ 0.5 GeV$^2$ from CLAS EG4~\cite{EG4} is currently under analysis. A more detailed summary of the quality of the published $g_1(x,Q^2)$ data is found in Ref~\cite{g1global,g1JP}.

\section{Spin Polarizabilities and $\chi$PT}
As discussed in Chapter~\ref{Section:Moments}, the spin polarizabilities are observables that describe the interaction between a nucleon's spin and external magnetic and electric fields. Determining these polarizabilities over a wide range of momentum transfer maps out the transition from the hadronic to partonic interactions of the nucleon. The generalized virtual-virtual Compton scattering\footnote{The incoming and outgoing photons are virtual, as apposed to real Compton scattering (RCS) and virtual Compton scattering (VCS). In VCS only the incoming photon is virtual.} (VVCS) polarizabilites, $\gamma_0(Q^2)$ and $\delta_{LT}(Q^2)$, are particularly interesting because the extra $x^2$ weighting causes the integrals to converge faster, with the bulk of the integral strength coming from the resonance region. At low $Q^2$, the experimental results can be compared directly to $\chi$PT calculations. The longitudinal-transverse spin polarizability, $\delta_{LT}(Q^2)$, is an ideal test of $\chi$PT because it is less sensitive to the $\Delta$-resonance\footnote{From Chapter~\ref{sec:ChiPT}, the nucleon resonances are included in the calculations as low energy constants because their energies are above the chiral length scale.}; the spin structure functions are approximately equal and opposite at this kinematic point. The additional $x^2$ weighting for $g_2(x,Q^2)$ in the forward spin polarizability, $\gamma_0 (Q^2)$, suppresses the $g_2(x,Q^2)$ contribution at typical Jefferson Lab kinematics, but $\gamma_0(Q^2)$ is more sensitive to the $\Delta$-resonance parameterization~\cite{Kao}.

\begin{figure}[htp]
\begin{center}
\includegraphics[scale=.80]{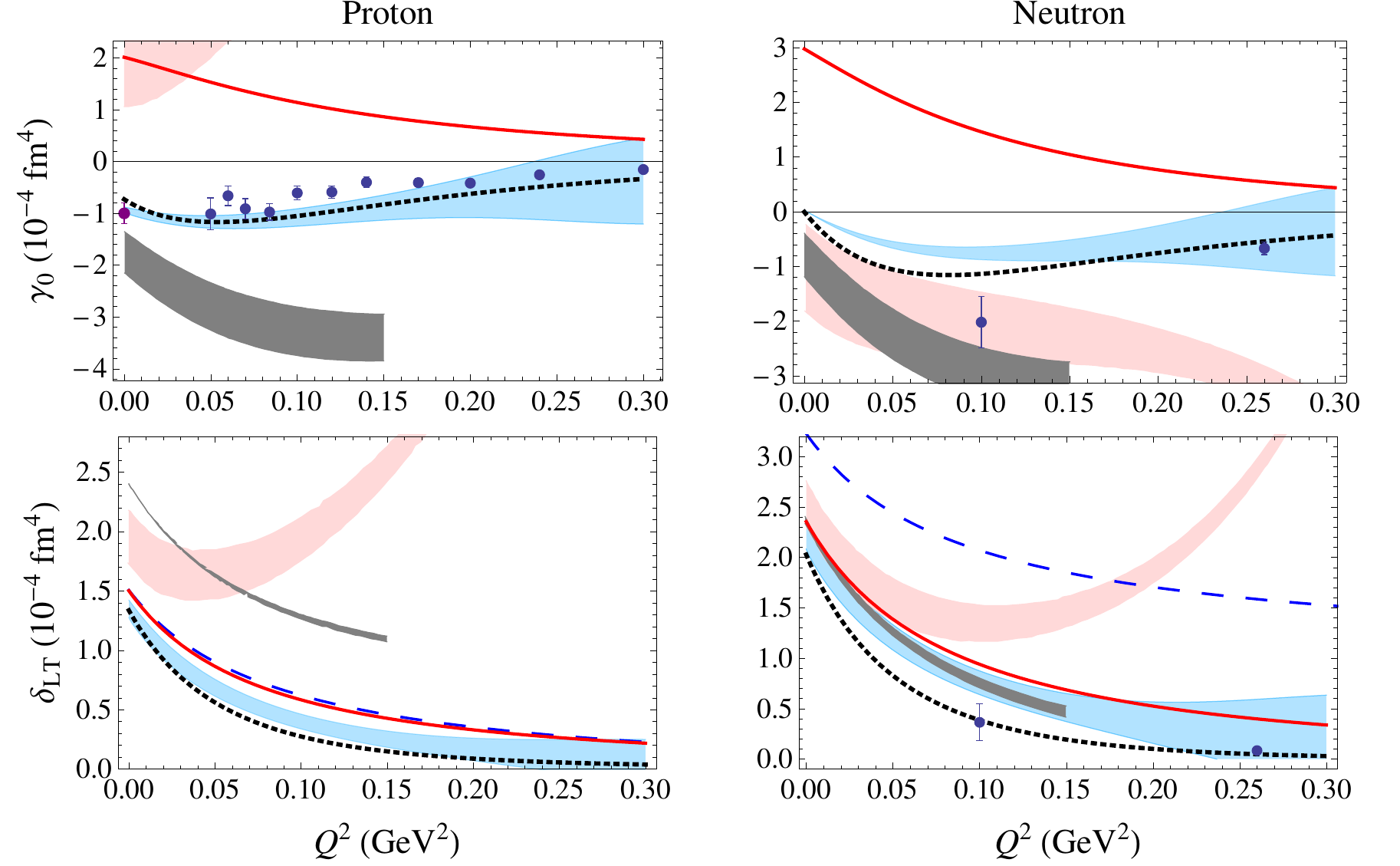}
\caption{\label{dlt_xpt}$\chi$-PT predictions of the spin polarizabilities. Reproduced from Ref~\cite{Gold}. Neutron data is from Ref~\cite{E94010} and the proton data is from Ref~\cite{EG1b} (blue points) and Ref~\cite{ELSA} (purple point at $Q^2 =0$). No measurement exists for $\delta_{LT}(Q^2)$ because there is currently no $g_2(x,Q^2)$ data at these kinematics.}  
\end{center}
\end{figure}

The current state of the polarizability measurements and corresponding $\chi$PT calculations is shown in Figure~\ref{dlt_xpt}. The experimental neutron results are from Jefferson Lab Hall A experiment E94-010~\cite{E94010}. The proton data is a combination of Ref~\cite{EG1b} (blue points) and Ref~\cite{ELSA} (purple point at $Q^2 =0$). The black dotted line is the MAID2007 prediction. Heavy Baryon $\chi$PT calculations~\cite{Kao} are the blue dashed lines, and are off scale for the forward spin polarizability. Both it and the Relativistic Baryon $\chi$PT calculation~\cite{Bernard2} (red band) show a large discrepancy between the theoretical predictions and measured data. This disagreement initiated further calculations to try and resolve the difference. The most recent calculations in RB$\chi$PT from Ref~\cite{Krebs} (grey band) and Ref~\cite{Gold} (leading order: red line and next-to-leading order: blue band) show much better agreement with the data. But there is still no proton $\delta_{LT}(Q^2)$, so the comparison is not complete. The $g_2(x,Q^2)$ structure function is not suppressed by kinematics in $\delta_{LT}(Q^2)$ and its contribution to the integral must be measured. 

\section{The Burkhardt Cottingham Sum Rule}
\begin{figure}[htp]
\begin{center}
\includegraphics[scale=.60]{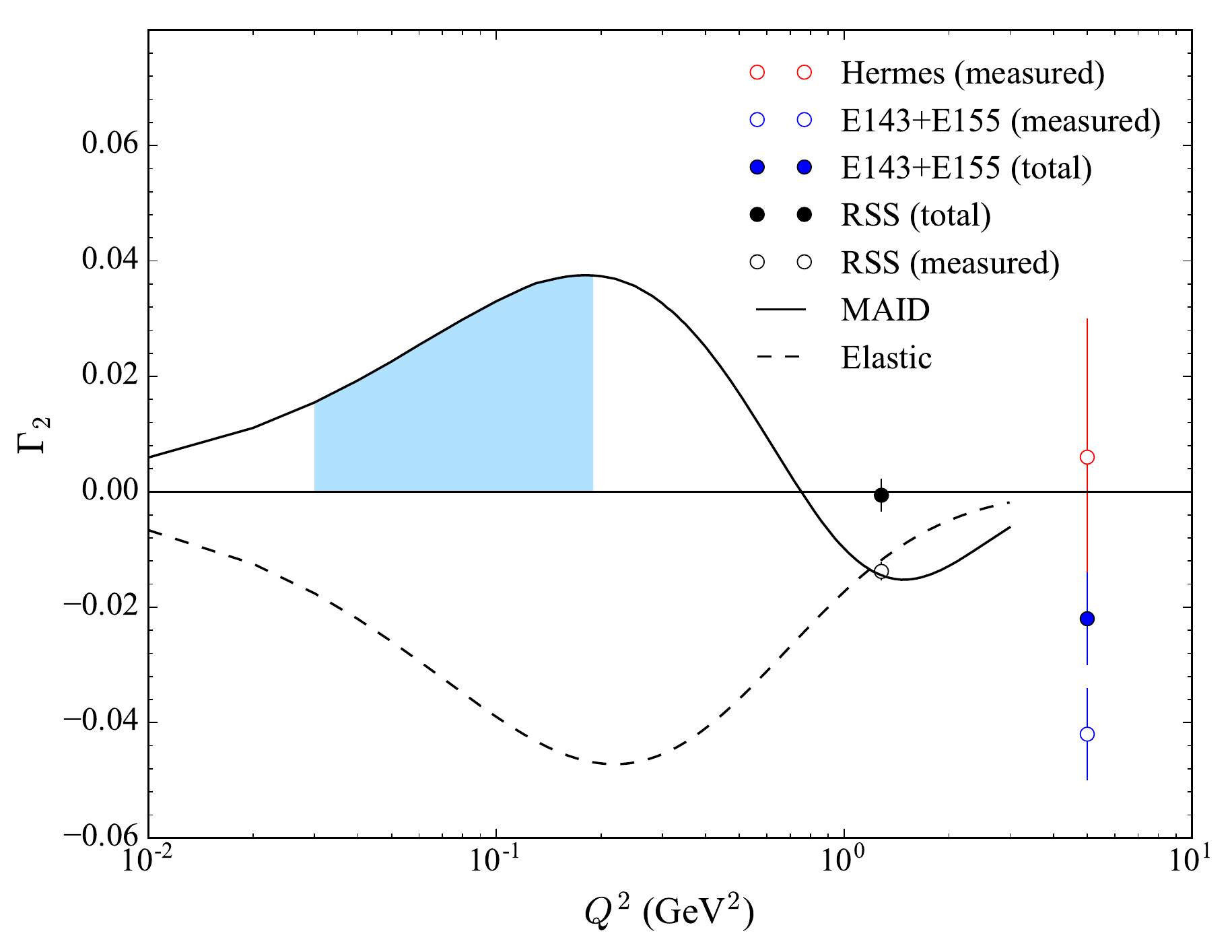}
\caption{\label{BCdata}Existing $\Gamma_2^p(Q^2)$ data and expected E08-027 measured contribution. The measured data points only include the integrated kinematic region accessible to the experiment; the total data points include contributions from the elastic and other unmeasured regions. }  
\end{center}
\end{figure}

The Burkhardt Cottingham Sum Rule integration covers all possible $x$ for a proton:
\begin{equation}
\Gamma_2(Q^2) = \int_0^1 g_2(x,Q^2) \mathrm{d}x = 0\,.
\end{equation} 
The full $x$ range is typically not accessible in a single experiment. Jefferson Laboratory experiments measure the resonance contribution to the sum rule. They rely on existing elastic form factor measurements for the $x=1$ component and assume leading twist behavior in the low-$x$ region. The elastic contribution to the sum rule is~\cite{BCElastic}
\begin{equation}
\Gamma_2^{\mathrm{el}} (Q^2) = \frac{\tau}{2}G_M(Q^2)\frac{G_E(Q^2)-G_M(Q^2)}{1+\tau}\,,
\end{equation}
where $G_E(Q^2)$ and $G_M(Q^2)$ are the electric and magnetic form factors as determined from elastic scattering experiments\footnote{This is derived from $g_2^{\mathrm{el}} = \frac{\tau}{2}G_M(Q^2)\frac{G_E(Q^2)-G_M(Q^2)}{1+\tau}\delta(x-1)$.}. The full integral exhibits a significant cancellation of the inelastic (resonance and DIS) and elastic contributions as shown in Figure~\ref{BCdata}. The elastic form factor parameterization is from Ref~\cite{ArringtonFit}.

The current world data for $\Gamma_2(Q^2)$ is also shown in Figure~\ref{BCdata}. The open circles are the measured contributions, while the filled circles represent the total integral. The SLAC~\cite{E155x,E143} data are at $Q^2$ = 5 GeV$^2$ and hint at a possible deviation from the predicted null result. It should also be noted that this discrepancy is only slightly over 2$\sigma$ from the expected result. A more recent measurement at HERMES~\cite{Hermes} at the same $Q^2$ as the SLAC experiments found agreement with the sum rule but also with significantly larger error bars. The only other published result is from RSS~\cite{KarlBC} at $Q^2$ = 1.3 GeV$^2$, which agrees with the sum rule. At best the current measurements are inconclusive, highlighting the need for further data. The resonance region coverage of E08-027 is highlighted in blue in Figure~\ref{BCdata}.

\section{The First Moment of $g_1(x,Q^2)$}
Similar to the B.C. Sum Rule, the first moment of $g_1(x,Q^2)$,
\begin{equation}
\label{gamma1eq}
\Gamma_1(Q^2) = \int_0^{x_{\mathrm{th}}} g_1(x,Q^2)dx\,,
\end{equation}
covers a large $x$-range that is not covered entirely by a single experiment. In the literature, the elastic contribution at $x=1$ is typically ignored and the first moment focuses solely on inelastic scattering contributions. The upper limit, $x_{th}$, is the pion production threshold. The Jefferson Lab experiments use a parameterization of the large amounts of world data in the DIS region to provide the low-$x$ extrapolation. For completeness, the elastic contribution to the moment is~\cite{BCElastic}
\begin{equation}
\Gamma_1^{\mathrm{el}} (Q^2) = \frac{\tau}{2}G_M(Q^2)\frac{G_E(Q^2)+\tau G_M(Q^2)}{1+\tau}\,,
\end{equation}
where $G_E(Q^2)$ and $G_M(Q^2)$ are the electric and magnetic form factors as determined from elastic scattering experiments.
 \begin{figure}[htp]
\begin{center}
\includegraphics[scale=0.43]{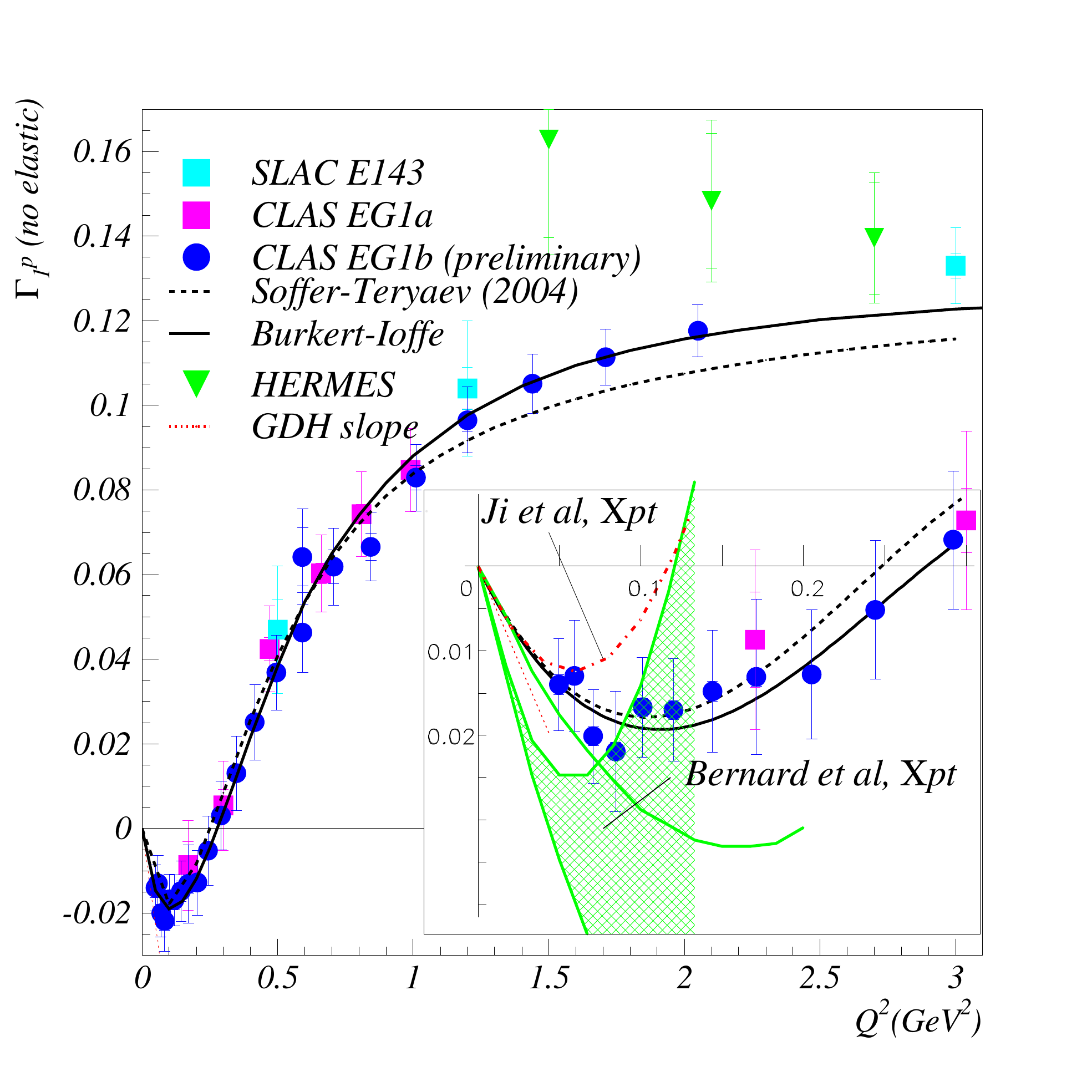}
\caption{\label{Gamma1JP}Existing $\Gamma_1^p(Q^2)$ data at low to moderate $Q^2$ compared with several theoretical predictions. Figure Reproduced from Ref~\cite{JPGamma1Plot}.}  
\end{center}
\end{figure}

The current world data on $\Gamma_1(Q^2)$ is shown in Figure~\ref{Gamma1JP}. The solid and dashed lines are phenomenological model predictions by Volker Burkert and Borris Ioffe~\cite{Burkert}, and Jacques Soffer and Oleg Teryaev~\cite{Soffer}, respectively. The low momentum transfer results show a change in sign of the slope of the moment, which is consistent with the predicted negative slope of the GDH sum rule. The two $\chi$PT calculations of Veronique Bernard $et\, al.$~\cite{Bernard2} and Xiangdong Ji $et\, al.$~\cite{Ji2} agree reasonably well with the measured data, but some large discrepancies begin to appear at the low $Q^2$ end of the published data. The constraint $\Gamma_1(0) = 0$ makes the first moment of $g_1(x,Q^2)$ a less stringent test of $\chi$PT as compared to the polarizabilities but comparisons between data and theory are still useful. The E08-027 longitudinal data point will be a high precision measurement at momentum transfer just below the published data and will help to distinguish between the $\chi$PT calculations, and should also further approach the predicted GDH slope.

\section{Hydrogen Hyperfine Splitting}
\label{sec:Hyper}
The energy scales of electron orbitals in atomic physics and excited nuclei in scattering experiments differ by many orders of magnitude ( a few eV to MeV and above) and, for most purposes, permit a separation of experimental and theoretical techniques to deal with each phenomena. But what happens if the experimental or theoretical precision reaches a level where there is a discernible link between the two? Data and calculations this precise would inform on the coupling between the atomic and nuclear regimes, and the effect of the finite size of the nuclei on atomic energy levels.

The interaction of the proton's magnetic dipole moment with the electron's magnetic dipole moment results in a shift of the energy levels of hydrogen based on the total angular momentum of the atom.  Experimental measurements of the hyperfine splitting\footnote{The shift is on the order of $\frac{m_e}{m_p} \simeq$ 2000 smaller than the fine structure correction.} in the hydrogen ground state boast accuracies at the 10$^{-13}$ MHz level, but theoretical calculations of the same splitting are only  accurate to $\sim$ 10$^{-6}$ MHz. The current experimental value is~\cite{hyper_data}
\begin{equation}
E_{\mathrm{hfs}} = 1420.405\,751\, 76\, 7(9)\,.
\end{equation}

Proton structure corrections,  $\Delta_p^{\mathrm{structure}}$, are the main theoretical uncertainty~\cite{Hyperfine}. The theoretical prediction for the hydrogen hyperfine splitting is written as
\begin{align}
\Delta E_{\mathrm{hfs}} &= (1 + \Delta^{\mathrm{R}} + \Delta^{\mathrm{small}} + \Delta^{\mathrm{QED}} + \Delta_p^{\mathrm{structure}} )E_F^p\,,\\
E_F^p &= 1418.840 \,\mathrm{MHz}\,,
\end{align}
where E$_F^p$~\cite{Fermi} is the magnetic-dipole interaction energy, and $ \Delta^{\mathrm{small}} $ includes corrections due to the muonic and hadronic vacuum polarizations and the weak interaction.  
The contributions due to recoil effects and radiative quantum electrodynamics,  represented by  $\Delta^{\mathrm{R}}$ and  $\Delta^{\mathrm{QED}}$ respectively, are known to high accuracy. The current state of the calculation values and uncertainties is shown in Table~\ref{hypercontrib}.

\begin{table}[htp]
\centering
\begin{tabular}{ l  r r } \hline
   Quantity &  Value (ppm) & $\delta$ (ppm)\\ \hline
  $\Delta^{\mathrm{QED}}$ & 1139.19 & 0.001 \\
  $\Delta^{\mathrm{R}}$ & 5.85 & 0.07\\
  $\Delta^{\mathrm{small}}$ & 0.14 & 0.02 \\
  $\Delta_p^{\mathrm{structure}}$ & $-$39.55 & 0.78 \\
  \hline  
\end{tabular}
\caption{\label{hypercontrib} Contributions and uncertainties in the calculation of $\Delta E_{\mathrm{hfs}}$ ~\cite{Hyperfine3}.} 
\end{table}

Experimentally measured ground state and excited state properties of the proton are needed to fully characterize $\Delta_p^{\mathrm{structure}}$. Elastic scattering determines the ground state properties, while the resonance structure of inelastic scattering is useful for the excited state properties.  
The structure dependent correction is usually split into two terms,
\begin{equation}
\Delta_p^{\mathrm{structure}} = \Delta^{Z} + \Delta^{\mathrm{pol}}\,,
\end{equation}
where $\Delta^{Z}$ is the ground state term, first calculated by Charles Zemach~\cite{Zemach}. The contribution of each term to the structure uncertainty is listed in Table~\ref{hypercontrib_spin}; the spin components are discussed in the remainder of the chapter.

\begin{table}[htp]
\centering
\begin{tabular}{ l  r  r } \hline
   Quantity &  Value (ppm) & $\delta$ (ppm)\\ \hline
  $\Delta_{\mathrm{Z}}$ & $-$41.33 & 0.44 \\
  $\Delta_{\mathrm{pol}}$ & 1.88 & 0.64\\
  $\Delta_1$ & 2.00 & 0.62 \\
  $\Delta_2$ & $-$0.12 & 0.12 \\
  \hline  
\end{tabular}
\caption{\label{hypercontrib_spin} The proton structure contributions and their uncertainties in the calculation of the hydrogen hyperfine splitting ~\cite{Hyperfine3,Hyperfine2}.} 

\end{table}
 The excited state term, $\Delta^{\mathrm{pol}}$, is further split into two additional terms,
\begin{equation}
\Delta^{\mathrm{pol}} = \frac{\alpha m_e}{\pi g_p m_p}(\Delta_1 + \Delta_2)\,,
\end{equation}
where $\Delta_1$ involves the elastic form factor $F_2(Q^2)$ and $g_1(x,Q^2)$,
\begin{align}
\label{D1int}
\Delta_1 &= \frac{9}{4} \int_0^{\infty} \frac{\mathrm{d}Q^2}{Q^2}{\bigg[}{\bigg(}\frac{G_M(Q^2) + G_E^2(Q^2)}{1+\tau}{\bigg)}^2 + \frac{8m_p^2}{Q^2}B_1(Q^2){\bigg]}\,,\\
\label{B1int}
B_1(Q^2) &= \int_0^{x_\mathrm{{th}}}\mathrm{d}x \beta_1(\tau)g_1(x,Q^2)\,,\\
\beta_1(\tau) &= \frac{4}{9}{\bigg(}-3\tau + 2\tau^2 + 2(2-\tau) \sqrt{\tau(\tau+1)}{\bigg)}\,,
\end{align} 
and $\tau = \nu^2/Q^2$, $x_{\mathrm{th}}$ is the pion production threshold, $m_e$ and $m_p$ are the mass of the electron and proton, respectively and $g_p$ is the Lande g-factor. The $\Delta_2$ term is only dependent on $g_2(x,Q^2)$,
\begin{align}
\Delta_2 &= -24m_p^2 \int_0^{\infty} \frac{\mathrm{d}Q^2}{Q^4}B_2(Q^2)\,,\\
B_2(Q^2) &= \int_0^{x_\mathrm{{th}}}\mathrm{d}x \beta_2(\tau)g_2(x,Q^2)\,,\\
\beta_2(\tau) &= 1 + 2\tau - 2 \sqrt{\tau(\tau+1)}\,.
\end{align}

\begin{figure}[htp]
\begin{center}
\includegraphics[scale=.60]{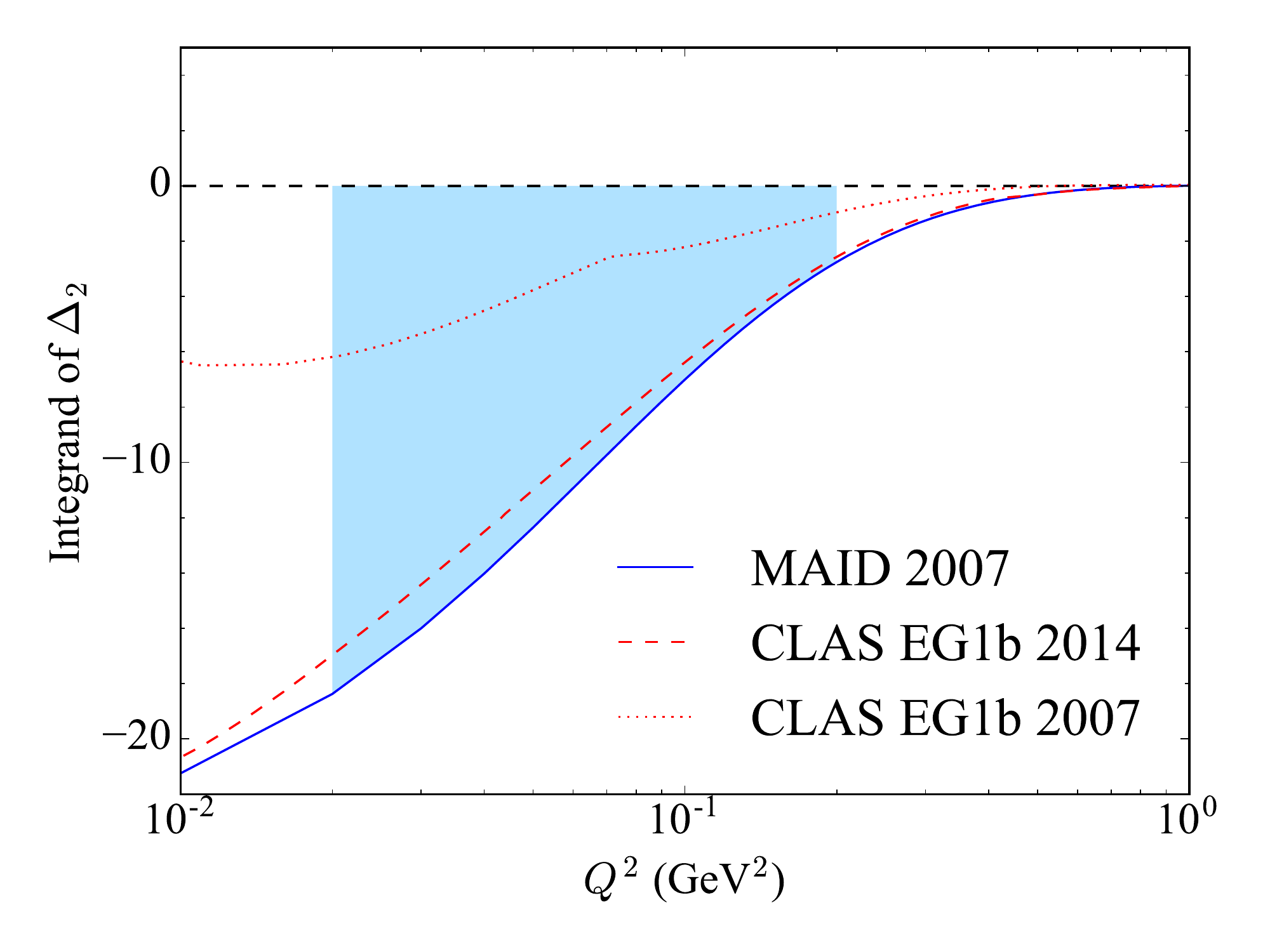}
\caption{\label{B2int}The quantity $B_2(Q^2)$ evaluated using phenomenological models, highlighting the current model dependence in determining the $g_2(x,Q^2)$ contribution to the hyperfine splitting of hydrogen.}  
\end{center}
\end{figure}

 Theorists have ample data to determine $\Delta_1$ but are forced to rely heavily on models to the calculate $\Delta_2$ contribution. The $\beta_1(\tau)$ term is very close to one, so the integral in equation~\eqref{D1int} is very similar to the first moment of $\Gamma_1(Q^2)$ and allows almost a one-to-one comparison for the $g_1(x,Q^2)$ contribution to the hyperfine splitting. The $\Delta_1$ contribution is further constrained at low momentum transfer by assuming the GDH slope, and the elastic form factors are well known. The models are less clear for $g_2(x,Q^2)$ where a lack precision data points puts few constraints on the models. The discrepancy between model results at low momentum transfer suggests that even 100\% uncertainty on $g_2(x,Q^2)$ may be optimistic~\cite{Hyperfine3,Hyperfine2}. This model discrepancy is highlighted in Figure~\ref{B2int}, and also shows the kinematic coverage of E08-027 in the shaded blue region. The CLAS EG1b 2007 model is used for the results in Ref~\cite{Hyperfine3} and Ref~\cite{Hyperfine2}.


\chapter{\sc The Experiment}
\label{ch:Experiment}
\begin{figure}[htp]
\begin{center}
\includegraphics[width=0.90\textwidth]{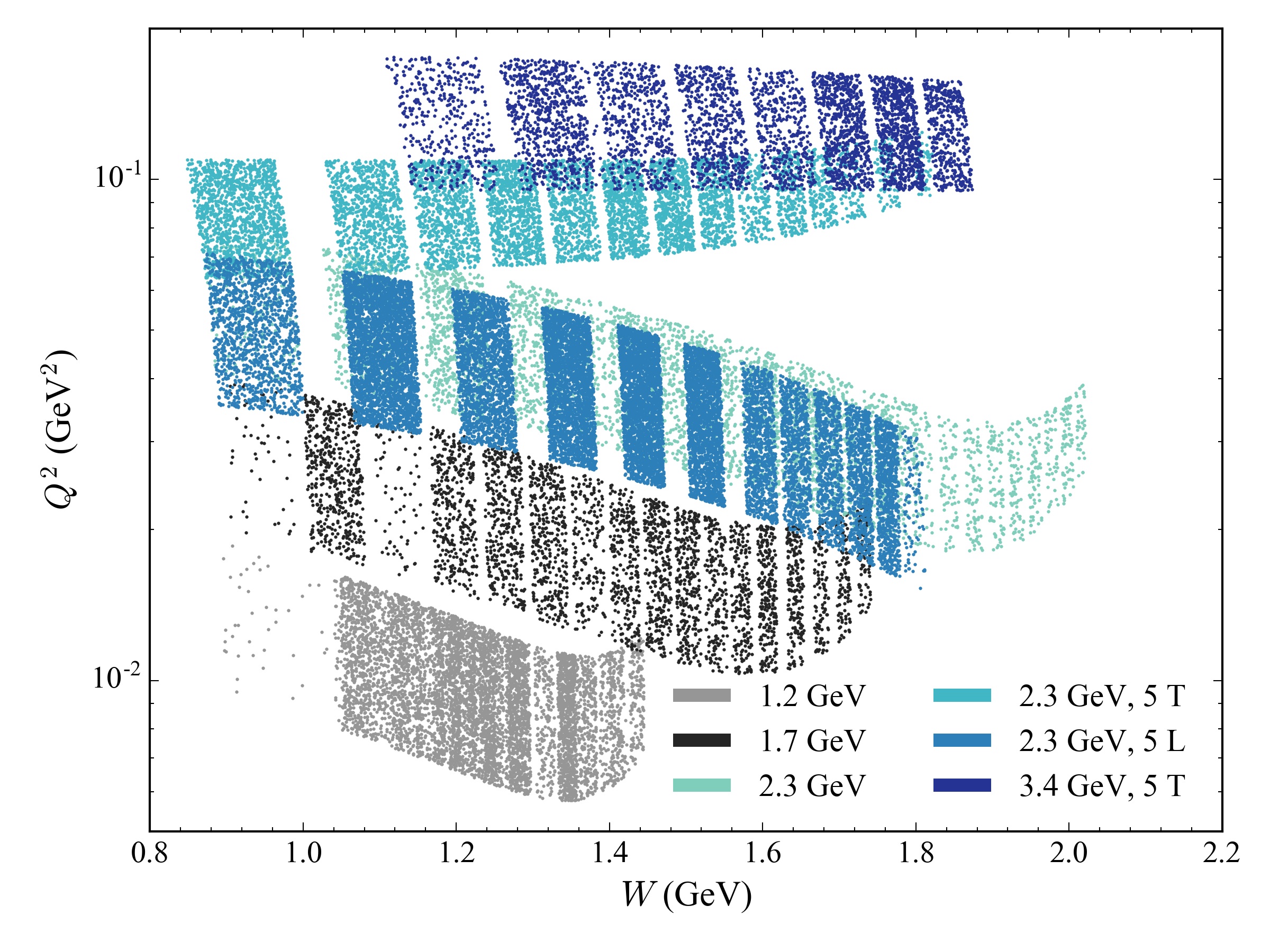}
\caption{\label{kin}The kinematic coverage of the E08-027 ($g_2^p$) experiment in $W$ vs. $Q^2$. }
\end{center}
\end{figure}

Experiment E08-027 ran in spring 2012 at Jefferson Lab's (JLab) Hall A. The experiment used inclusive electron-proton scattering to measure the proton's spin-dependent cross sections. The collected data focuses on the quasi-elastic and resonance regions at low momentum transfer (0.02 $< Q^2 <$ 0.20 GeV$^2$), with  the goal of extracting the proton's spin dependent structure functions, $g_{1,2}(x,Q^2)$, and calculating their related moments. The kinematic range of the experiment is shown in Figure~\ref{kin}. The figure is compiled from the collected data, and the plotted intensity is proportional to the number of events taken, scaled for the relative difference between the target polarizations of the 2.5 T and 5.0 T configurations. The spread in $Q^2$ at each beam energy is given by the approximate angular acceptance of the experiment: $\pm$ 1$^{\circ}$. The band structure of the data highlights the finite momentum acceptance of the experimental detectors.

Data was taken with longitudinally polarized electrons and a longitudinally and transversely polarized ammonia ($^{14}$NH$_3$) target.  The electron beam  energy ranged from 1.2 GeV to 3.4 GeV and its current was kept under 100 nA to minimize depolarization effects on the polarized target.  The transverse target field strength alternated between 2.5 T and 5.0 T in order to lessen the effect of the target field on the scattered electrons. At the lower beam energies, a transverse magnetic field increased the scattering angle of the detected electrons, and limited access to the low $Q^2$ values. This effect is seen in Figure~\ref{kin}, noting that higher values of $W$ indicate lower energy scattered electrons. Data was taken on a transversely polarized target at each kinematic setting, and for a longitudinally polarized target at one kinematic setting: 2.3 GeV, 5.0 T. In the remaining settings the parallel contribution to the data will be provided by the Hall B EG1~\cite{CLASg1,EG1b} and EG4~\cite{EG4} experiments.



The scattered electrons were detected using a pair of high-resolution momentum spectrometers (HRS).  The spectrometer detector package provided event track reconstruction, event triggering, and particle identification. Physical constraints limited the scattering angle of the spectrometers to 12.5$^{\circ}$. In order to to reach the lowest possible momentum transfer, a pair of septa magnets  were installed in front  of the HRS entrance. The magnets horizontally bent electrons scattered at $5.77^{\circ}$ (assuming no effect from the target field) into the spectrometers, which were located at 12.5$^{\circ}$.
\section{The Electron Accelerator}
\begin{figure}[htp]
\begin{center}
\subfigure[Diagram of the Jefferson Lab accelerator]{\label{fig:CebafSchem}\includegraphics[width=.80\textwidth]{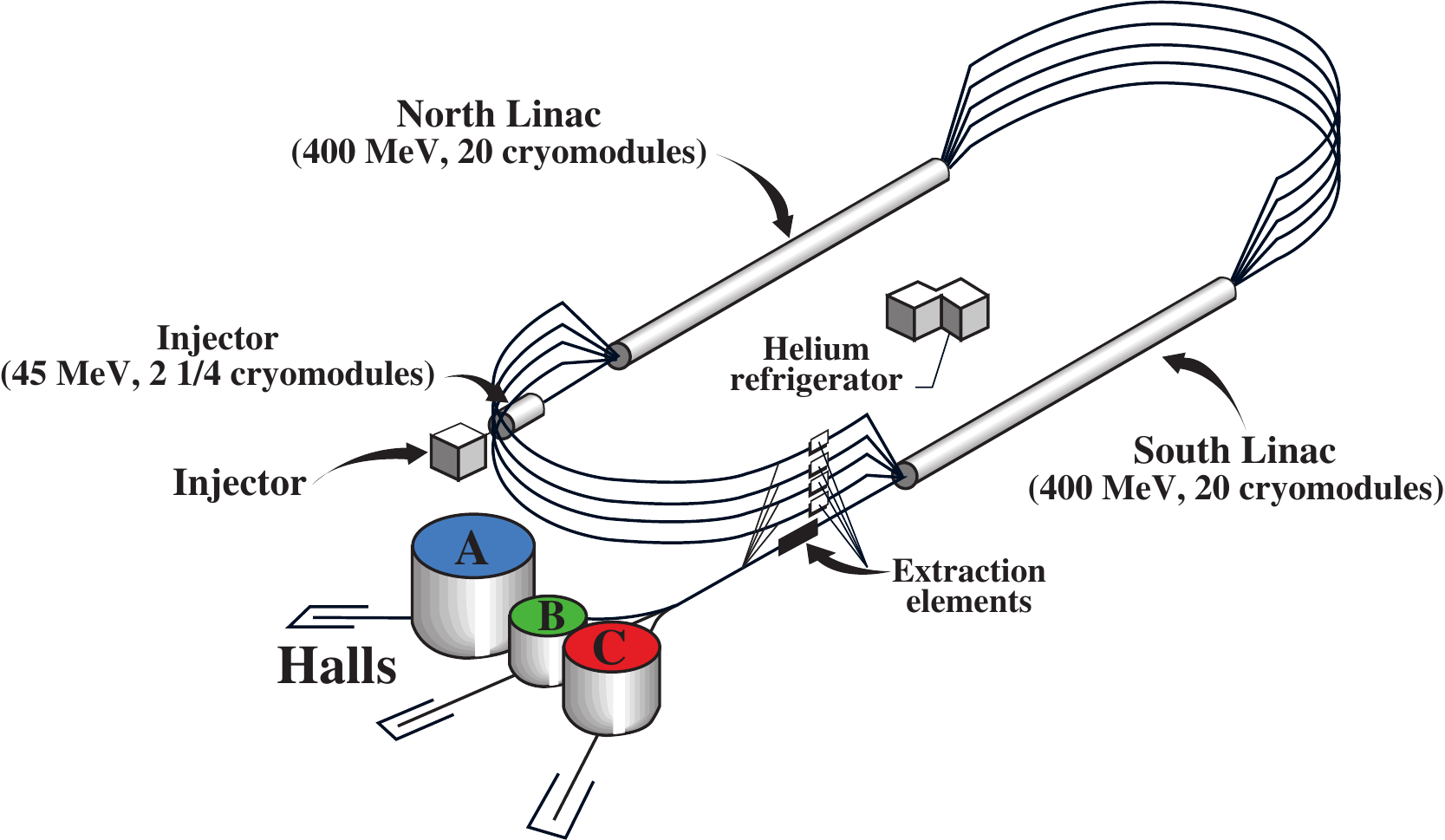}}
\qquad
\subfigure[Aerial view of Jefferson Lab]{\label{fig:AerialJLab}\includegraphics[width=.80\textwidth]{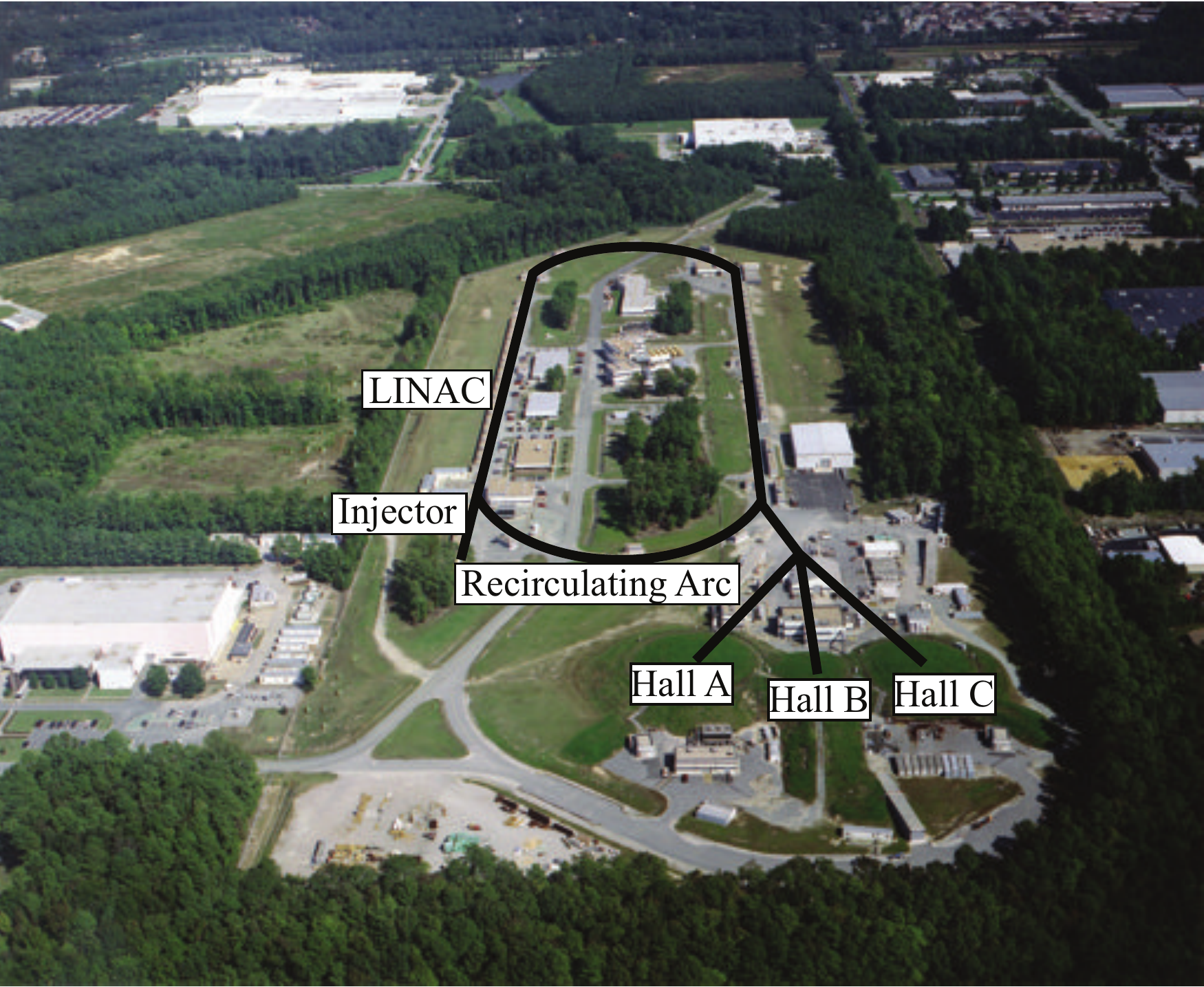}}
\caption{\label{CEBAF}CEBAF and the experimental halls. Reproduced from Ref~\cite{CEBAF}.}
\end{center}
\end{figure}
The Continuous Electron Beam Accelerator Facility (CEBAF) at Jefferson Lab is able to produce polarized electron beams up to 6 GeV in energy\footnote{After E08-027 the JLab accelerator was upgraded to maximum beam energies of 12 GeV and also added an additional experimental hall, Hall D.} and 200 $\mu A$ in current. The main components of the facility are shown in Figure~\ref{CEBAF} and include the polarized electron source, injector, two sets of linear accelerator cavities (LINACs), two recirculating arcs, and the three experimental halls. 

\subsection{Polarized Electron Source}
\begin{figure}[htp]
\begin{center}
\includegraphics[scale=.70]{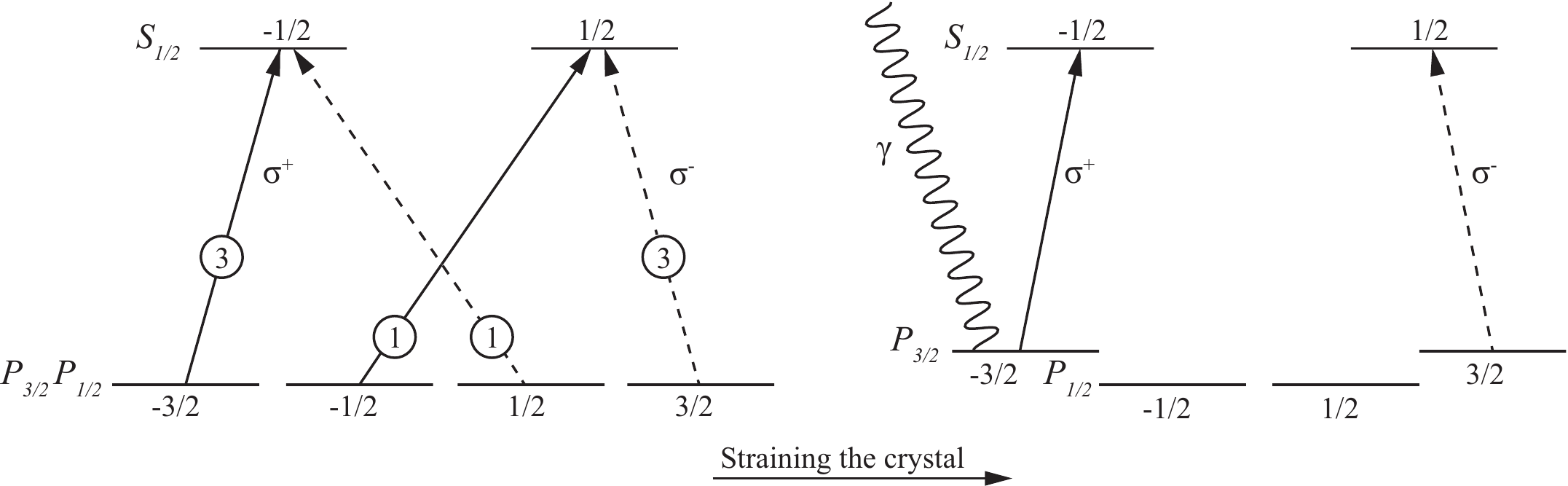}
\caption{\label{doped}Creating polarized electrons via stimulated photo-emission from a GaAsP crystal. Straining the crystal breaks the degeneracy and allows the polarized laser to excite electrons into a single polarization state. Based on a figure from Ref~\cite{doped}.}
\end{center}
\end{figure}

The electrons are created with a laser that stimulates photo-emission from a phosphorus doped gallium arsenide (GaAsP) cathode~\cite{Injector}.  The introduction of the phosphorous onto the gallium arsenide photo-cathode causes a mechanical strain on the GaAs crystal, which breaks a degeneracy present in the valence band. Before straining, three photons from the $P_{3/2}$ valence band are excited for every one from the $P_{1/2}$ state. Straining the GaAs crystal breaks the degeneracy, which leads to higher achievable polarizations. Illustrated in Figure~\ref{doped}, this process allows theoretical electron polarizations of 100\%. Real-world polarizations typically top-out at around 85\% at JLab, with crystalline imperfections and electron scattering during photo-emission being the limiting factors. The polarization of the electron is determined by the polarization of the 1497 MHz laser beam. Tuned to only stimulate the $P_{3/2}$ valence band, right-circularly polarized laser light excites electrons into the $m_j = -1/2$ conduction band state, and left-circularly polarized light excites electrons into the $m_j = +1/2$ state. The electrons are polarized longitudinally with respect to their momentum. 


The electron beam helicity is controlled by passing the injector laser through a voltage-controlled wave plate.  By changing the sign of the voltage applied to the wave plate the polarization of the incident photon is switched and causes the emission of an electron of opposite spin. The beam helicity is generated in a series of identical length helicity windows in a pseudo-random fashion at 960.02 Hz. The frequent flipping of the helicity limits time-dependent systematic effects. The systematic uncertainty is further reduced by generating the helicity in symmetric multi-window patterns.  For example, the helicity sequence in a quartet pattern is either (+ $-$ $-$ +) or ($-$ + + $-$), which limits any linear background.

The helicity signal is created from a programmable logic generator at the electron injector. The pseudo random nature of the helicity scheme eliminates systematic effects that could arise from the correlation between beam and helicity and data acquisition components in the experimental halls. Systematic effects are further minimized by delaying the helicity signal sent to the experimental data acquisition systems. The pseudo-random property of the algorithm means that the true electron helicity can still be extracted in analysis. The beam helicity is also flipped using an insertable half-wave plate. This wave plate allows for the study of helicity-dependent systematic effects and was flipped every few hours. A detailed overview of the helicity scheme for E08-027 is found in Ref~\cite{Chao}.
\subsection{Accelerating the Electrons}

The photo-emitted electrons are pulled from the conduction band into the injector through the application of a 100 kV bias voltage and organized into as many as three 499 MHz bunches~\cite{CEBAF}. The bias voltage accelerates the electrons  to 45 MeV prior to injection into the first superconducting LINAC.  After reaching energies of 400 MeV, magnets steer the accelerated electrons through a recirculation arc, where they then enter an identical second LINAC and are again accelerated. Following the second acceleration, the bunches are either separated and sent to the experimental halls or they enter another recirculation arc to return to the first LINAC. Any particular bunch can make up to five orbits before it enters a hall. Extraction into the experimental halls is carried out using radio-frequency (RF) separators and magnets. The bunch nature of the beam allows for each experimental hall to receive different (but correlated) beam energies and beam currents.

The two LINACs consist of 20 cryomodules each and accelerate the electrons at a rate exceeding 7 MeV/m. A single cryomodule consists of eight superconducting RF cavities powered at a frequency of 1497 MHz. The RF fields within the cavity create charge gradients that pull and accelerate the electrons as they move through it. The cavities themselves are made of niobium and kept at 2 K. High performance in the cavities allowed the accelerator to surpass the original design specification of 4 GeV electrons.

\section{Experimental Hall A}
\begin{figure}[htp]
\begin{center}
\includegraphics[scale=.60]{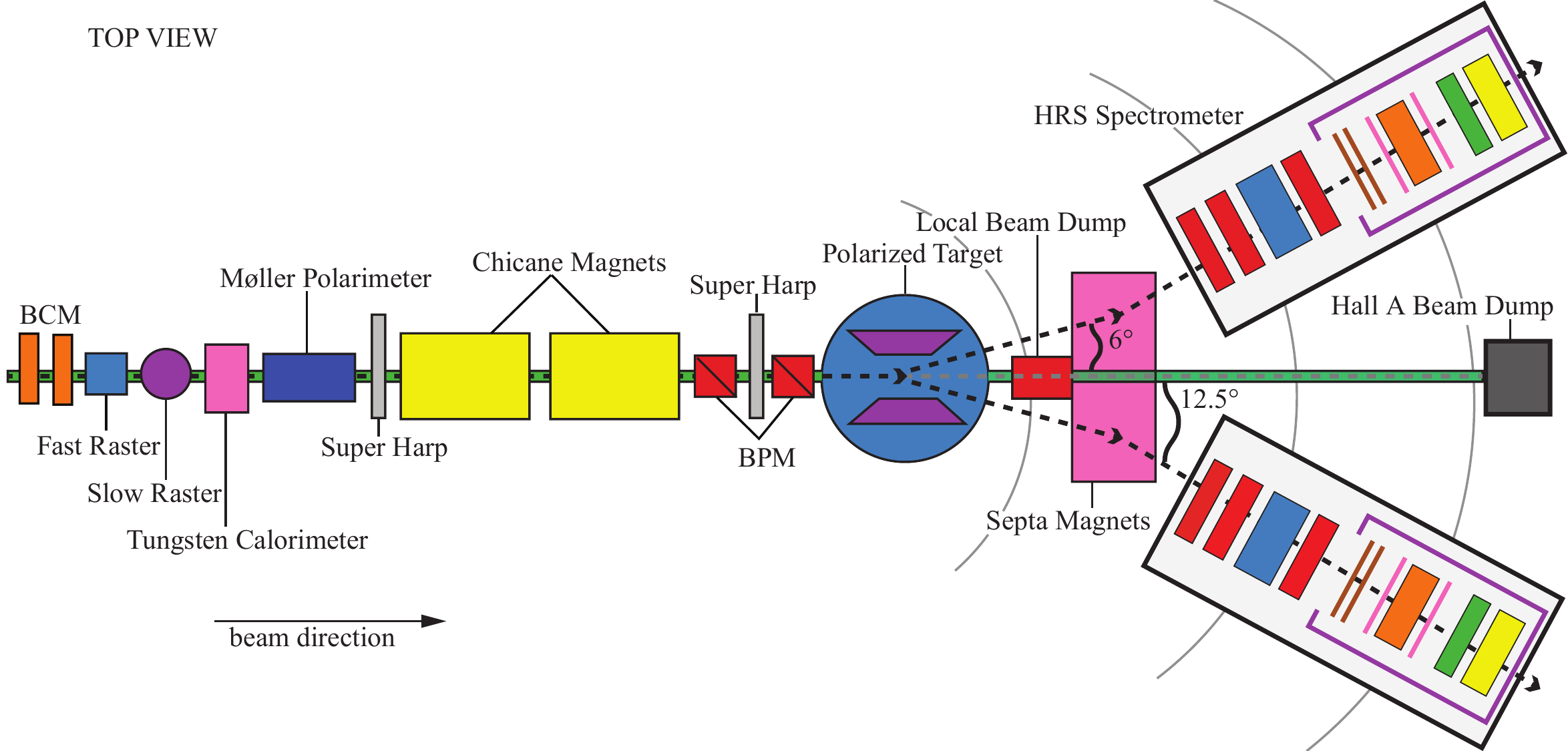}
\caption{\label{Beamline}Experimental Hall A for the E08-027 ($g_2^p$) experiment. Not to scale.}
\end{center}
\end{figure}

\begin{figure}[htp]
\begin{center}
\subfigure[Photo of the pivot prior to running the experiment.]{\label{fig:Cebaf111Schem}\includegraphics[width=.80\textwidth]{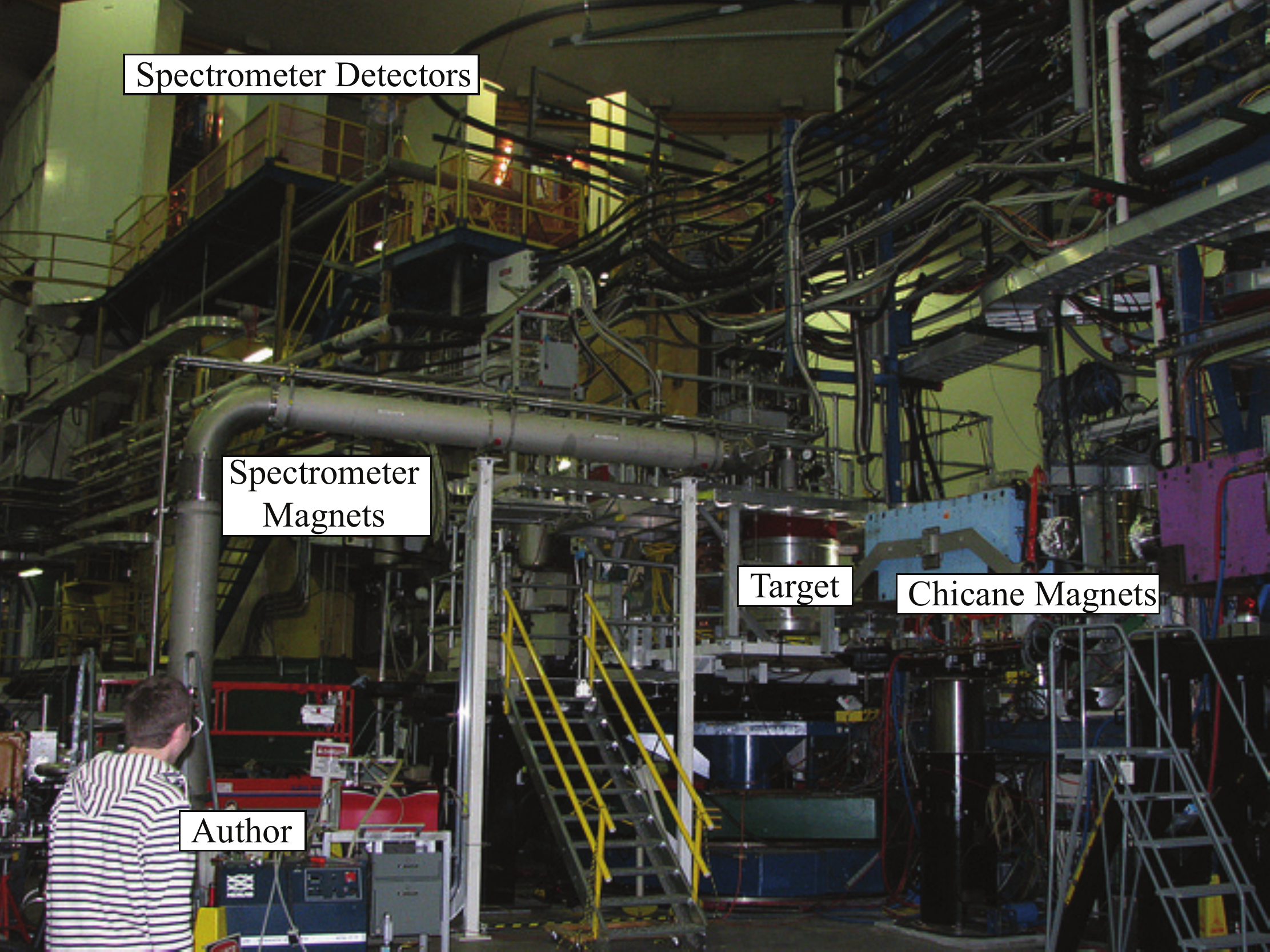}}
\qquad
\subfigure[CAD drawing of the Hall A pivot.]{\label{fig:AerialJLab111}\includegraphics[width=.80\textwidth]{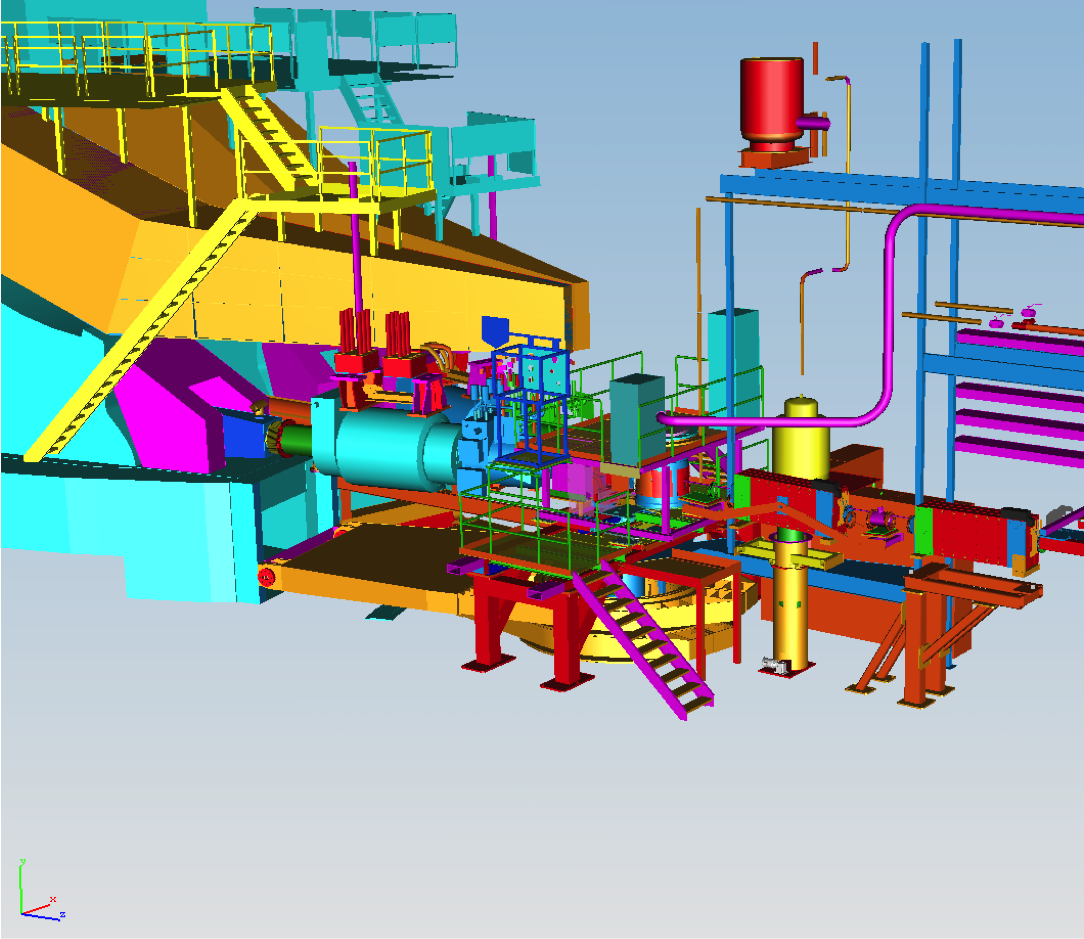}}
\caption{\label{HallA}The Hall A pivot  for the E08-027 ($g_2^p$) experiment.}
\end{center}
\end{figure}

The layout of Hall A along the beamline is shown in Figure~\ref{Beamline} and in Figure~\ref{HallA}, and includes several new components to deal with the difficult nature of operating a transversely polarized proton target in an electron beam.

\subsection{Beam Charge Monitors}
 Two beam current monitors (BCMs) are the first pieces of equipment along the Hall A beamline, and are 23 m upstream of the target. These monitors are resonant cavities in the TM$_{010}$ mode, tuned to the beam frequency of 1497 MHz~\cite{CEBAF}. An antenna inside the cavity picks up the resonant excitation of the passing electron beam. The output voltage of the antenna is proportional to the beam current, and this output signal is split to both the data acquisition system (DAQ) and experimental slow controls (EPICS). A voltage-to-frequency (V2F) converter converts the output into a signal suitable for the scaler counters, where the total counts is proportional to the beam charge accumulated.

Due to the low beam current ($<$ 100 nA) used during the experiment, the standard RMS-to-DC BCM receiver could not be used to produce the output signal. A new receiver was designed by the JLab instrumentation group. The new receiver uses analog amplifiers and filters to preprocess the 1497 MHz signal before digitizing the signal and applying more efficient digital filters to further reduce the signal to noise ratio. Finally, the output signal is converted back to analog to match the existing Hall A DAQ system setup.

 The BCM's are calibrated using a tungsten calorimeter. The calorimeter performs an invasive measurement of the beam current by placing a tungsten slug in the beam path. The energy absorbed by the slug is a function of the average beam current and is also given from the product between the increase in temperature of the calorimeter and its specific heat ($Q = KT$)~\cite{Tungsten}. Knowing the charge deposited in the tungsten calorimeter allows for the determination of the scaler proportionality constant. This calibration constant is used to determine the charge for runs without the use of the invasive calorimeter. Further details on both the BCM receiver and BCM calibration are found in Ref~\cite{Pengjia}.
 
\subsection{Beam Rasters}
\begin{figure}[htp]
\begin{center}
\includegraphics[scale=.75]{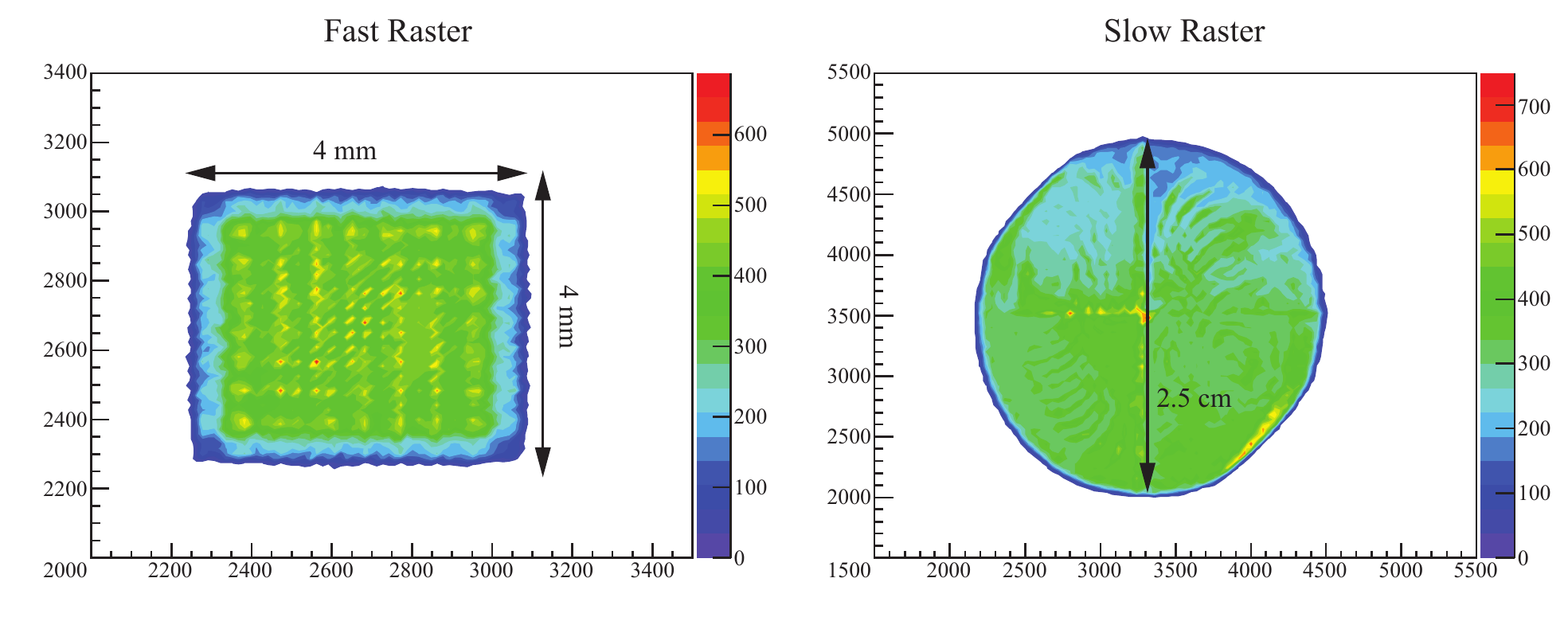}
\caption{\label{Raster}Fast and slow raster patterns as a function of the current (arb. units) applied to the dipole magnets.}
\end{center}
\end{figure}

 Located in between the BCMs and tungsten calorimeter are the fast and slow raster. The beam is rastered to minimize radiation damage and depolarization of the target material. The fast raster increases the initial 200 $\mu$m beam profile~\cite{CEBAF} to a 4 mm x 4 mm square beam spot. The circular slow raster pattern then ensures the beam moves over the full 1 inch (2.5 cm) diameter target cup and was new for experimental Hall A. Each raster consists of a time-varying magnetic dipole field controlled by a function generator. Both rasters use two dipole magnets, one for horizontal (X) motion and one for vertical (Y) motion. In the case of the fast raster, each magnet is driven by a 25 kHz triangle wave, which produces the raster pattern shown in Figure~\ref{Raster}.

The slow raster pattern is generated using a dual-channel function generator to create independent waveforms for the X and Y directions. The two waveforms used are
\begin{align}
x & = A_x t^{0.5} \mathrm{sin}(\omega t)\,,\\
y& = A_y (t+t_0)^{0.5} \mathrm{sin}(\omega t + \phi)\,,
\end{align}
where $A_x$ and $A_y$ are the amplitudes, $t$ is the amplitude modulation and $\phi$ is the phase difference between the X and Y waveforms. The frequency of the sine wave is the same for both X and Y at $\omega = 99.412$ Hz and the phase difference is locked at $\phi = \frac{\pi}{2}$. The amplitude modulation is cycled  between positive and negative values at a rate of 30 Hz, while the $t_0$ value is carefully adjusted to zero manually. This allows for the uniform slow raster pattern of Figure~\ref{Raster}.

\subsection{Beam Polarization}
Polarization of the electron beam is measured using the M${\o}$ller polarimeter, which uses the electron polarization dependence of the  (Christian) M${\o}$ller process ($ee \rightarrow ee$)  to determine the beam polarization~\cite{CEBAF}. The polarized M${\o}$ller cross section is~\cite{MollerXS}
\begin{align}
\frac{d\sigma}{d\Omega}  = \frac{d\sigma_0}{d\Omega} ( 1 + \sum_{i,j=x,y,z}P_i^b A_{ij}P_j^t)\,,
\end{align}
where $P_i^b$, $P_j^t$ are components of the beam and target polarizations, $A_{ij}$ are the asymmetry parameters and $d\sigma_0$/$d\Omega$ is the unpolarized  M${\o}$ller cross section. In the ultra-relativistic limit, the unpolarized cross section is
\begin{align}
\frac{d\sigma_0}{d\Omega}  = \frac{\alpha^2}{4E^2}\frac{(4-\mathrm{sin}^2\theta)^2}{\mathrm{sin}^4\theta}\,,
\end{align}
where $E$ and $\theta$ are the energy\footnote{In the laboratory frame, $E^2 = (E_0m + m^2)/2$ where $E_0$ is the beam energy and $m$ is the electron mass.}  and angle as measured in the center-of-mass frame. Assuming that the beam travels along the z-axis (the y-axis is normal to the zx-scattering plane), the longitudinal component of the incident electron polarization is extracted from $A_{zz}$,
\begin{align}
A_{zz} &= -\frac{(7+\mathrm{cos}^2\theta)\mathrm{sin}^2\theta}{(3+\mathrm{cos}^2\theta)^2}\,, \\
-A_{xx}&=A_{yy}=\frac{\mathrm{sin}^4\theta}{(3+\mathrm{cos}^2\theta)^2}\,,
\end{align}
and the measured target polarization. A transverse electron polarization also leads to an asymmetry, but its contribution is smaller than that of the longitudinal component: $A^{\mathrm{max}}_{xx} = A^{\mathrm{max}}_{zz}/7$~\cite{CEBAF}.

The polarization measurement is invasive and measures an asymmetry as opposed to a cross section. As a ratio of cross sections, the asymmetry cancels many of the sources of systematic error, but is still dependent on the beam and target polarizations. E08-027 made seven M${\o}$ller measurements and the results are shown in Table~\ref{g2pbeamPol}. The raw results are found in Ref~\cite{MollerRaw,Moller}.

\begin{figure}[htp]
\begin{center}
\includegraphics[scale=.60]{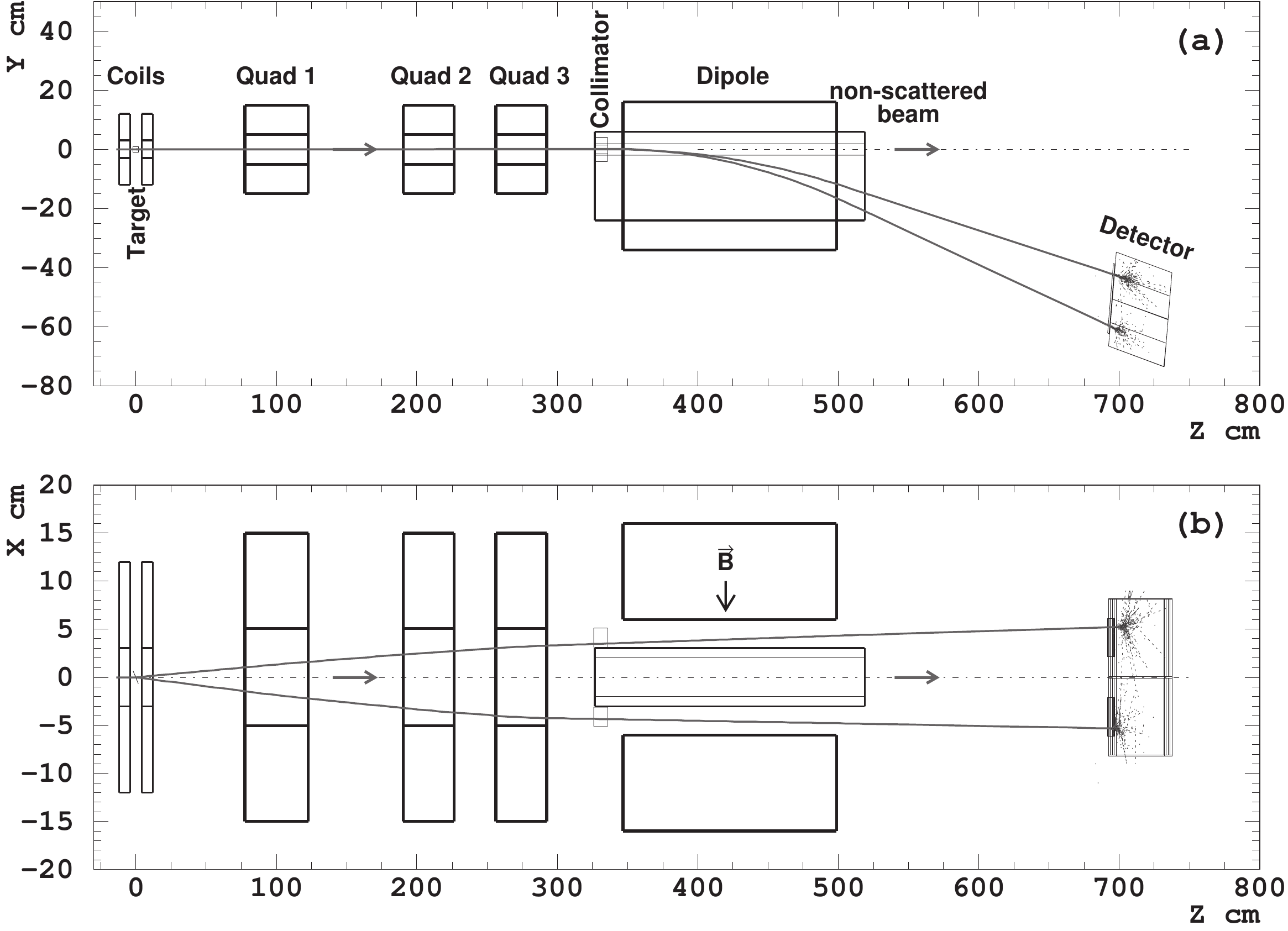}
\caption{\label{MollerSpect}M${\o}$ller spectrometer from the side (a) and from overhead (b). The three quadrupole magnets determine the acceptance of the spectrometer. The dipole magnet selects the electron momentum and bends the scattered electrons onto the lead-glass calorimeters. The polarized electrons in the iron and copper target foil are created using a 24 mT magnetic field. Reproduced from Ref~\cite{CEBAF}.}
\end{center}
\end{figure}

The schematic of the M${\o}$ller polarimeter is shown in Figure~\ref{MollerSpect}.  The spectrometer uses three quadrupole magnets to set the acceptance and a dipole magnet to select the electron momentum. The target foil is a combination of iron and copper and is oriented at an angle of $\pm$ 20$^{\circ}$ with respect to the beam in the horizontal plane in order to provide sensitivity to both longitudinal and transverse beam polarization components. The transverse components have opposite signs and cancel out during averaging to help reduce target angle uncertainties. The polarization of the target foil is created using a 24 mT magnetic field and is flipped between positive and negative polarizations to minimize false asymmetries.

M${\o}$ller scattering events are detected using a magnetic spectrometer with the detectors set to include scattering angles around the maximum analyzing power of $\theta_{CM}= 90^{\circ}$. The electrons are detected in coincidence using a lead-glass calorimeter. The spectrometer operates at beam energies ranging from 0.8 GeV to 6.0 GeV and can achieve statistical accuracy at the 0.2\% level for an hour of beam time.

\begin{table}[htp]
\begin{center}
\begin{tabular}{ l c  c  c  c r  }
\hline
  Date & $E_0$ (GeV) & Pol (\%) & Stat (Abs) & Sys (Rel) & Run Range  \\ \hline
  3/3/2012& 2.253 & 79.91&0.20 &1.7 & 2600 - 2777\\
  3/30/2012 & 2.253 & 80.16 & 0.52 & 1.7 & 2778 - 3956\\
  4/10/2012 & 1.710 & 88.52 & 0.30 & 1.7 & 3957 - 4697\\
  4/23/2012 & 1.157 & 89.72 & 0.29 & 1.7 & 4698 - 5483\\
  5/4/2012 & 2.253 & 82.65 & 0.58 & 1.7 & 5484 - 5642\\
   5/4/2012 & 2.253 & 80.40 & 0.45 & 1.7 & 5643 - 6100\\
   5/15/2012 & 3.350 & 83.59 & 0.31 & 1.7 & 6101 - 6216\\ \hline
\end{tabular}
\caption{\label{g2pbeamPol}M{\o}ller polarization results for the E08-027 experiment.}
\end{center}
\end{table}

\subsection{Beam Energy}
The incoming electron beam energy is measured using the arc-energy method~\cite{CEBAF,ArcE}. Before entering Hall A, the beam passes through a forty-meter arc, consisting of eight dipole magnets to bend the beam and nine quadrupole magnets  to focus the beam. The nominal bend of the arc is 34.3$^{\circ}$. The electrons are deflected as they pass through the dipoles' magnetic field, and this defection is related to the incoming beam energy:
\begin{equation}
p = \frac{c}{\theta} \int \vec{B} \cdot d\vec{l}\,,
\end{equation}
where $c$ is the speed of light. The equation is a consequence of the Lorentz force, $\frac{e}{c}\vec{v} \times \vec{B}$, for a magnetic field, and noting that the electron's path $l$ is related to the bend angle $\theta$ through $l = r\theta$. 

The magnetic field integral is measured using an identical reference dipole magnet. The deflection angle is measured using a set of ``Superharp" wire-scanners at the entrance and exit of the arc. During the bend angle measurement, the quadruples are turned off. This method provides an absolute measurement at the 2 x 10$^{-4}$ GeV level.
\begin{figure}[htp]
\begin{center}
\includegraphics[scale=.85]{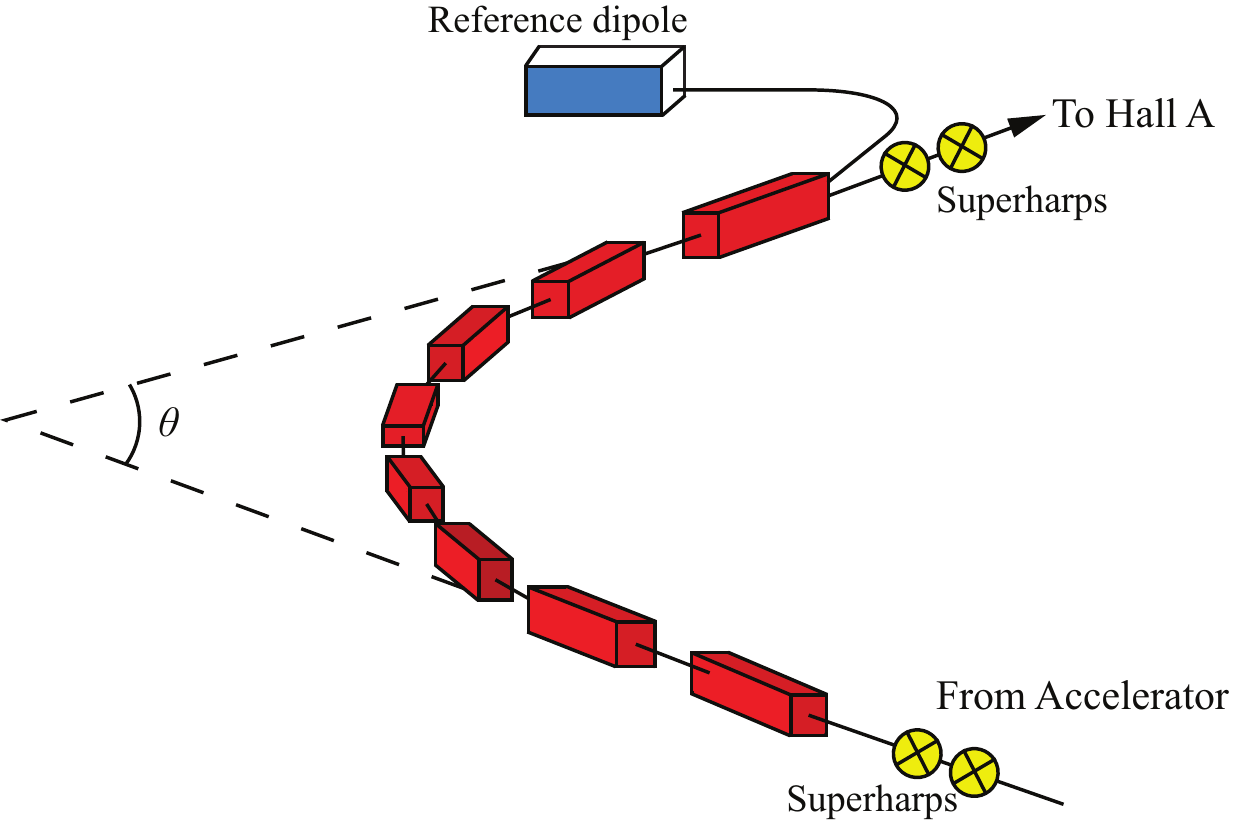}
\caption{\label{Arc}Beam arc method to calculate the energy of the incident electrons. The eight dipole magnets are in red and the nominal bend angle is $\theta$. }
\end{center}
\end{figure}

\begin{figure}[htp]
\begin{center}
\includegraphics[scale=.65]{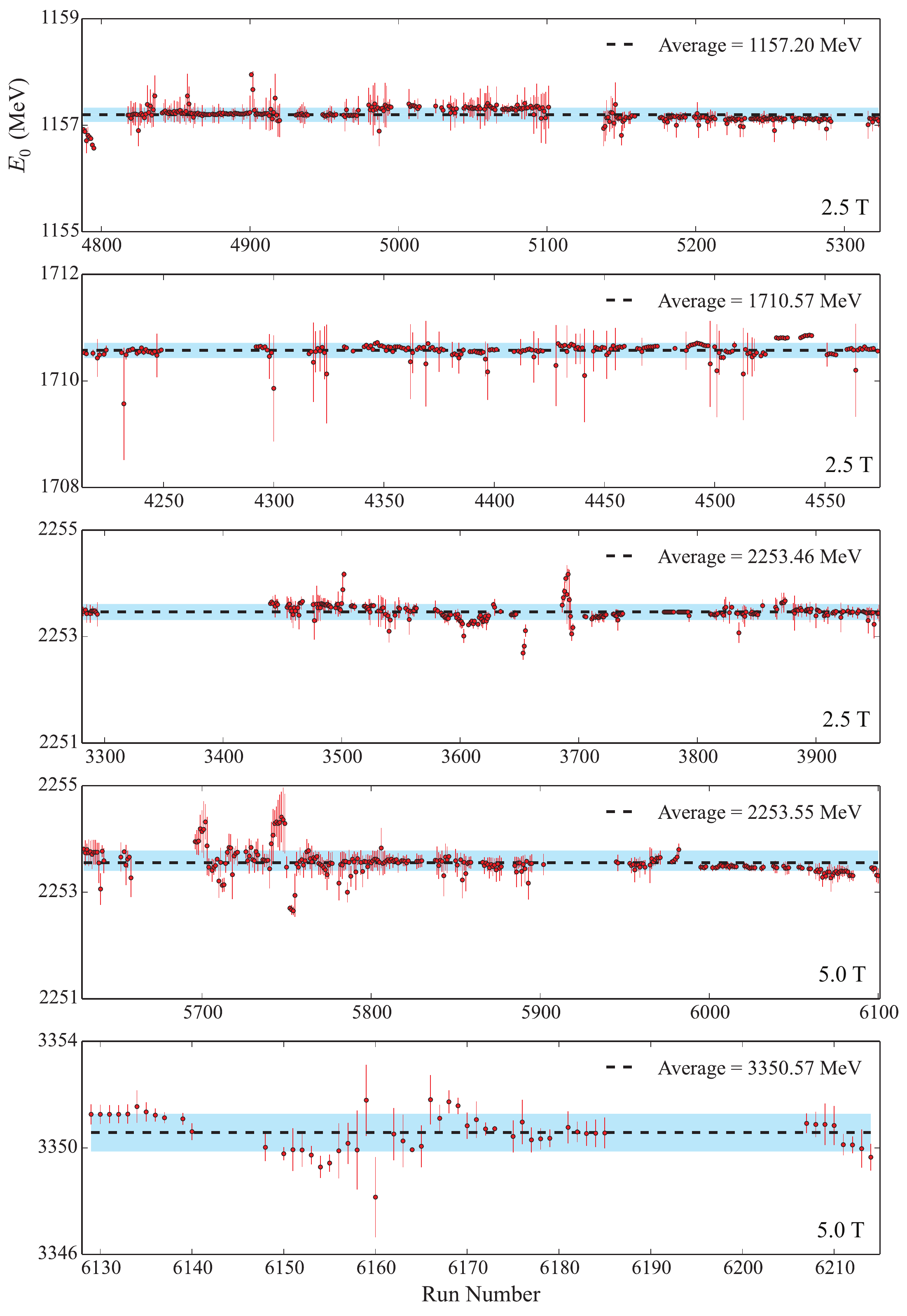}
\caption{\label{BeamE} Beam energy on a run-by-run basis for E08-027. The 2253 MeV plot at 5 T includes both transverse and longitudinal data. The dashed line is the average for a given setting and the blue band is the corresponding standard deviation. }
\end{center}
\end{figure}

The beam energy is also measured non-invasively and continuously using a procedure developed by Michael Tiefenback. In this method, the field integral value is determined by the set current of the dipoles within the arc. The conversion is done using a look-up table. Additional energy corrections are made depending on the beam position as determined from the arc beam-position monitors and Hall A beam line magnetic transfer functions.  The energy from this method is accurate at the 5 x 10$^{-4}$ GeV level. Beam energy measurement results for E08-027 are shown in Figure~\ref{BeamE}. The dotted line represents the average of all runs for a setting and the blue band is the corresponding standard deviation. The beam energy for each run is an average over the Tiefenback measurement (read out every few seconds) during the run, and the error bar is the standard deviation. 

\subsection{Dipole Chicane Magnets}
\begin{figure}[htp]
\begin{center}
\includegraphics[scale=.70]{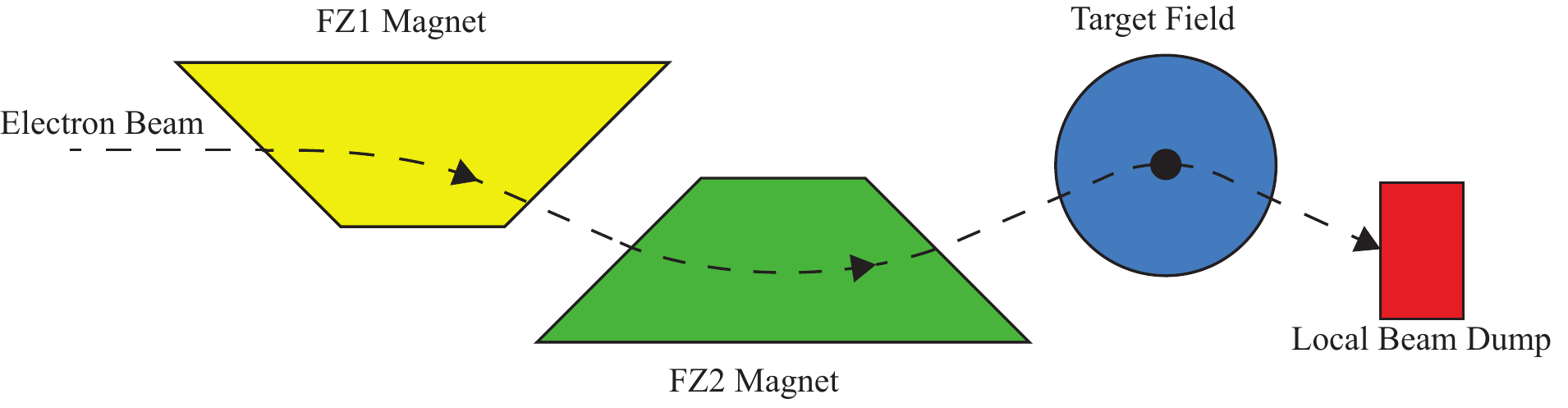}
\caption{\label{Chicane}Side-view of the chicane magnet operation for a transverse target field.}
\end{center}
\end{figure}

Two dipole chicane magnets are located after the M${\o}$ller polarimeter and were new equipment in Hall A. The magnets adjust the electron beam trajectory to compensate for the transverse target's magnetic field and ensure that the beam hits the center of the target, as shown in Figure~\ref{Chicane}. The right-hand rule dictates the electron is bent downwards from the magnetic field. The FZ1 dipole is located 5.92 m upstream of the target and  bends the beam downwards towards the FZ2 dipole. The FZ2 dipole is located 2.66 m upstream of the target, and is on a hydraulic support that can be moved vertically up and down. The FZ2 magnet bends the beam back up, towards the target. The chicane magnets are also used in a ``straight-through" configuration for a parallel target field where no bending is required. 
The chicane magnet configurations for each energy setting are listed in Table~\ref{chicaneConfig}~\cite{Roblin}. The incoming angle is the angle at which the electrons enter the target field, and the outgoing angle is the angle at which the electrons leave the target field. The tilt angle is the angle of the electrons at the target. The non-zero outgoing angle for the 5 T settings requires the use of a local beam dump.
\begin{table}[htp]
\begin{center}
\begin{tabular}{ l c  c  c  c r    }
\hline
 $E_0$ (GeV) & Target $\theta$ & Target B (T) & Incoming $\theta$ & Tilt Angle & Outgoing $\theta$  \\ \hline
1.157 & 90 & 2.5 &11.96 & 5.98 & 0.00\\
1.719 & 90 & 2.5 &8.06 & 4.03 & 0.00\\
2.253 & 90 & 2.5 &6.09 & 3.04 & 0.00\\
2.253 & 90 & 5.0 &6.09 & 0.00& $-$6.09\\
3.350 & 90 & 5.0 &4.09 & 0.00 & $-$4.09\\ \hline
\end{tabular}
\caption{\label{chicaneConfig}Chicane configurations for the E08-027 experiment.}
\end{center}
\end{table}



\subsection{Beam Position and Angle}
\begin{figure}[htp]
\begin{center}
\includegraphics[scale=15.0]{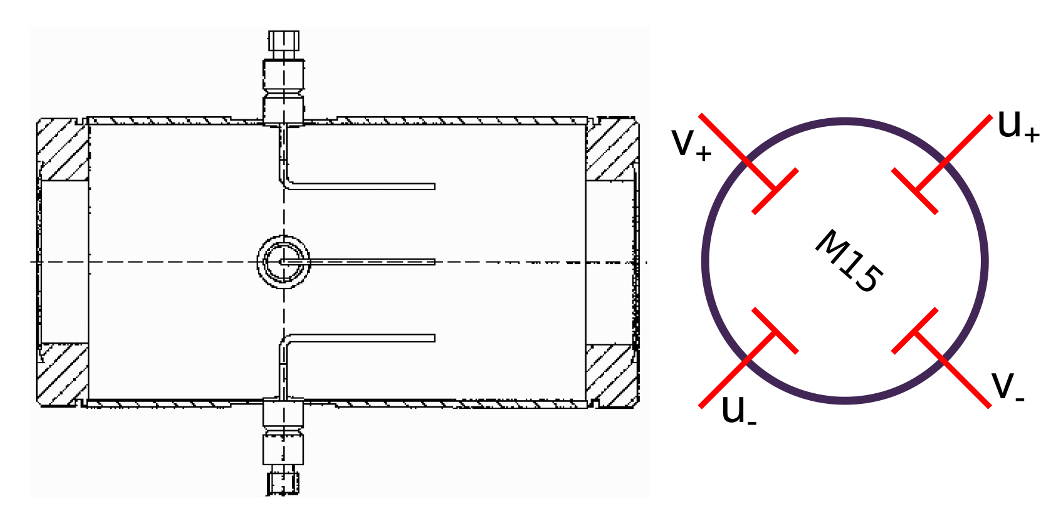}
\caption{\label{BPMdiagram}Diagram of the BPM cavity. The side (front) view is on the left (right). The four BPM antennas are labeled as $v_+$,$v_-$,$u_+$ and $u_-$ in the right figure. M15 is the model designation of the BPM chamber.}
\end{center}
\end{figure}

The beam position and beam angle after the chicane magnets is measured using two beam position monitors (BPMs) and they are located 95.5 cm and 69 cm upstream of the target. Each BPM cavity consists of four open-ended antennas that run coaxial with the incoming electron beam, as seen in Figure~\ref{BPMdiagram}. When the beam passes through the BPM cavity a signal is induced that is inversely proportional to the beam's distance from the antennae. Like the BCM receiver, the low beam current used during the experiment meant that a new BPM receiver had to be designed by the JLab instrumentation group~\cite{Musson}.

\begin{figure}[htp]
\begin{center}
\includegraphics[scale=.40]{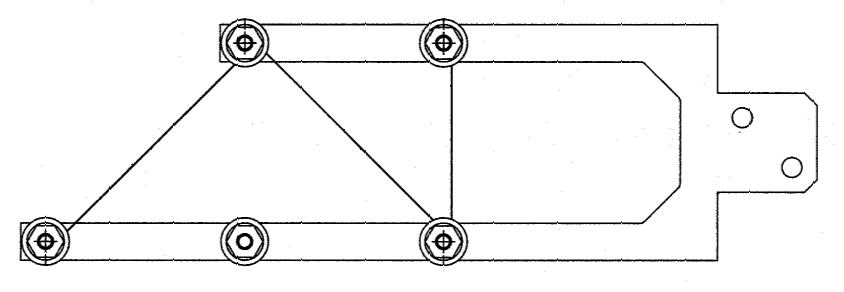}
\caption{\label{Harp}A diagram of the Superharp wire scanner showing the wire orientation.}
\end{center}
\end{figure}

The BPMs are calibrated using two Superharp wire scanners. Space limitations meant that only one scanner could be installed between the two BPMs; the other harp was placed before the first chicane magnet. The Superharps provide an invasive, but absolute measurement of the beam position. They consist of three 50 $\mu$m wires orientated vertically and at $\pm$ 45$^{\circ}$, with respect to the electron beam. A schematic of the harp is shown in Figure~\ref{Harp}. Scanning the wires across the beam profile creates a shower of radiation that is measured with photomultiplier tubes. A reference position for the wires is provided by a survey measurement.

The beam positions are transported from the last BPM to the target using a set of transport functions. The transverse target field breaks a linear projection of the positions to the target. Instead, the transport functions are created using thousands of simulated beam trajectories (from the  last BPM to the target) with different initial positions and angles.  A TOSCA model provides the target field map in the simulation and accounts for the deflection caused by the target's magnetic field. The TOSCA model is checked against measurements of the target field made before the experiment~\cite{Chao2,Jie2}, to confirm its accuracy.  

The BPM only provides the central beam position value of the rastered beam. The position of the rastered beam is determined from a combination of the BPM readout and the raster current information for each event. The magnet current of each raster is calibrated to determine the position of the raster with respect to the center of the raster pattern. 

The final position and angle uncertainties at the target are 1-2 mm and 1-2 mRad respectively. The uncertainty in the calibration is due to uncertainty in the harp calibration constant, survey data and electronic pedestal value.  Further details on the BPM calibration and reconstruction are found in Ref~\cite{Zhu}.

\subsection{Local Beam Dump}
\begin{figure}[htp]
\begin{center}
\subfigure[CAD drawing of the beam dump.]{\label{fig:dump_blue}\includegraphics[width=.40\textwidth]{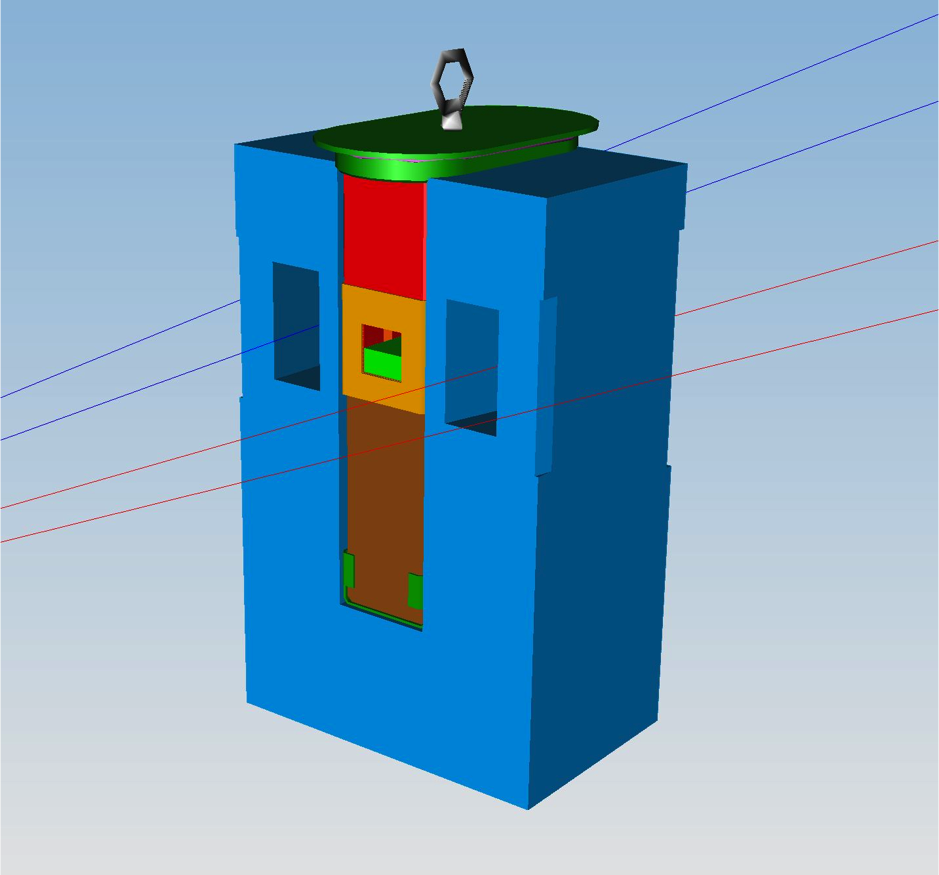}}
\qquad
\subfigure[CAD drawing of the target platform.]{\label{fig:dump_platform}\includegraphics[width=.5\textwidth]{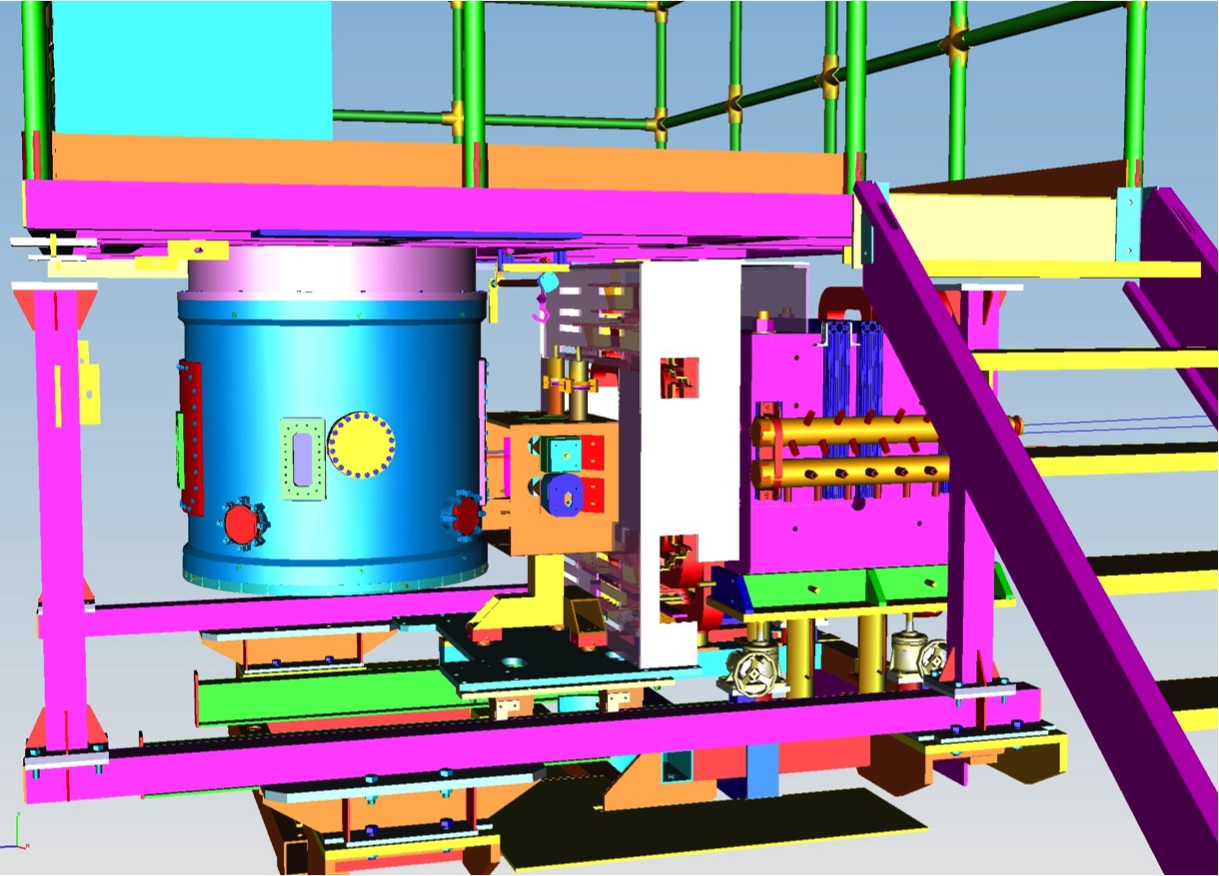}}
\qquad
\subfigure[Photo of the beam dump without insert. The septa magnets are in the background.]{\label{fig:Aeria234324lJLab111}\includegraphics[width=.50\textwidth]{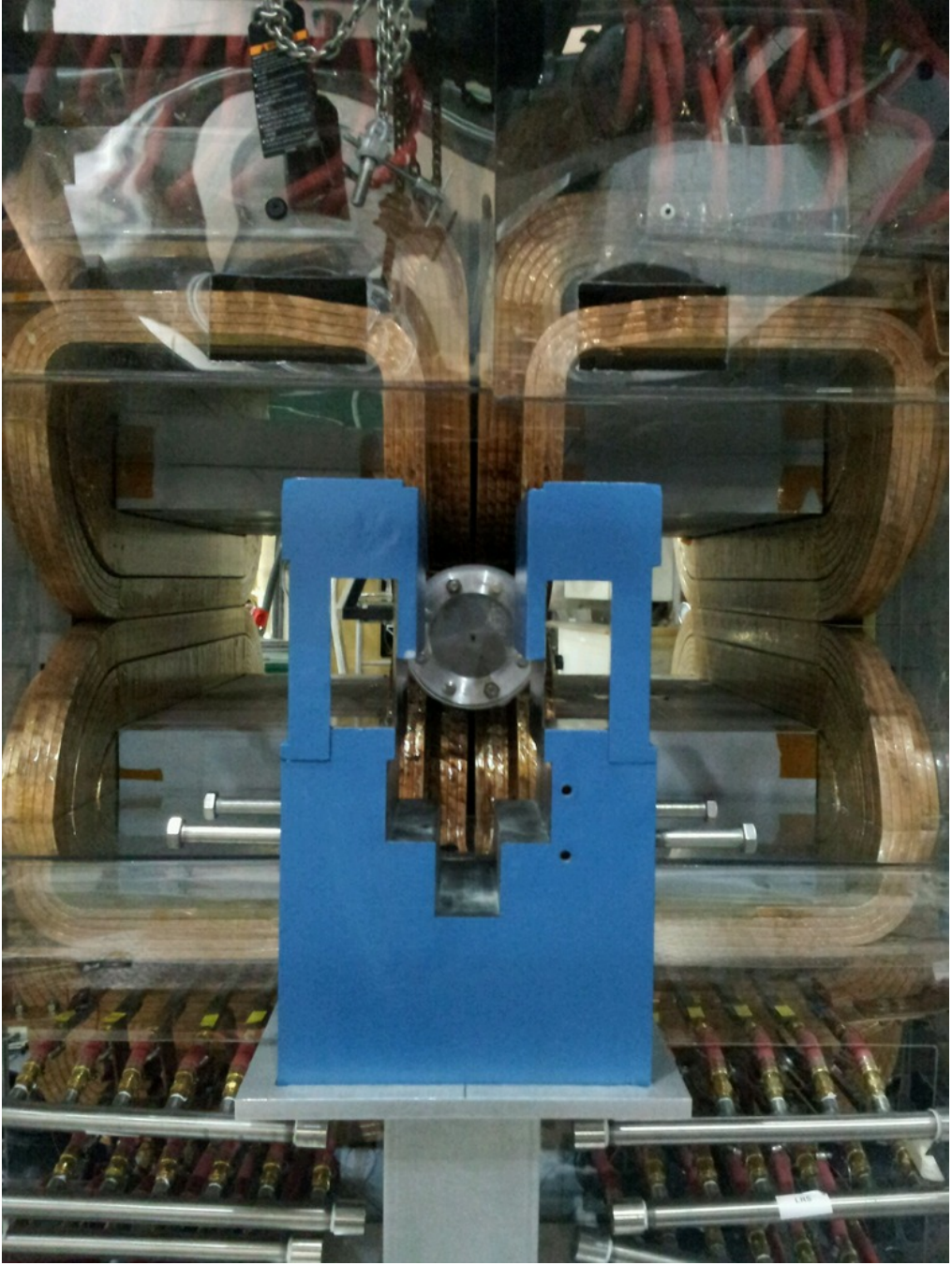}}
\caption{\label{Dump}The local beam dump for the E08-027 experiment.}
\end{center}
\end{figure}

The low beam current for the experiment allows the use of a copper and tungsten local beam dump. This is necessary for beam trajectories at the 5.0 T transverse target configuration, where the combination of beam energy and target field means that the electrons will not reach the standard Hall A beam dump. The chicane magnets insure that the 2.5 T beam trajectories exit the target field horizontally and are able to reach the standard Hall A beam dump. The local beam dump is installed downstream of the polarized target and contains a removable insert to control background radiation levels, see Figure~\ref{Dump}. The CAD drawing in Figure~\ref{fig:dump_blue} shows the insert (in brown, red and orange) and the blue support structure with holes to allow for passage of the scattered electrons. The CAD drawing in Figure~\ref{fig:dump_platform} shows the location of the local beam dump (inside the orange box) with respect to the blue polarized target can and pink septa magnets.

\section{The Polarized Target}
The E08-027 proton target consists of frozen ammonia ($^{\mathrm{14}}$NH$_{\mathrm{3}}$), which is polarized using dynamic nuclear
polarization (DNP) in a 5.0 or 2.5 T magnetic field at approximately 1 K. Similar targets have been used in Hall C and Hall B at Jefferson Lab previously, but this experiment is the first time such a target was installed in Hall A.

\subsection{Thermal Equilibrium Polarization}
\label{TEtalk}
\begin{figure}[htp]
\begin{center}
\includegraphics[scale=.50]{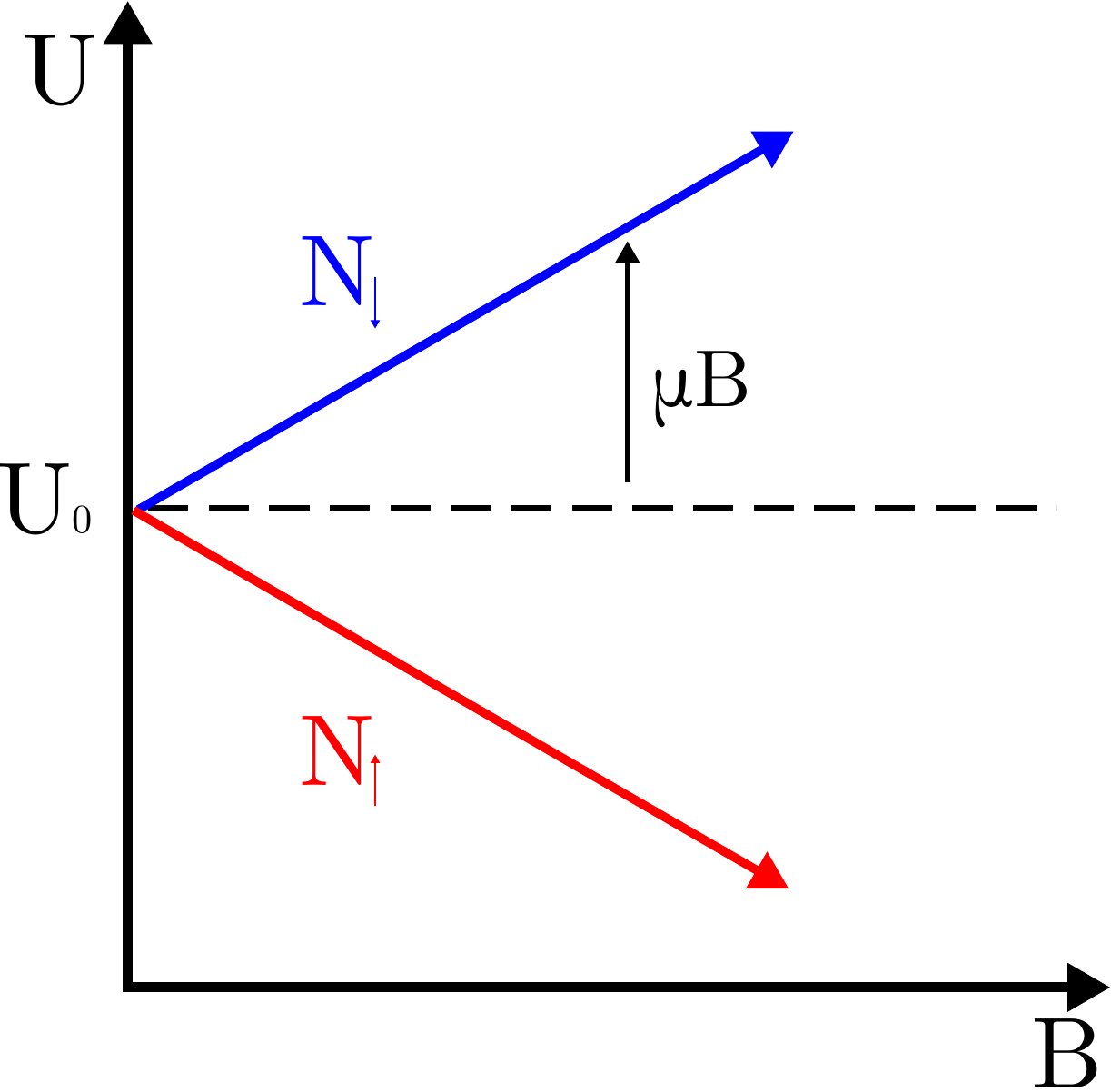}
\caption{\label{Zeeman}Energy states of a spin-$\frac{1}{2}$ particle in a magnetic field B. The magnetic field couples with the particles' spin to separate the energy levels according to spin state. }
\end{center}
\end{figure}

The starting point in the target polarization process is the thermal equilibrium (TE) polarization. At large magnetic fields and low temperatures, particles with a non-zero magnetic moment, $\mu$, tend to align themselves with the magnetic field. The Zeeman interaction (see Figure~\ref{Zeeman}) separates the nuclei into spin-dependent energy states. There are two such states for spin-$\frac{1}{2}$ nuclei (spin-up and spin-down), and the corresponding Zeeman energy is $\pm\mu \cdot B$.  

The vector polarization is defined as the difference in the fraction of nuclei aligned with and against an applied magnetic field  such that
\begin{equation}
\label{Pol}
P = \frac{N^{\uparrow} - N^{\downarrow}}{N^{\uparrow} + N^{\downarrow}}\\,
\end{equation}
where $N^{\uparrow}$ is the number of states aligned with the field and $N^{\downarrow}$ is the number of states anti-aligned with the field.

At thermal equilibrium, the relative state populations are expressed using Boltzmann statistics as~\cite{Pathria}
\begin{equation}
\label{Bolt}
\frac{N_1}{N_2} =  \mathrm{exp}\left(\frac{-\Delta E}{k_B T}\right) = \frac{N^{\uparrow}}{N^{\downarrow}} =  \mathrm{exp}\left(\frac{2\mu B}{k_B T}\right) \,,
\end{equation}
where $k_B$ is the Boltzmann constant and $T$ is the temperature. Combining equations~\eqref{Pol} and~\eqref{Bolt} leads to an expression for the thermal equilibrium polarization
\begin{equation}
\label{TE}
P_{\mathrm{TE}} = \frac{e^{\frac{\mu B}{k_B T}} - e^{\frac{-\mu B}{k_B T}}}{e^{\frac{\mu B}{k_B T}} + e^{\frac{-\mu B}{k_B T}}}\\.
\end{equation}
At 5 T and 1 K, this thermal equilibrium polarization is 99.8\%~\cite{Averett} for an electron but less than 1\% for a proton. The small magnetic moment of the proton  ($\mu_p \approx \frac{\mu_e}{660}$) results in a magnetic energy that is much less than the proton's thermal energy and thus a very low level of thermal equilibrium  polarization.

\subsection{The DNP Process}
Small equilibrium polarizations are enhanced through dynamic nuclear polarization (DNP). DNP leverages the large electron polarization with electron-proton spin coupling to increase the proton polarization. The enhanced polarization is driven using microwave transitions.

\begin{figure}[htp]
\begin{center}
\includegraphics[scale=.72]{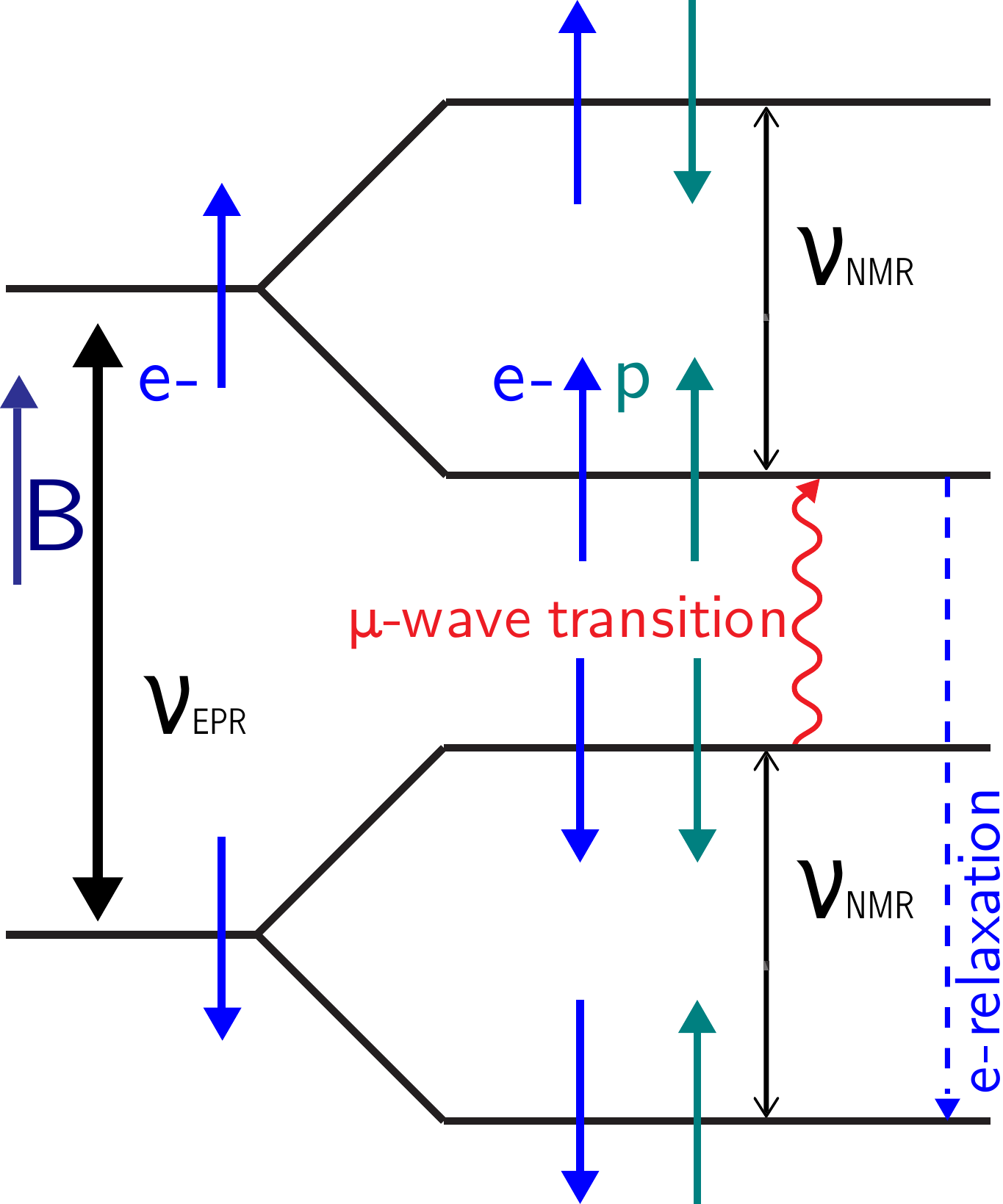}
\caption{\label{Hyper}Inducing the transition, $e_{\downarrow} p_{\downarrow} \rightarrow e_{\uparrow} p_{\uparrow}$, via microwave radiation. The four possible combinations of electron and proton spins determine the four hyperfine energy levels. The EPR frequency is determined from the Zeeman splitting of the electron. Based on a figure from~\cite{PolHyper}.  }
\end{center}
\end{figure}

A simple description of the solid-state effect and DNP process is given as follows: the solid target material sample, containing a large number of polarizable protons, is doped with paramagnetic radicals\footnote{Paramagnetic materials have an odd number of electrons and radicals have an unpaired valence electron.} to provide uniformly distributed, unpaired electron spins throughout the sample. A strong magnetic field (either  2.5 T and 5 T for E08-027) is applied at low temperature to create a large thermal equilibrium polarization of the unpaired electrons. The spin-spin interaction in the magnetic field of the electron and proton causes hyperfine splitting and four distinct energy levels based upon the alignment of the proton and electron spins, as shown in Figure~\ref{Hyper}. The corresponding Hamiltonian is
\begin{equation}
H = \vec{\mu}_e \cdot \vec{B} + \vec{\mu}_p \cdot \vec{B} + H_{ss}\\,
\end{equation}
where the first two terms are the Zeeman interaction of the electron and proton, respectively and $H_{ss}$ is the spin-spin interaction term~\cite{SE}.

Electron spins are flipped by applying an RF-field at the electron paramagnetic resonance (EPR) frequency, $\nu_{\mathrm{EPR}}$, which corresponds to a Zeeman energy of $\vec{\mu}_e \cdot \vec{B}$. In a similar manner, the proton spin is flipped with an RF-field at the nuclear magnetic resonance (NMR) frequency, $\nu_{\mathrm{NMR}}$, corresponding to a Zeeman energy of $\vec{\mu}_p \cdot \vec{B}$. The NMR frequency is not used to drive polarization, but is used to measure the resulting polarization.

Traditional dipole selection rules forbid the simultaneous flipping of both spins, but the presence of the spin-spin interaction creates mixing between electron and proton states, allowing access to the previously forbidden transitions~\cite{Tranny}; the electron polarization is now transferable to the proton. Microwaves tuned to the frequency 
\begin{align}
\nu_{\mu} &= \nu_{\mathrm{EPR}} - \nu_{\mathrm{NMR}}\, :\, e_{\downarrow} p_{\downarrow} \rightarrow e_{\uparrow} p_{\uparrow}\,, \\
\nu_{\mu} &= \nu_{\mathrm{EPR}} + \nu_{\mathrm{NMR}}\, :\,  e_{\downarrow} p_{\uparrow} \rightarrow e_{\uparrow} p_{\downarrow}\,,
\end{align}
induce either positive or negative polarization depending on the frequency. The frequency, $\nu_{\mu}$, is 140.1 (140.5) GHz at 5T and 1 K for positive (negative) polarization states. The electrons relax to the lower energy state in only a few seconds, but the relaxation time of the proton is tens of minutes. This allows a single electron to polarize multiple protons, driving a positive proton polarization state that is maintained with the use of microwaves.

\subsection{Spin Relaxation and Polarization Decay}
The vibrational and rotational motion of the proton and electron create a lattice structure within the sample~\cite{Levitt}. Irradiating microwaves add spin energy to the proton-electron system in order to create polarization. The additional spin energy is in thermal contact with the lattice, which provides a mechanism for the electron or proton to ``give up" its spin energy gained through polarization. They can return to their respective thermal equilibrium states by emitting a lattice phonon~\cite{MRI}. This process is referred to as spin-lattice relaxation. The spin coupling strength to the lattice is proportional to the magnetic moment of the polarized spin and explains the relatively shorter relaxation time of the electron ($\mu_e \approx 660 \mu_p$). 

The average lifetime of nuclei in the higher energy state is referred to as the relaxation time, $T_1$. This $T_1$ time is used to estimate the polarization decay rate of a DNP sample after terminating the microwave irradiation and ceasing the polarizing spin transitions. Changes in temperature and field strength change the vibrational and rotational energy of the lattice, which has a direct impact on $T_1$.

\subsection{Target Setup}
\begin{figure}[htp]
\begin{center}
\includegraphics[scale=1.2]{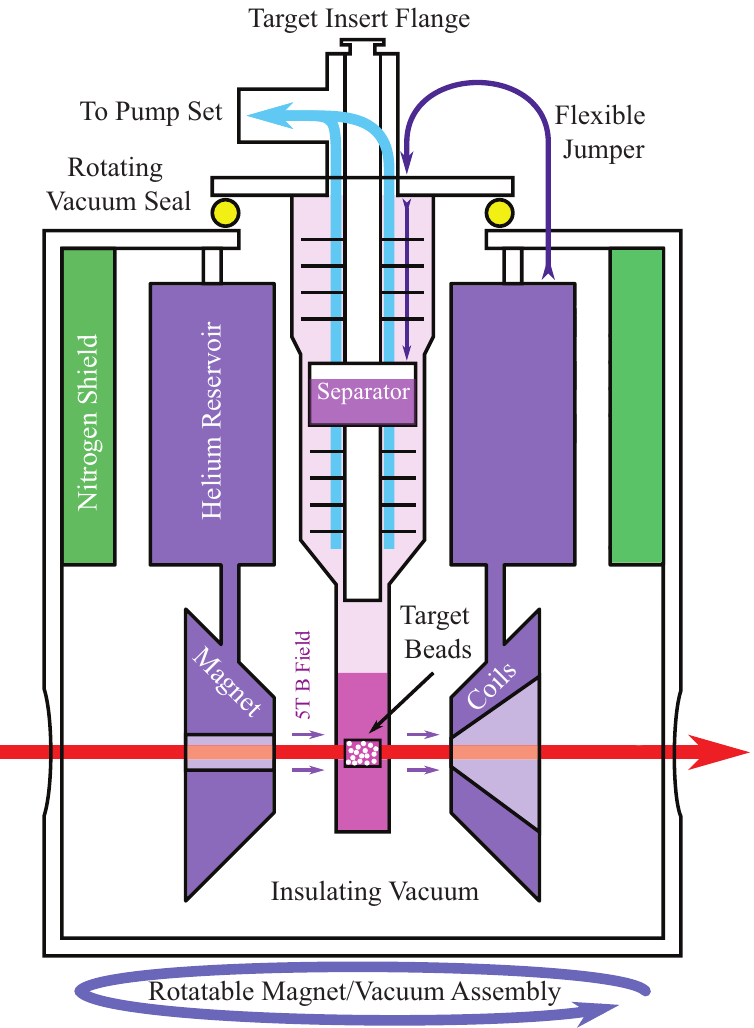}
\caption{\label{Target}Cut-away of the polarized NH$_\mathrm{3}$ target system for E08-027 in the parallel configuration.  Figure reproduced from Ref~\cite{TargetPol}.}
\end{center}
\end{figure}

The E08-027 polarized target system is shown in Figure~\ref{Target}. A superconducting split-pair magnet wound with niobium-titanium wire and kept in a bath of liquid helium at 4K provides the magnetic field. The open geometry of the split-pair design allows for the electron beam to pass through in both longitudinal and transverse target field configurations. Rotations of the magnetic field are carried out via a rotating vacuum seal. The uniformity of the field is 10$^{-4}$ or better over a cylindrical volume 2 cm in diameter and 2 cm in length~\cite{TargetPol}.  The magnet, along with the rest of the target system, is housed within an aluminum cryostat that is kept under an insulating vacuum. To minimize liquid helium loss, the helium reservoir is kept in thermal contact with a liquid nitrogen shield.

The target material is cooled to 1K using a liquid helium evaporation refrigerator. This refrigerator is isolated from the rest of the cryostat using a vacuum jacket. The interior refrigerator space is continuously pumped on using a set of Roots pumps to remove the higher temperature helium vapor. In addition, the liquid helium passes through a phase separator and a series of heat exchangers to reach 1K in the target nose.  The target nose holds the target materials in the uniform field region of the superconducting magnet. The temperature in the nose is determined by monitoring the $^{\mathrm{4}}$He vapor pressure.

\begin{figure}[htp]
\begin{center}
\includegraphics[width=0.8\textwidth]{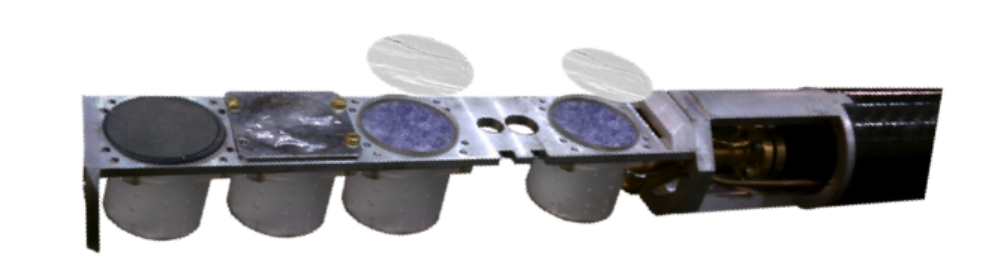}
\caption{\label{TargetStick}Target insert section of the E08-027 target stick. See text for a description of the target materials.}
\end{center}
\end{figure}

The E08-027 target stick is a carbon fiber tube with an aluminum insert (or ladder) at the end.  The ladder holds several different target materials as shown in Figure~\ref{TargetStick}. Starting from the left, there is a carbon foil, dummy cell, production target cell filled with solid ammonia beads, two empty holes for additional carbon foils and CH$_{\mathrm{2}}$ targets, and finally, another production target cell.  The dummy and production cells have thin aluminum end caps and also an embedded Cu-Ni coil for the NMR measurement. The absence of polarizable free protons in Kel-F plastic makes it an ideal material for the target cups. The target cell has a radius of 1.361 cm and a length of 2.827 cm. For a portion of the production runs at $E_0$ = 1.2 GeV, the length of the ammonia cell was shortened to 1.295 cm in order to limit radiation length seen by the incoming and scattered electrons.

\begin{figure}
\centering     
\subfigure[Pre-irradiation]{\label{fig:b}\includegraphics[width=.20\textwidth]{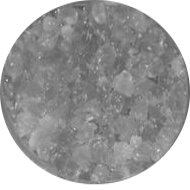}}
\qquad
\subfigure[Post-irradiation]{\label{fig:a}\includegraphics[width=.20\textwidth]{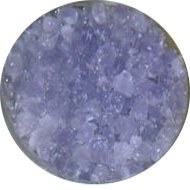}}
\caption{Ammonia samples before and after irradiation in the NIST accelerator}
\label{Sample}
\end{figure}

Ammonia is a suitable material for DNP and for use within an electron beam for several reasons. NH$_\mathrm{3}$ can be polarized to upwards of 90\% in a 5 T magnetic field in under 30 minutes. It also holds its polarization when subjected to the radiation damage from an electron beam.  Ammonia has a moderate dilution factor (ratio of free polarizable protons to the total number of nucleons) of approximately 17\%. The material is pre-irradiated at $\approx$ 90 K using a linear accelerator at the National Institute of Standards and Technology (NIST).  This ``warm'' dose produces paramagnetic centers that aid in the DNP polarization and changes the ammonia to a deep-shade of purple, see Figure~\ref{Sample}. The amount of time the material spends in a NIST accelerator is carefully monitored to produce an ideal number of paramagnetic centers. This corresponds to a density of approximately 10$^{19}$ spins per milliliter~\cite{PolHyper}. Increasing the number of centers also increases the number of spin-relaxation paths. The ``cold" dose radiation damage from the JLab electron beam adds additional paramagnetic centers beyond the ideal number causing a gradual decay in the polarization.  The effect of the cold dose is mitigated by warming up the ammonia to allow some of the extra paramagnetic centers recombine. This anneal process can be done a few times  before the material must be replaced.

The microwaves necessary for DNP are provided via  an extended interaction oscillator (EIO) tube and carried to the target material through a rectangular wave guide that impinges on the ammonia from a microwave horn. Two EIO tubes were used during the experiment: one for the 2.5 T field strength configuration at 70 GHz and one for the 5 T field strength configuration at 140 GHz. Each tube has a tunable frequency range of about 1\%~\cite{TargetPol} and this frequency is adjusted to polarize the protons positively or negatively. Smaller adjustments to the microwave frequency must be made constantly to maintain the polarization as the paramagnetic centers accumulate from the electron beam. Additional technical details on the E08-027 target cryostat are found in Ref~\cite{TargetPol}.

\subsection{Measuring Polarization}
The Zeeman interaction causes a particle with spin $S$ to split into 2$S$ + 1 energy levels when placed in a magnetic field $\vec{B}$. Each level is split by the energy 
\begin{equation}
h\omega_L = \vec{\mu} \cdot \vec{B}/S = g\mu_N B\,, 
\end{equation}where $g$ is the g-factor for the particle of spin $S$ and $\mu_N$ is the nuclear magneton. A RF-field applied at the Larmor frequency, $\omega_L$, flips the spin of the particle, as it absorbs or emits energy interacting with the field. The response of the polarized sample to the RF is characterized by its magnetic susceptibility and is a function of the RF-field frequency, $\omega$,
\begin{equation}
\chi(\omega) = \chi'(\omega) - i\chi''(\omega)\\,
\end{equation}
where $\chi'(\omega)$ is the dispersive term and $\chi''(\omega)$ is the absorptive part of the susceptibility~\cite{PolHyper}. The absorptive term is proportional to the absolute polarization of the sample through the following integral relation~\cite{Goldman}
\begin{equation}
\label{int}
P \propto \int^{\infty}_0 \chi''(\omega) d\omega\\.
\end{equation}
This absorptive term is measured by embedding a RF transmission coil in the polarized ammonia sample. The transmission coil is an LRC circuit with resonance frequency equal to that of the Larmor frequency of the proton.  The circuit is driven by an RF field that is swept through the Larmor frequency, creating a change in the inductance of the coil as the sample absorbs and emits energy. Measuring this change gives access to the absorptive term~\cite{PolHyper}. The proportionality constant in equation~\eqref{int} is found by comparing the NMR signal, in the absence of microwaves, to the calculated thermal equilibrium polarization, shown in equation~\eqref{TE}.

\section{The Hall A Spectrometers}
\begin{figure}[htp]
\begin{center}
\includegraphics[width=0.9\textwidth]{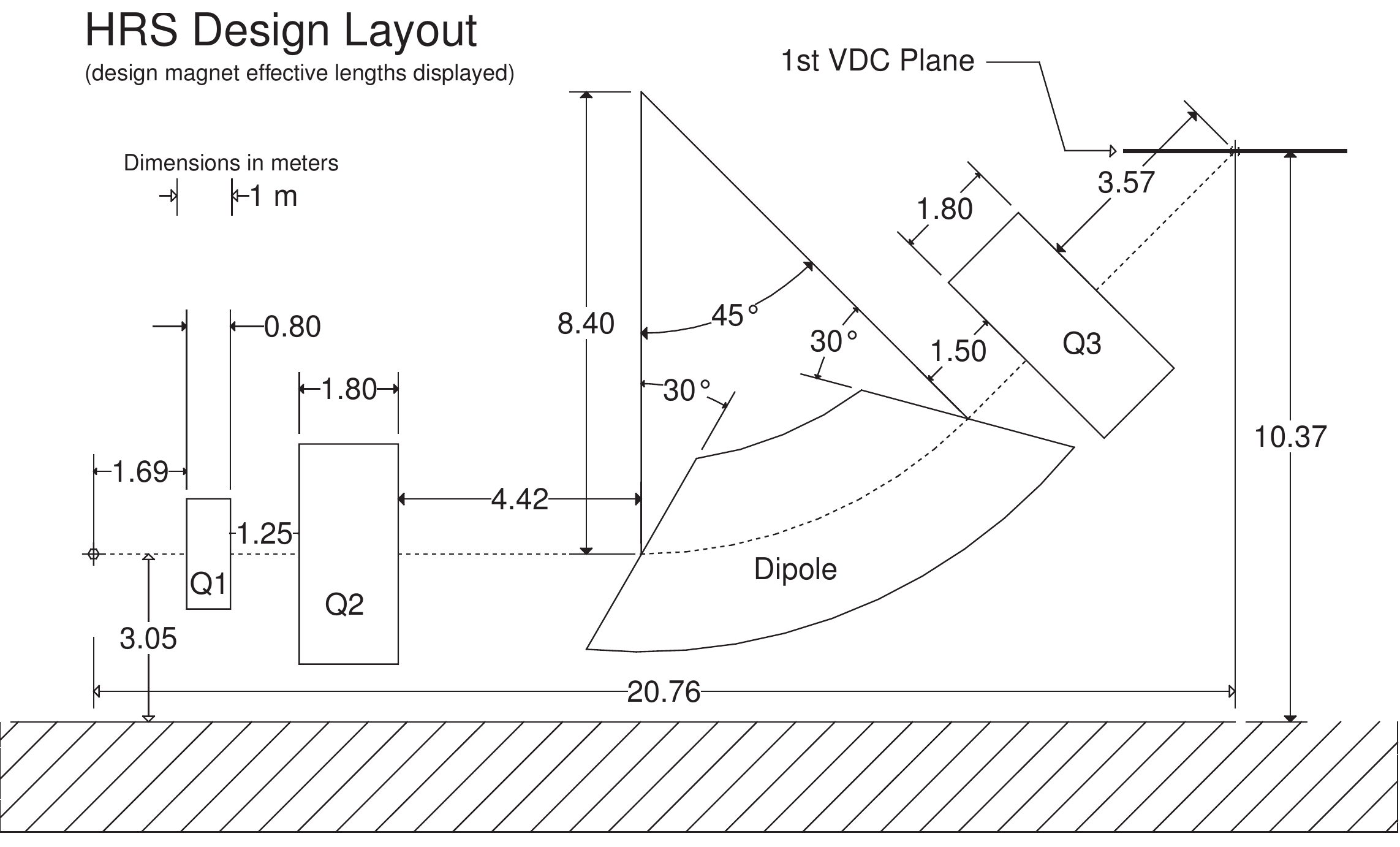}
\caption{\label{HRS}Spectrometer QQDQ magnet configuration, showing the trajectory of the scattered electrons (dotted line) into the detector stack, marked by the first VDC plane. The dipole magnet bends the electrons at 45$^{\circ}$ into the HRS detectors. Reproduced from Ref~\cite{CEBAF}. }
\end{center}
\end{figure}

Hall A contains two nearly identical magnetic spectrometers known as the high resolution spectrometers. Each spectrometer, referred to as LHRS and RHRS, consists of three superconducting quadrupole magnets and one superconducting dipole magnet in a QQDQ configuration as shown in Figure~\ref{HRS}\footnote{Not included in this diagram are the two septa magnets installed at the entrance to the Q1 quadrupole on each HRS.}. The quadrupoles focus the transported electrons, while the dipole determines the electron momentum that reaches the detector stack. The Q1 quadrupole focuses in the vertical plane and Q2 and Q3 in the transverse plane. The momentum resolution is at the 10$^{-4}$ level~\cite{CEBAF}, and the momentum acceptance is $\pm$4\% around the central value.

\subsection{Septa Magnets}
Located in front of the entrance to each spectrometer are two dipole septa magnets. Each septa consists of a set of top and bottom magnet coils, which bend electrons scattered  at approximately 6$^{\circ}$ into the spectrometers set at their minimum angular acceptance of 12.5$^{\circ}$.
\begin{figure}[htp]
\begin{center}
\subfigure[Septa diagram highlighting the damaged coils.]{\label{fig:sep_dia}\includegraphics[width=.50\textwidth]{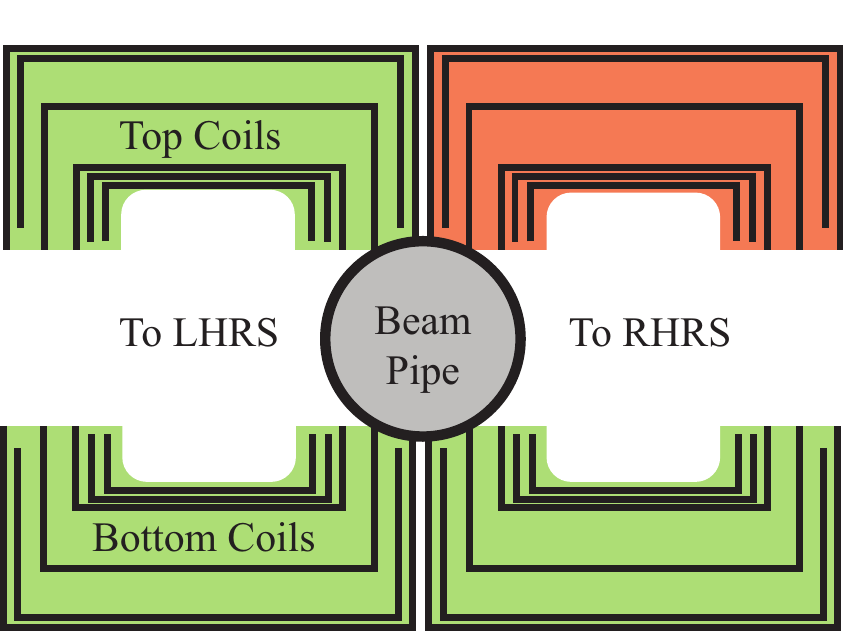}}
\qquad
\subfigure[Photo of the septa before installation in Hall A.]{\label{fig:sp_photo}\includegraphics[width=.35\textwidth]{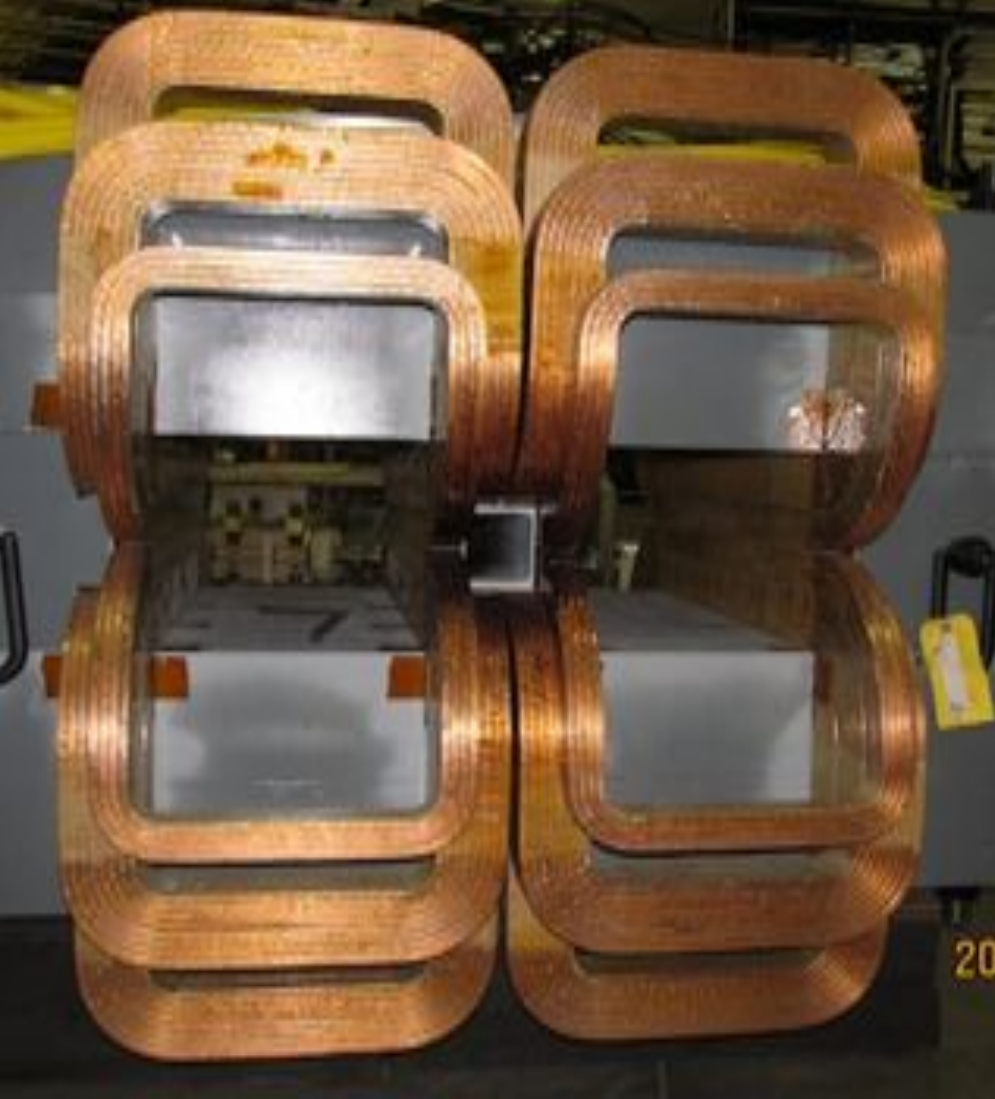}}
\caption{\label{septa_headon} The septa magnets for the E08-027 experiment. Each magnet had two sets of three wound coil packages. The design point is 48-48-16 turns of coil.}
\end{center}
\end{figure}
The scattered electron angle selected by the septa magnets is controlled by the field generated by the magnets and is produced by a single power supply to each HRS's septa. The septa current is also tuned with the dipole current to minimize acceptance effects. During the running of the experiment, the RHRS septa magnet was twice damaged, which lead to three different configurations of the magnet coils. This is shown in Figure~\ref{septa_headon}, where the damaged set of coils are highlighted in orange. Each coil set is a combination of three separate coils, as seen in Figure~\ref{fig:sp_photo}.   
\begin{table}[htp]
\begin{center}
\begin{tabular}{ l  c  c  r  }
\hline
Septa & Field (T) & Config & $E_0$ (GeV)   \\ \hline
 48-48-16& 2.5 &90$^{\circ}$ &2.2 \\
40-32-16& 2.5 & 90$^{\circ}$& 2.2 \\
40-00-16& 2.5 & 90$^{\circ}$ & 1.7 \\
40-00-16& 2.5 & 90$^{\circ}$ & 1.1 \\
40-00-16& 5.0 & 0$^{\circ}$& 2.2 \\
40-00-16& 5.0 & 90$^{\circ}$ & 2.2\\
40-00-16 & 5.0 & 90$^{\circ}$ & 3.3 \\ \hline
\end{tabular}
\caption{\label{septa_config}Septa coil configuration for the E08-027 experiment.}
\end{center}
\end{table}

The correspondence between kinematic setting and septa setting is shown in Table~\ref{septa_config}. The ideal situation is 48-48-16 turns of coil on the top and bottom coil packages and is valid for any run taken before 03/18/2012. The 40-32-16 configuration is valid from 03/18/2012 to 04/11/2012, and the 40-00-16 configuration is valid from 04/11/2012 until the end of the experiment.

\subsection{Detector Stack}
\begin{figure}[htp]
\begin{center}
\subfigure[RHRS calorimeters.]{\label{fig:shower}\includegraphics[width=.30\textwidth]{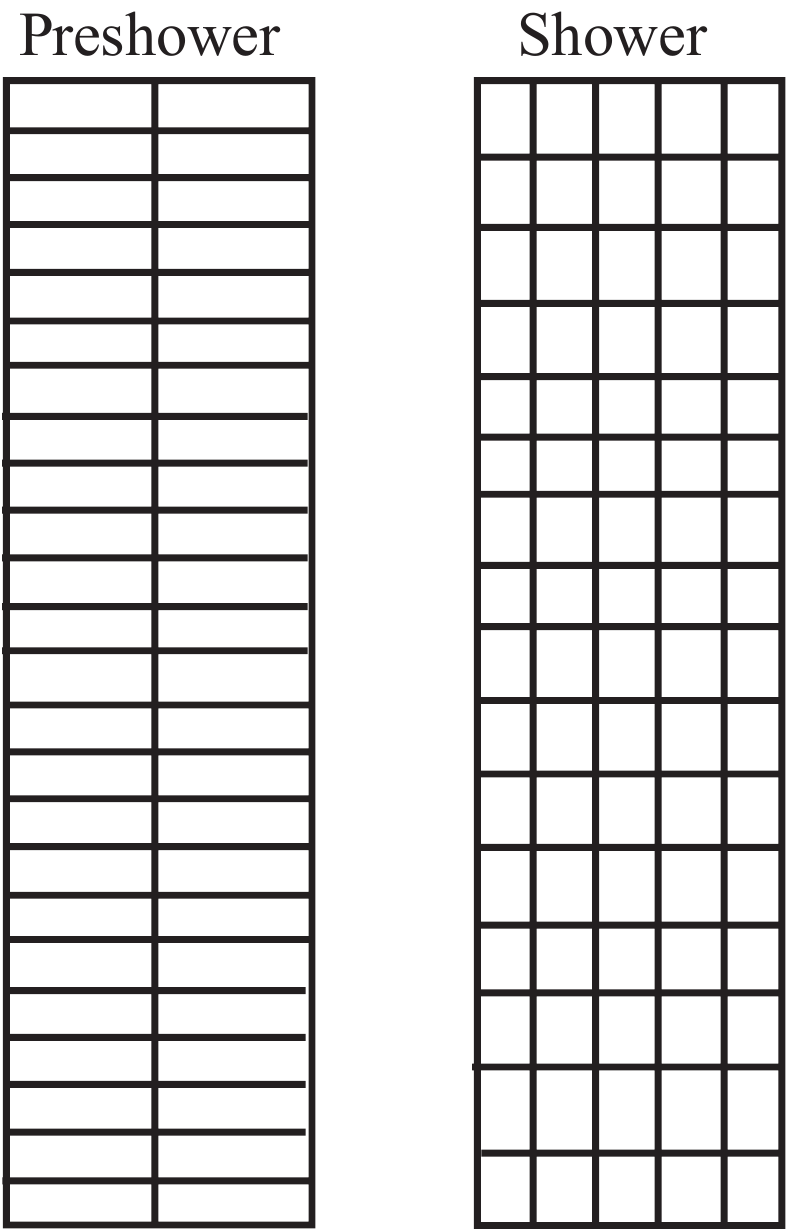}}
\qquad
\subfigure[RHRS detector stack.]{\label{fig:detector}\includegraphics[width=.60\textwidth]{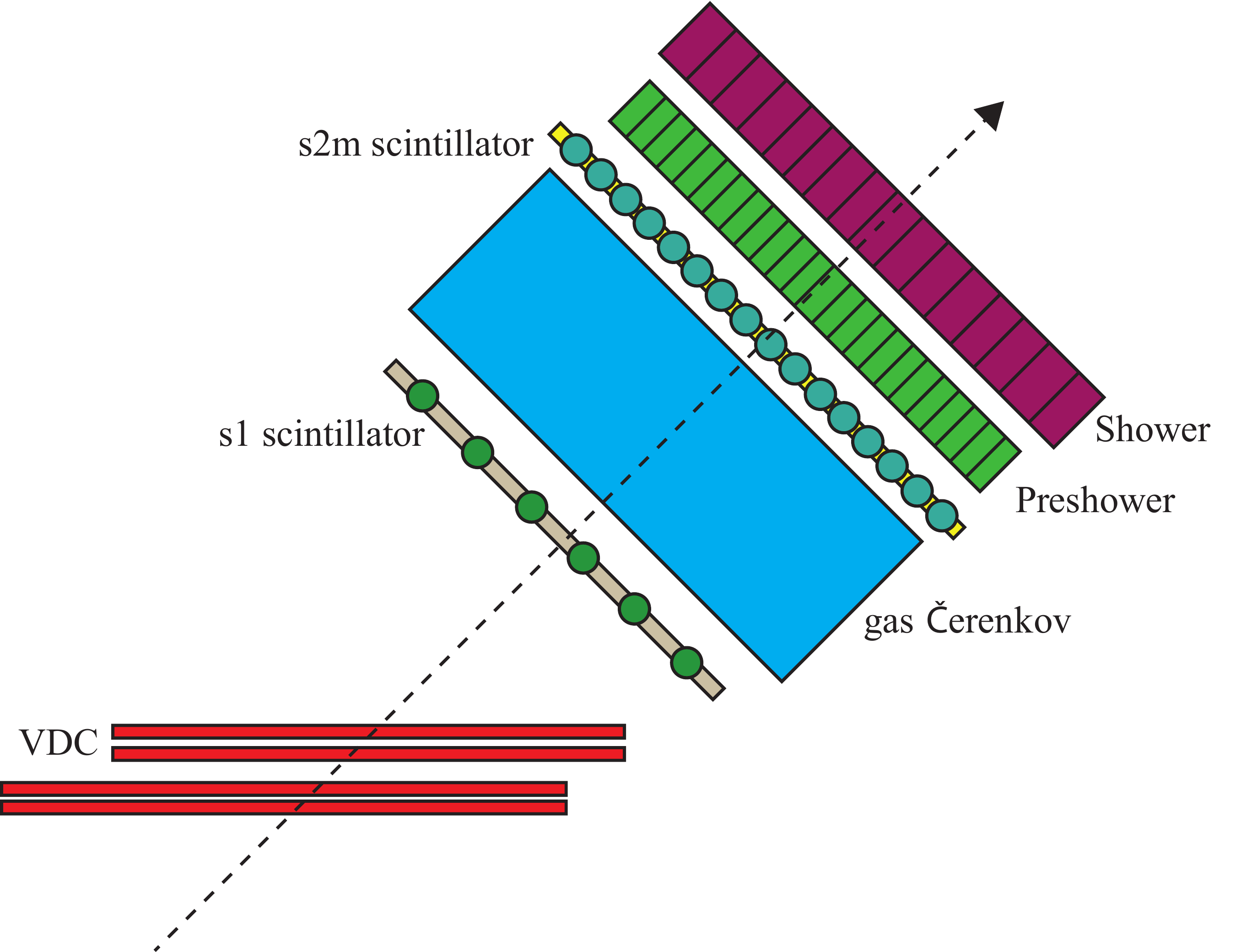}}
\caption{\label{Detectors}Frontal view of the lead glass blocks and side-view of the RHRS detector stack. The electron's trajectory is represented by the dotted line.}
\end{center}
\end{figure}

The electron detector stack for the RHRS is shown in Figure~\ref{Detectors}\footnote{The LHRS stack only differs in the layout of its electromagnetic calorimeters.}. Scattered electrons first pass through a pair of vertical drift wire-chambers (VDCs). The VDC's  offer good position and angular resolution for electron trajectory (track) reconstruction back to the target. Used in conjunction with the HRS dipole, the VDC's also provide electron momentum reconstruction. Next the electrons pass through a pair of segmented plastic scintillators (s1 and s2m), which form the data acquisition trigger. Particle identification is provided by a gas \v{C}erenkov detector and a two-layer electromagnetic calorimeter (Preshower and Shower). With the exception of the VDCs, the detector signals are read out using photomultiplier tubes.

\begin{figure}[htp]
\begin{center}
\includegraphics[scale=1.0]{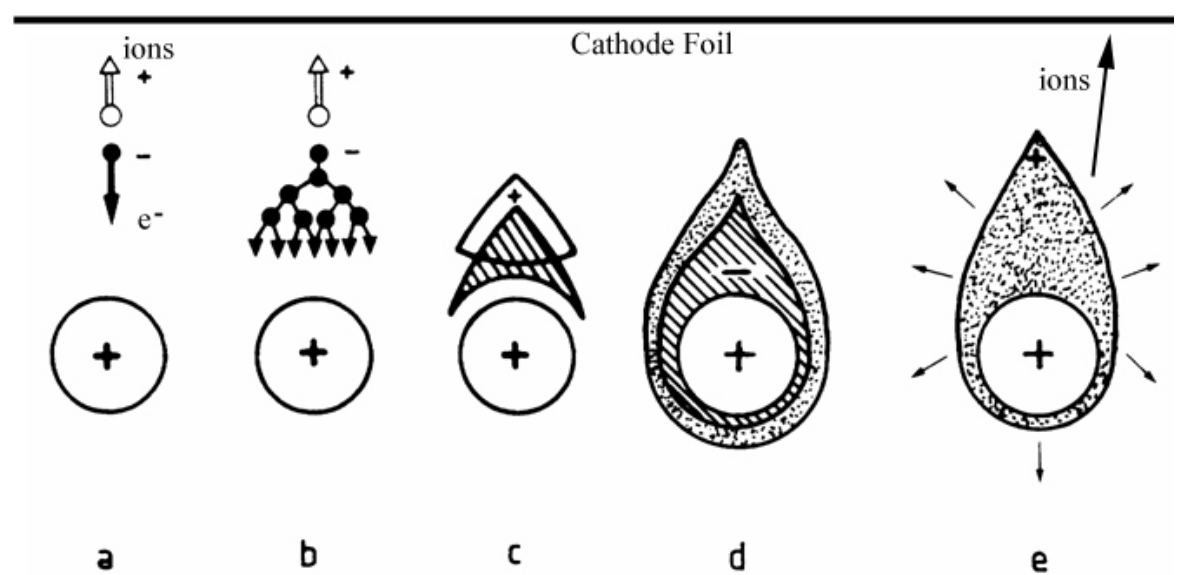}
\caption{\label{avalanche}Avalanche effect in drift chambers, and the creation of a detected signal. See text for a description. Figure reproduced from Ref~\cite{Charpak}.}
\end{center}
\end{figure}

\noindent{\bf Vertical Drift Chambers}\\
The vertical drift chambers are gas-filled cells containing an array of wires and a large electric field. The sense-wires are held at ground and as particles pass through drift chamber they ionize the gas. The process of creating a signal on the wires is shown in Figure~\ref{avalanche}. At (a) the primary electron travels towards the grounded sense wire. On its way to the wire, the electron, due to the increasing electric field, undergoes more ionizing collisions (b). The avalanche begins to develop (c) and as more electrons are rapidly collected (d) the remaining cloud of positive ions is stretched out towards the cathode (e).

The scattered electrons traverse several sense wires as they travel through the VDC. Typically the scattered electrons produce a cluster of four to six wires per plane. The smallest drift time as read-out via time-to-digital converters (TDC) corresponds to the shortest drift distance and thus the path of closest approach for the scattered electron and a given sense wire.  Using a linear fit of drift distances versus wire position the electron's trajectory is accurately reconstructed.

In order to keep the drift velocity as stable as possible a mixture of Argon (62\%) and Ethane (38\%) gas is used. Although any ionizable gas can be used, noble gases provide avalanche multiplication at lower electric fields. A polyatomic molecule such as ethane is used as a quencher. Due to the large number of non-radiated states in such molecules they absorb secondary emissions that would otherwise create spurious avalanches. Secondary emission occurs as the ions approach the cathode foil and gain an electron from it. In this process either a photon is emitted or an electron is re-emitted~\cite{Sauli}. 

\begin{figure}[htp]
\begin{center}
\includegraphics[scale=0.55]{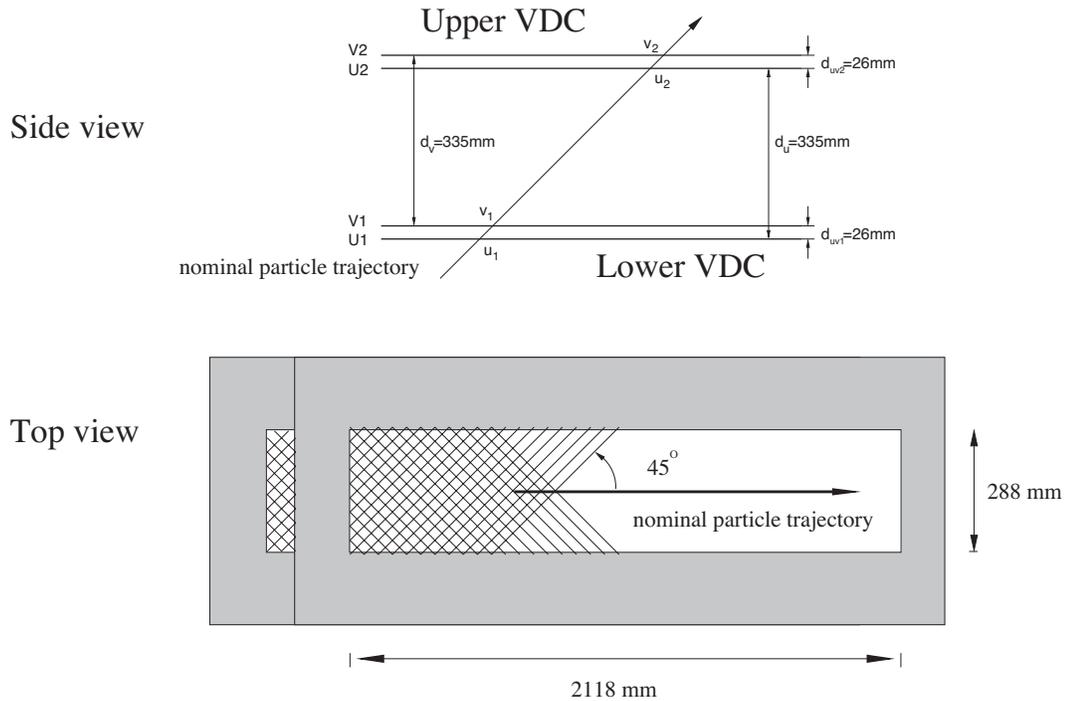}
\caption{\label{VDC}Layout of the HRS vertical drift chambers. Reproduced from Ref~\cite{CEBAF}.}
\end{center}
\end{figure}

The Hall A vertical drift chambers consist of four wire chambers in a U-V configuration as shown in Figure~\ref{VDC}. The UV chambers  consist of an upper and lower VDC.  The upper and lower chambers' sense wires are orientated 90$^{\circ}$ with respect to one another. The U and V planes lie within the horizontal plane of the hall and are at a 45$^{\circ}$ angle to the dispersive and non-dispersive directions. Each plane has 368 sense wires, spaced 4.24 mm apart. The spacing between the upper and lower UV planes is 335 mm. The VDC electric field is provided by a gold-plated Mylar foil held at $-$4 kV~\cite{CEBAF}.

\noindent{\bf Trigger Scintillators}\\
The s1 and s2m scintillators consist of segmented plastic scintillator readout on either side via photomultiplier
tubes (PMTs). The s1 detector has six 0.5 cm x 30 cm x 36 cm pieces of scintillator in a 1 x 6 arrangement
and the s2m scintillator has sixteen 43.2 cm x 5.1 cm x 14 cm scintillators in a 1 x 16 arrangement. In the
detector hut, the two scintillating planes are separated by about two meters~\cite{CEBAF}.  A logical AND between the two scintillator planes forms the data acquisition trigger.

Photomultiplier tube voltages for the scintillators are determined using cosmic rays. The flux
of cosmic ray muons at sea level is approximately 1 cm$^{-2}$ min$^{-1}$~\cite{PDG}; PMT voltages are set to give the
appropriate rate on the scintillator given its dimensions. The raw PMT signals are split: one copy is
sent to the DAQ analog-to-digital converters (ADC) and the other is sent to a discriminator for use in the
trigger logic and DAQ time-to-digital converters.

\begin{figure}[htp]
\begin{center}
\includegraphics[scale=0.45]{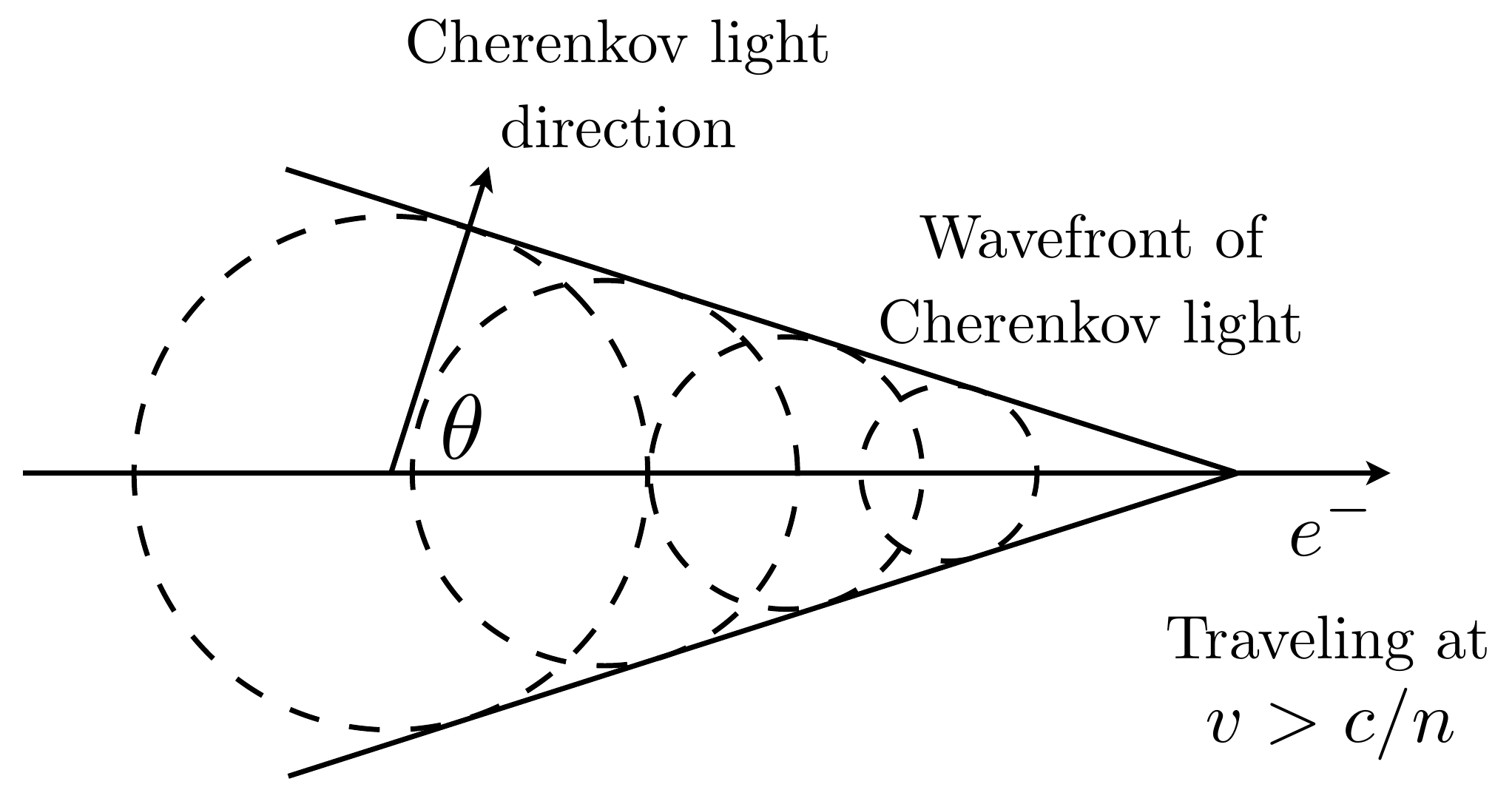}
\caption{\label{GasCerRad}\v{C}erenkov light is emitted at an angle of cos$\theta$ = $\frac{1}{n\beta}$.}
\end{center}
\end{figure}

\noindent{\bf Gas \v{C}erenkov}\\
Located between the two scintillator planes is the gas \v{C}erenkov detector, which uses the production of \v{C}erenkov light in CO$_2$ to distinguish electrons from other negatively charged particles. \v{C}erenkov radiation occurs when a particle moves through a medium faster than the phase velocity of light in the same medium. The threshold for the creation of \v{C}erenkov light is
\begin{align}
\beta  &\ge 1/n\,, \\
p &= \frac{mc}{\sqrt{n^2-1}}\,,
\end{align}
where $n = 1.00041$ is the index of refraction for CO$_2$. This corresponds to a momentum threshold of 0.017 GeV/c for electrons and 4.8 GeV for pions~\cite{GASC}. Given the momentum acceptance of the HRS, this means that only electrons will produce a signal in the detector. The \v{C}erenkov light is emitted in a conical shape as seen in Figure~\ref{GasCerRad}.

\begin{figure}[htp]
\begin{center}
\includegraphics[scale=0.45]{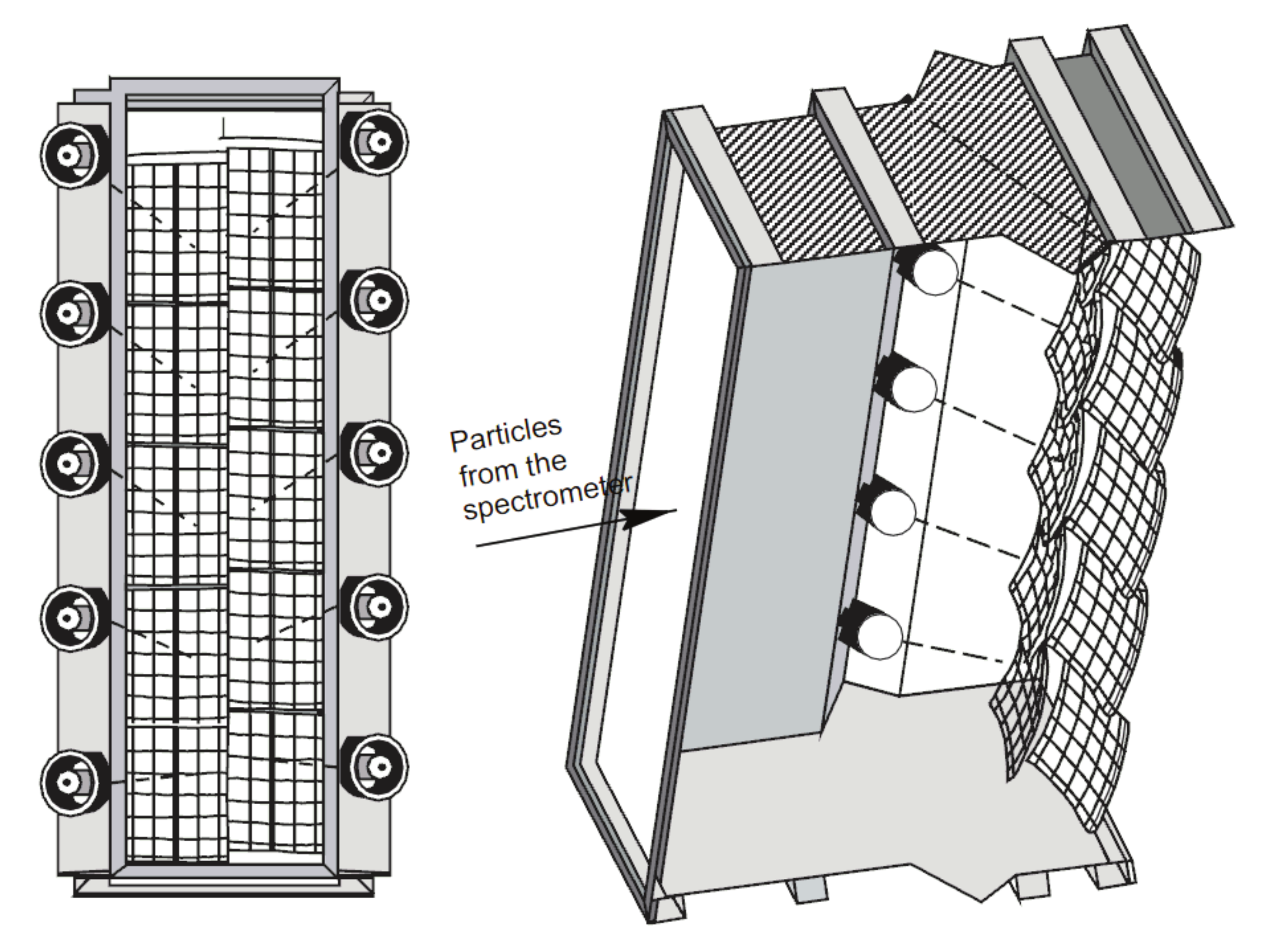}
\caption{\label{GasCer}Hall A HRS gas \v{C}erenkov detector. Reproduced from Ref~\cite{GASC}.}
\end{center}
\end{figure}

The gas \v{C}erenkov detector is shown in Figure~\ref{GasCer} and consists of ten spherical mirrors placed at one end of the detector\footnote{The angle of emitted radiation is cos$\theta$ = $\frac{1}{n\beta}$, so the majority of radiation travels in the forward direction.} to focus the  \v{C}erenkov radiation onto ten PMTs (dotted lines in Figure~\ref{GasCer}). There is a partial over-lap of the mirrors to fully capture all of the produced light. The signal on each PMT is run through an analog-to-digital converter (ADC) and then summed to produce the total light output from a single electron event. While pions do not directly produce a signal in the detector, they can produce electrons through ionizing collisions with the CO$_2$. These potential background events are removed with a lead-glass calorimeter.

\noindent{\bf Electromagnetic Calorimeters}\\
The electromagnetic calorimeters use a collection of lead glass blocks to induce pair production and bremsstrahlung radiation. Energetic particles create a cascade of photons and electron-positron pairs when they enter the calorimeters. The light from this cascade is detected using PMTs and is linearly proportional to the energy deposited.

The lay-out and number of lead glass blocks  differs slightly between the LHRS and RHRS. The LHRS calorimeter is two layers of thirty-four blocks, with the face of each block perpendicular to the particle's trajectory. The blocks in the first layer are 14.5 cm $\times$ 14.5 cm  $\times$ 30 cm and the blocks in the second layer are 14.5 cm  $\times$ 14.5 cm  $\times$ 35 cm. The RHRS calorimeter is two layers of forty-eight and eighty blocks, respectively. The blocks in the first layer are 10 cm $\times$ 10 cm  $\times$ 33 cm and the blocks in the second layer are 14.5 cm  $\times$ 14.5 cm  $\times$ 35 cm. The larger number of lead glass blocks on the RHRS means that it is a ``total energy absorber", i.e a passing particle deposits its energy fully in the calorimeter.

\section{Data Acquisition System and Trigger}
\label{DAQTRIGGER}
E08-027 used a few different singles triggers and each HRS operated its own data acquisition system (DAQ). For the experiment each HRS added an additional  hardware crate (four total on each arm) for the ADCs, TDCs and scalers as compared to previous experiments.  This helped to improve the DAQ deadtime, and is discussed in Appendix~\ref{app:Appendix-A}. The experiment also integrated  the HAPPEX DAQ into the standard HRS DAQ, for an additional and higher precision measurement of the beam charge and BPM information. Originally designed for the HAPPEX experiment~\cite{happex}, the DAQ used 18-bit ADCs as compared to the 12/13 bit FASTBUS ADCs.

\subsection{Data Trigger}
An electron passing through the s1 and s2m scintillator planes defines the main, T1 (T3) RHRS (LHRS) spectrometer trigger. It is formed as the logical AND of the following:
\begin{itemize}
\item The left or right PMTs of a scintillator segment of s1 fire
\item The left or right PMTs of a scintillator segment of s2m fire
\end{itemize}
The event must cause both s1 and s2m to fire, and no restriction is made on which scintillator segments fire.

\begin{figure}[htbp]
\begin{center}
\includegraphics[scale=.72]{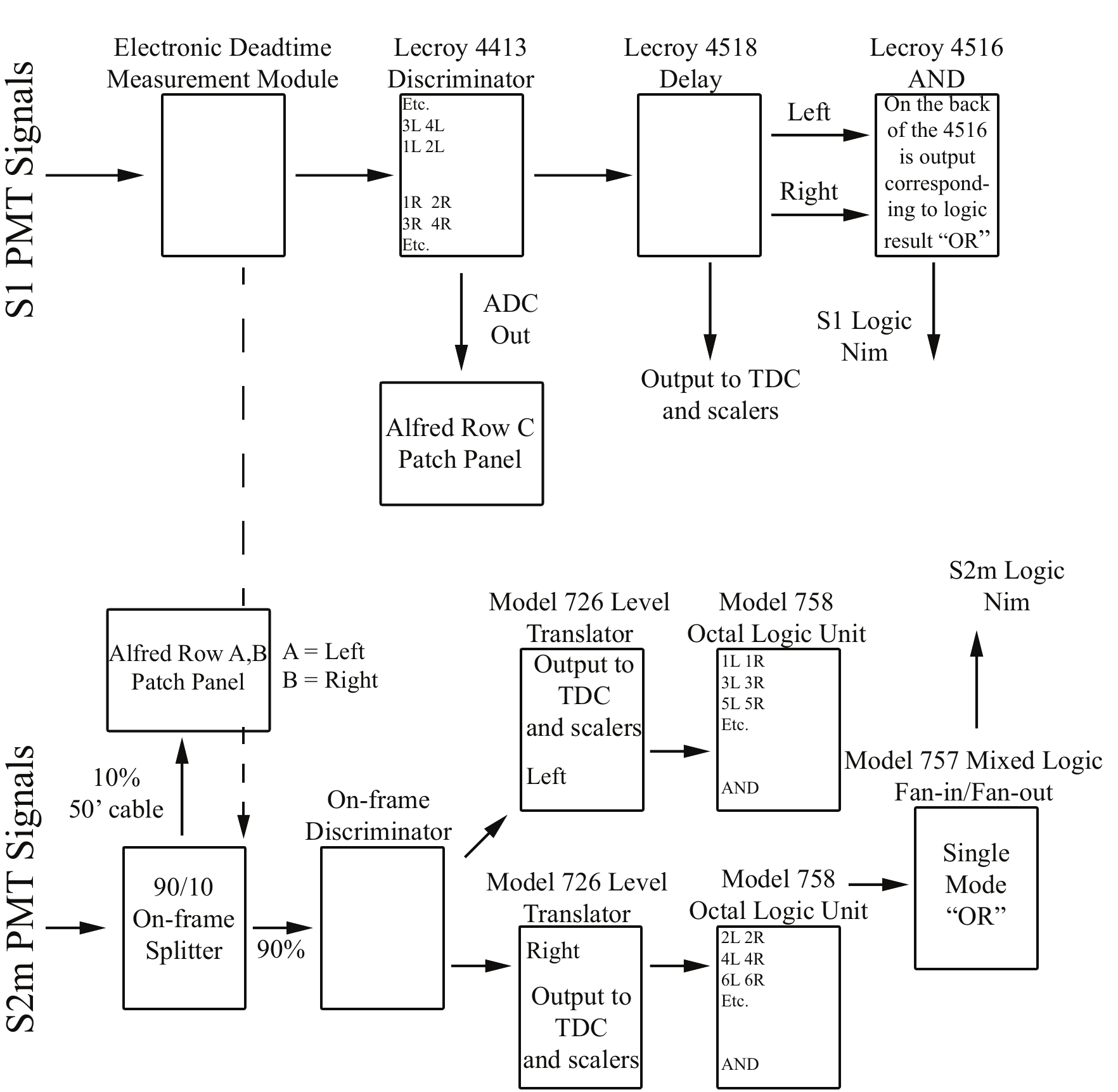}
\caption{Trigger logic for the s1 and s2m scintillators on the LHRS. The discriminator modules create the logic pulses for trigger formation. Delay is added to the s1 logic in order to align the detector signals at the trigger level.}
\label{fig:Trigger}
\end{center}
\end{figure}
The trigger logic of the scintillators on the LHRS is shown in Figure~\ref{fig:Trigger}.  The trigger modules are a combination of NIM and CAMAC electronics crates. The patch panels run the detector signals from the trigger modules to the data acquisition modules. The logic is identical on the RHRS, but the module layout is slightly different. The s1 logic defines the timing of the trigger; delay is added to the s2m logic signal to make it arrive $\approx$ 40 ns after the leading edge of the s1 signal. The Electronic Deadtime Measurement (EDTM) Module inserts a constant frequency signal into the raw s1 and s2m PMT signals. The EDTM signal is also sent to a TDC. The electronic deadtime is estimated by comparing EDTM events seen versus expected.

A secondary trigger, T$_2$ (T$_4$) for the RHRS (LHRS) measures the efficiency of the main trigger. Formed exclusive to the main trigger, the efficiency trigger is defined as the logical AND of the following:

\begin{itemize}
\item Either the s1 OR s2m scintillator planes fire but not both
\item The event also led to a signal being detected in the gas \v{C}erenkov
\end{itemize}

\begin{figure}[htbp]
\begin{center}
\includegraphics[scale=.70]{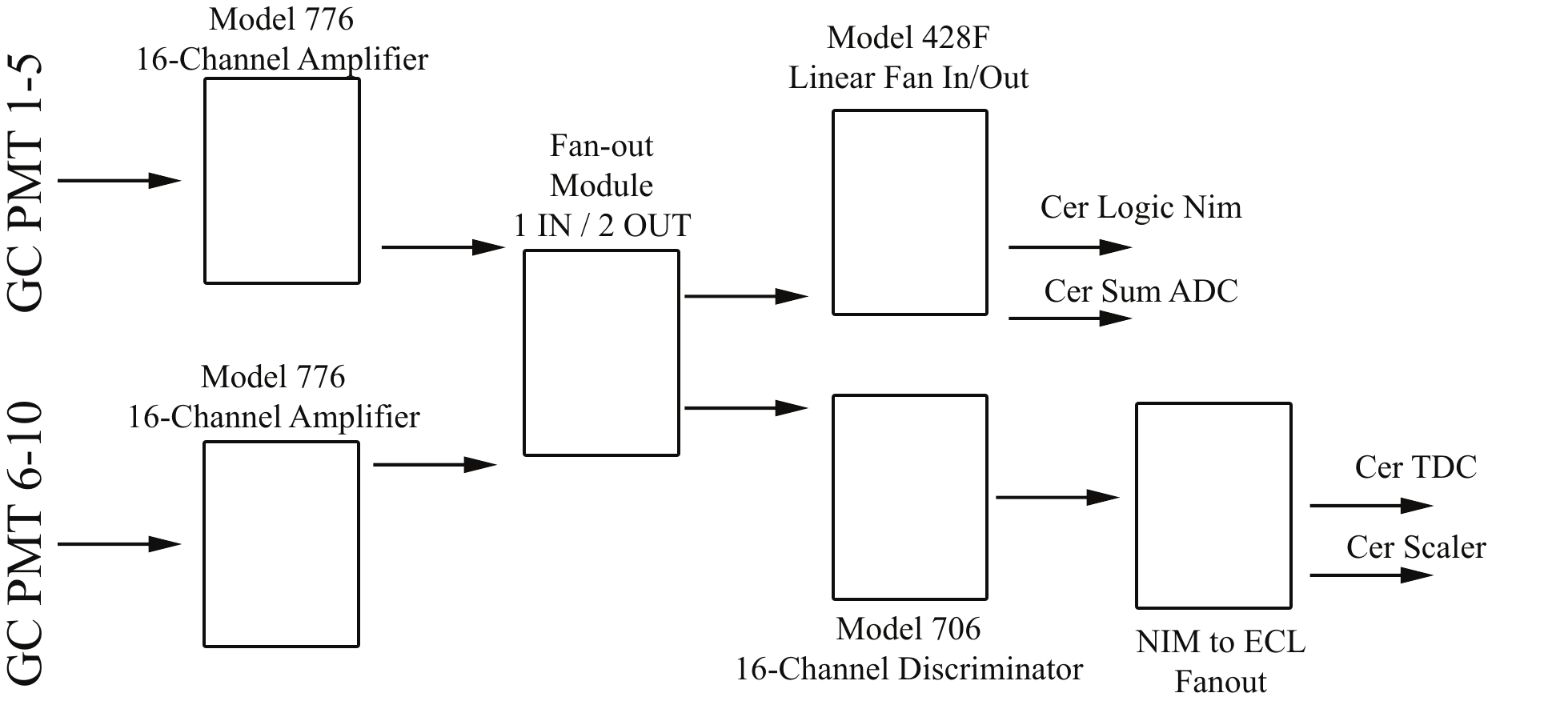}

\caption{NIM logic for the gas \v{C}erenkov trigger, TDC and scaler signals.}
\label{fig:Cer}
\end{center}
\end{figure}

\noindent The first requirement excludes main triggers while the second defines events that should have been been detected by both scintillator planes. The creation of the gas \v{C}erenkov logic is shown in Figure~\ref{fig:Cer}. The logic also creates a sum of all the \v{C}erenkov ADC signals, which is used for particle identification during analysis.

\begin{sidewaysfigure}[htbp]
\begin{center}
\includegraphics[scale=0.8]{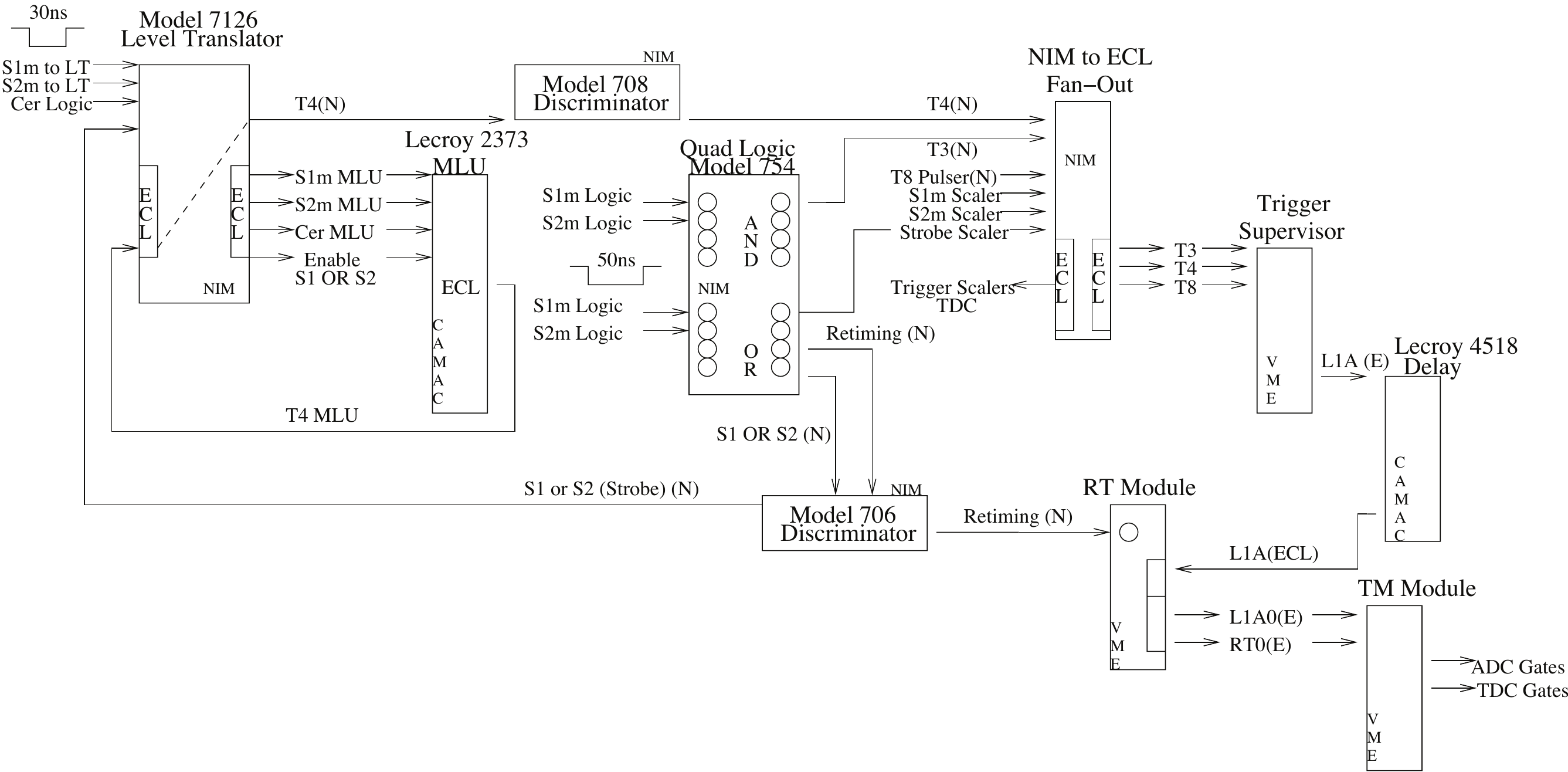}
\caption{E08-027 DAQ logic modules and trigger schematic.}
\label{fig:Tri}
\end{center}
\end{sidewaysfigure}

Two additional pulser triggers, T7 and T8, represent a 104 kHz and 1024 Hz clock, respectively. The slower clock calculates the scaler rates and the fast clock creates a time-stamp for each event. The layout and formation of the rest of the  trigger logic is shown in Figure~\ref{fig:Tri} and uses a combination of NIM, CAMAC and VME modules. A programable Lecroy 2373 Majority Logic Unit (MLU) creates the efficiency trigger. A graphical user interface (GUI) controls the programming of the MLU.

The formed triggers are sent to the trigger supervisor (TS), which decides whether or not the data acquisition system records the event. This decision depends on whether or not the DAQ is busy processing a previous trigger/event, and the prescale ($ps$) factor\footnote{ For every $ps$ event, only one is processed by the DAQ.}. In this manner, the TS maintains system busy, and acts as the link between the trigger system and hardware crates. If the event is accepted, the TS issues a level-one accept (L1A) signal. The TS can accept multiple triggers during the generation of the L1A signal. A trigger bit pattern for accepted events provides a map of which triggers co-existed for a given event. A TDC records the bit-pattern, with one trigger per channel.

The use of a re-time signal\footnote{The re-time signal is usually referred to as the ``strobe."} in Figure~\ref{fig:Tri} helps align the time signatures of the different triggers. The signal is the OR of the s1 and s2m detector signals. The Re-timing (RT) module ensures that there is an RT signal for every L1A, keeping the DAQ synchronized. Under normal operation, the timing of the L1A signal is adjusted to arrive $\approx$ 60 ns before the RT signal. If the L1A arrives after the RT or the L1A arrives greater then 200 ns before, the timing of the output (RTO) is shifted. This shift is detectable in the TDC spectrum. The gates and stops generated in the Transition Module (TM) for the ADCs, scalers and TDCs are based on an overlap from the L1AO and RTO signals. The arrival of the gates and stops to the front-end detector modules from the TM, signals the front-end electronics to take data.

\subsection{Data Acquisition}

The Hall A DAQ software is referred to as CODA (CEBAF online data acquisition)~\cite{CODA}. Data is digitized from the front-end hardware modules and then is built up into events and recorded by CODA. For the two spectrometers, the detector digitization occurs in FASTBUS (ADCs and TDCs) and VME (scalers and trigger supervisor) modules that are housed in hardware crates, referred to as Read-Out Controllers (ROCs). The ROC is actually a software routine that runs in a CPU located in the same crate as the digitizing electronics.

The FASTBUS ADCs and TDCs in the crates are a collection of Lecroy 1881/1881M and LeCroy 1877s modules, respectively. The scaler modules are SIS3801. Each HRS has three FASTBUS crates and one VME crate. The VME crate also runs a ring buffer server to store the scaler results of 1000 helicity windows, which is read-out every 50 physical events to reduce DAQ deadtime.  The data from the HAPPEX DAQ is stored in its own ring buffer with read-out controlled by the trigger supervisor. 
\begin{figure}[htbp]
\begin{center}
\includegraphics[scale=.80]{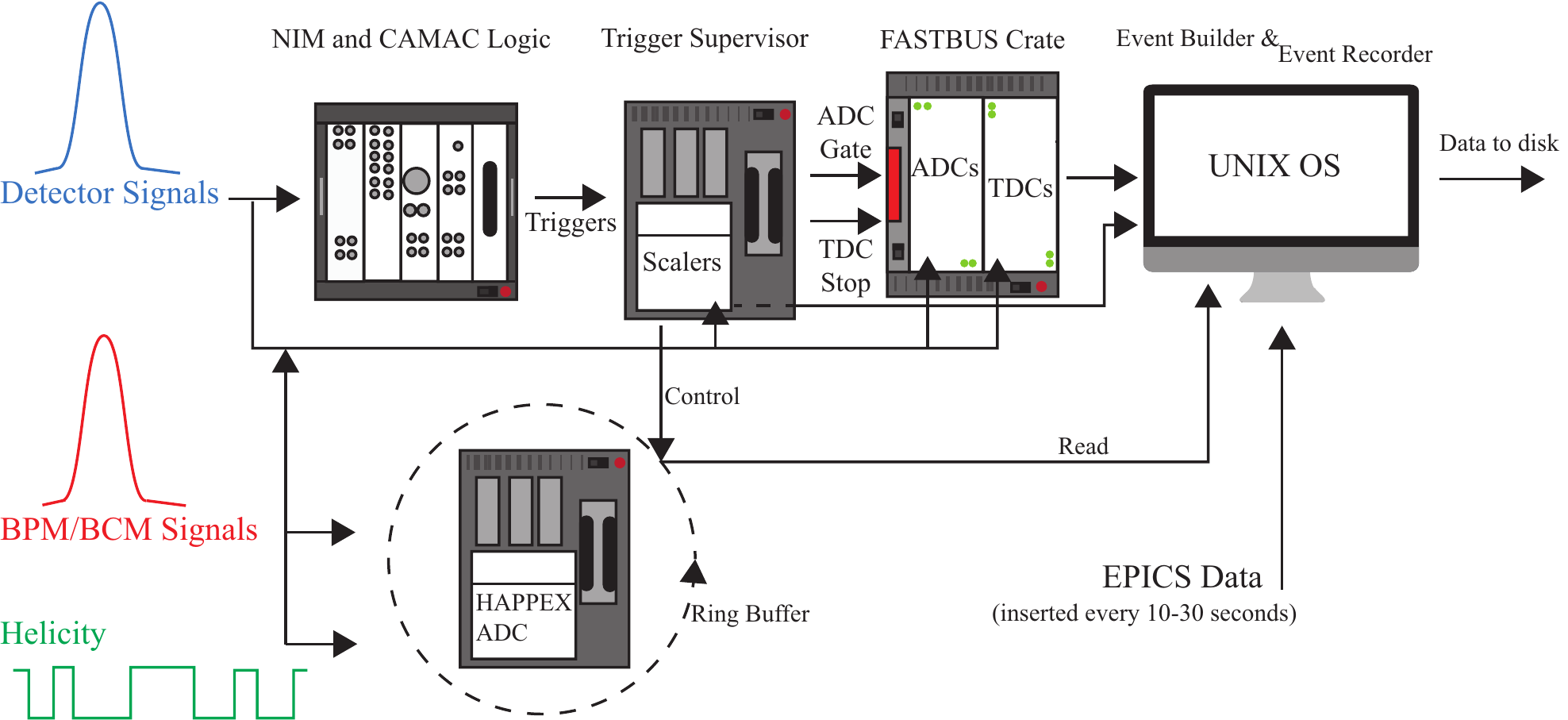}

\caption{DAQ Process: From raw detector signals to recorded events. The incoming electron information and scattered (detected) electron information is combined in software via the event builder and recorder. The trigger supervisor coordinates the readout of the two information streams.}
\label{DAQ}
\end{center}
\end{figure}

The data are processed in parallel between each ROC and the HAPPEX DAQ. The data fragments are collected, ordered and merged together via the CODA Event Builder. The Event Builder passes this formation to the Event Recorder for writing to disk. The whole data acquisition process is shown in Figure~\ref{DAQ}. EPICS\footnote{The Experimental Physics Industrial Controls software is an open source real-time control system for experiments~\cite{EPICS}.} data is information related to Hall A and acceleration instrumentation, and includes things such as the beam energy and HRS momentum.

\chapter{Analysis}
\label{ch:Analysis}
To extract the proton's spin-dependent cross sections, both asymmetries and unpolarized cross sections were measured. The experimental data was taken on the scattering of longitudinally polarized electrons from a transversely and longitudinally polarized NH$_3$ target.

\section{Decoding the Raw Data}
The standard Hall A analysis software is the Hall A Analyzer~\cite{Analyzer}. The Analyzer is a set of C++ libraries that sit on top of ROOT (itself a collection of C++ libraries specific to particle physics experimental analysis)~\cite{ROOT} and contains modules to perform analysis of standard Hall A equipment.  This analysis includes:

\begin{itemize}
\item Decoding raw CODA files into events
\item Particle tracking and reconstruction
\item Scaler decoding
\item ADC software gain for detector calibration
\end{itemize}
The translation between physical location of a detector ADC or TDC and location within the raw data file is done via a simple text file crate map input into the Analyzer.  This file also contains information on the scalers. General information on each run, such as beam energy and spectrometer angle is stored in a run database file. The variables written to the output ROOT file are controlled with an output definition file. For E08-027 an additional set of libraries called ``g2plib'' are added to the standard Analyzer and insert the BPM, BCM, raster, and helicty results into the replayed ROOT file.

\section{MySQL Analysis Database}
Additional static information on each run is stored in a MySQL database. The database allows for the sorting and selection of runs for the analysis and is split into tables for each HRS. The HRS tables are further divided into ``RunInfo" and ``AnaInfo" tables. The ``RunInfo" tables are snapshots of the start-of-run and end-of-run files on the Hall A electronic log book~\cite{HALOG}, while the newer ``AnaInfo" tables include information on run type and run quality, as well as a more accurate calculation of fields included in the original tables. The run-type field in the database sorts through the different calibration, optics, and production runs from the run period and each run is assigned a quality factor. The quality factor sorts through junk runs, runs that have potential problems, and good runs. Some criteria for junk runs include very short runs (less than a minute), runs where the DAQ malfunctioned or runs where the HRS magnet currents were not settled. The junk runs are avoided in the analysis. Runs with potential problems ($i.e$ one of the two BCMs was malfunctioning) are very likely OK to use in the analysis but should be checked for any outstanding issues. These analysis tables also provide a central location for the cross section normalization variables and the good-electron cuts. The database is accessible through a C++/Python library for analysis scripts and a web-interface~\cite{g2pmysql} for quick sorting. More information on the MySQL database is found in Ref.~\cite{MySQL}.

\section{Experimental Asymmetries}
\label{EAsymm}
The electron asymmetries are defined as the ratio of the difference in polarized cross sections to the sum of the polarized cross sections such that
\begin{equation}
\begin{aligned}
A_{\parallel} &=\frac{\frac{d^2\sigma}{d\Omega dE'} (\downarrow \Uparrow - \uparrow \Uparrow)}{\frac{d^2\sigma}{d\Omega dE'} (\downarrow \Uparrow + \uparrow \Uparrow)}\,,\\
A_{\perp} &=\frac{\frac{d^2\sigma}{d\Omega dE'} (\downarrow \Rightarrow - \uparrow \Rightarrow)}{\frac{d^2\sigma}{d\Omega dE'} (\downarrow \Rightarrow + \uparrow \Rightarrow)}.
\end{aligned}
\label{test69}
\end{equation}
The direction of the target polarization ($\Uparrow$ or $\Rightarrow$) with respect to the electron spin ($\uparrow$ and $\downarrow$) define the parallel and transverse asymmetries. The cross section normalization is the same for the numerator and denominator in equation~\eqref{test69}, so each asymmetry is proportional to the ratio of the difference of the raw electron events corresponding to the electron's  positive or negative helicity state, 
\begin{align}
\label{eq:Asymm}
A^{\mathrm{meas}} &=\frac {Y_+ - Y_-}{Y_++Y_-}\,,\\
A^{\mathrm{exp}} &= \frac{1}{f \cdot P_t \cdot P_b}A^{\mathrm{meas}}\,,\\
Y_{\pm}& = \frac{N_{\pm}}{LT_{\pm}Q_{\pm}}\,,
\end{align}
where $P_t$ and $P_b$ are the target and beam polarization respectively,  and $f$ is the dilution factor. The dilution factor accounts for scattering from unpolarized material in the ammonia target, such as nitrogen, helium and the aluminum caps on the target cups. The raw electron counts, $N_+$ and $N_-$, must also be corrected for false asymmetries related to the charge, $Q_{\pm}$, and DAQ livetime, $LT_{\pm}$. Charge asymmetries occur when unequal numbers of electrons corresponding to each helicity state scatter off the target, and livetime asymmetries occur if one helicity state is preferentially accepted by the trigger supervisor.

\section{Experimental Cross Sections}
The unpolarized measured cross section is defined according to
\begin{equation}
\sigma^{\mathrm{meas}} = \frac{d\sigma^{\mathrm{meas}}}{d\Omega dE'} = \frac{ps N}{N_\mathrm{{in}}\rho LT \epsilon_{\mathrm{det}}} \frac{1}{\Delta \Omega \Delta E' \Delta Z}\,,
\end{equation}
where:
\singlespacing
\begin{itemize}
\item $N$ is the number of scattered electrons recorded in the detectors defined by acceptance/detector cuts
\item $ps$ is the prescale factor
\item $N_\mathrm{{in}}$ is the number of incident electrons determined from the total BCM charge accumulated over a run
\item $\rho$ is the target density
\item $LT$ is the DAQ livetime correction factor 
\item $\epsilon_{det}$ is the product of all hardware and software detector efficiencies
\item $\Delta Z$ is the target length seen by the spectrometer
\item $\Delta E'$ is the scattered electron energy spread
\item $\Delta \Omega$ is the angular acceptance of the spectrometer 
\end{itemize}
\doublespacing
The unpolarized proton cross section is formed by removing the diluting contributions from scattering from material other than protons in the raw cross section. Breaking the measured cross section down into components gives
\begin{equation}
\frac{d\sigma^{\mathrm{exp}}}{d\Omega dE'} = \sigma^{\mathrm{exp}} = \sigma^{\mathrm{^1H}} = f\sigma^{\mathrm{meas}}= f( \sigma_{\mathrm{^1H}} +  \sigma_{\mathrm{^{14}N}} +  \sigma_{\mathrm{^4He}} +  \sigma_{\mathrm{^{27}Al}})\,,
\end{equation}
where $f$ is the dilution factor, and it is the same dilution factor defined in the asymmetries
\begin{equation}
f = \frac{\sigma_{\mathrm{^1H}}}{\sigma_{\mathrm{^1H}} +  \sigma_{\mathrm{^{14}N}} +  \sigma_{\mathrm{^4He}} +  \sigma_{\mathrm{^{27}Al}}}\,.
\end{equation}

The polarized cross section differences are calculated from the product of the experimental asymmetries and unpolarized cross sections,
\begin{equation}
\label{eq:poldiff}
\Delta\sigma^{\mathrm{exp}}_{\parallel, \perp} = 2 \cdot A^{\mathrm{exp}}_{\parallel, \perp} \cdot \sigma^{\mathrm{exp}}\,.
\end{equation}
The Born cross section, used in the evaluation of physics quantities such as the polarizabilities and sum rules, is determined after applying the radiative corrections,
\begin{equation}
\Delta\sigma^{\mathrm{Born}}_{\parallel, \perp} = \Delta\sigma^{\mathrm{exp}}_{\parallel, \perp}  - \Delta\sigma^{\mathrm{tail}}_{\parallel, \perp} + \delta(\Delta\sigma^{\mathrm{RC}}_{\parallel, \perp})\,,
\end{equation}
where $ \Delta\sigma^{\mathrm{tail}}_{\parallel, \perp}$ is the polarized elastic tail and $\delta(\Delta\sigma^{\mathrm{RC}}_{\parallel, \perp})$ are the polarized inelastic radiative corrections. The radiative corrections are discussed in detail in Chapter~\ref{ch:RC}.

\section{Detector Efficiency and Calibration Studies}
The detector efficiency correction to the cross section is a product of three separate detector efficiencies. These are the VDC efficiency, trigger efficiency and particle identification (PID) efficiency. The PID efficiency is broken down into a contribution from the gas \v{C}erenkov and lead glass calorimeters.

\subsection{Gas \v{C}erenkov Calibration}
\begin{figure}[htp]
\begin{center}
\includegraphics[scale=.55]{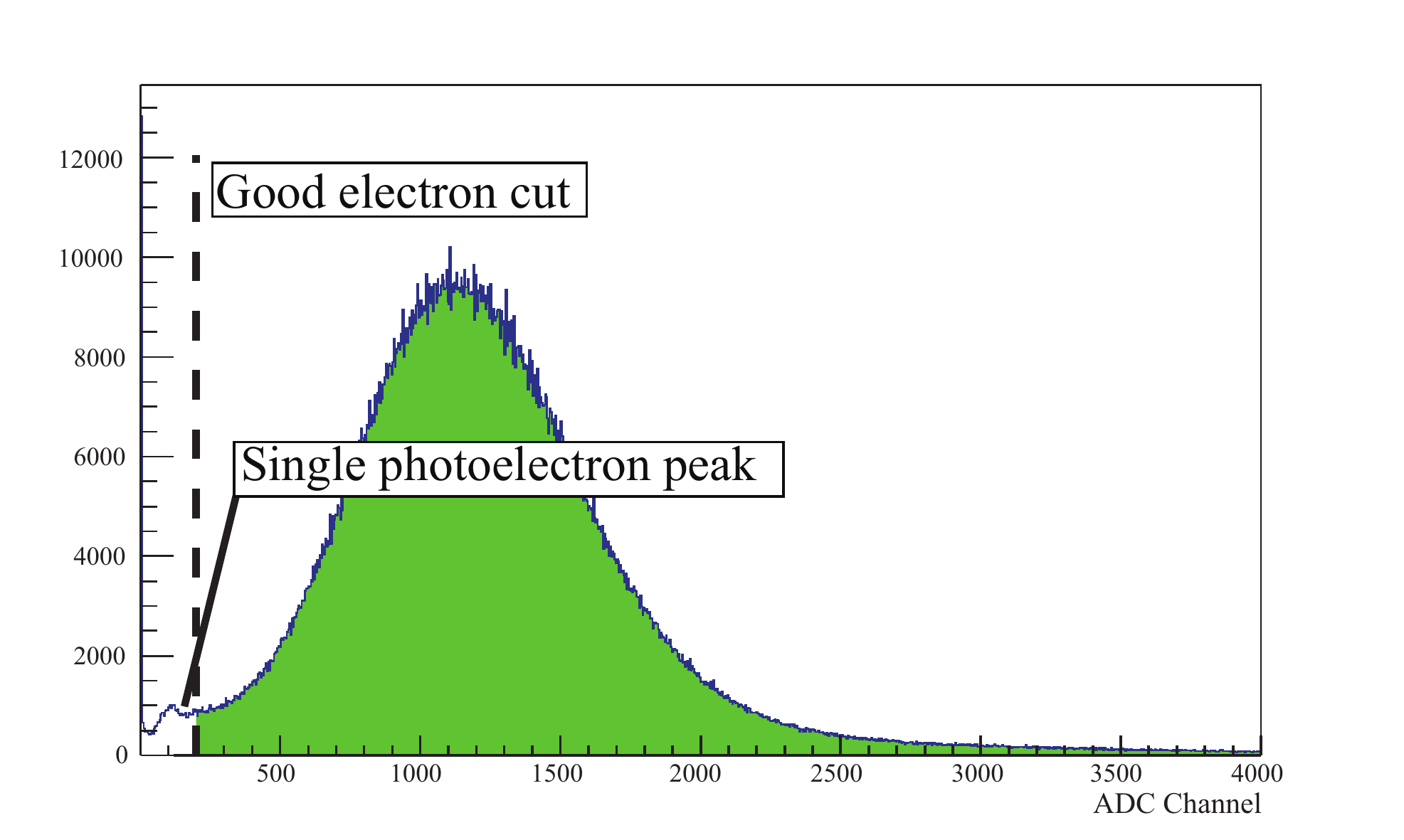}
\caption{Gas \v{C}erenkov ADC spectrum with PID cut. }
\label{fig:HRS_GC}
\end{center}
\end{figure}

Each of the 10 PMTs within the gas \v{C}erenkov have their own single photoelectron peak in the ADC spectrum. Single photoelectrons are events that result from secondary scattering within the detector or from noise (dark current) in the photomultiplier tube itself. Good electron events will produce multiple detected photoelectrons, so the single photoelectron peak is removed from the final analysis. To enhance the efficiency of removing these bad events, the ADC channels of the gas \v{C}erenkov have a software gain applied to align the single photoelectron peaks of each PMT. An example of the ADC spectrum, single photoelectron peak, and PID cut is shown in Figure~\ref{fig:HRS_GC}. Good electrons are located within the green shaded region in Figure~\ref{fig:HRS_GC}. Additional details on the gas \v{C}erenkov analysis are found in Ref.~\cite{Melissa}.

\subsection{Lead Glass Calorimeters Calibration}
To calibrate the lead glass calorimeters, the raw ADC signals from each PMT are converted into the energy deposited by the event. Each event creates a shower of secondary particles that is spread over multiple adjacent blocks, so the total energy of the event is proportional to the integration over the total detected signal. The differences in the layout of the lead glass calorimeters on the LHRS and RHRS leads to differences in the calibration technique for each spectrometer.
\begin{figure}[htp]
\centering     
\subfigure[LHRS]{\label{fig:a}\includegraphics[width=.6\textwidth]{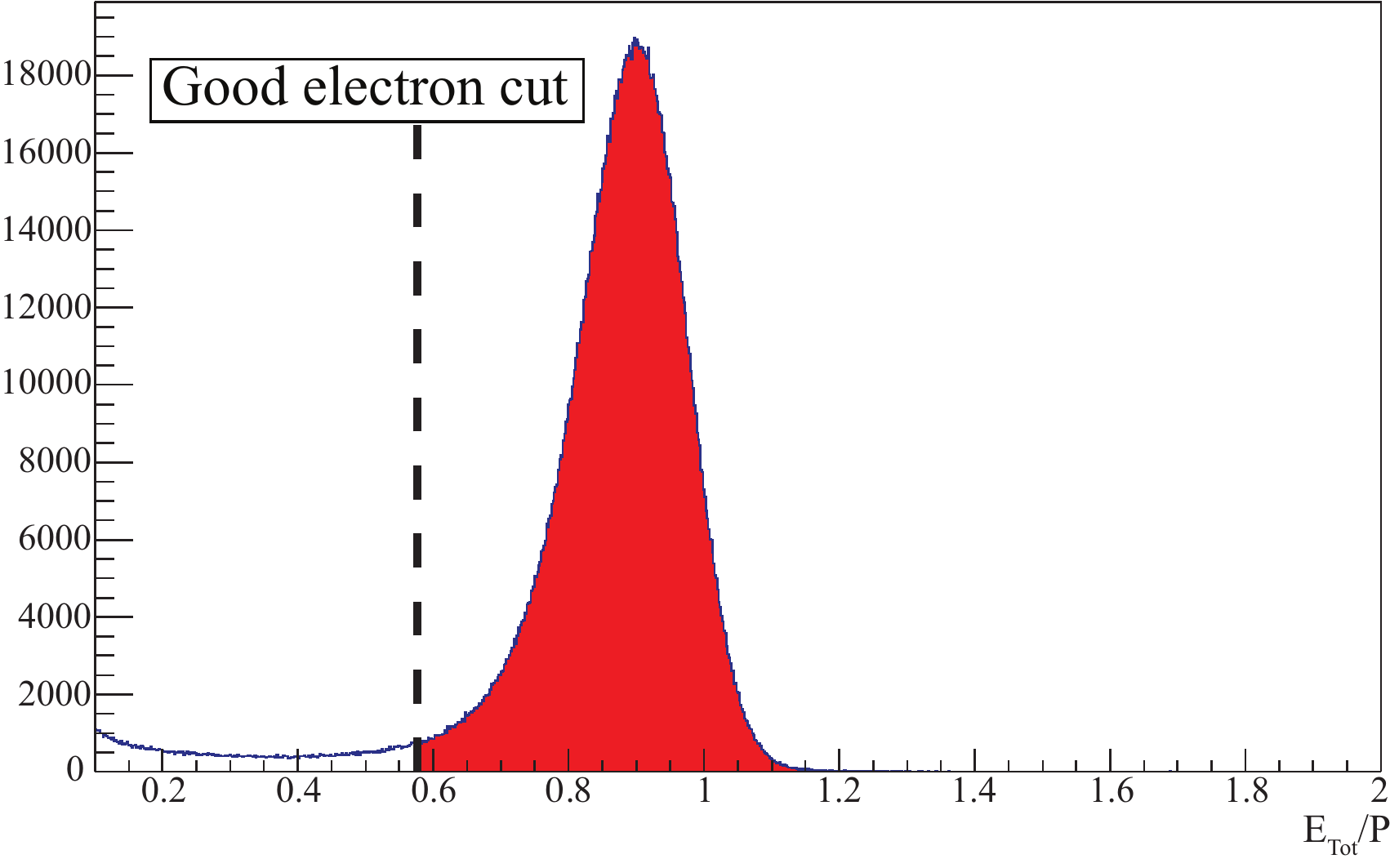}}
\qquad
\subfigure[RHRS]{\label{fig:b}\includegraphics[width=.6\textwidth]{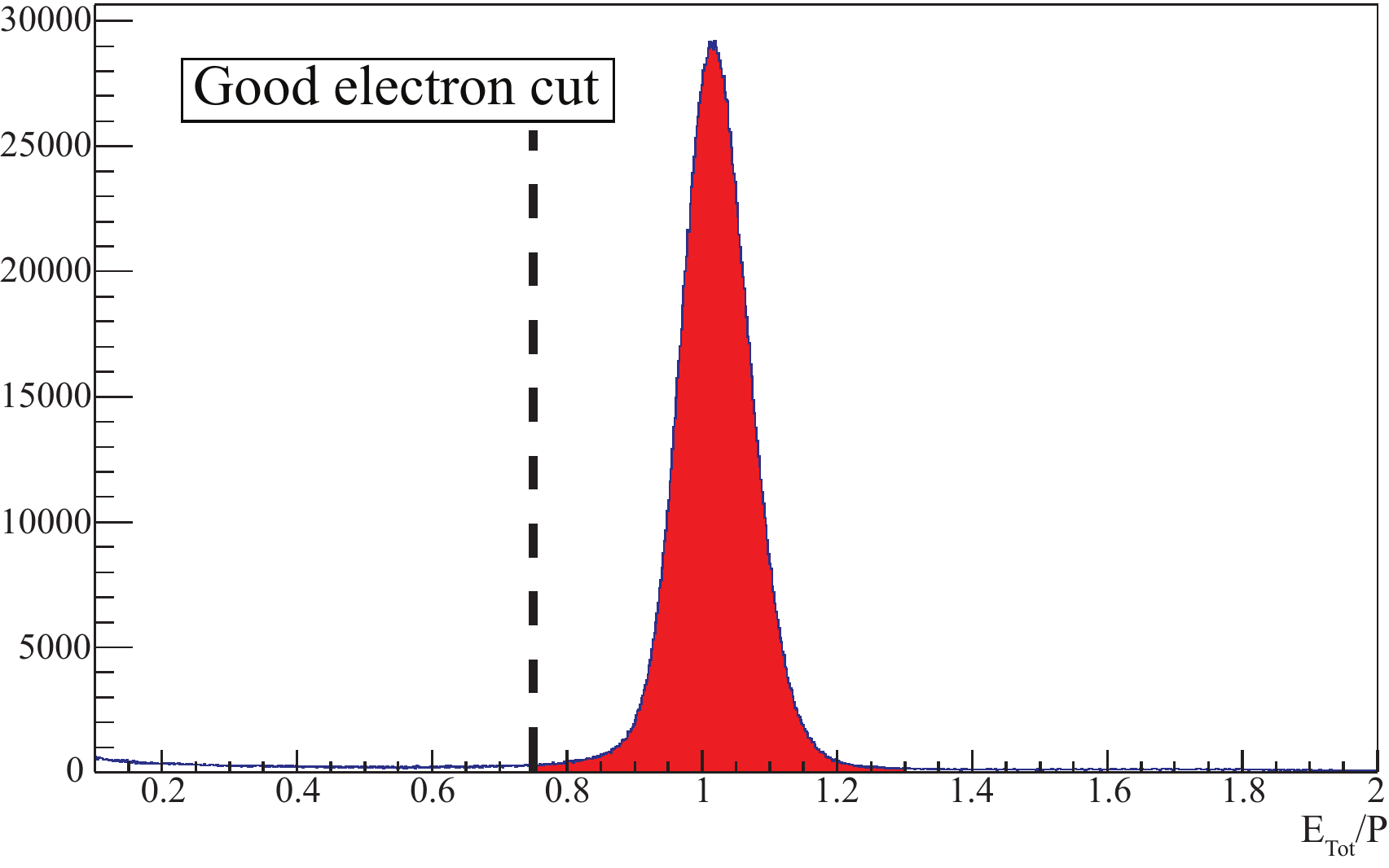}}
\caption{Energy deposited in the calorimeters with PID cut. The LHRS is not a total energy absorber so its signal peak is not centered around one.}
\label{fig:HRS_Sum}
\end{figure}

For the RHRS, the layer of lead glass is thick enough for an electron event to deposit all of its energy in the calorimeter. That means that for a pure electron sample the ratio of the total energy deposited in the calorimeter to the spectrometer central momentum, $P_0$, should be centered around one. The width of the distribution is the energy resolution. A $\chi^2$ minimization determines the calibration constants.

The LHRS is not a total energy absorber, and instead it is assumed that the energy shower obeys a gamma distribution
\begin{align}
\label{PRL}
\frac{dE}{dt} &= E_0\beta(\beta t)^{\alpha-1}\frac{e^{-\beta t}}{\Gamma(\alpha)}\,,\\
\frac{\alpha -1}{\beta} &= \mathrm{ln}(E_0) - \mathrm{ln}(E_c) - 1\,,
\end{align}
where $E_0$ is the electron momentum set by the spectrometer; $\beta$ is a free parameter, $E_c$ = 15.8 MeV is the critical energy, and $t$ is the radiation thickness of the calorimeter. The energy deposited in the LHRS calorimeters is proportional to integration of equation~\eqref{PRL}, where the proportionality constants, $\alpha$ and $\beta$, are the calibration constants. Additional details on the LHRS and RHRS lead glass calorimeter analysis are found in Ref.~\cite{Melissa}.


Good electron events are selected by placing a cut on the first calorimeter layer and also the sum of the two layers. The sum cut is shown in Figure~\ref{fig:HRS_Sum} for $E_0$ = 2253 MeV and $P_0$ = 2004 MeV.  Good electrons are located within the red shaded region in Figure~\ref{fig:HRS_Sum}. The calorimeter cuts are momentum dependent, in contrast with the gas \v{C}erenkov cut, which is constant across all kinematics. All of the PID cuts are uploaded to the MySQL database on a run-by-run basis for the analysis. The residual pion contamination after applying the three sets of PID cuts is negligible, with $\pi/e$ $<$ 0.0052 for all kinematic settings on both spectrometers~\cite{Melissa}.

\subsection{PID Efficiency}
The detection efficiency of the lead glass calorimeters is determined by selecting electron events that had a signal in the gas \v{C}erenkov and then counting how many of those same events are present in the calorimeters. The efficiency of the lead glass calorimeters is greater than 98\% and 98.8\% for the LHRS and RHRS respectively. A similar procedure determines the efficiency of the gas \v{C}erenkov: selecting electron events that have a signal in the calorimeters and then counting the number of events that also fire the  gas \v{C}erenkov. This efficiency is greater than 99.8\% for all kinematics on both spectrometers.

\subsection{VDC Track Efficiency}
\begin{figure}[htp]
\begin{center}
\includegraphics[scale=.45]{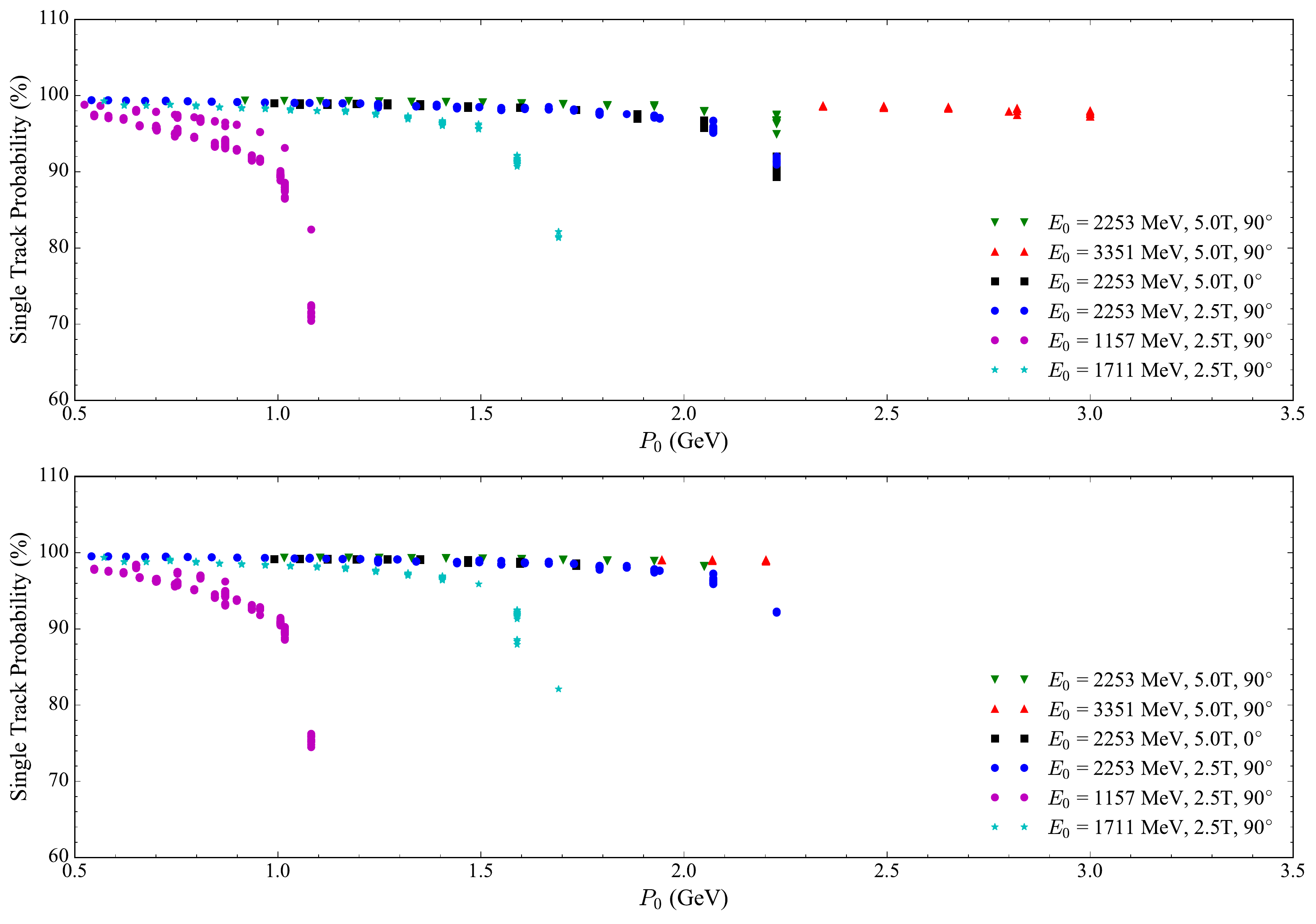}
\caption{VDC single track probability based upon the collected data. The LHRS (RHRS) is on top (bottom).}
\label{fig:HRS_1Track}
\end{center}
\end{figure}

The VDC efficiency is ratio of the ``good" electron events with a successful track reconstruction to the total number of ``good" events with no specification on the track reconstruction:
\begin{equation}
\epsilon_{\mathrm{VDC}} = \frac{N_{\mathrm{good}}}{N_{\mathrm{total}}}\,.
\end{equation}
The good events are a clean selection of electrons that pass PID and acceptance cuts. In the past, Hall A experiments only considered events that generated one track in the quantity $N_{\mathrm{good}}$. According to Figure~\ref{fig:HRS_1Track}, for some kinematics in E08-027 up to 30\% of events have multiple VDC tracks. The increase in the probability of multiple track events is driven by the increase in trigger rate as the electron momentum approaches elastic scattering. In order to not artificially cut out good electrons, the multiple track events are carefully studied to determine which events have at least one good track that passes the selection criteria.

The selection criteria for the multiple track events requires that one track deposit energy equivalent to the spectrometer central momentum, $P_0$, in the lead glass  calorimeters. If two or more tracks satisfy this requirement then the distance between the tracks is examined. For example, if two tracks are more than a calorimeter block away, then it is highly likely that one or both of those tracks are good. If the two tracks are within one calorimeter block of each other they will share the deposited energy and it is difficult to distinguish between them. Events like this contribute to the systematic uncertainty of the VDC multi-track correction.  Events with up to seven tracks are studied and the corrected uncertainty of the VDC efficiency is greater than 99.5\% for almost all kinematic settings. More details on the VDC multi-track efficiency study are found in Ref.~\cite{Jie}.

\subsection{Trigger Scintillator Efficiency and DAQ Livetime}
The trigger efficiency is

\begin{equation}
\epsilon_{\mathrm{trig}} = \frac{T_\mathrm{{main}}}{T_\mathrm{{main}}+T_\mathrm{{eff}}},
\end{equation}
where $T_\mathrm{{main}}$ and $T_\mathrm{{eff}}$ are the total number of trigger counts for the main and efficiency triggers respectively. These counts are determined from either the trigger scalers or the trigger latch pattern. The trigger latch pattern, which is created by sending the trigger signals to a TDC, is favored over the scalers because it is tied directly to recorded events and allows for cuts to be made in analysis. This correlation with recorded events also makes the trigger latch pattern susceptible to deadtime effects.  The corrected trigger count for the latched triggers, after accounting for the deadtime is,

\begin{equation}
T_{\mathrm{corr}} = \frac{T_i ps_i}{1 - DT_i},
\end{equation}
where $i$ is the trigger type (1-4), $T_i$ is the accepted trigger count, $ps_i$ is the prescale factor for the trigger and $DT_i$ is the deadtime associated with the trigger. The deadtime is determined from the livetime ($LT$) with $DT$ = 1 $-$ $LT$. The livetime is the ratio of accepted triggers to total triggers adjusted by the prescale factor such that

\begin{equation}
LT = \frac{ps_i T_i^{acc}}{T_i^{tot}},
\end{equation}
where the accepted triggers are determined from the latch pattern, and the total trigger count is from the trigger scalers.

\begin{figure}[h]
\begin{center}
\includegraphics[scale=.6]{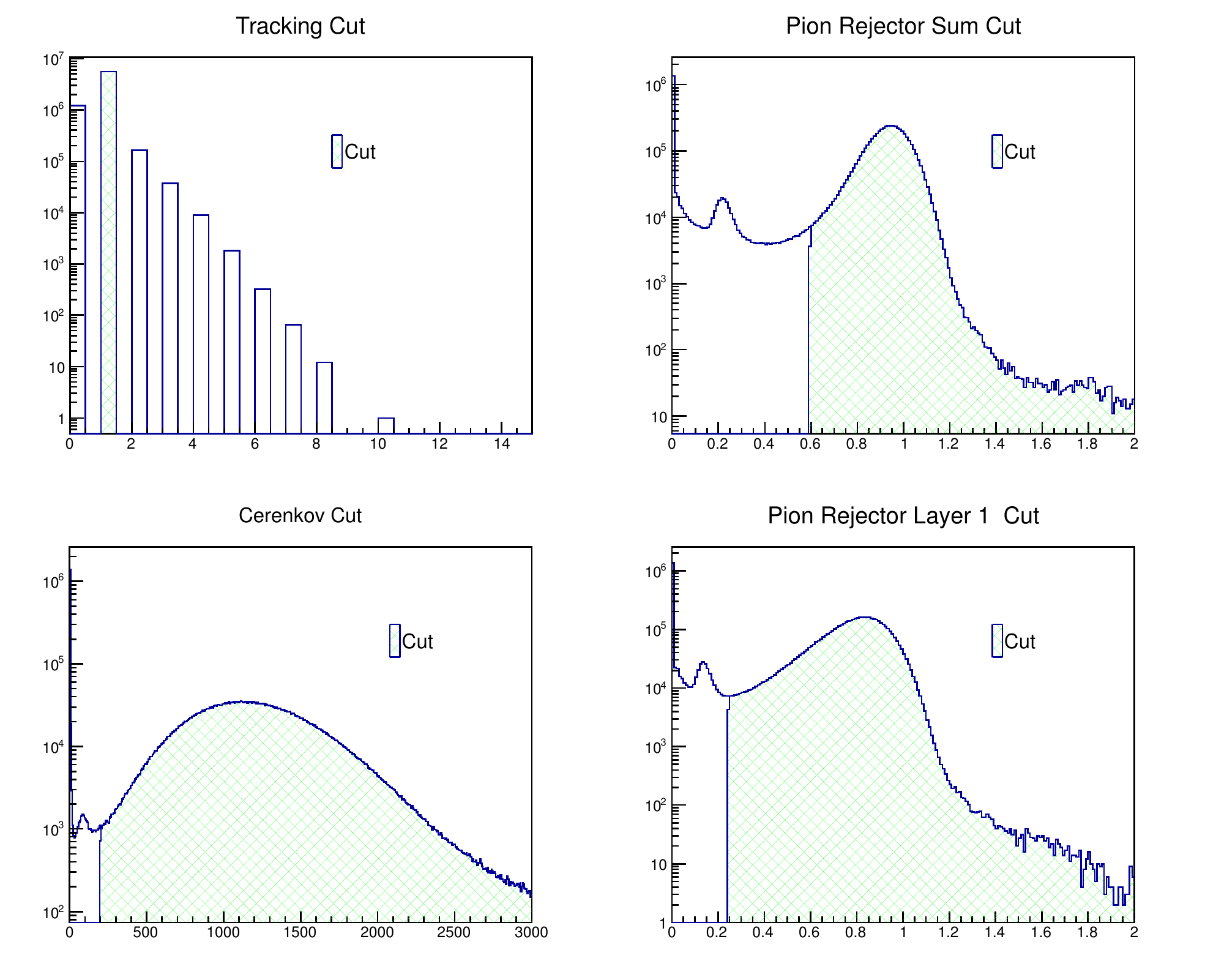}
\caption{ Good electron cuts for run 3888 on LHRS at $P_0$ = 0.898 GeV. The shaded green region is the good electron selection. The pion rejector cuts are on the total energy deposited and not the $E_{\mathrm{tot}}/p$ ratio.}
\label{fig:Cut}
\end{center}
\end{figure}

Analysis cuts ensure that only electron events are included in the efficiency calculations. The first cut is on the trigger latch pattern and selects the appropriate trigger. The trigger information is stored in a bit-wise manner. Each trigger is denoted as 2$^i$ where $i$ is the desired trigger. The rest of the cuts are shown in Figure~\ref{fig:Cut} for LHRS run 3288 at  $E_0$ = 2253 MeV and $P_0$ = 899 MeV. The tracking cut requires that the event creates a single track in the VDC. The particle identification cuts on the gas \v{C}erenkov and pion rejector layers eliminate  pion contamination and are a result of the analysis of Ref~\cite{Melissa}.

\begin{figure}[htp]
\begin{center}
\includegraphics[scale=.45]{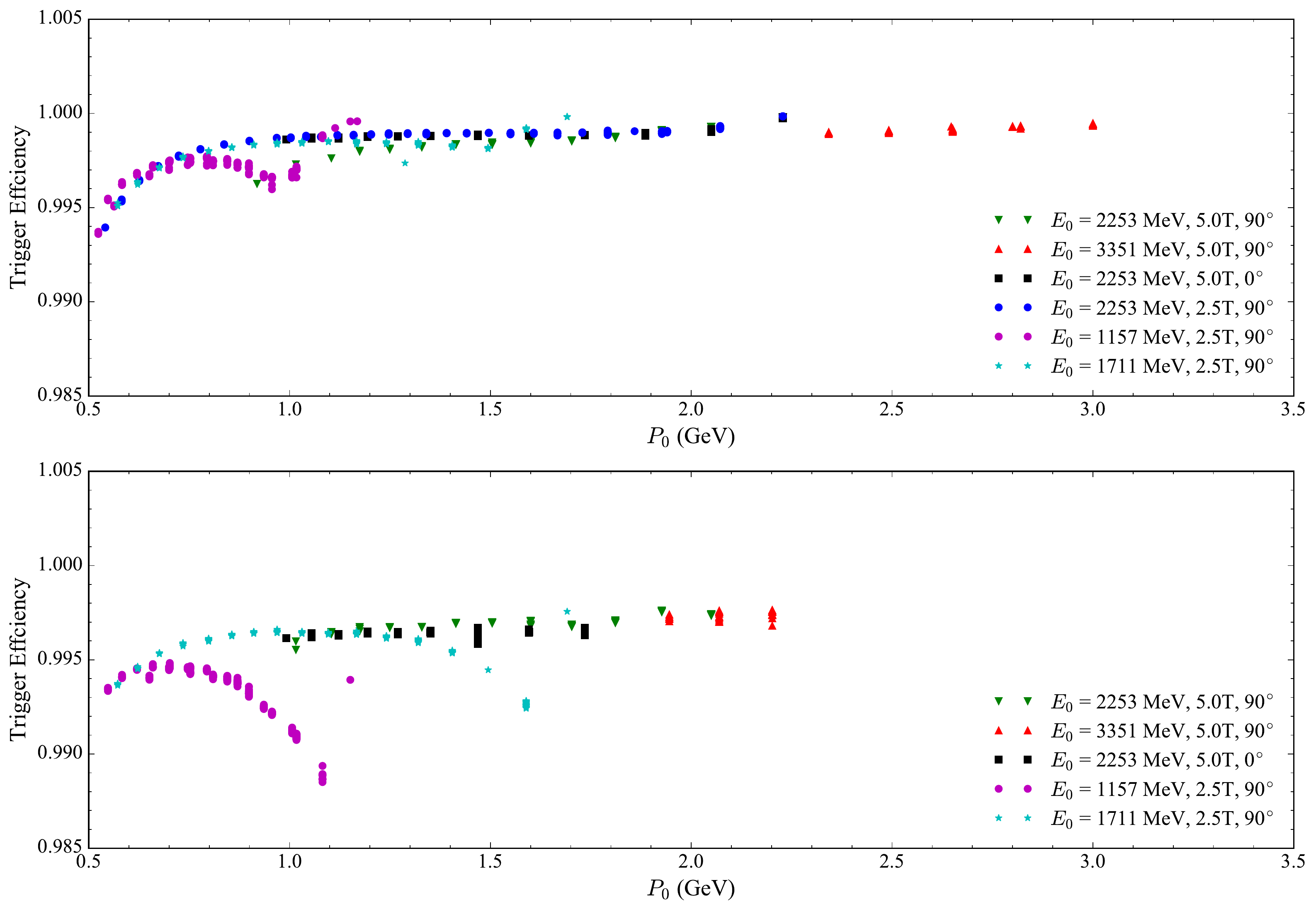}
\caption{Trigger efficiency plotted against $P_0$. The LHRS (RHRS) is on top (bottom).}
\label{fig:HRS_Tot}
\end{center}
\end{figure}

The results of the trigger efficiency calculation are shown in Figure~\ref{fig:HRS_Tot}. All LHRS runs have a trigger efficiency greater than 99$\%$ while there are 26 RHRS runs with trigger efficiencies below this number. These RHRS runs are listed in Ref~\cite{RyanTRIG}. It should also be noted that no efficiencies are calculated for the RHRS at $E_0$ = 2.253 GeV and a 2.5 T target field. The efficiency trigger, $T_2$, was broken during this run period. 

\subsection{Charge Asymmetry}
\begin{figure}[htp]
\begin{center}
\includegraphics[scale=.45]{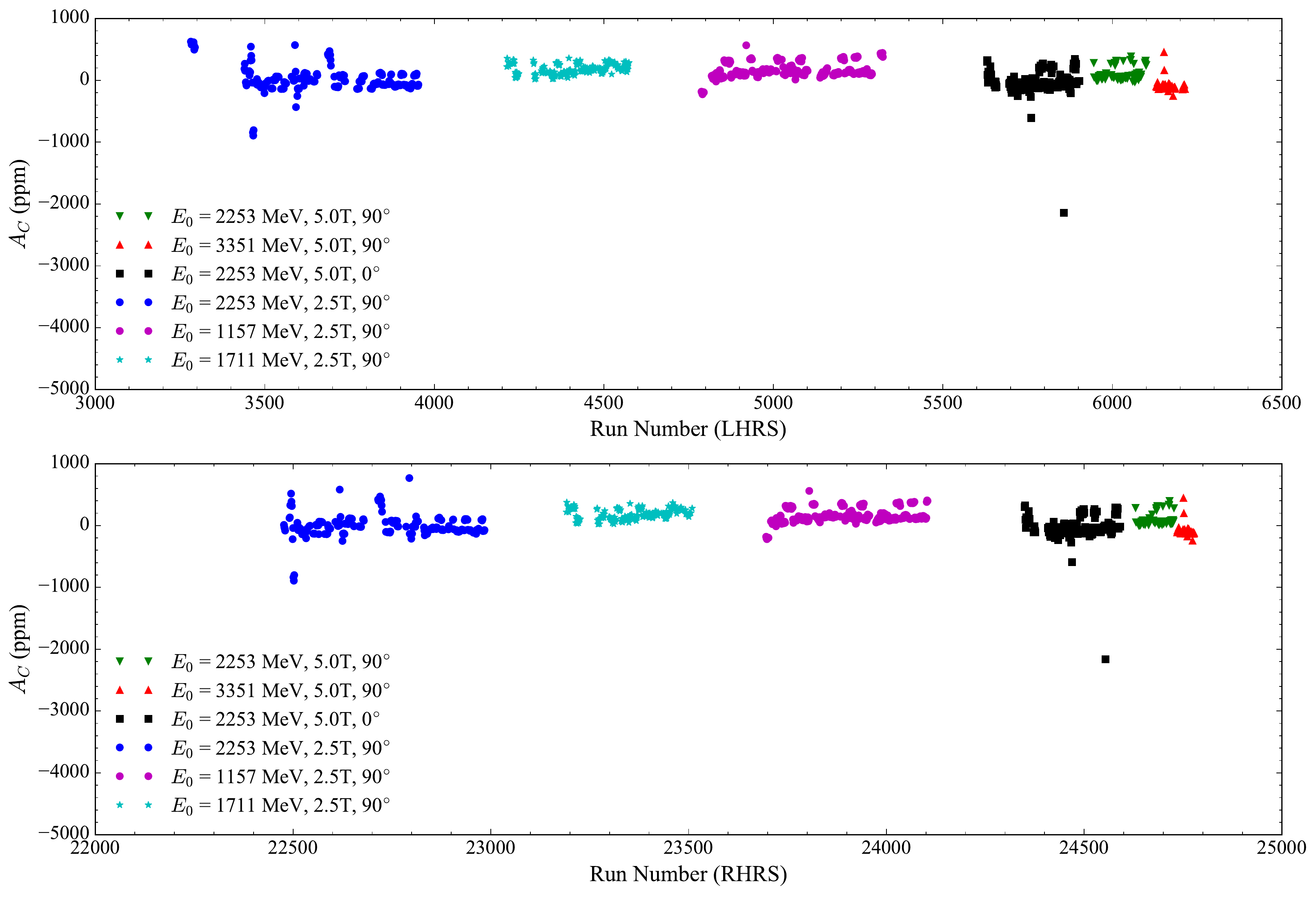}
\caption{Charge asymmetry plotted against run number for both spectrometers.}
\label{fig:HRS_CAsym}
\end{center}
\end{figure}

The distribution of charge between plus and minus helicity states is not perfectly uniform. This leads to a non-physics asymmetry of the form
\begin{equation}
A_Q = \frac{Q^+ - Q^-}{Q^+ + Q^-}\,,
\end{equation}
where $Q^{\pm}$ is the total accumulated charge during a run for the $\pm$ helicity state. The charge asymmetry is generally sourced to the accelerator injector, with imperfections in the Pockels cell of the half-wave plate as causes. This asymmetry is corrected for in the analysis by normalizing by the charge of each helicity state independently.  The presence of a large charge asymmetry is potentially indicative of other problems within a run, so it is still a useful quantity to study. The charge asymmetry for the entire run period is shown in Figure~\ref{fig:HRS_CAsym}, and,  with the exception of a few runs, is small.
\subsection{Livetime Asymmetry}
\begin{figure}[htp]
\begin{center}
\includegraphics[scale=.45]{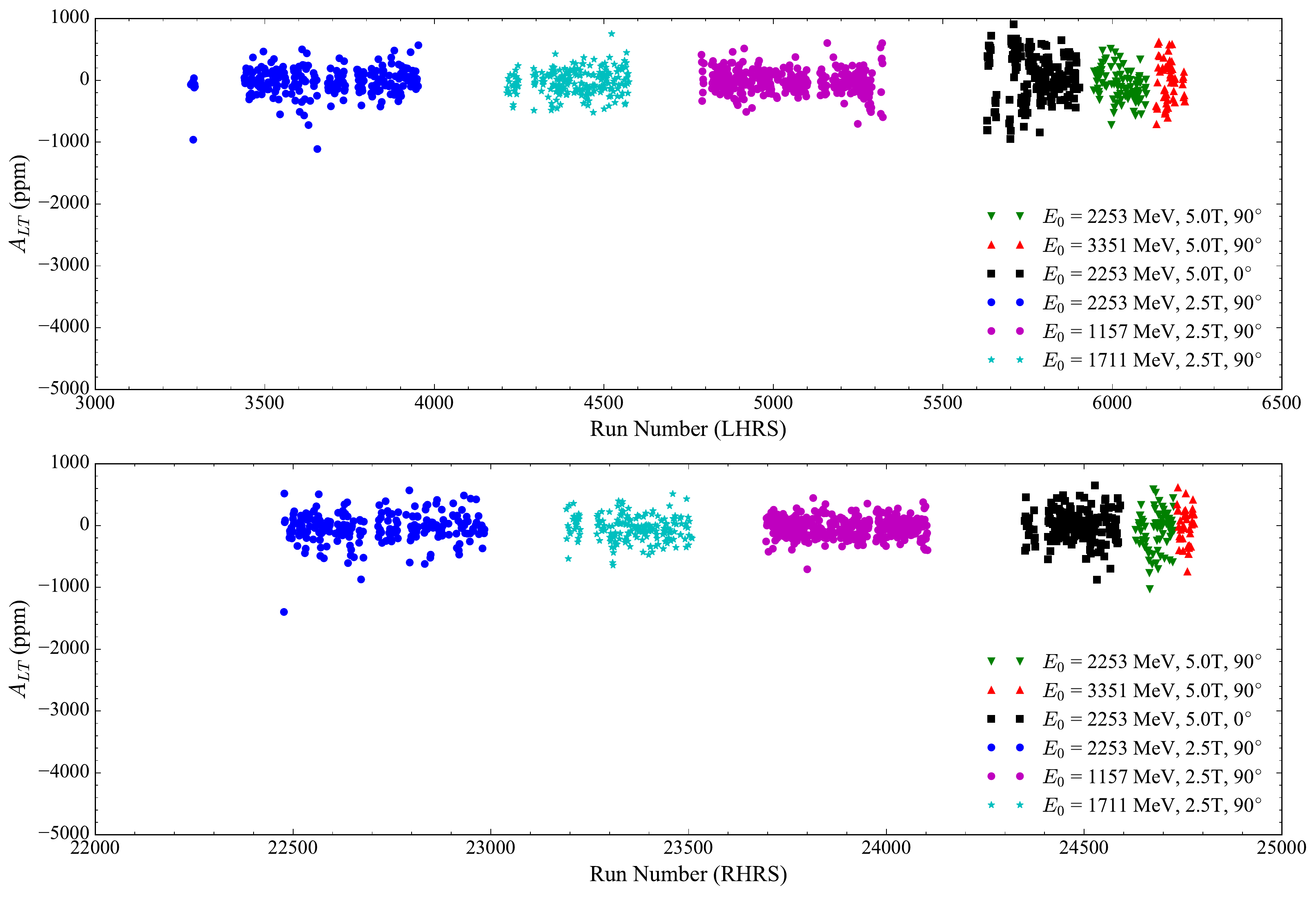}
\caption{Livetime asymmetry plotted against run number for both spectrometers.}
\label{fig:HRS_LTAsym}
\end{center}
\end{figure}

Similar to the charge, the DAQ livetime is not perfectly helicity independent. This creates a livetime asymmetry of the form
\begin{equation}
A_{LT} = \frac{LT^+ - LT^-}{LT^+ + LT^-}\,,
\end{equation}
where $LT^{\pm}$ is the livetime sorted by $\pm$ helicity state for a given run. A large physics asymmetry combined with highly correlated trigger and DAQ deadtime rate can cause a non-zero livetime asymmetry. Additional sources include varying event sizes between helicity states and also general deadtime fluctuations throughout a run that do not cancel on average. Again, this asymmetry is corrected for by normalizing by the livetime of each helicity state independently, but a large asymmetry potentially highlights other problems with a run. The livetime asymmetry for the entire run period is shown in Figure~\ref{fig:HRS_LTAsym}, and,  with the exception of a few runs, is small.

\section{Spectrometer Optics Studies}
\begin{figure}[htp]
\begin{center}
\includegraphics[scale=.45]{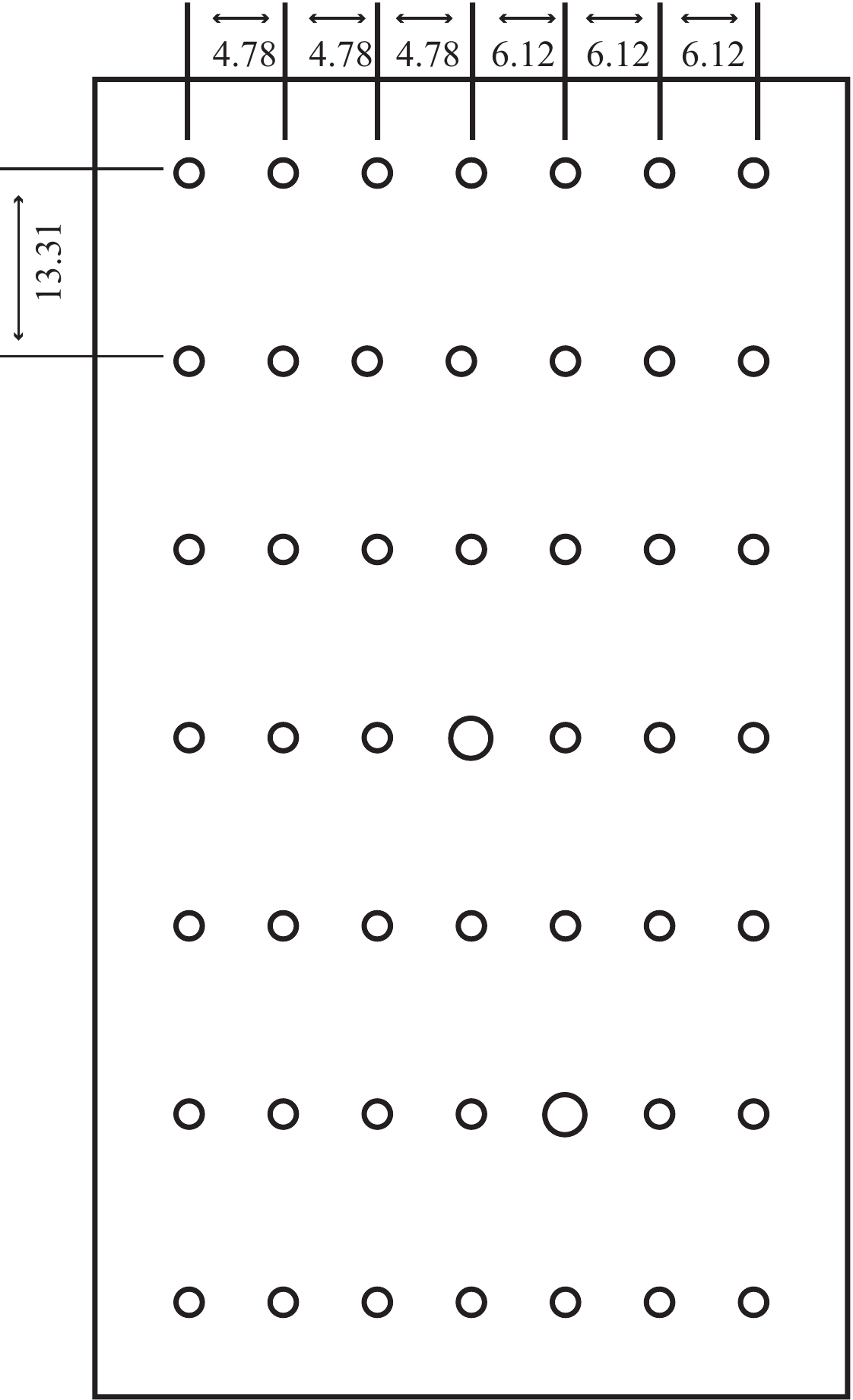}
\caption{Layout of the sieve slit used during E08-027. All measurements are in mm.}
\label{fig:Sieve}
\end{center}
\end{figure}

An optics matrix reconstructs the detected electrons back to the location of the electron-proton interaction at the target. The matrix elements are determined using elastic scattering  from a thin carbon foil target and a sieve slit collimator placed in front of the entrance to the HRS. The survey of the thin target allows for determination of the interaction vertex, and the surveyed sieve slit defines the electron's trajectory. The geometry of the sieve slit is show in Figure~\ref{fig:Sieve}. The two larger holes help determine the orientation of the sieve slit in the optics calibration. 

The addition of the septa magnets at the entrance of the HRS and presence of a strong transverse target field necessitate changes to the basic optics optimization procedure from previous Hall A experiments. Fortunately,  Hall A experiment E97-110 also used a pair of septa magnets with the HRS, so their optics matrix is used as a starting point in the analysis~\cite{Vince}. The effect of the target magnetic field on the scattered electrons is considered by first transporting the electrons to the entrance of the HRS using the equations of motion for charged particles in a magnetic field and then using the optics matrix to reconstruct the path through the spectrometer magnets~\cite{Chao4}. 
\subsection{Coordinate Systems}
\begin{figure}[htp]
\begin{center}
\includegraphics[scale=.45]{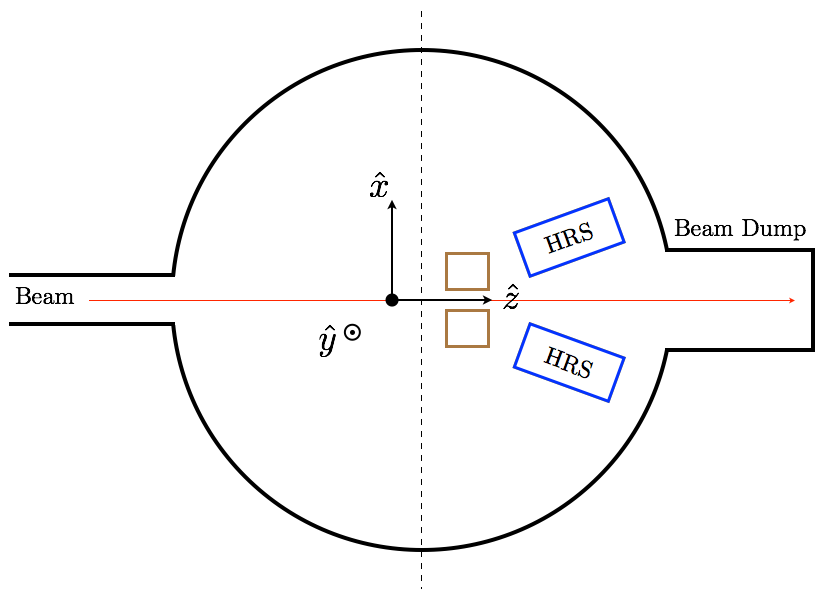}
\caption{Top view of the Hall A coordinate system. Reproduced from Ref~\cite{Chao4}.}
\label{fig:HRS_Hall}
\end{center}
\end{figure}

There are multiple coordinates systems used within Hall A. The following section will discuss only those needed for a basic understanding of the optics procedure. A full description of all the coordinate systems is found in Ref~\cite{Optics}. All angular coordinates refer to the tangent of the angle, even if not explicitly stated.
\newline \newline
{\bf Hall Coordinate System} \newline
The origin of the Hall A (lab) coordinate system is the intersection of the electron beam with the vertical axis (at the physical center of the target) of the polarized target. Along the beam line is the z-axis with $\hat{z}$ pointing downstream of the target. The x-axis is in the horizontal plane and $\hat{x}$ points left from looking along $\hat{z}$. The y-axis is in the vertical plane and $\hat{y}$ points up when viewed from the center of the hall. The coordinate system is shown in Figure~\ref{fig:HRS_Hall}.

\begin{figure}[htp]
\begin{center}
\includegraphics[scale=.70]{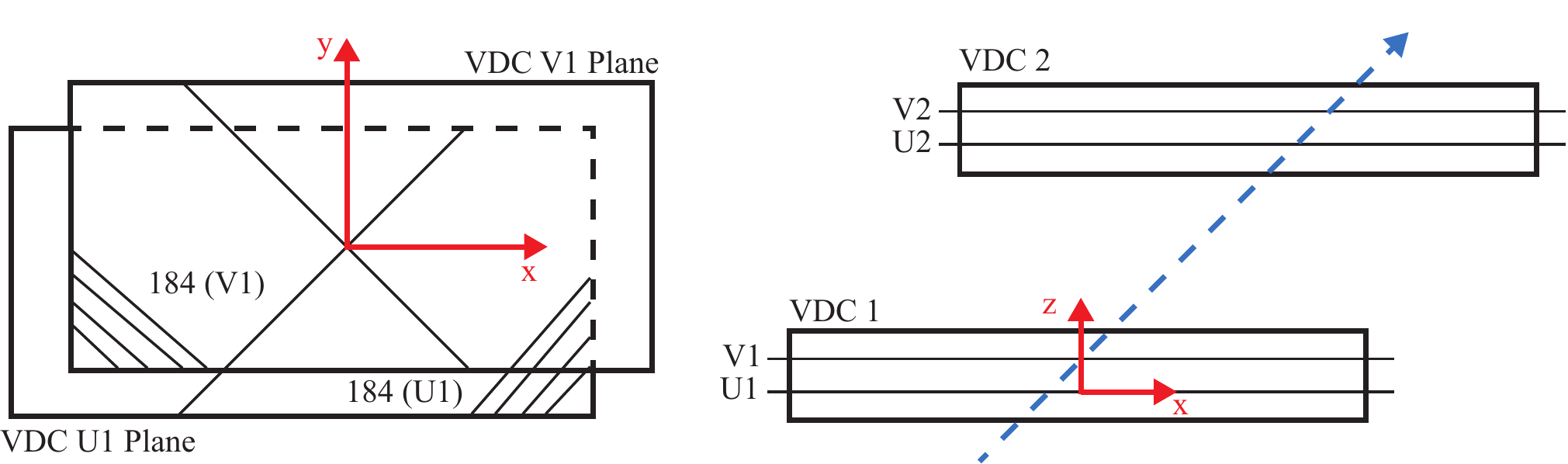}
\caption{Hall A HRS detector coordinate system. The origin of the coordinate system is the intersection of VDC wire 184 from the U1 and V1 planes.}
\label{fig:HRS_DEC}
\end{center}
\end{figure}

\begin{figure}[htp]
\begin{center}
\includegraphics[scale=.99]{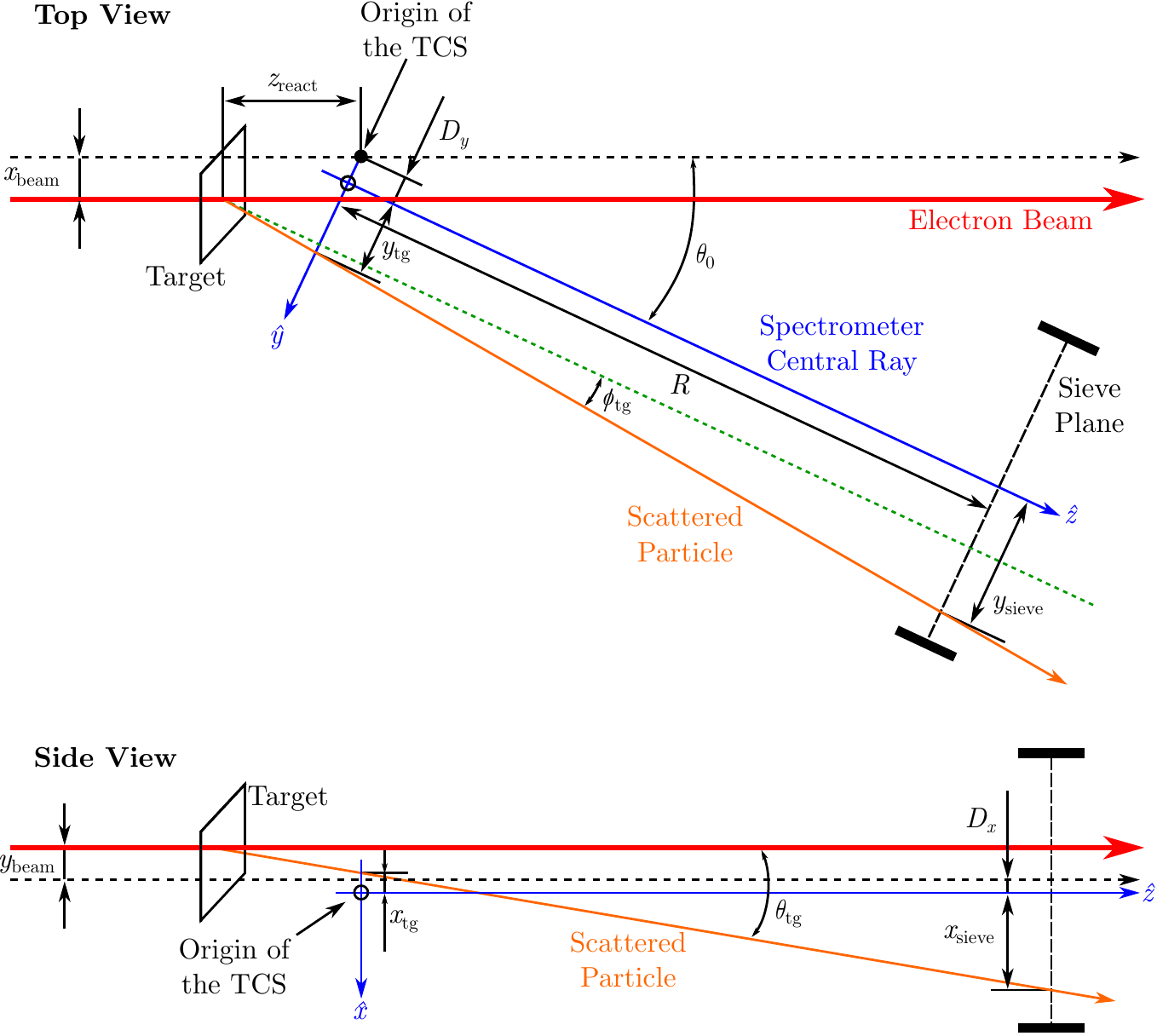}
\caption{Hall A target coordinate system. See text for a description. }
\label{fig:HRS_Tg}
\end{center}
\end{figure}

\noindent{\bf Detector Coordinate System} \newline
The vertical drift chambers define the detector coordinate system shown in Figure~\ref{fig:HRS_DEC}. The detected electron's trajectory is given by a pair of angular and spatial coordinates for the dispersive and non-dispersive axes. The dispersive axis runs along the length of the VDC and the electron's position is given by $x_{\mathrm{det}}$, while the tangent of the angle made by its trajectory is $\theta_{\mathrm{det}}$. The non-dispersive axis runs parallel to the width of the VDC and the corresponding detector variables are $y_{\mathrm{det}}$ and $\phi_{\mathrm{det}}$. The origin of the detector coordinate system is the intersection of VDC wire 184 from the U1 and V1 planes. The focal plane coordinates used to determine the optics matrix, are calculated by taking into account the off-set between the spectrometer central ray (defined as a trajectory passing through the geometrical center of the spectrometer) and the origin of the detector coordinate system~\cite{Optics}.
\\\\
\noindent{\bf Target Coordinate System} \newline
The optics matrix elements relate the focal plane coordinates to the target coordinates ($\theta_{\mathrm{tg}}$, $\phi_{\mathrm{tg}}$, $x_{\mathrm{tg}}$, $y_{\mathrm{tg}}$) defined from the target coordinate system shown in Figure~\ref{fig:HRS_Tg}. The $\hat{z}$ axis is defined from a line passing perpendicular through (and towards) the central sieve slit hole. The $\hat{x}$  axis is parallel to the sieve slit surface and points vertically down; $\hat{y}$ is parallel to the sieve slit surface in the transverse plane~\cite{Optics}.  The tangent of the in-plane and out of plane angles with respect to the central ray trajectory are given by $\phi_{\mathrm{tg}}$ and $\theta_{\mathrm{tg}}$ respectively. The central scattering angle of the spectrometer is represented by $\theta_{\mathrm{0}}$ and is the angle between $\hat{z}_{\mathrm{tg}}$ and $\hat{z}_{\mathrm{Hall}}$. The distance from the sieve slit to the origin of the target coordinate system is R. $D_x$ and $D_y$ are the vertical and horizontal deviations from the central ray of the spectrometer to the origin of the Hall coordinate system. 

\subsection{The Optics Matrix}
The optics matrix relates focal plane coordinates to the target coordinates and, to first order, is expressed as
\begin{equation}
\begin{pmatrix}
\delta \\
\theta \\
y \\
\phi \\
\end{pmatrix}_{\mathrm{tg}}
= \begin{pmatrix}
\langle \delta | x \rangle & \langle \delta | \theta \rangle& \langle \delta | y \rangle &\langle \delta | \phi \rangle\\
\langle \theta | x \rangle & \langle \theta | \theta \rangle& \langle \theta | y \rangle &\langle \theta| \phi \rangle\\
\langle y | x \rangle & \langle y | \theta \rangle& \langle y| y \rangle &\langle y| \phi \rangle\\
\langle \phi | x \rangle & \langle \phi | \theta \rangle& \langle \phi| y \rangle &\langle \phi| \phi \rangle\\ \end{pmatrix} 
\begin{pmatrix}
x \\
\theta \\
y \\
\phi \\
\end{pmatrix}_{\mathrm{fp}}\,.
\end{equation}
The relative momentum of the particle is defined as 
\begin{equation}
\delta = \frac{P-P_{\mathrm{0}}}{P_{\mathrm{0}}}\,,
\end{equation}
where $P$ is the particle's measured momentum and $P_{\mathrm{0}}$ is the spectrometer central momentum determined by the HRS dipole's magnetic field. 

\begin{figure}[htp]
\centering     
\subfigure[Before]{\label{fig:a}\includegraphics[width=.45\textwidth]{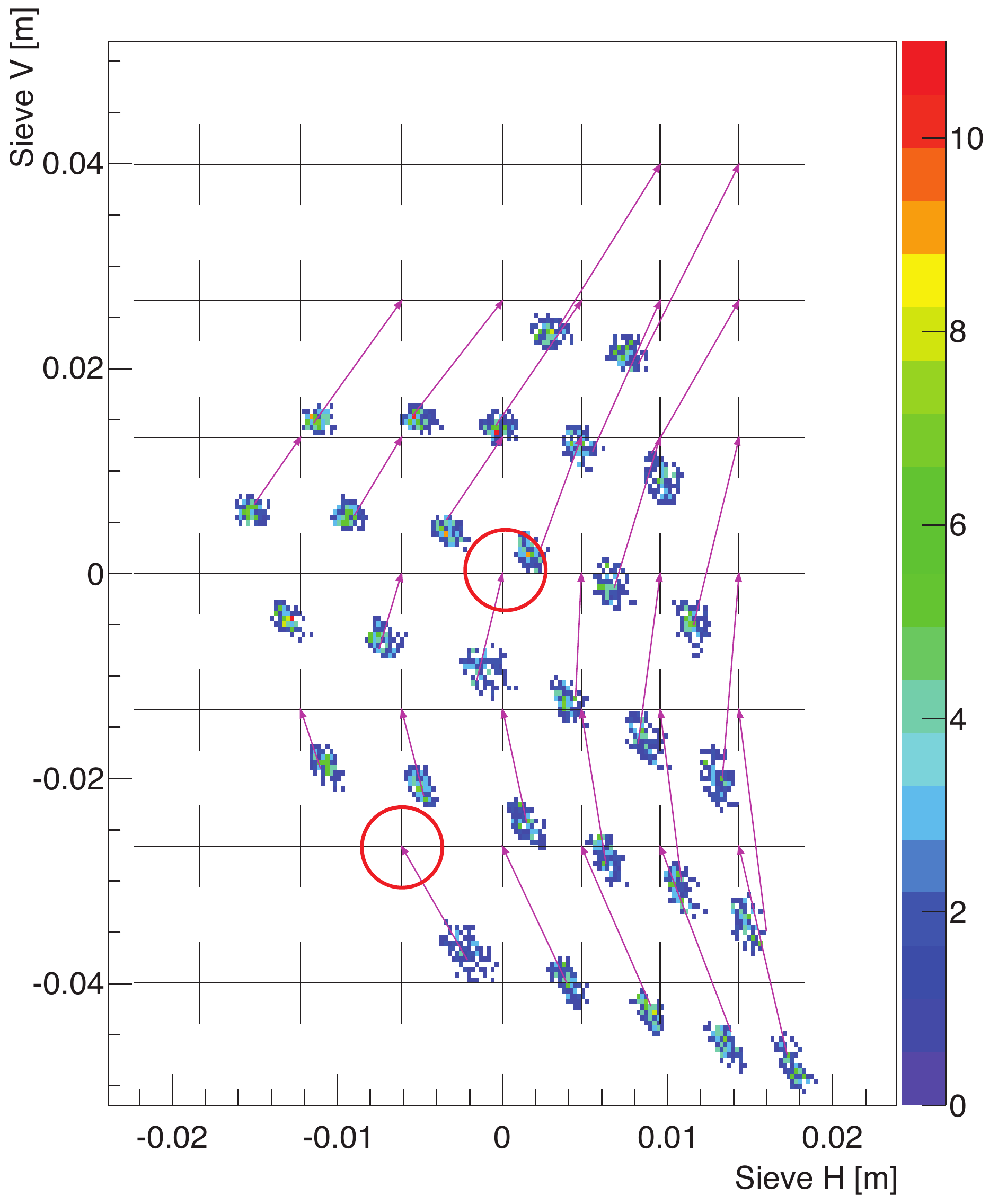}}
\qquad
\subfigure[After]{\label{fig:b}\includegraphics[width=.45\textwidth]{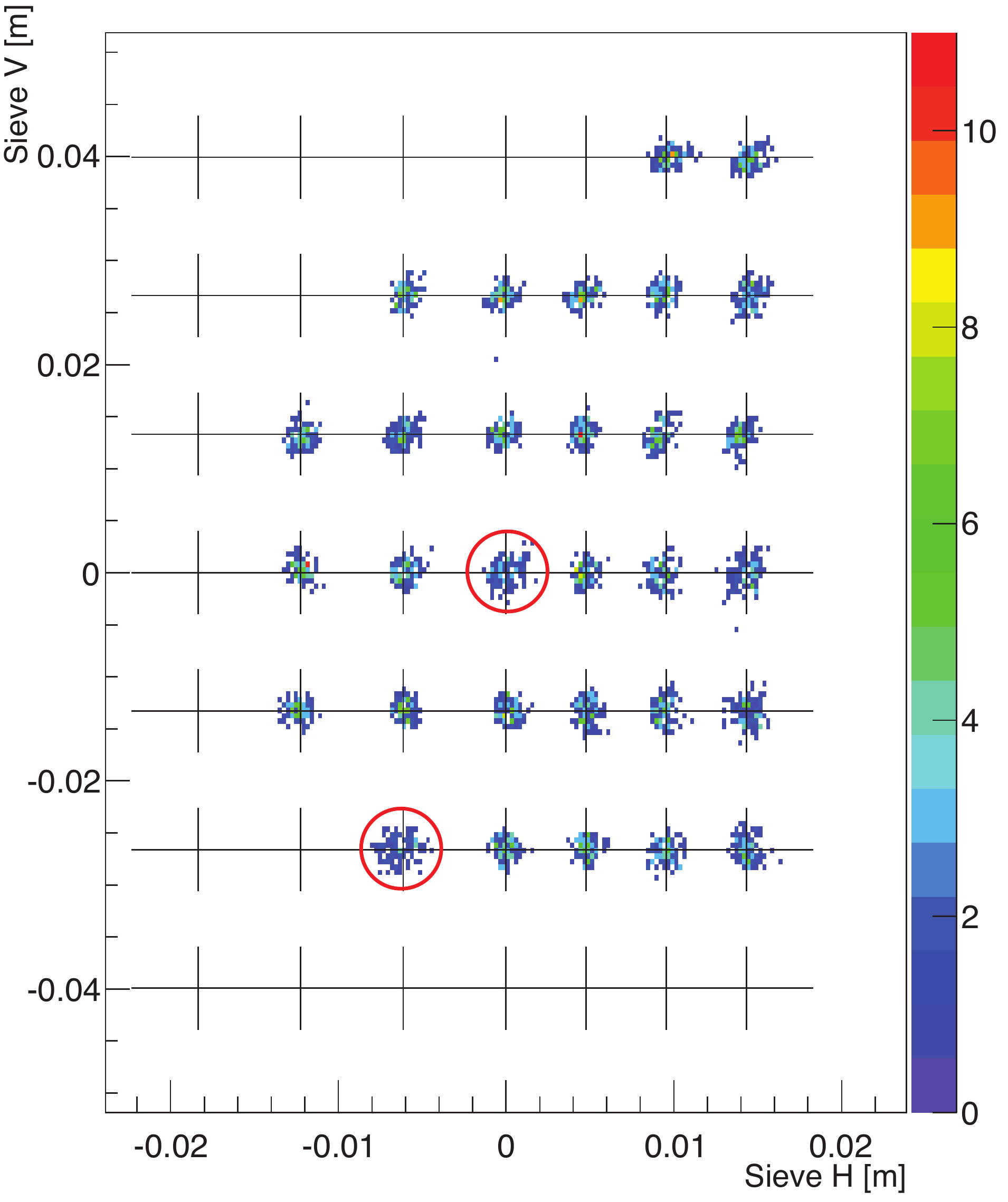}}
\caption{The $\chi^{2}$ minimization process aligns the electron events with the sieve holes. The orientation of the sieve slit is give by the two bigger holes circled in red.}
\label{Sieve}
\end{figure}

At higher order, the optics matrix is a tensor. Each optics matrix element is determined from a $\chi^{2}$ minimization process~\cite{Chao4}, which compares reconstructed events at the sieve plane to the actual surveyed sieve slit positions. In the process, only elastically scattered focal plane events corresponding to sieve slit holes are optimized. These events are selected using a series of graphical cuts. The results for the optimization of $\theta_{\mathrm{tg}}$ and $\phi_{\mathrm{tg}}$ for the RHRS are shown in Figure~\ref{Sieve} for $E_0$ = 2254 MeV and a longitudinal target field. Each cross represents the location of a sieve slit hole, and the bigger holes are circled in red. The left sieve plot is before optimization and uses an initial optics matrix.  In this plot, the arrows point to which hole these events actually came from, determined using the graphical cuts. The right plot is the optimized sieve pattern, using the new matrix generated from the optimization. This optimization is for a single beam energy and target field configuration; there is a separate optics matrix and optimization for each kinematic setting. There is also a separate optimization for $y_{\mathrm{tg}}$. 

The momentum calibration is done by using the elastic scattering conditions to precisely determine the momentum of the scattered particle,
\begin{equation}
E' = \frac{E}{1+E/M(1-\mathrm{cos}\theta)}\,,
\end{equation}
where $E$ is the energy of the incident electron beam, $M$ is the mass of the target and $\theta$ is the scattering angle. The entire momentum acceptance of the spectrometer is covered by performing a ``delta-scan" and taking data with several different $P_0$ values. The range of the delta scan is typically $\pm$4\% around the $E'$ determined from elastic scattering.

\subsection{Reconstructing the Scattering Angle}
\label{Sec:RecScatAngle}
The behavior of the scattered electrons in the transverse target field is determined using the equations of motion of the electron integrated over the magnetic field. The magnitude of the magnetic field is determined via the Biot-Savart law using the current density of the magnet coils and is experimentally confirmed with a field map~\cite{Chao2}.  The uncertainty in the reconstructed quantities, $\delta_{\mathrm{res}}$,  $\theta_{\mathrm{res}}$ and  $\phi_{\mathrm{res}}$ are approximately 2$\times$10$^{-4}$, 2 mrad (0.11$^{\circ}$) and 1.5 mrad (0.09$^{\circ}$) respectively. The precise details of the target field transportation and electron reconstruction is found in Ref~\cite{Chao4}. A detailed description of the uncertainties is found in Ref~\cite{Chao3}.

The calculation of the scattering angle on an event-by-event basis is complicated by the target field because it causes the beam to deviate from a straight line trajectory. The incident angle of the beam as it hits the target is readout by the BPM. The vector corresponding to the incident electron direction is
\begin{equation}
\vec{A} = [\mathrm{sin(\theta_{\mathrm{beam}}})\mathrm{cos(\phi_{\mathrm{beam}}}),\mathrm{sin(\theta_{\mathrm{beam}}})\mathrm{sin(\phi_{\mathrm{beam}}}),\mathrm{cos(\theta_{\mathrm{beam}}}) ]\,,
\end{equation}
where  $\theta_{\mathrm{beam}}$ and $\phi_{\mathrm{beam}}$ are the polar and azimuthal angles with respect to the beam in the Hall coordinate system. The translation between the BPM coordinate system and the Hall coordinate system is
\begin{align}
\theta_{\mathrm{beam}} & = \mathrm{tan^{-1}}{\bigg(}[\mathrm{tan}^2(\theta_{\mathrm{BPM}}) + \mathrm{tan}^2(\phi_{\mathrm{BPM}})]^{-1/2}{\bigg)}\,,\\
\phi_{\mathrm{beam}}    &=  \mathrm{tan^{-1}}{\bigg(}\mathrm{tan}(\theta_{\mathrm{BPM}})/ \mathrm{tan}(\phi_{\mathrm{BPM}}){\bigg)}\,.
\end{align}

The vector of the scattered electron direction is
\begin{equation}
\vec{B} = [ \phi_{\mathrm{rec}}\mathrm{cos}(\theta_0) + \mathrm{sin}(\theta_0),-\theta_{\mathrm{rec}},\mathrm{cos}(\theta_0) - \phi_{\mathrm{rec}}\mathrm{sin}(\theta_0)]\,,
\end{equation}
where $\theta_0$ is the central scattering angle of the spectrometer. The correlation between ROOT tree variable and the variables in the previous equations is given in Appendix~\ref{app:Appendix-E}. The angle between $\vec{A}$ and $\vec{B}$ is the scattering angle
\begin{equation}
\mathrm{cos}\theta_{\mathrm{scat}} = \frac{\vec{A}\cdot \vec{B}}{|\vec{A}||\vec{B}|}\,.
\end{equation}
The central scattering angle of the spectrometer is determined from survey and the results are shown in Table~\ref{table:theta}. The survey result is confirmed using elastic scattering. The results of the elastic scattering pointing study are in Ref~\cite{Min}.
\begin{table}
\begin{center}
\begin{tabular}{ l r }\hline
 HRS Arm & $\theta_0$ (rad)  \\[.2cm]  \hline 
 LHRS & 0.1007 $\pm$ 0.0007  \\[.2cm] 
 RHRS & 0.1009 $\pm$ 0.0007 \\[.2cm]\hline
\end{tabular}
\caption{The central scattering angle as determined from survey is approximately 5.73$^{\circ}$. }
\label{table:theta}
\end{center}
\end{table}

\section{Target Polarization}
The main focus of the offline target polarization analysis is a careful study of the target material calibration constants. Taken at thermal equilibrium, these calibration constants relate the integrated area under the NMR curve  to the thermal equilibrium polarization, which was described in Chapter~\ref{TEtalk}.

\subsection{Determining Calibration Constants}
The NMR curve is generated using a Liverpool Q-Meter. The Q-Meter is an LRC circuit at its core and the resonance of the circuit is tuned to the resonant frequency of the proton,
\begin{equation}
\omega_0 = 1/\sqrt{LC_0}\,,
\end{equation}
where the capacitance, $C_0$, is adjusting by changing the length of cable and the inductance $L$ is given by the loop of coil embedded in the sample. A tuned Q-Meter signal has the peak of the parabola of the ``Q-curve" centered at the proton resonance frequency. The curve itself represents the LCR circuits impedance as a function of frequency. When the frequency passes through the proton resonance the impedance changes, which is seen as either a dip or bump in the Q-curve.
\begin{figure}
\centering     
\subfigure[Raw NMR]{\label{fig:aNMR}\includegraphics[width=.80\textwidth]{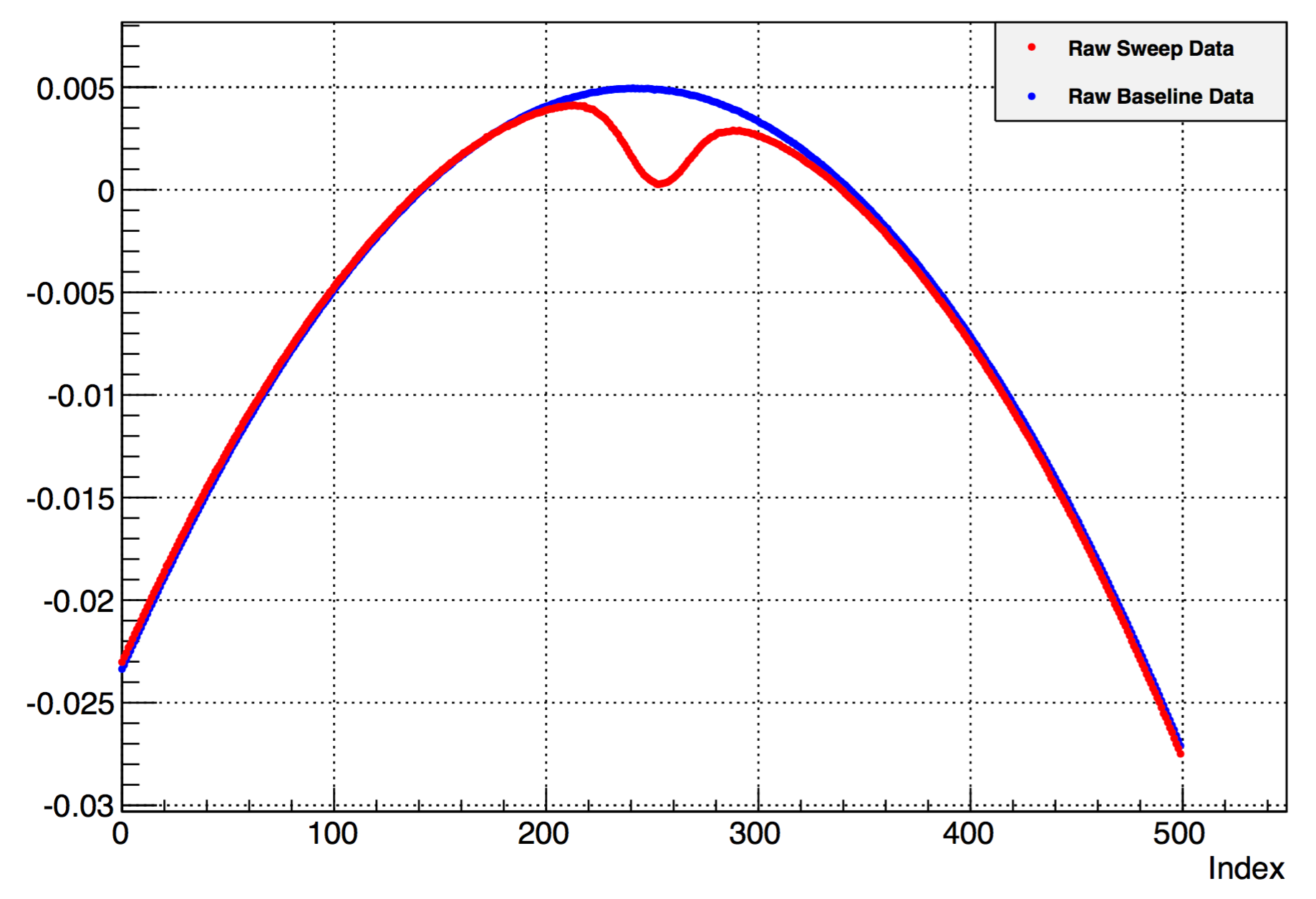}}
\qquad
\subfigure[Fit-Subtracted NMR]{\label{fig:bNMR}\includegraphics[width=.80\textwidth]{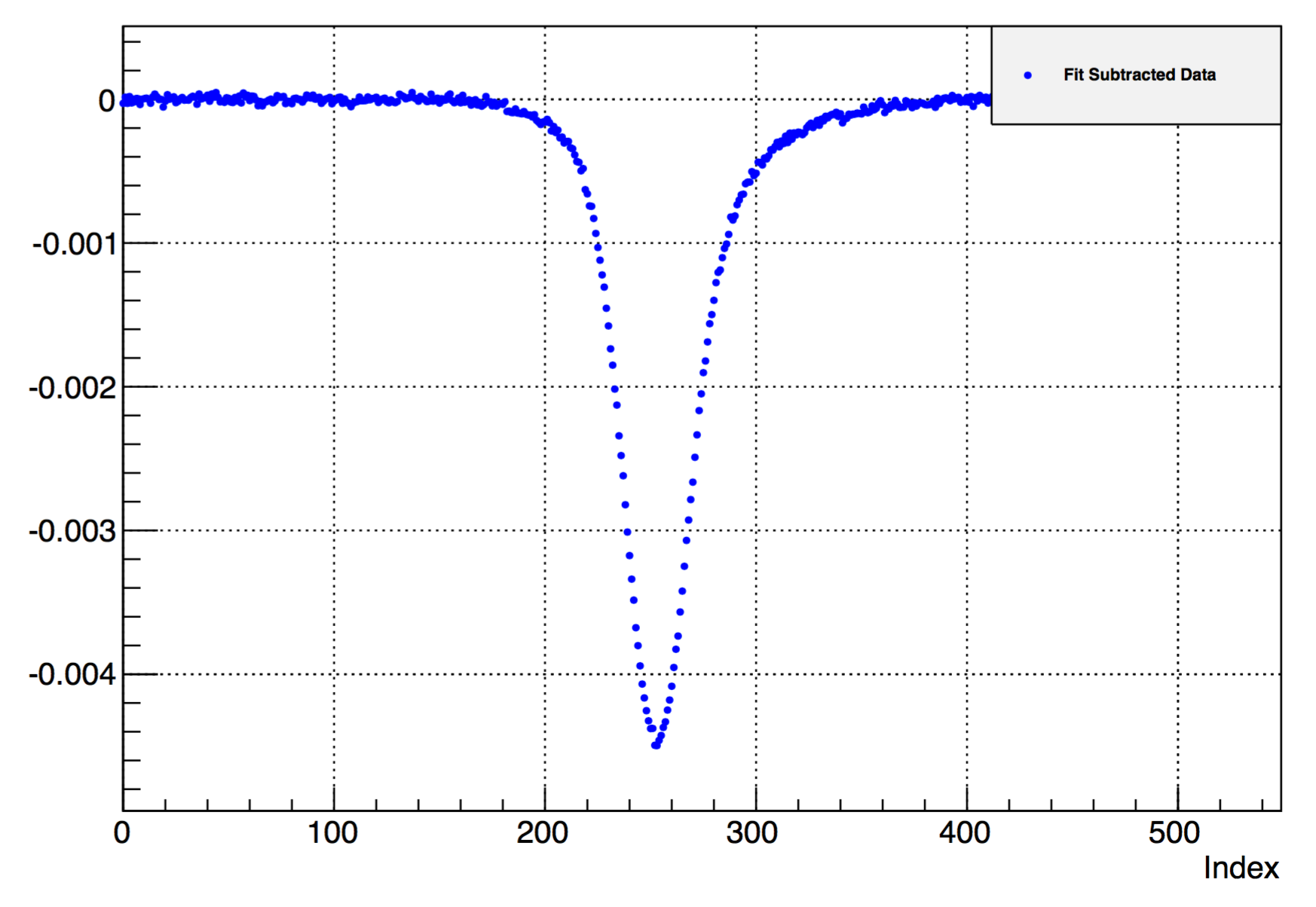}}
\caption{Q-Meter polarization signal. The proton couples with an LRC circuit to change the inductance of the circuit at the NMR frequency and produce a signal. Effects not related to the proton polarization are removed through a combination of background subtraction and fitting.}
\label{Fig:QCurve}
\end{figure}

The raw Q-curve and polarization signal is shown in Figure~\ref{Fig:QCurve}. The x-axis of Figure~\ref{fig:aNMR} is the frequency sweep of the RF generator in arbitrary units. The non-resonant background of the Q-meter is subtracted out via a baseline measurement and is the Q-curve when the proton is moved off resonance. The final polarization signal is the raw signal subtracted from the baseline signal, with a 3$^{rd}$ order polynomial fit to the wings. The fit removes any leftover background. An example of the subtracted and fit polarization signal is show in Figure~\ref{fig:bNMR}.



\begin{figure}[htp]
\begin{center}
\includegraphics[scale=.70]{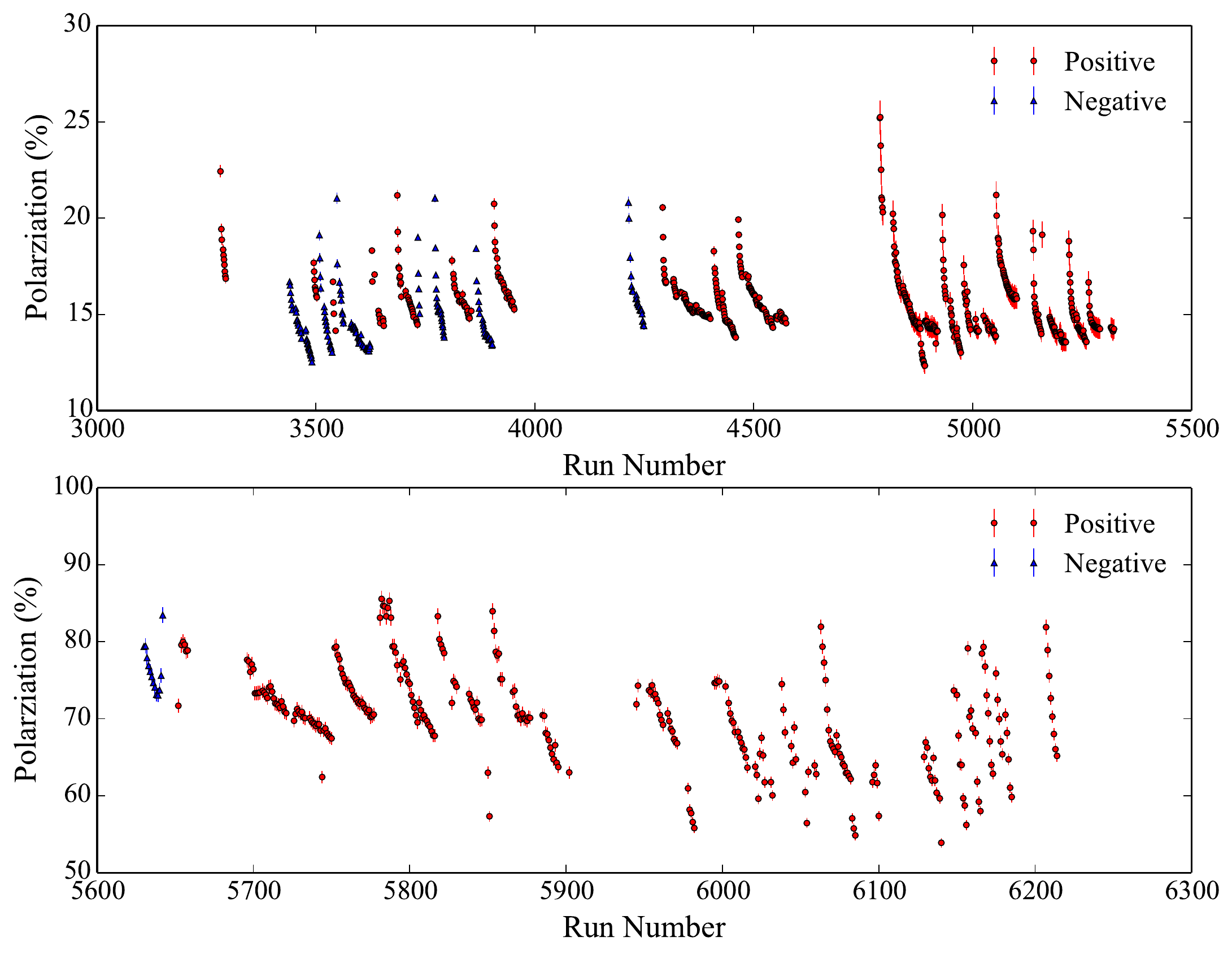}
\caption{\label{Fig:TargetPol}Target polarization for 2.5 T (top) and 5.0 T (bottom) on a run-by-run basis. Switching between positive and negative polarizations aids in minimizing systematic effects in the polarized physics quantities.}
\end{center}
\end{figure}

The area under the curve of the subtracted data is proportional to the polarization. The constant of proportionally is determined at thermal equilibrium where the polarization is known exactly from equation~\eqref{TE}. The enhanced polarization is the enhanced NMR area multiplied by this calibration constant. For each thermal equilibrium measurement (TE) on the ten different target materials there is a different calibration constant. A target material is considered different when it is removed from target cryostat or after it has been annealed. The final calibration constant for a material is the average of that material's TE measurements. The average polarization on a run-by-run basis is shown in Figure~\ref{Fig:TargetPol}. The average polarization for each setting is 15\% (70\%) for the 2.5 T (5.0 T) runs. 

\subsection{Target Polarization Uncertainty}
The uncertainty in the final calibration constants is separated into two categories. The  first is the uncertainty in the fit of the raw NMR signal. The second is the uncertainty in the determination of the absolute thermal equilibrium polarization and is directly related to the uncertainty in the target field and temperature measurements. The fit uncertainty is estimated by generating a Gaussian with the same amplitude as the TE on a baseline signal and the comparing the calculated area to the known area.  Each TE measurement on a given material is treated as a repeatable measurement so that total uncertainty is reduced if multiple TEs were taken on a material. The final relative uncertainties ranged from $\sim$3\% to 5\%. More information on the target polarization analysis is found in Ref~\cite{Toby}.

\section{Dilution Factor and Packing Fraction}
\label{sec:Dil}
Scattering from background material dilutes the polarized proton signal in the detected events. The measured asymmetries are a mix of proton scattering events and unpolarized scattering from nitrogen, helium and aluminum present within the target scattering chamber. Separating the detected events into proton and background components gives the following measured asymmetry,
\begin{equation}
A^{\mathrm{meas}} =\frac {Y_+ - Y_-}{Y_++Y_- +Y_{\mathrm{bg}}}\,,\\
\end{equation}
where $Y_{\mathrm{bg}}$ is the background contribution. This is a slightly altered form of equation~\eqref{eq:Asymm}. The background contribution cannot be separated from production runs on ammonia, and is instead determined from data taken on scattering from just the background materials. Combining the individually measured contributions reproduces the background contribution which is divided out in the measured data to produce a polarized proton yield. 

The dilution factor is expressed as
\begin{align}
\label{eq:Dilution}
f &= 1 - \frac{Y_{\mathrm{bg}}}{Y_{\mathrm{prod}}}\,,\\
Y_{\mathrm{bg}} & = Y_{\mathrm{N}} + Y_{\mathrm{He}} +Y_{\mathrm{Al}}\,, 
\end{align}
where $Y_{\mathrm{prod}}$ is the yield of a production run on the ammonia target. The aluminum contribution is determined from runs taken on a dummy (not-filled) ammonia cell and the helium contribution is determined from runs taken on pure liquid helium in the target nose. The nitrogen background is approximated by scaling carbon scattering data taken during the experiment. This scaling is done using ratios of the Bosted-Mamyan-Christy empirical cross section fit~\cite{N2Scale}. The background yields are also radiatively scaled to match the radiation lengths of the production data. 

In terms of cross sections the background and production components are
\begin{align}
Y_{\mathrm{bg}} &= A N_A{\bigg(} \frac{\rho_{\mathrm{NH}_3}L_{\mathrm{cup}}pf}{M_{\mathrm{NH}_3}}\sigma_{\mathrm{N}}  + \frac{\rho_{\mathrm{He}}(L_{\mathrm{out}}+(1-pf)L_{\mathrm{cup}})}{M_{\mathrm{He}}}\sigma_{\mathrm{He}} +\frac{\rho_{\mathrm{Al}}L_{\mathrm{Al}}}{M_{\mathrm{Al}}}\sigma_{\mathrm{Al}}   {\bigg)}\,,\\
\label{Dilution2}
Y_{\mathrm{prod}} &= Y_{\mathrm{bg}} +  A N_A{\bigg(} \frac{\rho_{\mathrm{NH}_3}L_{\mathrm{cup}}pf}{M_{\mathrm{NH}_3}}3\sigma_{\mathrm{H}}{\bigg)}\,,
\end{align}
where $N_A$ is Avogadro's number, $A$ is the acceptance correction, $L_{\mathrm{cup}}$ is the length of the target cup, $pf$ is the packing fraction, $L_{\mathrm{out}}$ is the length of liquid helium outside the cup, $L_{\mathrm{Al}}$ is the thickness of the aluminum target cup end caps, and $\rho_x$ and $M_x$ are the density and mass of background material $x=(\mathrm{NH}_3,\mathrm{Al},\mathrm{He}$), respectively. The measured yield is proportional to the cross section via $Y \propto \sigma \frac{\rho L}{M}$. The relevant lengths, masses, and densities are shown in Table~\ref{DilParam}.

\begin{table}[htp]
\begin{center}
\begin{tabular}{ l c  c  r }
\hline
  Material & $L$ (cm) & $M$ (u) & $\rho$  (g/cm$^3$)   \\ \hline
  NH$_3$& 2.8307 & 17.031&0.817\\
  He & 0.8738 & 4.0026 & 0.145\\
  C  & 0.1016 & 12.011 & 2.267\\
  Al & 0.0036 & 26.981 & 2.70 \\ \hline
\end{tabular}
\caption{\label{DilParam}Density and lengths of the dilution parameterization.}
\end{center}
\end{table}

The size of the liquid helium and nitrogen background is a function of the packing fraction of the ammonia beads. Ideally the target cup would be filled completely with solid ammonia, but the size and shape of the ammonia beads allows for some fraction of the target cup to be filled with superfluid helium, which aids in cooling the target. The ratio of the volume of the target cup to the volume occupied by the ammonia is the packing fraction. A larger packing fraction means there is more ammonia in the cup. The packing fraction is determined from the data by solving equation~\eqref{Dilution2} for the $pf$ to give
\begin{align}
\label{eq:pf}
pf = \frac{Y_{\mathrm{prod}}-Y_{\mathrm{dummy}}}{\frac{M_C\rho_{\mathrm{NH_3}}L_{\mathrm{cup}}}{M_{\mathrm{NH_3}}\rho_{C}L_{C}}{\bigg (}Y_\mathrm{C} - \frac{L_{\mathrm{cup}}+L_{\mathrm{out}}-L_{\mathrm{C}}}{L_{\mathrm{cup}}+L_{\mathrm{out}}}Y_{\mathrm{empty}}{\bigg)}{\bigg(}b+3\frac{\sigma_H}{\sigma_C}{\bigg)}-\frac{L_{\mathrm{tg}} } {L_{\mathrm{tg}}+L_{\mathrm{out}}}Y_{\mathrm{empty}} }\,,
\end{align}
where $Y_{\mathrm{prod}}$, $Y_{\mathrm{dummy}}$, $Y_{\mathrm{empty}}$, and $Y_{\mathrm{C}}$ are the yields from a production, dummy, empty and carbon runs, respectively. Simulation determines the carbon to nitrogen scaling factor, $b$, and the ratio $\sigma_H$/$\sigma_C$. The uncertainty in the packing fraction is minimized by extracting the $pf$ using a linear fit to equation~\eqref{eq:pf} beyond the resonance region. The validity of this fit is subject to good agreement between the normalized yields in the data. Large spreads in the yields create ambiguity in the fit and result in a larger uncertainty. The experimentally determined packing fractions for the 5 T settings are shown in Table~\ref{PackingFraction}. 

\begin{table}[htp]
\begin{center}
\begin{tabular}{ l c  c r }
\hline
  Setting & Material & $pf$ & $\delta_{pf}$ (\%)   \\ \hline
  3350 5T Tran.& 20 & 0.6009 &5.97\\
  3350 5T Tran. & 19 & 0.7091 & 5.78\\
  2254 5T Tran.  & 20 & 0.6351 & 4.49\\
  2254 5T Tran. & 19 & 0.6231 & 4.10 \\ 
  2254 5T Long.  & 18 & 0.6535 & 3.37\\
  2254 5T Long. & 17 & 0.5924 & 3.75 \\ \hline
\end{tabular}
\caption{\label{PackingFraction}Packing fractions for the 5 T settings. The 2.5 T settings are still under analysis.}
\end{center}
\end{table}

\begin{figure}
\centering     
\subfigure[$E_0$ = 2254 MeV 5T Transverse ]{\label{fig:Dil2254T}\includegraphics[width=.85\textwidth]{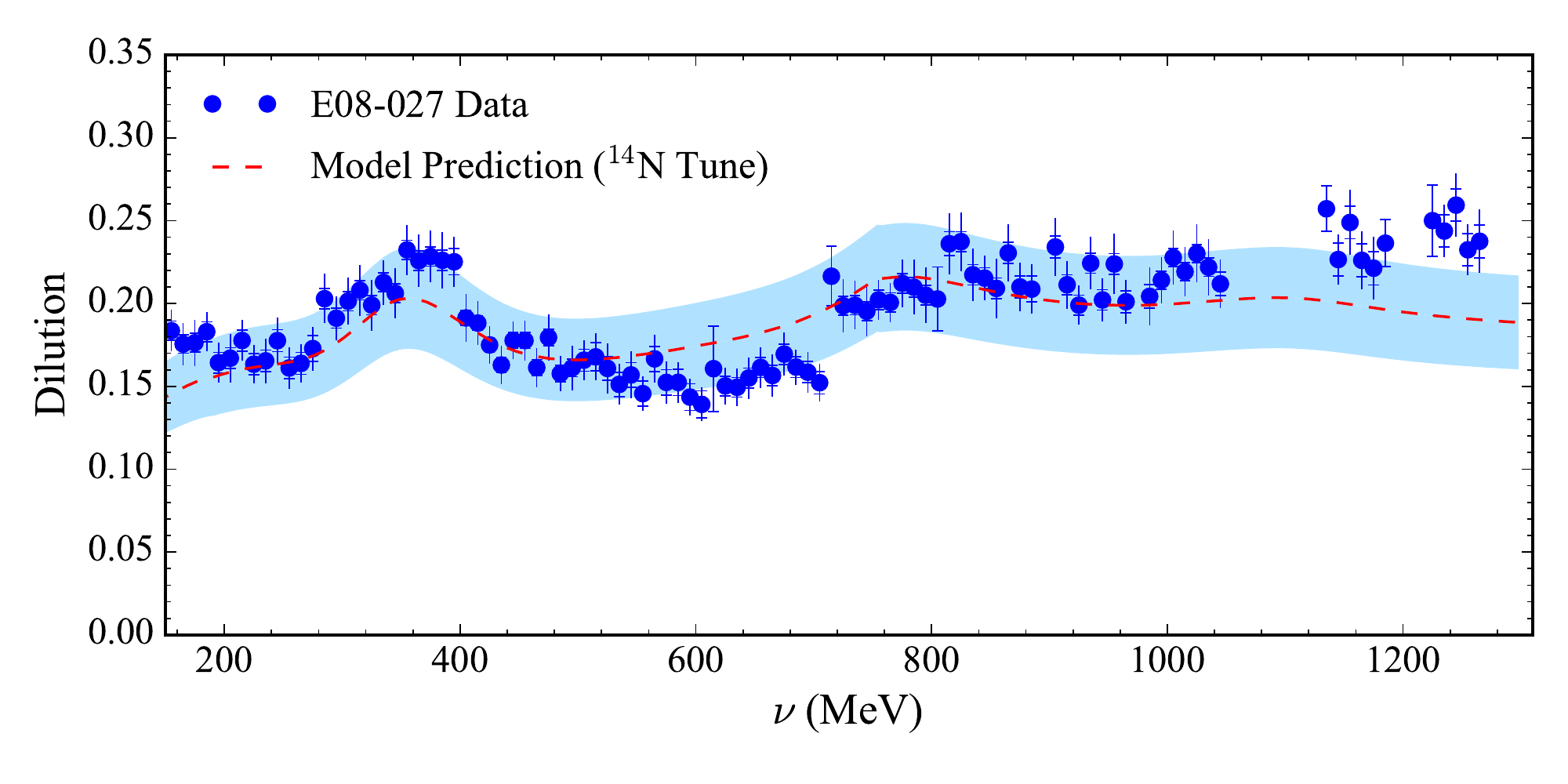}}
\qquad
\subfigure[$E_0$ = 2254 MeV 5T Longitudinal]{\label{fig:Dil2254L}\includegraphics[width=.85\textwidth]{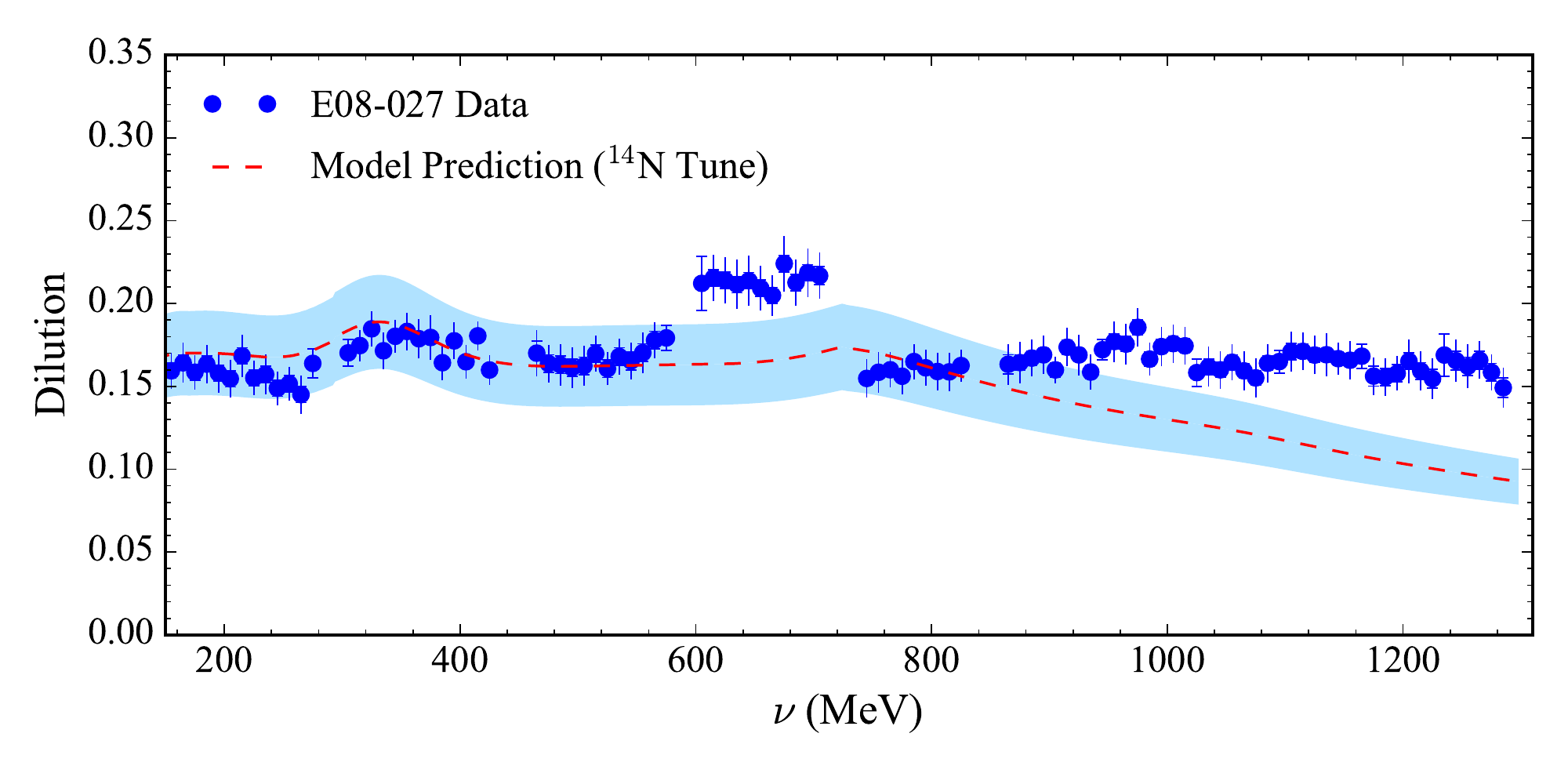}}
\qquad
\subfigure[$E_0$ = 3350 MeV 5T Transverse]{\label{fig:Dil3350}\includegraphics[width=.85\textwidth]{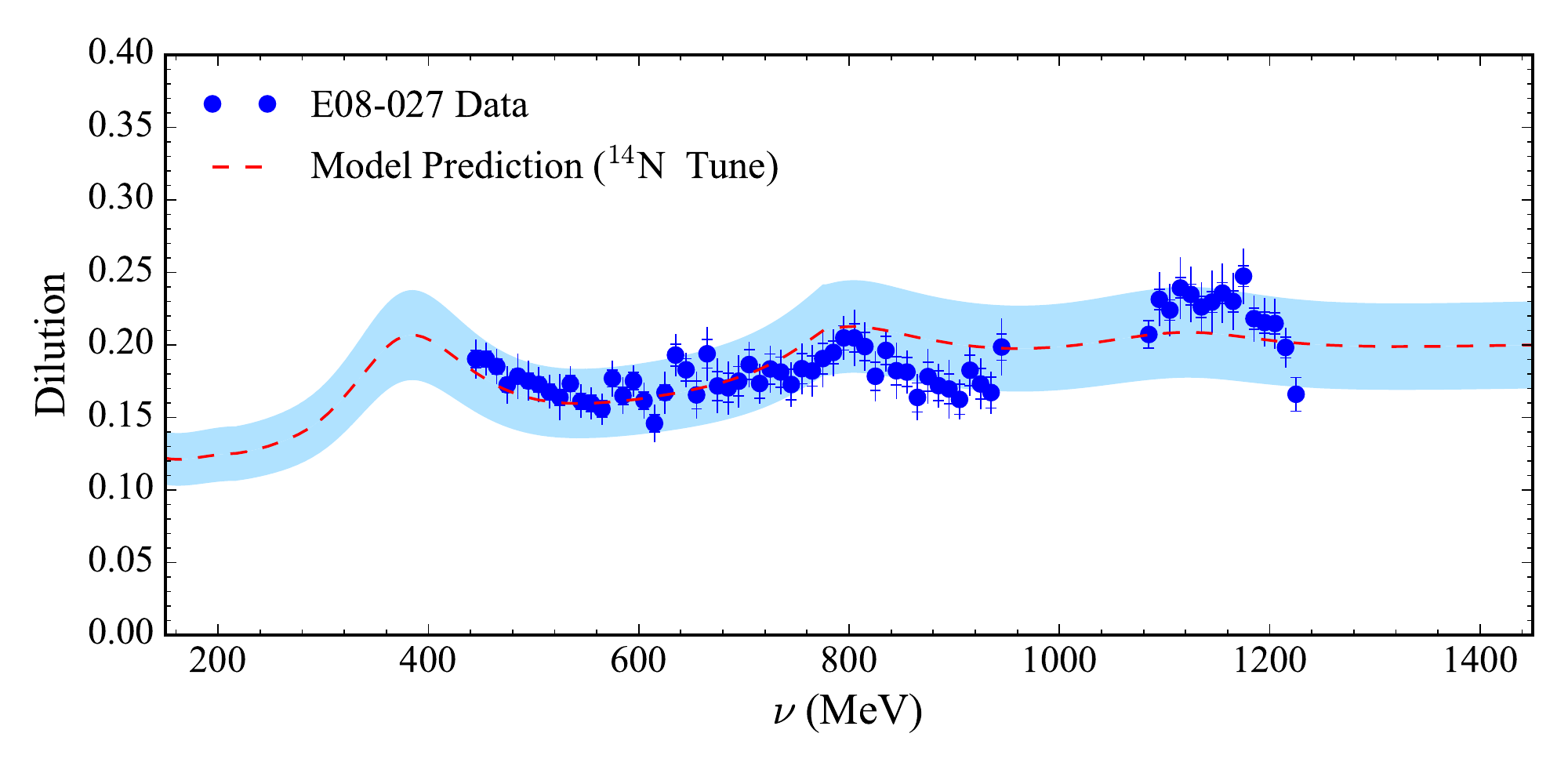}}
\caption{Experimentally determined dilution factors for the 5 T settings. The missing data at low $\nu$ at $E_0$ = 3350 MeV is due to an unresolved yield drift issue.}
\label{Fig:Dilution}
\end{figure}

The experimental dilution factors for the 5 T settings are shown in Figure~\ref{Fig:Dilution}. The inner error bars are statistical and the outer are the statistical and systematic errors added in quadrature. The model prediction is an application of equation~\eqref{eq:Dilution} using the Bosted-Mamyan-Christy fit and the radiation formalism of RADCOR, with the packing fraction values from Table~\ref{PackingFraction}. The model is tuned to better reproduce new nitrogen data from the Small Angle GDH experiment. This tune is discussed in Appendix~\ref{app:Appendix-C}. The blue bands are the systematic error of the model prediction and are approximately $\pm$15\%, and represents model uncertainties of 7\%, 10\%, 20\%, and 20\% for nitrogen, hydrogen, helium and aluminum respectively. Small gaps in the dilution coverage are filled in with a linear interpolation routine. Bigger gaps, such as $\nu$ $<$ 450 MeV at $E_0$ = 3.3 GeV 5T Transverse, are filled in using the tuned model dilution. More information on the dilution analysis is found in Ref~\cite{TobyDil}. 

\chapter{\sc Radiative Corrections}
\label{ch:RC}

 The difference between the Born cross section and the measured cross section are radiative corrections, and take into account all methods for radiative energy loss from the incoming and scattered electron. The goal of the analysis is to calculate a radiated cross section that can be subtracted from the experimental data to obtain a cross section that is more representative of the Born process of Figure~\ref{fig:aBorn}. Both elastic and inelastic processes contribute to the radiated cross section, but only states with a smaller invariant mass ($W$) radiatively affect larger invariant mass states. Elastic scattering events  have the lowest possible invariant mass, causing the elastic radiative tail to be present throughout the entire energy spectra. The first step in the radiative correction analysis is to remove this elastic tail, which is then followed by the inelastic radiative corrections. The analysis in this chapter is guided by theoretical work of Ref~\cite{Polrad,Polrad2,Polrad3,Polrad4} for polarized radiative corrections (POLRAD) and Ref~\cite{TSAI,MT,MillerT} for the unpolarized radiative corrections (RADCOR). The details of the radiative corrections analysis for the E08-027 data are discussed in Chapter~\ref{result:RC}.

\section{Radiative Processes}
\begin{figure}
\centering     
\subfigure[Born scattering process ]{\label{fig:aBorn}\includegraphics[width=.45\textwidth]{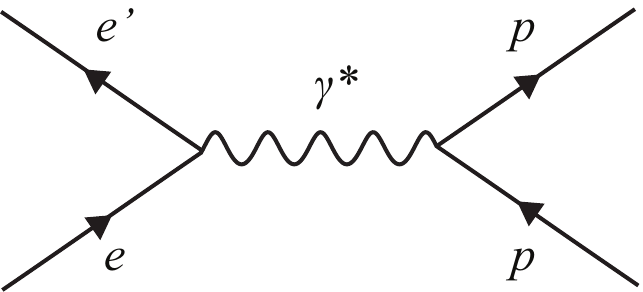}}
\qquad
\subfigure[External and internal bremsstrahlung]{\label{fig:bExt}\includegraphics[width=.45\textwidth]{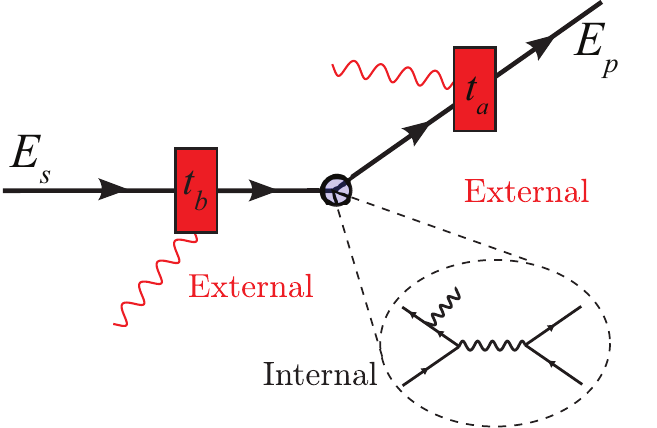}}
\caption{Born process and bremsstrahlung photon radiation, highlighting the distinction between external and internal bremsstrahlung.}
\end{figure}

The leading order Feynman diagram representing (elastic) $ep \rightarrow e'p$ scattering is shown in Figure~\ref{fig:aBorn}.  Theoretical analyses assume the physics of this lowest order diagram, but the likelihood of the process in Figure~\ref{fig:aBorn} occurring experimentally is virtually zero~\cite{TSAI}. There are higher order (in $\alpha$, the fine structure constant) loop corrections to Figure~\ref{fig:aBorn}; the electron can  also ``externally" radiate a bremsstrahlung photon as it passes through material before and after scattering, or ``internally" radiate a photon during the scattering process itself. The internal contribution is further divided into polarized and unpolarized contributions: the former occurring with a polarized beam and target and the later with both the beam and target unpolarized. 

The distinction between the external and internal radiative effects for an incoming electron of energy, $E_s$, and scattered energy, $E_p$, is shown in Figure~\ref{fig:bExt}. The external bremsstrahlung processes are characterized by the quantities $t_b$ and $t_a$, which are the radiation thicknesses before and after scattering, respectively. The calculation of these radiation thicknesses for E08-027 is discussed in Appendix~\ref{app:Appendix-D}. The incident and scattered electrons are also subject to ionizing collisions from materials in the beam path. 
The effects of the ionization and external bremsstrahlung are sometimes combined and referred to as electron straggling.


\section{Applicability of the First Born Approximation}
The formalism of the radiative corrections process is constructed under the first Born approximation, which implies single photon exchange during the scattering interaction. The error made in using this assumption is estimated as 
$\delta = 2Z\alpha\,\mathrm{sin}\frac{\theta}{2}$~\cite{Maximon,Barreau,Borie}. For the scattering angles around six degrees and target nuclei of $Z$=1, the error introduced by this assumption is approximately 0.15\% and is safely treated as a systematic error.

\subsection{Loop Corrections to the Scattering}
\begin{figure}
\centering     
\subfigure[Vacuum polarization]{\label{fig:Loop}\includegraphics[width=.45\textwidth]{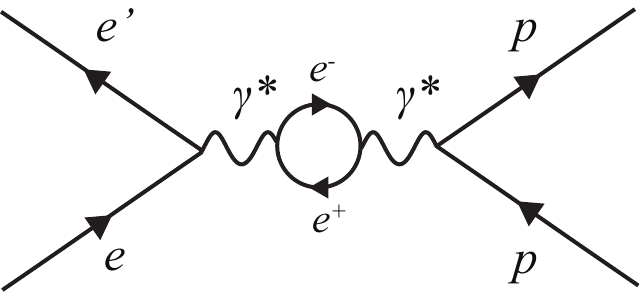}}
\qquad
\subfigure[Vertex correction]{\label{fig:Vertex}\includegraphics[width=.45\textwidth]{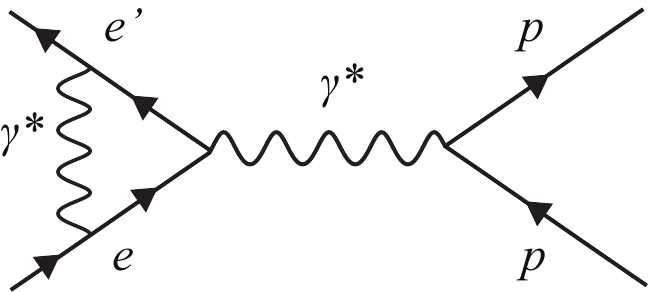}}
\qquad
\subfigure[Electron self-energy]{\label{fig:SelfA}\includegraphics[width=.45\textwidth]{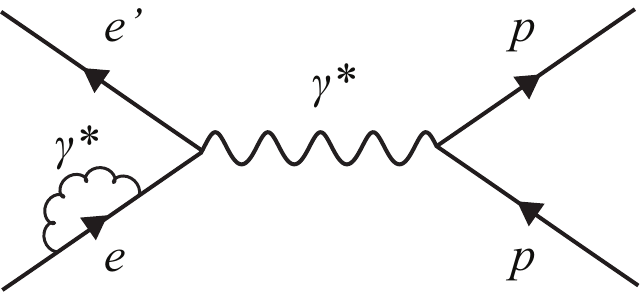}}
\qquad
\subfigure[Electron self-energy]{\label{fig:SelfB}\includegraphics[width=.45\textwidth]{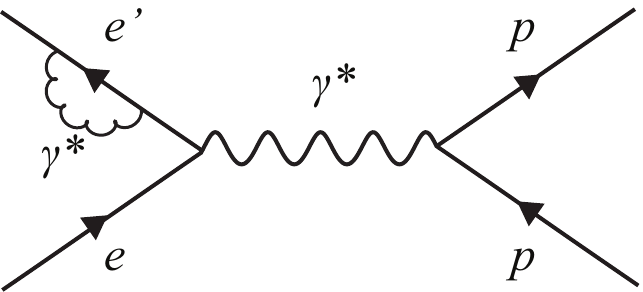}}
\qquad
\subfigure[Bremsstrahlung before scattering]{\label{fig:BremA}\includegraphics[width=.45\textwidth]{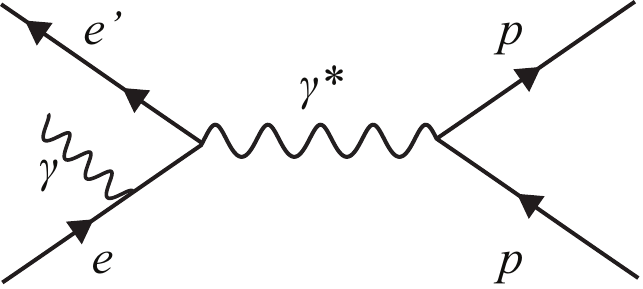}}
\qquad
\subfigure[Bremsstrahlung after scattering]{\label{fig:BremB}\includegraphics[width=.45\textwidth]{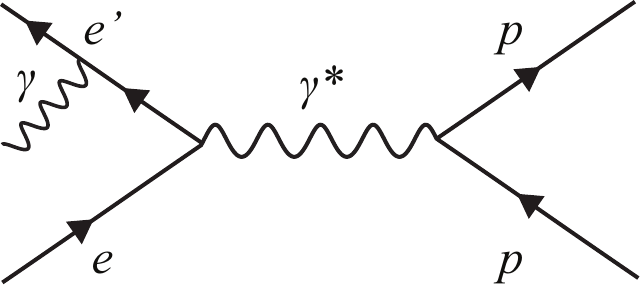}}
\caption{Next to leading order Feynman diagrams for the internal radiative corrections.}
\label{fig:LoopCorrections}
\end{figure}

The next to leading order processes considered in the radiative correction analysis are shown in Figure~\ref{fig:LoopCorrections}.  These loop corrections account for virtual photon effects on the scattering cross section.  The Feynman diagram in Figure~\ref{fig:Loop} is the vacuum polarization correction, where the virtual photon spontaneously splits into a $e^-/e^+$ pair. The charged pairs act as an electric dipole and their reorientation in the electron's electromagnetic field create a partial screening effect on the field~\cite{Loops}. The vertex correction, in Figure~\ref{fig:Vertex}, is the lowest order correction to the anomalous magnet moment of the electron; this diagram causes a small deviation in the electron's magnetic moment as predicted for a Dirac particle. The electron self-energy Feynman graphs are shown in Figure~\ref{fig:SelfA} and~\ref{fig:SelfB}. These diagrams contribute to the electron's mass renormalization and reflect the influence of the vacuum fluctuations on the ground state energy of the electron~\cite{Loops}.
Diagrams corresponding to two-photon exchange, target vertex and target bremsstrahlung\footnote{An approximate form for the target bremsstrahlung is found in Ref~\cite{Miller}.} are considered small and ignored. There are also contributions from muon, tau lepton and quark loops to the vacuum polarization diagram. Their inclusion is dependent on the radiative corrections scheme used; in the analysis of this thesis, the polarized formalism takes the higher order loops into account, while the unpolarized formalism does not.


\section{Bremsstrahlung Radiation}
The use of an accelerated (and charged) probe in electron scattering experiments causes real photon bremsstrahlung emission during the electron-proton interaction as shown in Figures~\ref{fig:BremA} and~\ref{fig:BremB}. Bremsstrahlung is also possible as the electron passes through material in the beam path before and after the target interaction.  For the purposes of doing the radiative corrections, internal bremsstrahlung occurs within the Coulombic field of the target nucleus and external bremsstrahlung occurs within the Coulombic field of the surrounding electrons and nuclei. 


 Photons have zero mass, and so the total energy of an emitted bremsstrahlung photon can be arbitrarily small. The bremsstrahlung process radiates a single hard photon consisting of most of the radiated energy, and the remaining energy is spread over an infinite number of soft photons~\cite{Soft}.  Being soft, these soft photons lack sufficient energy to be detected by the experiment, but they still have a real contribution to the bremsstrahlung cross section and radiative corrections. They must be taken into account in the radiative corrections analysis.




\subsection{Radiation Thickness and Length}\noindent  The amount of energy lost to bremsstrahlung photons by an electron passing through a material is parameterized by the materials radiation length\footnote{The radiation length of a material is the thickness required for an electron to lose 1-1/$e$ of its energy as it travels through the material.}. Differences in the physical thicknesses and densities of the various beam line materials are accounted for by weighting a material's radiation length by its thickness and density to produce a radiation thickness (see Appendix~\ref{app:Appendix-D}). In the calculation of the radiation thicknesses, it is assumed that the electron scattering occurs in the center of the target material, but not necessarily the center of all the material the electron traverses. If the total radiation thickness of the target before and after scattering is less than 0.10, the error on this assumption is very small~\cite{TSAI}.

\section{Ionization Energy Loss}
\label{ILOSS}
In addition to bremsstrahlung emission, it is possible for the electron to elastically scatter with atomic electrons from the materials in the beam path. This generally results in the ionization of the struck atom and causes the electron to lose a few MeV of energy. Multiple ionizing collisions affect the incident, $E_{\mathrm{inc.}}$, and final, $E_{\mathrm{fin.}}$, electron energies and are related to the beam and spectrometer momentum such that,
\begin{align}
E_{\mathrm{inc.}} &= E_{\mathrm{beam}} - \Delta_s\,,\\
E_{\mathrm{fin.}} &= E' + \Delta_p\,,\\
E' & \simeq p_{\mathrm{spect}}\,,
\end{align} 
where $\Delta_{s,p}$ is energy loss due to the ionizing collisions; for a highly relativistic electron $p \simeq E$. The exact values of the ionization energy loss parameters are calculated in Appendix~\ref{app:Appendix-D}, but, in general, the terms are independent of the electron energy and usually on the order of a few MeV.

\section{Scattering Amplitudes and Cross Sections}
Taking into account the first-order QED corrections and  real photon emission, the schematic expression for the inclusive electron scattering cross section is
\begin{align}
\sigma (e^-p) \approx   |\mathcal{M}_{\mathrm{Born}}|^2 &+ 2\mathrm{Re}[ \mathcal{M}_{\mathrm{Born}} (\mathcal{M}_{\mathrm{vert}} +  \mathcal{M}_{\mathrm{vac}} +  \mathcal{M}_{\mathrm{self}} )] \\ 
&+ | \mathcal{M}^{E_s}_{\mathrm{1\gamma}} +\mathcal{M}^{E_p}_{\mathrm{1\gamma}}  |^2 + \mathcal{O}(\alpha^4)\,, \nonumber
\end{align}
where $\mathcal{M}_{\mathrm{Born}}$ is the Born amplitude associated with Figure~\ref{fig:aBorn}, and $\mathcal{M}_{\mathrm{vert}}$, $\mathcal{M}_{\mathrm{vac}}$ and $\mathcal{M}_{\mathrm{self}}$ are the vertex, vacuum and self energy diagram amplitudes, respectively. The diagram for internal bremsstrahlung by the incoming (outgoing) electron is $\mathcal{M}^{E_s}_{\mathrm{1\gamma}}$ ( $\mathcal{M}^{E_p}_{\mathrm{1\gamma}}$). The above equation neglects two-photon exchange diagrams and hadron bremsstrahlung, but is representative of the processes considered in the remaining analysis.

There is a well-known infrared divergence\footnote{The expression approaches infinity as photon energy approaches zero.} in the vertex diagram. This divergence is canceled by a corresponding divergence of opposite sign in the internal bremsstrahlung cross section~\cite{Gramolin}. Photon emission always occurs and the divergence is always canceled.  Any experimental detector has a non-zero energy resolution, which makes the bremsstrahlung and vertex processes indistinguishable. Only the sum of the two processes is observable, which is finite. 

In terms of the radiative corrections process, this infrared divergence is treated differently by the unpolarized and polarized formalism. In the unpolarized case, the divergence is controlled with the introduction of infrared cut-off energy. The cut-off is the distinction between hard and soft photons, and is the maximal energy of an emitted photon that is not detectable in the experiment. Only contributions due to a single hard photon process are considered. Soft-photon effects are determined and applied separately. The choice of this cut-off parameter is some-what arbitrary and its choice should not effect the final radiatively corrected result. The polarized formalism uses the results from Ref~\cite{Bardin,Akhundov} to cancel out the infrared divergence exactly and without the use of any arbitrary parameters. 
\section{Elastic Radiative Corrections }\label{sec:Theory}
The cross section for the elastic radiative tail is 
\begin{equation}
\label{etail}
\sigma_{\mathrm{ el. tail}} = (\sigma_{\mathrm{int. b}} + \sigma_{\mathrm{ext. b}} + \sigma_{\mathrm{coll.}})\cdot F_{\mathrm{soft}}\,,
\end{equation}
where $\sigma_{\mathrm{int. b}}$, $\sigma_{\mathrm{ext. b}}$ and  $\sigma_{\mathrm{coll.}}$ represent the tail contributions from internal and external bremsstrahlung, and collisional loss processes, respectively. The term $F_{\mathrm{soft}}$ is an overall scale correction, accounting for soft photon emission. The external radiation and collisional loss processes are not polarization dependent. 

\subsection{Unpolarized Internal Radiative Tail}

The internal bremsstrahlung cross section for inclusive scattering experiments is calculated exactly (to lowest order of $\alpha$) with the following assumptions: one-photon exchange, and the much larger mass of the hadron system suppresses target bremsstrahlung so it can be ignored~\cite{MT}.  In inclusive scattering experiments, the scattered electrons are detected at different angles and energy,  but the emitted photons remain undetected. The internal radiative cross section is defined using the form factors (in the first Born approximation), as deduced from inclusive electron scattering, integrated over the unmeasured photon's energy and angle,

\begin{equation}
\label{brem}
\sigma_{\mathrm{Int}} = \int B_{\mu \nu} T_{\mu \nu} \, d\Omega_k \,,
\end{equation}
where $B_{\mu \nu}$ is the internal bremsstrahlung tensor and $T_{\mu \nu}$ is the leptonic tensor which includes the elastic form factors, $W_1(Q^2)$ and $W_2(Q^2)$, as written in Ref~\cite{MT}. 
\begin{figure}[htp]
\begin{center}
\includegraphics[width=0.60\textwidth]{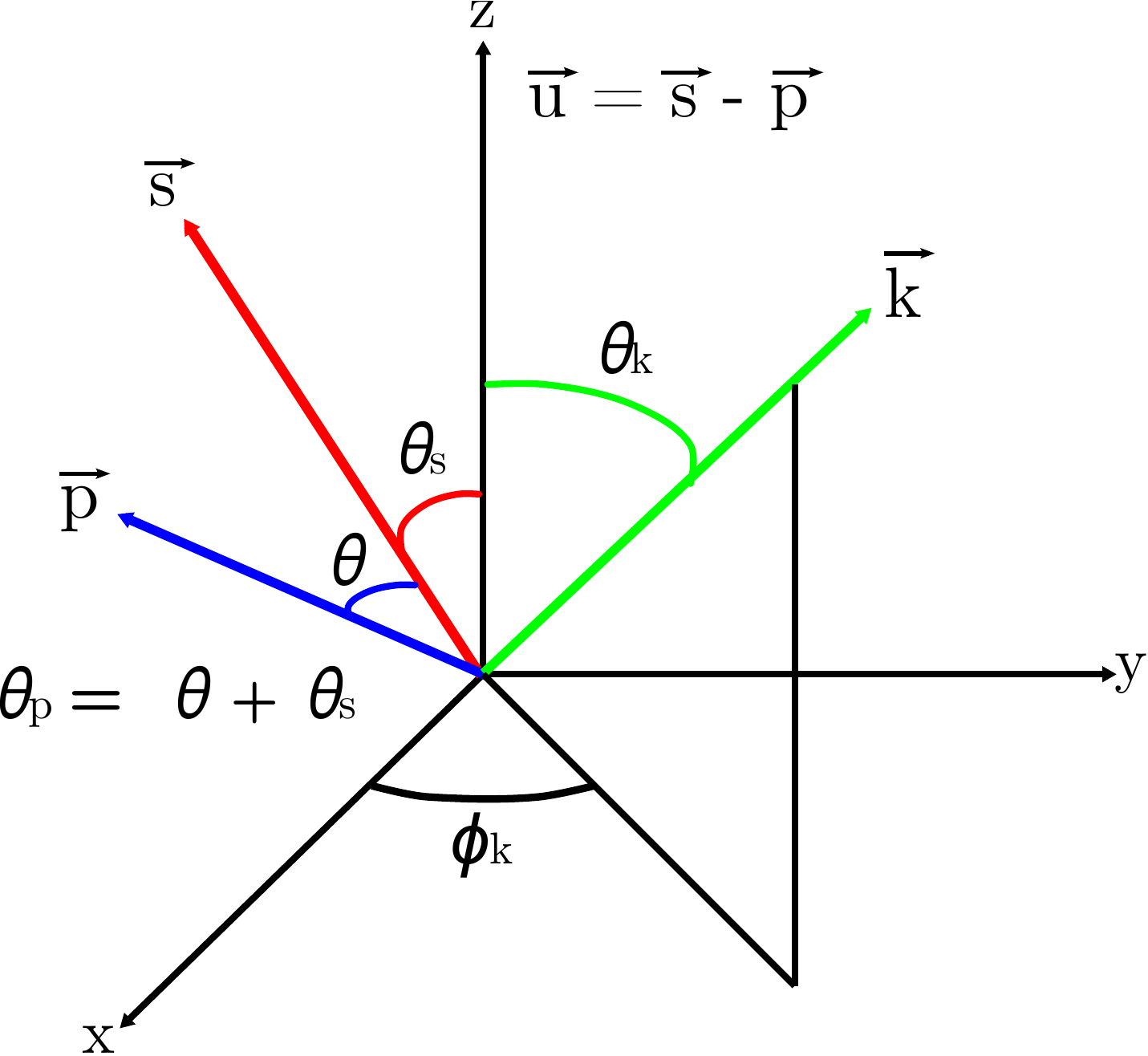}
\caption{\label{Coord}Internal bremsstrahlung coordinate system. Reproduced from Ref~\cite{TSAI}.}
\end{center}
\end{figure}

The integration of equation~\eqref{brem} is calculated in the lab frame of reference and makes the following variable definitions: four-vectors $s$, $p$, $t$ and $k$ refer to the incident electron, outgoing electron, target particle and real emitted photon, respectively.  Two additional four-vectors used are the missing mass in the absence of radiation, $u = s + t - p$ and the final state of the target nucleus $P_f = u - k$. The angle between $u$ and $k$ is $\theta_k$. A diagram of coordinate system is show in Figure~\ref{Coord}. The ${\bf z}$ axis is along the ${\bf u}$ direction and the electron momenta $s$ and $p$ are in the $x$-$z$ plane.

The exact cross section for the elastic radiative tail is 
\begin{equation}\label{Exact}
\begin{split}
\sigma_{\mathrm{exact}} &= \left( \frac{d^2\sigma}{d\Omega dE_p} \right)_{\mathrm{ex}} = \frac{\alpha^3}{2\pi} \left(\frac{E_p}{E_s}\right)  \int_{-1}^1 \frac{2 M_T \omega d(\mathrm{cos}\theta_k)}{q^4(u_0 - |\vec{u}| \mathrm{cos}\theta_k)} \\
& \times {\Bigg (}  \tilde{W}_2 (Q^2) {\Bigg \{}  \frac{-am^2}{x^3} \left[2E_s(E_p + \omega) + \frac{Q^2}{2}\right]  - \frac{a' m^2}{y^3}\left[ 2E_p(E_s + \omega) + \frac{Q^2}{2}\right]  \\
& -2 + 2 \nu(x^{-1} - y^{-1})\{m^2(s  \cdot p - \omega^2) + (s \cdot p)[2E_sE_p - (s \cdot p ) + \omega (E_s - E_p)]\} \\
& +x^{-1}\left[2(E_sE_p + E_s\omega + E_p^2) + \frac{Q^2}{2} - (s \cdot p) - m^2 \right] \\
& - y^{-1}\left[2(E_pE_s + E_p\omega + E_s^2) + \frac{Q^2}{2} - (s \cdot p) - m^2 \right]  {\Bigg \}}  \\
& + \tilde{W}_1(Q^2){\Bigg [}{\Bigg (}\frac{a}{x^3} + \frac{a'}{y^3} {\Bigg )}m^2(2m^2 + Q^2) + 4 +4\nu(x^{-1} - y^{-1})(s \cdot p)(s \cdot p - 2m^2)  \\
& + (x^{-1} - y^{-1})(2s \cdot p + 2m^2 - Q^2) {\Bigg ]} {\Bigg )}\,,  
\end{split}
\end{equation}
where the numerous kinematic factors ($a, b, \nu \ldots$ ) are defined in Ref~\cite{Stein} A25 through A41. The incoming and outgoing electron energies are $E_s$ and $E_p$, respectively. The radiated photon energy is $\omega$. The two terms $\tilde{W}_2 (Q^2)$ and $\tilde{W}_1(Q^2)$ contain the elastic form factors and are defined as
\begin{align}
\tilde{W}_1 (Q^2) &= \tilde{F} (Q^2) W_1^{\mathrm{el}} (Q^2)\,,  \\ 
\tilde{W}_2 (Q^2) &= \tilde{F} (Q^2) W_2^{\mathrm{el}} (Q^2) .
\end{align}

The integration of equation~\eqref{Exact} has the potential for a divide-by-zero error when $a' = a$. This is caused by the choice of factorization for the $\phi_k$ integration and is not an actual problem with the form of the exact internal tail~\cite{TSAI}. In order to keep the integral finite, a small point near the uncertainty is avoided in the integration. This creates the possibility for error depending on the  numerical integration routine used. A comparison of FORTRAN and Python calculations of the exact internal nitrogen tail estimates the systematic error to be approximately 0.4\% of the internal tail, which contributes 0.2\% of the total tail. Although not used in this analysis, Leonard Maximon and Steven Williamson show that the numerical systematic can be avoided by carrying out a piecewise analytic evaluation of the internal elastic tail~\cite{Williamson}.

\subsection{Polarized Internal Elastic Tail}

The exact polarized internal elastic tail is given as
\begin{equation}
\label{poltail}
\frac{d^2\sigma}{dxdy} = -\alpha^3y\int_{\tau_{\mathrm{min}}}^{\tau_{\mathrm{max}}}d\tau \sum_{i=1}^{8} \sum_{j=1}^{k_i} \theta_{ij}(\tau) \frac{2M^2R^{j-2}_{\mathrm{el}}}{(1+\tau)(Q^2+R_{\mathrm{el}}\tau)^2}\mathcal{F}_i^{\mathrm{el}}(R_{\mathrm{el}},\tau)\,,
\end{equation}
where the numerous kinematic factors are defined in Ref~\cite{Polrad}. The above equation is similar to equation~\eqref{Exact}, noting that integration of the photon phase space, $\tau$, is related to the emitted photon's angle and energy:  (1$+\tau$) =  ($u_0$ - $|\vec{u}|\mathrm{cos}\theta_k$) and $\omega = R$. The $\theta_{i,j}(\tau)$ are analytic functions of the bremsstrahlung and hadron tensor contraction. The summation indexes $j, k_i$ are functions of the eight form factor contributions, $\mathcal{F}_i^{\mathrm{el}}(R_{\mathrm{el}},\tau)$. In spin-1/2 scattering $i$ runs from one to four and represents combinations of $F_1(Q^2),\, F_2(Q^2),\, g_1(Q^2)$, and $g_2(Q^2)$. The Jacobian converts the cross section into more experimentalist friendly units
\begin{equation}
\frac{d^2\sigma}{dxdy} = {\bigg(}\frac{M\nu}{E_p}{\bigg)}\frac{d^2\sigma}{dE_pd\Omega}\,,
\end{equation}
where $x$ is  the Bjorken-x scaling variable and y= $\nu/E_s$.
\subsection{Virtual Photon Corrections}
The function $\tilde{F} (Q^2)$ or ``fbar" represents corrections which are independent of the infrared cut-off.  The terms in fbar exist even in the absence of hard photon emission and are included as part of the form factors.
The vacuum polarization and (non-infrared divergent) vertex  Feynman graph contributions are
\begin{align}
\delta_{\mathrm{vac}} &= \frac{2\alpha}{\pi}{\bigg [}-\frac{5}{9} + \frac{1}{3}\mathrm{ln}{\bigg(}\frac{Q^2}{m_e^2}{\bigg )}{\bigg]} \label{mtvac}\,, \\
\delta_{\mathrm{vertex}} &= \frac{2\alpha}{\pi}{\bigg [}-1 + \frac{3}{4}\mathrm{ln}{\bigg(}\frac{Q^2}{m_e^2}{\bigg )}{\bigg]}\,,
\end{align}
where $m_e$ is the mass of the electron. The non-infrared soft-photon contribution is sometimes referred to as the (Julian) Schwinger~\cite{TSAI,Schwinger} term and is given as
\begin{equation}
\delta_s =  \frac{\alpha}{\pi}{\Bigg [}\frac{1}{6}\pi^2 - \Phi\bigg{(}\mathrm{cos}^2\frac{\theta}{2}{\bigg)}{\Bigg ]}\,, 
\end{equation}
where $\Phi(x)$ is the Spence function
\begin{equation}
  \Phi(x) = \int_0^x \frac{-\mathrm{ln}|1-y|}{y}\mathrm{d}y\,.
\end{equation}
Adopting the convention to exponentiate the non-infrared divergent terms, fbar is
\begin{equation}
\begin{split}
\label{FBAR}
 \tilde{F} (Q^2)   &= \frac{\mathrm{exp}(\delta_{\mathrm{vac}} + \delta_{\mathrm{vertex}} + \delta_{\mathrm{s}})}{\Gamma(1 + bt_a)\Gamma(1+ bt_b)}\,,\\
&\approx 1 + (\delta_{\mathrm{vac}} + \delta_{\mathrm{vertex}} + \delta_{\mathrm{s}}) + 0.5772 (bt_a + bt_b)\,,\\
& = 1 +( 0.5772 \cdot b(t_a + t_b)) + \frac{2\alpha}{\pi} {\Bigg [}\frac{-14}{9} + \frac{13}{12}\mathrm{ln}\frac{Q^2}{m_e^2}{\Bigg]} + \frac{\alpha}{\pi}{\Bigg [}\frac{1}{6}\pi^2 - \Phi(\mathrm{cos}^2\frac{\theta}{2}){\Bigg ]}\,,
\end{split}
\end{equation}
where $b \simeq 4/3$, and $t_b$ and $t_a$ are the radiation lengths before and after scattering, respectively. The gamma functions are due to the normalization of the external bremsstrahlung straggling function. Equation~\eqref{FBAR} varies slightly from Ref~\cite{Stein} A44, in that it drops the third term of A44. The third term is related to the angle peaking approximation, which is not used in this elastic radiative tail analysis.

The inclusion of the muon and tau lepton loops results in the following form for the vacuum polarization correction~\cite{Badalek,Dasu}
\begin{equation}
\label{fullvac}
\delta^{e,\mu,\tau}_{\mathrm{vac}} = \frac{2\alpha}{\pi}{\bigg[}-\frac{5}{9} + \frac{4m_l^2}{3Q^2} + \frac{1}{3} \sqrt{1+\frac{4m_l^2}{Q^2}}{\bigg(}1 - \frac{2m_l^2}{Q^2}{\bigg)} \mathrm{ln}{\bigg(}\frac{\sqrt{1 + 4m_l^2/Q^2} +1}{\sqrt{1 + 4m_l^2/Q^2} -1}{\bigg)}{\bigg]}\,,
\end{equation}
where $m_l$ is the mass of lepton $l$. In the limit $Q^2 >> m_e^2$, equation~\eqref{fullvac} reduces to equation~\eqref{mtvac}. Quark loops also contribute to the vacuum diagram. Quark terms are sensitive to the parameterization of the quark masses and, in this analysis, the quark loops are treated as a systematic error, along with the muon and tau lepton loops. A comparison of fbar using equation~\eqref{fullvac} and equation~\eqref{mtvac} gives an error of ~0.4\% at the kinematics of the experiment detailed in this analysis.

\subsection{The Angle Peaking Approximation}
Described in detail in Appendix C of Ref~\cite{MT}, the angle peaking approximation simplifies the integral of equation~\eqref{Exact} for the unpolarized internal bremsstrahlung. The approximation assumes that the spectrum of emitted photons is strongly peaked at angles corresponding to the incoming and outgoing electrons. This assumption holds when the majority of bremsstrahlung photons are emitted in the direction of the electron momentum. The approximation results in two terms: one for photon emission related to the incoming electron and one for photon emission related to the outgoing electron. The purpose of the angle peaking approximation is to save computing resources, and also limit the uncertainty arising from a limited kinematic knowledge of the form factors. Advances in computing power and an accumulation of elastic scattering data since the original radiative corrections papers were released in the 1970s alleviate the need for such an approximation in the calculation of the elastic radiative tail. The full integral calculation is completed in a manner of seconds on modern hardware and the form factors cover a wide momentum transfer range. 

The formalism of the inelastic radiative corrections does make use of the angle peaking approximation though, so the result is quoted below:
\begin{equation}
\begin{split}
\sigma_{\mathrm{int}}^{\mathrm{pk}} (E_s,E_p)&= \frac{bt_r\phi(v_p)}{\omega_p} \tilde{F}(q_p^2)\sigma_{\mathrm{el}}(E_s)\,,\\
&+  \frac{M_T + 2(E_s - \omega_s)\mathrm{sin}^2\frac{\theta}{2}}{M_T - 2E_p\mathrm{sin}^2\frac{\theta}{2}} \frac{bt_r\phi(v_s)}{\omega_s} \tilde{F}(q_s^2)\sigma_{\mathrm{el}}(E_s - w_s)\,,
\end{split}
\end{equation}
where $M_T$ is the mass of the target, and $\omega_s$ and $\omega_p$ are the maximum energies of the incoming and outgoing emitted photons, such that
\begin{align}
\omega_s &= E_s - \frac{E_p}{1-2\tfrac{E_p}{M_T}\mathrm{sin}^2\frac{\theta}{2}}\,, \\ 
\omega_p &=  \frac{E_s}{1+2\tfrac{E_s}{M_T}\mathrm{sin}^2\frac{\theta}{2}} - E_p\,.
\end{align}
The $bt_r\phi(v)$ term is evaluated using the method of equivalent radiators. This defines the effective internal radiation length as
\begin{equation}
bt_r = \frac{\alpha}{\pi}{\Bigg[}\mathrm{ln}{\Bigg (}\frac{Q^2}{m_e^2}{\Bigg )} -1 {\Bigg]}\\,
\end{equation}
and uses the following functional form for the bremsstrahlung spectrum
\begin{equation}
\phi(v) = 1 -v + \frac{3}{4}v^2\,,
\end{equation}
where $\nu$ is defined in equation~\eqref{vs} and equation~\eqref{vp}.
\subsection{External Radiative Corrections}
\label{ExtElastic}

The incident and scattered electrons lose energy as they travel through materials via ionizing collisions with atomic electrons and via bremsstrahlung within an atom's Coulomb field. There is also the possibility that the direction of the electron's momentum will change. The change in direction and scattering angle is small for large electron energies and is neglected from further analysis. The external radiative corrections are not polarization dependent. 

The amount of energy lost by the electron is a statistical quantity. Ignoring ionization, the probability for an electron to lose energy $E_0 - E$ in a thickness $t$ is~\cite{TSAI} 
\begin{align}
I^{MT}(E_0,E,t) &= \frac{tx_0}{\mathrm{\Gamma}(1+bt)} \left(\frac{E_0 -E}{E_0}\right)^{bt} W_b(E_0,E_0-E) \label{tsaiI}\,, \\
W^{MT}_b(E_0,E_0-E) & = \frac{1}{x_0}\frac{b}{E_0 -E}\phi{\bigg(}\frac{E_0 - E}{E_0}{\bigg)} \label{tsaiB}\,,
\end{align}
where $\mathrm{\Gamma}(x)$ is the gamma function, $x_0$ is the unit radiation length for material of atomic number $Z$, $E_0$ is the energy of the electron before any loss occurs and $W_b(E_0,E_0-E)$ is the probability that the electron loses energy, $E_0-E$, per unit radiation length due to bremsstrahlung. The shape of the bremsstrahlung spectrum, $\phi(v)$, is  approximated as $\phi(v) = 1 - v + \tfrac{3}{4}v^2$. The approximation for $\phi(v)$  holds for $E > $ 100 MeV and $v <$ 0.8~\cite{TSAI}. The term $[(E_0 - E) / E_0]^{bt}/\Gamma (1+bt) $ is a correction due to multiple photon emission, and is discussed in the following section.

In his thesis, Guthrie Miller presents an alternative form of equation~\eqref{tsaiI} and equation~\eqref{tsaiB}~\cite{Miller,MillerT}
\begin{align}
I^{M}(E_0,E,t) &= \mathrm{ln}{\bigg(}\frac{E_0}{E}{\bigg)}^{bt}\frac{1+btP(z)}{\mathrm{\Gamma}(1+bt)}\, t \,W_b(E_0,E) \label{MillerI}\,, \\
W^{M}_b(E_0,E) & = \frac{1}{E_0 - E} {\bigg [} 1 + {\bigg (}\frac{E}{E_0}{\bigg)}^2 - \frac{2E}{3E_0} {\bigg ]} \label{MillerB}\,, \\
P(z) & = 0.5388z - 2.1938 z^2 + 0.9634z^3\,,
\end{align}
where $z = (E_0 - E)/E_0$. The term $a$ is dropped from equation A7 in Ref~\cite{Miller} (the above equation~\eqref{MillerB}) because it is negligible. Miller states that the uncertainty in his energy loss function is less than 0.2\% for $t < 0.1$, $Z < 30$ and $(E_0 - E) > 0.002E_0$. The terms $[\mathrm{ln}(E_0/E)]^{bt}/\Gamma (1+bt)$ and $1 + btP(z)$ are corrections due to multiple photon emission.
\begin{figure}[htp]
\begin{center}
\includegraphics[width=0.50\textwidth]{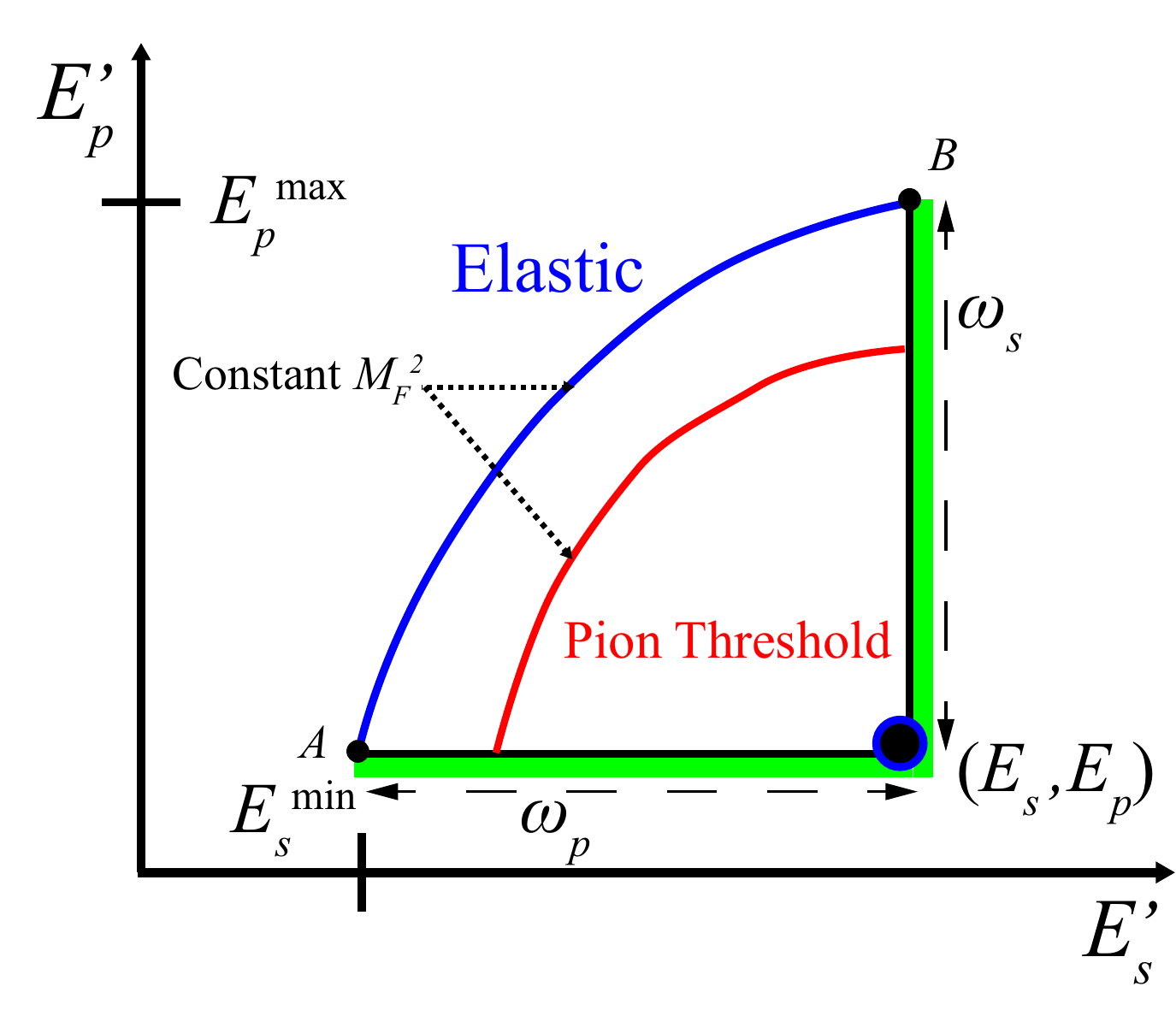}
\caption{\label{Triangle}External bremsstrahlung two-dimensional integration area and the energy peaking approximation. The triangular integration region is approximated as two line integrals (shaded in green).}
\end{center}
\end{figure}

The raw experimental cross section is a convolution of the internal bremsstrahlung cross section and the energy loss functions over the incoming and outgoing electron energies and possible radiation lengths ($T$ is the total radiation length),
\begin{equation}\label{EXT}
\sigma_{\mathrm{ext}} (E_s,E_p) = \int_0^T \frac{\mathrm{d}t}{T}  \int_{E_s^{ \mathrm{min}}}^{E_s} \mathrm{d}E_s' \int_{E_p}^{E_p^{\mathrm{max}}} \mathrm{d} E_p' I(E_s,E'_s,t)\sigma_r(E'_s,E'_p,\theta)I(E_p,E'_p,T-t) \,,
\end{equation} 
where the limits of integration are set by the kinematics of elastic scattering 
\begin{align}
\label{IntBounds1}
E_p^{\mathrm{max}} & = \frac{E'_s}{1 + \tfrac{E_s}{M_T}(1-\mathrm{cos}\theta)}\,, \\
\label{IntBounds2}
E_s^{\mathrm{min}} & = \frac{E_p}{1 + \tfrac{E_p}{M_T}(1-\mathrm{cos}\theta)}\,.
\end{align}

A few approximations simplify equation~\eqref{EXT}.  Dropping the radiation length integral and making the replacement $t \rightarrow t_b$ and $T-t \rightarrow t_a$ introduces a systematic error of less than 0.01\% when $bT < 0.1$~\cite{TSAI}.  For discrete states of constant $M_F^2$, the surface integral is replaced by two line integrals via a delta function\footnote{The following identity is useful in evaluating the delta function: $\int_{-\infty}^{\infty} f(x) \delta(ax - b)dx = f(b/a)/|a|$.}. The resulting integration region is shown along the green lines in Figure~\ref{Triangle}. The two line integrals are heavily peaked near the points A and B, $i.e.$ the majority of the energy is lost to a single photon before or after scattering. In the energy-peaking approximation, the external elastic tail is approximated by  evaluating equation~\eqref{EXT} at points A and B:
\begin{equation}
\sigma(E'_s,E'_p,\theta) \rightarrow \sigma(E'_s,\theta) [ 2M_T + 2E'_s(1-\mathrm{cos}\theta)]\delta(2M_T(E'_s -E'_p) - 2E'_sE'_p(1-\mathrm{cos}\theta))\,,
\end{equation}
which leads to the following expression in the energy-peaking approximation
\begin{equation}
\begin{split} 
\label{peakingapprox2}
\sigma_{\mathrm{ext}}(E_s,E_p) &\approx\tilde{\sigma}(E_s,\theta) I(E_p + \omega_p,E_p,t_a)\\
&+ \tilde{\sigma}(E_s - \omega_s,\theta) \frac{M_T+(E_s +\omega_s)(1-\mathrm{cos}\theta)}{M_T-E_p(1-\mathrm{cos}\theta)}I(E_s,E_s - \omega_s, t_b)\,, 
\end{split}
\end{equation}
where $\tilde{\sigma}(E,\theta)$ is the elastic cross section multiplied by fbar. The gamma functions of the soft photon terms are absorbed into fbar.  It is worth pointing out that there is a typo in the delta function of Ref~\cite{TSAI} equation C.4. The correct form is found in equation C.10 of Ref~\cite{MT}, which is reproduced above. The external tail due solely to bremsstrahlung  (and neglecting multiple photon radiation) is
\begin{equation}\label{External}
\begin{split}
\sigma^{MT}_{\mathrm{ext}}(E_s,E_p)  &= \frac{M_T + 2(E_s - \omega_s)\mathrm{sin}^2\frac{\theta}{2}}{M_T - 2E_p\mathrm{sin}^2\frac{\theta}{2}} \left\{ \tilde{\sigma}_{\mathrm{el}} (E_s - \omega_s) \left[\frac{bt_b}{\omega_s} \phi(v_s) \right] \right\} \\
& +  \tilde{\sigma}_{\mathrm{el}} (E_s ) \left[\frac{bt_a}{\omega_p}\phi(v_p) \right] \,,\\
\end{split}
\end{equation} 
where the Tsai form of the bremsstrahlung loss function, $W^{MT}_b(E_0,E_0 - E)$, is used and
\begin{align}
\label{vs}
v_s &= \omega_s / E_s\,,\\
\label{vp}
v_p & = \omega_p/( E_p + \omega_p)\,.
\end{align}
The correction due to ionization energy loss is determined from the M$\o$ller cross section
 \begin{align}
 W_i(E_0-E,\xi) &= \frac{\xi}{(E_0-E)^2}\,,\\
 \xi &= \frac{\pi m_e}{2 \alpha} \frac{t_b+t_a}{(Z+\eta)\mathrm{ln}(183Z^{-1/3})}\,,\\\
 \xi_a &= \xi_b = \xi/2\,,\\
 \eta &= \mathrm{ln}(1440Z^{-2/3})/\mathrm{ln}(183Z^{-1/3})\,.
 \end{align}

Adding back in the terms due to ionization energy loss results in Equation A49 of Ref~\cite{Stein}. This form of $\xi$ assumes that the effective collisional loss thickness is the same before and after scattering and that the collisional thickness is directly proportional to the radiation thickness. Both are not always true. Ref~\cite{RadLength} shows an alternate and more accurate calculation for $\xi$ and represents what was used in this analysis. In general, the ionization is largest closest to the elastic peak, and the bremsstrahlung is more important for tails far away from the elastic peak.

The uncertainty in the energy-peaking approximation is understood by considering a known integral with a structure similar to the full external radiative tail integration~\cite{TSAI}
\begin{equation}
\label{EXACTpeak}
\int_0^1 x^{T_i -1} (1-x)^{T_f -1} dx = \frac{\Gamma(T_i)\Gamma(T_f)}{\Gamma(T_i + T_f)}\,,
\end{equation}
which under the peaking approximation is
\begin{equation}
\label{peakingapprox}
\int_0^1 x^{T_i -1} (1-x)^{T_f -1} dx \approx \frac{1}{T_f} + \frac{1}{T_i}\,,
\end{equation}
where $T_f$ and $T_i$ are equivalent to the final and initial radiation lengths. For small radiation lengths, $T_i + T_f =  bt <  0.1$,  the energy-peaking approximation is a very good approximation. The absolute value of the uncertainty is found by comparing the results of equation~\eqref{EXACTpeak} and equation~\eqref{peakingapprox}.

\subsection{Multiple Photon Processes}
Up to this point all formulas presented assume that a single photon carries all of the radiation energy in the radiative process.  This is not actually what happens. When an electron passes through material it always experiences multiple scatterings accompanied by soft photon emission and energy loss due to ionization, which is proportional to the target thickness~\cite{TSAI}. The small effect of the ionization loss is accounted for by having $E \rightarrow E - \Delta$, as shown in Chapter~\ref{ILOSS}.

The soft-photon correction terms in equation~\eqref{tsaiI} and equation~\eqref{MillerI} are found by solving the diffusion equation
\begin{equation}
\label{diffusion}
\frac{\partial I(E_0,E,t)}{\partial t} = -I(E_0,E,t) \int_0^E d\epsilon W_b(E,\epsilon) + \int_{0}^{E_0-E} d\epsilon I(E_0,E+\epsilon,t) W_b(E+\epsilon,\epsilon)\,,
\end{equation}
where $\epsilon$ is the energy lost by the electron. A numerical solution to this equation, subject  to the boundary condition
\begin{equation}
I(E_0,E,0) = \delta(E_0 - E)\,,
\end{equation}
is presented in both Ref~\cite{TSAI} and Ref~\cite{Miller}. The differences in the two results are a result of the differences in the choice of the bremsstrahlung straggling function, and Miller includes ionization in his energy loss probability function ($W \rightarrow W_b + W_i$).

To determine the effect of multiple photon radiation, Mo and Tsai compute equation~\eqref{EXT} using a form of $I(E_0,E,t)$ that only includes the soft photon terms of the straggling function, and the target length approximation
\begin{align}
 I^{MT}_{\mathrm{soft}}(E_p + \omega_p,E_p,t_a)& \int_{E_s - \omega_s}^{E_s} dE'_s I^{MT}_{\mathrm{soft}}(E_s,E'_s,t_b) \\&\approx  \frac{1}{\Gamma(1 + bt_b +bt_a)}{\bigg(}\frac{\omega_s}{E_s} {\bigg)}^{bt_b}{\bigg(}\frac{\omega_p}{\omega_p + E_p}{\bigg)}^{bt_a} \nonumber\,, \\
 I^{MT}_{\mathrm{soft}}(E_s,E_s - \omega_s,t_b) &\int_{E_p}^{E_p + \omega_p} dE'_p I^{MT}_{\mathrm{soft}}(E'_p,E_p,t_a) \\&\approx \frac{1}{\Gamma(1 + bt_b + bt_a)}{\bigg(}\frac{\omega_p}{\omega_p + E_p}{\bigg)}^{bt_a} {\bigg(}\frac{\omega_s}{E_s} {\bigg)}^{bt_b} \nonumber\,,
\end{align}
where the integration is computed using equation (B36) of Ref~\cite{TSAI}. The gamma function is absorbed into fbar and the remaining terms are the multiple photon correction factor. Both Ref~\cite{TSAI} and Ref~\cite{Stein} and the polarized formalism of Refs~\cite{Polrad,Polrad2,Polrad3} use this factor
\begin{equation}\label{Soft}
F_{\mathrm{soft}} = \left(\frac{\omega_s}{E_s}\right)^{b(t_b + t_r)} \left(\frac{\omega_p}{E_p + \omega_p}\right)^{b(t_a+t_r)}\,,
\end{equation}
where $t_r = b^{-1}(\alpha / \pi)[\mathrm{ln}(Q^2/m_e^2)-1]$ is the thickness of an ``equivalent radiator" used to take into account internal soft-photon emission. The term is a an overall scale factor that is applied to the single emission cross sections from the internal and external radiative effects.

In Ref~\cite{Miller}, Miller details an alternative calculation of the soft-photon correction, which is outlined below.  Since photons have zero mass, the total energy of a photon can be arbitrarily small. It also means that photon radiation always occurs.  For sufficiently soft photons, the net result is
\begin{equation}
d\sigma = \sigma_0 \,\frac{2t_r}{\omega} {\bigg (}\frac{\omega}{\sqrt{E_sE_p}}{\bigg)}^{2t_r} d\omega\,,
\end{equation}
where $\sigma_0$ is the cross section without radiation, $\omega$ is the (small) photon energy and $t_r$ is the equivalent radiator. The factor of two in front of $t_r$ is for consistency between the Miller and Mo and Tsai definitions of the equivalent radiator. Making the substitution $\sigma_0 \rightarrow \sigma_{1\gamma}$ and allowing for arbitrary soft photon radiation gives
\begin{equation}
d\sigma = \sigma_{1\gamma} \,\int_0^k \frac{2t_r}{\omega} {\bigg (}\frac{\omega}{\sqrt{E_sE_p}}{\bigg)}^{2t_r} d\omega = \sigma_{1\gamma} {\bigg(}\frac{k}{E_sE_p}{\bigg)}^{t_r} \,,
\end{equation}
where $k$ is the total, limiting, energy of the soft photons and $\sigma_{1\gamma}$ is the cross section for emitting a single, hard photon,  $eN \rightarrow eN\gamma$. The effective soft photon limit, $k$, is found by solving
\begin{equation}
\begin{split}
\int_{E_s - \omega_s}^{E_s} I(E,E_1,t_b) &\sigma_0(E_1, \theta) I(E_{1,p},E_p,t_a) dE_1 = {\bigg(}\frac{k}{\sqrt{E_sE_p}}{\bigg )}^{b(t_b+t_a)} \frac{1}{\Gamma(1 + bt_a + bt_b)}\\
{\bigg[} t_b &W_b(E_s,E_s-\omega_s) \frac{M_T + 2(E_s - \omega_s)\mathrm{sin}^2\frac{\theta}{2}}{M_T - 2E_p\mathrm{sin}^2\frac{\theta}{2}} \sigma_0(E_s - \omega_s,\theta)\\
&+ t_a W_b(E_p + \omega_p, E_p)\sigma_0(E_s,\theta) {\bigg]}\,,
\end{split}
\end{equation}
where the left-hand side is equation~\eqref{EXT} under the target length approximation, and assuming $t_r \ll t_b,t_a$. The integral over $E_{1,p}$ is reduced to a delta function for elastic scattering and is subject to the condition
\begin{equation}
E_{1,p} =  \frac{E_1}{1 + \frac{E_1}{M_T}(1-\mathrm{cos}\theta)}\,.
\end{equation}
Adding back in the internal multiple photon correction through an equivalent radiator gives 
\begin{equation}
F_{\mathrm{soft}} = {\bigg (} \frac{k}{\sqrt{E_sE_p}}{\bigg)}^{2bt_r + bt_a +bt_b}\,.
\end{equation}

\begin{figure}
\centering     
\subfigure[Nitrogen ]{\label{fig:softN2}\includegraphics[width=.8\textwidth]{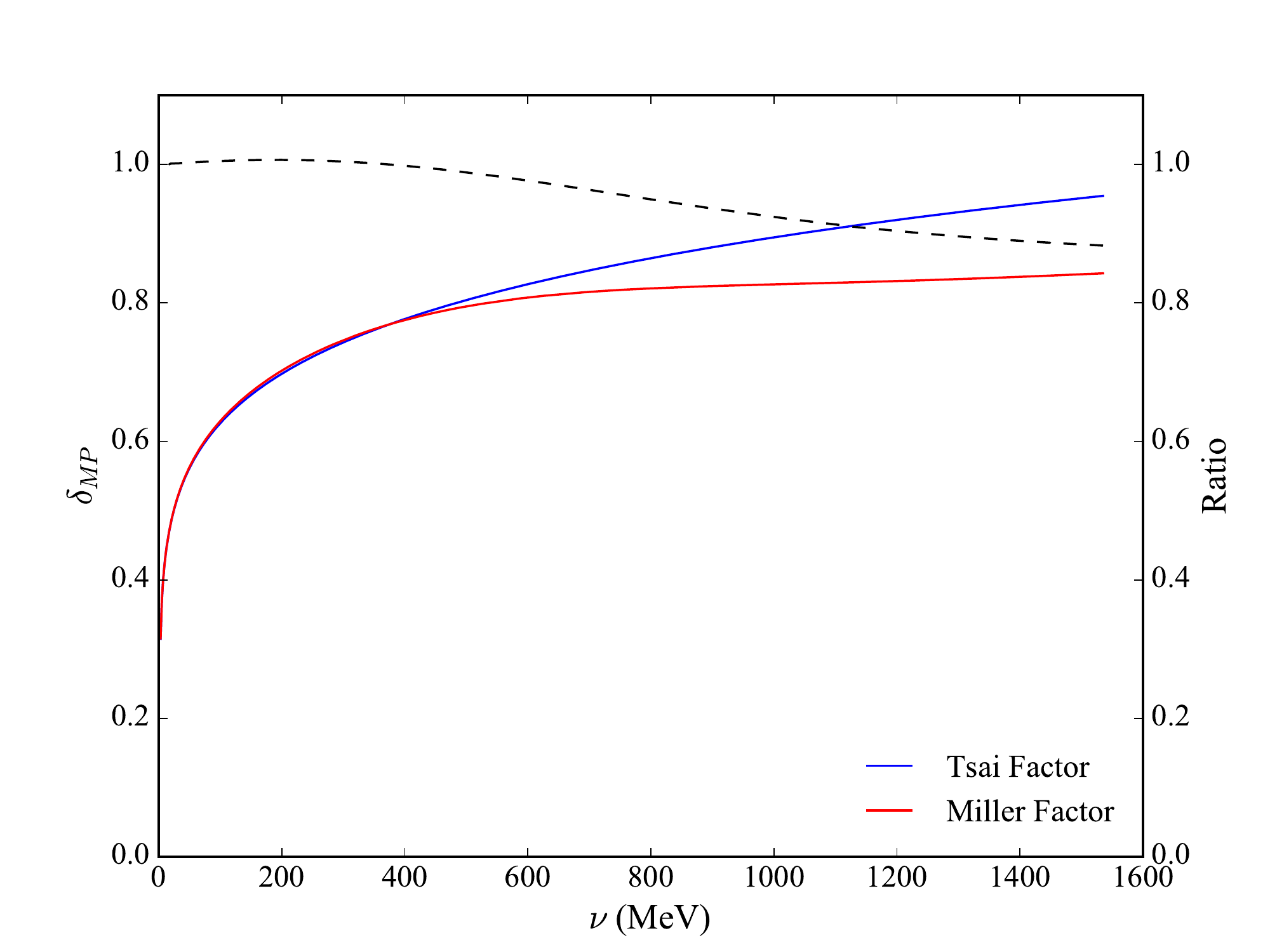}}
\qquad
\subfigure[Proton]{\label{fig:softP}\includegraphics[width=.8\textwidth]{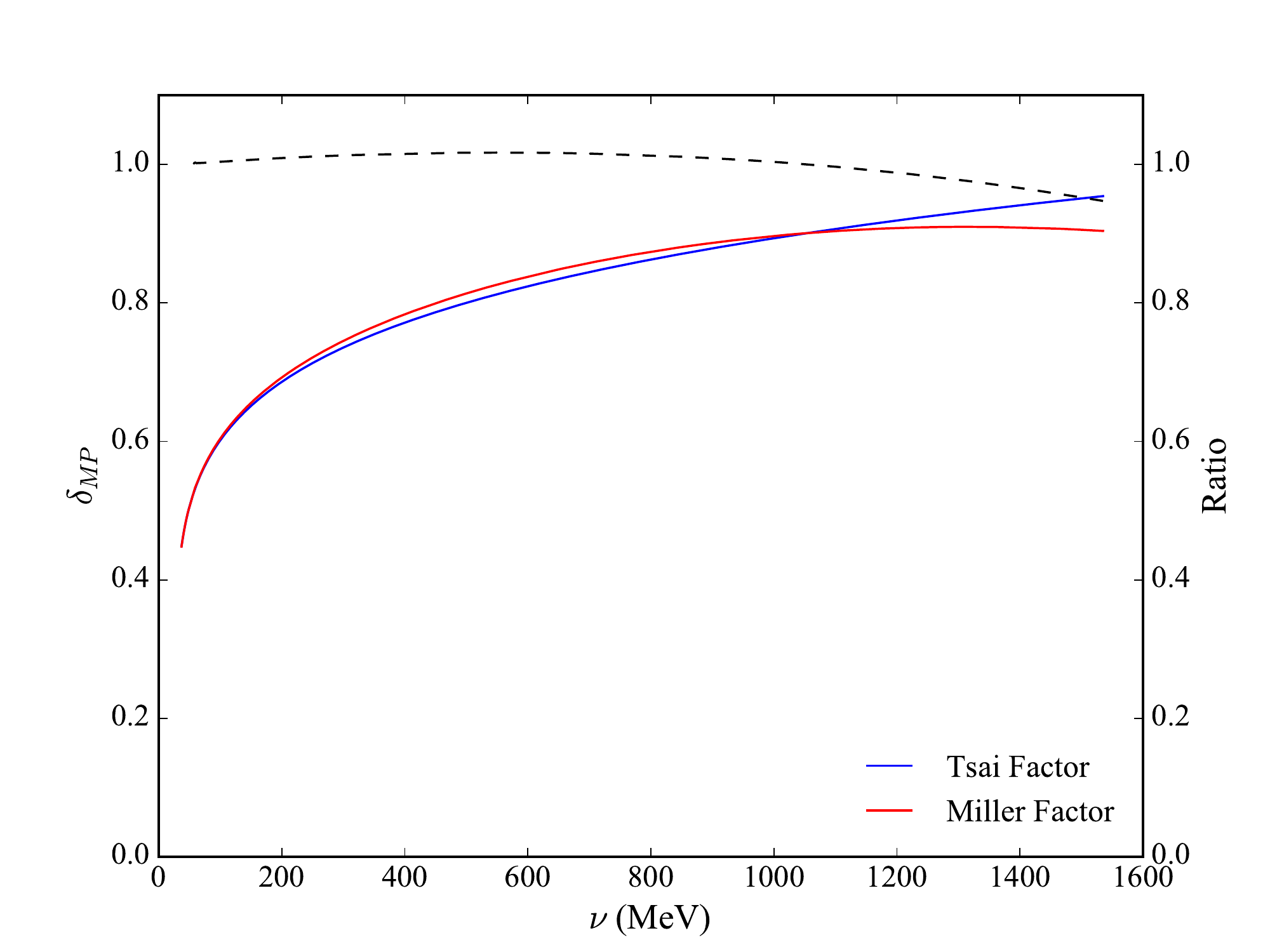}}
\caption{Comparison of the Mo and Tsai and Miller soft photon factors. The difference is most pronounced in nitrogen scattering at small electron energies and is because of the logarithmic $Q^2$ dependence of the nitrogen elastic form factors.}
\label{SoftP}
\end{figure}

The similarities and differences between the Miller and Mo and Tsai multiple photon corrections are highlighted in Figure~\ref{SoftP} for $E_0$ = 2.2 GeV and $\theta_{\mathrm{sc}}$ = 6.0$^{\circ}$. The radiation thicknesses are from Appendix~\ref{app:Appendix-D}. Both terms approach zero at low energy transfer, which is to say that all photons are soft near the elastic peak. Differences between the two formulations appear at larger energy losses.  The reason for the difference, most strongly seen in nitrogen, is because Miller's soft photon limit $k$ depends on how strongly the cross section varies across the kinematics. The nitrogen elastic form factors are logarithmic with $Q^2$; the proton discrepancy is smaller because the proton's dipole-like elastic form factors vary less severely with $Q^2$. The Mo and Tsai result falsely assumes that the cross section does not change significantly.  The rest of this analysis will use the Miller soft photon factor.

The uncertainty in the soft photon correction is two-fold: how dependent is the correction on the form factor fit and how well does the factor apply to the internal radiative tail. The former is straightforward to investigate and is studied by varying the fit and noting the change in $F_{\mathrm{soft}}$. The latter is more complicated, but is estimated by studying the dependence of the correction on $t_r$. Fortunately this dependence is rather small. A 10\% change in $t_r$ leads to a 0.5\% change in $F_{\mathrm{soft}}$, for the proton results given in Figure~\ref{SoftP}.

\subsection{Elastic Cross Section and Asymmetry}
The elastic, non-radiative cross section is defined from the elastic form factors as
\begin{equation}\label{elastic}
\frac{d\sigma}{d\Omega} (E_s,E_p,\theta) = {\bigg (} \frac{d\sigma}{d\Omega}{\bigg )}_0 F^2(Q^2,\theta)\,,
\end{equation}
with
\begin{equation}
 {\bigg (} \frac{d\sigma}{d\Omega}{\bigg )}_0= \frac{\alpha^2 \mathrm{cos}^2\tfrac{\theta}{2}}{4E^2\mathrm{sin}^4\tfrac{\theta}{2}} \frac{E_p}{E_s} ^2\,,
\end{equation}
and
\begin{align}
F^2(Q^2,\theta) &= W^{\mathrm{el}}_2(Q^2) + 2\mathrm{tan}^2\frac{\theta}{2}W_1^{\mathrm{el}}(Q^2)\,, \\
W_1(Q^2) & = \tau G_M^2(Q^2)\,,\\
W_2(Q^2) & = \frac{G_E^2(Q^2)+ \tau G_M^2(Q^2)}{1+\tau}\,,\\
\tau &= \frac{Q^2}{4M_T^2}\,,
\end{align}
where  $\alpha$ is the fine structure constant. The Mott cross section term is divided by $4Z^2$ as in Ref~\cite{MT} equation (B.3). Sometimes the full internal bremsstrahlung integral is written in terms of $F(Q^2)$ and $G(Q^2)$, which obey the following relation
\begin{align}
F(Q^2) &= 4[G_E(Q^2) + \tau G_M(Q^2)]/(1+\tau)\,,\\
G(Q^2)&= Q^2G_M(Q^2)\,.
\end{align}
For polarized elastic scattering the spin structure functions are written in terms of the electric and magnetic form factors as 
\begin{align}
g_1^{\mathrm{el}} (x,Q^2) &= \frac{1}{2}G_M(Q^2)\frac{G_E(Q^2) + \tau G_M(Q^2)}{1+\tau}\delta(x-1)\,,\\
g_2^{\mathrm{el}} (x,Q^2) &= \frac{\tau}{2}G_M(Q^2)\frac{G_E(Q^2) - G_M(Q^2)}{1+\tau}\delta(x-1)\,,
\end{align}
where the delta function restricts their validity to elastic scattering. This results in the following expression for the elastic polarized cross section~\cite{Puckett,JonesRSS}
\begin{align}
\label{pol_elastic}
\frac{1}{2}{\bigg[}\sigma^+ - \sigma^-{\bigg]} &= -2\sigma_{\mathrm{Mott}}\frac{E'}{E}\sqrt{\frac{\tau}{1+\tau}}\mathrm{tan}\frac{\theta}{2}{\bigg[}\sqrt{\tau{\bigg(}1+(1+\tau)\mathrm{tan}\frac{\theta}{2}{\bigg)}}\mathrm{cos}\theta^*G_M^2(Q^2)\nonumber \\ 
&+\mathrm{sin}\theta^* \mathrm{cos}\phi^*G_M(Q^2)G_E(Q^2) {\bigg]}\,, 
\end{align}
where $\theta^*$ and $\phi^*$ are the polar and azimuthal angles between the momentum transfer vector $\vec{q}$ and the proton's spin vector, $\vec{S}$. The azimuthal angle is the angle between the scattering plane defined by $\vec{k} \times \vec{k'}$ and the plane defined by $\vec{q} \times \vec{S}$. The elastic asymmetry is the ratio of equation~\eqref{pol_elastic} and equation~\eqref{elastic}.

\subsection{Fortran Analysis Code}
The previously described theoretical calculations are carried out in a collection of Fortran subroutines and functions that are collectively referred to as ROSETAIL~\cite{RTAIL}.  Equation numbers from the Ref~\cite{MT} and Ref~\cite{Stein} papers are referenced throughout the code, and equation numbers from this thesis are referenced in parentheses. 

The radiative corrections analysis is run from within the subroutine ``rtails", which takes $E_s$, $E_p$, $\theta$, $t_b$, $t_a$ and target type as arguments and returns the external and internal cross section corrections.  The nomenclature used for the target type can be found by searching for the `ctarg' variable, but two examples are `nitr' for nitrogen and `prot' for the proton. A `do' statement loops through the entire energy spectrum.

Within the``rtails" subroutine, the code first calls the function ``fbar" to calculate $\tilde{F}(Q^2)$, equation~\eqref{FBAR}, and next  calls the subroutine ``externl" to calculate the external radiative corrections, equation~\eqref{External}. The elastic cross section and form factors needed to complete the external correction are calculated in the functions ``sigbar" and ``fmfac" respectively. Also calculated in the ``externl" subroutine is the soft-photon correction factor $F_{\mathrm{soft}}$, equation~\eqref{Soft}. ``Externl" returns the cross-section corrections before and after the target as xextb and xexta, in units of nb $\cdot$ sr$^{-1}$ $\cdot$ MeV$^{-1}$.

After calculating the external corrections, the code moves onto the internal corrections. The integration of equation~\eqref{Exact} is done using a Simpson integration routine. The integrand is put together in the function ``xsect", which follow's Mo and Tsai's~\cite{MT} equation B.3 and B.5. There is also a preparation function for ``xsect" called ``xsectp" that calculates parameters for equation~\eqref{Exact} that do not depend on $\theta_k$. This eliminates the need to redefine certain variables every time ``xsect" is called but $E_s$, $E_p$ and $\theta$ are unchanged.

The results of the internal and external corrections are passed back to the main program and then are multiplied by the soft-photon correction term. The corrections are printed for each kinematic setting individually and as a sum.

A version of ROSETAIL modified by K. Slifer and S. Choi includes the POLRAD formalism to calculate the polarized elastic tails. This modified code supports polarization vectors at the target of 180$^{\circ}$ and 270$^{\circ}$ for proton and helium-3 scattering. A combination of the elastic cross section and elastic asymmetry from Ref~\cite{Polrad4} create the polarized cross sections for the external tail. The calculation of the polarized internal tail is described below.

Within the ``polsig\_el'' subroutine the code first calls function ``conkin'' to calculate the necessary kinematic constants and then ``deltas'' for the computation of the infrared divergence, vacuum polarization and vertex diagram contributions. The structure functions are calculated next in ``bornin'', followed by the analytic tail integration in equation~\eqref{poltail}. The results are combined and multiplied by the Jacobian to return the total polarized elastic tail. It is very important to note that the polarized version of ROSETAIL returns the polarized cross section and not the polarized cross section difference. The polarized ROSETAIL result must be multiplied by two for polarized cross section differences.

\section{Inelastic Radiative Corrections}
\label{InelasticRC}
The inelastic radiative corrections exist on a continuum of states, which contrasts with the discrete spectrum of the elastic radiative tail. A continuum is reached by summing up many discrete levels, so the inelastically (and unpolarized) radiated cross section is found by integrating equation~\eqref{etail} over all states above the elastic peak. For the internal bremsstrahlung tail this results in\footnote{There is a missing factor of $1/2\pi$ in Ref~\cite{MT}.}~\cite{MT} 
\begin{equation}
\label{InelasInt}
\frac{d\sigma_r}{d\Omega dE'} (\omega > \Delta E) = \frac{\alpha^3}{4\pi^2}\frac{E_p}{M_T E_s} \int_{-1}^{1} d (\mathrm{cos}\theta_k)\int_{\Delta E}^{\omega_{\mathrm{max}} }\frac{\omega d\omega}{q^4}\int_0^{2\pi} B^c_{\mu\nu}T_{\mu\nu}d\phi_k\,,
\end{equation}
where the continuum tensor, $B^c_{\mu\nu}$, is the same as in the elastic case except that $F(Q^2)$ and $G(Q^2)$ are replaced by $F(Q^2,M_f^2)$ and $G(Q^2,M_f^2)$. Virtual photon corrections are accounted for with the same fbar as in the elastic tail. The integral over the final states, $M_f^2$ ($W^2$), is replaced with an integral over radiated photon energy using
\begin{align}
F(Q^2) \rightarrow &\int_{u_c^2}^{u^2} F(Q^2,M_f^2) dM_f^2 =\int_0^{\omega_{\mathrm{max}} } F(Q^2,M_f^2) 2(u_0 - | u | \mathrm{cos}\theta_k)d\omega\,, \\
\label{omegaF}
\omega &= \frac{1}{2} (u^2 - M_f^2)/(u_o - |u|\mathrm{cos}\theta_k\,,\\
\omega_{\mathrm{max}} & = \frac{1}{2} (u^2 - u_c^2)/(u_o - |u|\mathrm{cos}\theta_k)\,,\
\end{align}
where the start of the inelastic spectrum is denoted by $u_c^2$, and is the pion threshold ($u_c^2 = M_p + m_{\pi}$) for a proton and the two body breakup threshold for nuclear targets.

Equation~\eqref{InelasInt} approaches infinity as the radiated photon energy goes to zero. This infrared divergence is treated by replacing the lower-bound of the the integral with an IR cut-off parameter, $\Delta E$. The full expression for the internal inelastic radiative tail is~\cite{Badalek}
\begin{align}
\label{FullInt}
\frac{d\sigma_r}{d\Omega dE'} (E_s,E_p)&= \frac{d\sigma}{d\Omega dE'} (E_s,E_p) e^{-\delta_r({\Delta E})}\tilde{F}(Q^2) + \frac{d\sigma_r}{d\Omega dE'} (\omega > \Delta E)\,, \\
\delta_r (\Delta E) &= \frac{\alpha}{\pi}{\bigg(} \mathrm{ln}\frac{E_s}{\Delta E} + \mathrm{ln}\frac{E_p}{\Delta E} {\bigg)}{\bigg (} \mathrm{ln}\frac{Q^2}{m_e^2} -1 {\bigg )}\,,
\end{align}
which is slightly different from equation B.6 of Ref~\cite{MT}, but is consistent with the previous choice of fbar.  For the purposes of radiatively correcting data, equation~\eqref{FullInt} is not of much use because it requires knowledge of the inelastic structure functions over a wide kinematic range. 

In practice, the internal bremsstrahlung spectrum is approximated using the angle-peaking approximation and method of equivalent radiators. The full inelastically radiated spectrum is given by
\begin{equation}
\begin{split}
\sigma_{\mathrm{In. Rad}} = \int F_{\mathrm{soft}} &\cdot {\Bigg (} \frac{M_T + 2(E_s - \omega_s)\mathrm{sin}^2\frac{\theta}{2}}{M_T - 2E_p\mathrm{sin}^2\frac{\theta}{2}} \left\{ \tilde{\sigma} (E_s - \omega_s) \left[\frac{b(t_b +t _r)}{\omega_s} \phi(v_s) + \frac{\xi}{2\omega_s^2}\right] \right\} \\
& +  \tilde{\sigma} (E_s ) \left[\frac{b(t_a +t_r)}{\omega_p}\phi(v_p) + \frac{\xi}{2\omega_p^2}\right] {\Bigg)}dM_F^2\,,
\end{split}
\end{equation}

\noindent where the terms in the set of big parentheses  are equation~\eqref{External} with the equivalent radiator thickness, $t_r$, added to the radiation thickness before and after scattering.

Following Appendix C of Ref~\cite{TSAI}, the integral over the final hadron states, $M_F^2$, is expressed as a two-dimensional integral over $E_s'$ and $E_p'$ (see Figure~\ref{Triangle}). The area integral is reduced to (using the energy peaking approximation) the sum of three separate integrals: 
 a line integral over $E_p'$ for fixed $E_s$ ($\sigma_{\mathrm{after}}$), a line integral over $E_s'$ for fixed $E_p$ ($\sigma_{\mathrm{before}}$) and a small area integral near $E_s$ and $E_p$ defined as $\Delta E$ by $R\Delta E$ ($\sigma_{\mathrm{low}}$), where $R$ is
\begin{equation}
R = \frac{M_T + E_s (1-cos\theta)}{M_T - E_p (1-cos\theta)}\,.
\end{equation}
The small area integration corresponds to the emission of very low energy (soft) photons  and is calculated under the assumption that the cross section is constant in the region. Ref~\cite{TSAI} states that the bin size, $\Delta E$, should obey $\Delta E/\xi \geq 10$ but still be small enough that cross section remains approximately constant. The result should be independent of the choice of  $\Delta E$. For this analysis, $\Delta E$ is 10 MeV.

The low energy photon term is given by:
\begin{equation}
\label{low}
\sigma_{\mathrm{low}} = {\Bigg (} \frac{R\Delta E}{E_s}{\Bigg )}^{b(t_b + t_r)} {\Bigg (} \frac{\Delta E}{E_p}{\Bigg )}^{b(t_a+ t_r)} {\Bigg [}1 - \frac{\xi/\Delta E}{1-b(t_a + t_b + 2t_r)}{\Bigg]}{\tilde \sigma}(E_s,E_p)\,.
 \end{equation}
 The integral due to energy loss before scattering is given by
 \begin{equation}
 \label{before}
 \begin{split}
\sigma_{\mathrm{before}} = \int_{E_{s\, \mathrm{min}}}^{E_s - R\Delta E} &{\tilde \sigma}(E_s',E_p){\bigg \{}{\Bigg (}\frac{E_s - E_s'}{E_pR}{\Bigg )}^{b(t_a + t_r)}{\Bigg (}\frac{E_s - E_s'}{E_s}{\Bigg )}^{b(t_b + t_r)} {\bigg \}}\\
& \times{\Bigg [} \frac{b(t_b + t_r)}{E_s - E_s'} \phi {\Bigg (}\frac{E_s - E_s'}{E_s}{\Bigg )} + \frac{\xi}{2(E_s-E_s')^2}{\Bigg ]} dE'_s \,,
\end{split}
 \end{equation}
and the integral due to energy loss after scattering is given by
 \begin{equation}
 \label{after}
 \begin{split}
\sigma_{\mathrm{after}} = \int_{E_{p+\Delta E}}^{E_{p\, \mathrm{max}}} &{\tilde \sigma}(E_s,E_p'){\bigg \{}{\Bigg (}\frac{E_p'- E_p}{E_p'}{\Bigg )}^{b(t_a + t_r)}{\Bigg (}\frac{(E_p' - E_p)R}{E_s}{\Bigg )}^{b(t_b + t_r)}{\bigg \}} \\
& \times{\Bigg [} \frac{b(t_a + t_r)}{E_p' - E_p} \phi {\Bigg (}\frac{E_p' - E_p}{E_p'}{\Bigg )} + \frac{\xi}{2(E_p'-E_p)^2}{\Bigg ]} dE'_p \,.
\end{split}
\end{equation}
The soft photon correction is located within the curly brackets in the above equations. The low energy cross section does not have a specific soft photon correction because the entire term itself represents soft photon emission. The integration bounds are the same as in equations~\eqref{IntBounds1} and~\eqref{IntBounds2}. 

\subsection{Internal Bremsstrahlung and the Peaking Approximation}
Replacing the full internal bremsstrahlung integral with an equivalent radiator introduces a potential source of systematic error. The relative size of this error is determined by comparing an internally radiated cross section using the full integral of equation~\eqref{FullInt} and the angle peaking approximation given by equation~\eqref{low}, equation~\eqref{before} and equation~\eqref{after}. Setting $t_a = t_b = \xi =  0 $ isolates the internal contribution. The soft-photon terms are also neglected in the $E_s$ and $E_p$ integrals.

To facilitate the comparison a model for $F(Q^2,M_f^2)$ and $G(Q^2,M_f^2)$ is used. The model is an empirical fit to electron-nucleus form factors, $F_1(Q^2,W^2)$ and $ F_2(Q^2,W^2)$ from Bosted-Mamyan-Christy~\cite{Bosted3,Bosted1,Bosted2}.  In the model, the inelastic cross section is defined as
\begin{align}
\frac{d\sigma}{d\Omega dE'} &= \frac{\alpha\mathrm{cos}\frac{\theta}{2}}{4 E^2 \mathrm{sin}^4\frac{\theta}{2}} {\bigg (} \frac{2}{\nu}F_2 (Q^2, W^2) + \frac{2}{M_T}\mathrm{tan}^2\frac{\theta}{2}F_1(Q^2,W^2){\bigg )}\,,  \\
Q^2 &= 4E_sE_p \mathrm{sin}^2\frac{\theta}{2}\,, \\
W^2 &= M_p^2 + 2M_p\nu - Q^2\,, \\
\nu  &= E_s - E_p \,.
\end{align}
The internal bremsstrahlung formalism defines all the kinematic variables with respect to the mass of the target nucleus and at the most probable momentum transfer, $i.e$ $E_s \rightarrow E_s - \omega$. This means that for the invariant mass input of the model, $W^2 \equiv M_f^2$ for the proton only, with the $M_f^2$ defined in equation~\eqref{omegaF}. For $Z >1$, the effective invariant mass for the model  is given by
\begin{equation}
W^2_{\mathrm{eff}} = M_p^2 + 2M_p(E_s - \omega - E_p) - Q^2\,.
\end{equation}
The relationship between the Mo and Tsai and traditional structure functions is
\begin{align}
F(Q^2,M_f^2) &= \frac{2}{M_T \nu_{\mathrm{eff}}}F_2(Q^2,M_f^2)\,,\\
G(Q^2,M_f^2) &= 2F_1(Q^2,M_f^2)\,, 
\end{align}
where $\nu_{\mathrm{eff}} = E_s - \omega - E_p $ corresponds to the most probable energy loss. 

\begin{figure}
\centering     
\subfigure[Nitrogen ]{\label{fig:softN2}\includegraphics[width=.85\textwidth]{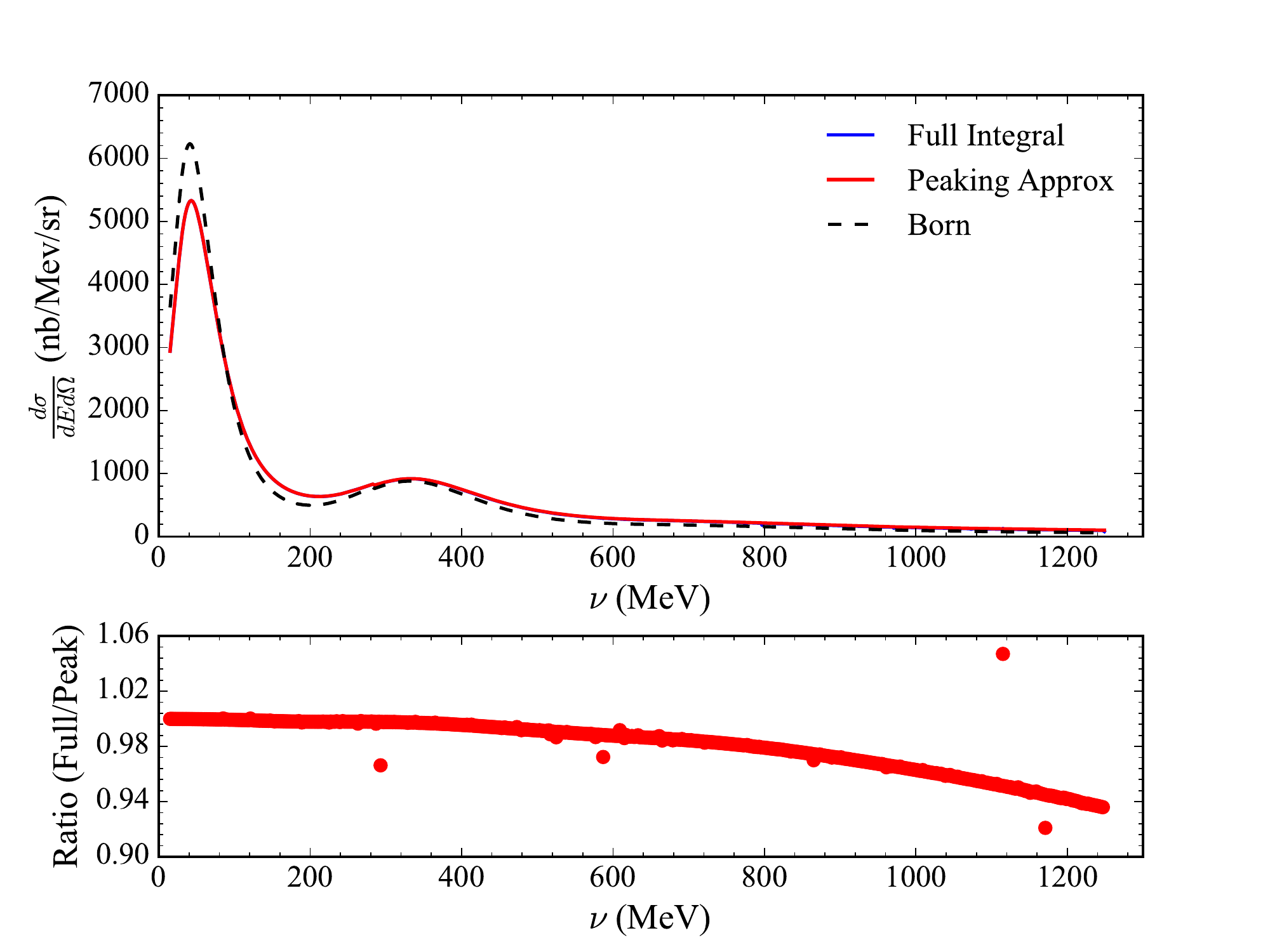}}
\qquad
\subfigure[Proton]{\label{fig:softP}\includegraphics[width=.85\textwidth]{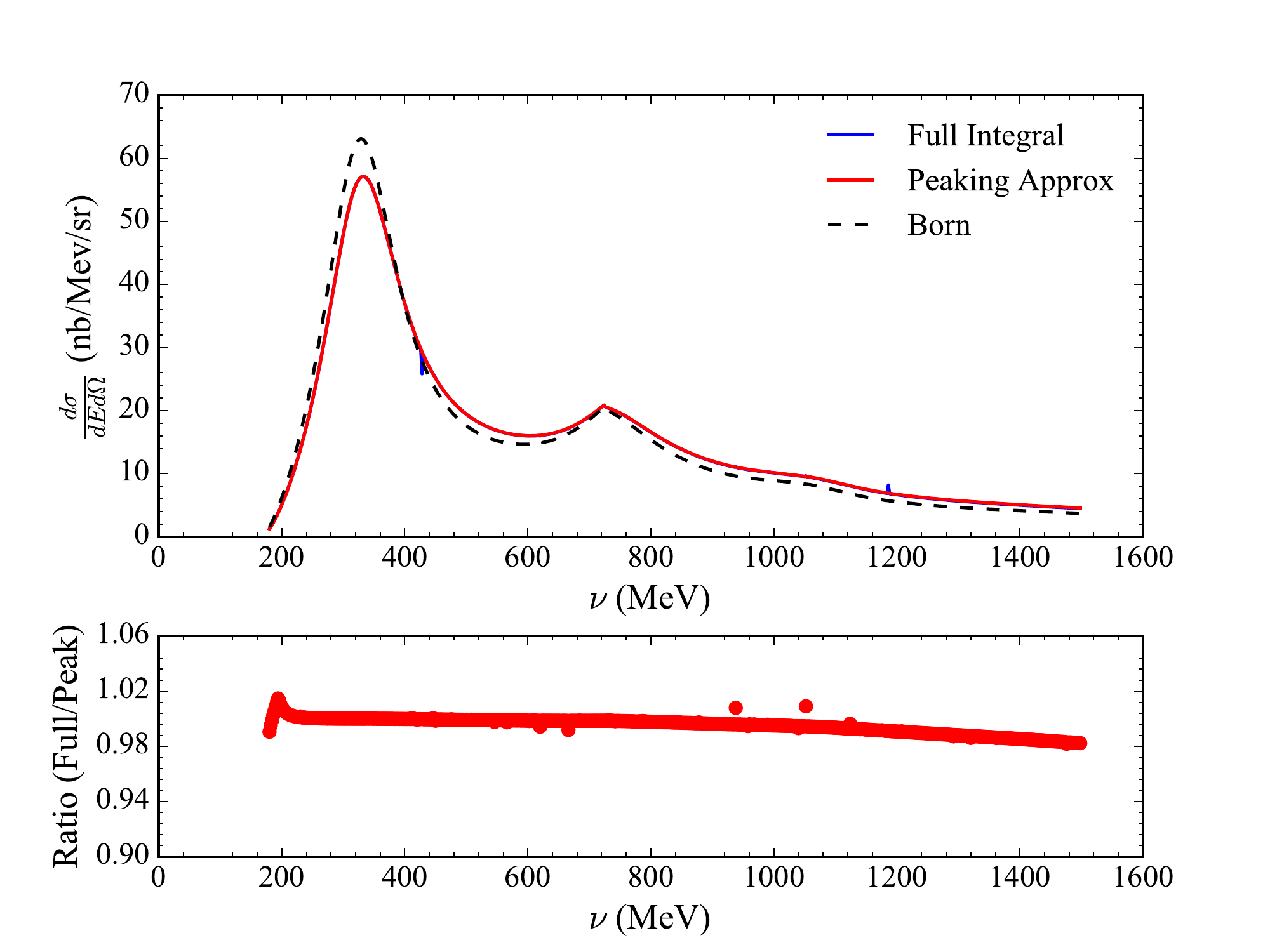}}
\caption{Comparison of the angle peaking approximation and full integral for the inelastic internal unpolarized bremsstrahlung.}
\label{IntBComp}
\end{figure}

The results of the comparison for the proton and nitrogen nucleus are shown in Figure~\ref{IntBComp} for $E_0$ = 2.2 GeV and $\theta_{\mathrm{sc}}$ = 6.0$^{\circ}$. In general, the approximation is very good. The difference between the proton is less than 2\% maximum at the kinematics relevant to this analysis. The case is similar for nitrogen with the total contribution to the fully radiated cross section (internal plus external plus low energy) at the 3\% level maximum. A few points in Figure~\ref{IntBComp} stray from the general trend. The cause of this is from two potential sources and is not a problem with the formalism. The first is a discontinuity in the model at the stitching together of the different kinematic channels, which then propagates through the integral. The second is a potential lack of accuracy in the integration method itself at certain kinematic points.

The angle-peaking approximation for the inelastic bremsstrahlung is used for the rest of this analysis. The level of error it introduces is tolerable and it is also significantly faster. For example, the full nitrogen integral took approximately one week to complete, while the angle-peaking calculation finished in a few minutes. The time difference was smaller for the proton but still existed. 

\subsection{Inelastic Polarized Internal Bremsstrahlung}
As with the unpolarized internal bremsstrahlung, the polarized inelastic internal cross section is given as an integral over the final states such that
\begin{align}
\label{polin}
\sigma^{\mathrm{in}} = \frac{d^2\sigma}{dxdy} = -\alpha^3y\int_{\tau_{\mathrm{min}}}^{\tau_{\mathrm{max}}}d\tau \sum_{i=1}^{8} {\bigg \{} &\theta_{i1}(\tau)\int_0^{R{\mathrm{max}}}\frac{dR}{R}{\bigg[}\frac{\mathcal{F}_i(R,\tau)}{(Q^2 +R\tau)^2} - \frac{\mathcal{F}_i(0,0)}{Q^4}{\bigg ]} \nonumber \\
&+ \sum_{j=2}^{k_i} \theta_{i,j}(\tau) \int_0^{R_\mathrm{max}} dR \frac{R^{j-2}}{(Q^2+R\tau)^2}\mathcal{F}_i(R,\tau){\bigg \}}\,,
\end{align}
where the $\mathcal{F}_i(R,\tau)$ are the inelastic structure functions, and $R_{\mathrm{max}} = \omega_{\mathrm{max}}$~\cite{Polrad}. The polarized formalism does not make use of an angle peaking approximation and evaluates the bremsstrahlung processes in the full kinematic triangle. The infrared divergence as $R\rightarrow 0$ is dealt with according to the prescription in Ref~\cite{Bardin}. 


\subsection{Inelastic Soft Photon Correction}
\label{InelasticSoft}
The uncertainty in the inelastic soft photon correction is checked by making a comparison between the Mo and Tsai factor given in the equation~\eqref{before} and equation~\eqref{after} and an equivalent term by G. Miller~\cite{Miller}. The Miller inelastic soft photon term is similar to the term for the elastic tail and is
\begin{equation}
F^M_{\mathrm{soft}} = {\bigg (} \frac{k_1}{E_s}{\bigg)}^{(bt_r + bt_b)}{\bigg (} \frac{k_1'}{E_p}{\bigg)}^{(bt_r + bt_a)}\,,
\end{equation}
where $k_1, k_1'$ are the soft-photon limiting energies. In general, they  should be a small fraction of $E_s$, $E_p$ so the photons are actually soft,
\begin{align}
k_1 &= \mathrm{min}[(1/3)E,\omega_1]\,, \\ 
k_1' &= \mathrm{min}[(1/3)E,\omega_1']\,. 
\end{align}
 The definitions of $\omega_1$ and $\omega_1'$ are different for the $E_s$ and $E_p$ integrals:
\begin{equation}
\begin{aligned} [c]
&\mathrm{d}E_s\,\mathrm{integral}: \nonumber \\
&\omega_1 = E_s - E_s' \\
&\omega_1' = \omega_p /R 
\end{aligned}
\qquad \qquad
\begin{aligned}[c]
&\mathrm{d}E_p\,\mathrm{integral}: \nonumber \\
&\omega_1 = R\omega_p \\
&\omega_1' = E_p' - E_p
\end{aligned}
\end{equation}
The choice of $\frac{1}{3}$ is arbitrary and uncertain within the range 0.2 to 0.8~\cite{Miller}, so any variation in the radiated cross section based upon a different fraction is a measure of the systematic uncertainty.  A comparison between the Mo and Tsai factor and the Miller factor is an additional check on the uncertainty. On average, the systematic difference between the two procedures is at the 1-2\% level.


\subsection{Correcting the Inelastic Data}
The cross sections for $E_s' < E_{s\,\mathrm{min}},E_p$ and $E_s,E_p' > E_{p\,\mathrm{max}}$ are zero after the elastic tail is subtracted. This leaves only inelastically radiated data in the experimental cross section. A direct calculation of equation~\eqref{low}, equation~\eqref{before}, equation~\eqref{after} and equation~\eqref{polin} to remove the inelastic radiated events requires prior knowledge of the non-radiative inelastic cross sections (or structure functions for the full internal bremsstrahlung calculation) at the desired kinematics. There are two possible methods for getting around this problem. The first involves using a model to reconstruct a radiated cross section that matches the data.   A comparison between the radiated and unradiated model corrects the data. The second method is an iterative/`unfolding' procedure that extracts the non-radiative inelastic cross section from the radiated experimental cross section. The unfolding method is preferred because it is naturally less model dependent. Although, the unfolding procedure requires a spectrum that starts at the threshold of the continuum and is relatively gap free, so in some cases it is still necessary to revert to the first method.

In the unfolding process, the first iteration uses the measured cross section as input to create a model of the unfolded (non-radiative) cross section. The model is radiated according to equation~\eqref{low}, equation~\eqref{before}, equation~\eqref{after} and equation~\eqref{polin}.  The radiated model is compared to the actual measured input cross section. A new model is created by subtracting the differences between the old model and the measured cross section. The process is repeated on the new model and the results converge after a few iterations. The necessary cross section values needed for the integration are evaluated using extrapolation and interpolation on the measured data.

\begin{figure}[htp]
\begin{center}
\includegraphics[width=0.80\textwidth]{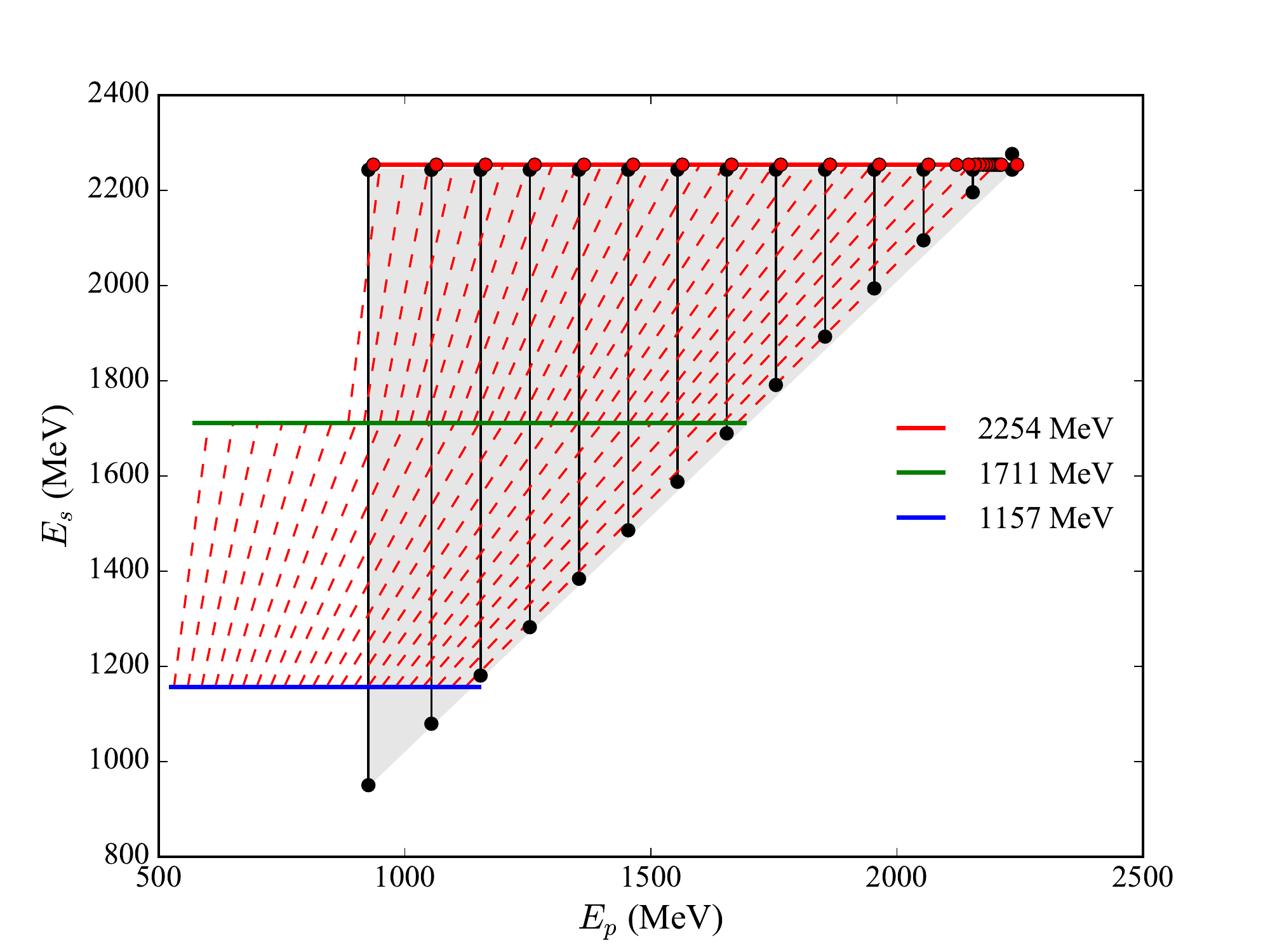}
\caption{\label{ExpTri}Experimental radiative corrections triangle.}
\end{center}
\end{figure}

The experimental cross sections are measured at many different values of the outgoing electron energy, $E_p$ but only a few values incident electron energy $E_s$. This means that the $dE'_p$ integration is calculated using the already unfolded spectrum via an interpolation at constant $E_s$, assuming that the measured spectrum starts below the continuum and is continuous across energy loss/invariant mass. Extrapolation to $E_s' < E_s$ is needed in order to perform the $dE_s'$  integral~\cite{TSAI}. The $dE_s'$  extrapolation is aided by providing the unfolding procedure a range of spectra at different incident energies. For spectra at the lowest beam energy, model input provides the necessary cross sections for extrapolation.

The extrapolation and interpolation procedure is sketched out in Figure~\ref{ExpTri} for radiatively correctly data at $E_0$ = 2.2 GeV. The gray shaded region represents the full kinematic phase space. Black (red) points are the minimum and maximum values for the $E_s$ ($E_p$) integrals at a single $\nu$ point, while the black (red) lines represent the range of necessary values for the integration. The red and blue lines represent data taken at lower values of $E_0$ and aid the extrapolation to different values of $E_s$ at the red dotted lines of constant $W$. The small uncovered region as low $E_p$ and $E_s$ is determined from a Mott cross section scaling of the lowest available data.

\subsection{Fortran Analysis Code}
The unfolding process is done in a set of Fortran analysis codes known as RADCOR~\cite{RADCOR} for the unpolarized corrections and POLRAD~\cite{POLCODE} for the polarized processes. For polarized data sets, typically the unpolarized external corrections are calculated with RADCOR first and then input into POLRAD for the final internal polarized contributions. The results are insensitive to the order of operations.

For RADCOR, an input file, $input\_radcor.txt$, supplies the incident energy spectra and other necessary experimental constants to the main program radcor.f. The input spectra  already have their elastic tail subtracted and are `smoothed' so they contain no gaps or discontinuities. 

There are three nested control loops in radcor.f: the iterations loop, the beam-energy spectra loop and the $\nu$-bin loop. Working outwards, the $\nu$-bin loop loops over the energy spectrum for a single beam energy and radiates the model cross section. Equation~\eqref{low} is defined as SIGLOW. Two functions, FES and FEP, prepare the integrands for equation~\eqref{before} and equation~\eqref{after} respectively and the function TERP does any necessary interpolations. The end of the loop updates the model and records any differences between the current and previous cross section models. The beam-energy spectra loop does this process for each beam energy, while the iterations loop defines the number of total iterations for each beam energy.

The output file contains the initial (radiated) and final (unfolded) smoothed energy spectra (cross sections are in units of nb $\cdot$ sr$^{-1}$ $\cdot$ MeV$^{-1}$) and the difference between the last two iterated spectra versus $\nu$ (MeV). The correction to the unsmoothed experimental data is applied, on a bin-by-bin basis as either the difference or ratio between the initial and final smoothed energy spectra in a separate procedure.

The input file in POLRAD is $input\_polrad.txt$ and supplies the incident energy spectra and the lower energy spectra for extrapolation for the main program jintpolrad.f and polsig.f. These inputs have their elastic tail subtracted and are `smoothed'. POLRAD computes the full internal bremsstrahlung integral and thus uses the spin structure functions (as opposed to measured cross sections in RADCOR) in its calculations. This requires the the input file include both perpendicular and parallel polarized cross section differences. The general structure and output of the POLRAD code is the same as that of RADCOR: it iterates over the data to produce a born cross section that when radiated matches the data input. The output file contains the initial and unfolded smoothed cross sections.

\chapter{\sc Results}
\label{ch:Results}
The main analysis focus of this thesis is the study of the 5 T and high $Q^2$ settings at $E_0$ = 2254 MeV (longitudinal and transverse) and $E_0$ = 3350 MeV (see Figure~\ref{kin}). A preliminary 2.5 T analysis is in Appendix~\ref{app:Appendix-I}. This includes analysis  of the proton asymmetries, polarized cross section differences, and the radiative corrections. After obtaining the polarized cross section differences, it is possible to evaluate the spin structure functions and their moments. This work includes the full statistics of E08-027 and combines LHRS and RHRS data.   

\section{Asymmetry}
\label{Sec:Asymm}
The measured raw asymmetries are formed on a run-by-run and kinematic bin-by-bin basis as described in equation~\eqref{EAsymm}. A statistically weighted combination of similar kinematic bins from different runs gives the full analyzing power of the measured result. The statistical error bars are given by Poisson statistics as $\delta \approx N^{-1/2}$, for $N \approx 2N^+ \approx 2N^-$. The use of a prescale factor in the DAQ changes the event rate and count rate behavior from pure Poisson fluctuations. This is accounted for by introducing the correction factor~\cite{Qiang}

\begin{equation}
S = \sqrt{1-LT\cdot f_A{\bigg(}1-\frac{1}{ps}{\bigg)}}\,,
\end{equation}
where $LT$ is the livetime, $ps$ is the prescale factor and $f_A$ is the ratio of accepted events to the total number of events: $f_A$ = $N_{\mathrm{accepted}}/N_{\mathrm{total}}$. The total statistical uncertainty on a given bin in a given run is
\begin{equation}
\delta_A \simeq \frac{1}{2}\sqrt{\frac{S^2_+}{N_+}+\frac{S^2_-}{N_-}}\,.
\end{equation}
The final measured raw asymmetry is the statistically weighted average such that
\begin{align}
A &= \frac{ \sum_i A_i/\delta A_i^2}{\sum_i 1/\delta A_i^2}\,,\\
\delta A &= \sqrt{\frac{1}{\sum_i1/\delta A_i^2}}\,,
\end{align}
where $A_i$ is the asymmetry calculated for run $i$ and $\delta A_i$ is its corresponding statistical uncertainty.

The sign of the asymmetry is dependent on the position of the insertable half-wave plate (IHWP) and target spin direction. In the case of transverse target asymmetries, the sign also flips between LHRS and RHRS data. The following conventions are used in this analysis to fix the asymmetry sign:
\begin{itemize}
\item IHWP is ``IN" [``OUT"]: +1 [$-$1] (`OUT" corresponds to a MySQL value of 0)
\item Target polarization greater [less] than 0: +1 [$-$1]
\item LHRS (RHRS) transverse asymmetry: +1 [$-$1]
\end{itemize}
The final asymmetry sign is a product of the above values. For example, a RHRS run on a target transversely polarized to 80\% with the IHWP ``OUT" has the sign +1.

\subsection{Analysis Cuts}
\label{Sec:Cuts}
\begin{figure}[htp]
\centering     
\subfigure[LHRS: Transverse]{\label{fig:LHRSAcc}\includegraphics[width=.80\textwidth]{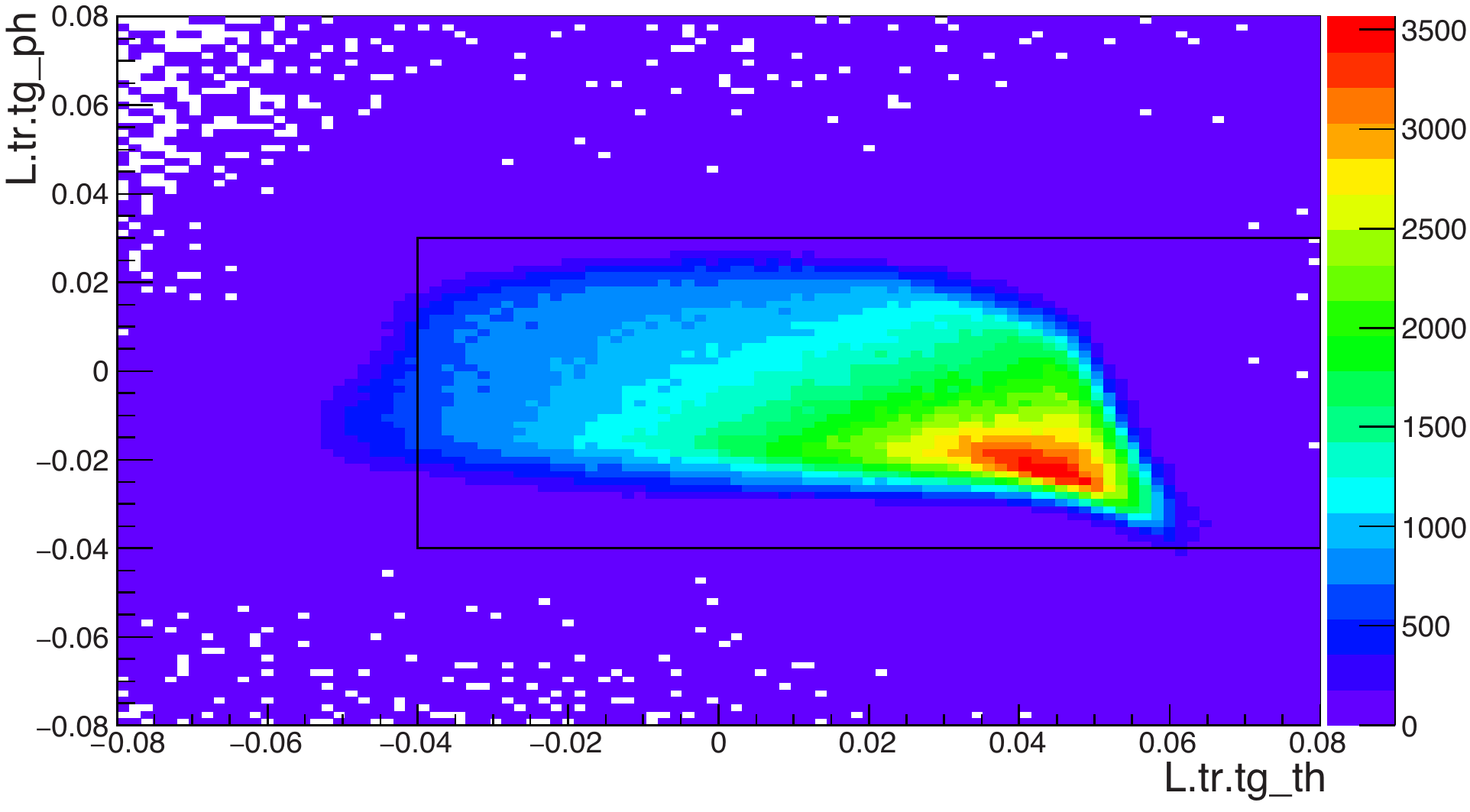}}
\qquad
\subfigure[RHRS: Transverse]{\label{fig:RHRSAcc}\includegraphics[width=.80\textwidth]{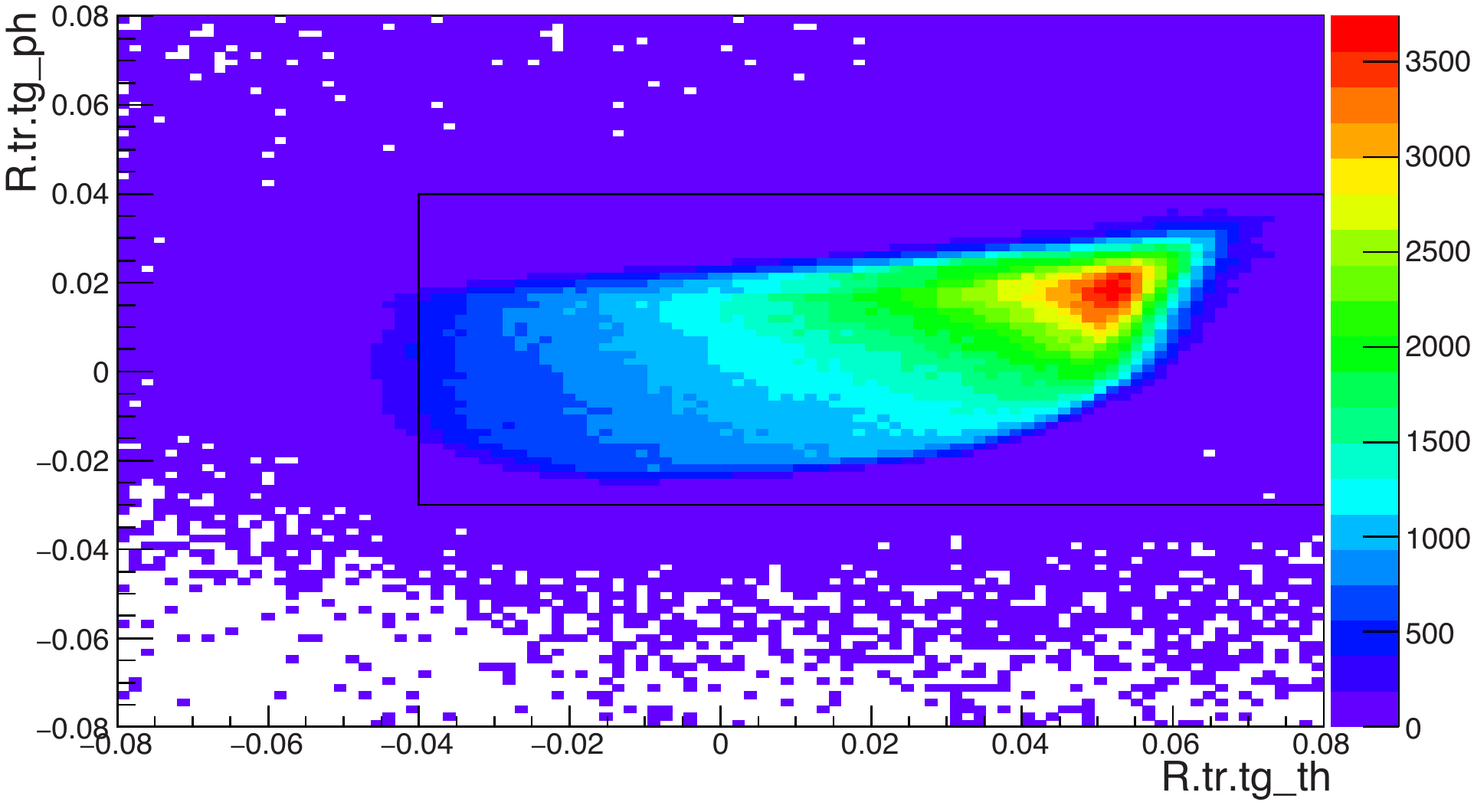}}
\qquad
\subfigure[LHRS: Longitudinal]{\label{fig:LHRSAccLong}\includegraphics[width=.80\textwidth]{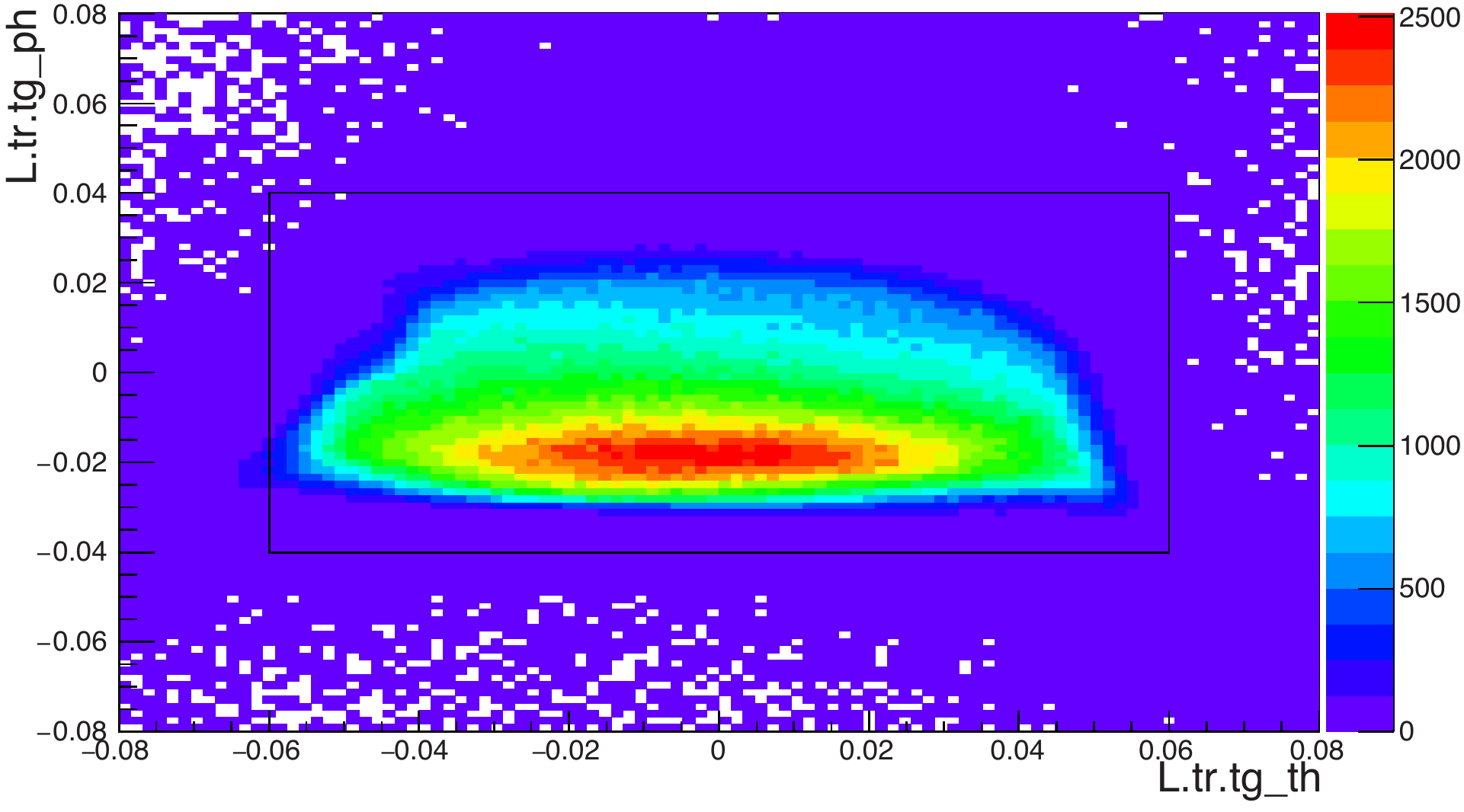}}
\caption{Acceptance cuts for $E_0$ = 2254 MeV 5T and $P_0$ = 1.8 GeV.}
\label{AccCuts}
\end{figure}

The asymmetry is a ratio of cross sections, so the spectrometer acceptance correction cancels to first order. This allows the use of a larger kinematic phase space in the asymmetry, which is beneficial because of the statistics required for an asymmetry measurement. For the asymmetry analysis, the following loose acceptance cuts are placed on the data\footnote{The RHRS cuts are formed by making the substitution L$\rightarrow$R.} :
\begin{itemize}
\item $|$L.rec.dp$|$ $<$ 4\%  
\item $|$L.tr.r\_y$|$ $<$ 80 mm 
\item $|$L.tr.r\_x$|$ $<$ 800 mm 
\item  $-$40 mrad $<$ L.tr.tg\_ph $<$ 40 mrad  
\item  $-$40 mrad $<$ L.tr.tg\_th $<$ 80 mrad : transverse
\item  $-$60 mrad $<$ L.tr.tg\_th $<$ 60 mrad : longitudinal
\end{itemize}
The physical target phi and target theta cuts are shown in Figure~\ref{AccCuts}; the 2-D histograms are resonstructed target phi and theta without target field corrections. The ``hot-spot" in the transverse acceptance is a direct result of the transverse target field and represents small scattering angle events. The correlation between the ``hot-spot" and scattering angle is the reason for its change in sign between the LHRS and RHRS. The longitudinal data is more symmetric for both the LHRS and RHRS, as expected. The ``hot-spot" is again small scattering angle events and switches signs between the two spectrometers. The ``dp" cut removes edge events from the electron's reconstructed momentum spectrum. The ``x" and ``y" cuts are very loose cuts on the electron track and act as low-level electron event filter.

\begin{figure}[htp]
\centering     
\subfigure[Good electron cuts]{\label{fig:LHRSSel}\includegraphics[width=.80\textwidth]{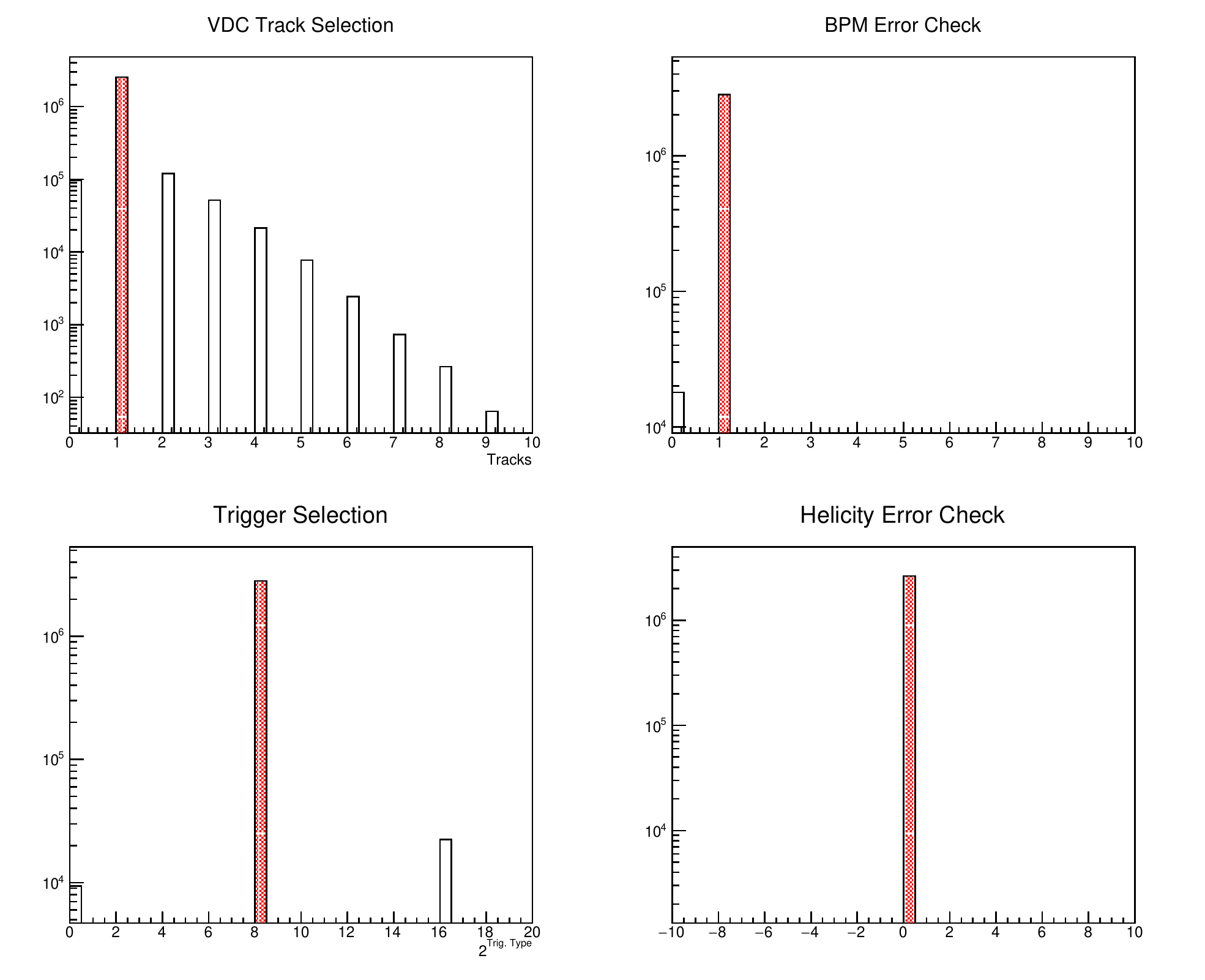}}
\qquad
\subfigure[Reconstructed momentum and PID cuts.]{\label{fig:LHRSSelc2}\includegraphics[width=.80\textwidth]{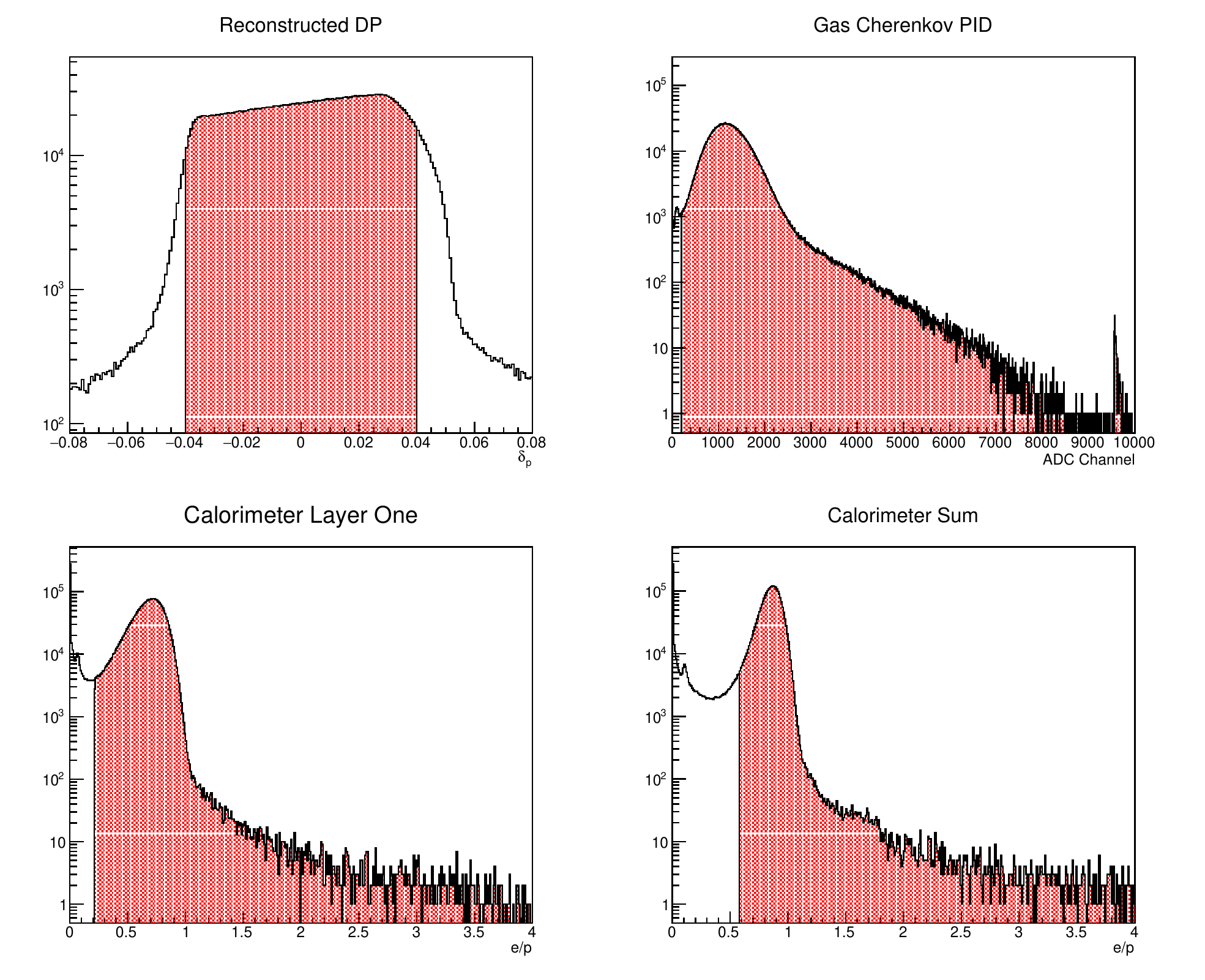}}
\caption{Good electron cuts for $E_0$ = 2254 MeV 5T and $P_0$ = 1.8 GeV. The trigger selection cut is applied as $2^i$ where $i=1,2,3,4$ is the trigger type defined in Chapter~\ref{DAQTRIGGER}.}
\label{OtherCuts}
\end{figure}

The remaining analysis cuts are\footnote{R.ps.e+R.sh.e/(p0\_MeV*(1.0+R.rec.dp)) and  DR.evtypebits\&(1$<<$1)$>$0 for the RHRS.}:

\begin{itemize}
\item hel.L.error\&0x2f0f = 0 
\item Lrb.bpmavail = 1
\item L.tr.n = 1
\item L.cer.asum\_c $>$ {\it \v{C}erenkov cut}
\item L.prl1.e/(p0\_MeV*(1.0+L.rec.dp)) $>$ {\it calorimeter layer one cut}
\item L.prl1.e+L.prl2.e/(p0\_MeV*(1.0+L.rec.dp)) $>$ {\it calorimeter sum cut}
\item DL.evtypebits\&(1$<<$3)$>$0
\end{itemize}
The helicity error and BPM cuts ensure that there is a valid reconstruction of that quantity for the event. The tracking cut of $n=1$ selects events with only one electron track in the VDC and the ``evtypebits" cut selects only events caused by a $T_3$ trigger. The MySQL database provides the PID cut values for the gas \v{C}erenkov and calorimeters. These additional cuts are shown in Figure~\ref{OtherCuts}.

\subsection{Scattering Angle Fit}
\label{sec:Angfit}
The acceptance cuts define a range of scattering angles admitted into the spectrometers, and at the transverse settings the central value of this range is a function of the spectrometers dipole momentum. Understanding the scattering angle dependence as a function of the electron's energy is critical to accurately reconstructing the kinematics of each event. The scattering angle is reconstructed on an event-by-event basis as described in Chapter~\ref{Sec:RecScatAngle}, weighted by $\sigma_{\mathrm{Mott}}^{-1}$ to remove physics dependence to first order, and fit with the following functional form
\begin{equation}
\label{FitEQ}
\theta_{\mathrm{rec}} = \mathrm{exp}(p_0 + p_1 E_p) + p_2 +p_3E_p\,, 
\end{equation} 
where $\theta_{\mathrm{rec}}$ is the scattering angle in degrees, $E_p$ is the electron momentum in GeV, and $p_0$-$p_3$ are the fit parameters~\cite{JixieAngle}.

\begin{figure}[htp]
\centering     
\subfigure[$E_0$ = 2254 MeV]{\label{fig:2254Ang}\includegraphics[width=.80\textwidth]{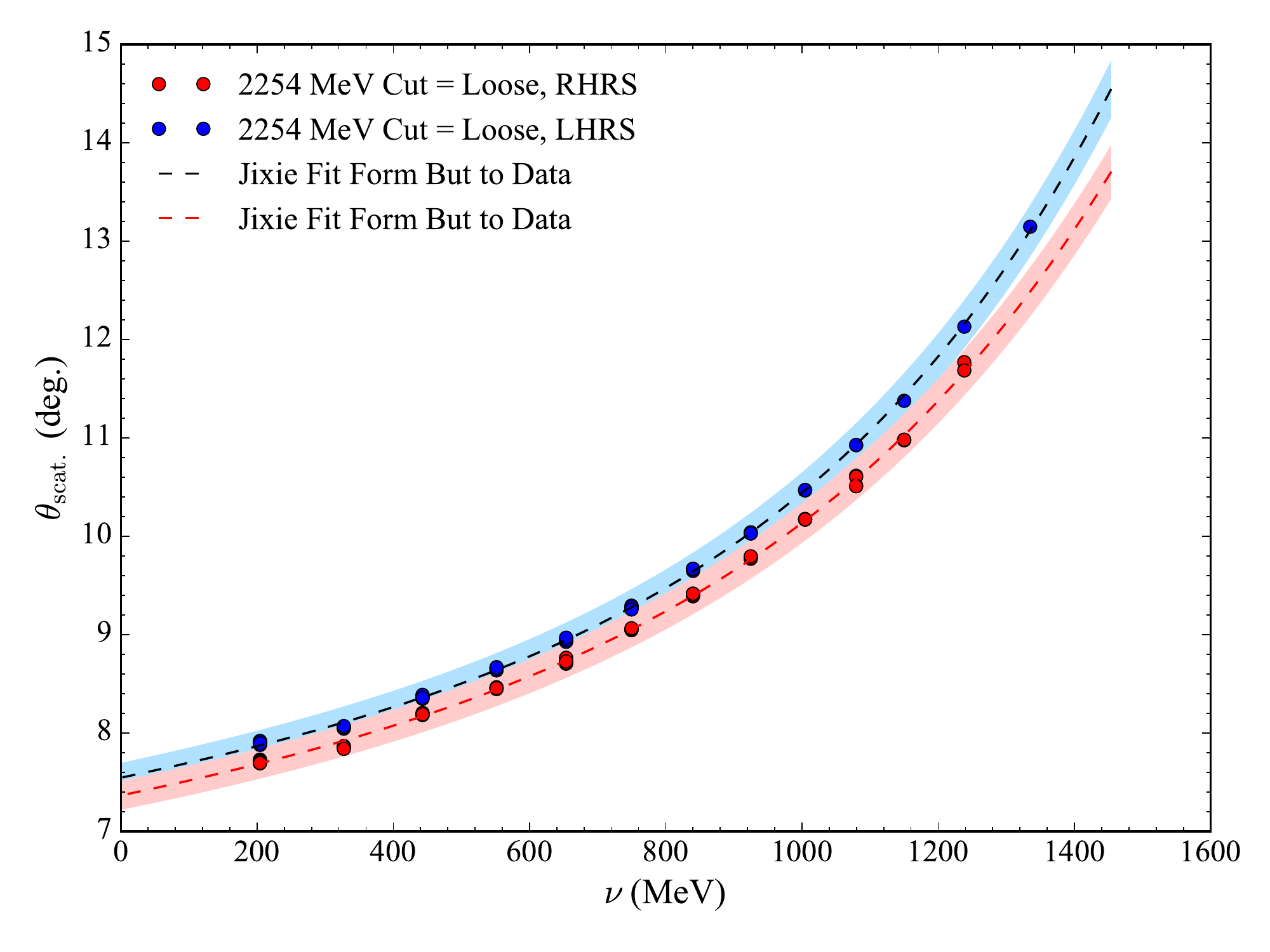}}
\qquad
\subfigure[$E_0$ = 3350 MeV]{\label{fig:3350Ang}\includegraphics[width=.80\textwidth]{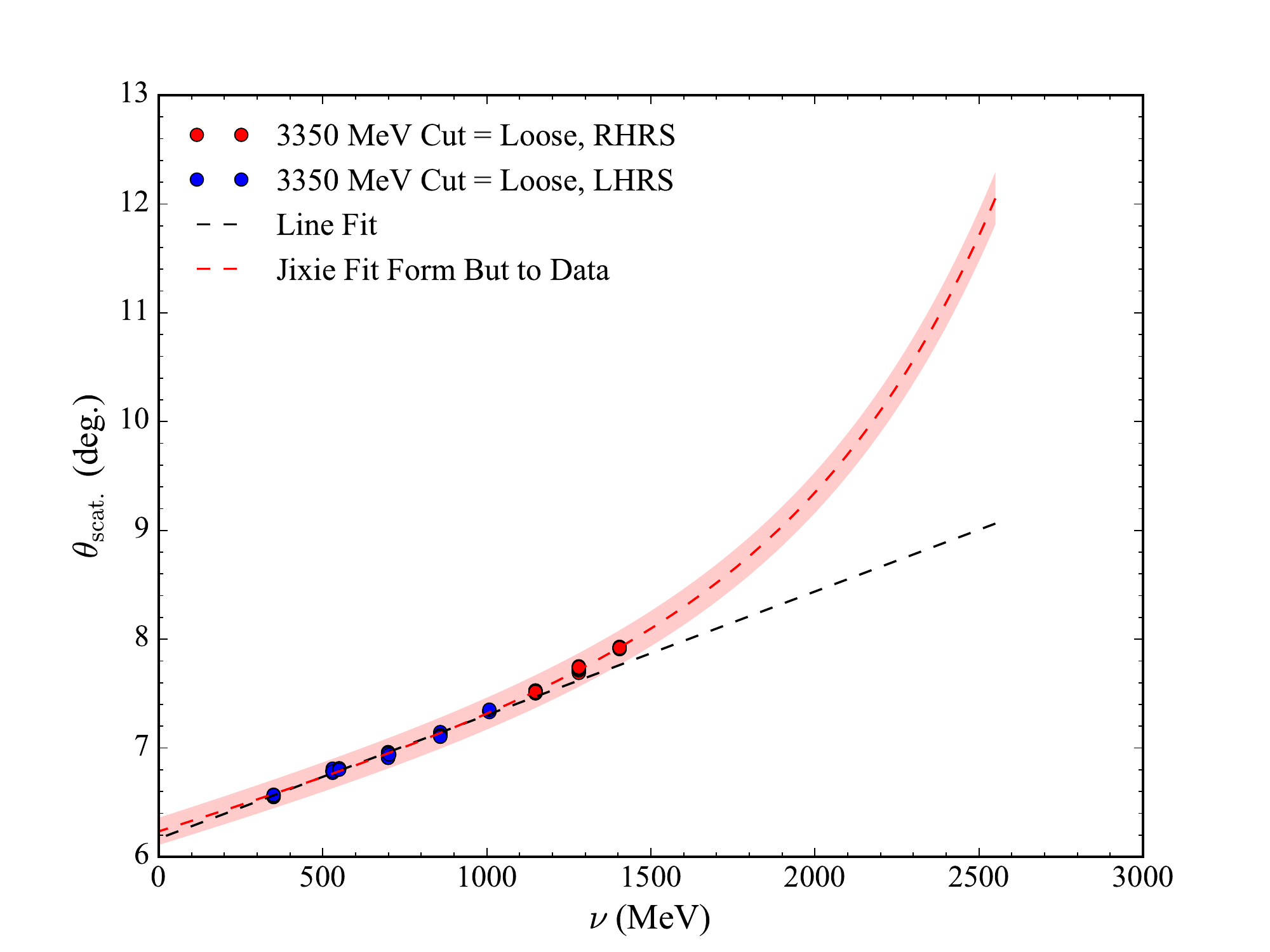}}
\caption{Reconstructed scattering angle for the 5 T transverse target field settings. The loose cut is defined in Chapter~\ref{Sec:Cuts} and the ``Jixie" fit form is from Ref~\cite{JixieAngle}.}
\label{ScatRecon}
\end{figure}

 The central scattering angle as a function of electron energy and with the acceptance cuts of Chapter~\ref{Sec:Cuts} are shown in Figure~\ref{ScatRecon} for the 5 T transverse settings. At the $E_0$ = 2254 MeV setting the difference between the LHRS and RHRS reconstruction is $\approx$ 0.3$^{\circ}$ which is within the systematic uncertainty of $\approx$ 2\% given by the colored bands around the dotted-line fit. The uncertainty is determined from a simple Monte-Carlo using Gaussian distributions centered around the central value of the input quantities of Chapter~\ref{Sec:RecScatAngle} and width equivalent to the quantity's uncertainty. The beam position uncertainties are $\pm$ 1 mrad and $\pm$ 2 mrad for the reconstructed target quantities.   The fit parameters are found in Table~\ref{table:fitparam}. For the $E_0$ = 3350 MeV setting, the fit is to both the LHRS and RHRS data, where the RHRS data is shifted down by 0.3$^{\circ}$. The $E_0$ = 2254 MeV fit parameters are for the LHRS data.  

\begin{table}[htp]
\begin{center}
\begin{tabular}{ l c c c c } \hline
 $E_0$ (MeV) & $p_0$ & $p_1$ & $p_2$ & $p_3$  \\[.2cm]  \hline 
 2254 & 3.485 & $-$2.113 & 9.231 & $-$0.871  \\[.2cm] 
 3350 & 2.667 & $-$1.723 & 9.124 & $-$0.876\\[.2cm]\hline
\end{tabular}
\caption{Fit parameters for scattering angle reconstruction at 5 T for equation~\eqref{FitEQ}.}
\label{table:fitparam}
\end{center}
\end{table}

\begin{figure}[htp]
\centering     
\includegraphics[width=.80\textwidth]{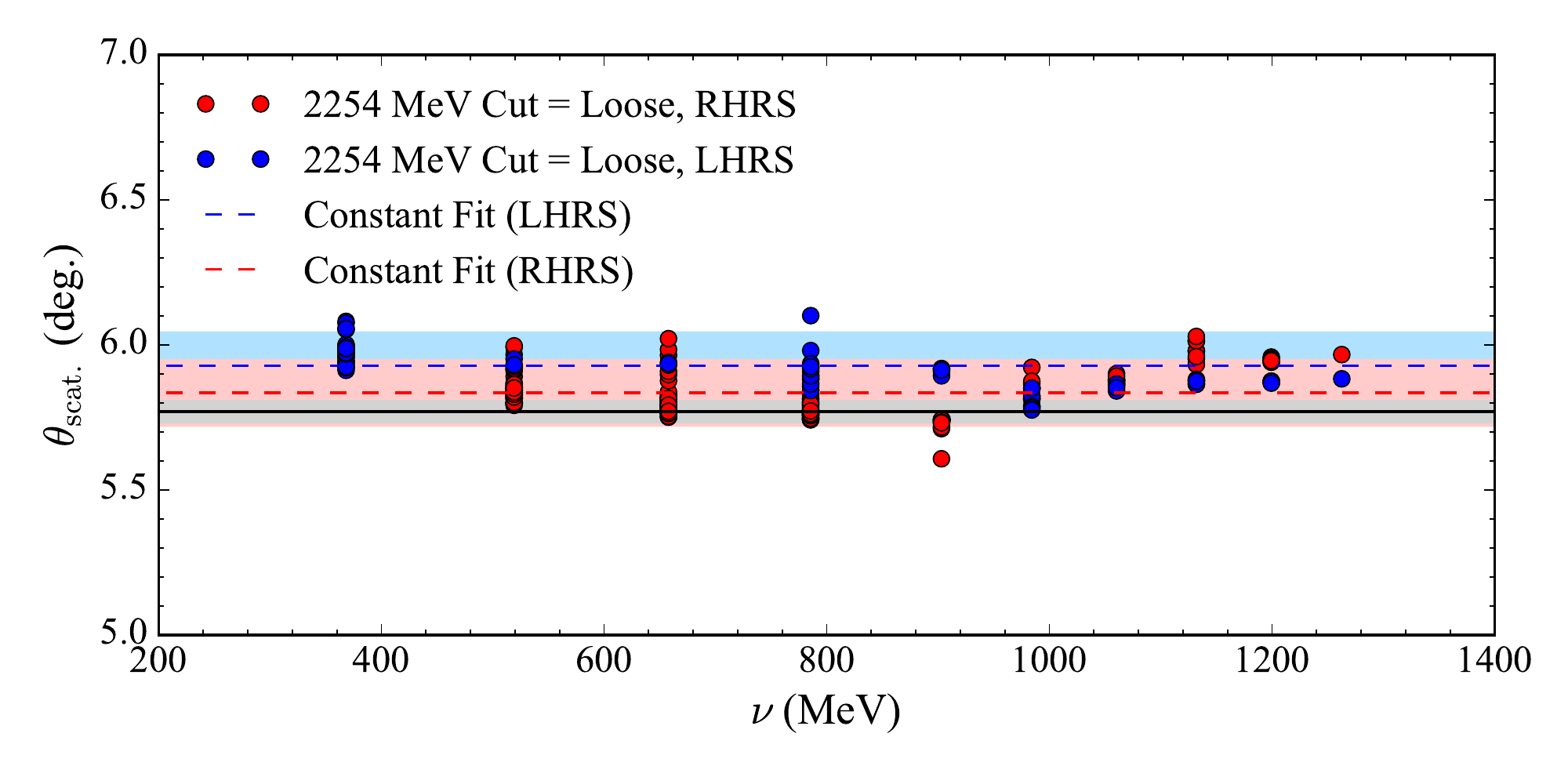}
\caption{Reconstructed scattering angle for the 5 T longitudinal target field setting. The loose cut is defined in Chapter~\ref{Sec:Cuts}. }
\label{ScatReconLong}
\end{figure}

The accuracy of the angle reconstruction and Mott cross section weighting is checked against the longitudinal data with a horizontal line fit to the reconstructed results. The target field has a minimal effect at this setting, so the reconstructed angle should be the spectrometer pointing angle (see Table~\ref{table:theta}), within uncertainty. The results of this study are shown in Figure~\ref{ScatReconLong}. The black solid line is the point resulting at 5.77$^{\circ}$ and with an uncertainty of $\pm$ 0.04$^{\circ}$ given by the gray band. The LHRS and RHRS one parameter fits agree with the pointing result within uncertainty, and the pointing result is used for the rest of the analysis. 

\subsection{Out of Plane Polarization Angle}
The transverse asymmetries have an extra dilution factor in the out-of-plane polarization angle. This angle, defined in equation~\eqref{DiffXSPerp}, is the azimuthal angle between the polarization plane and scattering plane~\cite{Leader}. In terms of the reconstructed quantities from Appendix~\ref{app:Appendix-E}, the relevant vectors are
\begin{align}
\vec{k} &= [ \mathrm{tan}(\phi_{\mathrm{BPM}}), \mathrm{tan}(\theta_{\mathrm{BPM}}),1]\,,\\
\vec{k'} &= [ \mathrm{sin}(\theta_{\mathrm{rec\_l}})\mathrm{cos}(\phi_{\mathrm{rec\_l}}),\mathrm{sin}(\theta_{\mathrm{rec\_l}})\mathrm{sin}(\phi_{\mathrm{rec\_l}}), \mathrm{cos}(\theta_{\mathrm{rec\_l}})]\,,\\
\vec{s} & = [\pm1,0,0]\,,
\end{align}
for the incoming electron, outgoing electron and proton spin, respectively. The sign of the proton spin vector flips with the sign of the polarization. The angle between two planes is defined as the angle between their normal vectors. The normal vectors for the scattering and polarization plane are defined as
\begin{align}
\vec{n_1} & = \vec{k} \times \vec{k'}\,,\\
\vec{n_2} & = \vec{k} \times \vec{s}\,,
\end{align} 
respectively. The angle between $\vec{n_1}$ and $\vec{n_2}$ is the out-of-plane polarization angle. The correction is applied on a bin-by-bin basis to the data as
\begin{equation}
\delta_{\mathrm{OPP}} = [\mathrm{cos}{\theta_{\mathrm{OPP}}}]^{-1}\,,
\end{equation} 
and the angle ranges from $\approx$ 40$^{\circ}$ to 65$^{\circ}$ at $E_0$ = 2254 MeV and from $\approx$ 25$^{\circ}$ to 35$^{\circ}$ at $E_0$ = 3350 MeV. The angle increases with decreasing electron momentum. The uncertainty is calculated in the same Monte-Carlo as the scattering angle and is $\pm$ 0.5$^{\circ}$ about the central value.
\subsection{Asymmetry Studies}
The asymmetries for the LHRS and RHRS are compared before they are combined. The comparison is a statistical statement that the spectrometers are measuring the same asymmetry and is done using the reduced $\chi^2$ metric. The ideal value is found by computing the reduced $\chi^2$ for a given HRS asymmetry compared to that asymmetry with each point varied randomly within its statistical error bar. This process is repeated for one million trials and the average reduced $\chi^2$ is the benchmark value for the LHRS and RHRS comparison. This value is approximately: $<$ red. $\chi^2$ $>$ $\approx$ 1.3.

\begin{figure}[htp]
\centering     
\subfigure[$E_0$ = 2254 MeV 5T Transverse]{\label{fig:TranComp}\includegraphics[width=.80\textwidth]{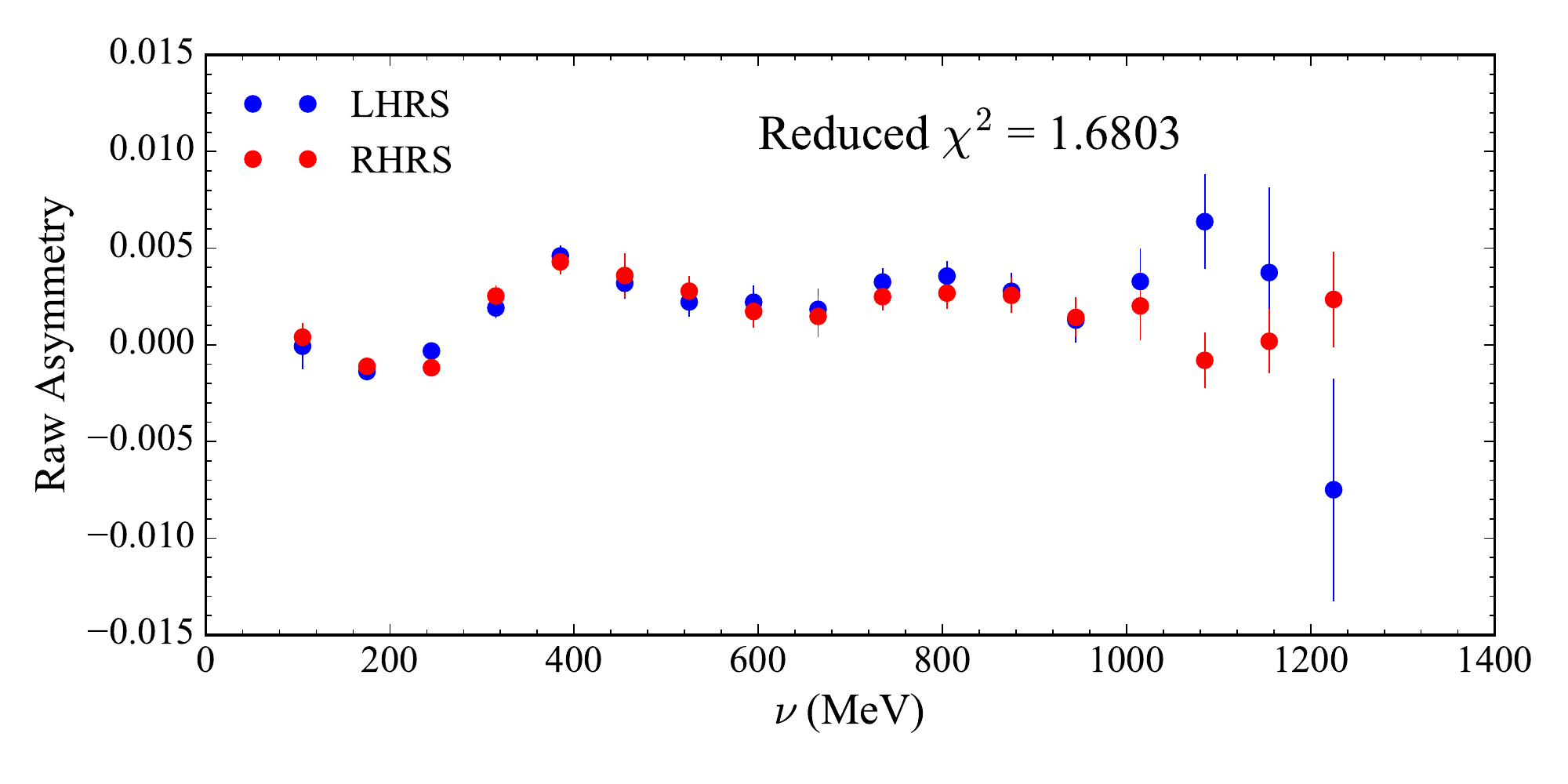}}
\qquad
\subfigure[$E_0$ = 3350 MeV 5T Transverse]{\label{fig:3350TranComp}\includegraphics[width=.80\textwidth]{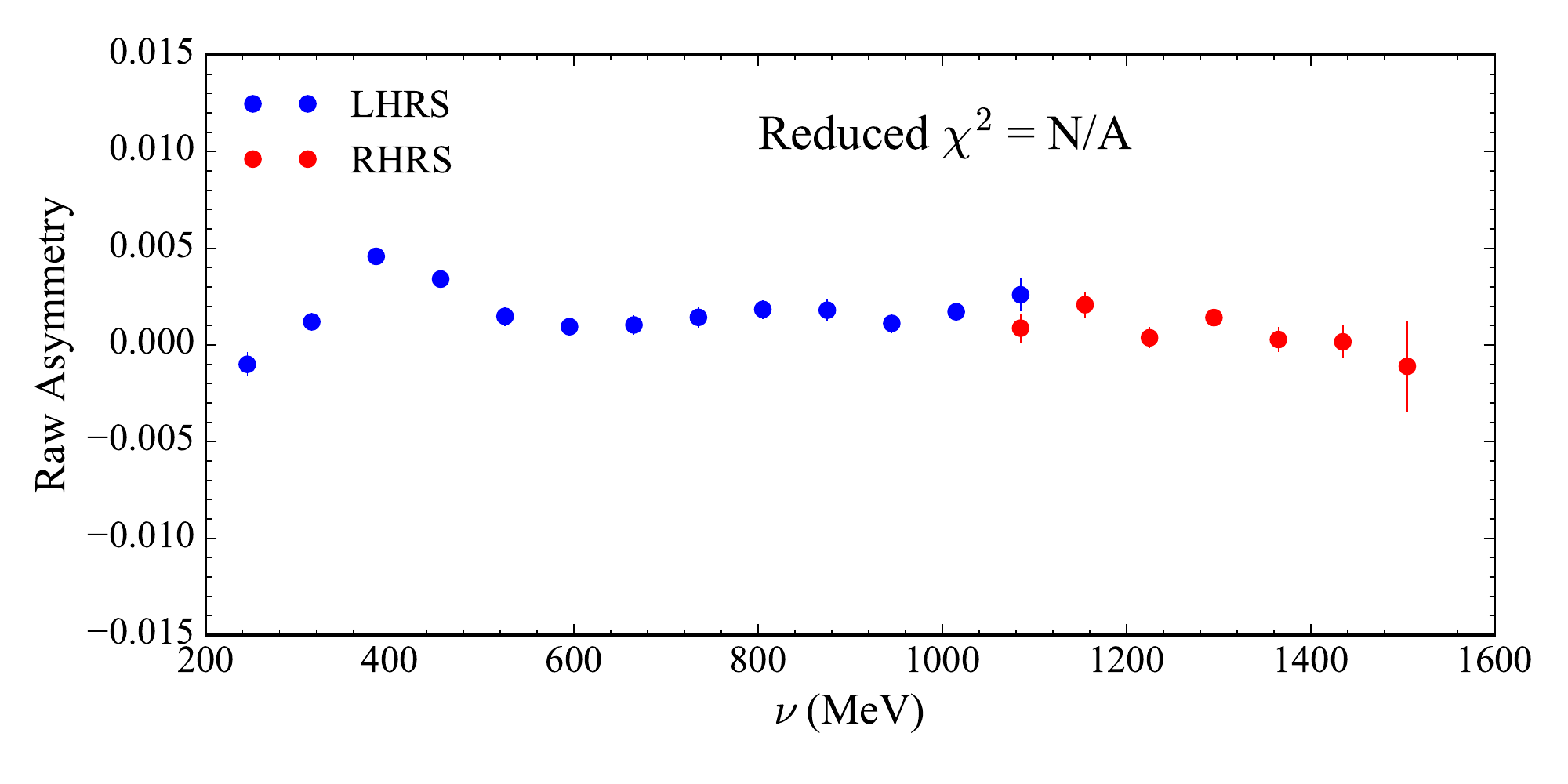}}
\qquad
\subfigure[$E_0$ = 2254 MeV 5T Longitudinal]{\label{fig:LongComp}\includegraphics[width=.80\textwidth]{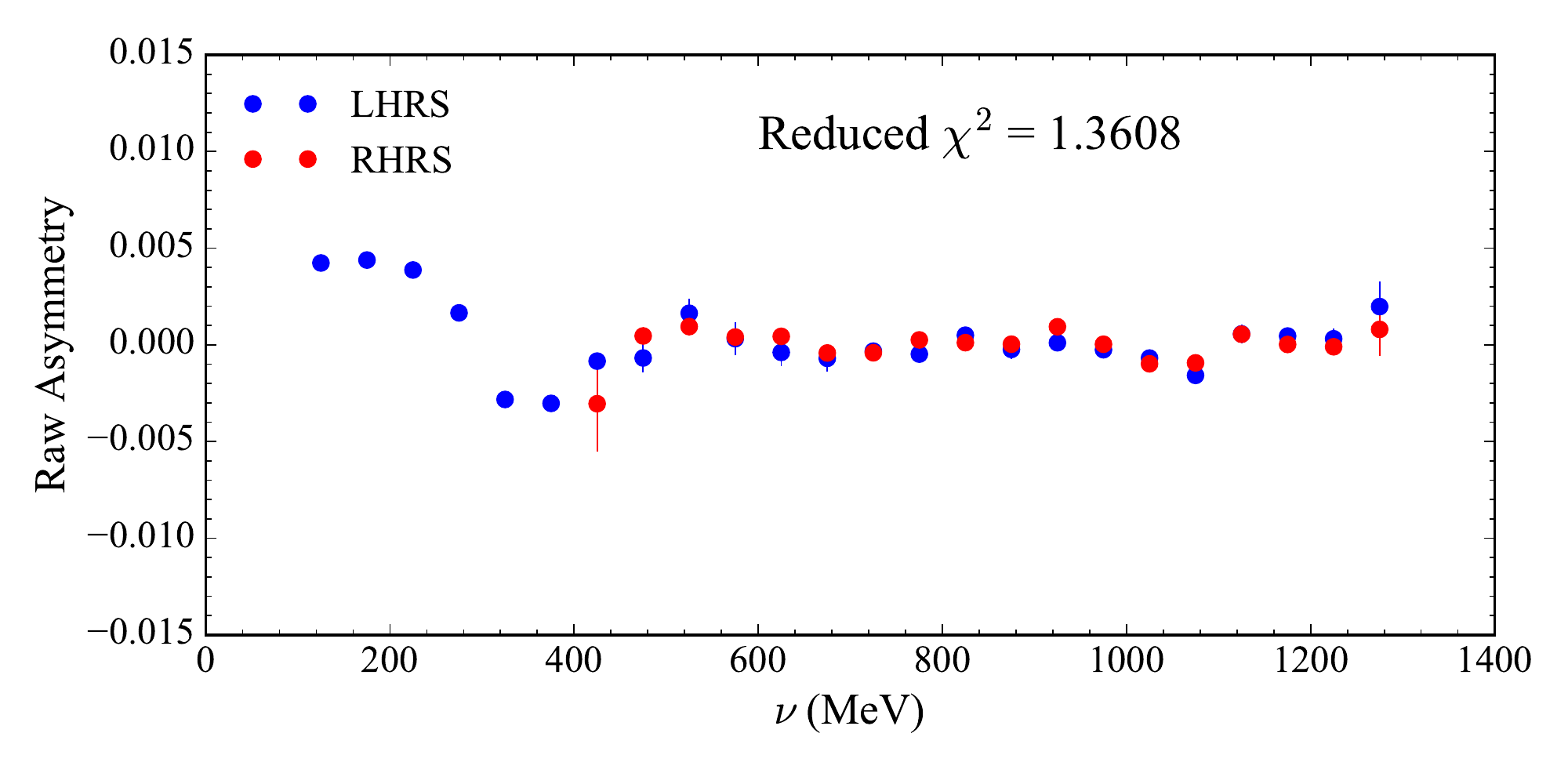}}
\caption{Asymmetry comparison between the LHRS and RHRS.}
\label{AsymComp}
\end{figure}

The comparison of the asymmetries between the two spectrometers is shown in Figure~\ref{AsymComp}. These asymmetries are corrected for target and beam polarization, half-wave plate status, HRS status, and out-of-plane polarization angle. The results agree well with the ideal reduced $\chi^2$ of 1.3, but no comparison is made for $E_0$ = 3350 MeV, since the data only overlap at single kinematic point at this setting. The slightly larger than expected value for $E_0$ = 2254 MeV 5T Transverse is because of a single point at approximately $\nu = 1100$ MeV. If this point is removed, the reduced $\chi^2$ is comparable to that of the longitudinal setting. 

\begin{figure}[htp]
\centering     
\includegraphics[width=.80\textwidth]{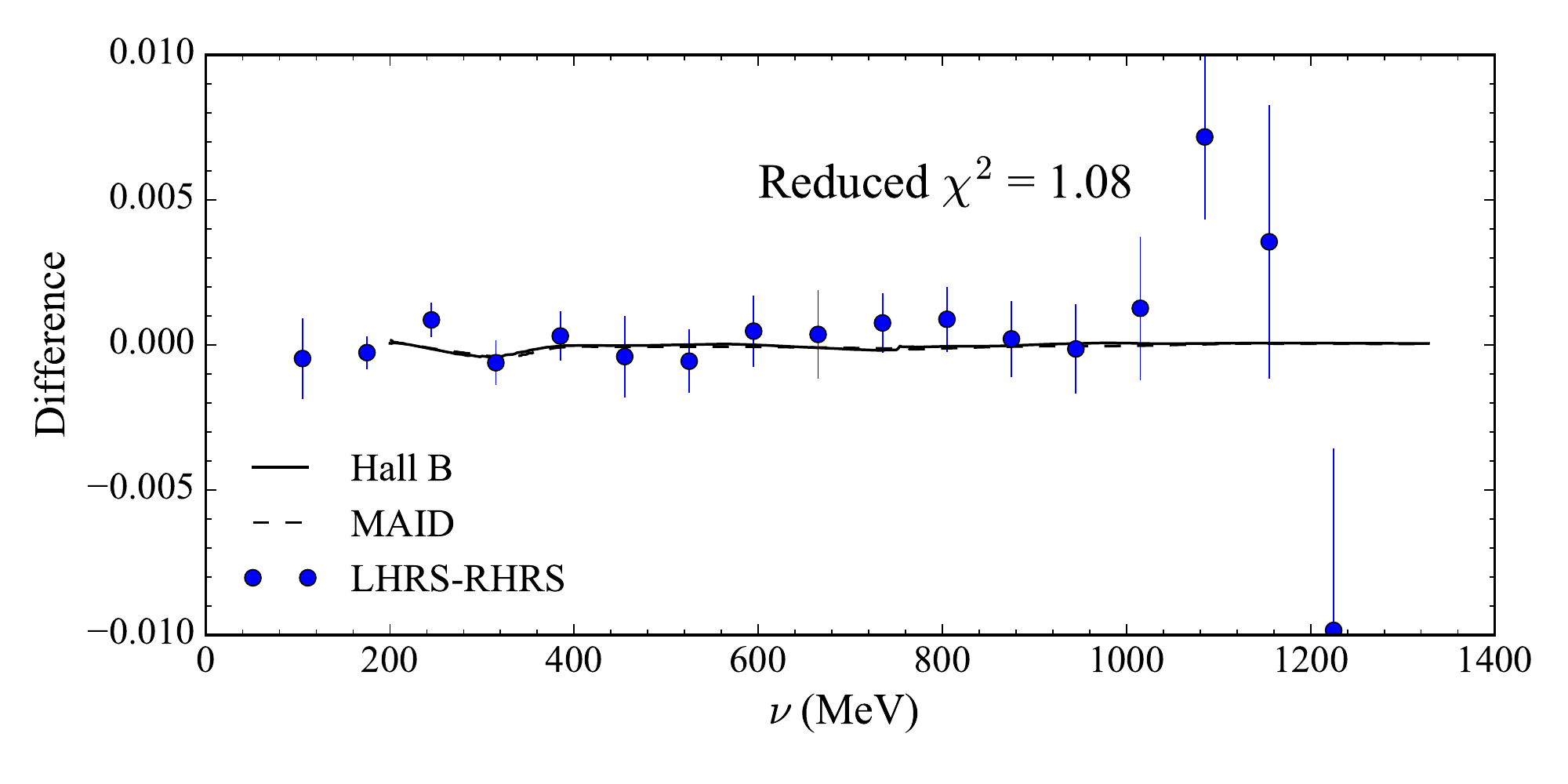}
\caption{Effect of the angular difference on the asymmetry between the two spectrometers. The effect is not statistically significant.}
\label{LMinusR}
\end{figure}

As an additional check on this data, the difference between the LHRS and RHRS asymmetries are calculated point-by-point and compared to an equivalent difference calculated using polarized models at the $E_0$ = 2254 MeV transverse setting. The models are generated using the angular fits in Figure~\ref{fig:2254Ang} for each HRS to mimic the data. The comparison gives a measure on the size of the effect of the angle difference on the asymmetries between the LHRS and RHRS. From Figure~\ref{LMinusR}, it is clear that this difference is negligible, with the reduced $\chi^2$ of the difference and zero very close to one. 

The loose cut used in the asymmetry analysis has an angular acceptance of approximately $\pm$2$^{\circ}$ about the central value. If there is a strong physics preference to scatter at a specific angle, then the measured quantity is not accurately described by the central scattering angle. Effects of this nature are referred to as bin-centering effects, and the measured quantity is corrected back to the central value of the bin. In general, this is a bigger issue for cross sections with their strong Mott cross section angular dependence and good statistics. 

\begin{figure}[htp]
\centering     
\includegraphics[width=.80\textwidth]{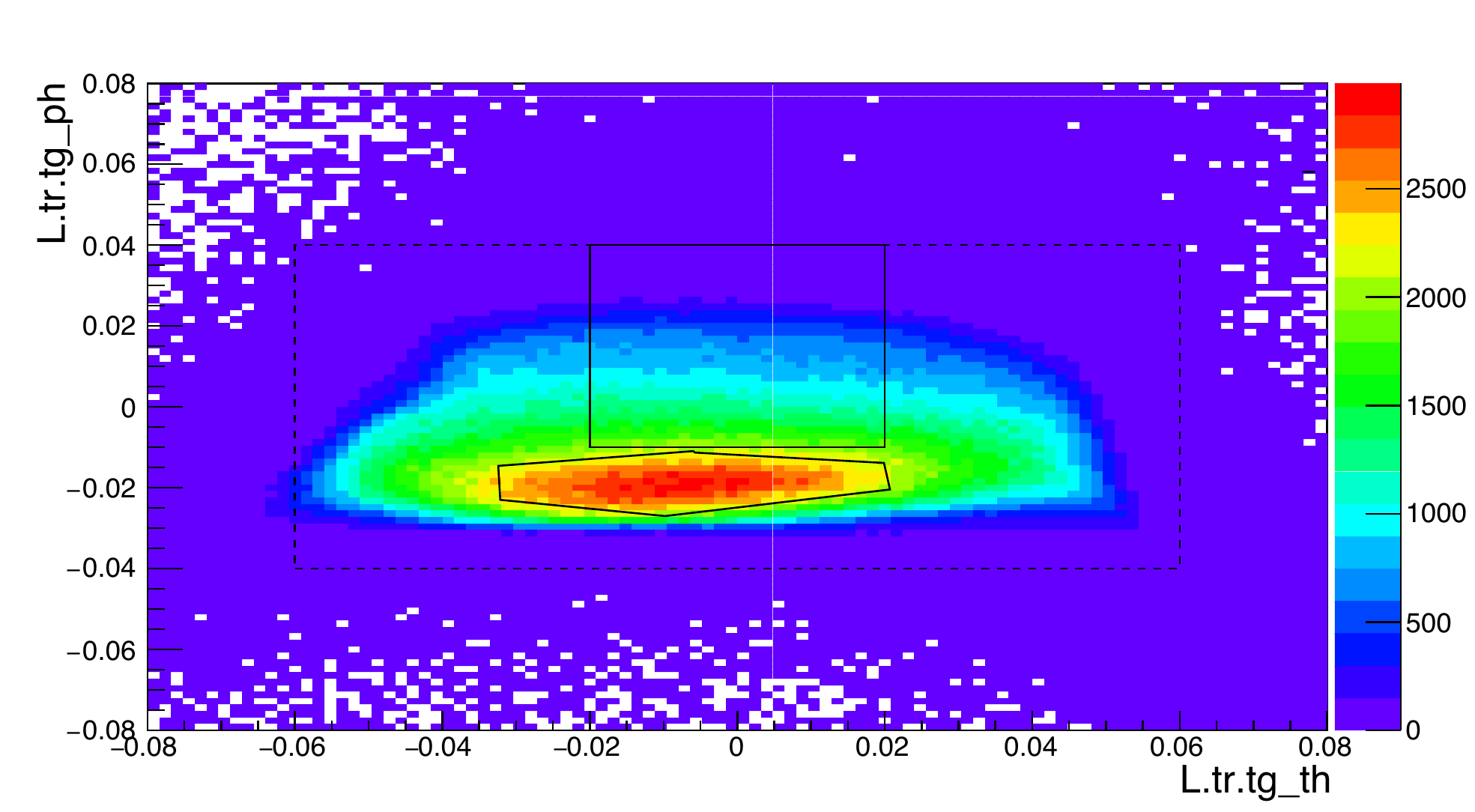}
\caption{Longitudinal asymmetry cut selection study. The three sections are defined as follows: ``Hot" is the graphical cut on the red region, ``Loose" is the dotted-line box cut (defined in Chapter~\ref{Sec:Cuts}), and ``Not-Hot" is the solid-line box cut. }
\label{CutCompare}
\end{figure}

The bin-centering effect for the asymmetries is studied at the $E_0$ = 2254 MeV longitudinal setting. This setting is picked for the study because it has the best statistics of the three 5 T settings. The angular acceptance is broken down into three sections (``Hot", ``Not-Hot", and ``Loose") with three different cuts, as shown in Figure~\ref{CutCompare}. The dotted line is the loose cut described in Chapter~\ref{Sec:Cuts} and corresponds to a central scattering angle of 5.77$^{\circ}$. The tight box cut excluding the hot-spot (solid line) has a central scattering angle of $\approx$ 6.3$^{\circ}$.  The graphical cut on the hot-spot has a central scattering angle of $\approx$ 4.9$^{\circ}$. The areas of the tight box cut and hot-spot graphical cut are selected to give a similar number of events in each region.

\begin{figure}[htp]
\centering     
\includegraphics[width=.90\textwidth]{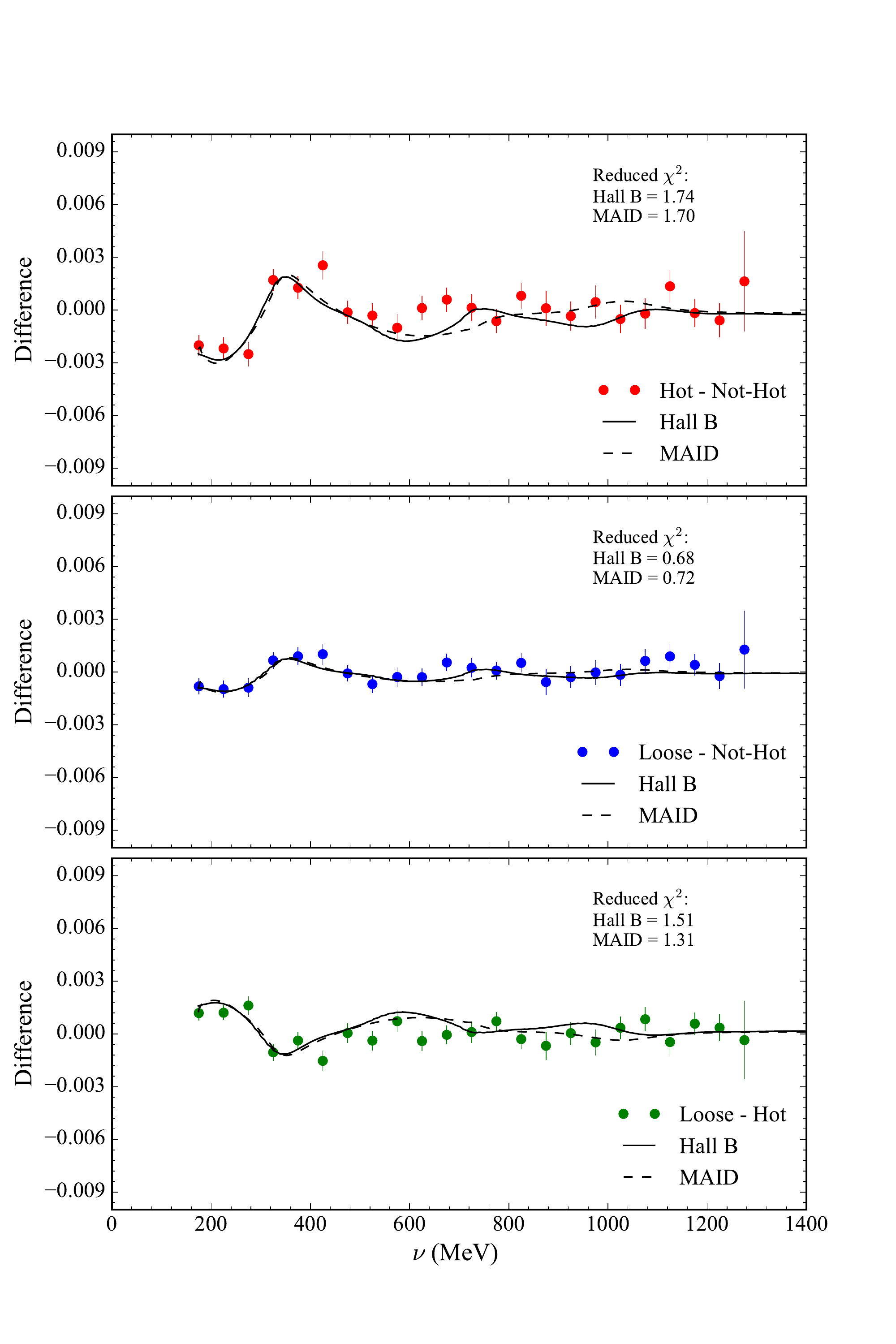}
\caption{Effect of different acceptance cuts on the asymmetry given as the difference of the asymmetries. The difference is computed for the three permutations of cuts and compared to the same quantity calculated using polarized models.}
\label{DiffCompare}
\end{figure}

\begin{figure}[htp]
\centering     
\includegraphics[width=.90\textwidth]{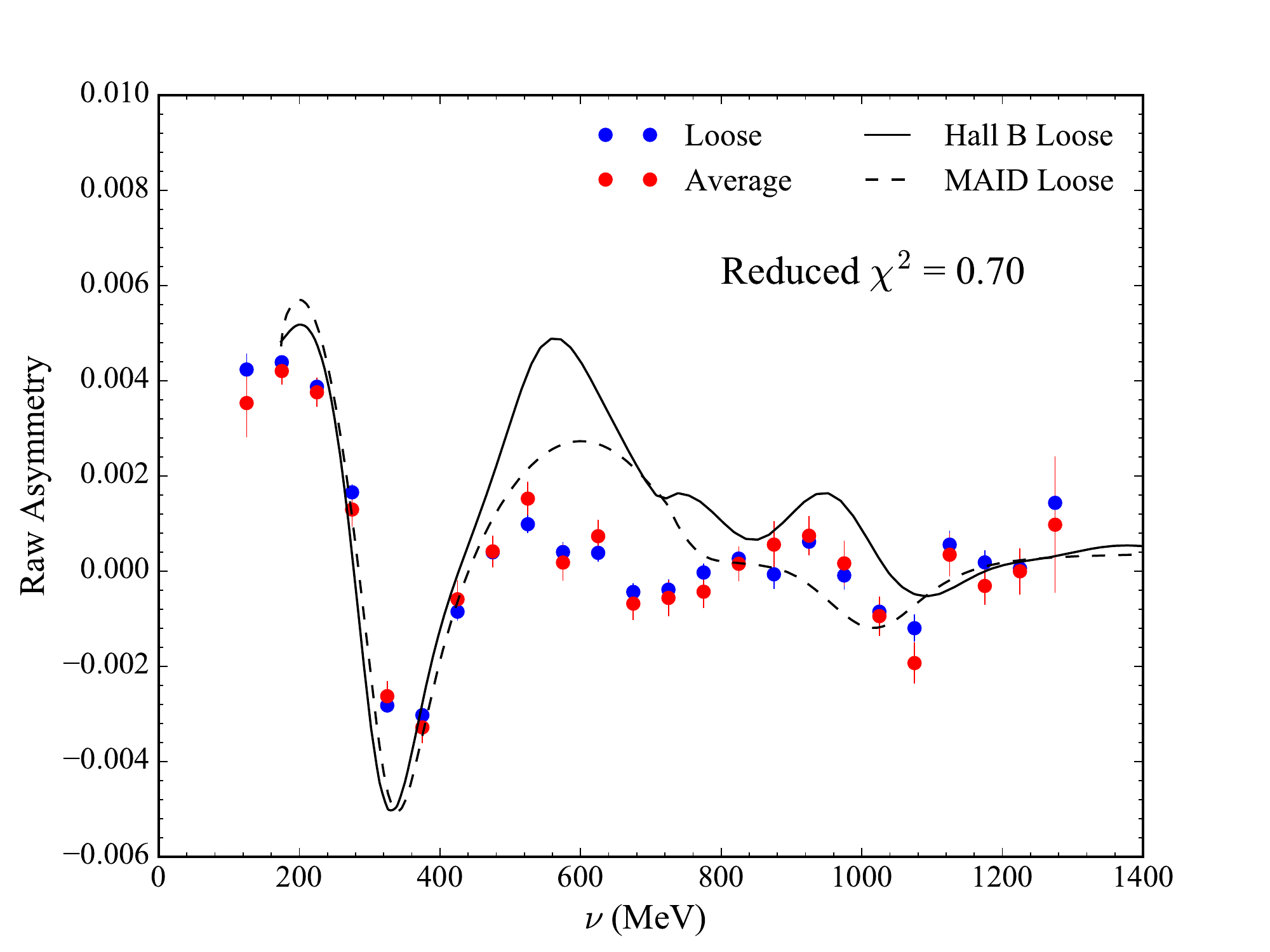}
\caption{Estimate of the bin centering effect on the longitudinal asymmetry.}
\label{BinCenter}
\end{figure}

The difference between the asymmetries with the different cuts is shown in Figure~\ref{DiffCompare}. These differences are compared to polarized asymmetry models run at the angles of each cut and multiplied by the corresponding model dilution to match the data. The angular effect seen in the asymmetries is reproduced well with the models. The bin-centering effect is tested by comparing the average asymmetry of the hot-spot cut and tight box cut to the asymmetry of the loose cut. The average angle of the two cuts is 5.6$^{\circ}$ $\pm$ 0.16$^{\circ}$, and is the same (within uncertainty) as the loose cut angle. If there is a noticeable bin-centering effect, then the average asymmetry of the two cuts should be different from the asymmetry of the loose cut. This is not seen in Figure~\ref{BinCenter}, and the reduced $\chi^2$ of the averaged and loose asymmetries is close to one. A similar study for the transverse field produces the same result. Going forward, it is assumed that there is no bin-centering effect in the asymmetries.  

\subsection{Results and Systematic Errors}
\label{AsymSys}
\begin{figure}[htp]
\centering     
\subfigure[$E_0$ = 2254 MeV 5T Transverse]{\label{fig:FinalTran}\includegraphics[width=.80\textwidth]{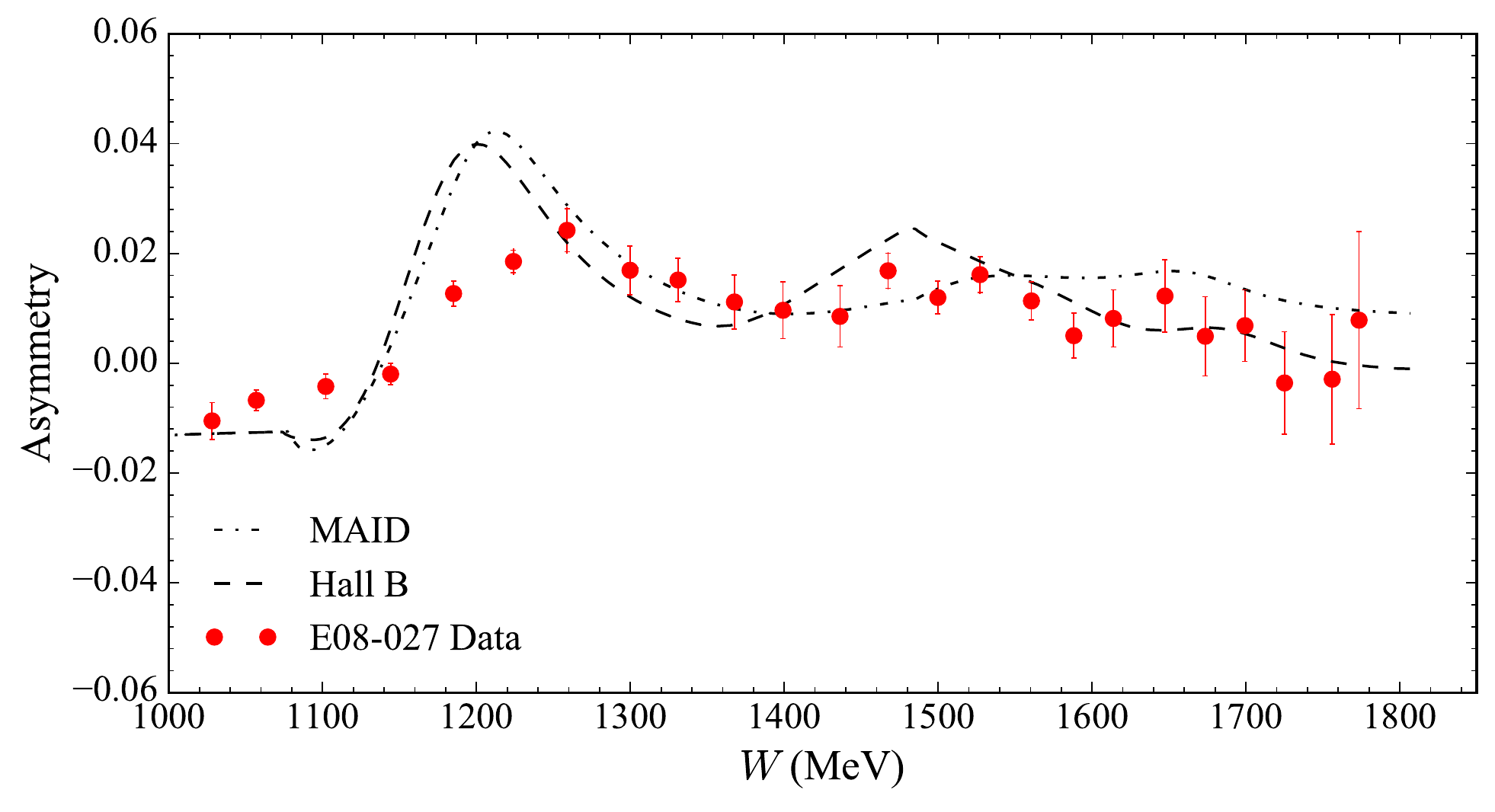}}
\qquad
\subfigure[$E_0$ = 3350 MeV 5T Transverse]{\label{fig:Final3350}\includegraphics[width=.80\textwidth]{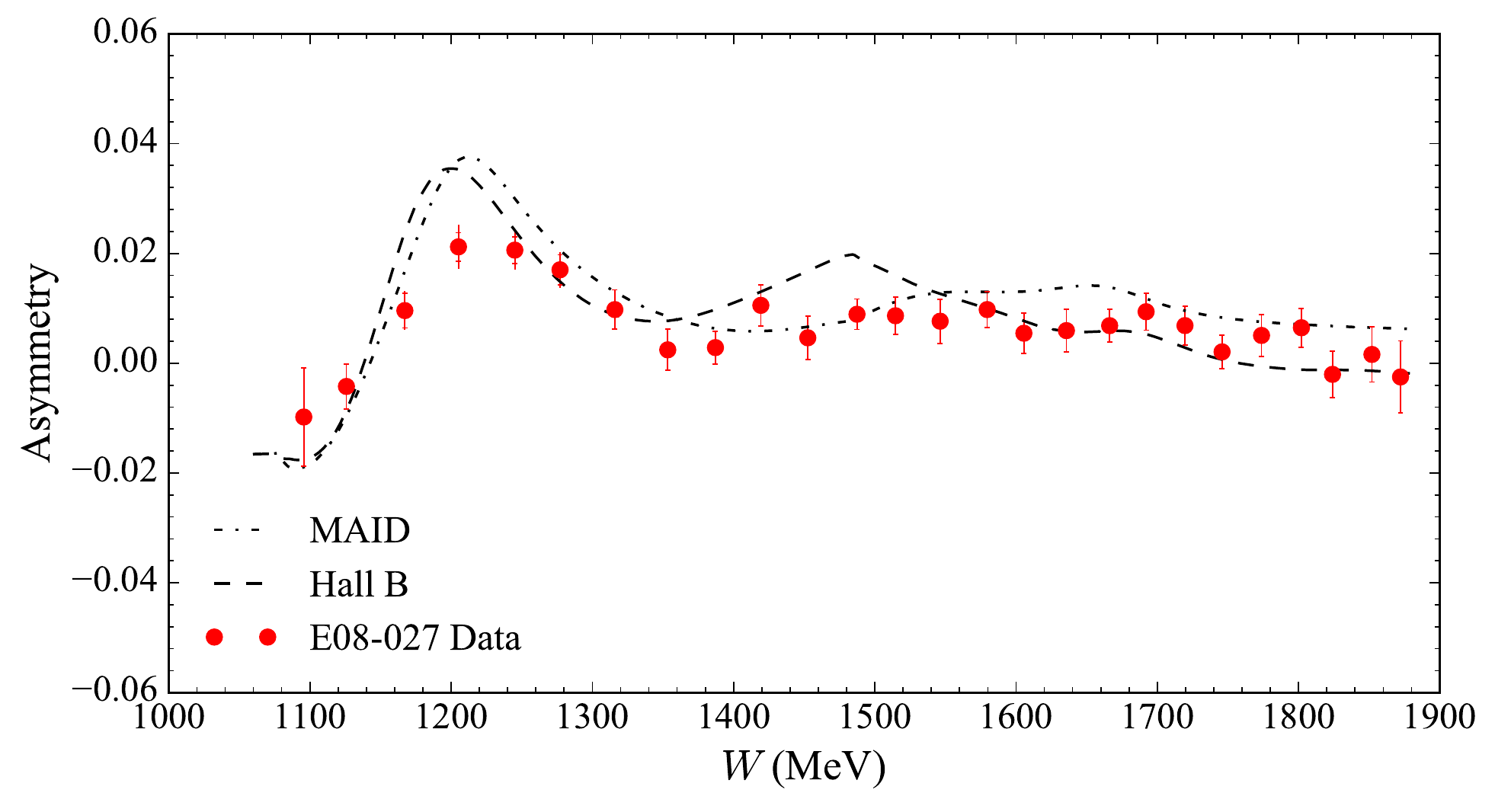}}
\qquad
\subfigure[$E_0$ = 2254 MeV 5T Longitudinal]{\label{fig:FinalLong}\includegraphics[width=.80\textwidth]{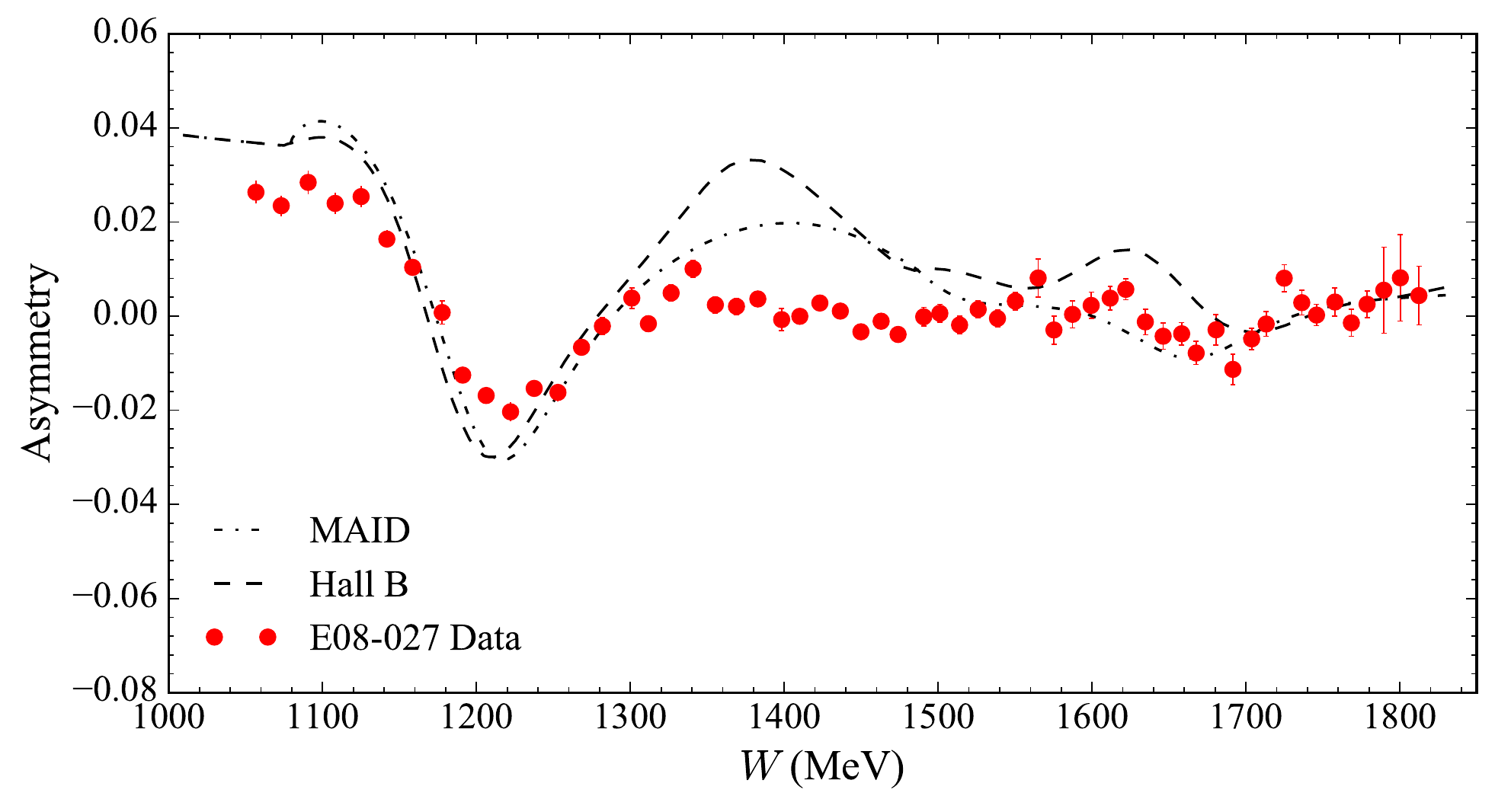}}
\caption{Final asymmetries for the 5 T settings.}
\label{FinalAsym}
\end{figure}

The final asymmetries before radiative corrections for the 5 T settings are shown in Figure~\ref{FinalAsym} and include the experimental dilution factors from Chapter~\ref{sec:Dil}. The $W/\nu$ values are statistically weighted averages based upon the asymmetry error bars. This helps to minimize bin-centering effects from the larger binning used in an asymmetry analysis. The inner (and dominant) error bars are statistical, while the outer error bars are the statistical and systematic errors added in quadrature. The statistical error includes the error on the counts as described in Chapter~\ref{Sec:Asymm}, with the statistical error from the dilution analysis and beam and target polarization analysis also propagated through. The statistical uncertainty arising from the counts is by far the largest contributor to this uncertainty. 
\begin{table}[htp]
\begin{center}
\begin{tabular}{ l  c  c  c c c | r }
\hline
$E_0$ (MeV)& Config & $\delta_{\mathrm{dil}}$ (\%) & $\delta_{\mathrm{e\,pol}}$ (\%) & $\delta_{\mathrm{p\,pol}}$ (\%) & $\delta_{\mathrm{OoP}}$ (\%) & $\delta_{\mathrm{tot}}$ (\%)\\ \hline
 2254& 90$^{\circ}$&5.0 $-$ 7.0 &1.7 & 2.0 $-$ 4.5 &0.5 $-$ 1.5&7.0 $-$ 8.5\\
2254& 0$^{\circ}$ & 5.0 $-$ 7.0& 1.7 & 2.0 $-$ 4.5&N/A& 7.0 $-$ 8.5\\
3350& 90$^{\circ}$ & 7.0 $-$ 15.0 & 1.7 & 2.0 $-$ 4.5&0.5 $-$ 1.5&8.0 $-$ 15.0\\
 \hline
\end{tabular}
\caption{\label{asym_sys}Systematic errors for the asymmetry analysis of the 5 T settings.}
\end{center}
\end{table}

The largest systematic error comes from the dilution factor correction and is 5-7\% at both $E_0$ = 2254 MeV settings and 7-15\% at the $E_0$ = 3350 MeV setting. The increased dilution systematic error at $E_0$ = 3350 MeV is caused by unresolved yield drifts in the production data. For settings with a significant yield drift a model dilution is used instead and a 15\% systematic error is applied. These regions appear as a gap in the dilution coverage in Figure~\ref{fig:Dil3350}. The remaining systematic errors are: 1.7\% from the beam polarization measurement, 2-4.5\% from the target polarization analysis and 0.5-1.5\% from the out-of-plane polarization angle (only applicable at the transverse settings). The residual pion contamination to the asymmetry is discussed in Ref~\cite{MelissaT}, and is a few orders of magnitude smaller than the measured asymmetry, so it is neglected in the analysis. The total systematic error on the asymmetry is the previously mentioned sources of error added in quadrature. The systematic errors are compiled in Table~\ref{asym_sys}.

While the dilution factor takes into account scattering from any unpolarized target material, this does not necessarily preclude the scattering from polarized nitrogen atoms. This polarized nitrogen would contribute to any polarized asymmetry measurement and alter the results. Fortunately, this effect is small (on the order of 2\%) and is safely included as a systematic error in the target polarization. Details of the calculation of the polarized nitrogen contribution to the NMR result is found in Appendix~\ref{app:Appendix-G}.

The finer binning and better statistics of the longitudinal data arises because the experiment took approximately three times the number of runs here as compared to the two other 5 T kinematic points. With its higher precision, the longitudinal data shows three resonances past the large $\Delta$(1232) resonance. These include the $N^*$ (Roper) at $W=$ 1440 MeV, the $N_1^*$ at $W\approx$ 1500 MeV, and the $N_2^*$ $W\approx$ 1700 MeV. The $Q^2$ values at the $\Delta$(1232) resonance are approximately 0.043 GeV$^2$, 0.086 GeV$^2$ and 0.13 GeV$^2$ for the $E_0$ = 2254 MeV 5 T longitudinal, 2254 MeV 5 T transverse and 3350 MeV 5 T transverse settings, respectively. An additional study of two separate but parallel asymmetry analyses is detailed in Appendix~\ref{app:Appendix-H}.

\section{Polarized Cross Section Differences}
The polarized cross section differences are formed from the experimental unpolarized cross sections and transverse and parallel asymmetries, according to equation~\eqref{eq:poldiff}. For this analysis, an updated Bosted-Mamyan-Christy model (see Chapter~\ref{BostedModSec}) for hydrogen is used in place of an experimental cross section because the spectrometer acceptance analysis is still ongoing~\cite{ChaoCom}. This use of a model has a few advantages and disadvantages. The model can be generated at the necessary kinematics and minimizes the systematic effect from combining an asymmetry and cross section generated with different acceptance cuts\footnote{The asymmetry is statistics limited and uses a much looser acceptance cut. The cross section is systematic limited and uses a much tighter acceptance cut.}. The model does not require a significant time investment to produce a result. The accuracy of an empirical model is beholden to the data available for fitting; it is difficult to estimate the uncertainty in a region where there is no past data. In Hall A,  past uncertainties in the cross section acceptance corrections are are no more than 5\%~\cite{VinceT}. Even the best empirical model will struggle to match that level of uncertainty.

The updated model includes fits to low $Q^2$ proton data that is not included in the published fit in Ref~\cite{Bosted3}. The additional data comes from the SLAC E61 and SLAC ONEN1HAF experiments~\cite{HallCRes}, and the new fit is provided by Eric Christy~\cite{ChristyCom}. The model is generated to mimic the experimental conditions and run at the central scattering angle determined from data. The model is also fully radiated according to the analysis of Chapter~\ref{ch:RC}, with the radiation lengths from Appendix~\ref{app:Appendix-D}.
\begin{figure}[htp]
\centering     
\subfigure[$E_0$ = 4.499 GeV, $Q^2 \approx$ 0.094 GeV$^2$ ]{\label{fig:FinalTran222}\includegraphics[width=.47\textwidth]{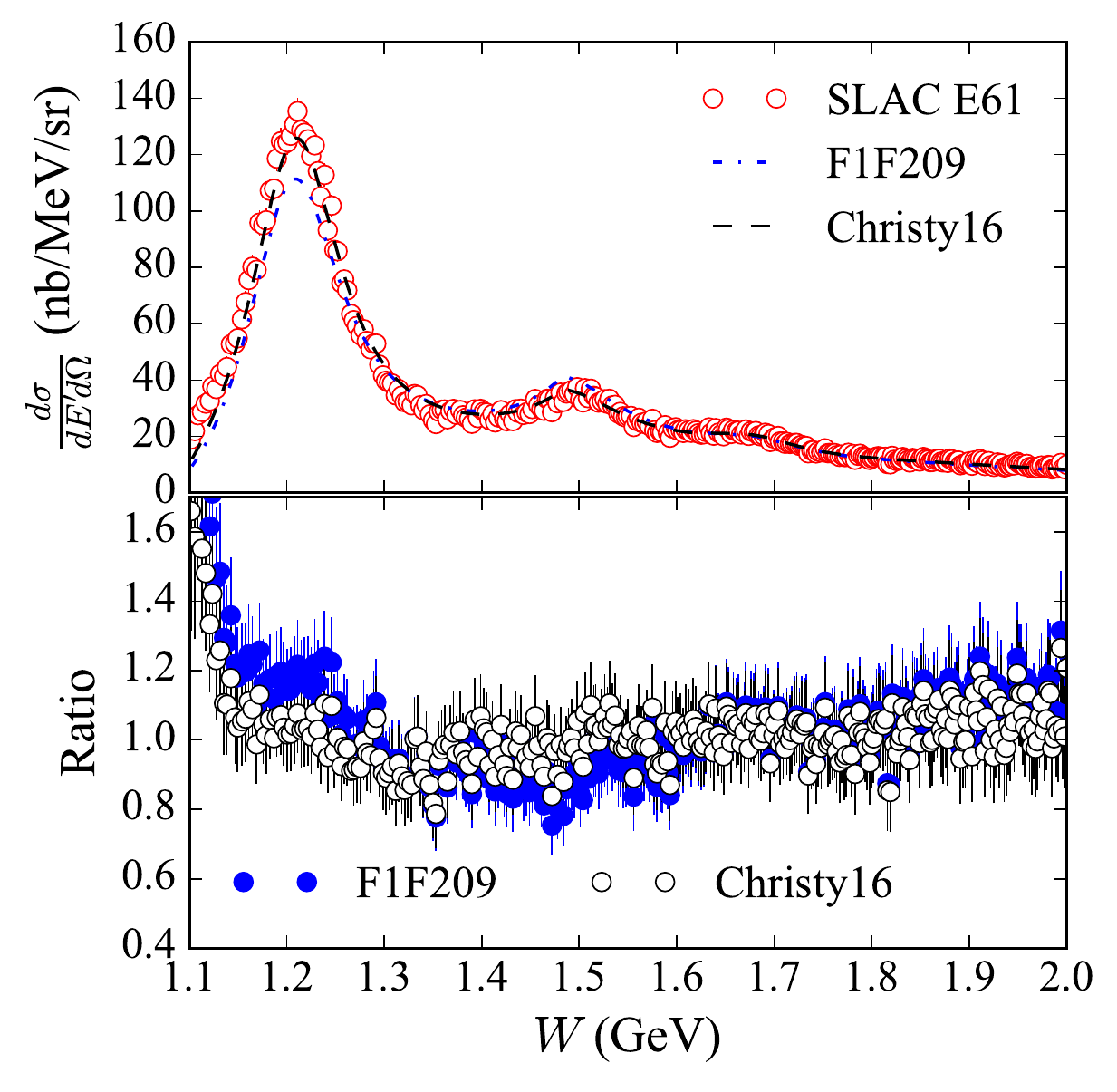}}
\qquad
\subfigure[$E_0$ = 9.301 GeV, $Q^2 \approx$ 0.056 GeV$^2$]{\label{fig:Final3350222}\includegraphics[width=.47\textwidth]{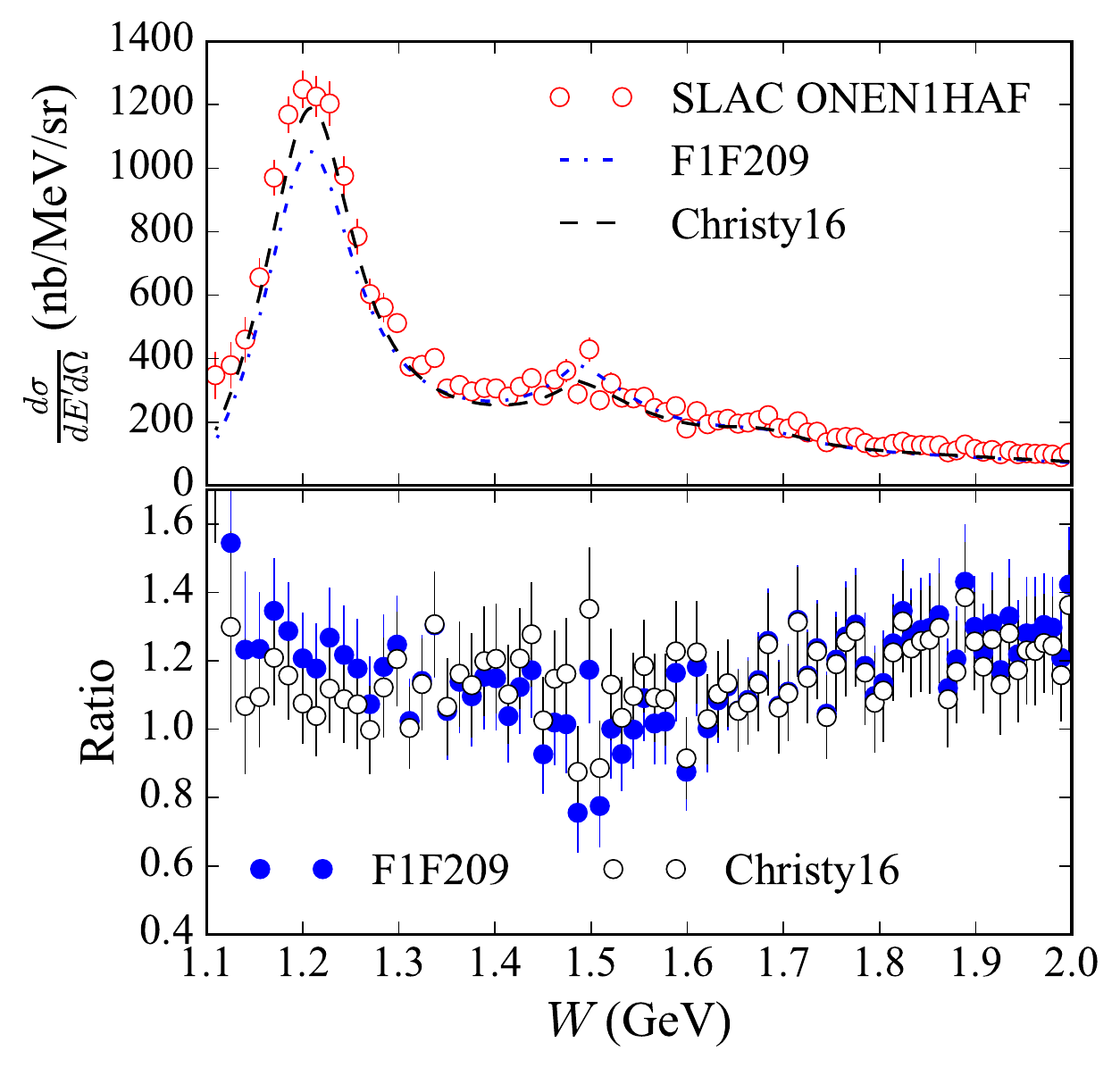}}
\qquad
\subfigure[$E_0$ = 11.799 GeV, $Q^2 \approx$ 0.09 GeV$^2$]{\label{fig:FinalLon222g}\includegraphics[width=.47\textwidth]{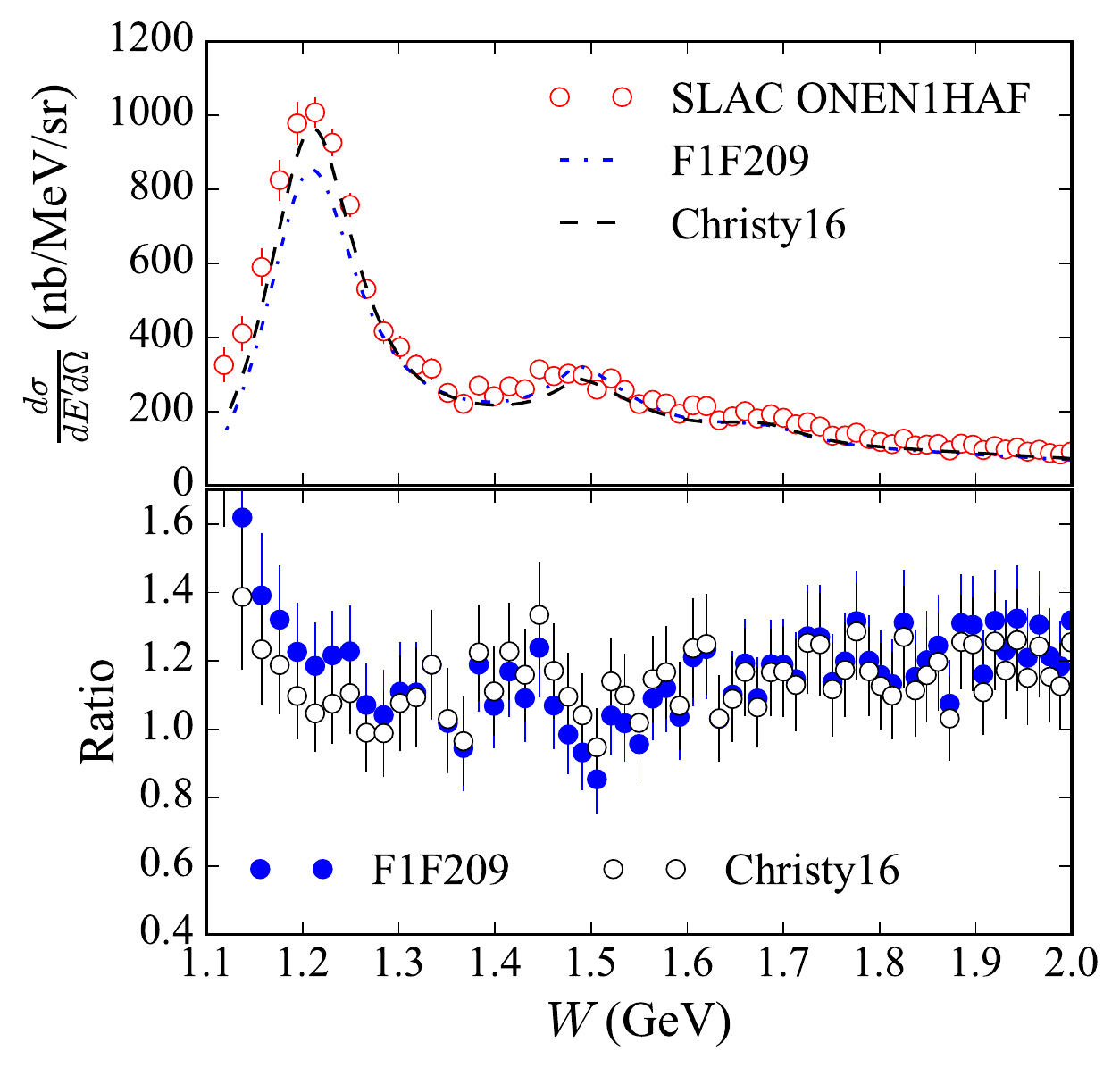}}
\qquad
\subfigure[$E_0$ = 13.805 GeV, $Q^2 \approx$ 0.124 GeV$^2$]{\label{fig:FinalLon22werwer2g}\includegraphics[width=.47\textwidth]{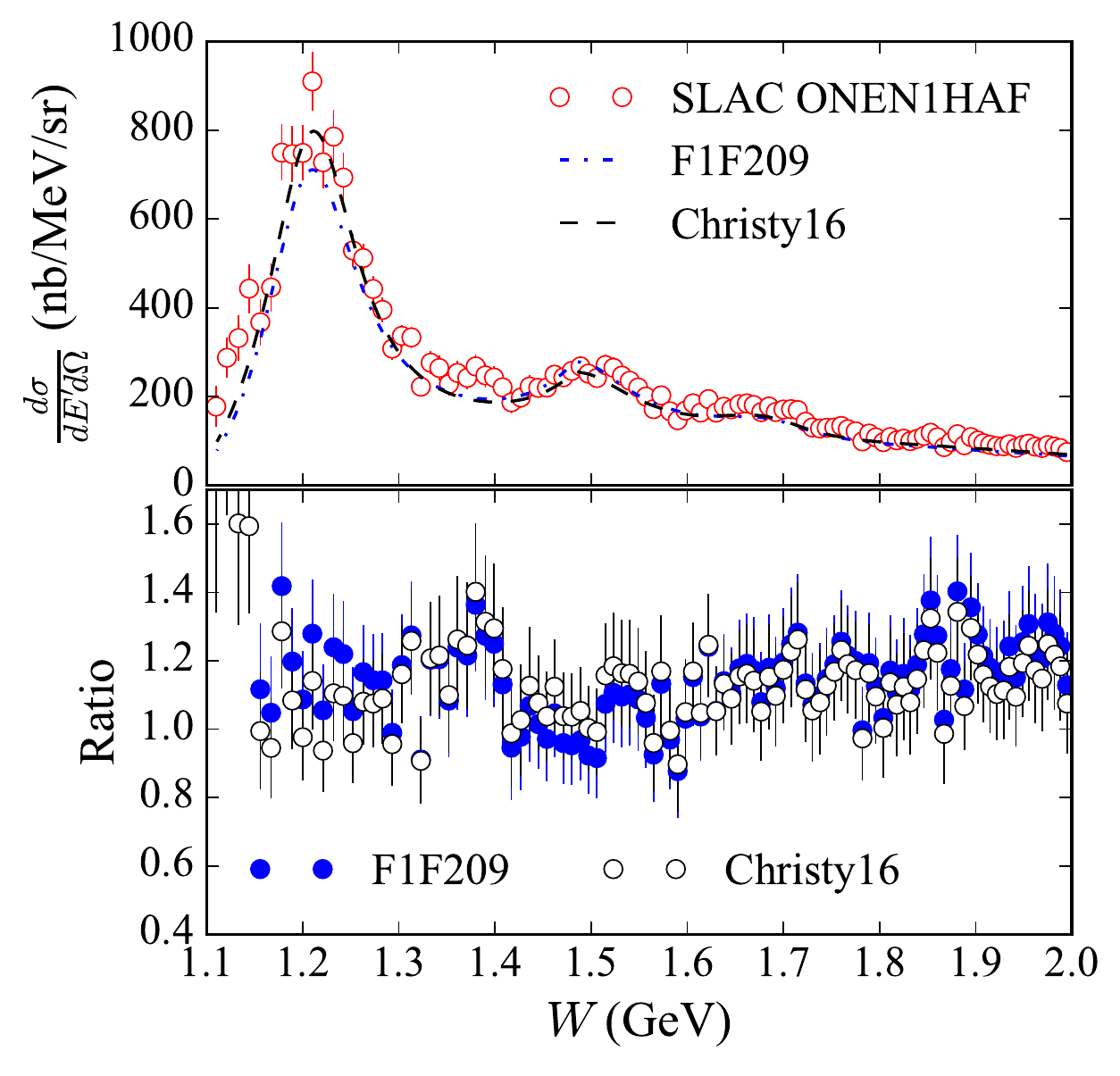}}
\caption{Low $Q^2$ improvement of the unpolarized model fit at the E08-027 kinematics. The advantage of the new fit is highlighted by comparing the ratio of the data to the model generated at the same kinematics. The F1F209 fit is from Ref~\cite{Bosted3,Bosted2}.}
\label{Christy16}
\end{figure}

The improvement in the fit is shown in Figure~\ref{Christy16} for the $Q^2$ range of E08-027, where the reported $Q^2$ values represent the value at the $\Delta$(1232) resonance. Only statistical error bars are shown. The new fit (labeled ``Christy16") greatly improves agreement between the model and existing data at the first resonance peak. The difference is smaller for the rest of the data spectrum at the Born level, but for radiation purposes the $\Delta$(1232) resonance is important to the entire inelastically radiated spectrum.

\subsection{Unpolarized Model Input}

\begin{figure}[htp]
\centering     
\includegraphics[width=.80\textwidth]{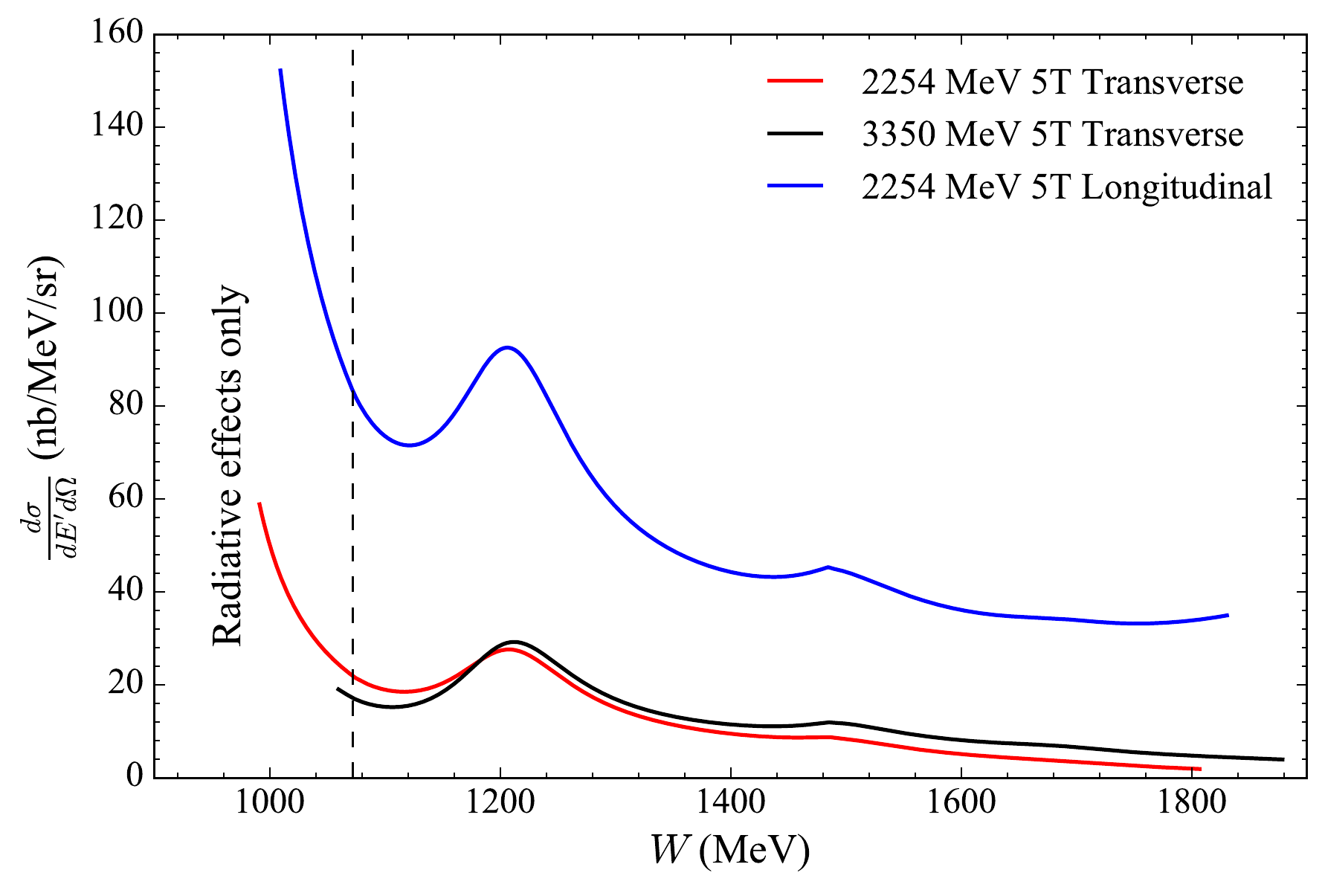}
\caption{Model generated radiated unpolarized cross sections.}
\label{InelasticRadXSBosted}
\end{figure}

The radiated unpolarized inelastic cross section models are shown in Figure~\ref{InelasticRadXSBosted}. The rapidly increasing cross section to the left of the pion production threshold (dotted line) is almost solely from the radiative elastic tail. The elastic tail is also causing the rise in the $E_0$ = 2254~MeV longitudinal cross section at large invariant mass. This is not seen in the transverse settings because the increase in scattering angle cancels out the increase in the elastic cross section at smaller electron energies.  

\begin{figure}[htp]
\centering     
\includegraphics[width=0.90\textwidth]{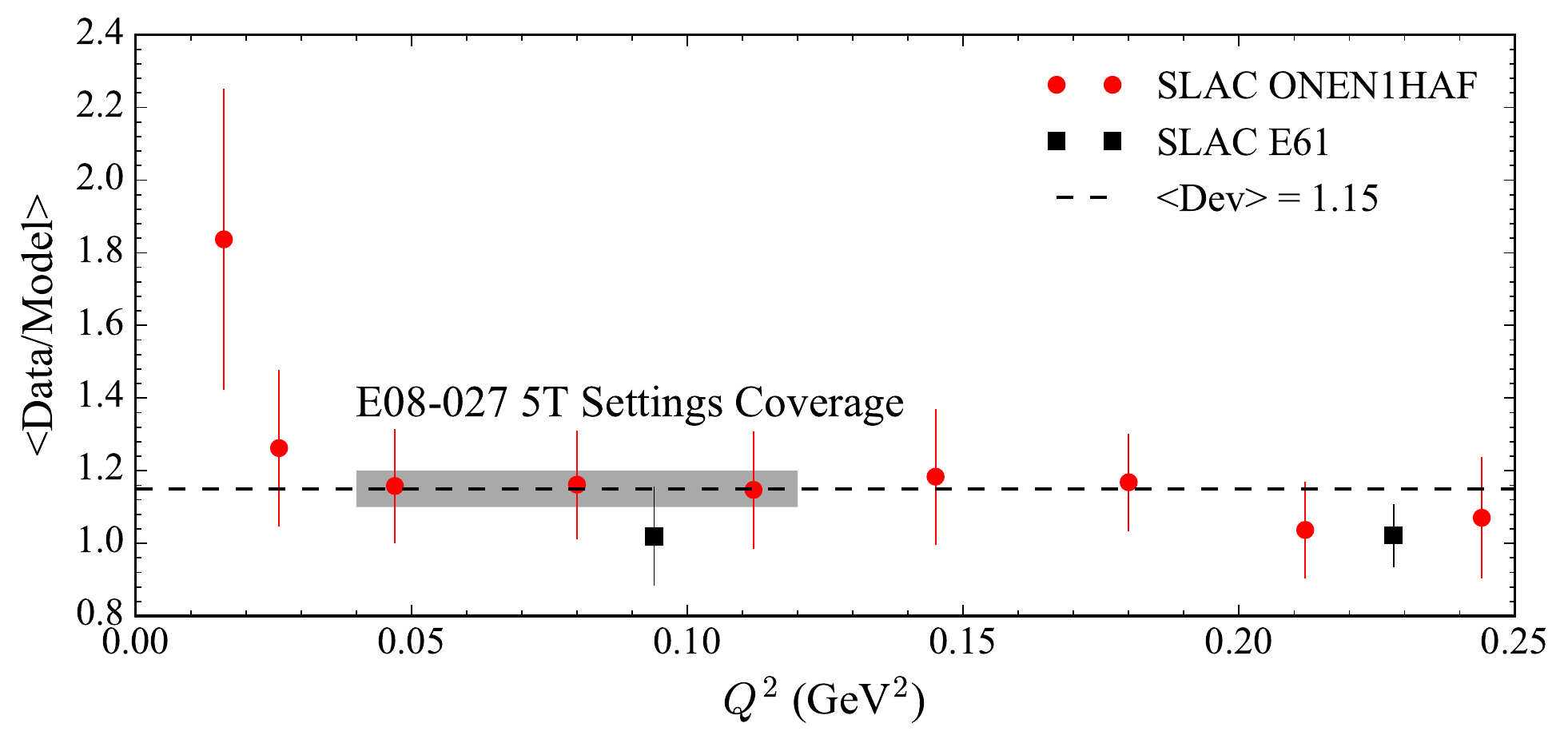}
\caption{Comparison between the Bosted-Mamyan-Christy model and existing data. See Figure~\ref{Christy16} for the individual spectra in the E08-027 kinematic region.}
\label{BostedModelComp}
\end{figure}

The uncertainty in the unpolarized cross section model before radiation is estimated by comparing the model prediction to available data at the E08-027 kinematics. The data used for the model comparison is from the Hall C Resonance Data Archive~\cite{HallCRes}. The results are shown in Figure~\ref{BostedModelComp}. The points represent the average ratio between the model and the data at a single beam energy and the error bar represents the standard deviation in this average. The reported $Q^2$ value corresponds to the $\Delta$(1232) resonance, but the data cover most of the resonance region. At the kinematics of the 5 T settings, the difference is approximately 1.15 with a standard deviation of 0.15. If the model is scaled by 1.15, the average difference is centered around 1.0 with standard deviation of 0.15. For this analysis, the systematic error on unpolarized model is taken as 15\%.

 Very preliminary cross section studies on the E08-027 data agree with the 15\% systematic error~\cite{TobyCom}. In these studies, the complex acceptance shape is approximated, geometrically, as a box via very tight cuts on the acceptance. The area of the box is the acceptance correction. The uncertainty on this area is large  (on the order of 20-30\%) for very tight cuts.

\subsection{Results and Systematic Error}
\begin{figure}[htp]
\centering     
\subfigure[$E_0$ = 2254 MeV 5T Transverse]{\label{fig:DSTran}\includegraphics[width=.80\textwidth]{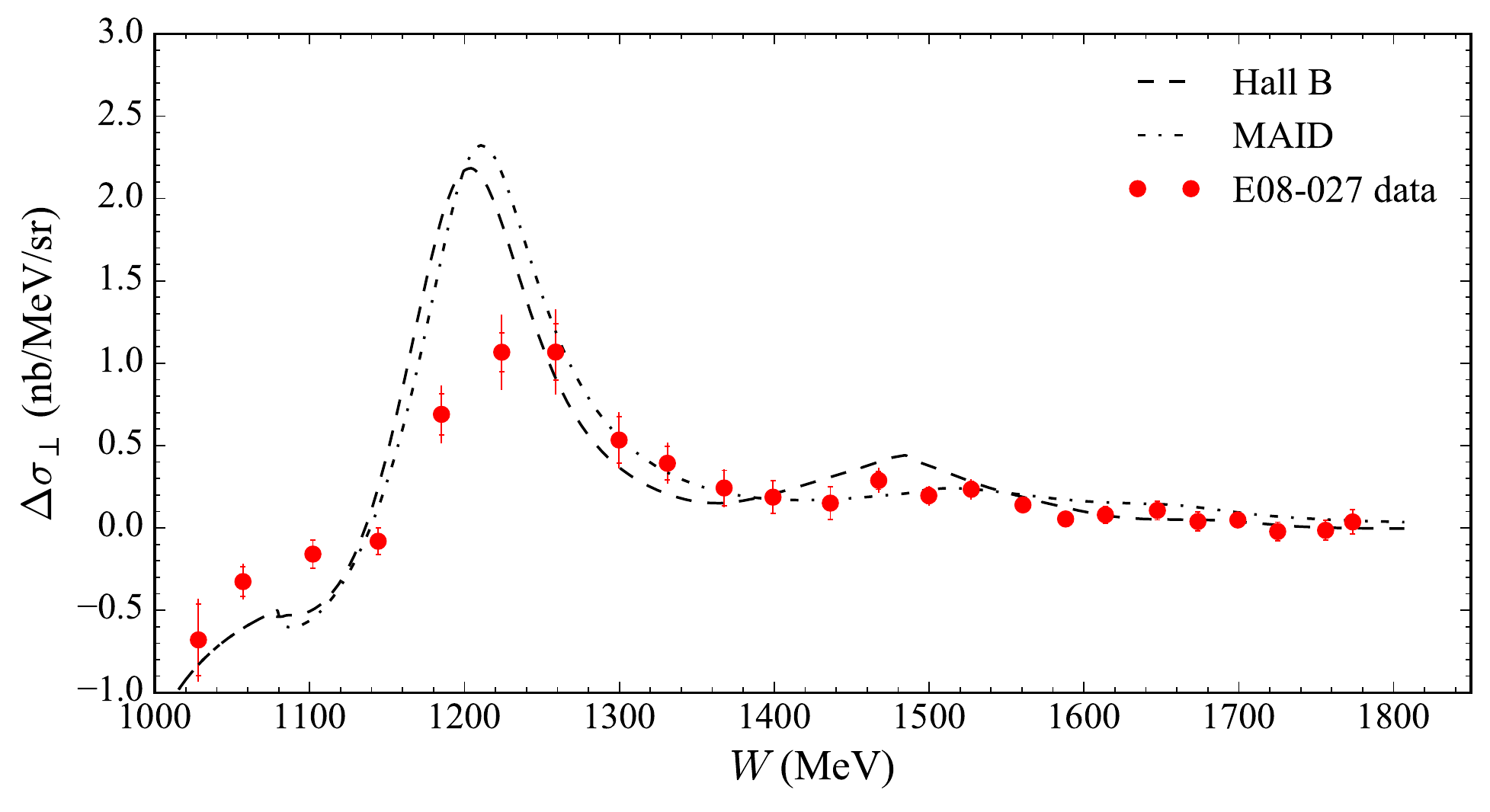}}
\qquad
\subfigure[$E_0$ = 3350 MeV 5T Transverse]{\label{fig:DSTran3350}\includegraphics[width=.80\textwidth]{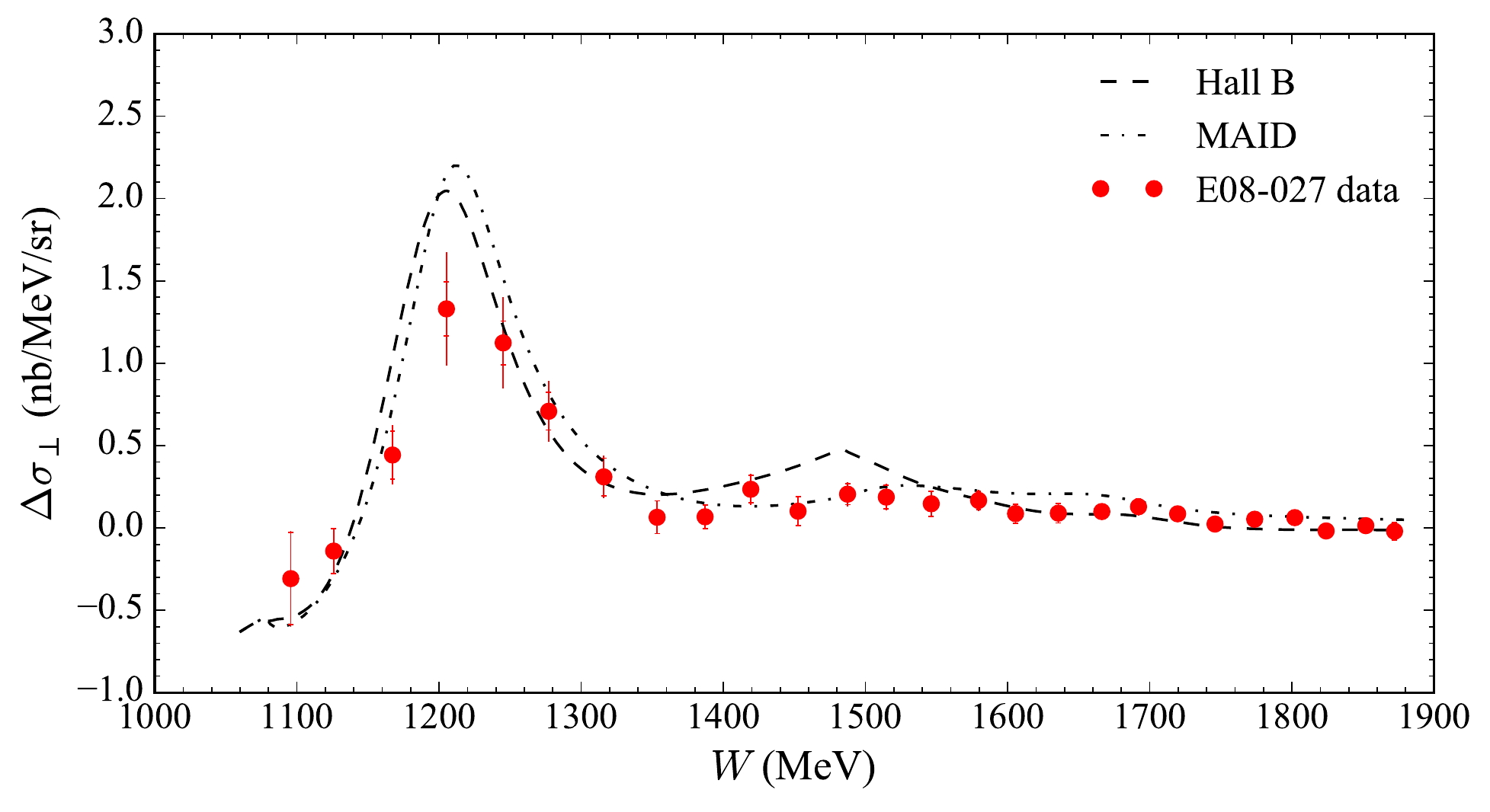}}
\qquad
\subfigure[$E_0$ = 2254 MeV 5T Longitudinal]{\label{fig:DSLong}\includegraphics[width=.80\textwidth]{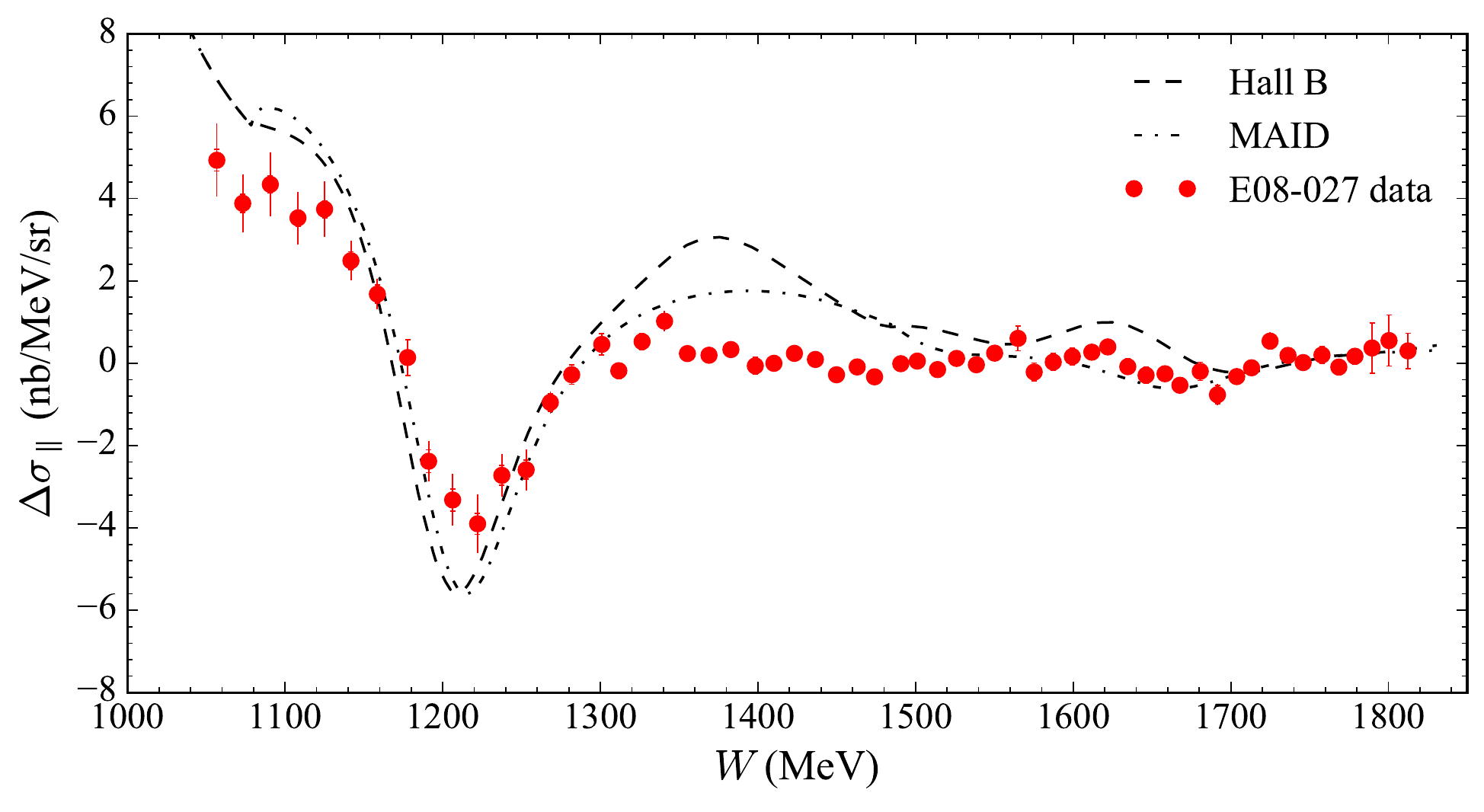}}
\caption{Polarized cross section differences before radiative corrections for the 5 T settings.}
\label{PolDS}
\end{figure}

The experimental (before radiative corrections) polarized cross section differences are shown in Figure~\ref{PolDS}. The inner error bars are statistical, and the outer error bars are the statistical and systematic errors added in quadrature. The systematic error on the polarized cross section differences is the asymmetry systematic error added in quadrature with the unpolarized model error. There are three components to the model error: how accurate is the radiation procedure for the model, how well does the model represent data at similar kinematics, and how does the scattering angle uncertainty correlate with the generated models. The accuracy of the radiation formalism is discussed in detail in Chapter~\ref{ch:RC} and the result is an uncertainty of approximately 3\%. The unpolarized model comparison to data gives a systematic error of 15\%. The total systematic error is the three sources added in quadrature. The angle uncertainty propagation is discussed below.

The largest source of angle dependence in the unpolarized cross section is the Mott cross section, with $\sigma_{\mathrm{Mott}} \propto \mathrm{cos}^2\frac{\theta}{2}/\mathrm{sin}^4\frac{\theta}{2}$. The smaller the angle the more sensitive the Mott cross section is to variations in that angle. Using the results from Chapter~\ref{sec:Angfit} for both the angle and uncertainty corresponds to a Mott uncertainty of 4-8\% (high $\nu$ to low $\nu$) at $E_0$ = 2254 MeV 5 T transverse, 7-9\% at  $E_0$ = 3350 MeV 5 T transverse and 2.8\% at $E_0$ = 2254 MeV 5 T longitudinal. The smaller uncertainty at the longitudinal setting is due to the use of the survey angle and its resulting uncertainty in this analysis. The final systematic errors on the radiated polarized cross section differences are listed in Table~\ref{ds_sys}, and are dominated by the 15\% error assigned to the unpolarized model input. 

\begin{table}[htp]
\begin{center}
\begin{tabular}{ l c c c c  c |r }
\hline
$E_0$ (MeV)& Config & $\delta_{\mathrm{asym}}$ & $\delta_{\mathrm{mod}}$ (\%) & $\delta_{\mathrm{rad}}$ (\%) & $\delta_{\mathrm{Mott}}$ (\%) & $\delta_{\mathrm{ds\,rad}}$ (\%) \\ \hline
 2254& 90$^{\circ}$& 7.0 $-$ 8.5&15.0& 3.0 &4.0 $-$ 8.0 & 17.3 $-$ 19.3 \\
2254& 0$^{\circ}$ & 7.0 $-$ 8.5& 15.0& 3.0& 2.8 & 17.0 $-$ 18.0\\
3350& 90$^{\circ}$ & 8.0 $-$ 15.0& 15.0 & 3.0& 7.0 $-$ 9.0 & 18.0 $-$ 22.5\\
 \hline
\end{tabular}
\caption{\label{ds_sys}Systematic errors for the radiated (measured) polarized cross section differences. The asymmetry systematic errors are from Table~\ref{asym_sys}.}
\end{center}
\end{table} 

\section{Experimental Radiative Corrections}
\label{result:RC}
The polarized elastic radiative tail is subtracted from the measured polarized cross section difference and inelastic radiative corrections are applied to obtain the Born polarized cross section difference:
\begin{equation}
\Delta \sigma^{\mathrm{Born}}_{\perp,\parallel} = \Delta \sigma^{\mathrm{exp}}_{\perp,\parallel} - \Delta \sigma^{\mathrm{tail}}_{\perp,\parallel} + \delta(\Delta \sigma^{\mathrm{RC}}_{\perp,\parallel})\,.
\end{equation}
The analysis of this section follows closely the formalism and procedures laid out in Chapter~\ref{ch:RC}, but includes details specific to the E08-027 data.

\subsection{Polarized Elastic Tail Subtraction}
\begin{figure}[htp]
\centering     
\includegraphics[width=0.90\textwidth]{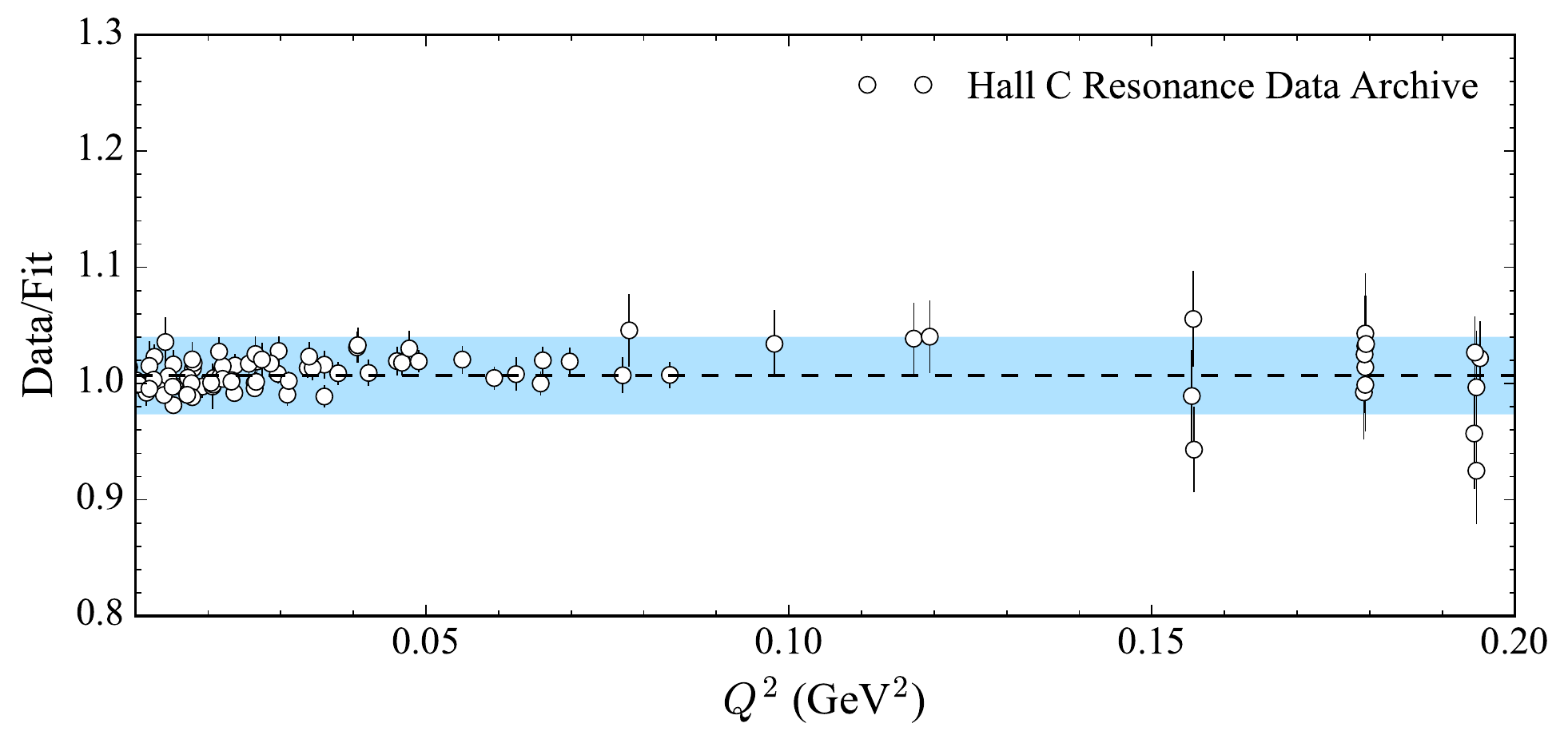}
\caption{Comparison between the Arrington proton form factor fit and existing data. The existing data is from Ref~\cite{HallCRes}.}
\label{FFComp}
\end{figure}

The first step in the radiative corrections analysis is removing the polarized elastic tail. The code ROSETAIL~\cite{RTAIL} was updated to include the proton elastic form factor parameterization of John Arrington~\cite{JAFit} via the Rosenbluth separation technique. The updated fit includes data over the range 0.007 $< Q^2 <$ 31 GeV$^2$. The quoted uncertainty on the new fit is 2\% on $G_E(Q^2)$ and 1\% on $G_M(Q^2)$ at the kinematics of E08-027. A comparison of the calculated elastic cross section, using the fit and equation~\eqref{RosieQ}, to measured data from the Hall C Resonance Data Archive~\cite{HallCRes} is in agreement with the quoted uncertainties and is shown in Figure~\ref{FFComp}. The dotted line is the average model-data ratio and the blue band is the standard deviation on that average at $\pm$ 3\%.




\begin{figure}[htp]
\centering     
\includegraphics[width=0.80\textwidth]{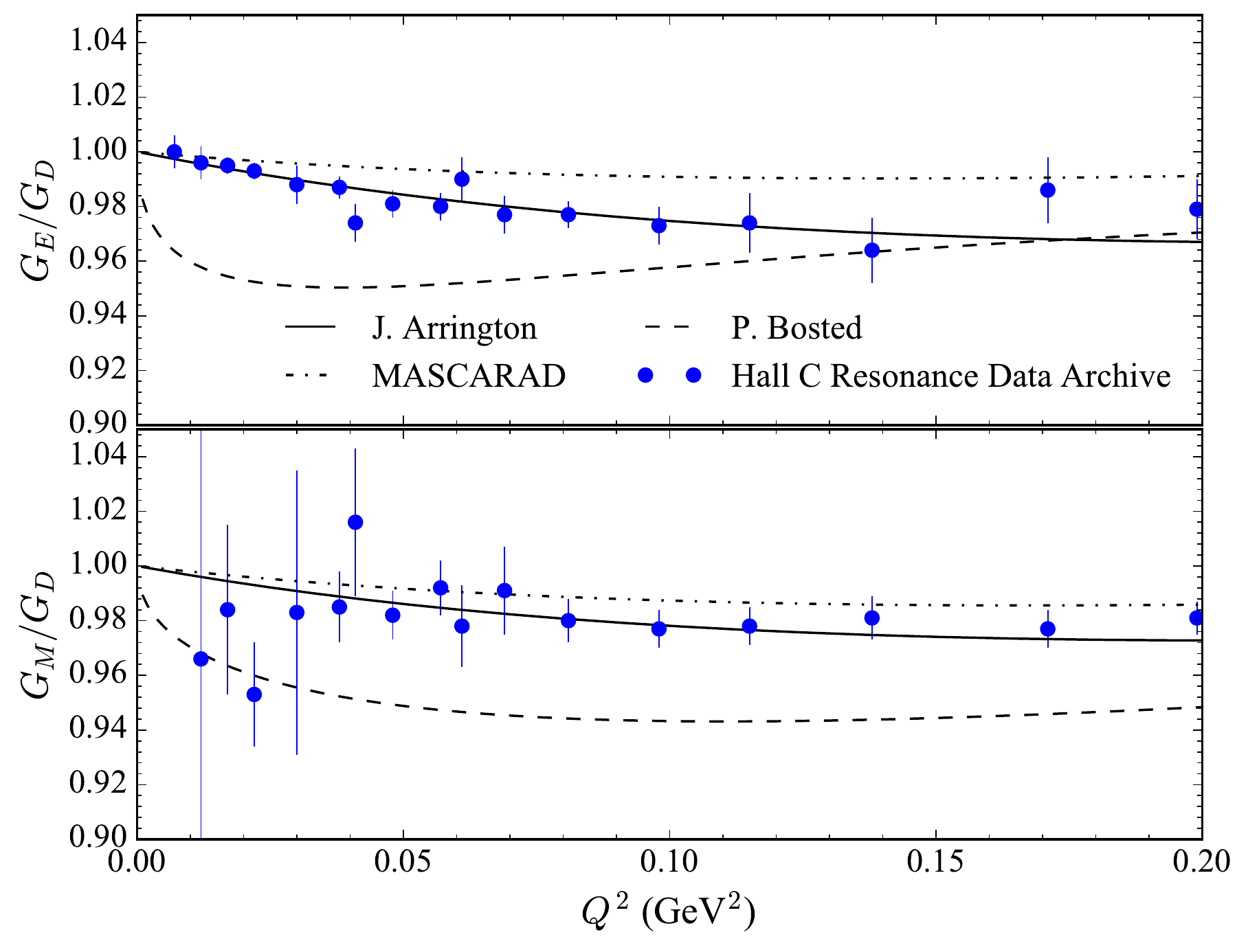}
\caption{Different proton form factor fits normalized to the dipole parameterization (see equation~\eqref{dipoleparam}). The best agreement is given by the fit of Ref~\cite{JAFit}.}
\label{FFFComp}
\end{figure}

Previous versions of ROSETAIL used a different fit for the unpolarized elastic cross section, MASCARAD elastic asymmetry, and POLRAD polarized elastic cross section. The proton form factor function call is now centralized and all subroutines use the same parametrization. The difference between the elastic form factor fits is shown in Figure~\ref{FFFComp}; the new Arrington fit does a much better job of reproducing the data at the E08-027 kinematics. Details on the Peter Bosted and MASCARAD fit are found in Ref~\cite{BostedGE} and Ref~\cite{Polrad4}, respectively. The polarized elastic cross sections were previously formed from the dipole parameterization
\begin{equation}
\label{dipoleparam}
G_E(Q^2),G_M(Q^2)/\mu_p = (1 + Q^2/ 0.0756)^{-2}\,,
\end{equation}
where the momentum transfer squared is given in units of GeV$^2$ and $\mu_p = 2.793$ is the proton magnetic dipole moment.

\begin{figure}[htp]
\centering     
\subfigure[$E_0$ = 2254 MeV 5T Transverse]{\label{fig:Tail2254}\includegraphics[width=.80\textwidth]{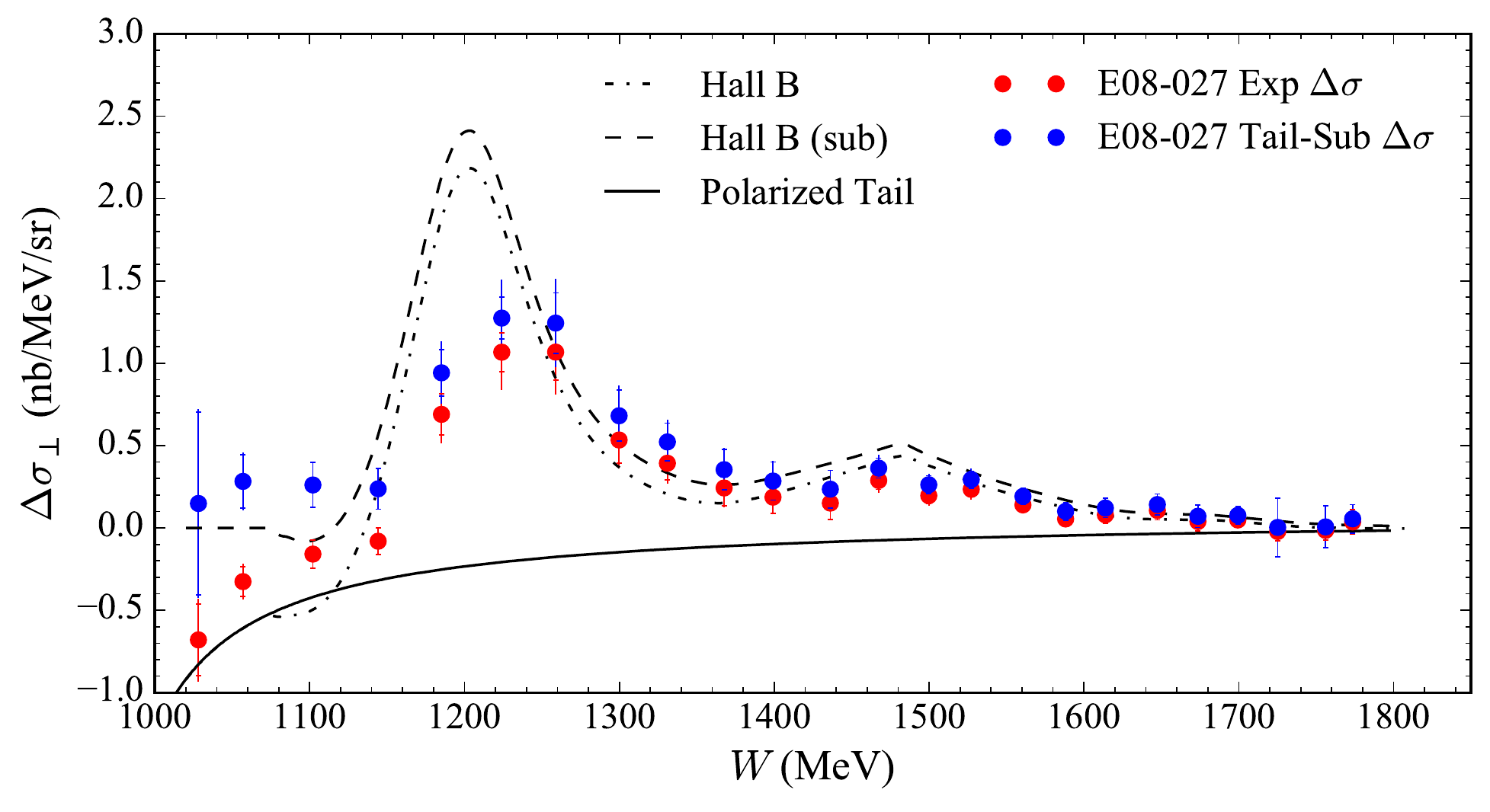}}
\qquad
\subfigure[$E_0$ = 3350 MeV 5T Transverse]{\label{fig:Tail3350}\includegraphics[width=.80\textwidth]{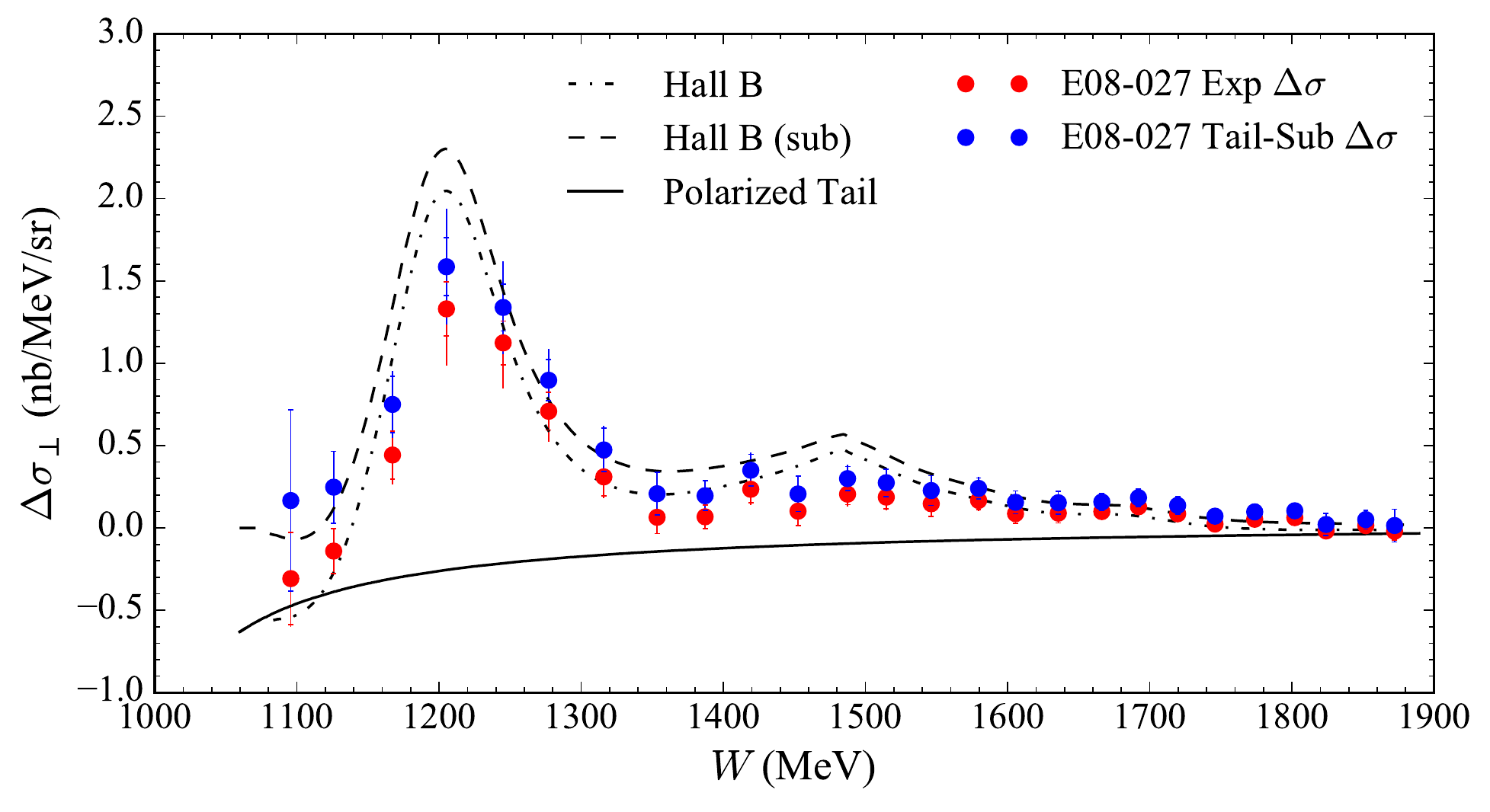}}
\qquad
\subfigure[$E_0$ = 2254 MeV 5T Longitudinal]{\label{fig:TailLong}\includegraphics[width=.80\textwidth]{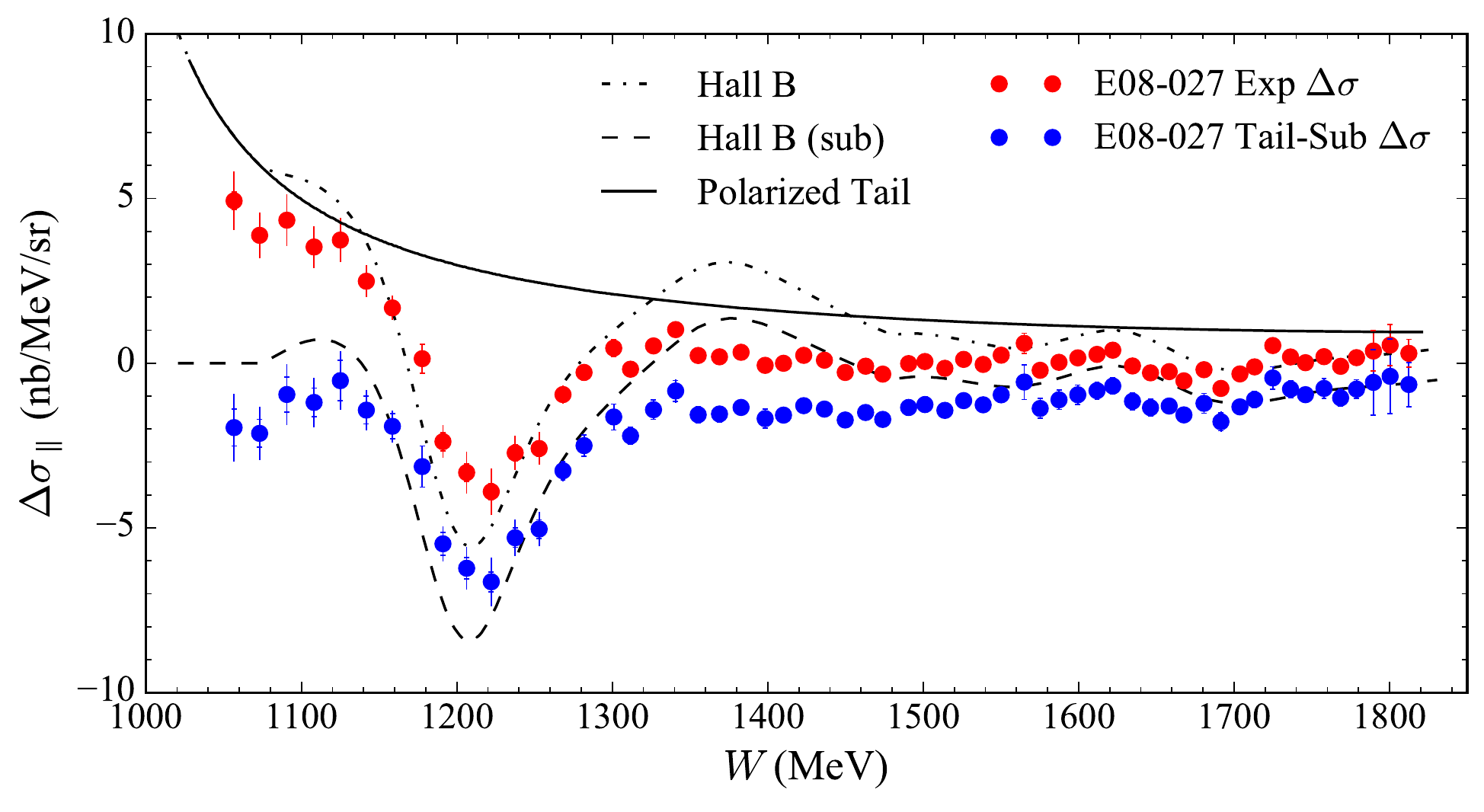}}
\caption{Polarized tail and tail subtraction for the 5 T settings.}
\label{TailSub}
\end{figure}

The calculated polarized elastic radiative tails and tail-subtracted cross section differences are shown in Figure~\ref{TailSub}. The inner error bars are statistical and the outer are the statistical and systematic errors added in quadrature. The statistical error on the tail subtracted results is modified to account for the unwanted tail events such that
\begin{equation}
\label{stat_sub}
\delta_{\mathrm{sub}} = {\bigg [}{\bigg(}\frac{1}{\delta_{\mathrm{exp}}}{\bigg)}^2 {\bigg(}\frac{\sigma_{\mathrm{sub}}}{\sigma_{\mathrm{tail}}+\sigma_{\mathrm{sub}}} {\bigg )} {\bigg ]}^{-\frac{1}{2}}\,,
\end{equation}  
where $\delta_{\mathrm{exp}}$ is the statistical error on the non-subtracted polarized cross section difference. 

The systematic errors follow the discussion from Chapter~\ref{ch:RC} and are listed in Table~\ref{TailSys}. Sensitivity to assumptions made in the theoretical calculations of the tails (such as the choice of fbar or soft-photon term) are summed linearly and reported as $\delta_\mathrm{int,ext}$, and are then added in quadrature with the uncertainty arising from the form factor parameterization, $\delta_{\mathrm{FF}}$, and angle reconstruction uncertainty, $\delta_{\mathrm{\theta}}$. The systematic error is calculated on a bin-by-bin basis to account for the $Q^2$ dependence of the proton elastic form factors, and is propagated with the systematic uncertainty on the experimental polarized cross section difference to get the final systematic error on the tail subtracted result. 
\begin{table}[htp]
\begin{center}
\begin{tabular}{ l c c  c  c  r}
\hline
  $E_0$ (MeV) &Config& $\delta_{\mathrm{int}}$ (\%)& $\delta_{\mathrm{ext}}$ (\%) & $\delta_{\mathrm{FF}}$  (\%) & $\delta_{\mathrm{\theta}}$  (\%)  \\ \hline
  3350 & 90$^{\circ}$& 1.0 & 1.5 &3.0$-$5.0&3.5\\
  2254 & 90$^{\circ}$  & 1.0 & 1.5 & 3.0$-$5.0&3.5\\
  2254 & 0$^{\circ}$ & 1.0 & 1.5 & 2.0&1.5 \\ \hline
\end{tabular}
\caption{\label{TailSys}Systematic errors for the polarized radiative elastic tail calculation at the 5 T settings.}
\end{center}
\end{table}

The sign of the elastic tail is confirmed by making sure the rise in the tail at low $W$ is matched with the data. Below the pion production threshold of $W$ = 1072 MeV, the data should consist of only polarized tail events and the resulting tail subtraction should lead to a zero result. This is seen in Figure~\ref{fig:Tail2254}. At the longitudinal setting, the subtraction is slightly non-zero but does trend to a constant off-set. It is possible that this is an artifact of using a model for the unpolarized cross section and should be investigated again when the cross section data is available. The kinematic coverage at the $E_0$ = 3350 MeV setting is not sufficient to see this effect. 

\subsection{Polarized Inelastic Radiative Corrections}
The existing inelastic radiative corrections code and interpolation and extrapolation procedure assume a constant scattering angle for the input spectra. This assumption is not valid for the E08-027 transverse target field configurations, and the existing code has been updated to reflect this. In both RADCOR and POLRAD, the input files now include a column for the scattering angle, which is read into an array of the same size and shape as the $W/\nu$ variables. The arrays are filled upon execution of the FORTRAN program.
\begin{figure}[htp]
\centering     
\subfigure[Radiating and unfolding at a constant scattering angle.]{\label{fig:RCConst}\includegraphics[width=.80\textwidth]{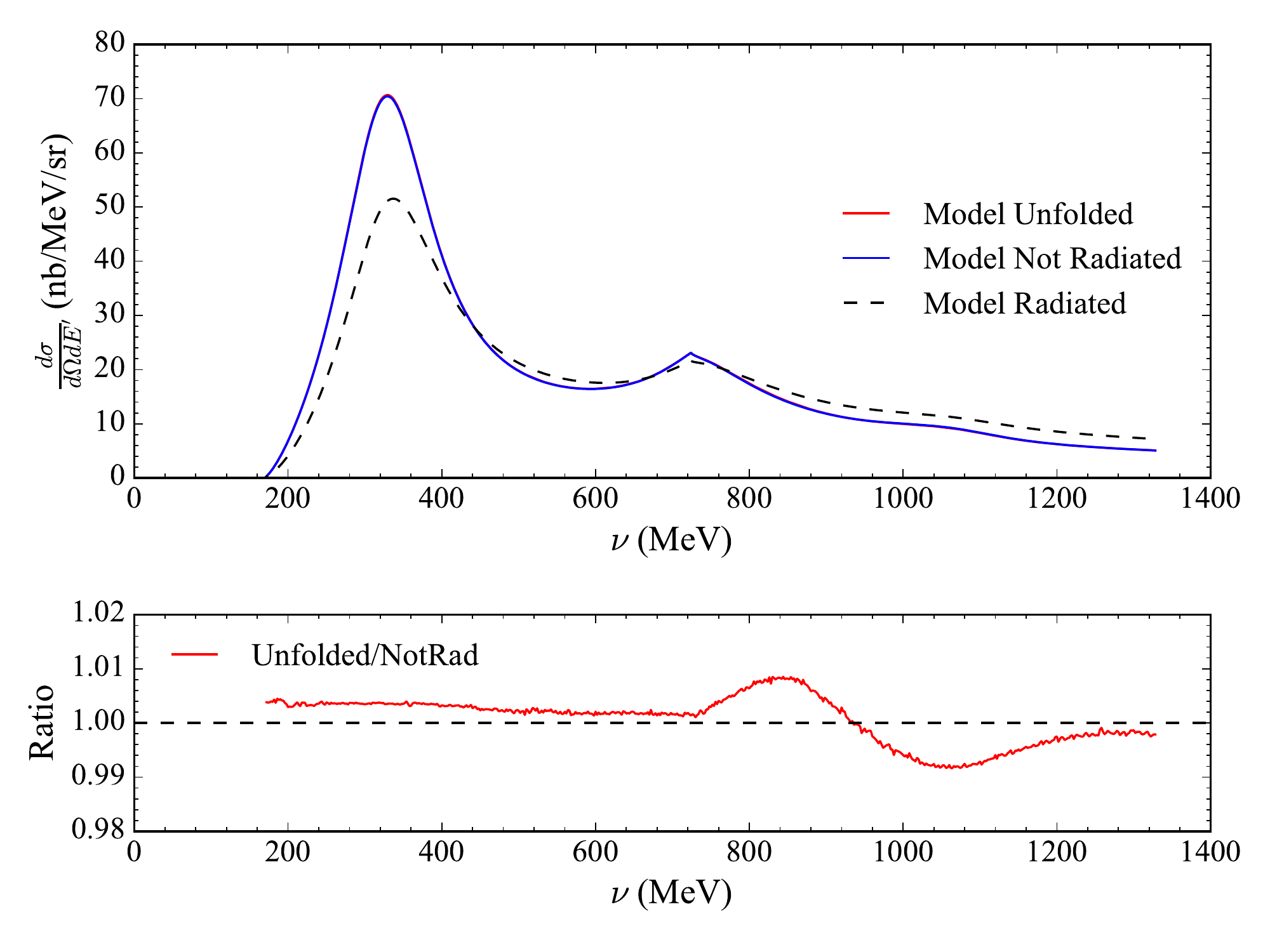}}
\qquad
\subfigure[Error introduced in the unfolding procedure from a changing angle.]{\label{fig:RCInFit}\includegraphics[width=.80\textwidth]{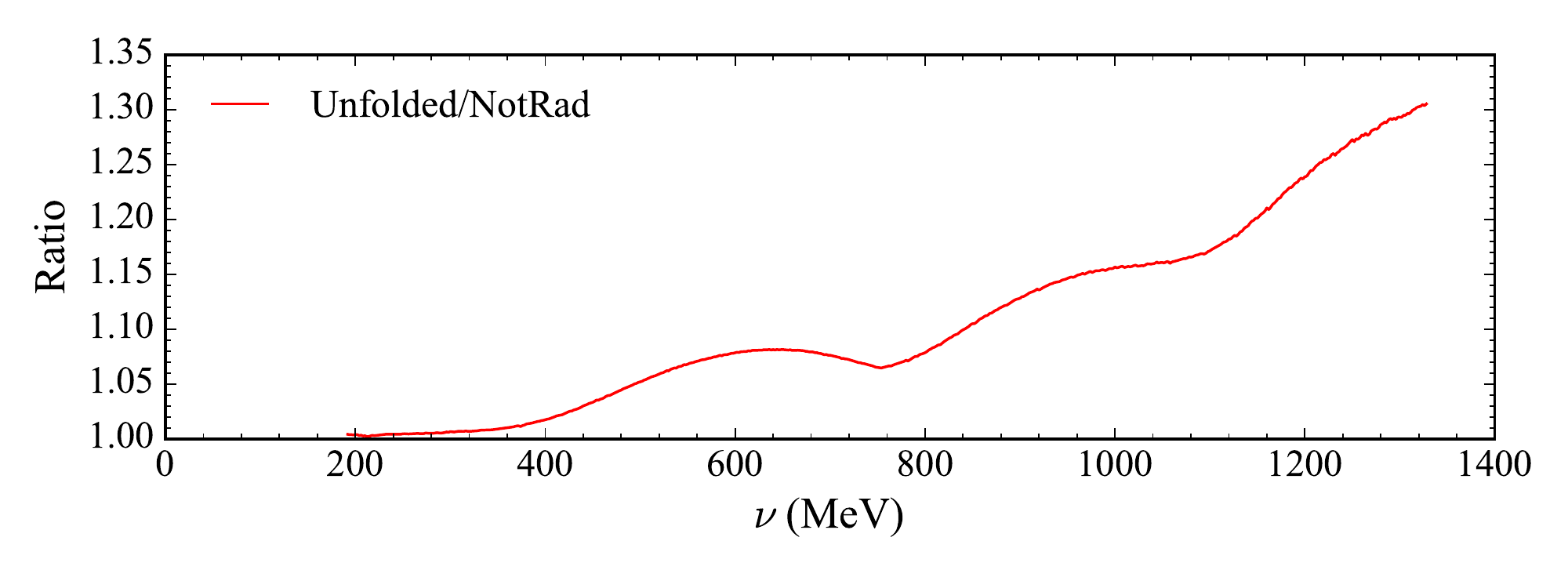}}
\qquad
\subfigure[Unfolding error if the (non-constant) angle dependence is the same for all spectra.]{\label{fig:RCSameFit}\includegraphics[width=.80\textwidth]{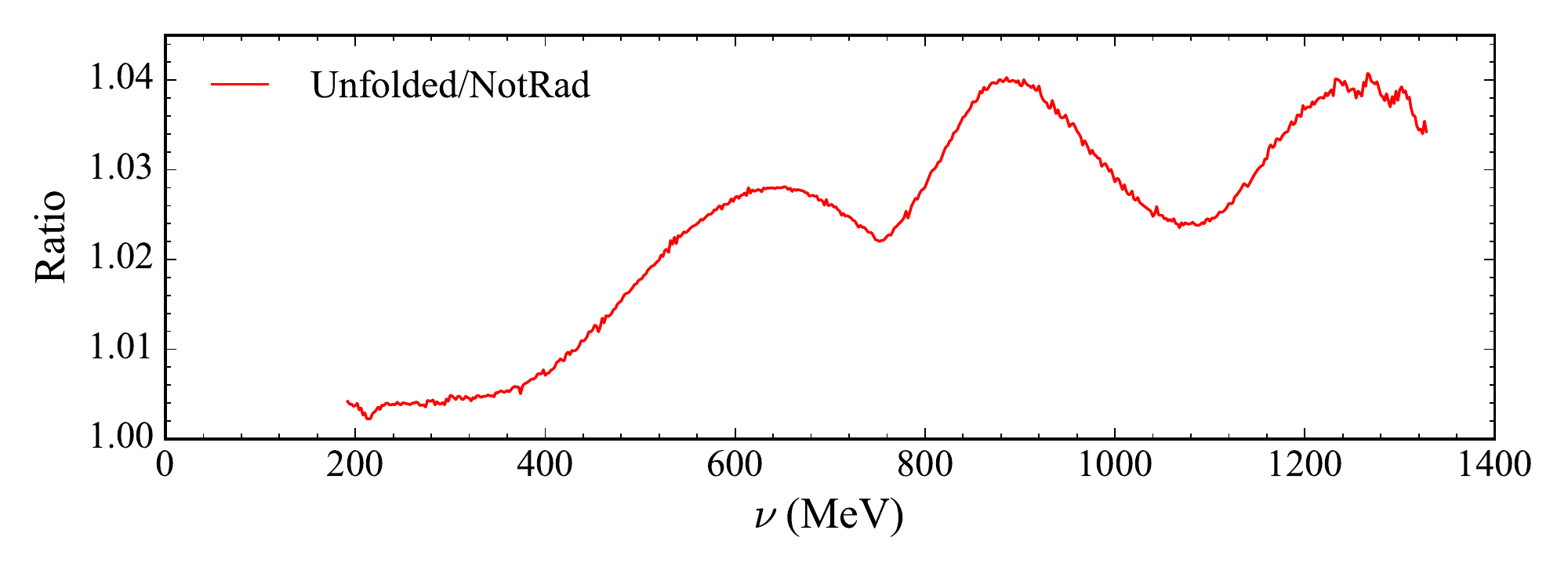}}
\caption{Effect of a non-constant scattering angle on the unfolding procedure in the inelastic radiative corrections.}
\label{RCProc}
\end{figure}
 
 Testing of the updated code revealed that the extrapolation procedure is very sensitive to the change in angle. This is seen in Figure~\ref{RCProc} for an unpolarized cross section model. For this study, the beam energy is set to $E_0$ = 2254 MeV and $\theta_{\mathrm{sc}}$ is set to a constant angle in order to mimic the longitudinal setting and also to an energy-dependent angle to simulate the transverse setting. Two spectra at $E_0$ = 1157 MeV and $E_0$ = 1711 MeV are inputs to aid the extrapolation, and their $W/\nu$ coverage is representative of the data taken during E08-027. The Born model run at the $E_0$ = 2254 MeV kinematics is radiated and put through the unfolding procedure of RADCOR. The result is compared directly to the original Born model and any differences between the two are a systematic effect of the inelastic radiative corrections process.

  At a constant angle, the extrapolation introduces a systematic difference on the order of 1\%. If the scattering angle is varied and the input spectra are allowed to have their own independent angular dependence\footnote{This is the case that corresponds to the measured data.}, then the extrapolation uncertainty is as large as 30\%. This uncertainty is reduced to approximately 4\% if the input spectra are generated with the same angle fit (henceforth called the ``same-fit" method) as the $E_0$ = 2254 MeV transverse data. For the remainder of the analysis the inelastic radiative corrections use the ``same-fit" method for the extrapolation spectra. The uncertainty from using an input model is less than the uncertainty from using data taken with a different angular dependence. Very similar results were found in a polarized cross section model study and also using the POLRAD code. 

\begin{figure}[htp]
\centering     
\subfigure[$E_0$ = 2254 MeV 5T Transverse]{\label{fig:Interp2254}\includegraphics[width=.80\textwidth]{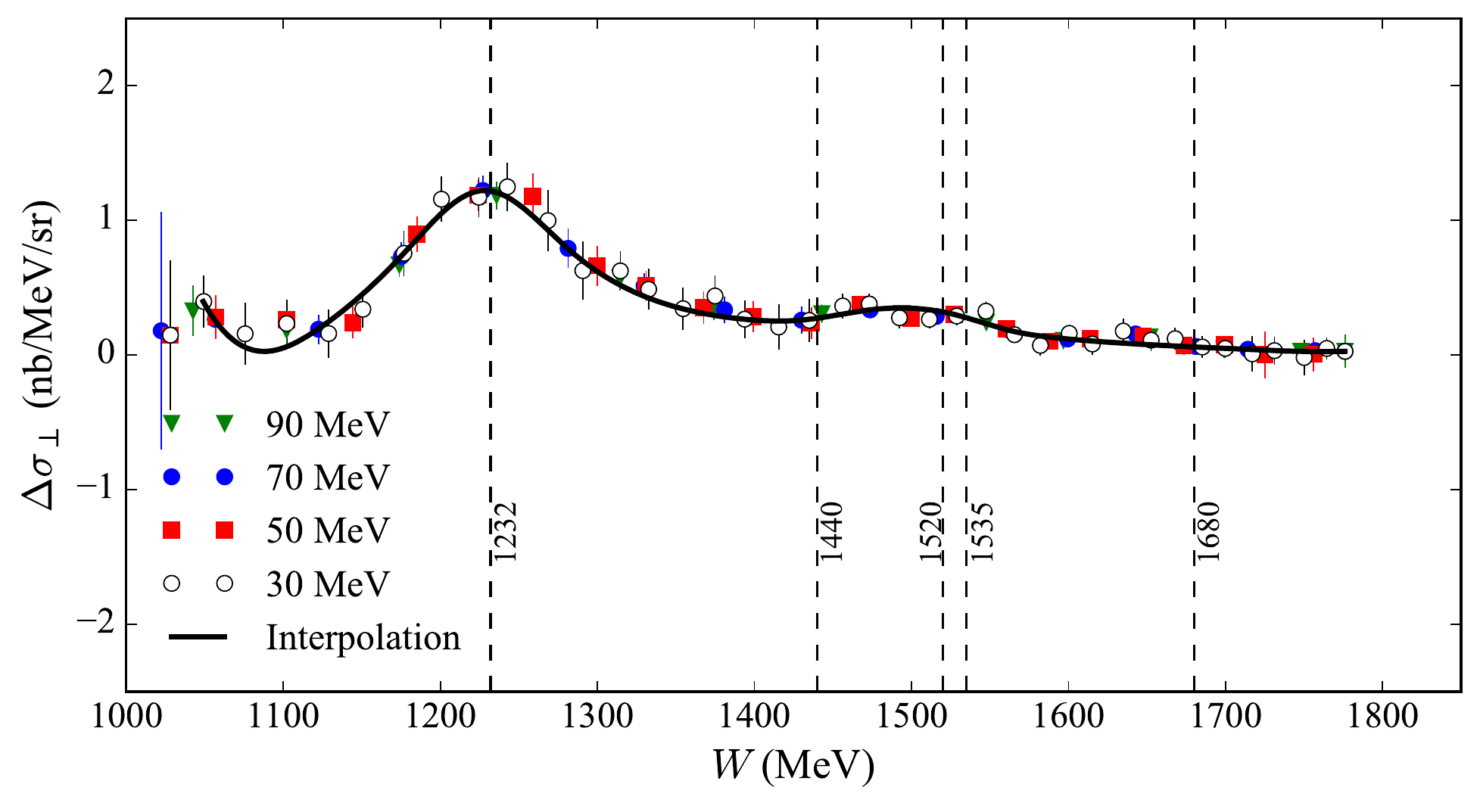}}
\qquad
\subfigure[$E_0$ = 3350 MeV 5T Transverse]{\label{fig:Interp3350}\includegraphics[width=.80\textwidth]{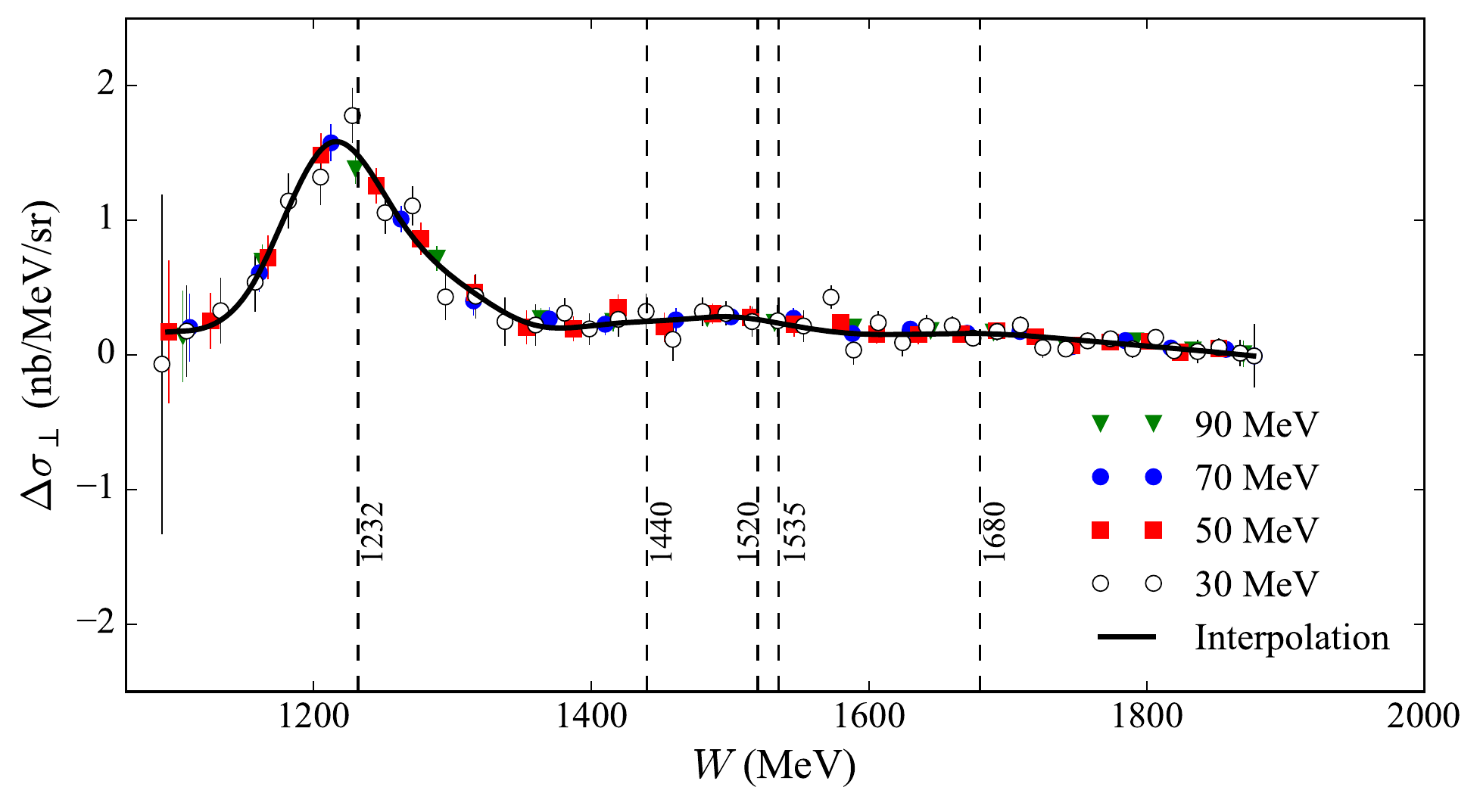}}
\qquad
\subfigure[$E_0$ = 2254 MeV 5T Longitudinal]{\label{fig:InterpLong}\includegraphics[width=.80\textwidth]{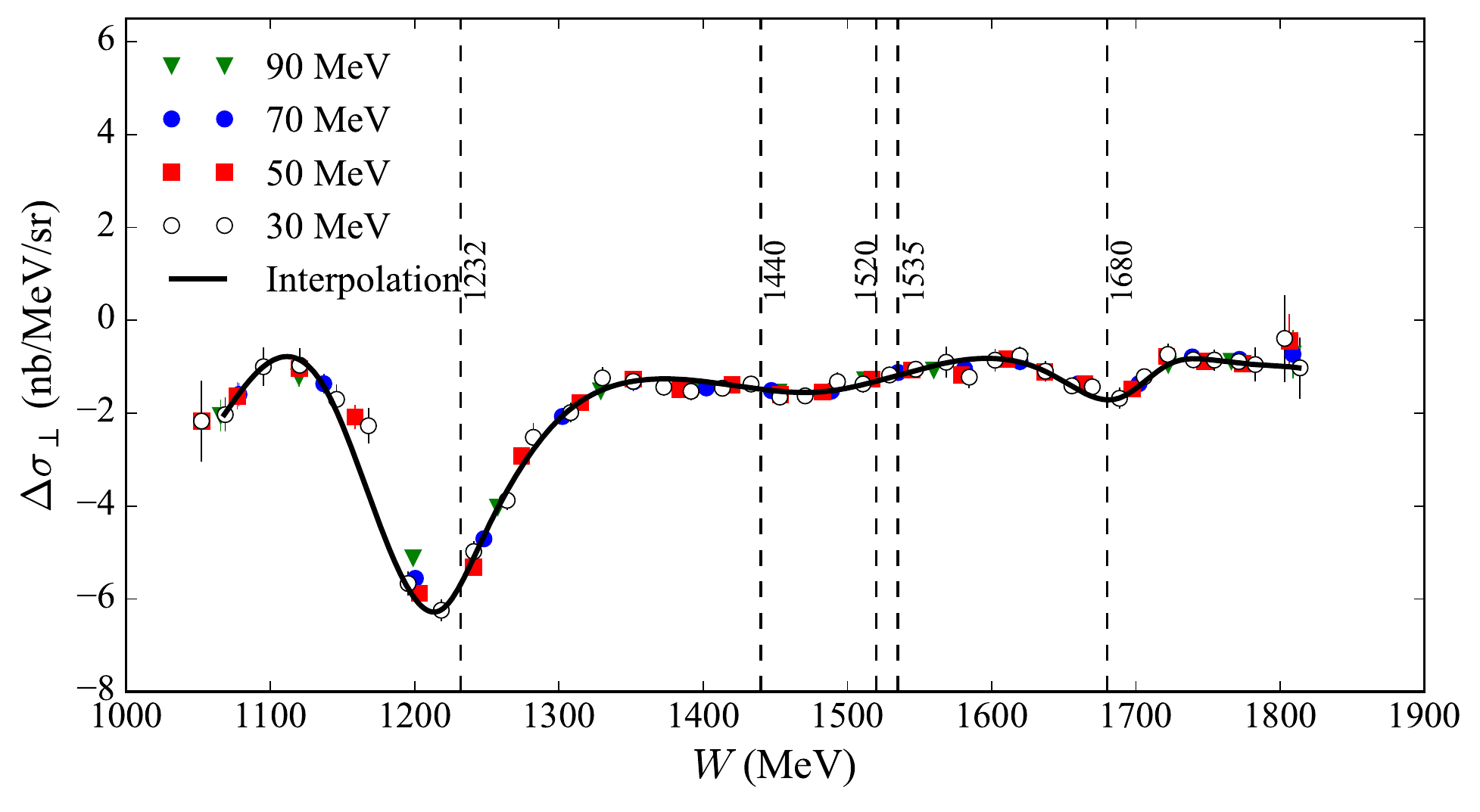}}
\caption{Smoothing of the tail-subtracted polarized cross sections differences.}
\label{RCInterp}
\end{figure}

The first step in the inelastic radiative corrections is to `smooth' the tail-subtracted cross sections in order to fill in small gaps in the data and to remove discontinuities and fluctuations that would be enhanced by the iteration process. The smoothing is completed using a cubic interpolation to the data in Python. The interpolation is done versus the invariant mass to more easily distinguish between resonance structures and statistical fluctuations. A super-set of the data is also created at 90, 70, 50 and 30 MeV binning and is used to locate the absolute maximum of the $\Delta$(1232) resonance peak. The cubic interpolation result is shown in Figure~\ref{RCInterp}, and only data points above the pion threshold are considered. 

The inelastic radiative corrections factor is applied to the data as 
\begin{equation}
\delta(\Delta \sigma^{\mathrm{RC}}_{\perp,\parallel}) = \Delta  \sigma^{\mathrm{Born\, interp.}}_{\perp,\parallel} -\Delta  \sigma^{\mathrm{sub.\, interp.}}_{\perp,\parallel}\,,
\end{equation}
where $\Delta  \sigma^{\mathrm{Born\, interp.}}_{\perp,\parallel}$ is the unfolded result after running POLRAD. The external radiative effects are corrected for first in RADCOR and those results are fed into POLRAD to produce the Born polarized cross section difference. This order of operations is not a one-to-one correspondence with what occurs experimentally: first the incoming electron experiences unpolarized radiative effects characterized by $t_b$, then the electron internally radiates bremsstrahlung photons with the polarized proton, and finally the scattered electrons undergoes external radiative effects from $t_a$.

 The order of the radiative corrections procedure used in the analysis introduces a possible systematic error. This systematic error is checked by comparing the radiatively corrected results of the following: running POLRAD first and then RADCOR, running RADCOR external before followed by POLRAD internal and then finally RADCOR external after, and running RADCOR and then POLRAD. The difference between these three options is minimal ($<$ 0.5\%) and the systematic error is neglected.

\begin{table}[htp]
\begin{center}
\begin{tabular}{ lc c c  c  r }
\hline
  $E_0$ (MeV) &Config& $\delta_{\mathrm{strag}}$ (\%)& $\delta_{\mathrm{soft}}$ (\%) & $\delta_{\mathrm{other}}$  (\%) \\ \hline
  3350 & 90$^{\circ}$& $<$ 0.5& 1.0$-$3.0 &0.8\\
  2254 & 90$^{\circ}$ & $<$ 1.0 & 1.0$-$2.5 & 0.8\\
  2254 & 0$^{\circ}$ & $<$ 1.0 & 1.0$-$2.5  & 0.8\\ \hline
\end{tabular}
\caption{\label{IRCSys}Theoretical systematic errors for the polarized inelastic radiative corrections at the 5 T settings.}
\end{center}
\end{table}

Some additional sources of systematic error in the inelastic radiative corrections carry over from the polarized elastic tail analysis. These include the evaluation of the cross sections under the first Born approximation, higher order virtual photons loops in fbar and the energy peaking approximation for the external radiative corrections. They are summed together as $\delta_{\mathrm{other}}$ in Table~\ref{IRCSys}. The systematic related to the choice of external straggling function and soft photon term is also included in Table~\ref{IRCSys}. These systematic errors are summed linearly on a bin-by-bin basis to give the theoretical systematic error in the inelastic radiative corrections.

\begin{figure}[htp]
\centering     
\includegraphics[width=0.80\textwidth]{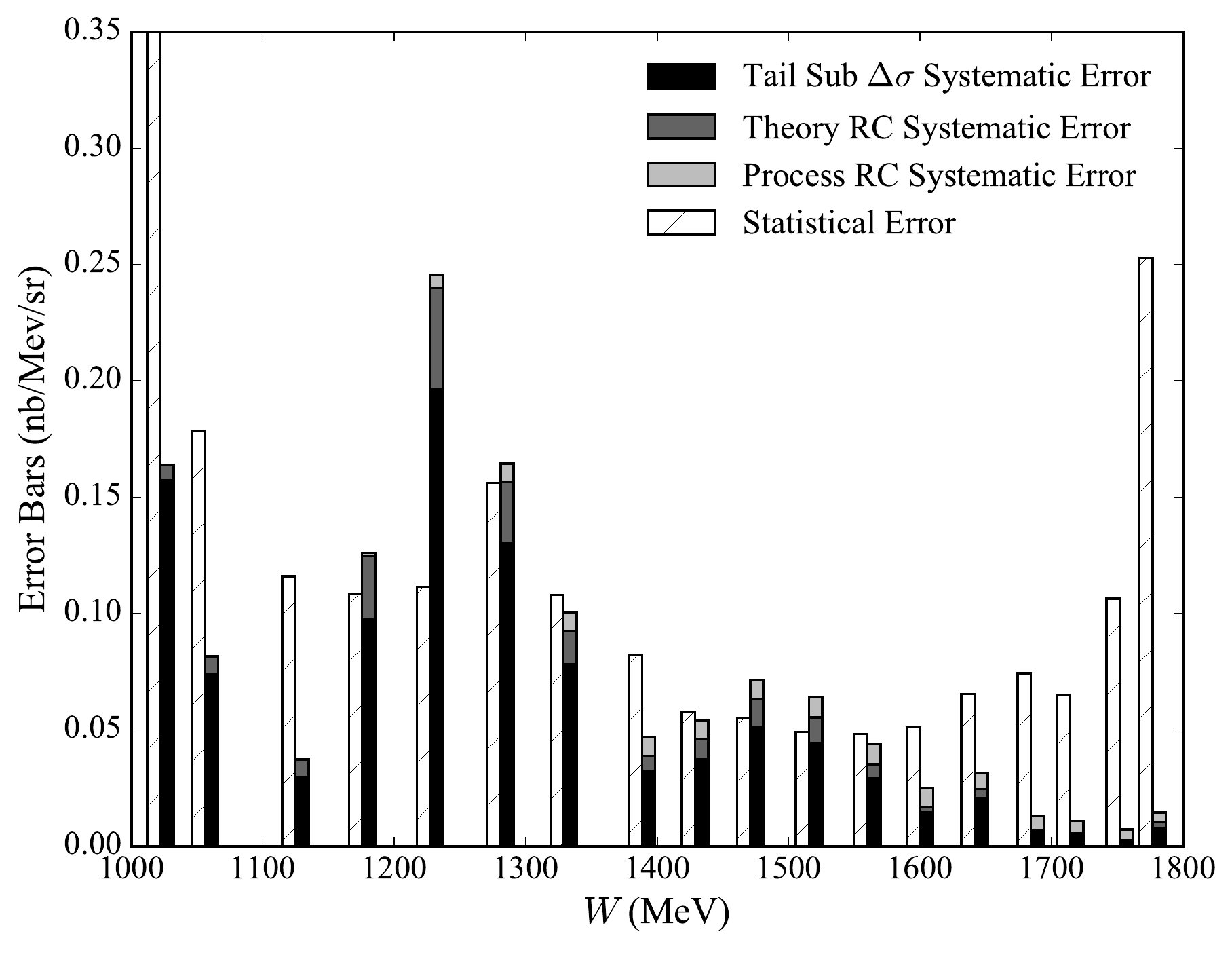}
\caption{Sources of error in the $E_0$ = 2254 MeV 5T transverse kinematic setting.}
\label{ErrorBarComp}
\end{figure}

\begin{figure}[htp]
\centering     
\subfigure[$E_0$ = 2254 MeV 5T Transverse]{\label{fig:Born2254}\includegraphics[width=.78\textwidth]{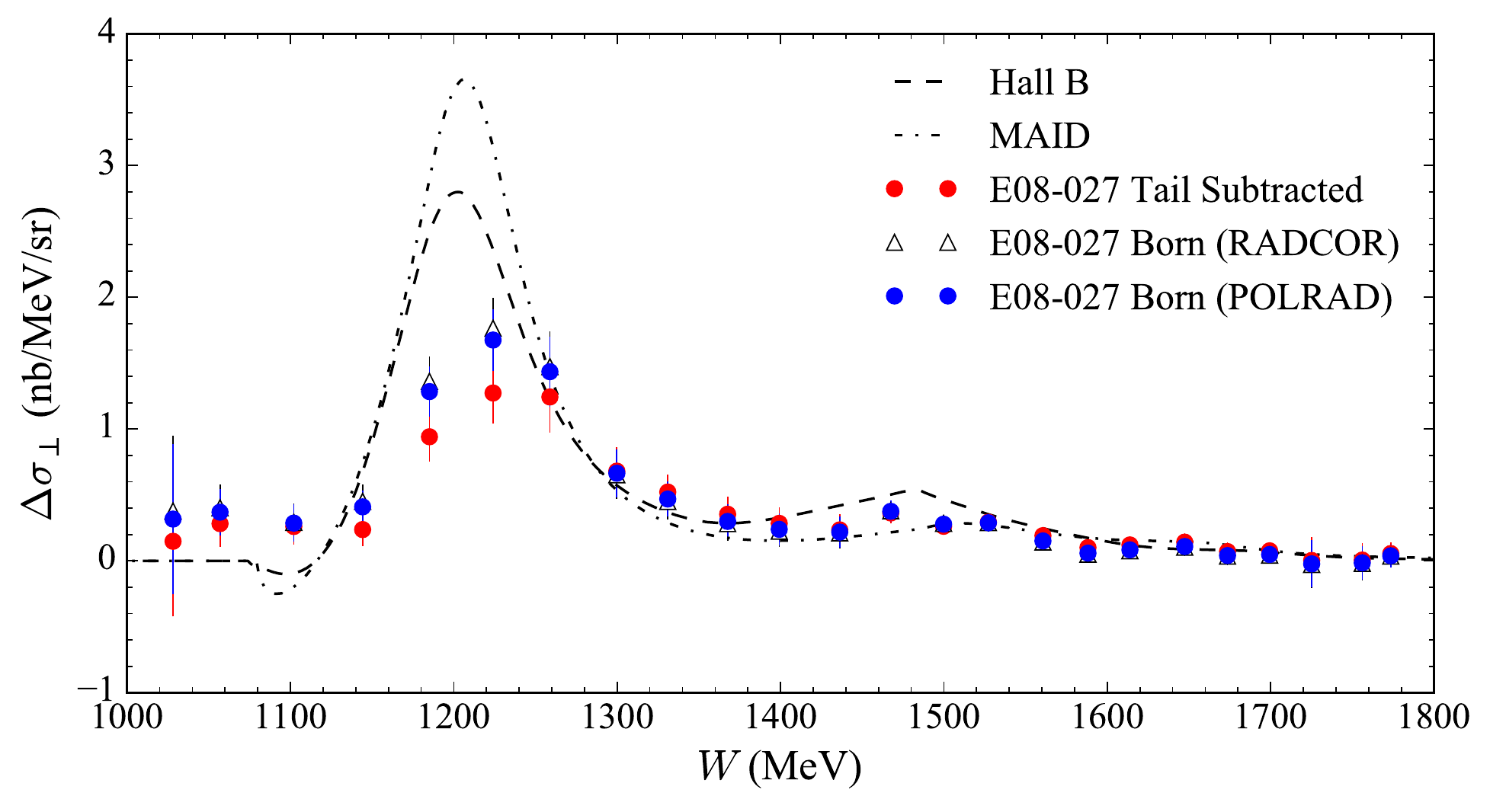}}
\qquad
\subfigure[$E_0$ = 3350 MeV 5T Transverse]{\label{fig:Born3350}\includegraphics[width=.78\textwidth]{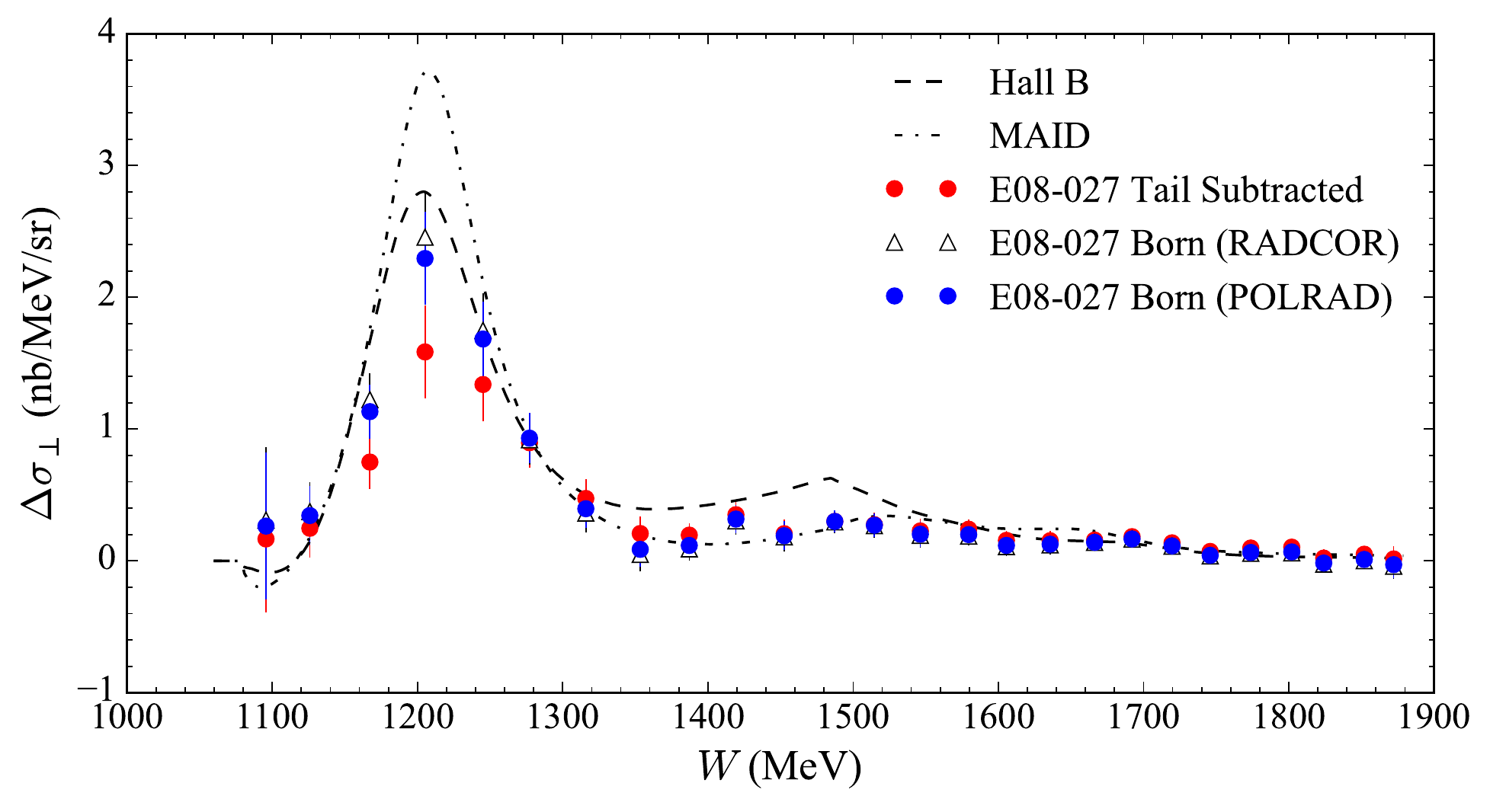}}
\qquad
\subfigure[$E_0$ = 2254 MeV 5T Longitudinal]{\label{fig:BornLong}\includegraphics[width=.78\textwidth]{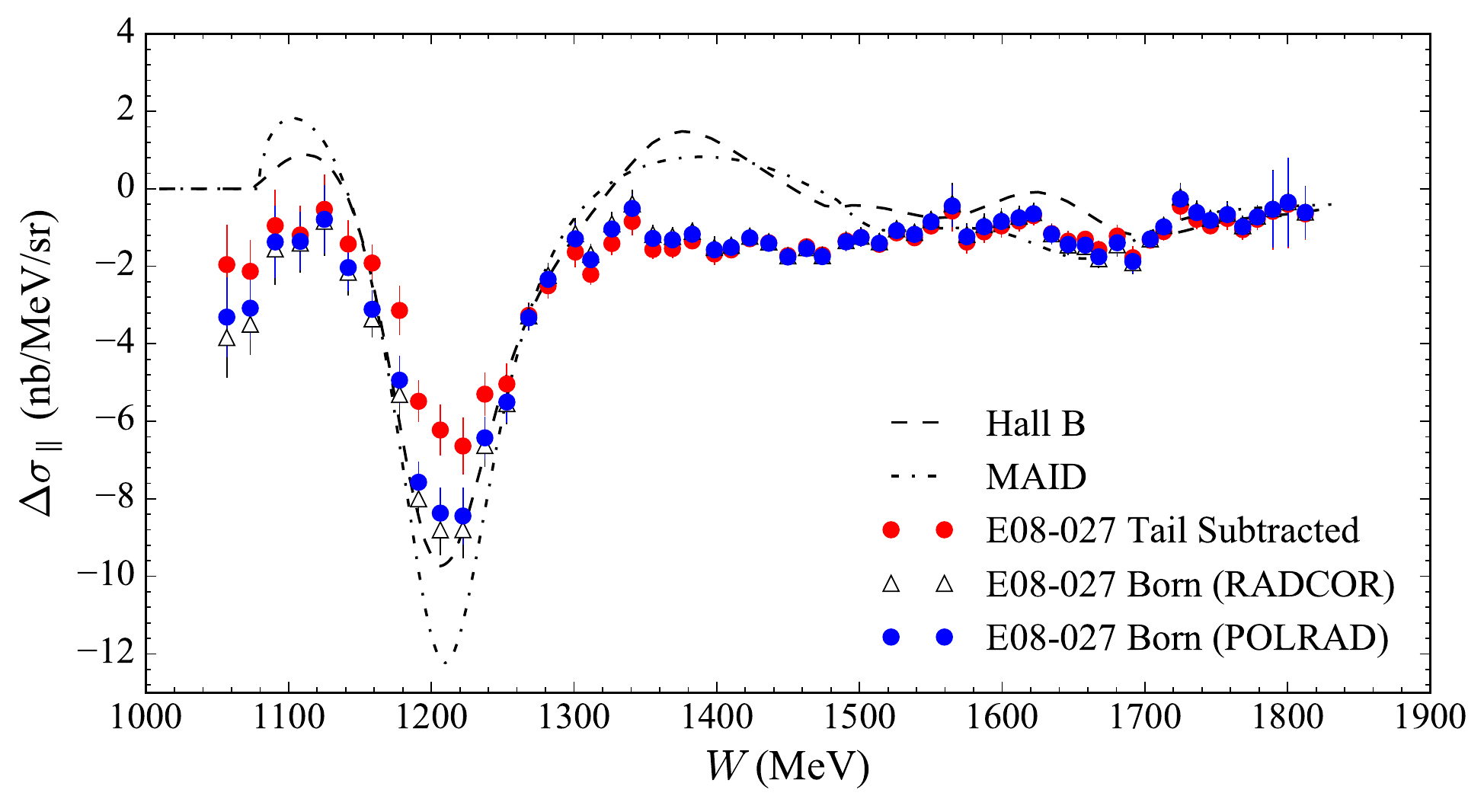}}
\caption{Born polarized cross section differences for the 5 T kinematic settings. The RADCOR formalism is an approximation; the POLRAD points represent the correct result.}
\label{BornDSResult}
\end{figure}

There is an additional source of systematic error in the inelastic radiative corrections and that is the unfolding procedure itself. The first component of this is shown in Figure~\ref{RCProc} and reflects how well the procedure can radiatively correct a known result, $i.e$ a radiated model cross section. The second component to the systematic error is how sensitive are the radiatively corrected results to the input model spectra. In this analysis, the input spectra are treated like pseudo-data and are inelastically radiated. This allows the unfolding procedure to modify the extrapolation spectra directly, which produces better iterative results\footnote{As opposed to providing the procedure static Born cross sections.}. The overall scale of the input model is varied as well as the specific model used to determine the systematic error. Varying the scale has a larger effect on the results, so the Hall B model is used for all the radiatively corrected results in this analysis. 

The size and variation of the scaling is determined by taking the average and standard deviation of the ratio between the inelastically radiated data and an inelastically radiated model. This results in a scale of 0.70 and a variation of $\pm$ 0.25. In general, the model and equivalent extrapolation have the largest effect at large $W/\nu$ and where the polarized cross section crosses zero. At the $\Delta$(1232) resonance this systematic is $<$ 0.5\% but increases to around 20\% with increasing $W/\nu$. Even at large $W/\nu$ the total inelastic radiative corrections systematic error (procedural combined with theoretical) is never larger than the statistical error bars. This is shown in Figure~\ref{ErrorBarComp} for $E_0$ = 2254 MeV 5 T transverse; the results are very similar for the other 5 T settings.

The Born polarized cross section differences are shown in Figure~\ref{BornDSResult}. The error bars shown represent the total error of the statistical and systematic errors added in quadrature. The total error is mostly statistics dominated as seen in Figure~\ref{ErrorBarComp}, except for around the $\Delta$(1232) resonance where the systematic errors take over.  The dominant systematic error is the 15\% assigned to the unpolarized model input, in conjunction with the error assigned to the uncertainty in the Mott cross section. Figure~\ref{BornDSResult} also shows the comparison between using a combination of RADCOR and POLRAD (labeled POLRAD) and just RADCOR (labeled RADCOR) for the external and internal inelastic radiative corrections. The difference between the two formalisms is around 5\%  and so the correct method is used.

\section{Spin Structure Function Extraction}
The spin structure functions are extracted from the Born polarized cross section differences following the analysis of Chapter~\ref{ch:epScatter} and equations~\eqref{DiffXSPara} and~\eqref{DiffXSPerp}. The result is
\begin{align}
g_1(x,Q^2)  &= \frac{MQ^2}{4\alpha}\frac{y}{(1-y)(2-y)}{\bigg[}\Delta \sigma_{\parallel} + \mathrm{tan}\frac{\theta}{2}\Delta \sigma_{\perp}{\bigg]}\,,\\
g_2(x,Q^2)  &= \frac{MQ^2}{4\alpha}\frac{y^2}{2(1-y)(2-y)}{\bigg[}-\Delta \sigma_{\parallel} + \frac{1+(1-y)\mathrm{cos}\theta}{(1-y)\mathrm{sin}\theta}\Delta \sigma_{\perp}{\bigg]}\,,
\end{align}
where $y=\nu/E$. At the E08-027 kinematics, the parallel cross section difference dominates the contribution to $g_1(x,Q^2)$ and the perpendicular cross section difference is the main contributor to $g_2(x,Q^2)$.  For $g_1(x,Q^2)$, the maximum contribution from the perpendicular polarized cross section is 5\% at the longitudinal kinematic setting scattering angle of 5.77$^{\circ}$. Taking into consideration the change in scattering angle and kinematic dependence of $y$, the maximum contribution to $g_2(x,Q^2)$ from the unmeasured parallel polarized cross section is 6\% (5\%) at $E_0$ = 2254 MeV (3350 MeV). The quoted contributions assume that the perpendicular and parallel polarized cross sections are roughly equal in magnitude.

 In this analysis, the unmeasured component to the spin structure functions is provided by a polarized model, and so as to minimize the model uncertainty, the structure functions are written in terms of the measured polarized cross section and the opposing spin structure function such that
\begin{align}
g_1(x,Q^2) & = K_1\bigg{[}\Delta \sigma_{\parallel}{\bigg(}1+\frac{1}{K_2}\mathrm{tan}\frac{\theta}{2}{\bigg)}{\bigg]}  + \frac{2g_2(x,Q^2)}{K_2y}\mathrm{tan}\frac{\theta}{2}\,,\\
\label{g2fromg1}
g_2(x,Q^2) & = \frac{K_1y}{2}\bigg{[}\Delta \sigma_{\perp}{\bigg(}K_2+\mathrm{tan}\frac{\theta}{2}{\bigg)}{\bigg]} - \frac{g_1(x,Q^2)y}{2}\,,
\end{align}
where the kinematic terms, $K_1$ and $K_2$, are defined as
\begin{align}
K_1 & = \frac{MQ^2}{4\alpha}\frac{y}{(1-y)(2-y)}\,,\\
K_2 & = \frac{1+(1-y)\mathrm{cos}\theta}{(1-y)\mathrm{sin}\theta}\,. 
\end{align}

\begin{figure}[htp]
\centering     
\subfigure[CLAS EG1b : $Q^2$ = 0.084 GeV$^2$]{\label{fig:eg1b1}\includegraphics[width=.47\textwidth]{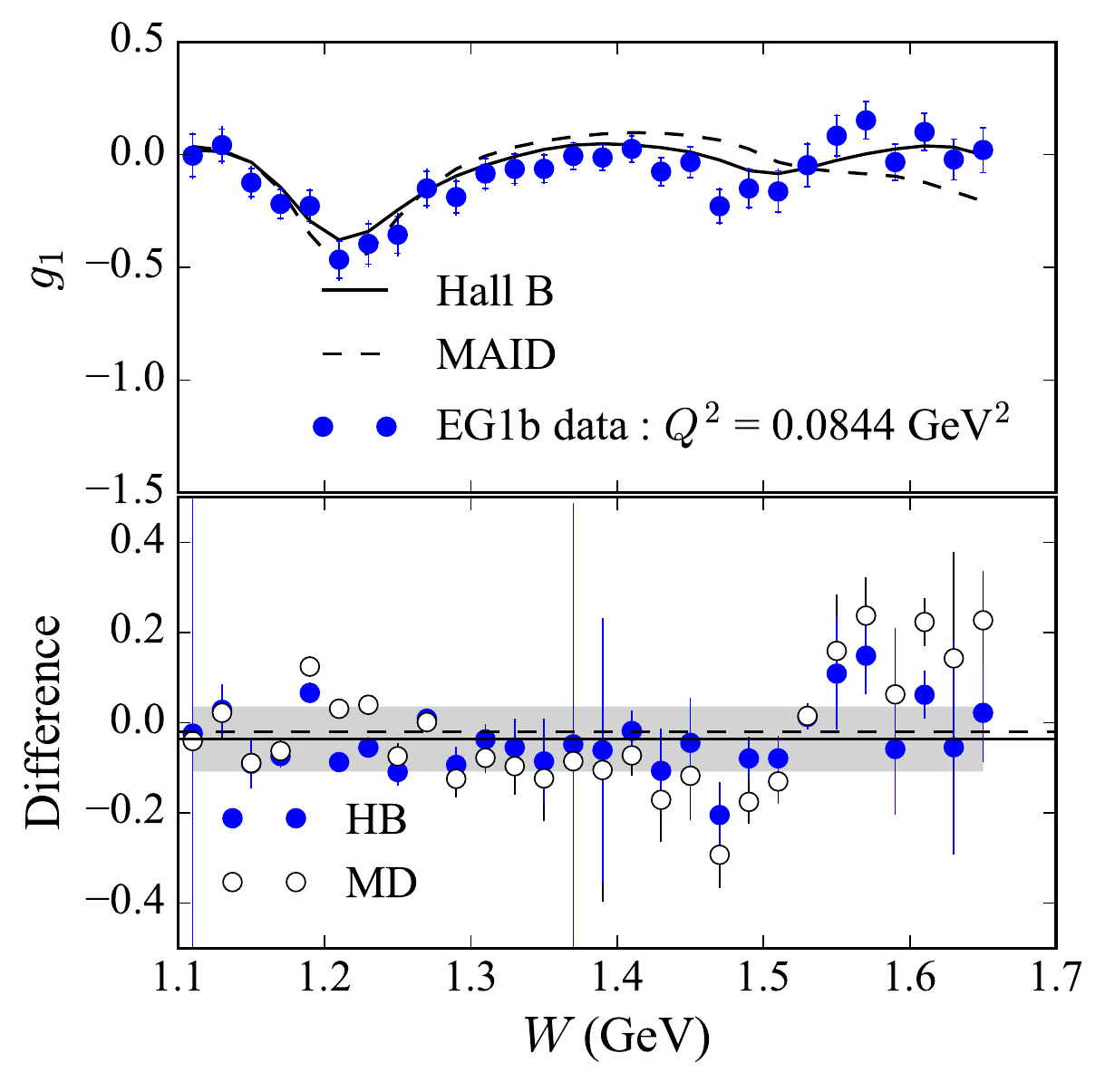}}
\qquad
\subfigure[CLAS EG1b : $Q^2$ = 0.101 GeV$^2$]{\label{fig:eg1b2}\includegraphics[width=.47\textwidth]{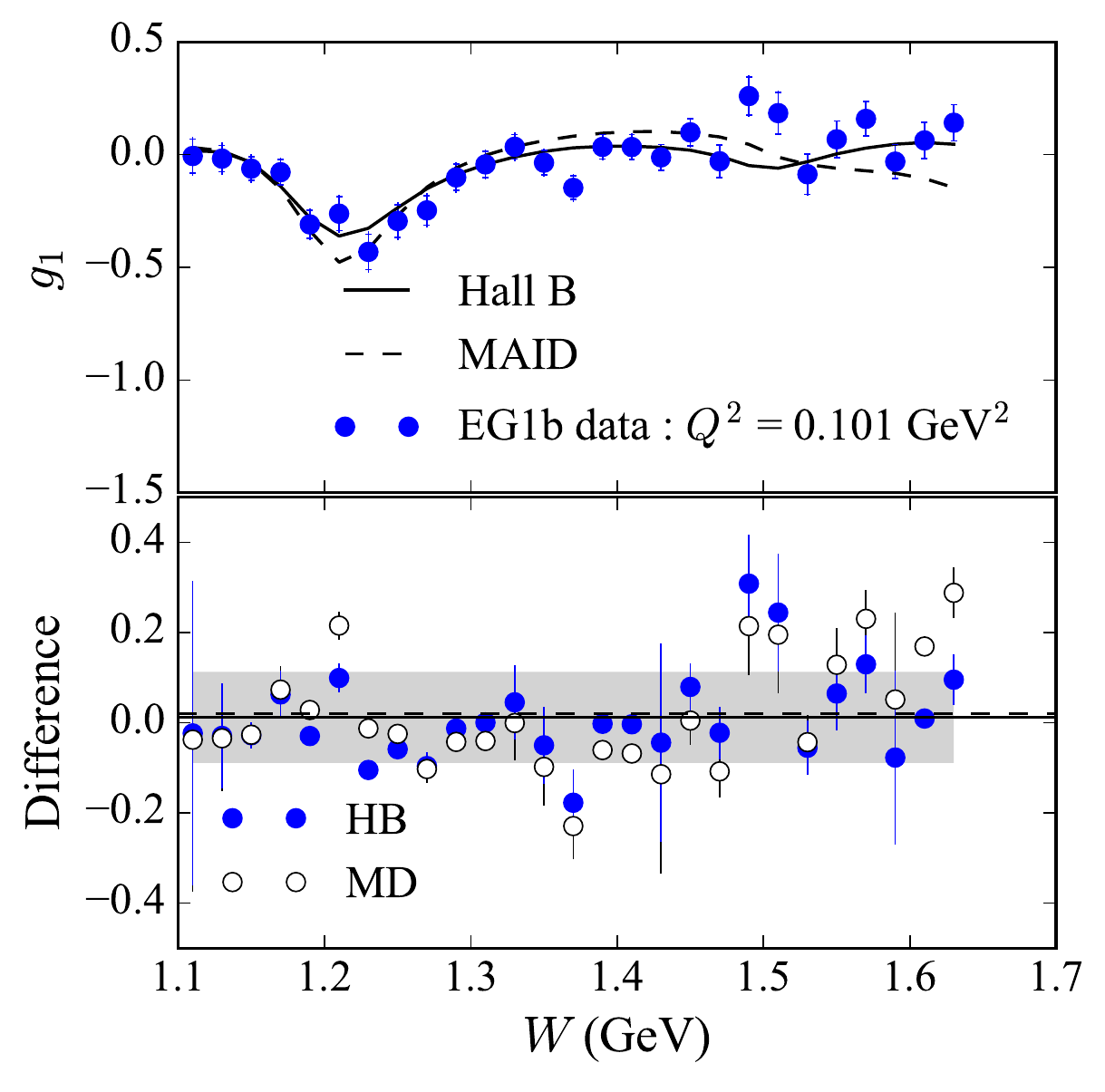}}
\qquad
\subfigure[CLAS EG1b : $Q^2$ = 0.120 GeV$^2$]{\label{fig:eg1b3}\includegraphics[width=.47\textwidth]{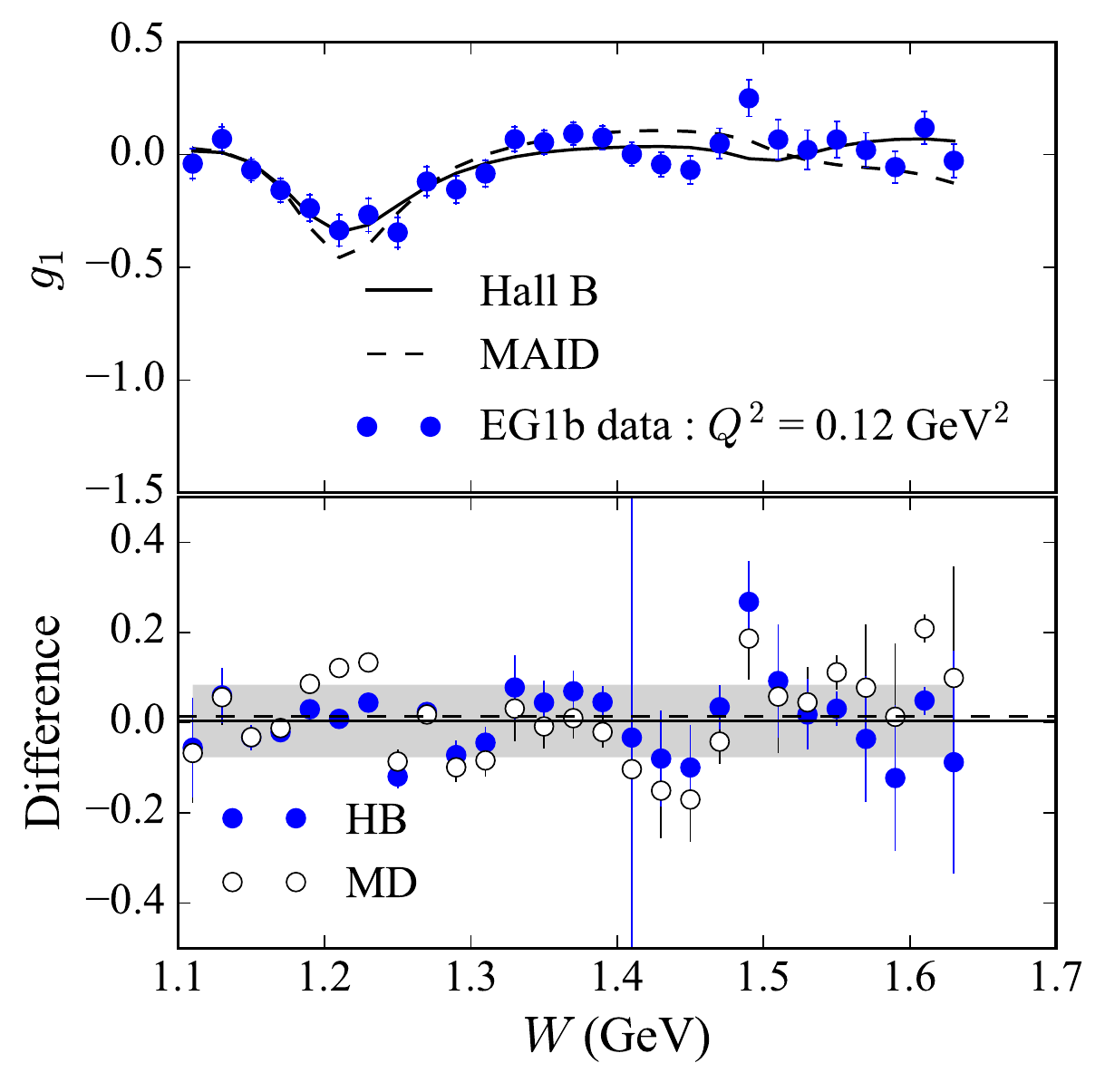}}
\qquad
\subfigure[CLAS EG1b : $Q^2$ = 0.144 GeV$^2$]{\label{fig:eg1b3}\includegraphics[width=.47\textwidth]{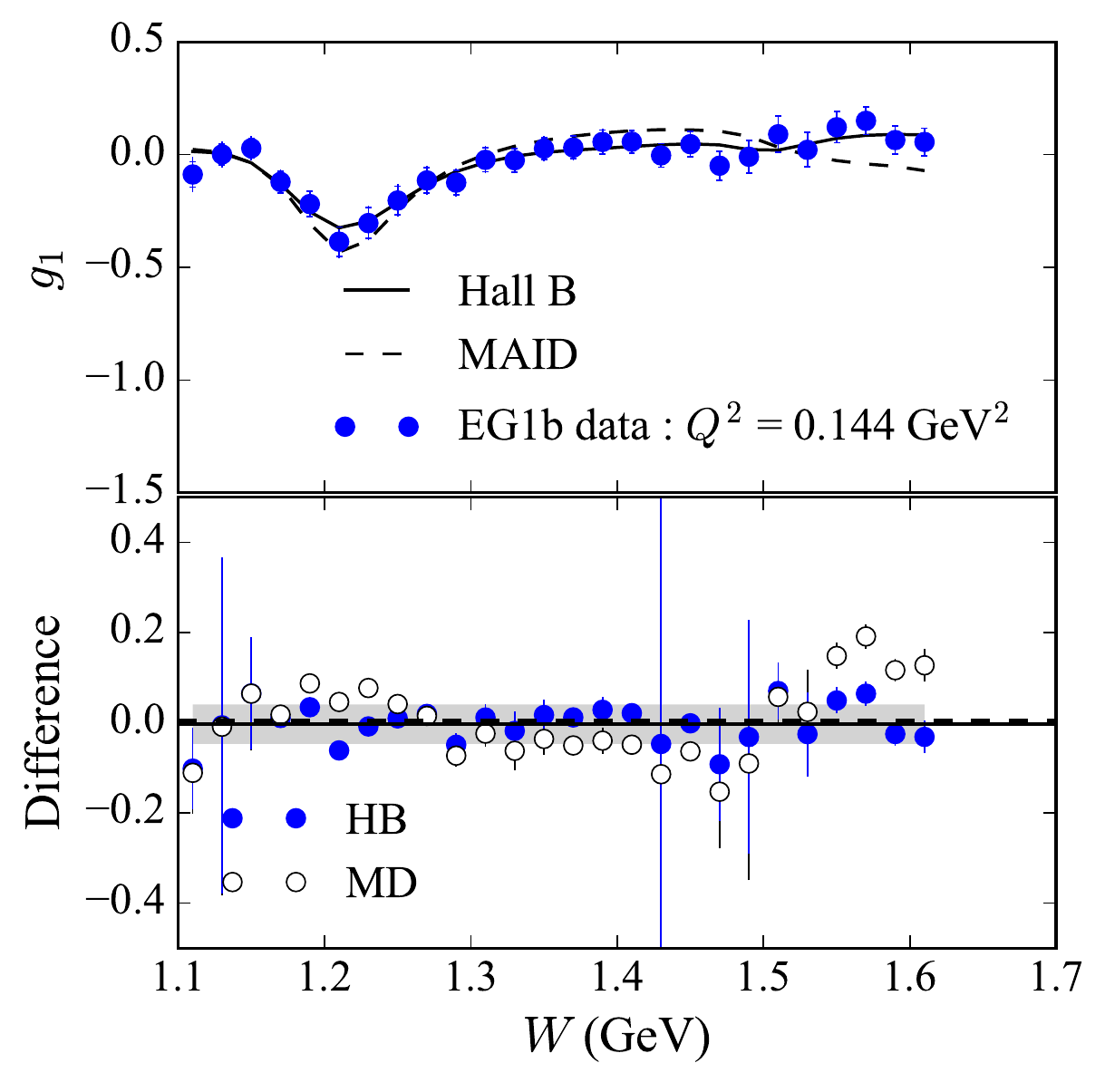}}
\caption{Comparison between the CLAS EG1b data and polarized models for $g_1(x,Q^2)$ at the E08-027 5 T transverse kinematics. The standard deviation of the model difference (grey band) is approximately 0.04. Across the four settings shown, $<$reduced $\chi^2$$>$ = 1.1 for the Hall B model and $<$reduced $\chi^2$$>$ = 2.5 for MAID, indicating the Hall B model is a better fit to the CLAS EG1b data. }
\label{HallBcompg1}
\end{figure}

For the $g_1(x,Q^2)$ component of $g_2(x,Q^2)$ in equation~\eqref{g2fromg1}, the CLAS EG1b~\cite{HallBRes} data set provides an additional check on the model uncertainty. The EG1b experiment ran in Hall B at Jefferson Lab and extracted $g_1(x,Q^2)$ for the proton and deuteron covering 0.05 $<$ $Q^2$ $<$ 5.0 GeV$^2$ and $W$ $<$ 3 GeV. The range in $Q^2$ values for the E08-027 5 T transverse kinematic settings are 0.086 $<$ $Q^2$ $<$ 0.104 GeV$^2$ and 0.124 $<$ $Q^2$ $<$ 0.133 GeV$^2$ for the $E_0$ = 2254 MeV and $E_0$ = 3350 MeV settings, respectively. The comparison between EG1b data and Hall B and MAID polarized models in the E08-027 kinematic region is shown in Figure~\ref{HallBcompg1}. The error weighted average difference between the data and model is very close to zero (solid and dashed lines for Hall B and MAID, respectively), and so the systematic error is determined by comparing extracted values of $g_2(x,Q^2)$ with the Hall B model shifted by the standard deviation of the average (grey band).

\begin{figure}[htp]
\centering     
\subfigure[$E_0$ = 2254 MeV 5T Transverse]{\label{fig:SSF2254}\includegraphics[width=.80\textwidth]{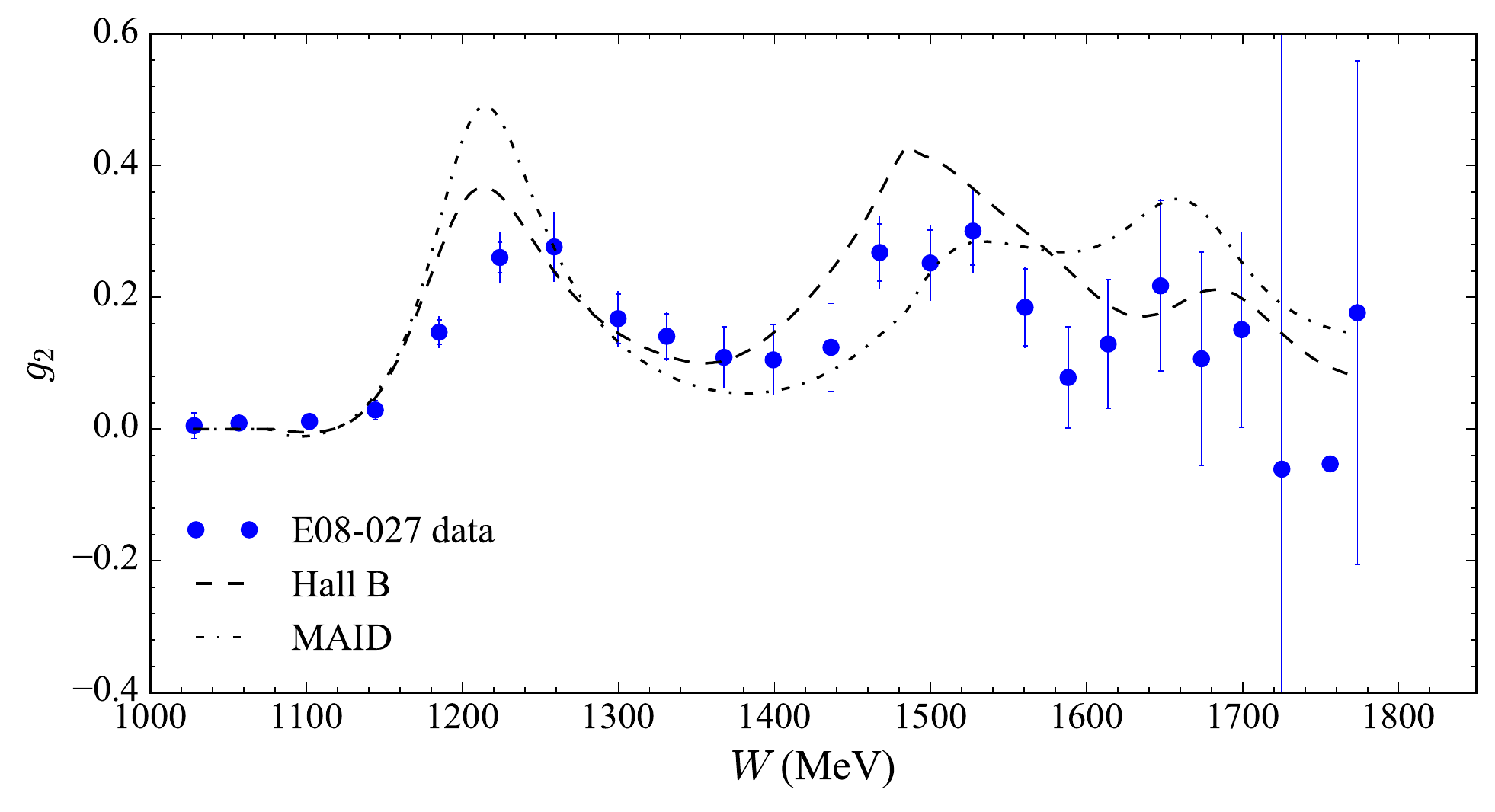}}
\qquad
\subfigure[$E_0$ = 3350 MeV 5T Transverse]{\label{fig:SSF3350}\includegraphics[width=.80\textwidth]{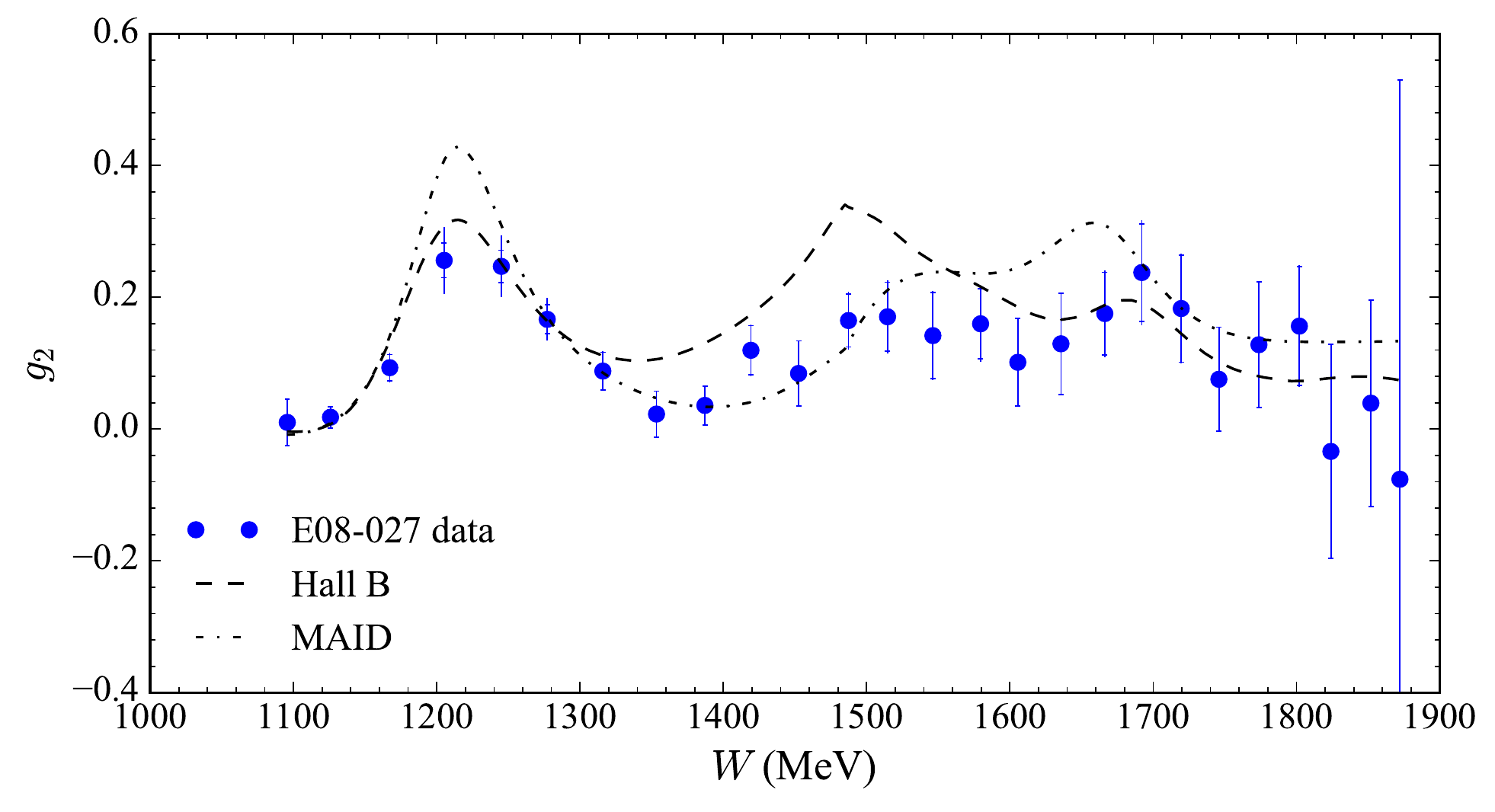}}
\qquad
\subfigure[$E_0$ = 2254 MeV 5T Longitudinal]{\label{fig:SSFLong}\includegraphics[width=.80\textwidth]{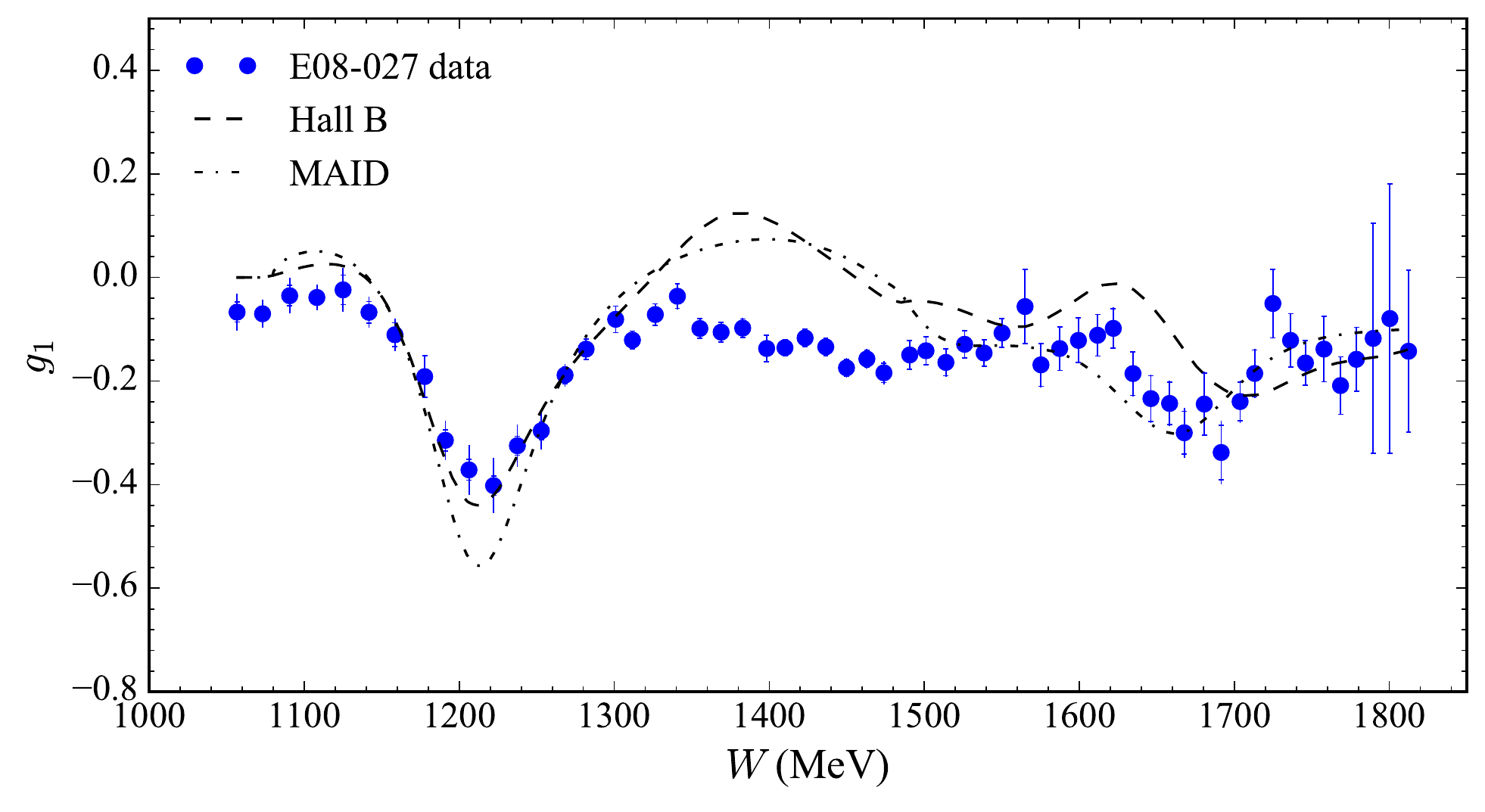}}
\caption{Born spin structure functions for the 5 T kinematic settings.}
\label{SSFResult}
\end{figure}

The results for the E08-027 structure functions at the 5 T kinematic settings are shown in Figure~\ref{SSFResult}. The inner error bars are statistical and the outer are the total error: systematic error added in quadrature with the statistical uncertainty. The statistical uncertainty in the structure functions is propagated from the polarized cross section differences via
\begin{equation}
\delta_{g_{1,2}} = \sqrt{{\bigg(}\frac{\partial g_{1,2}}{\partial \Delta\sigma_{\parallel,\perp}}{\bigg)}^2 \delta^2_{\Delta\sigma_{\parallel,\perp}} + {\bigg(}\frac{\partial g_{1,2}}{\partial g_{2,1}}{\bigg)}^2 \delta^2_{g_{2,1}}  }\,,
\end{equation}
where the statistical uncertainty on the model provided structure function is zero. At the longitudinal setting, the additional systematic error from the model input is the standard deviation of $g_1(x,Q^2)$ calculated using the Hall B and MAID models, and also assuming $g_2(x,Q^2)$ is zero. This standard deviation is added linearly to the systematic error propagated from the parallel polarized cross section difference.
\section{Moments and Sum Rules}
\label{SumRulesMoments}

The data is measured at a constant beam  and varied scattered electron energies. This results in a non-constant $Q^2$ across the $W/\nu$ spectrum of a given incident energy. The calculated moments in Chapter~\ref{Section:Moments} are integrations of the spin structure functions at constant momentum transfer, so the data must first be corrected to a constant $Q^2$ before evaluating the physics quantities.

\subsection{Model Evolving to Constant Momentum Transfer}
\label{EvolveMeth}

Typically the adjustment to a constant $Q^2$ is done via extrapolating and interpolating among measured data taken over a wide range of beam energies\footnote{See Ref~\cite{KarlT} for example.}. The limited scope of the E08-027 5 T kinematic settings makes this procedure more difficult and instead a polarized model method is used. In this method, a model calculation of the spin structure function is done at the measured $Q^2$ value and the desired, constant $Q^2$. The difference between the two model calculations is applied to the data in order to correct it to a constant momentum transfer:
\begin{align}
\label{evoeq}
\delta_{\mathrm{evolve}} &= g_{1,2}^\mathrm{mod} (x_{\mathrm{data}},Q_{\mathrm{data}}^2) -g_{1,2}^\mathrm{mod} (x_{\mathrm{const}},Q_{\mathrm{const}}^2)\,,\\
x_{\mathrm{const}} &= Q_{\mathrm{const}}^2/ (W^2 - M^2 + Q_{\mathrm{const}}^2)\,,
\end{align}
where the evolution is also done at a constant $W$. The constant $Q^2$ value is chosen as the momentum transfer closest to the $\Delta$(1232) resonance in the measured data.
\begin{figure}[htp]
\centering     
\includegraphics[width=1.0\textwidth]{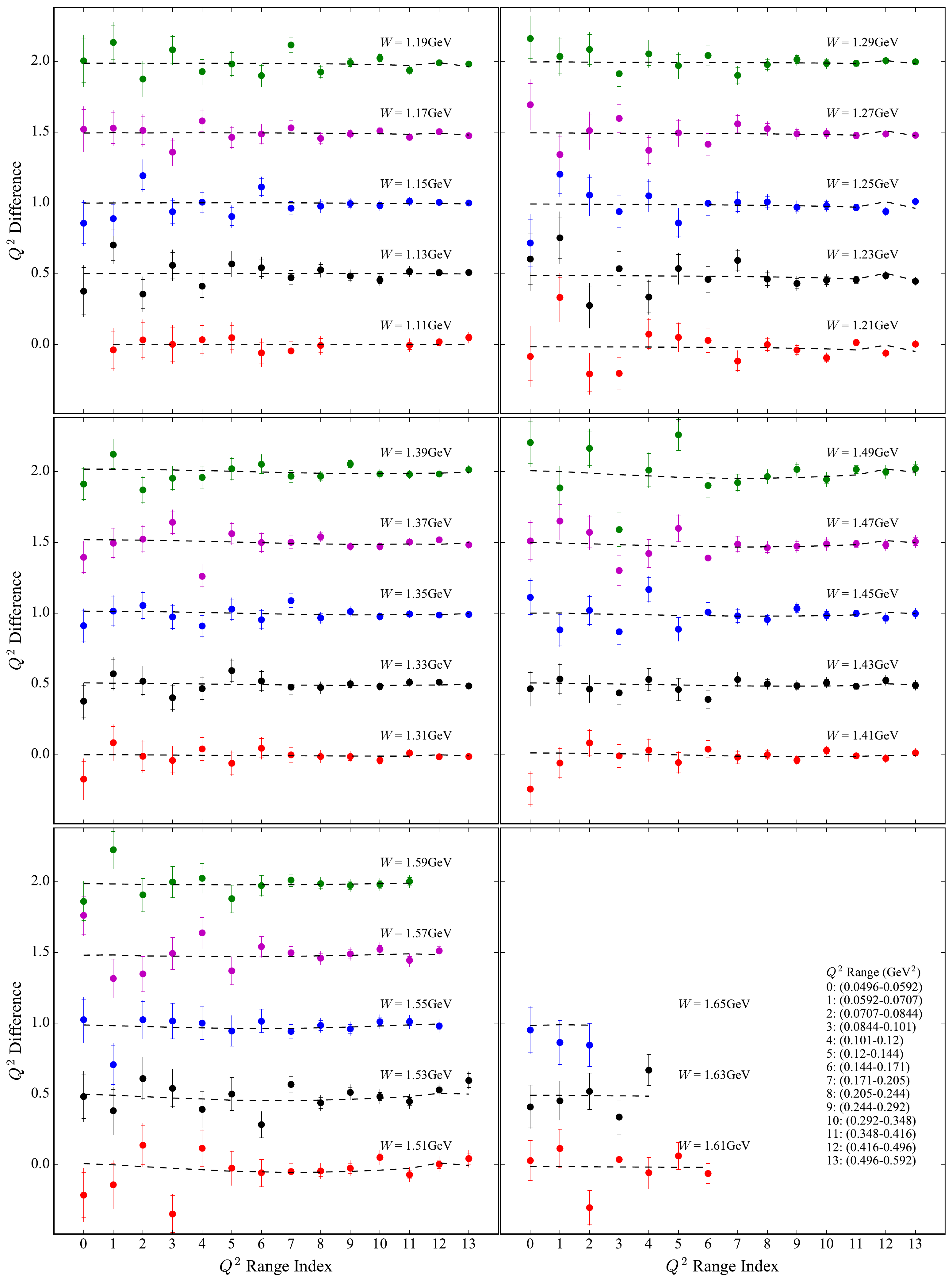}
\caption{Momentum transfer evolution of the CLAS EG1b data as compared to the CLAS EG1b model for 0.0496 GeV$^2$$<$ Q$^2$ $<$ 0.592 GeV$^2$ . The data is broken up into thirteen $Q^2$ pairs for a given $W$ bin.}
\label{Q2evolveHallB}
\end{figure}

The applicability of this method is tested against the CLAS EG1b~\cite{HallBRes} data set. The data is broken down into fourteen different $Q^2$ ranges that cover 0.0496 $< Q^2 <$ 0.592 GeV$^2$. The difference between the data is calculated at a given $W$ bin for adjacent $Q^2$ values. The results of the comparison are shown in Figure~\ref{Q2evolveHallB} and the $Q^2$ ranges are given by an index that is mapped out in the plot legend. In a given plot, increasing $W$ bins are offset by 0.5 for clarity; the inner error bars are statistical and the outer are the statistical and systematic errors added in quadrature. The reduced $\chi^2$ on the entire data set for both MAID and Hall B is approximately 1.1. This is similar to the results seen in the asymmetry cut study in Figure~\ref{DiffCompare}. Because the reduced $\chi^2$ does not prefer one model over the other, the correction is applied as the average of the $\delta_\mathrm{evolve}$ in equation~\eqref{evoeq} for both MAID and Hall B, with the difference between the two taken as the systematic error.
\begin{figure}[htp]
\centering     
\subfigure[Evolving the $E_0$ = 2254 data set.]{\label{fig:g2evo}\includegraphics[width=.80\textwidth]{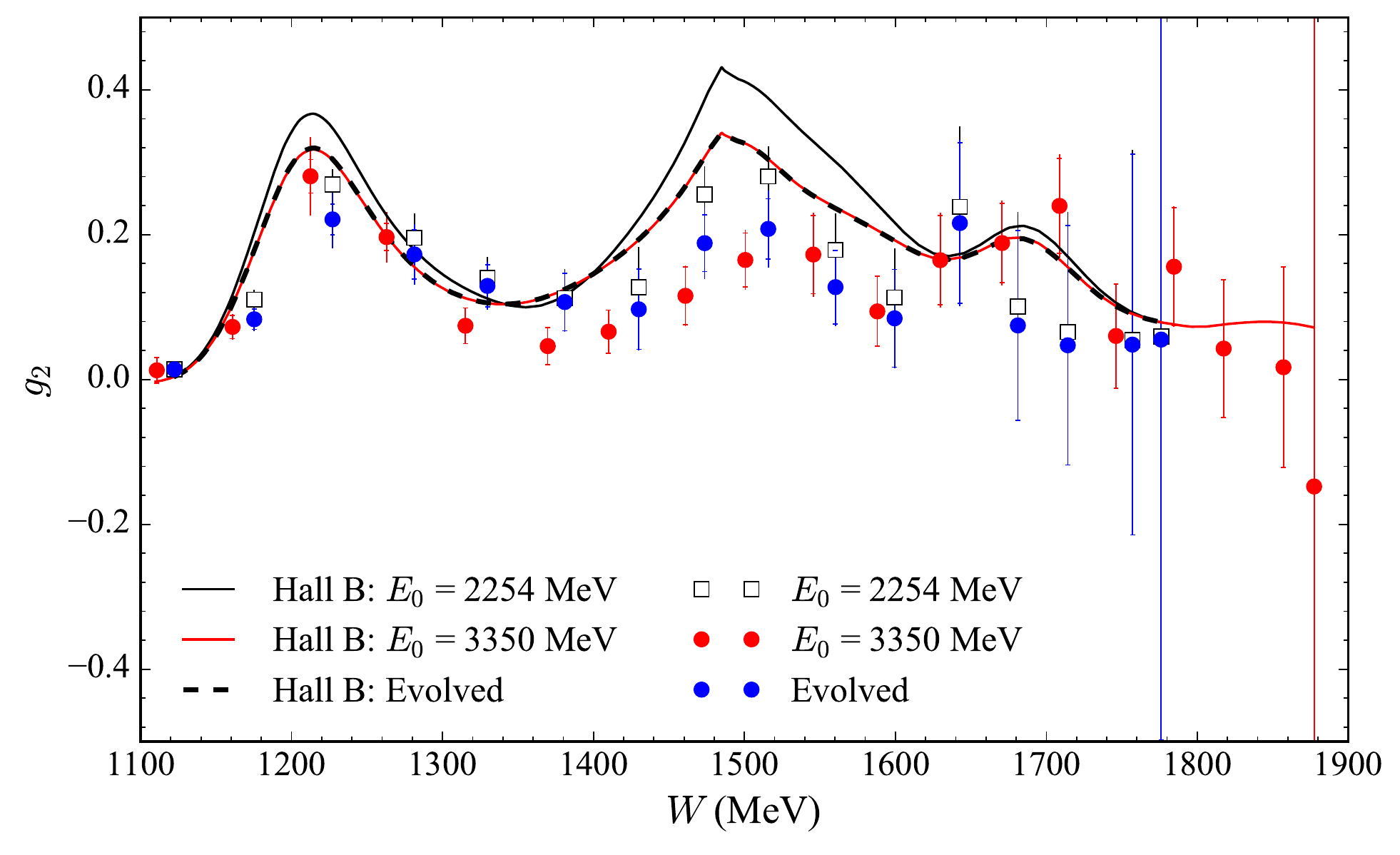}}
\qquad
\subfigure[Difference between the evolved and $E_0$ = 3350 MeV data points]{\label{fig:g2evodiff}\includegraphics[width=.80\textwidth]{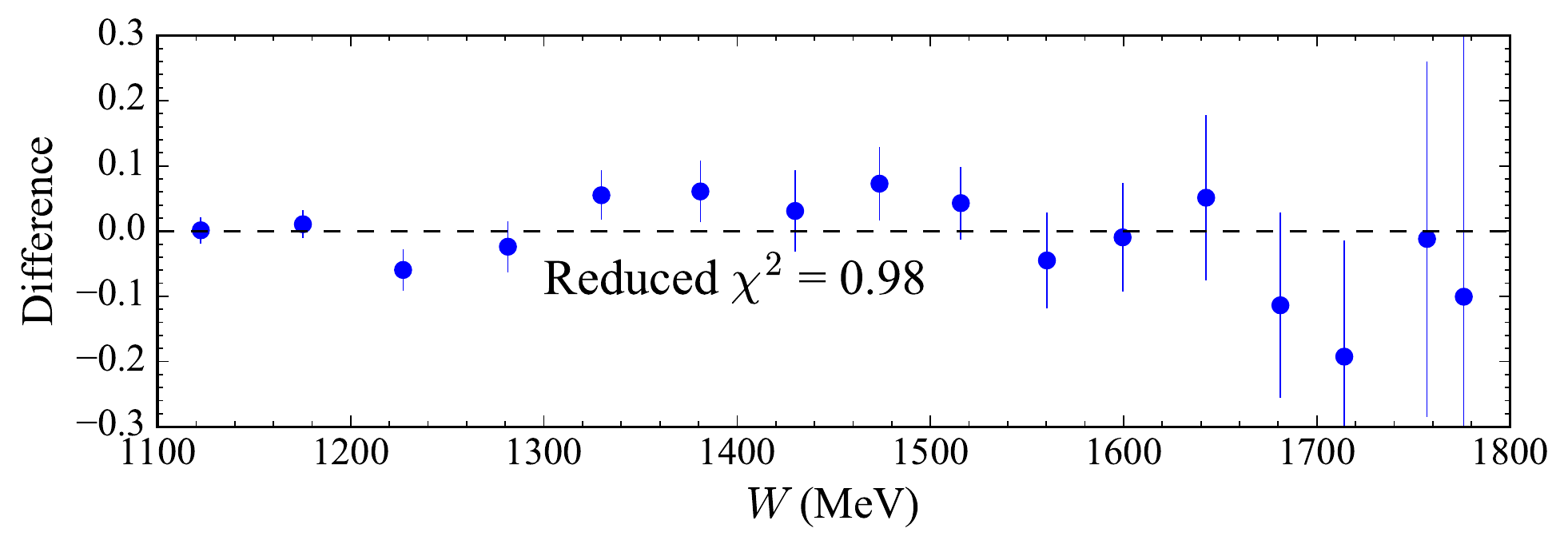}}
\caption{Testing the model evolution procedure for $g_2(x,Q^2)$ at the 5 T transverse kinematic settings. The adjustment in $Q^2$ is from approximately 0.086 GeV$^2$ to 0.13 GeV$^2$.}
\label{g2evolvecomp}
\end{figure}

\begin{figure}[htp]
\centering     
\subfigure[$E_0$ = 2254 MeV 5T Transverse]{\label{fig:g22254}\includegraphics[width=.80\textwidth]{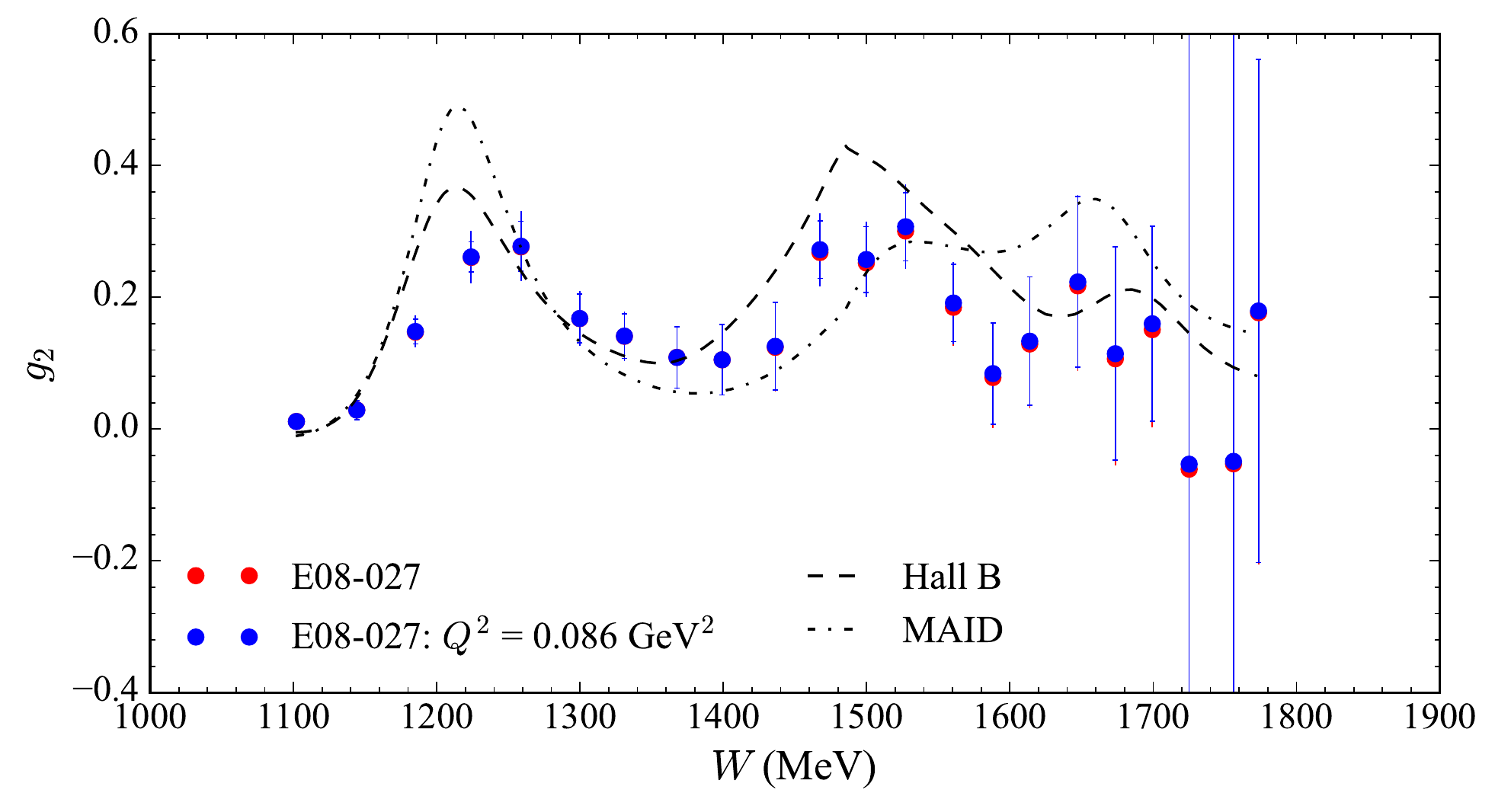}}
\qquad
\subfigure[$E_0$ = 3350 MeV 5T Transverse]{\label{fig:g23350}\includegraphics[width=.80\textwidth]{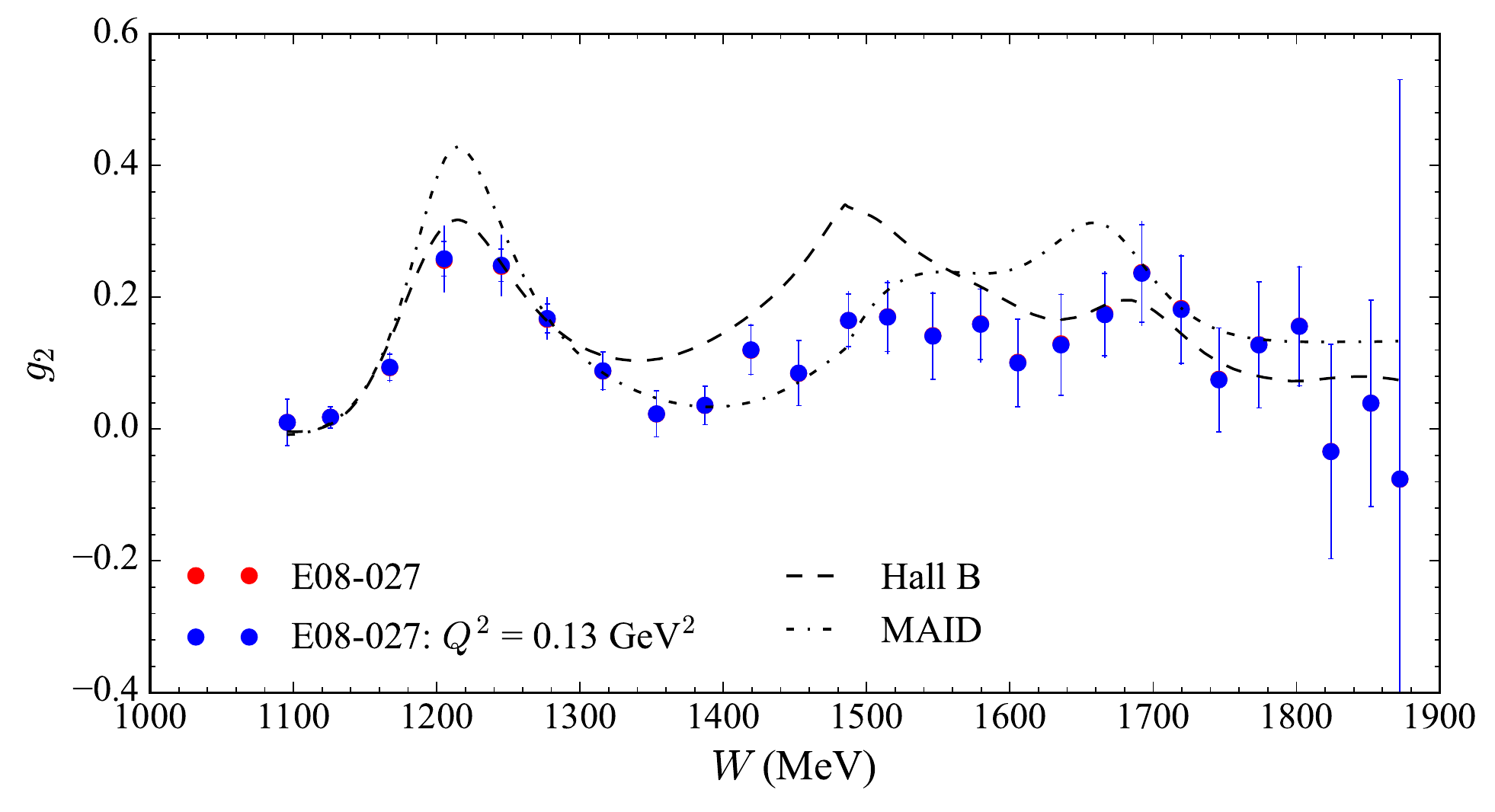}}
\qquad
\subfigure[$E_0$ = 2254 MeV 5T Longitudinal]{\label{fig:g1Long}\includegraphics[width=.80\textwidth]{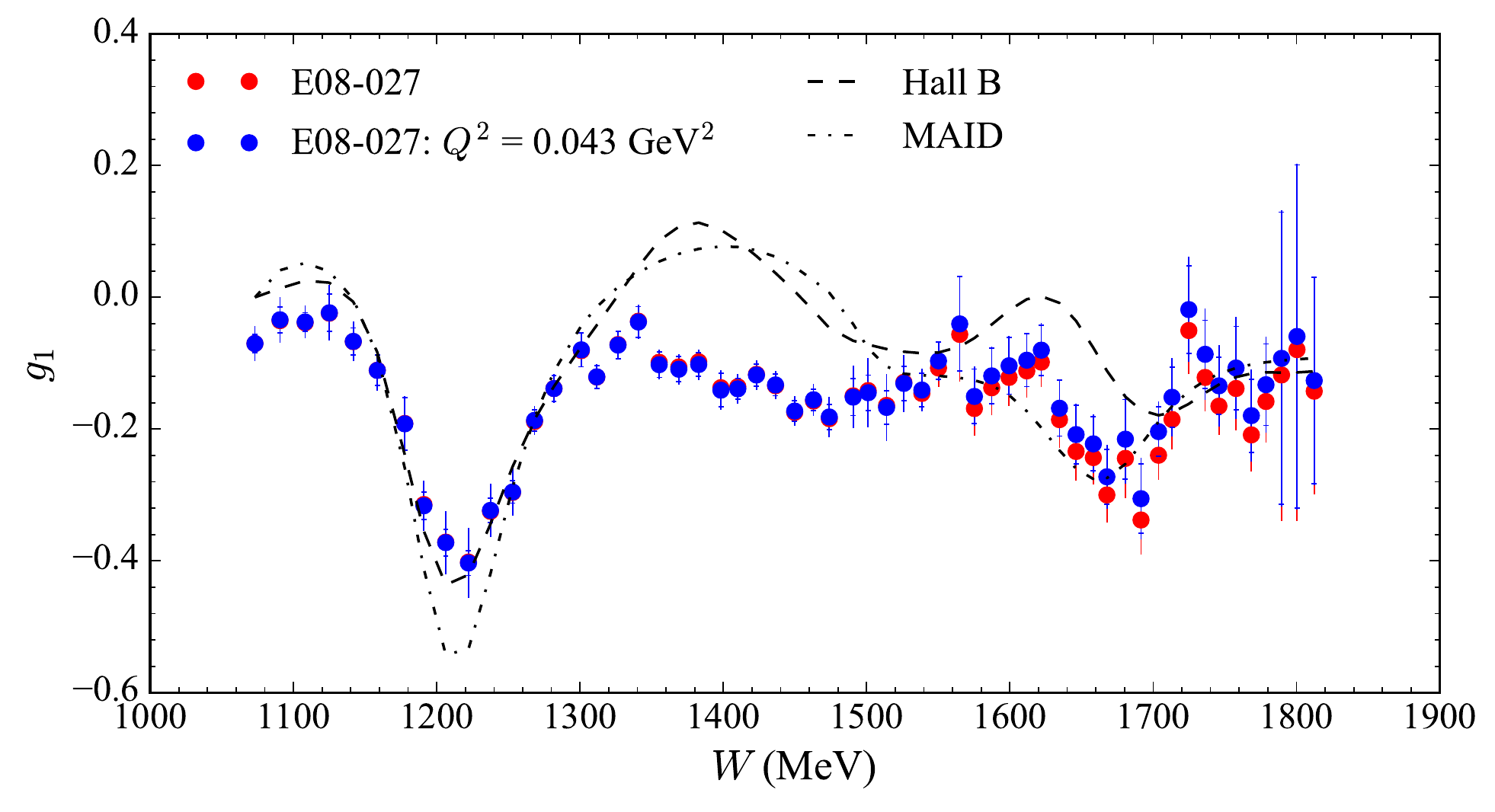}}
\caption{E08-027 spin structure functions evolved to a constant momentum transfer.}
\label{SSFCon}
\end{figure}
The EG1b data set only consists of $g_1(x,Q^2)$ measurements and is only applicable at the longitudinal setting. The validity of the model method at the two 5 T transverse kinematic settings is tested by attempting to evolve one data set to the other and then comparing the residual difference between the two data sets. The results of this study are shown in Figure~\ref{fig:g2evo}, where the $E_0$ = 2254 MeV data at non-constant $Q^2$ (white squares) is evolved to the non-constant $Q^2$ values of the $E_0$ = 3350 MeV data (red circles). The evolved result is the blue circles. Figure~\ref{fig:g2evodiff} highlights that the residual difference between the evolved $E_0$ = 2254 MeV and measured $E_0$ = 3350 MeV data points is consistent with zero.

The results of the momentum transfer evolution are shown in Figure~\ref{SSFCon}. The inner error bars are statistical and the outer are the total error: systematic error added in quadrature with the statistical error. The correction is very small at the transverse kinematic settings; the size of the correction is added linearly to the systematic error on the Born $g_2(x,Q^2)$ as an upper bound on the systematic uncertainty. The $Q^2$ increases by approximately 18\% across the $E_0$ = 2254 MeV 5 T transverse setting, but decreases by over 50\% at the same energy but longitudinal target field configuration. The change is even smaller at $E_0$ = 3350 MeV and only decreases by about 7\%. The uncertainty in the $Q^2$ determination from a reconstructed scattering angle uncertainty of $\pm$ 2\% is at the 5\% level for the transverse kinematic settings.
\begin{figure}[htp]
\centering     
\includegraphics[width=.80\textwidth]{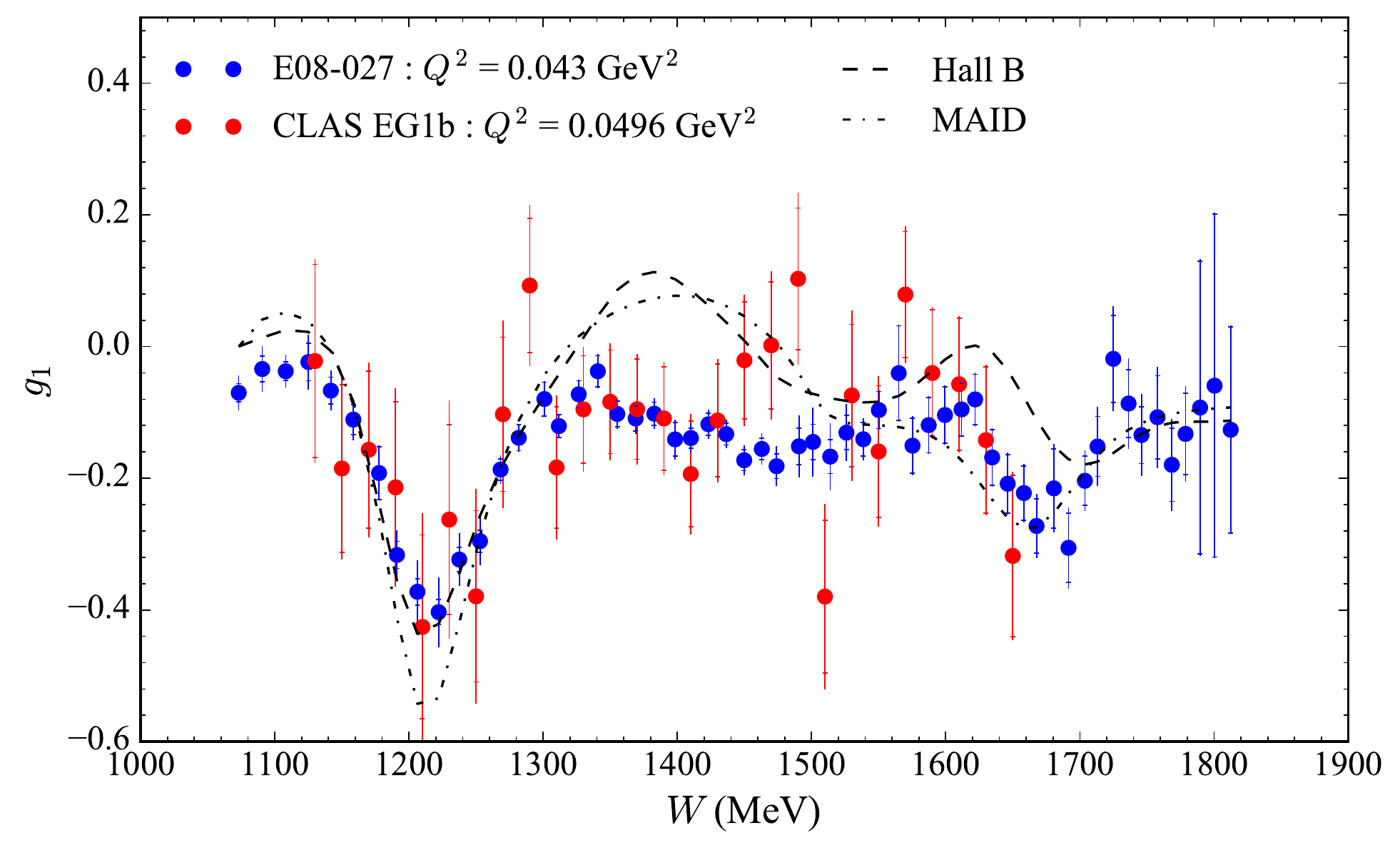}
\caption{Comparison of EG1b data and E08-027 longitudinal $g_1(x,Q^2)$.}
\label{g1HallBcomp}
\end{figure}

The E08-027 longitudinal kinematics sit right below the lowest $Q^2$ measured by CLAS EG1b. The two data sets agree very well, as shown in Figure~\ref{g1HallBcomp}. Both experiments do not show a sign flip in $g_1(x,Q^2)$ immediately after the $\Delta$(1232) resonance, which is predicted by both polarized models. It is worthwhile noting that the E08-027 data reaches the pion-production threshold, but the CLAS EG1b data does not. This is important in calculating the $x$-weighted integrals of the moments and sum rules because the E08-027 and polarized models differ by a sign in this kinematic region. Published EG1b results use the Hall B model to fill in this unmeasured high-$x$ region.

\subsection{Description of the Moment Plot Quantities and Uncertainties}
The presented figures of the subsequent moment calculations adhere to the following format:
\begin{itemize}
\item Inner error bars represent the statistical uncertainty
\item Outer error bars are the systematic error and statistical error added in quadrature
\item $\chi$PT calculations are given by the blue and grey bands and represent calculations by Vladimir Pascalutsa $et\, al.$~\cite{PascaCom} and Ulf Meissner $et\, al.$~\cite{Krebs}, respectively. The Pascaluta calculation is an unpublished update to their results in Ref~\cite{Gold}.
\item MAID model calculations are given by a dashed line and are limited to the resonance region contributions only. Hall B model calculations are given by the dashed-dotted line and include DIS contributions.
\item Published results from the EG1b experiment are given as red circles, and integration of the EG1b data from the CLAS database~\cite{HallBRes} by this author are presented as white squares.
\end{itemize} 

The systematic error on the calculated moments of the E08-027 data is calculated assuming the errors on the structure functions are highly correlated. The systematic error on the sampled integration is negligible, but is checked by comparing a sampled integration of a polarized model with the same sample frequency as the data to a Gaussian quadrature integration of the same moment. Additional systematic errors relevant to a given moment are discussed within the appropriate section.

The statistical uncertainty is calculated using a Monte Carlo method, where the mean value of each point to be integrated is adjusted by its statistical uncertainty. The results of this method are compared to an exact calculation of the error propagation as written in the FORTRAN analysis code ``BetterSimpson"~\cite{BS} and agree well.

\subsection{The First Moment of $g_1(x,Q^2)$}
\begin{figure}[htp]
\centering     
\subfigure[Total contribution]{\label{fig:Gamma1Mdata}\includegraphics[width=.80\textwidth]{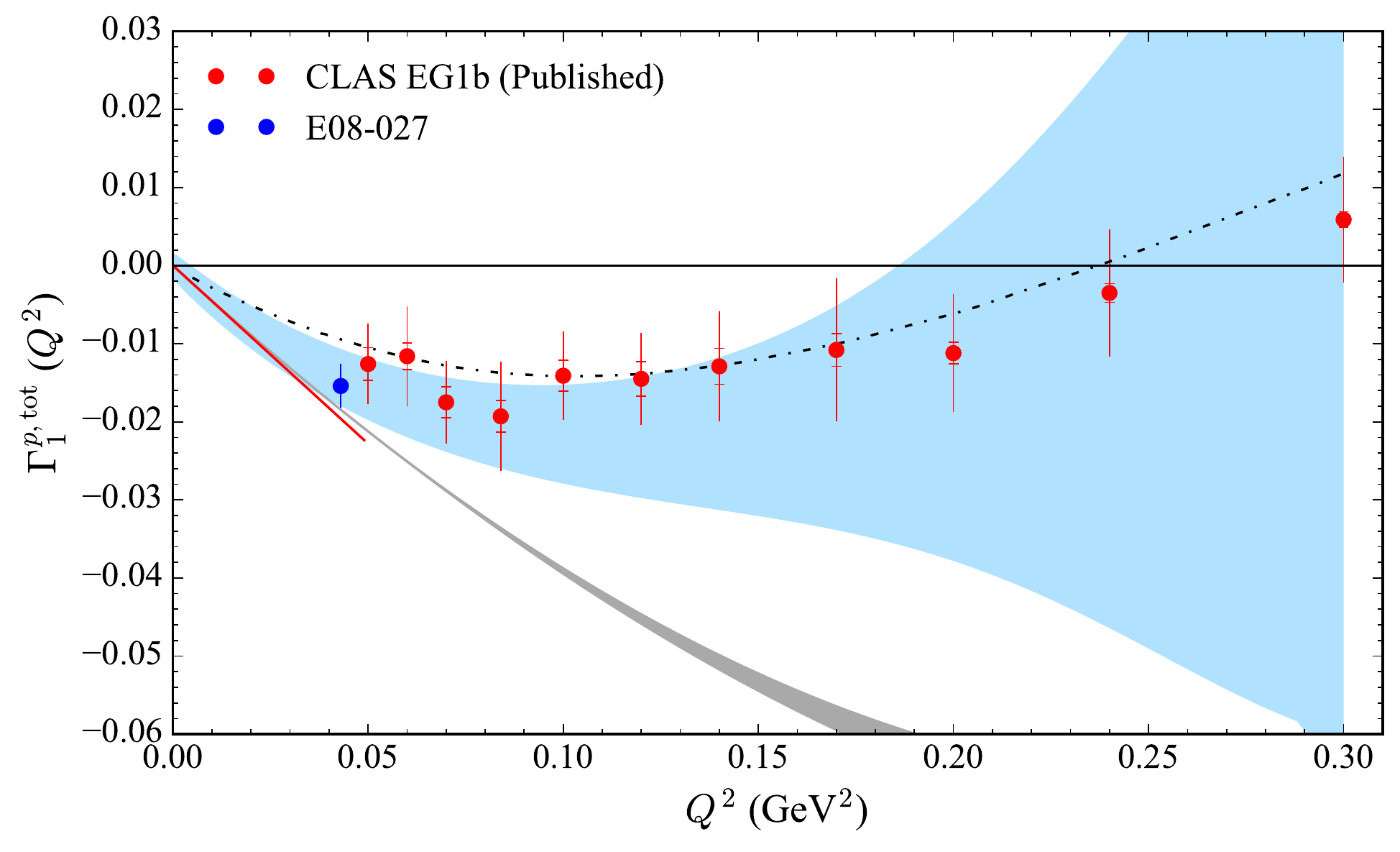}}
\qquad
\subfigure[Cumulative step-by-step integration of $\int_{x_{\mathrm{min}}}^{x_{\mathrm{max}}} g_1(x,Q^2)dx$.]{\label{gamma1integrand}\includegraphics[width=.80\textwidth]{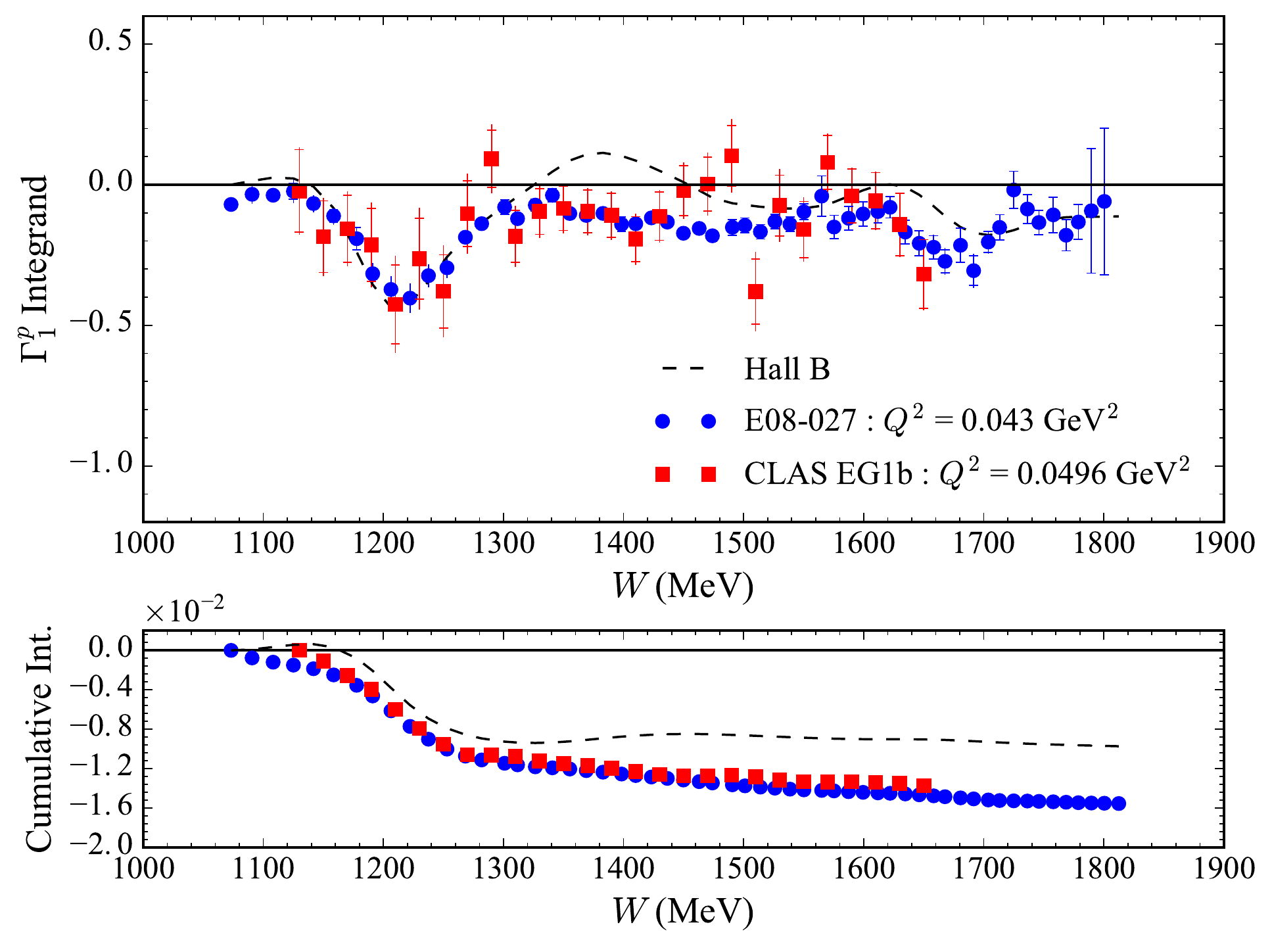}}
\caption{E08-027 results for the first moment of $g_1(x,Q^2)$. The colored bands are $\chi$PT calculations from Refs~\cite{Krebs,PascaCom}.}
\label{Gamma1Results}
\end{figure}

The results for $\Gamma_1(Q^2)$, defined in equation~\eqref{gamma1eq}, are shown in Figure~\ref{Gamma1Results} and Table~\ref{E08027Gamma1}. The solid red line in Figure~\ref{Gamma1Results} is the prediction given by the GDH slope. The Hall B model is used to calculate the low $x$ (and high $x$ for the CLAS EG1b data) contributions to the moment. Saturation of the integral occurs around the $\Delta$(1232) resonance, as shown in Figure~\ref{gamma1integrand}; the low $x$ input is not corrected for in the E08-027 data but added linearly to the systematic uncertainty.  

\begin{table}[htp]
\begin{center}
\begin{tabular}{ l  c  c  c  c  c r }
  Setting& $Q^2$ (GeV$^2$)&$\Gamma_1^{\mathrm{meas.}}$ &  $\Gamma_1^{\mathrm{low\, }x}$ & $\Gamma_1^{\mathrm{tot.}}$&$\delta^{\mathrm{tot.}}_{\mathrm{stat}}$  & $\delta^{\mathrm{tot.}}_{\mathrm{sys}}$    \\ \hline
  2254 5T Long. &  0.043 & $-$0.01541 & 0.0003& $-$0.01541  & 0.0006 & 0.0028
\end{tabular}
\caption{\label{E08027Gamma1}Results for the E08-027 $\Gamma_1(Q^2)$ integration. }
\end{center}
\end{table}

\begin{figure}[htp]
\centering     
\subfigure[Measured contribution]{\label{fig:Gamma1M}\includegraphics[width=.80\textwidth]{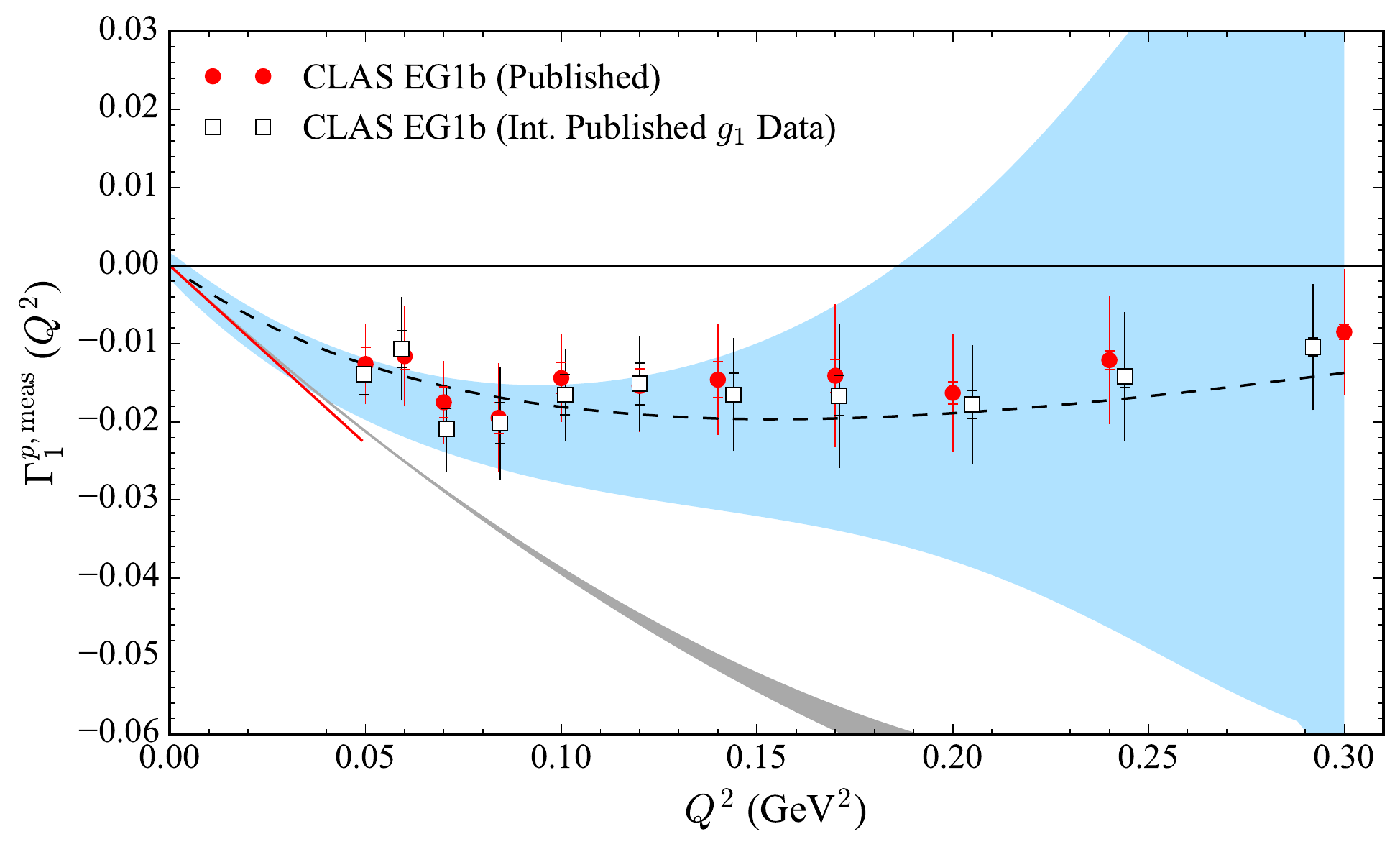}}
\qquad
\subfigure[Total contribution (measured + model extrapolations)]{\label{fig:Gamma1T}\includegraphics[width=.80\textwidth]{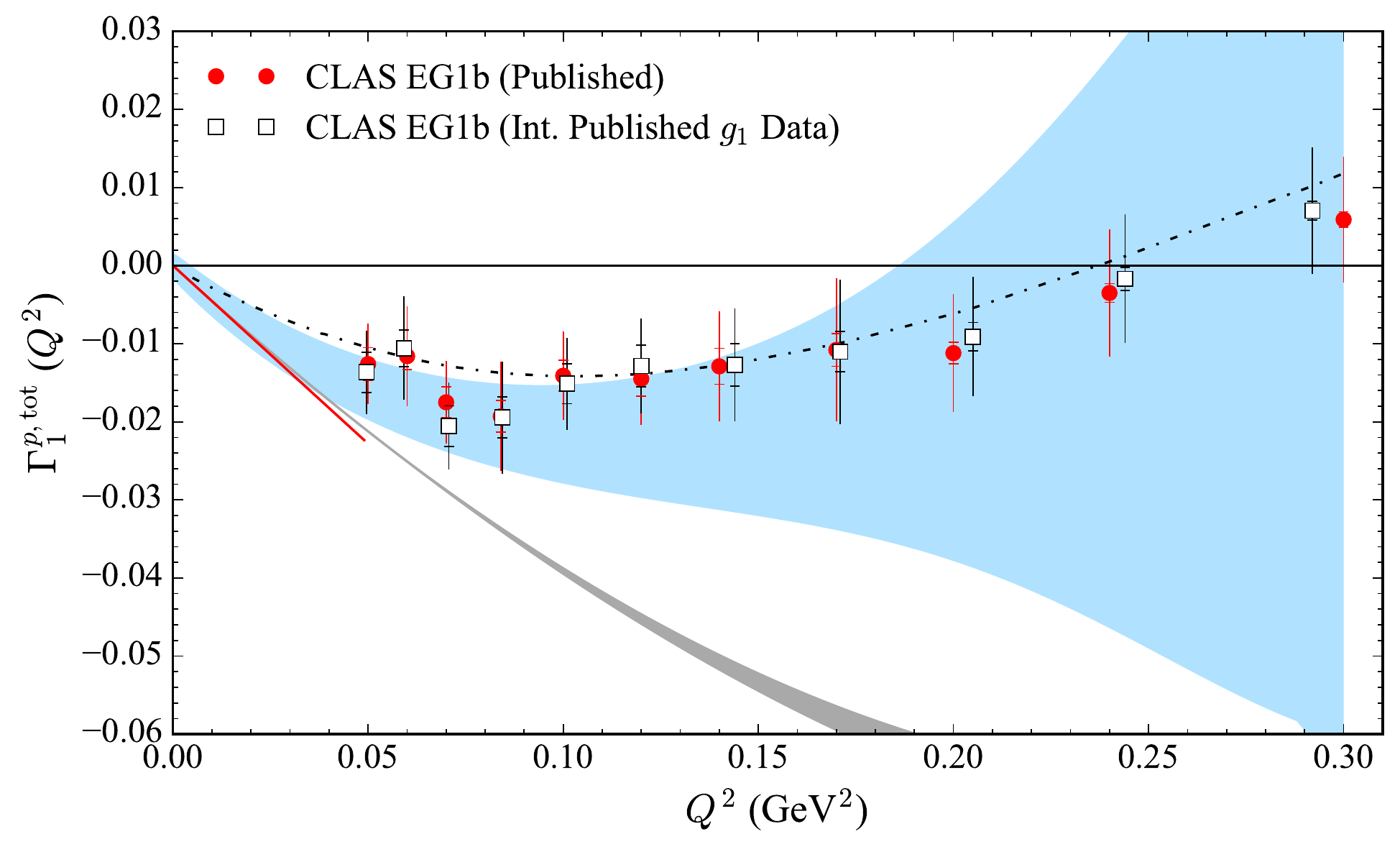}}
\caption{Reproducing the CLAS EG1b results for the first moment of $g_1(x,Q^2)$. The colored bands are $\chi$PT calculations from Refs~\cite{Krebs,PascaCom}.}
\label{Gamma1Results_repo}
\end{figure}

As an additional check on the integration method, uncertainty propegatation, and unmeasured extrapolation, the Hall B $g_1(x,Q^2)$ data is integrated and compared to the published results. The integration of this data agrees well with the published moments, including the propagation of the statistical uncertainty, as seen in Figure~\ref{Gamma1Results_repo}. Additionally, the unmeasured contribution to the moments is easily reproduced in going from Figure~\ref{fig:Gamma1M} to~\ref{fig:Gamma1T}. The systematic uncertainty is consistent with the assumption of fully correlated errors. It should also be noted that the off-set in $Q^2$ is a discrepancy in the published momentum transfer of the moments and momentum transfer of the $g_1(x,Q^2)$ values from the Hall B database. 

The E08-027 data is consistent with the CLAS EG1b results, but with significantly smaller error bars\footnote{The dominant contribution to the systematic error is from the unpolarized cross section model. This error should be reduced further when data becomes available.}. The data is not at a low enough $Q^2$ to confirm the GDH slope, but it does trend in the correction direction.
While the E08-027 data is consistent with both $\chi$PT calculations, the Pascalutsa group's calculation is in much better agreement with the entire data set. Meissner $et\, al.$ are aware of the increasing discrepancy with $Q^2$ and suggest potential improvements to their results by calculating to the next-to-next-to-leading order diagrams~\cite{Krebs}. 
\subsection{Extended GDH Sum}
\begin{figure}[htp]
\centering     
\includegraphics[width=.80\textwidth]{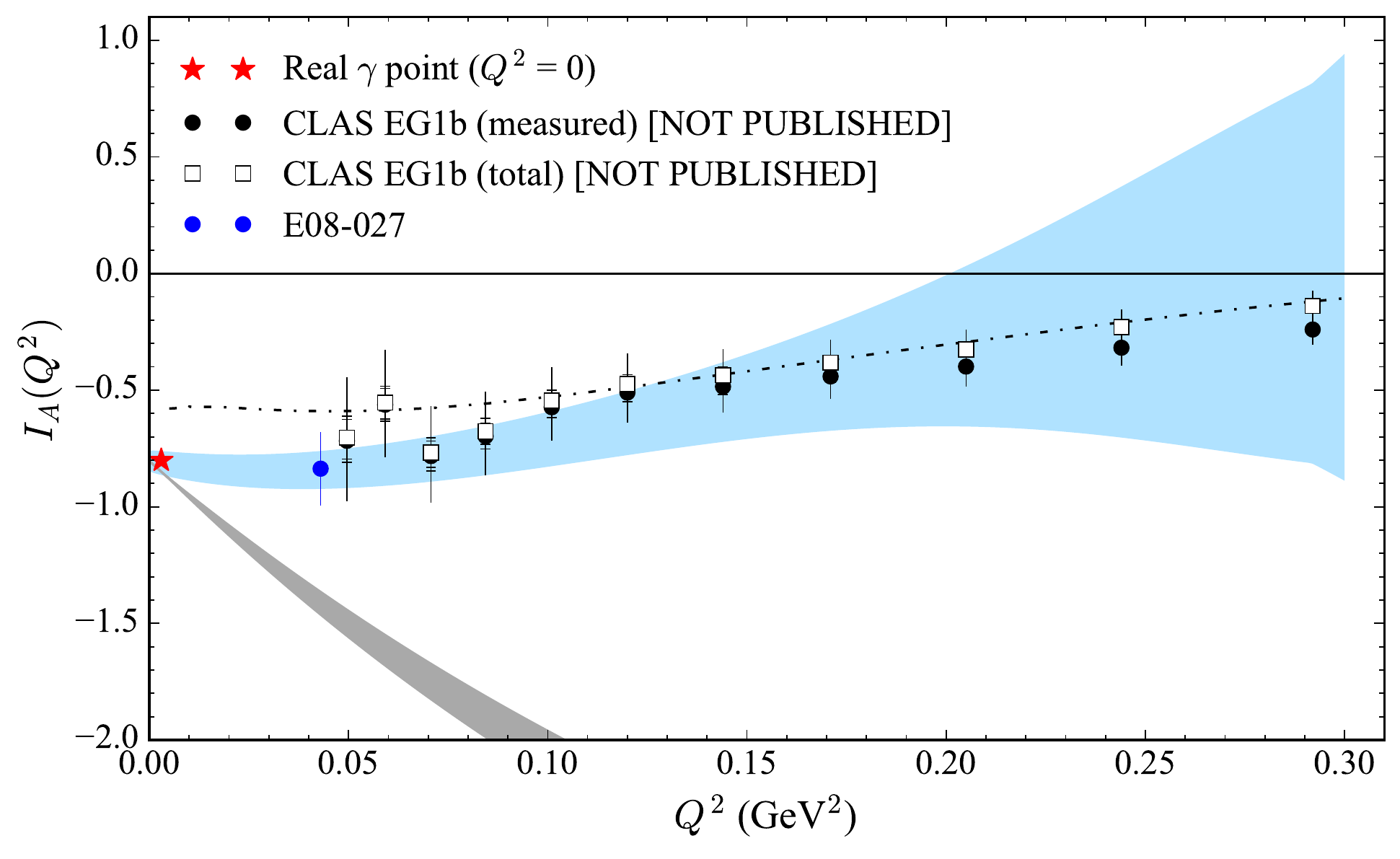}
\caption{E08-027 results for the extended GDH sum. The CLAS EG1b points are not published; they are integrations of their published $g_1(x,Q^2)$ data and use the Hall B model for the $g_2(x,Q^2$) component. The colored bands are $\chi$PT calculations from Refs~\cite{Krebs,PascaCom}.}
\label{GDHresults}
\end{figure}

\begin{figure}[htp]
\centering     
\includegraphics[width=.80\textwidth]{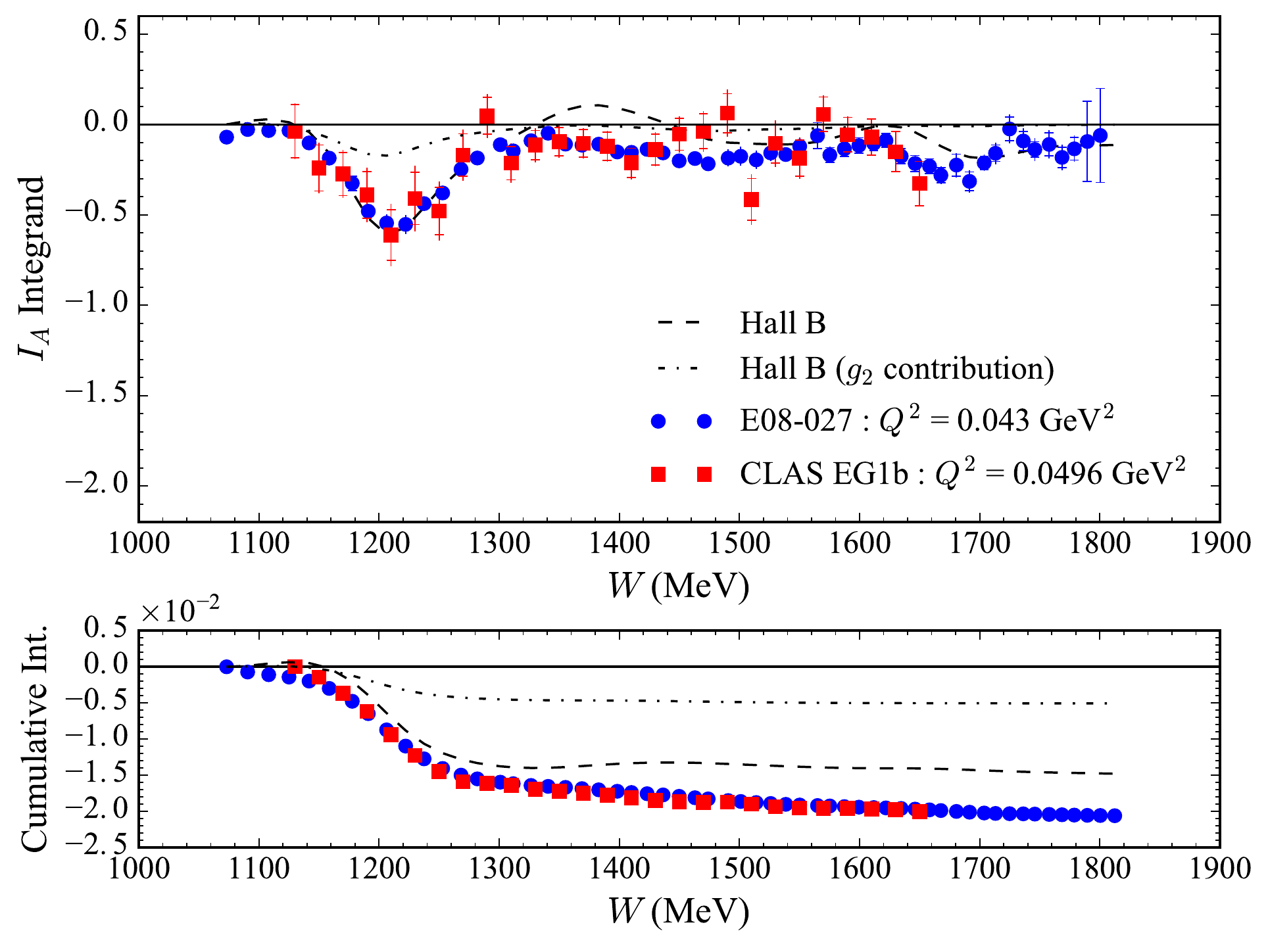}
\caption{Cumulative step-by-step integration of $I_A(Q^2)$.}
\label{GDHintegrand}
\end{figure}

Figure~\ref{GDHresults} and Table~\ref{E08027GDH} show the results for the extended GDH sum of equation~\eqref{GDHeq}. In Figure~\ref{GDHresults}, the GDH sum result for a real photon is
\begin{equation}
I_A(Q^2 = 0) = -\frac{1}{4}\kappa_p^2 = -0.8010\,,
\end{equation}
where $\kappa_p=1.79$ is the anomalous magnetic moment of the proton. The unmeasured $g_2(x,Q^2)$ component of the integral is provided by the Hall B model, with an assumed 50\% systematic uncertainty. This uncertainty is estimated by comparing the E08-027 $g_2(x,Q^2)$ data (both 5 T and  preliminary 2.5 T results) with the Hall B model.  The model structure function provides approximately 20\% of the strength of the integrand, as seen in Figure~\ref{GDHintegrand} and so a 10\% systematic uncertainty is added in quadrature with the previously described systematic errors, which are evaluated assuming they are fully correlated.

Similar to the $\Gamma_1(Q^2)$ result, the low $x$ contribution is approximately 1.5\% of the measured integral and is added linearly to the other systematic uncertainties. In the EG1b data, the extrapolated portions of the integral are not necessarily small and negligible. Their systematic uncertainty is estimated by varying the input parameters to the Hall B model and noting their effect on the resultant integral. These parameters are discussed in Appendix~\ref{app:Appendix-F}.

\begin{table}[htp]
\begin{center}
\begin{tabular}{ l cc  c  c  c r }
  Setting& $Q^2$ (GeV$^2$)&$I_A^{\mathrm{meas.}}$ & $I_A^{\mathrm{low\, }x}$ & $I_A^{\mathrm{tot.}}$&$\delta^{\mathrm{tot.}}_{\mathrm{stat}}$  & $\delta^{\mathrm{tot.}}_{\mathrm{sys}}$    \\ \hline
  2254 5T Long. &0.043& $-$0.83669 & 0.0119&  $-$0.83669 & 0.0240 & 0.1550
\end{tabular}
\caption{\label{E08027GDH}Results for the E08-027 $I_A(Q^2)$ integration. }
\end{center}
\end{table}

The calculated moments suggest a smooth and relatively flat approach to the $I_A(0)$ point, which is consistent with the Pascalutsa $et.\, al$ calculation. The systematic uncertainty in the E08-027 data point is dominated by the $g_2(x,Q^2)$ contribution and unpolarized cross section model systematic. This could be improved in the future with the addition of the $E_0$ = 2254 MeV 2.5 T data set and unpolarized cross sections from data. Referring back to Figure~\ref{kin}, the $E_0$ = 2254 MeV 2.5 T data is at approximately the same momentum transfer as the longitudinal data.

\subsection{Forward Spin Polarizability and Higher Order $\gamma_0$}
The CLAS EG1b results for $\gamma_0(Q^2)$ are calculated using a redefinition of equation~\eqref{GAMMAEQ}:
\begin{equation}
\label{EG1bgamma0}
\gamma_0 (Q^2) = \frac{16M^2\alpha}{Q^6}\int_0^{x_{\mathrm{th}}} x^2 A_1(x,Q^2)F_1(x,Q^2)dx\,.
\end{equation}
The spin structure function dependence is contained within the virtual photon asymmetry, $A_1(x,Q^2)$, and alleviates the need for a separate extraction of the $g_1(x,Q^2)$ and $g_2(x,Q^2$) terms. The perpendicular asymmetry dependence on $A_1(x,Q^2)$ is determined from a Rosenbluth-like separation of $A_1(x,Q^2)$ at different kinematic points. The CLAS setup in Hall B can only achieve parallel target polarization. More details on the $A_1(x,Q^2)$ extraction from the EG1 data are in Ref~\cite{FerschT}.

\begin{figure}[htp]
\centering     
\includegraphics[width=.80\textwidth]{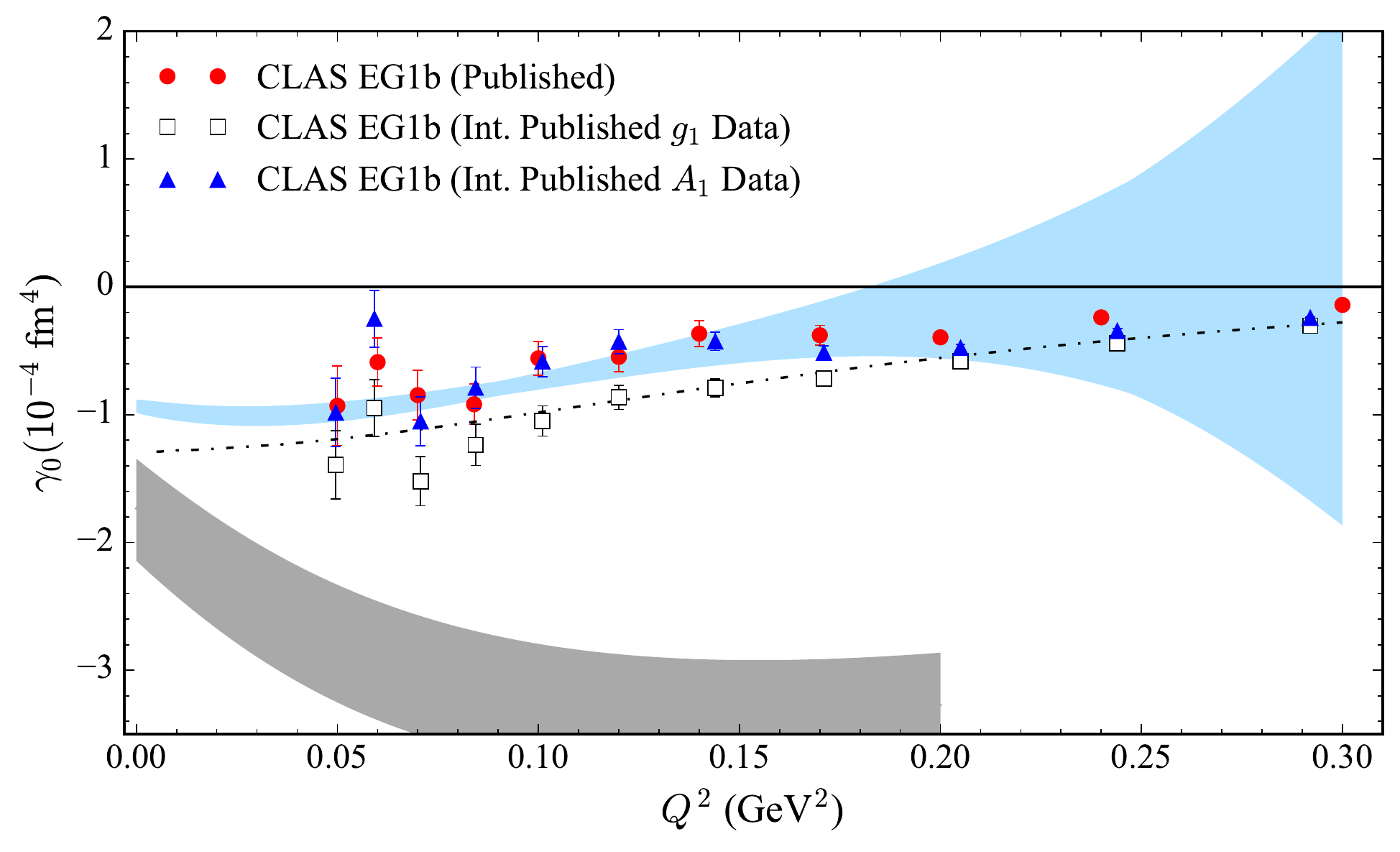}
\caption{Reproducing the CLAS EG1b $\gamma_0(Q^2)$ results. For clarity, only statistical error bars are shown. The colored bands are $\chi$PT calculations from Refs~\cite{Krebs,PascaCom}.}
\label{gamma0results_repo}
\end{figure}

The treatment of the $g_2(x,Q^2)$ contribution leads to discrepancies in reproducing the published CLAS EG1b results. The differences are highlighted in Figure~\ref{gamma0results_repo}; the integration of the CLAS $A_1(x,Q^2)$ data reproduces the published $\gamma_0(Q^2)$ data, but using a combination of the published $g_1(x,Q^2)$ data and the Hall B $g_2(x,Q^2)$ model gives a different result. The E08-027 data is crucial here to resolve the difference because it has both $g_1(x,Q^2)$ and $g_2(x,Q^2)$ data at a low $Q^2$ data point.

\begin{figure}[htp]
\centering     
\includegraphics[width=.80\textwidth]{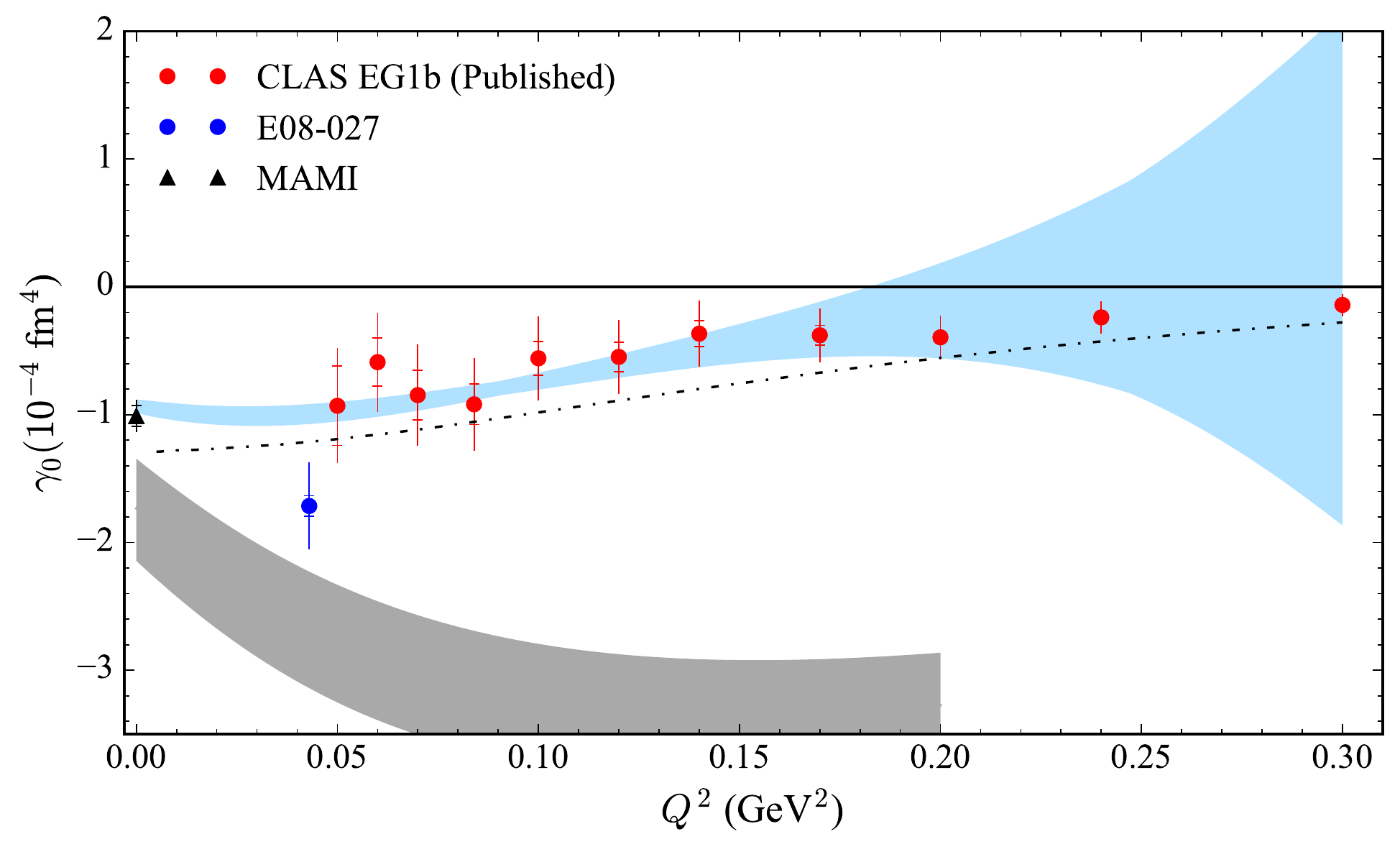}
\caption{E08-027 results for the forward spin polarizability. The colored bands are $\chi$PT calculations from Refs~\cite{Krebs,PascaCom}.}
\label{gamma0results}
\end{figure}

The E08-027 results are shown in Figure~\ref{gamma0results} and Table~\ref{E08027Gamma0}. The real photon point at $Q^2=0$ is from the ELSA-MAMI collaboration~\cite{ELSA} and is
\begin{equation}
\gamma_0(Q^2 = 0) = [-1.01 \pm 0.08\,(\mathrm{stat}) \pm 0.10\,(\mathrm{sys})] \cdot 10^{-4}\,\mathrm{fm}^4\,.
\end{equation}
As with the extended GDH sum, the Hall B model provides the $g_2(x,Q^2)$ component with an attached 50\% uncertainty that contributes a 10\% uncertainty to the integrated result. 

\begin{figure}[htp]
\centering     
\includegraphics[width=.80\textwidth]{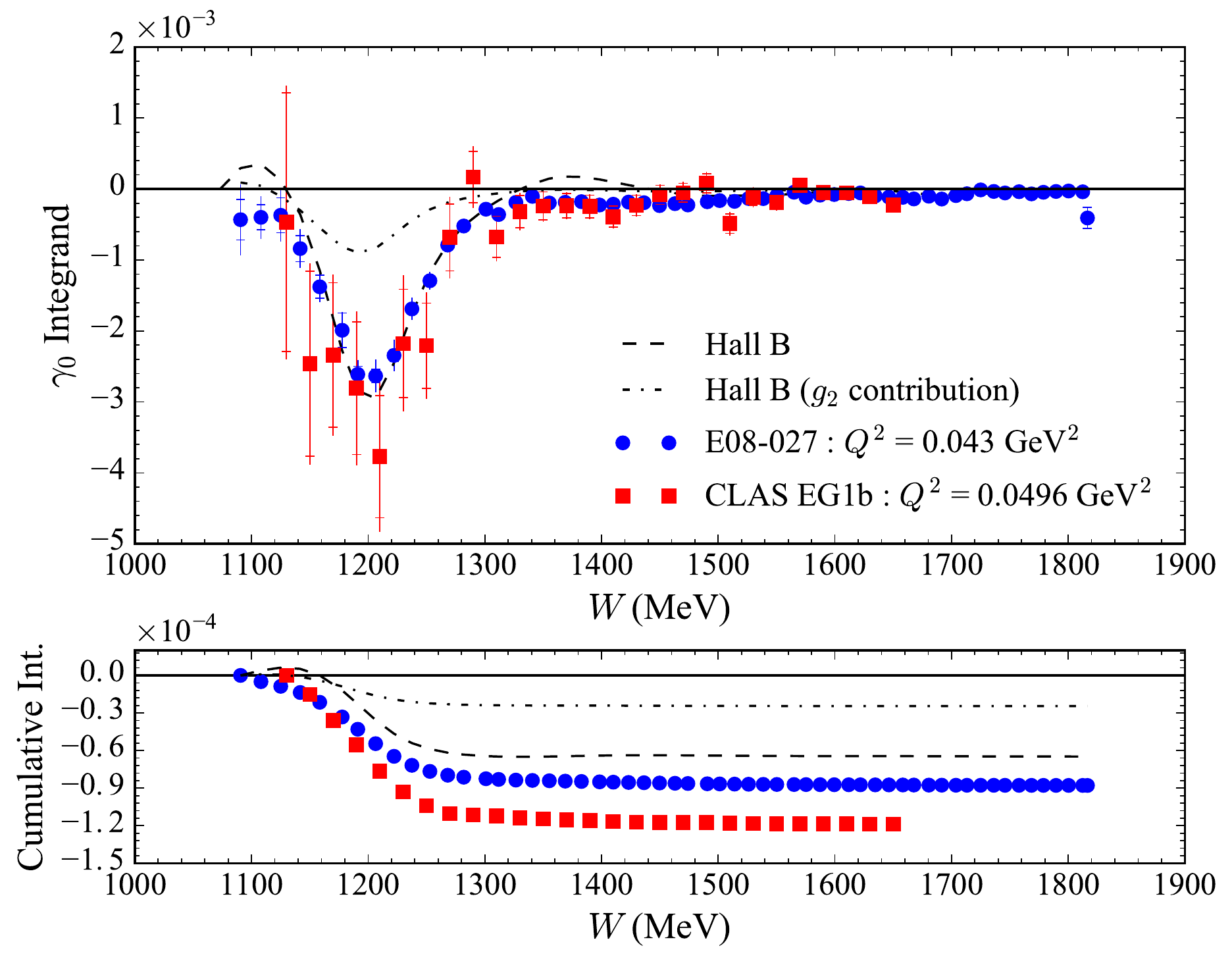}
\caption{Cumulative step-by-step integration of $\gamma_0(Q^2)$. For the CLAS EG1b data, the Hall B model is used to determine the contribution to the integrand below $W$ $\approx$ 1120 MeV.}
\label{gamma0int}
\end{figure}

Further discrepancies arise in the E08-027 result and EG1b results from the EG1b treatment of its unmeasured high $x$ region. Highlighted in Figure~\ref{gamma0int}, EG1b assumes Hall B model behavior between its largest $x$ value and the pion production threshold. The sign of the model is different from the sign of the E08-027 data, which does reach the pion production threshold. This sign difference leads to a larger negative contribution to the E08-027 result. The $x^2$ weighting of the moment enhances the discrepancy between the two data sets and is also why this effect is minimized in the first order moments previously discussed.
\begin{table}[htp]
\begin{center}
\begin{tabular}{ l  c c  c  c  c r }
  Setting& $Q^2$(GeV$^2$) &$\gamma_0^{\mathrm{meas.}}$ & $\gamma_0^{\mathrm{low\, }x}$ & $\gamma_0^{\mathrm{tot.}}$&$\delta^{\mathrm{tot.}}_{\mathrm{stat}}$  & $\delta^{\mathrm{tot.}}_{\mathrm{sys}}$    \\ \hline
  2254 5T Long.& 0.043& $-$1.7145 & $-$0.0007&  $-$1.7145 & 0.0808 & 0.3308
\end{tabular}
\caption{\label{E08027Gamma0}Results for the E08-027 $\gamma_0(Q^2)$ integration. Units are $10^{-4}$ fm$^4$. }
\end{center}
\end{table}
From the results in Figure~\ref{gamma0results}, a relatively sharp change in the slope of the moment is needed to agree with the ELSA-MAMI result. This change is currently not predicted by either $\chi$PT calculations presented. CLAS EG4 and the 2.5 T transverse settings of E08-027 can potentially fill in the low $Q^2$ gap of the current data and provide information on the suggested slope transition. Of course, the E08-027 results presented here are subject to change upon availability of a experimentally measured cross section.

The next term in the expansion of equation~\eqref{MomentSum} and equation~\eqref{MomentExpansion} is
\begin{equation}
\gamma_{0}^* (Q^2)  = \frac{64M^4\alpha}{Q^{10}}\int_{0}^{x_0} x^4{\bigg (} g_1(x,Q^2) - \frac{4M^2}{Q^2} x^2 g_2(x,Q^2){\bigg )} dx\,,
\end{equation}
and represents a higher order moment of the generalized forward spin polarizability. Continuing on to higher and higher order moments allows for the full construction of the forward Compton amplitude. From the optical theorem, this is equivalent to a reconstruction of the hadronic tensor and, when verified by theoretical predictions, is a calculation of the internal dynamics of the proton at all energy scales. The strong momentum transfer dependence at low $Q^2$ indicates that the polarizabilities are dominated by long-range effects and resonance structure and is more evidence of the complex proton internals.


\begin{figure}[htp]
\centering     
\includegraphics[width=.80\textwidth]{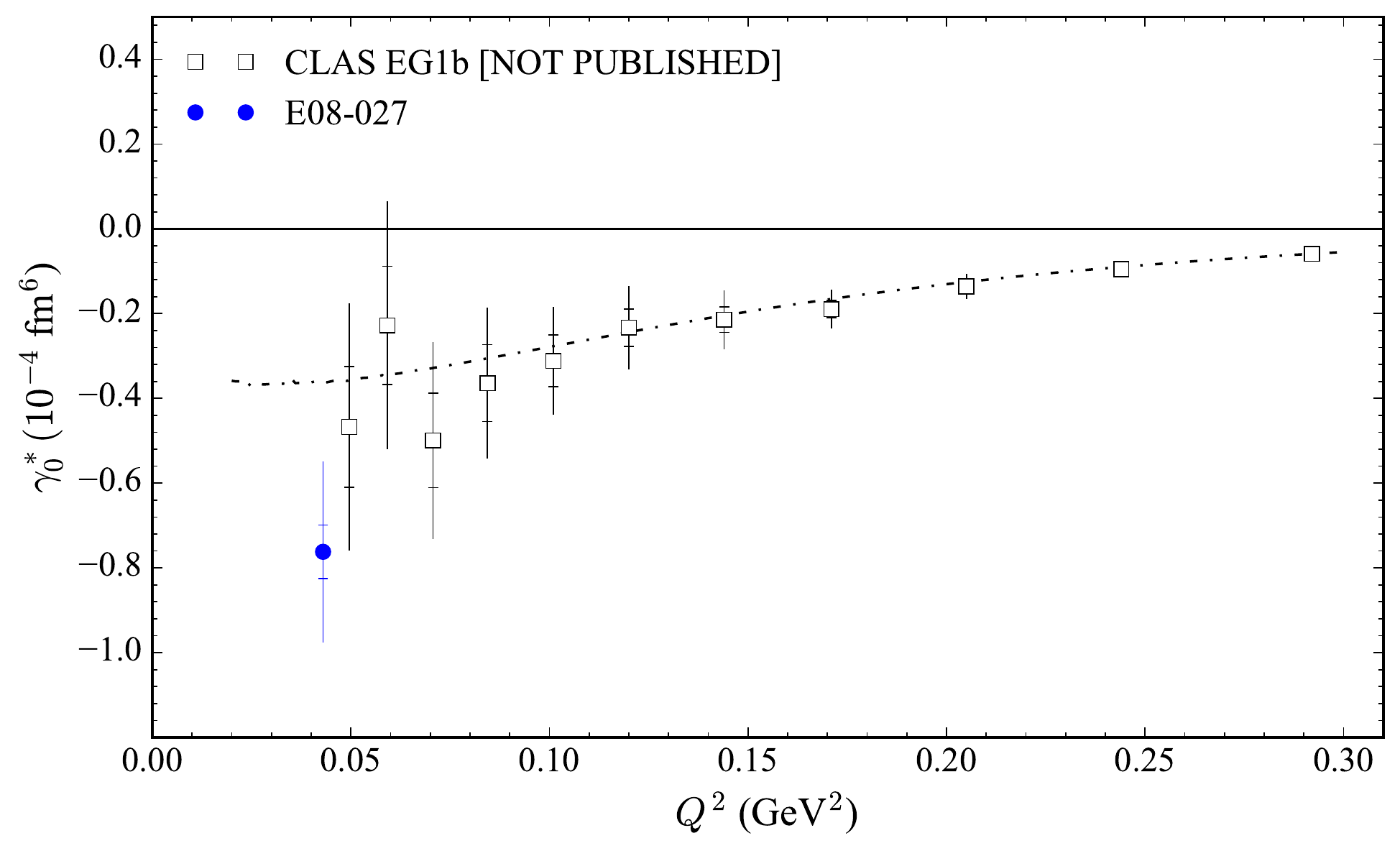}
\caption{E08-027 results for the $\mathcal{O}(\nu^5)$ forward spin polarizability. The CLAS EG1b results are integrations of their published $g_1(x,Q^2)$ data and the Hall B model for $g_2(x,Q^2)$.}
\label{gamma0starresults}
\end{figure}

\begin{figure}[htp]
\centering     
\includegraphics[width=.80\textwidth]{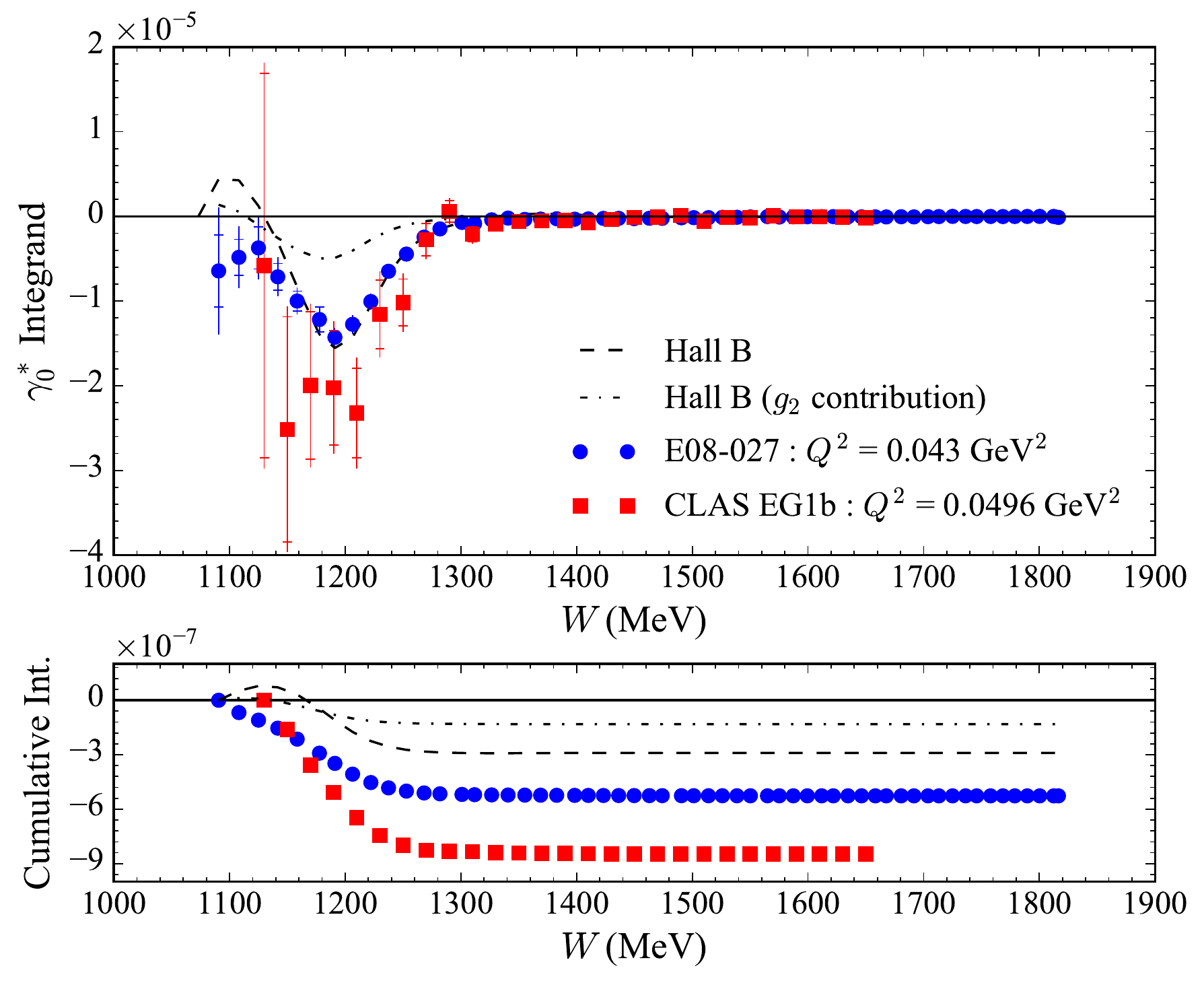}
\caption{Cumulative step-by-step integration of $\gamma^*_0(Q^2)$.}
\label{gamma0starint}
\end{figure}

The results for $\gamma^*_0(Q^2)$ are shown in Figure~\ref{gamma0starresults} and Table~\ref{E08027Gamma0star}. The $x^4$ weighting causes the integral to saturate immediately after the $\Delta$(1232) resonance as seen in Figure~\ref{gamma0starint}, and minimizes any contribution from the DIS regime. This weighting also magnifies the discrepancy between the E08-027 and EG1b data sets in the region of the pion-production threshold. The increased $1/Q^2$ weighting also drives an increase in the error bars at lower and lower momentum transfers. There are no current $\chi$PT calculations for the higher order polarizabilities. Those calculations, along with lower momentum transfer data, would be very useful in mapping out the moment as it approaches the real photon point.

\begin{table}[htp]
\begin{center}
\begin{tabular}{ l c c  c  c  c r } 
  Setting& $Q^2$ (GeV$^2$) &$\gamma_0^{*\,\mathrm{meas.}}$ & $\gamma_0^{*\,\mathrm{low\, }x}$ & $\gamma_0^{*\,\mathrm{tot.}}$&$\delta^{\mathrm{tot.}}_{\mathrm{stat}}$  & $\delta^{\mathrm{tot.}}_{\mathrm{sys}}$    \\ \hline
  2254 5T Long.& 0.043 & $-$0.7621 & $-$2$\times$10$^{-5}$&  $-$0.7621 & 0.0630 & 0.2037 
\end{tabular}
\caption{\label{E08027Gamma0star}Results for the E08-027 $\gamma^*_0(Q^2)$ integration. Units are $10^{-4}$ fm$^6$.}
\end{center}
\end{table}
\subsection{Hyperfine Splitting Contributions}
\label{HyperFineDATA}

\begin{figure}[htp]
\centering     
\includegraphics[width=.80\textwidth]{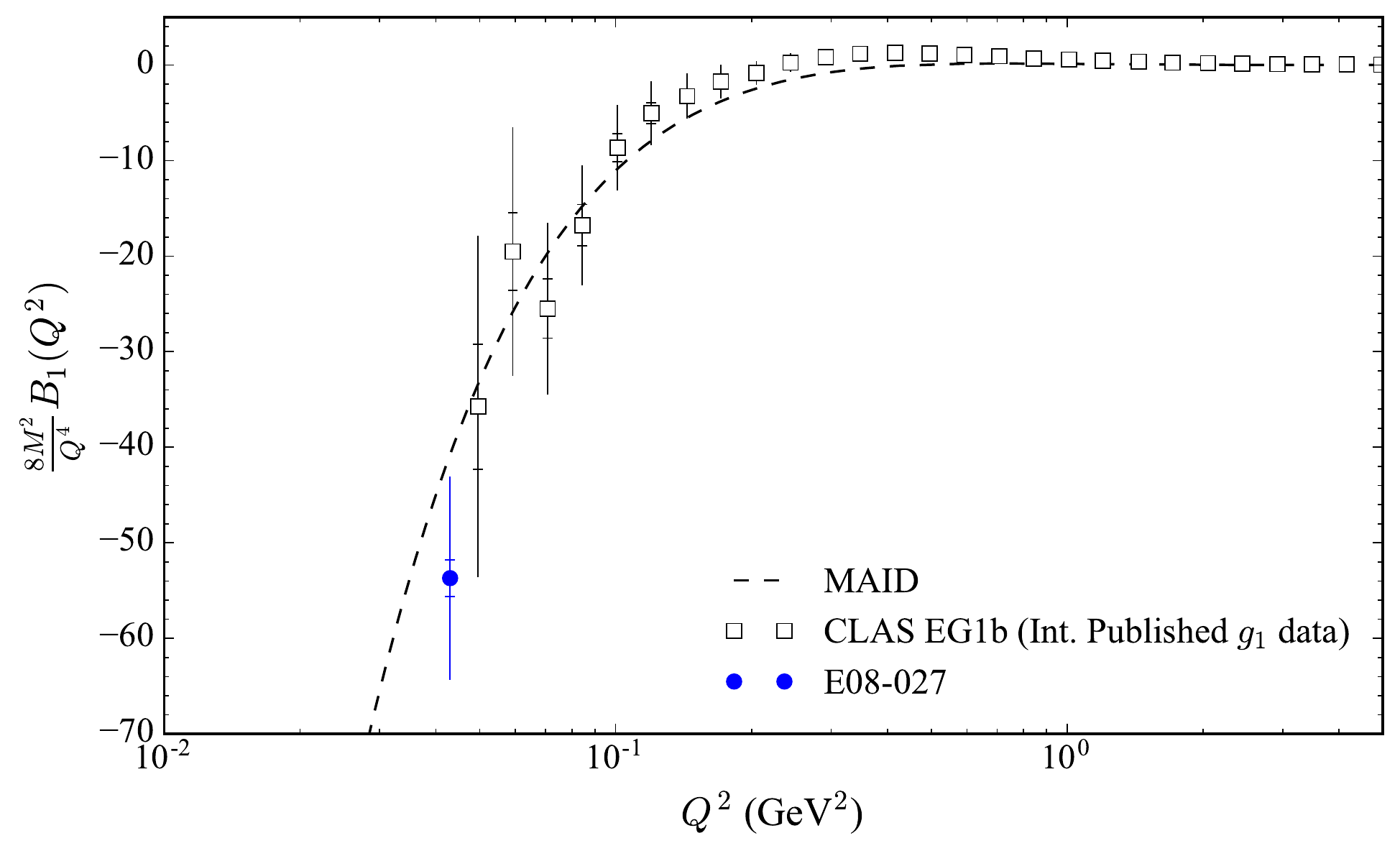}
\caption{$g_1(x,Q^2)$ contribution to the $\Delta_1$ integrand.}
\label{b1result}
\end{figure}

The $B_1(Q^2)$ integral in equation~\eqref{B1int} is very similar to the first moment of $g_1(x,Q^2)$ with $\beta_1(\tau)$ approximately equal to one. Adding in the $1/Q^4$ weighting and other terms, the $g_1(x,Q^2)$ contribution to the $\Delta_1$ integrand is shown in Figure~\ref{b1result}. Almost all of the integrand's strength comes from $Q^2<1 $GeV$^2$. The $F_2(Q^2)$ term is provided by the Arrington fit~\cite{ArringtonFit} to $G_E(Q^2)$ and $G_M(Q^2)$ where
\begin{equation}
F_2(Q^2) = \frac{G_M(Q^2)-G_E(Q^2)}{1+\tau}\,,
\end{equation} 
and $\tau$ = $Q^2/4M^2$. 

Integrating Figure~\ref{b1result} with respect to $Q^2$ gives the $g_1(x,Q^2)$ portion of $\Delta_1$ in the region of 0.43 $< Q^2 <$ 5.0 GeV$^2$. The unmeasured low $Q^2$ region of $\Delta_1$ is found by interpolating between the data higher at $Q^2$ and $Q^2=0$. The GDH sum rule constrains the slope\footnote{The GDH slope predicts the behavior of $g_1(x,Q^2)$ as $Q^2$ $\rightarrow$ 0: See Chapter~\ref{GDHSLOPECHAPTER}. The slope of $G_E(Q^2)$ as $Q^2\rightarrow 0$ give the charge radius of the proton~\cite{Povh}.} of the interpolation~\cite{Hyperfine}. The result is~\cite{Hyperfine2} 
\begin{equation}
\Delta_1[0,Q_i^2] = {\bigg(}-\frac{3}{4}\kappa_p^2r_p^2 + 18M^2c_1{\bigg)}Q_i^2\,,
\end{equation}
where $r_p$ is the charge radius of the proton and $c_1$ = 4.94 $\pm$ 0.30 (stat) $\pm$ 1.22 (sys) GeV$^{-4}$ is a constant of the fit. The CLAS EG1b parametrization provides the $Q^2>$ 5 GeV$^2$ contribution.

\begin{table}[htp]
\begin{center}
\begin{tabular}{ l c  c  r  r r } \hline
  Term& $Q^2$ (GeV$^2$) & Contribution & Result & Stat  & Sys    \\ \hline
  $\Delta_1$ & (0,0.043)& $F_2$ and $g_1$&  1.28 & 0.20 & 0.83 \\
             & (0.043,5.0)& $F_2$&  7.65 & $-$ & 0.45 \\
             & (0.043,5.0)& $g_1$&  $-$0.77 & 0.22 & 2.46 \\
             & (5.0,$\infty$)& $F_2$&  0.00 & $-$ & $-$ \\
             & (5.0,$\infty$)& $g_1$&  0.45 & $-$ & 0.45 \\ \hline
   Total $\Delta_1$ & & & 8.63 & 0.30 & 4.19 \\ \hline
\end{tabular}
\caption{\label{E08027Delta1}Results for the $\Delta_1$ contribution to the hydrogen hyperfine splitting.}
\end{center}
\end{table}

The results for $\Delta_1$ across the entire kinematic phase space are shown in Table~\ref{E08027Delta1}, and agree well with $\Delta_1$ = 8.85 $\pm$ 0.30 (stat) $\pm$ 3.57 (sys\footnote{The model error is added linearly with the systematic error here.}) from Ref~\cite{Hyperfine2}. The value for the charge radius of the proton used in this analysis is $r_p$ = 0.88 fm.

Combining the results of Chapter~\ref{sec:OPE} and Chapter~\ref{sec:Hyper}, the $g_2(x,Q^2)$ contribution to the hyperfine splitting is written to leading twist as
\begin{equation}
\label{B2WW}
B_2^{WW} (Q^2) = \int_0^{x_{\mathrm{th}}}{\bigg[}4\sqrt{\tau(\tau+1)}-4\tau -2\sqrt{\tau}\mathrm{ln}{\bigg(}\frac{\sqrt{\tau(\tau+1)}+\sqrt{\tau} }{\tau}{\bigg)}{\bigg]}g_1(x,Q^2)dx\,.
\end{equation}
Previously analyses in Ref~\cite{Hyperfine3} and Ref~\cite{Hyperfine2} use equation~\eqref{B2WW} to help estimate the $g_2(x,Q^2)$ effect on $\Delta_2$ and it provides a good reference for comparison to the E08-027 transverse data. It also quantifies the magnitude of higher twist effects on the $g_2(x,Q^2)$ extracted from the E08-027 data. 

\begin{figure}[htp]
\centering     
\includegraphics[width=.80\textwidth]{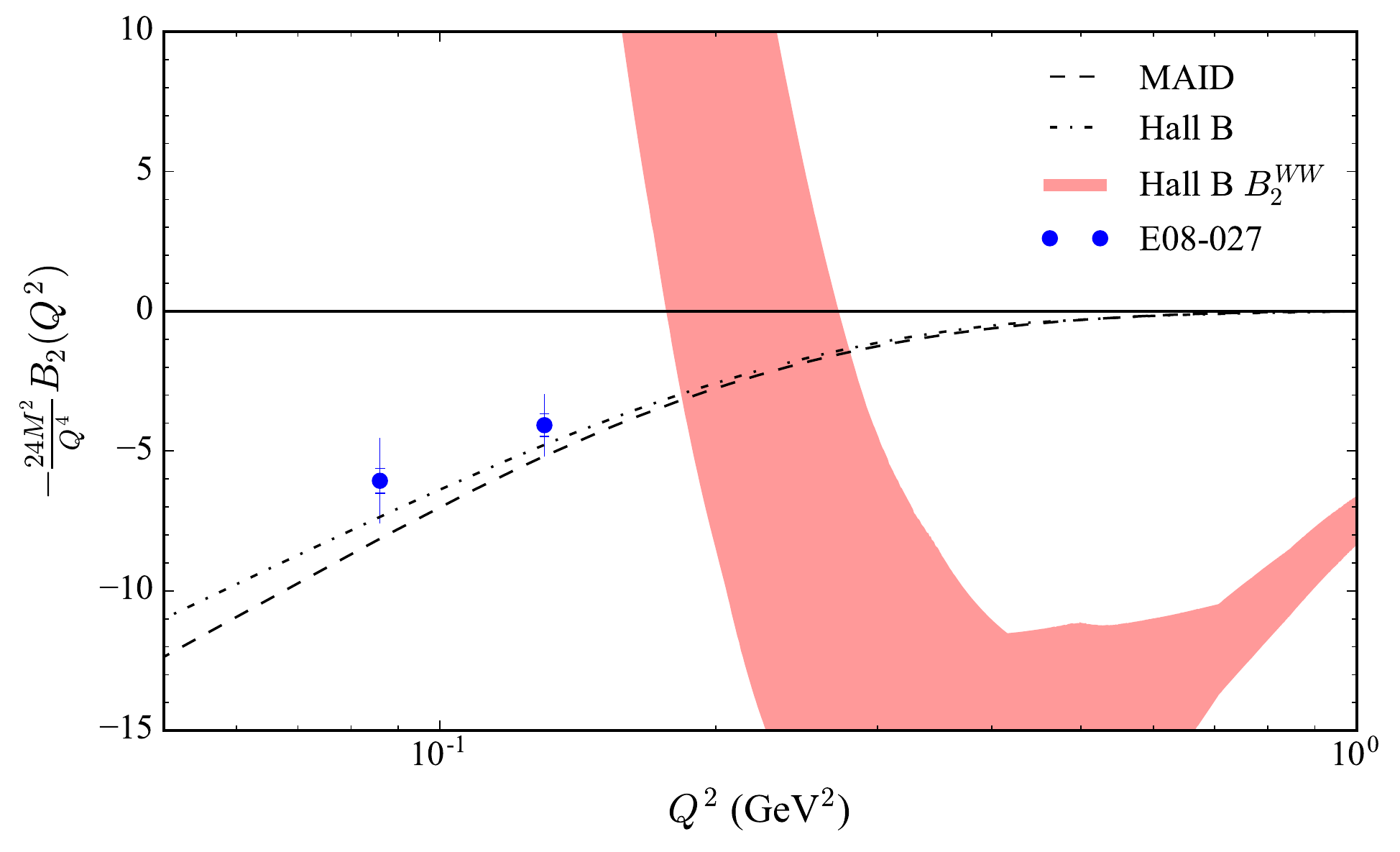}
\caption{$g_2(x,Q^2)$ contribution to the $\Delta_2$ integrand. The width of the colored band is the uncertainty in the measured $B_2^{WW}$ result from the EG1b data.}
\label{b2result}
\end{figure}

The $g_2(x,Q^2)$ and $g_2^{WW}(x,Q^2)$ contributions to the $\Delta_2$ integrand are shown in Figure~\ref{b2result}. Not surprisingly, there are significant higher twist effects at the low momentum transfer of the E08-027 data. The leading twist prediction is provided accurately by the Hall B model, which is expected because the model is a phenomenological fit that includes the CLAS EG1b data. Even at the largest $Q^2$ settings of E08-027, the data still represent a sizable portion of the $\Delta_2$ integrand.

The limited kinematic coverage of the $g_2(x,Q^2)$ does not a permit a comprehensive analysis of the $\Delta_2$ integrand, but it can still be checked against the model predictions. The results of the comparison are shown in Table~\ref{E08027Delta2}, and the data is consistent (within uncertainty) with both models. The uncertainty arising from using the trapezoidal rule on just two data points is 5\%. This is estimated by calculating a full numerical integration of the model in the data's momentum transfer range and then repeating the same integration with only the end points included.

\begin{table}[htp]
\begin{center}
\begin{tabular}{ l c   c  c r }\hline
  $\Delta_2$& $Q^2$ (GeV$^2$) & Result & Stat  & Sys    \\ \hline
  MAID &  (0.086,0.130)&   $-$0.29 & $-$  & $-$ \\
  Hall B           & (0.086,0.130)&   $-$0.26 & $-$ & $-$ \\
   E08-027   & (0.086,0.130)&   $-$0.22 & 0.03 & 0.06  \\ \hline
\end{tabular}
\caption{\label{E08027Delta2}Comparison of  the $\Delta_2$ contribution to the hydrogen hyperfine splitting.}
\end{center}
\end{table}

The remaining portions of the $\Delta_2$ term must be provided by a model, and this is where significant discrepancies arise. A comparison of MAID\footnote{MAID is limited to the resonance region and $Q^2<$ 5 GeV$^2$, but these are the dominant kinematics for the $\Delta_2$ integrand.} , Hall B and the Hall B (HB 2007) model version used in Ref~\cite{Hyperfine3} and Ref~\cite{Hyperfine2} is shown in Table~\ref{ModDelta2}. More recent model calculations disagree by more than 100\% from the previously published results and the newer models are in agreement with the E08-027 results. The lower momentum transfer E08-027 data will expand the data coverage even further into the low $Q^2$ region, providing additional confidence to the model calculations. Even still, the two data points shown here significantly impact the analysis of Ref~\cite{Hyperfine3,Hyperfine2}.
\begin{table}[htp]
\begin{center}
\begin{tabular}{ l c  r  r  r  } \hline
  Term& $Q^2$ (GeV$^2$) & MAID & Hall B & HB 2007\\ \hline
  $\Delta_2$ & (0,0.05)&-$0.87$& $-$0.80& $-$0.23  \\
             & (0.05,20)& $-$1.26 & $-$1.16 & $-$0.33 \\
             & (20,$\infty$)&0.00 &  0.00 & 0.00 \\ \hline
   Total $\Delta_2$ & &$-$2.13 & $-$1.96 & $-$0.56 \\ \hline
\end{tabular}
\caption{\label{ModDelta2}Model results for $\Delta_2$ in the the hydrogen hyperfine splitting calculation.}
\end{center}
\end{table}

\section{Summary of Results}
The moment results presented in this chapter are summarized in Table~\ref{MomentTable}. These results can be improved with an unpolarized cross section from the data. This would remove the largest source of model dependence on the results. The majority of the work remaining, in this regard, is on understanding the acceptance and determining the useful acceptance cuts. The acceptance correction is calculated from a Monte Carlo simulation, and is tuned to match the optics reconstructed variables. To first order, the uncertainty in the acceptance is a product of the uncertainties in the reconstructed variables. For a tight cut, the acceptance area is smaller and easier to reproduce in simulation, but has a larger systematic uncertainty because the smaller area is more sensitive to the reconstructed uncertainties. The difficulty is in optimizing the acceptance cuts and simulation to minimize the total uncertainty. This process is nearly complete for the longitudinal setting and work has begun on transferring the results to the transverse kinematic settings.   

\begin{table}[htp]
\begin{center}
\begin{tabular}{ l  c r  r  r r } \hline
  Quantity & $Q^2$(GeV$^2$) & Result&  Stat & Sys & Units  \\ \hline
  $\Gamma_1$& 0.043& $-$0.015 &  0.001 & 0.002 & N/A\\ 
  $I_A$& 0.043& $-$0.84&  0.024 & 0.155 & N/A\\ 
  $\gamma_0 $& 0.043& $-$1.71 &  0.081 & 0.331 & 10$^{-4}$ fm$^4$\\ 
  $\gamma_0^*$& 0.043& $-$0.76 &  0.063 & 0.204& 10$^{-4}$ fm$^6$\\ \hline
  $\Delta_1$& (0.043,$\infty$)& 8.63 &  0.30 & 4.19 & ppm\\ 
  $\Delta_2$& (0.086,0.130) & $-$0.22 &  0.03 & 0.06 & ppm\\  \hline
\end{tabular}
\caption{\label{MomentTable}Summary of the E08-027 moment results presented in Chapter~\ref{ch:Results}.}
\end{center}
\end{table}

 A preliminary analysis, following the procedures of this chapter, of the 2.5 T settings has also been carried out. The results of this analysis are described in detail in Appendix~\ref{app:Appendix-I}. The majority of the work left to finalize this data is: run-by-run data comparison to check for systematic effects, experimental dilution factor and packing fraction determination, and unpolarized cross section measurements. This work is complicated by the presence of unsolved yield drifts in the data. The yield drifts are differences above the 5\% level in normalized yields at a given spectrometer momentum setting. A summary of these yield drifts is found in Ref~\cite{Yield}. Attempts to resolve these drifts have focused on changes in beam position (scattering angle) and the BPM reconstruction~\cite{JieBPM}. Work on this issue is still in process, but the data at the 2.5 T settings is statistics limited so the additional systematic uncertainty from unresolved drifts is potentially not harmful to the calculated moments. More work needs to be done on the impact of unresolved yield drifts in the final results.




\chapter{\sc Conclusion}
\label{ch:Conclusion}

Any explanation of the proton is not complete if it cannot describe how the proton interacts with particles external to its internal quarks and gluons. Although, the nature of QCD is such that these internal interactions influence the external behavior. For example, the spin structure of the proton has a noticeable effect on the energy levels in atomic hydrogen. This is remarkable considering the separation in energy  and length scales: the energy of electron orbitals in hydrogen is on the order of eV, but the proton mass is almost a GeV, and the electron is, on average, 5$\times$10$^{-11}$ m away from the proton of approximate radius 10$^{-15}$ m. Broad interactions with external magnetic and electric fields are characterized by the proton's generalized polarizabilities and are also related back to complex internal structure of the proton.

The E08-027 experiment performed the first measurement of the proton spin structure function, $g_2^p(x,Q^2)$, in the low momentum transfer region of 0.02 $<$ $Q^2$ $<$ 0.20 GeV$^2$. The low $Q^2$ region is useful for investigating the global properties of the proton, where the observed phenomena largely result from the collective interaction of the intrinsic quarks and gluons. Theoretical predictions in this non-pertubative QCD region are provided by an effective QCD Lagrangian: Chiral Perturbation Theory. The inclusion of hadronic degrees of freedom in the  $\chi$PT momentum expansion is non-trivial and comparisons between the data and theory are crucial for a full understanding of the proton. Both the hydrogen hyperfine splitting contributions and generalized polarizabilties are determined from integrals of the proton structure functions.

The results of this thesis represent significant progress from the previous E08-027 theses~\cite{Chao3,MelissaT,JieT,MinT,PengiaT}, including a full characterization of the radiation corrections procedure on the unique angular dependence present in the data. The statistics presented are the full analyzing power of the 5 T kinematic settings. At the longitudinal setting, the statistical error bars are substantially smaller than the previously published data of CLAS EG1b~\cite{EG1b}. 

The integrated structure function moments of $g_1(x,Q^2)$ are largely in agreement with the $\chi$PT calculations from Pascalutsa $et\,al.$~\cite{Gold,PascaCom} and the CLAS EG1b data. The data also trend towards the real photon point of the GDH slope for $\Gamma_1(Q^2)$ and the GDH sum for $I_A(Q^2)$. There is, however, a bigger discrepancy in the results of the forward spin polarizability and is, in part, related to the treatment of the $g_2(x,Q^2)$ component. Future analysis of the 2.5 T data set and CLAS EG4~\cite{EG4} data could potentially help resolve or confirm the observed difference. Combining both data sets gives the $g_1(x,Q^2)$ and $g_2(x,Q^2)$ components to the forward spin polarizability and negates the need to resort to a Rosenbluth-like separation for the transverse component.


Analysis of the spin structure function contribution to the calculation of the hydrogen hyperfine splitting energy levels show agreement with the previously published results for the $g_1(x,Q^2)$ component, but significant differences in the $g_2(x,Q^2)$ terms. This is perhaps not surprising given the larger amount of $g_1(x,Q^2)$ data compared to the other spin structure function. The data also show significant higher twist effects in the transverse E08-027 results. Better agreement between data and model is achieved with the use of newer models, but these results differ by over 100\% from the previous model results~\cite{Hyperfine3,Hyperfine2}. The comparison range can be extended even further into the dominant low momentum transfer region with the 2.5 T data set. 

In conclusion, the E08-027 data provides an experimental determination of a previously unmeasured kinematic region for the proton's spin structure functions. The low momentum transfer data is especially useful for $g_2(x,Q^2)$ because the next lowest published data point is from RSS~\cite{RSS} at $Q^2$ = 1.3 GeV$^2$ and is an order of magnitude higher. Data at these low momentum transfers gives insights into the hadronic structure of the proton and when combined with the higher $Q^2$ data show the transition into partonic degrees of freedom. Knowledge of both kinematic regions is required for a complete understanding of the proton.

\addcontentsline{toc}{chapter}{\sc Appendices}  
\appendixtitletocon                             
\renewcommand\appendixpagename{\sc Appendices}  
\appendixpage                                   
\begin{appendices}
\chapter{\sc DAQ Deadtime Study}
\label{app:Appendix-A}

The deadtime is defined as the probability that any given trigger will be lost and not processed by the data acquistion system,
\begin{equation}
DT = \frac{N_{\mathrm{rej}}}{N_{\mathrm{acc}}+N_{\mathrm{rej}}}\,,
\end{equation}

where $N_{\mathrm{acc}}$ is the number of triggers accepted in a particular time period, and $N_{\mathrm{rej}}$ is the number of triggers rejected in the same time period. In a simple single-trigger, non-prescaled system the DAQ deadtime is approximately the time it takes for that trigger to be converted, processed, and transferred out of the front-end module

\begin{equation} \label{eq:DT}
DT \approx D_r + D_c  \,,
\end{equation}

\noindent where $D_r$ is the readout deadtime and $D_c$ is the conversion deadtime~\cite{RDT}. The conversion deadtime is correlated with the time it takes for a front-end ADC (TDC) module to complete an analog (time)-to-digital conversion, while the readout deadtime is correlated with the time it takes to transfer data out of the front-end modules. The two deadtimes are not entirely independent because first the front-end conversion occurs and then all the modules are read. Using Poisson probability theory, the two deadtime components are broken down into two infinite sums
\begin{equation} \label{eq:DT_tot}
\begin{aligned}
D_c &=  \sum_{n=1}^\infty \frac{\mu_c^n e^{-\mu_c}}{n!}\,, \\ 
D_r &=  \sum_{n=1}^\infty \frac{\mu_r^{b+n} e^{-\mu_r}}{(b+n)!} \,,
\end{aligned}
\end{equation}

\noindent where $\mu_c = R \tau_c$, $\mu_r = R (\tau_r -\tau_c)$, R is the trigger rate, $\tau_c$ is the conversion time, $\tau_r$ is the readout time and $b$ is the buffer factor~\cite{RDT}.  

The buffer factor is a feature of the trigger supervisor that allows it to store converted events in a buffer before read-out. Provided the buffer is not full, new events are processed while older events are being read out, decoupling the conversion and readout deadtimes. The conversion time is fixed and module dependent. The readout time depends on the number of modules being readout. When running the DAQ at low to moderate rate in buffered mode, b = 8, the buffer factor reduces the the readout contribution to the overall deadtime. The deadtime is then dominated by the conversion time. As the rate increases the probability of a full buffer also increases and readout begins to dominate the deadtime. 

\section{Improving Deadtime}
To limit the error deadtime corrections introduce into the cross section normalization, experiments typically aim for a maximum of 20$\%$ deadtime~\cite{Bob}. In the past, this translated into a maximum non-prescaled acceptable DAQ rate of 4 kHz. To increase the rate, a third FASTBUS crate was added to both spectrometer DAQs. The read out of each crate is done in parallel so by distributing the module population throughout the three crates, the readout time of each was decreased. This resulted in an improvement of the DAQ rate from 4 kHz to 6 kHz while maintaining comparable deadtime. More information on the read out of the FASTBUS crates is found in Ref~\cite{TS}.
\begin{figure}[hbt]
\includegraphics[scale=.75]{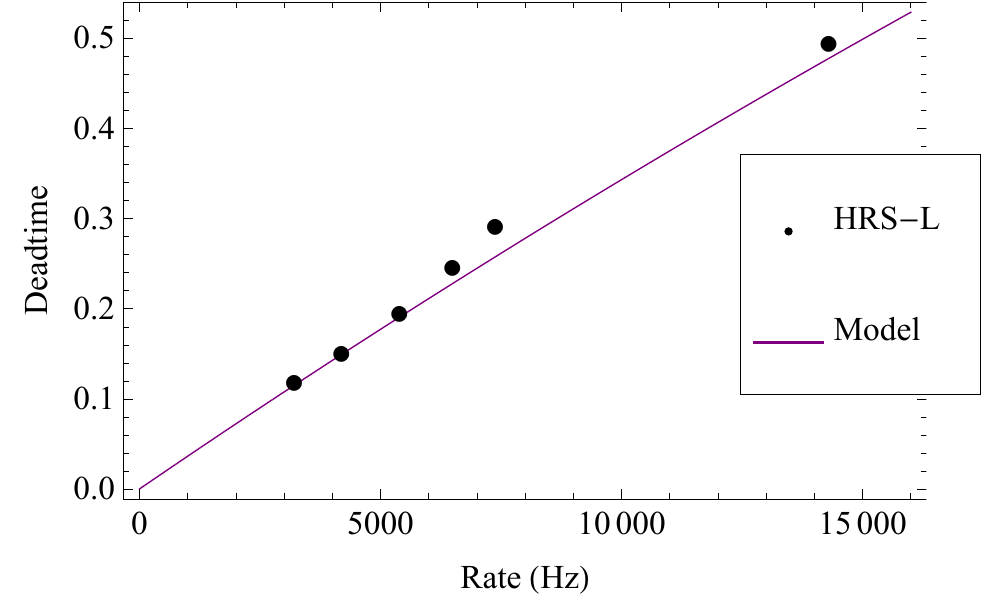}
\includegraphics[scale=.75]{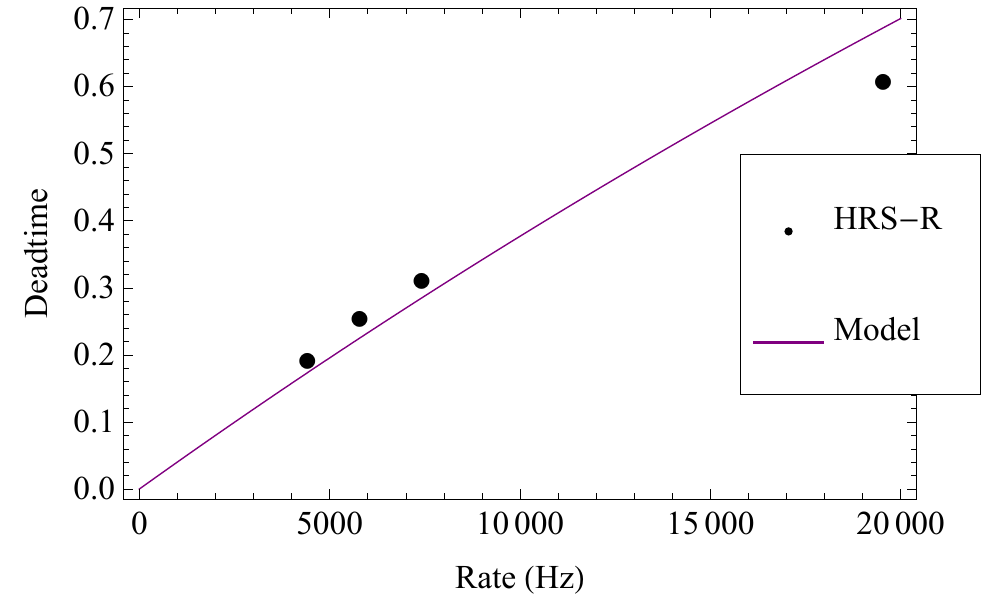}
\caption{Deadtime for both HRSs during commissioning.}
\label{fig:DTMod}
\end{figure}

\section{Deadtime Testing}

The updated DAQ system was initially tested during commissioning for the experiment in December 2011. During the test, the LHRS DAQ consisted of  three FASTBUS crates and a TS crate that included scaler modules. The total deadtime is a sum of the contributions from the TS crate and slowest FASTBUS crate. The data simulated experimental conditions, using an electron beam incident on a carbon-foil target. The results in Table~\ref{tab:DTresults} highlight the 6 kHz improvement, and include the conversion and readout times used in the model in Figure~\ref{fig:DTMod}. These times were measured with cosmic ray data. The TS  crates cannot be buffered. The busy time listed for these components is the total busy time and is labeled as the ``readout" time. The model from equation~\ref{eq:DT_tot}, plotted in Figure~\ref{fig:DTMod}, compares favorably with the experimental data. In addition, two-crate performance was tested on the RHRS because its third crate had not yet been installed. The RHRS results are shown in Table~\ref{tab:DTresults2}
\begin{table}[ht]
\begin{center}
\begin{tabular}[t]{ |c|c|c|c|c| } \hline
Rate (kHz) &   DT (\%) & Crate & $\tau_c$ ($\mu$s) &$\tau_r$($\mu$s) \\ \hline
14.2   &    49 & TS &   & 36  \\     
7.3 &   29  & FASTBUS & 12 & 90 \\ \cline{3-5}
6.5 & 24  \\  
5.3 & 20 \\ 
4.2 & 15 \\ 
3.2 & 12\\ \cline{1-2} 
\end{tabular}
\caption{Results of the LHRS commissioning deadtime study.}
\label{tab:DTresults}
\end{center}
\end{table}

\begin{table}[ht]
\begin{center}
\begin{tabular}[t]{ |c|c|c|c|c| } \hline
Rate (kHz) &   DT (\%) & Crate & $\tau_c$ ($\mu$s) &$\tau_r$($\mu$s) \\ \hline
19.5   &    60 & TS &   & 44  \\     
7.4&   31  & FASTBUS & 12 & 100 \\ \cline{3-5}
5.7 & 25  \\  
4.4 & 20 \\  \cline{1-2}
\end{tabular}
\caption{Results of the RHRS commissioning deadtime study. }
\label{tab:DTresults2}
\end{center}
\end{table}

\chapter{\sc Reconstructed ROOT Tree Variable Definitions}
\label{app:Appendix-E}

The following ROOT tree variable definitions are useful when reconstructing the scattering angle and out-of-plane polarization angle (RHRS substitution is given by L$\rightarrow$R):
\begin{itemize}
\item $\theta_{\mathrm{BPM}}$ = L[R]rb.tgt\_0\_theta
\item $\phi_{\mathrm{BPM}}$ = L[R]rb.tgt\_0\_phi
\item $\theta_{\mathrm{rec}}$ = L[R].rec.th
\item $\phi_{\mathrm{rec}}$ = L[R].rec.ph
\item $\theta_{\mathrm{rec\_l}}$ = L[R].rec.l\_th
\item $\phi_{\mathrm{rec\_l}}$ = L[R].rec.l\_ph
\end{itemize}
The ``BPM" variables are in the BPM coordinate system, and the reconstructed ``rec\_l" variables are in the Hall coordinate system. The ``rec" variables are reconstructed in the target coordinate system (TCS). Both sets of ``rec" variables incorporate target field effects, and use the ``tr.tg\_*" variables as a starting point. The following trigonometric identifies are useful in translating between the BPM coordinate system and the Hall coordinate system:
\begin{align}
\mathrm{cos}(\mathrm{atan}(x)) &= \frac{1}{\sqrt{1+x^2}}\\
\mathrm{sin}(\mathrm{atan}(x)) &= \frac{x}{\sqrt{1+x^2}}\,.
\end{align}

\chapter{\sc Nitrogen Model Tune}
\label{app:Appendix-C}

The Bosted-Mamyan-Christy  fit to inclusive electron scattering is considered to be a good (10-20\% level) fit to data for 0.2 $<$ $Q^2$ $<$ 5.0 GeV$^2$~\cite{Bosted3,Bosted1,Bosted2}. At lower $Q^2$, the large quasi-elastic and $\Delta$-resonance peaks are hard to model, and, in the case of nitrogen, there is lack of experimental data to constrain the fit. Given these the difficulties, the accuracy of the base fit is at best 20-30\% for nitrogen at the E08-027 kinematics as determined from the low $Q^2$ data of the Hall A experiment, Small Angle GDH (E97-110). The accuracy is improved by adjusting the free parameters in the fit to better match the data.

\section{The Small Angle GDH Experiment }
\begin{figure}[htp]
\begin{center}
\includegraphics[scale=.60]{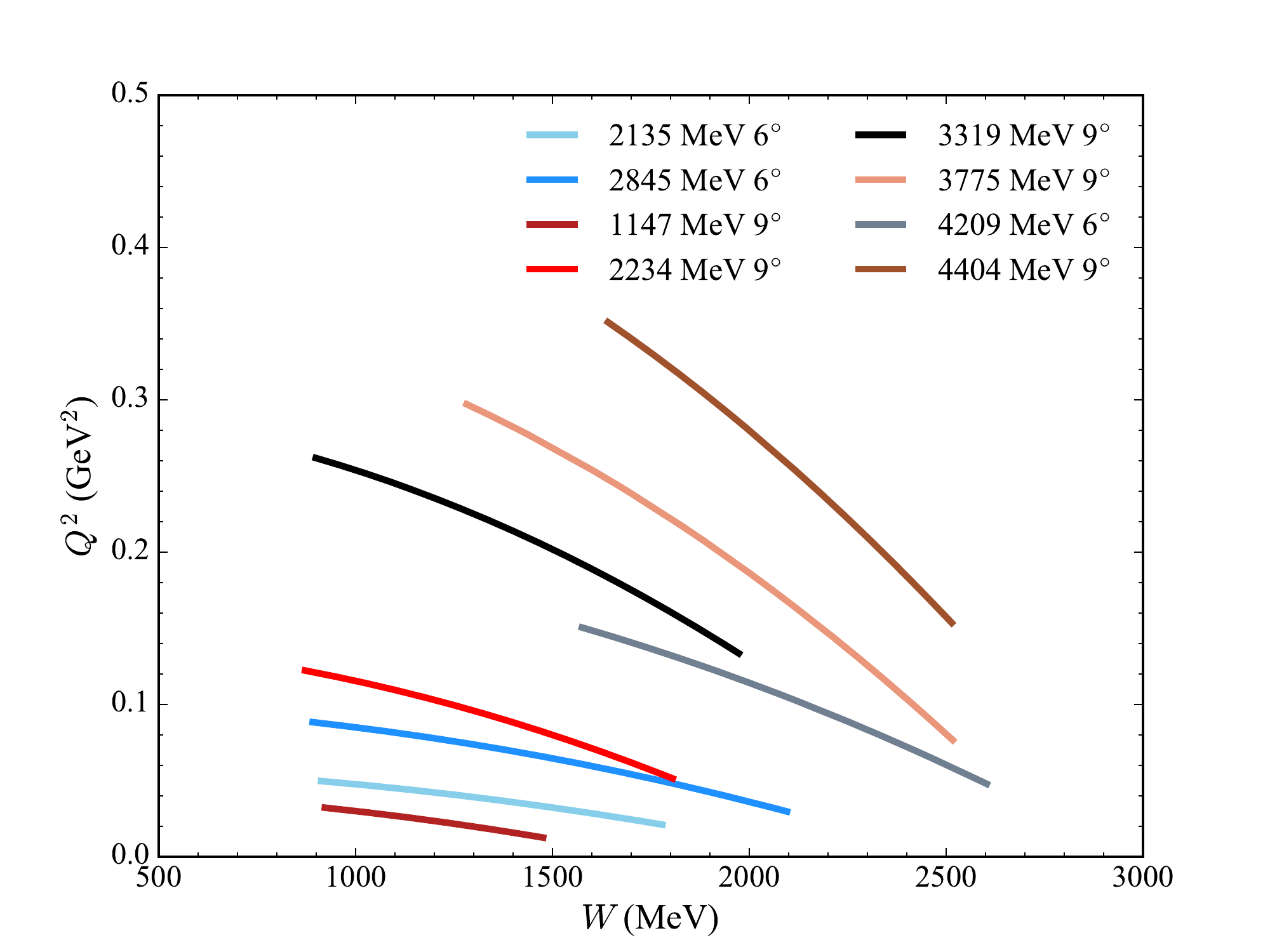}
\caption{Small Angle GDH experimental kinematics. }
\label{fig:saGDHkin}
\end{center}
\end{figure}

The Small Angle GDH (saGDH) experiment ran in Hall A at JLab~\cite{Vince}. The incoming electrons ranged in energy from from 1.1 GeV to 4.4 GeV and scattered  from a longitudinally or transversely polarized $^\mathrm{3}$He gas target. Small forward scattering angles of 6$^{\circ}$ and 9$^{\circ}$ meant the experiment probed a low four-momentum transfer squared region of 0.02 GeV$^{2}$ $<$ $Q^2$ $<$ 0.30 GeV$^2$. The gas target cell used in the experiment contained a small amount of nitrogen gas (in addition to the $^\mathrm{3}$He) to aid in the polarization.  In order to determine the nitrogen contribution to the  electron-helium scattering, the experiment took data with only the nitrogen gas inside a reference cell~\cite{PUBMAYBE}. The kinematics of the experiment are shown in Figure~\ref{fig:saGDHkin}.

\section{Tuning the Fit}
The Bosted-Mamyan-Christy fit is divided into three regions: the quasi-elastic, inelastic and dip regions. The dip region is in between the quasi-elastic and $\Delta$(1232) peaks. The total fit output is the sum of the three regions. Each region has three adjustable parameters: the Fermi-momentum ($k_F$), binding energy ($E_s$) and an overall scale factor in the quasi-elastic and inelastic channels. The dip channel is a Gaussian and its shaped is controlled with height, location and width parameters. The base model uses values from Ref~\cite{Superscale} for the Fermi-momentum and binding energy.
\begin{table}[htp]
\begin{center}
\begin{tabular}{ l c  c r }\hline
Region & $k_f$ (GeV) & $E_s$ (GeV) & Scale\\
\hline
Quasi-elastic & 0.196 (0.228) & 0.009 (0.0165) & 0.76 (0.0)\\
Inelastic &  0.330 (0.228) & 0.016 (0.0165) & 0.77(0.0)\\
&&&\\
& Width & Loc & Amp \\
\hline
Dip & 0.124 (0.046) & 0.861 (0.981) & 0.005 (0.005)\\ \hline
\end{tabular}
\caption{Free parameters for Bosted model saGDH tune.}
\label{Param}
\end{center}
\end{table}
The updated fit parameters are determined from a $\chi^2$ minimization on each channel and for each kinematic setting of the saGDH nitrogen data. The channels are not entirely independent so the minimization is iterated to get the best set of parameters in each region. The adjusted model parameters are an average of the new parameters at each setting. The results of the saGDH updated fit are shown in Table~\ref{Param}. Terms in parentheses are from the base model.

\section{Fit Performance}
\begin{figure}[htp]
\begin{center}
\includegraphics[scale=.70]{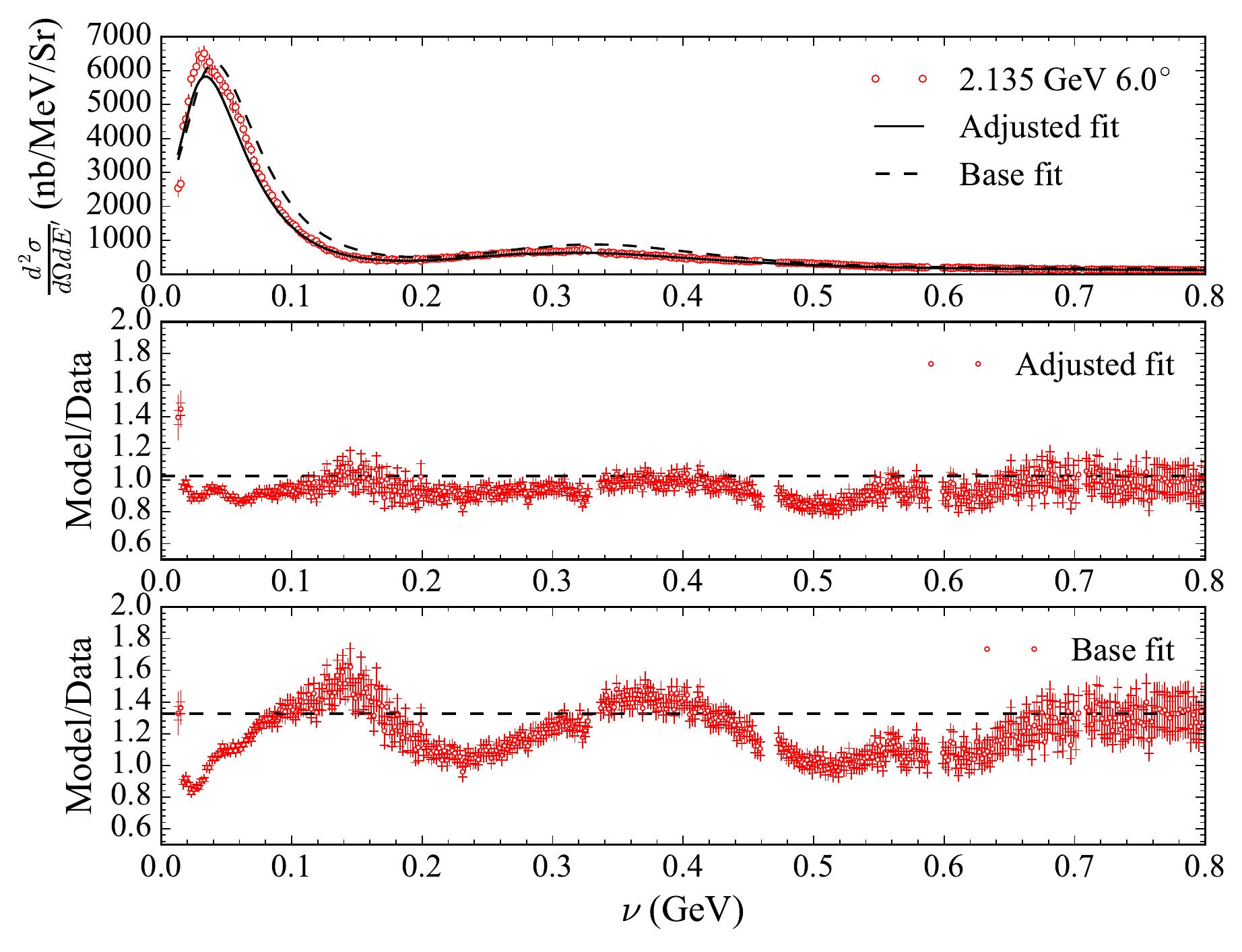}
\caption{Tuned Bosted-Mamyan-Christy fit for nitrogen at $E_0$ = 2135 MeV. }
\label{fig:saGDHtune}
\end{center}
\end{figure}

\begin{figure}[htp]
\begin{center}

\subfigure[$E_0$ = 2845 MeV, $\theta_{\mathrm{sc}}$ = 6$^{\circ}$ ]{\label{tune1}\includegraphics[width=.80\textwidth]{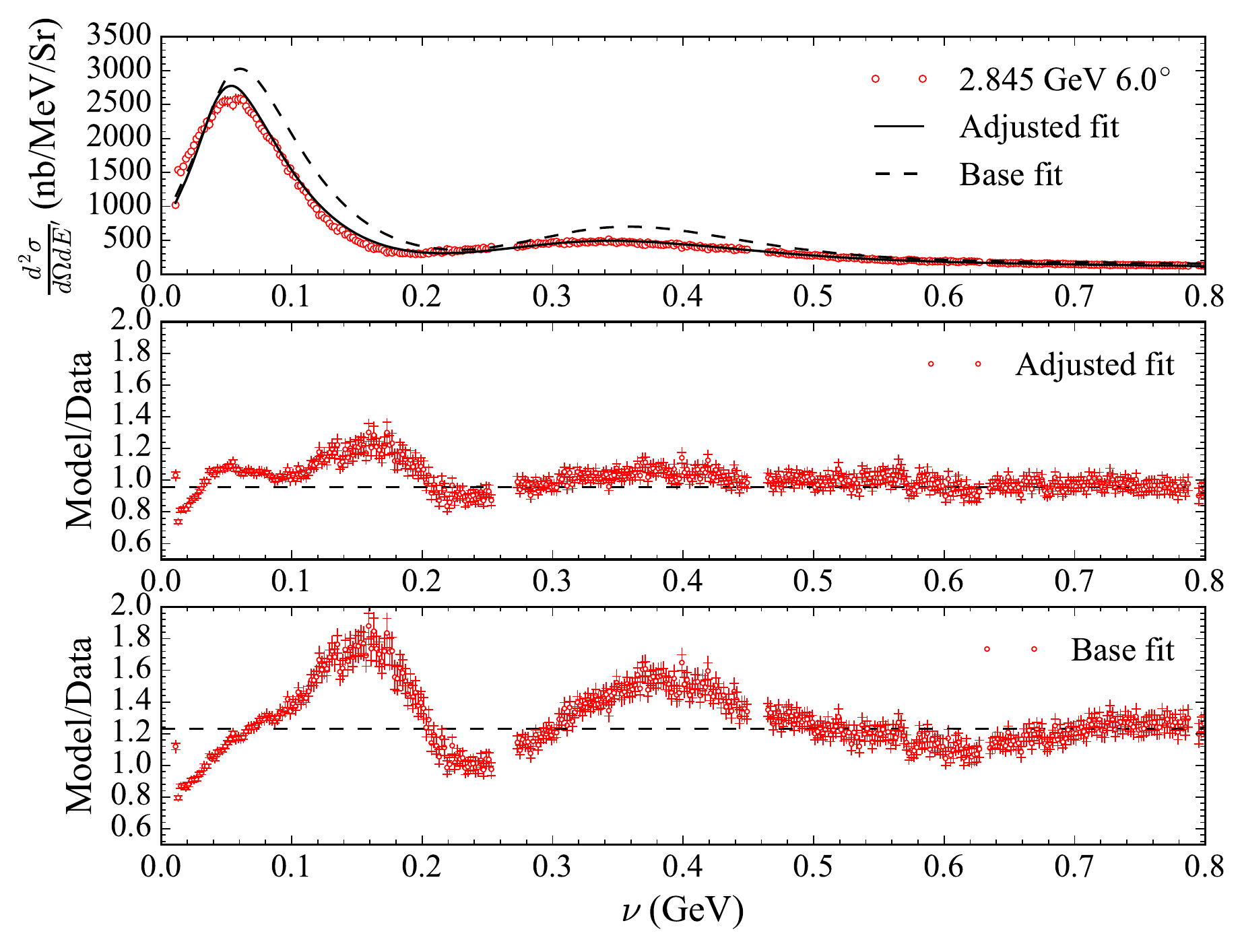}}
\qquad
\subfigure[$E_0$ = 1147 MeV, $\theta_{\mathrm{sc}}$ = 9$^{\circ}$ ]{\label{tune2}\includegraphics[width=.80\textwidth]{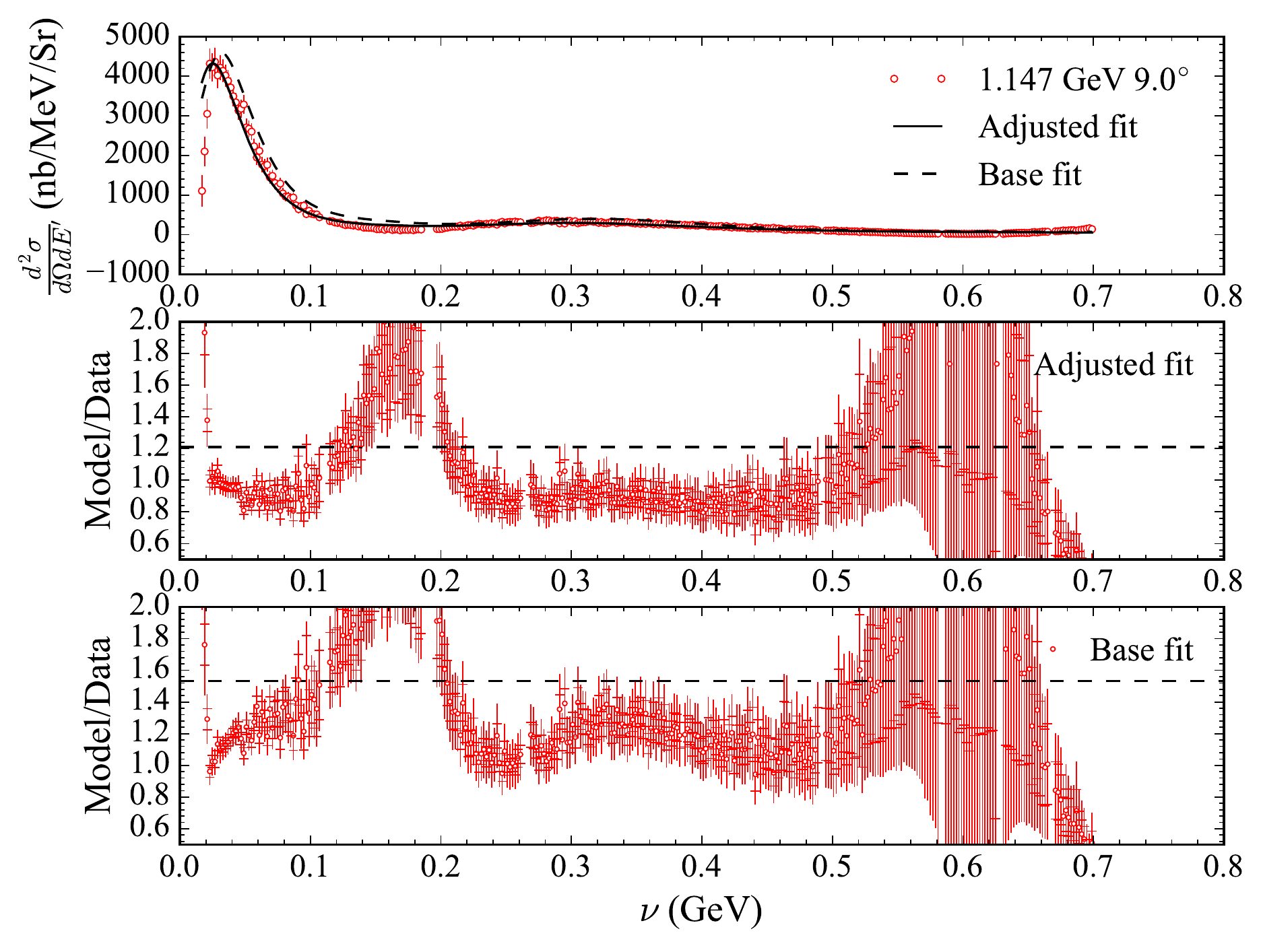}}
\caption{Tuned Bosted-Mamyan-Christy fit for nitrogen at $E_0$ = 2845 MeV, 1147 MeV. }
\label{fig:saGDHtune2}
\end{center}
\end{figure}

\begin{figure}[htp]
\begin{center}

\subfigure[$E_0$ = 2234 MeV, $\theta_{\mathrm{sc}}$ = 9$^{\circ}$ ]{\label{tune3}\includegraphics[width=.80\textwidth]{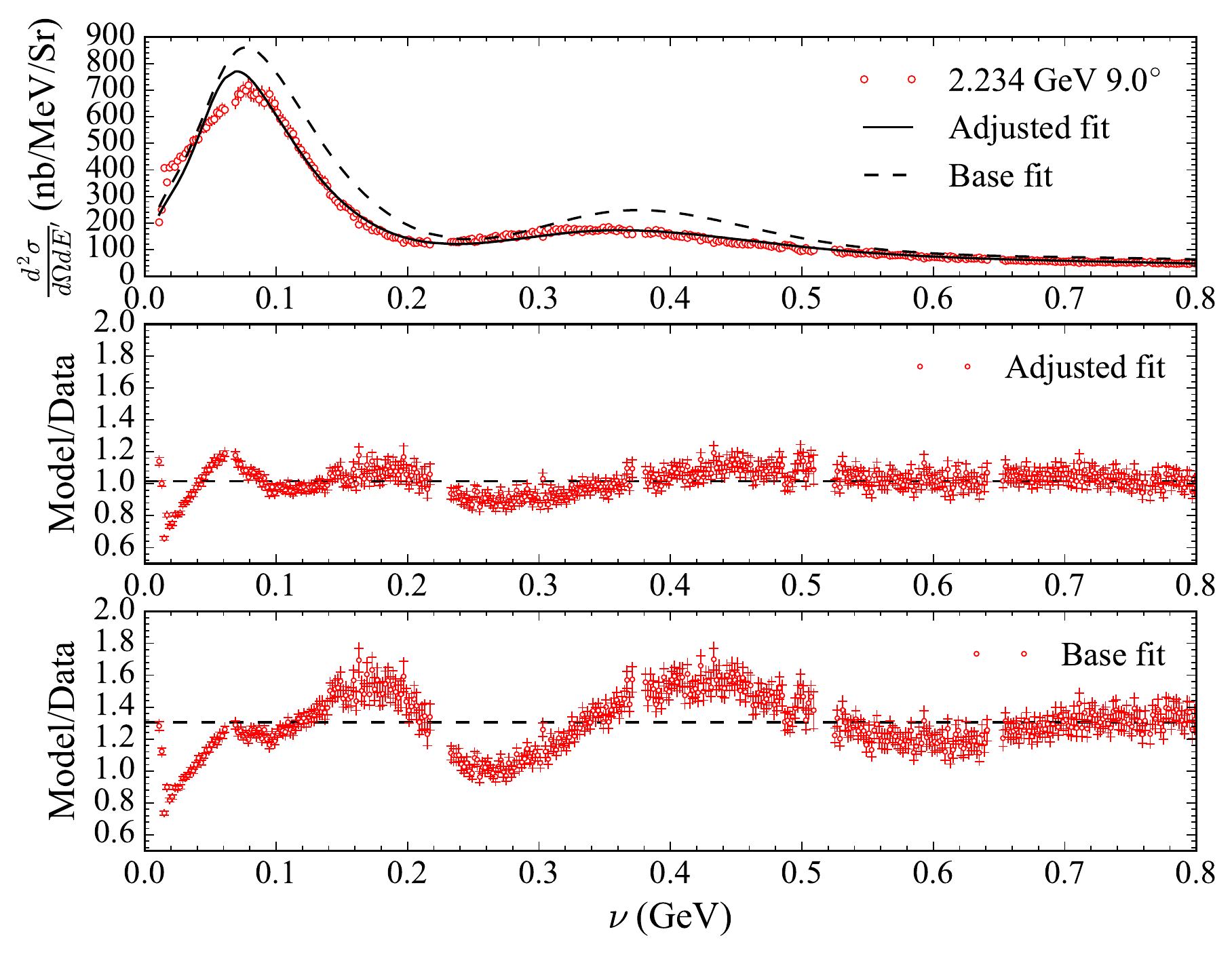}}
\qquad
\subfigure[$E_0$ = 3319 MeV, $\theta_{\mathrm{sc}}$ = 9$^{\circ}$ ]{\label{tune4}\includegraphics[width=.80\textwidth]{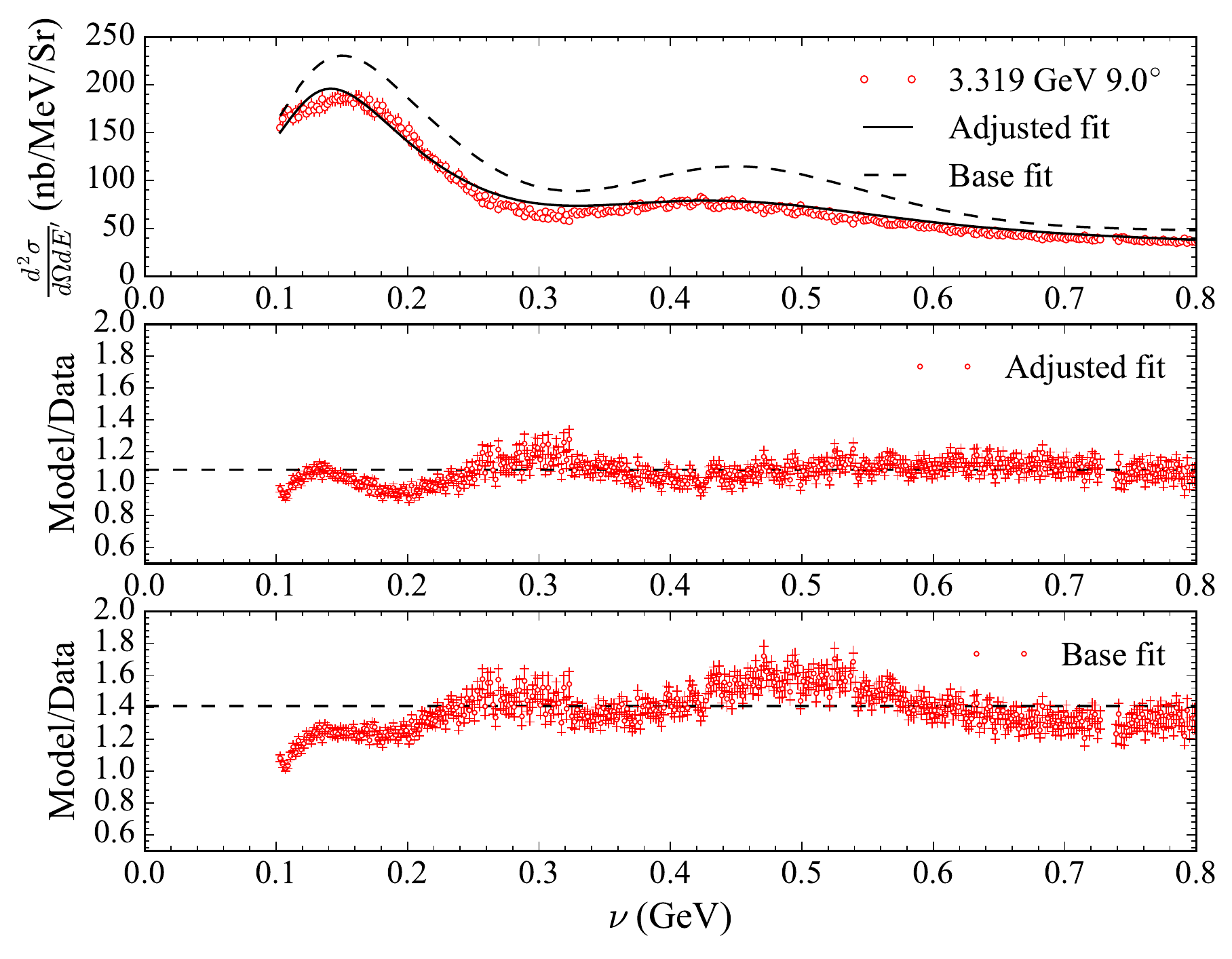}}
\caption{Tuned Bosted-Mamyan-Christy fit for nitrogen at $E_0$ = 2234 MeV, 3319 MeV. }
\label{fig:saGDHtune3}
\end{center}
\end{figure}

The updated fit is accurate at the 7\% level across the four-momentum transfer squared region of 0.02 GeV$^{2}$ $<$ $Q^2$ $<$ 0.30 GeV$^2$. The stated accuracy represents the average in the data-model ratio at each setting and each $\nu$-bin. An example of the improved fit is shown in Figure~\ref{fig:saGDHtune} through~\ref{fig:saGDHtune3}, along with a comparison to the base model. In the ratio plots, the dashed lines represent the average deviation between model and data. The adjustments to the parameters are purely empirical and no physics information should be derived from them. The applicability of this tune is limited to the $Q^2$ range fitted.


\chapter{\sc E08-027 Radiation Lengths}
\label{app:Appendix-D}

Real photon bremsstrahlung emission occurs as the electron beam interacts with the Coulombic fields of the materials it traverses before, after, and within the target. This bremsstrahlung emission resulting from material interaction is characterized by the material's radiation length. The radiation length of a material is formally defined as the thickness required for an electron to lose 1-1/$e$ of its energy as it travels through the material. 

For a single atomic species, the radiation length in mass per unit area is given by
\begin{equation}
	X_0 = \frac{A}{ N_A \sigma_\mathrm{rad}(Z)}\,, 
\end{equation}
and for a composite material it is given by the sum
\begin{equation}
	X_0^{-1} = \sum_k \frac{w_k N_A \sigma_\mathrm{rad}(Z_k)}{A_k} = \sum_k \frac{w_k}{X_0^k}\,,
\end{equation}
where $w_k$ represents the mass fraction for the $k$-th component of the total mass of the composite material. In order to accommodate differences in the physical thicknesses and densities of the various beam line materials, each material's radiation length is weighted by the material's thickness and density and then summed giving 
\begin{equation}
	t = \sum_j \frac{ \rho_j \ell_j }{X_0^j}\,,
\end{equation}
where the sum over $j$ runs over each composite material with physical thickness $\ell_j$ and mass density $\rho_j$. The weighted sum is referred to as the radiation thickness. The radiation lengths of some common materials are shown in Table~\ref{RadLengths}. The radiation lengths for the composite materials relevant to E08-027 are shown in Table~\ref{RadLengths2}. 

\begin{table} [t]
\begin{center} 
\begin{tabular}{ccccc}\hline
\bf $Z$ & {\bf atom} & \bf $A$ ($\mathrm{amu}$) & \bf $\sigma_\mathrm{rad}$ ($\mathrm{barns}$)& \bf $X_0$ ($\mathrm{g/cm^2}$ ) \\ \hline 
 & & & & \\ 
1	&	$\mathrm{H}$	&	1.00794	&	0.02655	&	63.0435 \\ 
2	&	$\mathrm{He}$	&	4.00260	&	0.07047	&	94.3224 \\ 
4	&	$\mathrm{Be}$	&	9.01218	&	0.22956	&	65.1900 \\ 
 & & & & \\ 
5	&	$\mathrm{B}$	&	10.81100	&	0.34073	&	52.6868 \\ 
6	&	$\mathrm{C}$	&	12.01070	&	0.46711	&	42.6969 \\ 
7	&	$\mathrm{N}$	&	14.00670	&	0.61226	&	37.9879 \\ 
8	&	$\mathrm{O}$	&	15.99940	&	0.77597	&	34.2382 \\ 
 & & & & \\ 
11	&	$\mathrm{Na}$	&	22.98977	&	1.37637	&	27.7362 \\ 
12	&	$\mathrm{Mg}$	&	24.30500	&	1.61235	&	25.0315 \\ 
13	&	$\mathrm{Al}$	&	26.98154	&	1.86596	&	24.0112 \\ 
14	&	$\mathrm{Si}$	&	28.08550	&	2.13706	&	21.8231 \\ 
 & & & & \\ 
19	&	$\mathrm{K}$	&	39.09830	&	3.74958	&	17.3151 \\ 
20	&	$\mathrm{Ca}$	&	40.07800	&	4.12249	&	16.1434 \\ 
33	&	$\mathrm{As}$	&	74.92160	&	10.41949	&	11.9401 \\ 
38	&	$\mathrm{Sr}$	&	87.62000	&	13.51897	&	10.7624 \\ 
56	&	$\mathrm{Ba}$	&	137.32700	&	27.45233	&	8.3066 \\ 
\hline 
\end{tabular}
\caption{Radiation lengths for atomic species~\cite{PDG}.}
\label{RadLengths}
\end{center}
\end{table}
\begin{table} [ht]
\begin{center} 
\begin{tabular}{ccc}\hline
\bf  Material &  Ammonia &  Kapton  \\ \hline 
 & & \\ 
\bf Formula & $\mathrm{^{14}NH_3}$ & $\left ( \mathrm{C_{22}H_{10}N_2O_5} \right )_n$ \\ 
 & & \\ 
 C by wt.	&	-	&	0.691133 \\ 
 O by wt. 	&	-	&	0.209235 \\ 
 N by wt.	&	0.822453	&	0.073270	\\ 
 H by wt.	&	0.177547	&	0.026362  \\ 
 & & \\ 
\bf $X_0\ \left ( \mathrm{g/cm^2} \right )$  & 40.87 	&	40.5761	\\ \hline
\end{tabular}
\caption{Radiation lengths for composite materials~\cite{PDG}.}
\label{RadLengths2}
\end{center}
\end{table}

In addition to the bremsstrahlung processes, electrons elastically scatter with atomic electrons from the materials in the beam path. This results in the ionization of the struck atom and energy loss for the electron. The energy lost to a single ionizing collision is given by~\cite{TSAI}
\begin{align}
\frac{\xi}{\rho x} = & \frac{Z a}{A\beta^2}\,,  \\
a = & 2 \pi N_A r_e^2 m_e c^2 = 0.15353747\ \mathrm{MeV \cdot \frac{cm^2}{mol}}\,, \\
\beta =& \frac{v}{c} = \frac{pc}{E} \,,  
\end{align}
where $\rho$ and $x$ are the materials mass density and thickness, $N_A$ is Avogadro's numbers and $r_e$ is the classical electron radius. The effects of the ionization of many atoms results in a net energy loss to the electron, which is equivalent to adjusting the incoming and outgoing electron energies such that
\begin{align}
E_{\mathrm{inc.}} &= E_{\mathrm{beam}} - \Delta_s\,,\\
E_{\mathrm{fin.}} &= E' + \Delta_p\,,\\
E' & \simeq p_{\mathrm{spect}}\,,
\end{align} 
where $\Delta_{s,p}$ is energy loss due to the ionizing collisions; for a highly relativistic electron $p \simeq E$.  The form of $\Delta_{s,p}$ was calculated by both Landau and Bethe-Bloch and is given by~\cite{Leo}
\begin{align}
\frac{\Delta}{\rho x} =&{\bigg(} \frac{\xi}{\rho x}{\bigg)}{\bigg[} 2 \log \left ( \frac{p c}{I} \right)  - \delta(X) + g {\bigg ]}\,,\\
\bar{g}       = & \log \left ( \gamma-1 \right )  - F(\gamma)\ \ \ \ \ \ \ \ \ \ \ \ \ \ \!\mathrm{mean\ energy\ loss\ (Bethe-Bloch)}\,,    \\
        g_\mathrm{mp}  = & \log\left [ \frac{2 \xi}{m_ec^2}\right ] -\beta^2 + 0.198\ \ \ \ \ \mathrm{most\ probable\ (Landau)}\,, \\
        F(\gamma) =  & \left [ 1 + \frac{2}{\gamma} -\frac{1}{\gamma^2}\right ] \log(2) - \frac{1}{8} \left [1 - \frac{1}{\gamma}\right ]^2 - \frac{1}{\gamma^2}\,,  \\
        \gamma  = & \frac{1}{\sqrt{1-\beta^2}} = \frac{E}{m_\mathrm{e} c^2}\,,
\end{align}
where $I$ is the mean excitation energy of the material and $\delta(X)$ is the density correction. The density correction is given as~\cite{Fermi22,Density}
\begin{eqnarray}
\delta(X) & = & \left \{ \begin{array}{ll}
                        \delta\! \left ( X_0^\delta \right ) \times 10^{2 \left ( X-X_0^\delta \right ) }   & X \le X_0^\delta \\
                        2\log(10) \left ( X - X_a^\delta \right ) + a_\delta \left ( X_1^\delta - X \right )^{m_\delta} & X_0^\delta < X < X_1^\delta \\
                        2\log(10) \left ( X - X_a^\delta \right )  & X \ge X_1^\delta
                \end{array} \right \} \\
	X & = & \log_{10} \left ( \frac{p}{m_e c} \right)\,, \\
	X_a^\delta & = & \frac{-C_\delta }{2\log(10)}\,, \\
	a_\delta & = & \frac{ \delta \left ( X_0^\delta \right ) + 2 \log(10) \left ( X_a^\delta - X_0^\delta \right ) }{\left ( X_1^\delta - X_0^\delta \right )^{m_\delta}}\,. 
\end{eqnarray}
where $C_{\delta}$, $X_0^\delta$, $X_1^\delta$, $m_\delta$  and $\delta(X_0^\delta)$ are material dependent and are shown in Table~\ref{DensityTabC} and Table~\ref{DensityTab}. The densities given are for the natural form of the element (gas, diatomic gas, liquid, solid) at 1 atm and 20$\mathrm{^oC}$. For all these materials in Table~\ref{DensityTabC}, $\delta(X_0)$ is zero and the density of ammonia listed in Table~\ref{DensityTabC} is for gaseous ammonia. For material densities different than those listed in Table~\ref{DensityTabC} and Table~\ref{DensityTab} the density correction parameters are adjusted as follows
\begin{eqnarray}
	{X_{a,0,1}^{\delta'}} & = & X_{a,0,1}^\delta - \frac{1}{2}\log_{10} \left ( \mathrm{\frac{N}{N_0}} \right )\,,
\end{eqnarray}
where $\mathrm{N_0}$ is the density from the table and $\mathrm{N}$ is the desired density~\cite{RadLength}.

\begin{table}[h]
\begin{center}
\begin{tabular}{ccccrcccc}\hline
 & $Z/A$ & $I$ ($e\mathrm{V}$) & $\rho$ ($\mathrm{g/cm^3}$)& $-C$ & $X_0$ & $X_1$ & $a$ &  $m$  \\ \hline
 & & & & & & & & \\ 
$\mathrm{^{14}NH_3}$	&	0.5972	&	53.69&	$\mathrm{8.2602E\!-\!04}$	&	9.8763	&	1.6822	&	4.1158	&	0.08315	&	3.6464 \\ 
$\mathrm{Kapton}$	&	0.5126	&	79.60	&	$\mathrm{1.4200E\!+\!00}$	&	3.3497	&	0.1509	&	2.5631	&	0.15971	&	3.1921 \\
& & & & & & & & \\  
\hline 
\end{tabular}
\caption{Density correction parameters for composite materials.}
\label{DensityTabC}
\end{center}
\end{table}

\begin{landscape}
\begin{table}
\begin{center}
\begin{tabular}{ccccccrccccc}\hline
 & & & & & & & & & & & \\ 
 &$Z/A$ &$I$ &$I'$ (g) &$I'$ (l/s) &$\rho$ &$-C$ &$X_0$ &$X_1$ &$a$ &$m$ &$\delta(X_0)$ \\ 
 & & $e\mathrm{V}$ & $e\mathrm{V}$ & $e\mathrm{V}$ & $\mathrm{g/cm^3}$ &  & & & & & \\ \hline
 & & & & & & & & & & & \\ 
$\mathrm{H}$	&	0.9922	&	19.2	&	19.2	&	19.2	&	$\mathrm{8.3748E\!-\!05}$	&	9.5835	&	1.8639	&	3.2718	&	0.14095	&	5.7273	&	0.00 \\ 
$\mathrm{He}$	&	0.4997	&	41.8	&	41.8	&	47.2	&	$\mathrm{1.6632E\!-\!04}$	&	11.1393	&	2.2017	&	3.6122	&	0.13443	&	5.8347	&	0.00 \\ 
$\mathrm{Be}$	&	0.4438	&	63.7	&	63.7	&	72.0	&	$\mathrm{1.8480E\!+\!00}$	&	2.7847	&	0.0592	&	1.6922	&	0.76146	&	2.4339	&	0.14 \\ 
 & & & & & & & & & & & \\ 
$\mathrm{B}$	&	0.4625	&	76.0	&	76.0	&	85.9	&	$\mathrm{2.3700E\!+\!00}$	&	2.8477	&	0.0305	&	1.9688	&	0.56221	&	2.4512	&	0.14 \\ 
$\mathrm{C}$	&	0.4995	&	78.0	&	70.0	&	81.0	&	$\mathrm{2.0000E\!+\!00}$	&	2.9925	&	-0.0351	&	2.4860	&	0.20489	&	3.0036	&	0.10 \\ 
$\mathrm{N}$	&	0.4998	&	82.0	&	82.0	&	82.0	&	$\mathrm{1.1653E\!-\!03}$	&	10.5400	&	1.7378	&	4.1323	&	0.15955	&	3.2125	&	0.00 \\ 
$\mathrm{O}$	&	0.5000	&	95.0	&	97.0	&	106.0	&	$\mathrm{1.3315E\!-\!03}$	&	10.7004	&	1.7541	&	4.3213	&	0.11778	&	3.2913	&	0.00 \\ 
 & & & & & & & & & & & \\ 
$\mathrm{Na}$	&	0.4785	&	149.0	&	149.0	&	168.4	&	$\mathrm{9.7100E\!-\!01}$	&	5.0526	&	0.2880	&	3.1962	&	0.07608	&	3.6452	&	0.08 \\ 
$\mathrm{Mg}$	&	0.4937	&	156.0	&	156.0	&	176.3	&	$\mathrm{1.7400E\!+\!00}$	&	4.5297	&	0.1499	&	3.0668	&	0.08162	&	3.6166	&	0.08 \\ 
$\mathrm{Al}$	&	0.4818	&	166.0	&	166.0	&	187.6	&	$\mathrm{2.6989E\!+\!00}$	&	4.2395	&	0.1708	&	3.0127	&	0.07934	&	3.6345	&	0.12 \\ 
$\mathrm{Si}$	&	0.4985	&	173.0	&	173.0	&	195.5	&	$\mathrm{2.3300E\!+\!00}$	&	4.4351	&	0.2014	&	2.8715	&	0.14840	&	3.2546	&	0.14 \\ 
 & & & & & & & & & & & \\ 
$\mathrm{K}$	&	0.4860	&	190.0	&	190.0	&	214.7	&	$\mathrm{8.6200E\!-\!01}$	&	5.6423	&	0.3851	&	3.1724	&	0.20027	&	2.9233	&	0.10 \\ 
$\mathrm{Ca}$	&	0.4990	&	191.0	&	191.0	&	215.8	&	$\mathrm{1.5500E\!+\!00}$	&	5.0396	&	0.3228	&	3.1191	&	0.15475	&	3.0745	&	0.14 \\ 
$\mathrm{As}$	&	0.4405	&	347.0	&	347.0	&	392.1	&	$\mathrm{5.7300E\!+\!00}$	&	5.0510	&	0.1767	&	3.5702	&	0.06725	&	3.4176	&	0.08 \\ 
$\mathrm{Sr}$	&	0.4337	&	366.0	&	366.0	&	413.6	&	$\mathrm{2.5400E\!+\!00}$	&	5.9867	&	0.4585	&	3.6778	&	0.07058	&	3.4435	&	0.14 \\ 
$\mathrm{Ba}$	&	0.4078	&	491.0	&	491.0	&	554.8	&	$\mathrm{3.5000E\!+\!00}$	&	6.3153	&	0.4190	&	3.4547	&	0.18267	&	2.8906	&	0.14 \\ 
 & & & & & & & & & & & \\ 
\hline 
\end{tabular}
\caption{Density correction parameters for a single atomic species~\cite{Density}. }
\label{DensityTab}
\end{center}
\end{table}
\end{landscape}

\section{Materials in the Beam Path}
\begin{figure}[htp]
\begin{center}
\includegraphics[width=1.0\textwidth]{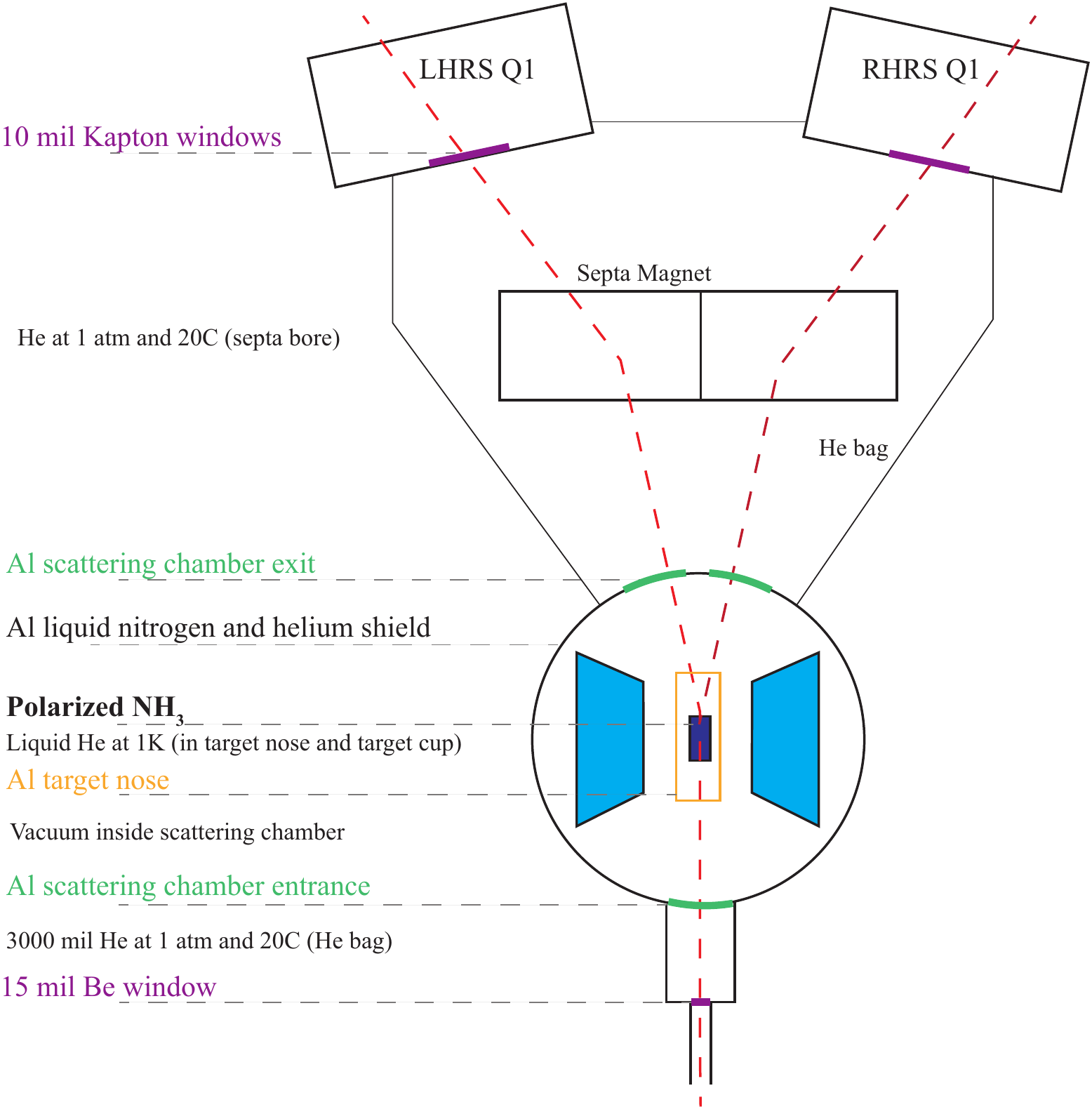}
\caption{\label{SetUp}Hall A Layout and material thicknesses for E08-027.}
\end{center}
\end{figure}

The materials encountered by the electron before and after scattering are shown in Figure~\ref{SetUp}. The incident electrons exit the beam pipe through a beryllium window and proceed through a helium bag before entering the aluminum scattering chamber. The scattering chamber is held under vacuum and contains the Helmoltz coils of the target magnetic field and the bottom of the target refrigerator. The bottom of the refrigerator or ``nose" is made out of aluminum and keeps the target materials immersed in $\sim$ 1 K super-fluid helium. 

The E08-027 the target cell has a radius of 1.361 cm and a length of 2.827 (1.295 for the short production cell) cm. The amount of liquid helium versus frozen ammonia beads in the target cell is dependent on the packing fraction of the ammonia beads. For this analysis the packing fraction is assumed to be 0.55. The NMR coil is not considered in the radiation lengths. The separation between radiation thicknesses before and after scattering is half-way through the target material. This does not coincide with the center of the target cup for the carbon foil targets. 


The scattered electrons exit through the other side of the aluminum target nose and continue through the vacuum of the scattering chamber. After leaving the scattering chamber, they continue onwards to the septa magnet and, ultimately, the Kapton entrance windows of the high resolution spectrometers. To prevent hazardous ionization of the air, a helium bag starts at the exit of the scattering chamber and ends at the entrance of the spectrometers, surrounding the bores of the septa magnets. 

\section{Results}
The E08-027 radiation lengths and energy loss parameterization results are shown in Table~\ref{RadNorm} through Table~\ref{EDummy}. The results apply to all production, packing fraction and dilution runs\footnote{Optics runs were frequently taken without liquid helium in the target nose, so these values may not apply there.}. The density of liquid helium plateaus below 2 K; a constant density of 0.145 g/cm$^3$ is assumed. 
 
\begin{table} 
\begin{center} 
\begin{tabular}{c|cccccc}\hline 
\multirow{4}{*}{\bf material} & \multirow{2}{*}{$X_0$} & \multirow{2}{*}{$\rho$} & \multirow{2}{*}{$\bar{X}_0$} & \multirow{2}{*}{$\ell$} & \multirow{2}{*}{$t$} & \multirow{2}{*}{\bf fraction} \\ 
& & & & & &  \\ 
& \multirow{2}{*}{$\mathrm{g/cm^2}$} & \multirow{2}{*}{$\mathrm{g/cm^3}$} & \multirow{2}{*}{$\mathrm{cm}$} & \multirow{2}{*}{$\mathrm{cm}$} & \multirow{2}{*}{$-$} & \multirow{2}{*}{(of total)} \\ 
& & & & & &  \\ \hline 
& & & & & &  \\ 
Beryllium Window  &65.190&1.85E+00&3.53E+01&3.81E$-$01&1.08E$-$03&0.048 \\ 
Helium Bag  &94.322&1.79E$-$04&5.28E+05&7.62E+00&1.44E$-$05&0.001 \\ 
Scat. Chamber  &24.011&2.70E+00&8.89E+00&1.78E$-$02&2.00E$-$03&0.089 \\ 
L$\mathrm{N_2}$ Shield  &24.011&2.70E+00&8.89E+00&3.80E$-$03&4.28E$-$04&0.019 \\ 
4K Shield  &24.011&2.70E+00&8.89E+00&1.30E$-$03&1.46E$-$04&0.006 \\ 
Target Nose  &24.011&2.70E+00&8.89E+00&1.27E$-$02&1.43E$-$03&0.063 \\ 
LHe in Nose &94.322&1.45E$-$01&6.50E+02&4.37E$-$01&6.72E$-$04&0.030 \\ 
Target End Cap  &24.011&2.70E+00&8.89E+00&1.80E$-$03&2.00E$-$04&0.009 \\ 
LHe Target Cup  &94.322&1.45E$-$01&6.50E+02&6.36E$-$01&9.78E$-$04&0.043 \\ 
Solid $\mathrm{NH_3}$   &40.872&8.17E$-$01&5.00E+01&7.77E$-$01&1.55E$-$02&0.691 \\ 
& & & & & &  \\ \hline 
& & & & & &  \\ 
\bf Total Before & \multicolumn{6}{c}{2.25E$-$02}  \\ 
& & & & & &  \\ \hline 
& & & & & &  \\ 
Solid $\mathrm{NH_3}$   &40.872&8.17E$-$01&5.00E+01&7.77E$-$01&1.55E$-$02&0.591 \\ 
LHe Target Cup  &94.322&1.45E$-$01&6.50E+02&6.36E$-$01&9.78E$-$04&0.037 \\ 
Target End Cap  &24.011&2.70E+00&8.89E+00&1.80E$-$03&2.00E$-$04&0.008 \\ 
LHe in Nose &94.322&1.45E$-$01&6.50E+02&4.37E$-$01&6.72E$-$04&0.026 \\ 
Target Nose  &24.011&2.70E+00&8.89E+00&1.27E$-$02&1.43E$-$03&0.054 \\ 
4K Shield  &24.011&2.70E+00&8.89E+00&1.30E$-$03&1.46E$-$04&0.006 \\ 
L$\mathrm{N_2}$ Shield  &24.011&2.70E+00&8.89E+00&3.80E$-$03&4.28E$-$04&0.016 \\ 
Scat. Chamber  &24.011&2.70E+00&8.89E+00&5.08E$-$02&5.71E$-$03&0.217 \\ 
Helium Bag  &94.322&1.79E$-$04&5.28E+05&1.71E+02&3.24E$-$04&0.012 \\ 
Kapton Window  &40.576&1.42E+00&2.86E+01&2.54E$-$02&8.89E$-$04&0.034 \\ 
& & & & & &  \\ \hline 
& & & & & &  \\ 
\bf Total After & \multicolumn{6}{c}{2.63E$-$02} \\  
& & & & & &  \\ \hline 
\end{tabular} 
 \caption{Radiation thicknesses for the normal g2p production target. }
  \label{RadNorm} 
\end{center} 
\end{table} 
 
\begin{table} 
\begin{center} 
\begin{tabular}{c|ccccc}\hline 
\multirow{4}{*}{\bf material} & \multirow{2}{*}{$\rho$} & \multirow{2}{*}{$\ell$} & \multirow{2}{*}{$\xi$} & \multirow{2}{*}{mp} & \multirow{2}{*}{$dE$} \\ 
& & & & &  \\ 
 & \multirow{2}{*}{$\mathrm{g/cm^3}$} & \multirow{2}{*}{$\mathrm{cm}$} & \multirow{2}{*}{MeV} & \multirow{2}{*}{MeV} & \multirow{2}{*}{MeV} \\ 
& & & & &  \\ \hline 
& & & & &  \\ 
Beryllium Window &1.85E+00&3.81E$-$01&4.80E$-$03&0.077&0.136\\ 
Helium Bag &1.79E$-$04&7.62E+00&1.04E$-$04&0.002&0.004\\ 
Scat. Chamber &2.70E+00&1.78E$-$02&3.56E$-$03&0.054&0.099\\ 
L$\mathrm{N_2}$ Shield &2.70E+00&3.80E$-$03&7.61E$-$04&0.010&0.021\\ 
4K Shield  &2.70E+00&1.30E$-$03&2.60E$-$04&0.003&0.007\\ 
Target Nose  &2.70E+00&1.27E$-$02&2.54E$-$03&0.038&0.070 \\ 
LHe in Nose &1.45E$-$01&4.37E$-$01&4.86E$-$03&0.090&0.149 \\ 
Target End Cap  &2.70E+00&1.80E$-$03&3.55E$-$04&0.005&0.010 \\ 
LHe Target Cup  &1.45E$-$01&6.36E$-$01&7.08E$-$03&0.138&0.217 \\ 
Solid $\mathrm{NH_3}$   &8.17E$-$01&7.77E$-$01&5.82E$-$02&1.148&1.675 \\ 
& & & & &  \\ \hline 
& & & & &  \\  
\bf Total Before & & &8.26E$-$02&1.565&2.387\\ 
 & & & & &  \\ \hline 
& & & & &  \\ 
Solid $\mathrm{NH_3}$   &8.17E$-$01&7.77E$-$01&5.82E$-$02&1.148&1.675 \\ 
LHe Target Cup  &1.45E$-$01&6.36E$-$01&7.08E$-$03&0.138&0.217 \\ 
Target End Cap  &2.70E+00&1.80E$-$03&3.55E$-$04&0.005&0.010 \\ 
LHe in Nose &1.45E$-$01&4.37E$-$01&4.86E$-$03&0.090&0.149 \\ 
Target Nose  &2.70E+00&1.27E$-$02&2.54E$-$03&0.038&0.070 \\ 
4K Shield  &2.70E+00&1.30E$-$03&2.60E$-$04&0.003&0.007\\ 
L$\mathrm{N_2}$ Shield &2.70E+00&3.80E$-$03&7.61E$-$04&0.010&0.021\\ 
Scat. Chamber &2.70E+00&5.08E$-$02&1.01E$-$02&0.165&0.282\\ 
Helium Bag &1.79E$-$04&1.71E+02&2.34E$-$03&0.057&0.087\\ 
Kapton Window &1.42E+00&2.54E$-$02&2.84E$-$03&0.044&0.080\\ 
& & & & &  \\ \hline 
 & & & & &  \\ 
 \bf Total After & & &8.94E$-$02&1.698&2.599\\ 
 & & & & &  \\ \hline 
\end{tabular} 
 \caption{Energy loss for the normal g2p production target. } 
\end{center} 
\end{table} 
 
\begin{table} 
\begin{center} 
\begin{tabular}{c|cccccc}\hline 
\multirow{4}{*}{\bf material} & \multirow{2}{*}{$X_0$} & \multirow{2}{*}{$\rho$} & \multirow{2}{*}{$\bar{X}_0$} & \multirow{2}{*}{$\ell$} & \multirow{2}{*}{$t$} & \multirow{2}{*}{\bf fraction} \\ 
& & & & & &  \\ 
& \multirow{2}{*}{$\mathrm{g/cm^2}$} & \multirow{2}{*}{$\mathrm{g/cm^3}$} & \multirow{2}{*}{$\mathrm{cm}$} & \multirow{2}{*}{$\mathrm{cm}$} & \multirow{2}{*}{$-$} & \multirow{2}{*}{(of total)} \\ 
& & & & & &  \\ \hline 
& & & & & &  \\ 
Beryllium Window  &65.190&1.85E+00&3.53E+01&3.81E$-$01&1.08E$-$03&0.080 \\ 
Helium Bag  &94.322&1.79E$-$04&5.28E+05&7.62E+00&1.44E$-$05&0.001 \\ 
Scat. Chamber  &24.011&2.70E+00&8.89E+00&1.78E$-$02&2.00E$-$03&0.148 \\ 
L$\mathrm{N_2}$ Shield  &24.011&2.70E+00&8.89E+00&3.80E$-$03&4.28E$-$04&0.032 \\ 
4K Shield  &24.011&2.70E+00&8.89E+00&1.30E$-$03&1.46E$-$04&0.011 \\ 
Target Nose  &24.011&2.70E+00&8.89E+00&1.27E$-$02&1.43E$-$03&0.105 \\ 
LHe in Nose &94.322&1.45E$-$01&6.50E+02&4.37E$-$01&6.72E$-$04&0.050 \\ 
Target End Cap  &24.011&2.70E+00&8.89E+00&1.80E$-$03&2.00E$-$04&0.015 \\ 
LHe Target Cup  &94.322&1.45E$-$01&6.50E+02&2.91E$-$01&4.48E$-$04&0.033 \\ 
Solid $\mathrm{NH_3}$   &40.872&8.17E$-$01&5.00E+01&3.56E$-$01&7.12E$-$03&0.526 \\ 
& & & & & &  \\ \hline 
& & & & & &  \\ 
\bf Total Before & \multicolumn{6}{c}{1.35E$-$02}  \\ 
& & & & & &  \\ \hline 
& & & & & &  \\ 
Solid $\mathrm{NH_3}$   &40.872&8.17E$-$01&5.00E+01&3.56E$-$01&7.12E$-$03&0.361 \\ 
LHe Target Cup  &94.322&1.45E$-$01&6.50E+02&2.91E$-$01&4.48E$-$04&0.023 \\ 
Target End Cap  &24.011&2.70E+00&8.89E+00&1.80E$-$03&2.00E$-$04&0.010 \\ 
LHe in Nose &94.322&1.45E$-$01&6.50E+02&1.97E+00&3.03E$-$03&0.153 \\ 
Target Nose  &24.011&2.70E+00&8.89E+00&1.27E$-$02&1.43E$-$03&0.072 \\ 
4K Shield  &24.011&2.70E+00&8.89E+00&1.30E$-$03&1.46E$-$04&0.007 \\ 
L$\mathrm{N_2}$ Shield  &24.011&2.70E+00&8.89E+00&3.80E$-$03&4.28E$-$04&0.022 \\ 
Scat. Chamber  &24.011&2.70E+00&8.89E+00&5.08E$-$02&5.71E$-$03&0.290 \\ 
Helium Bag  &94.322&1.79E$-$04&5.28E+05&1.71E+02&3.24E$-$04&0.016 \\ 
Kapton Window  &40.576&1.42E+00&2.86E+01&2.54E$-$02&8.89E$-$04&0.045 \\ 
& & & & & &  \\ \hline 
& & & & & &  \\ 
\bf Total After & \multicolumn{6}{c}{1.97E$-$02} \\  
& & & & & &  \\ \hline 
\end{tabular} 
\caption{Radiation thicknesses for the thin g2p production target. } 
\end{center} 
\end{table} 
 
\begin{table} 
\begin{center} 
\begin{tabular}{c|ccccc}\hline 
\multirow{4}{*}{\bf material} & \multirow{2}{*}{$\rho$} & \multirow{2}{*}{$\ell$} & \multirow{2}{*}{$\xi$} & \multirow{2}{*}{mp} & \multirow{2}{*}{$dE$} \\ 
& & & & &  \\ 
 & \multirow{2}{*}{$\mathrm{g/cm^3}$} & \multirow{2}{*}{$\mathrm{cm}$} & \multirow{2}{*}{MeV} & \multirow{2}{*}{MeV} & \multirow{2}{*}{MeV} \\ 
& & & & &  \\ \hline 
& & & & &  \\ 
Beryllium Window &1.85E+00&3.81E$-$01&4.80E$-$03&0.077&0.136\\ 
Helium Bag &1.79E$-$04&7.62E+00&1.04E$-$04&0.002&0.004\\ 
Scat. Chamber &2.70E+00&1.78E$-$02&3.56E$-$03&0.054&0.099\\ 
L$\mathrm{N_2}$ Shield &2.70E+00&3.80E$-$03&7.61E$-$04&0.010&0.021\\ 
4K Shield  &2.70E+00&1.30E$-$03&2.60E$-$04&0.003&0.007\\ 
Target Nose  &2.70E+00&1.27E$-$02&2.54E$-$03&0.038&0.070 \\ 
LHe in Nose &1.45E$-$01&4.37E$-$01&4.86E$-$03&0.090&0.149 \\ 
Target End Cap  &2.70E+00&1.80E$-$03&3.55E$-$04&0.005&0.010 \\ 
LHe Target Cup  &1.45E$-$01&2.91E$-$01&3.24E$-$03&0.061&0.099 \\ 
Solid $\mathrm{NH_3}$   &8.17E$-$01&3.56E$-$01&2.67E$-$02&0.505&0.767 \\ 
& & & & &  \\ \hline 
& & & & &  \\  
\bf Total Before & & &4.72E$-$02&0.845&1.362\\ 
 & & & & &  \\ \hline 
& & & & &  \\ 
Solid $\mathrm{NH_3}$   &8.17E$-$01&3.56E$-$01&2.67E$-$02&0.505&0.767 \\ 
LHe Target Cup  &1.45E$-$01&2.91E$-$01&3.24E$-$03&0.061&0.099 \\ 
Target End Cap  &2.70E+00&1.80E$-$03&3.55E$-$04&0.005&0.010 \\ 
LHe in Nose &1.45E$-$01&1.97E+00&2.19E$-$02&0.437&0.671 \\ 
Target Nose  &2.70E+00&1.27E$-$02&2.54E$-$03&0.038&0.070 \\ 
4K Shield  &2.70E+00&1.30E$-$03&2.60E$-$04&0.003&0.007\\ 
L$\mathrm{N_2}$ Shield &2.70E+00&3.80E$-$03&7.61E$-$04&0.010&0.021\\ 
Scat. Chamber &2.70E+00&5.08E$-$02&1.01E$-$02&0.165&0.282\\ 
Helium Bag &1.79E$-$04&1.71E+02&2.34E$-$03&0.057&0.087\\ 
Kapton Window &1.42E+00&2.54E$-$02&2.84E$-$03&0.044&0.080\\ 
& & & & &  \\ \hline 
 & & & & &  \\ 
 \bf Total After & & &7.11E$-$02&1.325&2.096\\ 
 & & & & &  \\ \hline 
\end{tabular} 
 \caption{Energy loss for the thin g2p production target. } 
\end{center} 
\end{table} 

\begin{table} 
\begin{center} 
\begin{tabular}{c|cccccc}\hline 
\multirow{4}{*}{\bf material} & \multirow{2}{*}{$X_0$} & \multirow{2}{*}{$\rho$} & \multirow{2}{*}{$\bar{X}_0$} & \multirow{2}{*}{$\ell$} & \multirow{2}{*}{$t$} & \multirow{2}{*}{\bf fraction} \\ 
& & & & & &  \\ 
& \multirow{2}{*}{$\mathrm{g/cm^2}$} & \multirow{2}{*}{$\mathrm{g/cm^3}$} & \multirow{2}{*}{$\mathrm{cm}$} & \multirow{2}{*}{$\mathrm{cm}$} & \multirow{2}{*}{$-$} & \multirow{2}{*}{(of total)} \\ 
& & & & & &  \\ \hline 
& & & & & &  \\ 
Beryllium Window  &65.190&1.85E+00&3.53E+01&3.81E$-$01&1.08E$-$03&0.076 \\ 
Helium Bag  &94.322&1.79E$-$04&5.28E+05&7.62E+00&1.44E$-$05&0.001 \\ 
Scat. Chamber  &24.011&2.70E+00&8.89E+00&1.78E$-$02&2.00E$-$03&0.141 \\ 
L$\mathrm{N_2}$ Shield  &24.011&2.70E+00&8.89E+00&3.80E$-$03&4.28E$-$04&0.030 \\ 
4K Shield  &24.011&2.70E+00&8.89E+00&1.30E$-$03&1.46E$-$04&0.010 \\ 
Target Nose  &24.011&2.70E+00&8.89E+00&1.27E$-$02&1.43E$-$03&0.101 \\ 
LHe in Nose &94.322&1.45E$-$01&6.50E+02&4.37E$-$01&6.72E$-$04&0.047 \\ 
$\mathrm{^{12}C}$ disc   &42.697&2.27E+00&1.88E+01&1.59E$-$01&8.43E$-$03&0.594 \\ 
& & & & & &  \\ \hline 
& & & & & &  \\ 
\bf Total Before & \multicolumn{6}{c}{1.42E$-$02}  \\ 
& & & & & &  \\ \hline 
& & & & & &  \\ 
$\mathrm{^{12}C}$ disc   &42.697&2.27E+00&1.88E+01&1.59E$-$01&8.43E$-$03&0.385 \\ 
LHe Target Cup  &94.322&1.45E$-$01&6.50E+02&2.51E+00&3.86E$-$03&0.176 \\ 
LHe in Nose &94.322&1.45E$-$01&6.50E+02&4.37E$-$01&6.72E$-$04&0.031 \\ 
Target Nose  &24.011&2.70E+00&8.89E+00&1.27E$-$02&1.43E$-$03&0.065 \\ 
4K Shield  &24.011&2.70E+00&8.89E+00&1.30E$-$03&1.46E$-$04&0.007 \\ 
L$\mathrm{N_2}$ Shield  &24.011&2.70E+00&8.89E+00&3.80E$-$03&4.28E$-$04&0.020 \\ 
Scat. Chamber  &24.011&2.70E+00&8.89E+00&5.08E$-$02&5.71E$-$03&0.261 \\ 
Helium Bag  &94.322&1.79E$-$04&5.28E+05&1.71E+02&3.24E$-$04&0.015 \\ 
Kapton Window  &40.576&1.42E+00&2.86E+01&2.54E$-$02&8.89E$-$04&0.041 \\ 
& & & & & &  \\ \hline 
& & & & & &  \\ 
\bf Total After & \multicolumn{6}{c}{2.19E$-$02} \\  
& & & & & &  \\ \hline 
\end{tabular} 
 \caption{Radiation thicknesses for the normal g2p carbon dilution target. } 
\end{center} 
\end{table} 
 
\begin{table} 
\begin{center} 
\begin{tabular}{c|ccccc}\hline 
\multirow{4}{*}{\bf material} & \multirow{2}{*}{$\rho$} & \multirow{2}{*}{$\ell$} & \multirow{2}{*}{$\xi$} & \multirow{2}{*}{mp} & \multirow{2}{*}{$dE$} \\ 
& & & & &  \\ 
 & \multirow{2}{*}{$\mathrm{g/cm^3}$} & \multirow{2}{*}{$\mathrm{cm}$} & \multirow{2}{*}{MeV} & \multirow{2}{*}{MeV} & \multirow{2}{*}{MeV} \\ 
& & & & &  \\ \hline 
& & & & &  \\ 
Beryllium Window &1.85E+00&3.81E$-$01&4.80E$-$03&0.077&0.136\\ 
Helium Bag &1.79E$-$04&7.62E+00&1.04E$-$04&0.002&0.004\\ 
Scat. Chamber &2.70E+00&1.78E$-$02&3.56E$-$03&0.054&0.099\\ 
L$\mathrm{N_2}$ Shield &2.70E+00&3.80E$-$03&7.61E$-$04&0.010&0.021\\ 
4K Shield  &2.70E+00&1.30E$-$03&2.60E$-$04&0.003&0.007\\ 
Target Nose  &2.70E+00&1.27E$-$02&2.54E$-$03&0.038&0.070 \\ 
LHe in Nose &1.45E$-$01&4.37E$-$01&4.86E$-$03&0.090&0.149 \\ 
$\mathrm{^{12}C}$ disc  &2.27E+00&1.59E$-$01&2.76E$-$02&0.500&0.770 \\ 
& & & & &  \\ \hline 
& & & & &  \\  
\bf Total Before & & &4.45E$-$02&1.056&1.256\\ 
 & & & & &  \\ \hline 
& & & & &  \\ 
$\mathrm{^{12}C}$ disc   &2.27E+00&1.59E$-$01&2.76E$-$02&0.500&0.770 \\ 
LHe Target Cup  &1.45E$-$01&2.51E+00&2.80E$-$02&0.564&0.857 \\ 
LHe in Nose &1.45E$-$01&4.37E$-$01&4.86E$-$03&0.090&0.149 \\ 
Target Nose  &2.70E+00&1.27E$-$02&2.54E$-$03&0.038&0.070 \\ 
4K Shield  &2.70E+00&1.30E$-$03&2.60E$-$04&0.003&0.007\\ 
L$\mathrm{N_2}$ Shield &2.70E+00&3.80E$-$03&7.61E$-$04&0.010&0.021\\ 
Scat. Chamber &2.70E+00&5.08E$-$02&1.01E$-$02&0.165&0.282\\ 
Helium Bag &1.79E$-$04&1.71E+02&2.34E$-$03&0.057&0.087\\ 
Kapton Window &1.42E+00&2.54E$-$02&2.84E$-$03&0.044&0.080\\ 
& & & & &  \\ \hline 
 & & & & &  \\ 
 \bf Total After & & &7.93E$-$02&1.189&1.896\\ 
 & & & & &  \\ \hline 
\end{tabular} 
 \caption{Energy loss for the normal g2p carbon dilution target. } 
\end{center} 
\end{table} 
 
\begin{table} 
\begin{center} 
\begin{tabular}{c|cccccc}\hline 
\multirow{4}{*}{\bf material} & \multirow{2}{*}{$X_0$} & \multirow{2}{*}{$\rho$} & \multirow{2}{*}{$\bar{X}_0$} & \multirow{2}{*}{$\ell$} & \multirow{2}{*}{$t$} & \multirow{2}{*}{\bf fraction} \\ 
& & & & & &  \\ 
& \multirow{2}{*}{$\mathrm{g/cm^2}$} & \multirow{2}{*}{$\mathrm{g/cm^3}$} & \multirow{2}{*}{$\mathrm{cm}$} & \multirow{2}{*}{$\mathrm{cm}$} & \multirow{2}{*}{$-$} & \multirow{2}{*}{(of total)} \\ 
& & & & & &  \\ \hline 
& & & & & &  \\ 
Beryllium Window  &65.190&1.85E+00&3.53E+01&3.81E$-$01&1.08E$-$03&0.128 \\ 
Helium Bag  &94.322&1.79E$-$04&5.28E+05&7.62E+00&1.44E$-$05&0.002 \\ 
Scat. Chamber  &24.011&2.70E+00&8.89E+00&1.78E$-$02&2.00E$-$03&0.236 \\ 
L$\mathrm{N_2}$ Shield  &24.011&2.70E+00&8.89E+00&3.80E$-$03&4.28E$-$04&0.051 \\ 
4K Shield  &24.011&2.70E+00&8.89E+00&1.30E$-$03&1.46E$-$04&0.017 \\ 
Target Nose  &24.011&2.70E+00&8.89E+00&1.27E$-$02&1.43E$-$03&0.169 \\ 
LHe in Nose &94.322&1.45E$-$01&6.50E+02&4.37E$-$01&6.72E$-$04&0.079 \\ 
$\mathrm{^{12}C}$ disc   &42.697&2.27E+00&1.88E+01&5.08E$-$02&2.70E$-$03&0.318 \\ 
& & & & & &  \\ \hline 
& & & & & &  \\ 
\bf Total Before & \multicolumn{6}{c}{8.47E$-$03}  \\ 
& & & & & &  \\ \hline 
& & & & & &  \\ 
$\mathrm{^{12}C}$ disc   &42.697&2.27E+00&1.88E+01&5.08E$-$02&2.70E$-$03&0.117 \\ 
LHe Target Cup  &94.322&1.45E$-$01&6.50E+02&2.73E+00&4.20E$-$03&0.182 \\ 
LHe in Nose &94.322&1.45E$-$01&6.50E+02&4.37E$-$01&6.72E$-$04&0.029 \\ 
Target Nose  &24.011&2.70E+00&8.89E+00&1.27E$-$02&1.43E$-$03&0.062 \\ 
4K Shield  &24.011&2.70E+00&8.89E+00&1.30E$-$03&1.46E$-$04&0.006 \\ 
L$\mathrm{N_2}$ Shield  &24.011&2.70E+00&8.89E+00&3.80E$-$03&4.28E$-$04&0.019 \\ 
Scat. Chamber  &24.011&2.70E+00&8.89E+00&5.08E$-$02&5.71E$-$03&0.247 \\ 
Helium Bag  &94.322&1.79E$-$04&5.28E+05&1.71E+02&3.24E$-$04&0.014 \\ 
Kapton Window  &40.576&1.42E+00&2.86E+01&2.54E$-$02&8.89E$-$04&0.038 \\ 
& & & & & &  \\ \hline 
& & & & & &  \\ 
\bf Total After & \multicolumn{6}{c}{2.31E$-$02} \\  
& & & & & &  \\ \hline 
\end{tabular} 
 \caption{Radiation thicknesses for the thin g2p carbon dilution target. } 
\end{center} 
\end{table} 
 
\begin{table} 
\begin{center} 
\begin{tabular}{c|ccccc}\hline 
\multirow{4}{*}{\bf material} & \multirow{2}{*}{$\rho$} & \multirow{2}{*}{$\ell$} & \multirow{2}{*}{$\xi$} & \multirow{2}{*}{mp} & \multirow{2}{*}{$dE$} \\ 
& & & & &  \\ 
 & \multirow{2}{*}{$\mathrm{g/cm^3}$} & \multirow{2}{*}{$\mathrm{cm}$} & \multirow{2}{*}{MeV} & \multirow{2}{*}{MeV} & \multirow{2}{*}{MeV} \\ 
& & & & &  \\ \hline 
& & & & &  \\ 
Beryllium Window &1.85E+00&3.81E$-$01&4.80E$-$03&0.077&0.136\\ 
Helium Bag &1.79E$-$04&7.62E+00&1.04E$-$04&0.002&0.004\\ 
Scat. Chamber &2.70E+00&1.78E$-$02&3.56E$-$03&0.054&0.099\\ 
L$\mathrm{N_2}$ Shield &2.70E+00&3.80E$-$03&7.61E$-$04&0.010&0.021\\ 
4K Shield  &2.70E+00&1.30E$-$03&2.60E$-$04&0.003&0.007\\ 
Target Nose  &2.70E+00&1.27E$-$02&2.54E$-$03&0.038&0.070 \\ 
LHe in Nose &1.45E$-$01&4.37E$-$01&4.86E$-$03&0.090&0.149 \\ 
$\mathrm{^{12}C}$ disc   &2.27E+00&5.08E$-$02&8.83E$-$03&0.150&0.246 \\ 
& & & & &  \\ \hline 
& & & & &  \\  
\bf Total Before & & &2.57E$-$02&0.424&0.732\\ 
 & & & & &  \\ \hline 
& & & & &  \\ 
$\mathrm{^{12}C}$ disc   &2.27E+00&5.08E$-$02&8.83E$-$03&0.150&0.246 \\ 
LHe Target Cup  &1.45E$-$01&2.73E+00&3.04E$-$02&0.615&0.930 \\ 
LHe in Nose &1.45E$-$01&4.37E$-$01&4.86E$-$03&0.090&0.149 \\ 
Target Nose  &2.70E+00&1.27E$-$02&2.54E$-$03&0.038&0.070 \\ 
4K Shield  &2.70E+00&1.30E$-$03&2.60E$-$04&0.003&0.007\\ 
L$\mathrm{N_2}$ Shield &2.70E+00&3.80E$-$03&7.61E$-$04&0.010&0.021\\ 
Scat. Chamber &2.70E+00&5.08E$-$02&1.01E$-$02&0.165&0.282\\ 
Helium Bag &1.79E$-$04&1.71E+02&2.34E$-$03&0.057&0.087\\ 
Kapton Window &1.42E+00&2.54E$-$02&2.84E$-$03&0.044&0.080\\ 
& & & & &  \\ \hline 
 & & & & &  \\ 
 \bf Total After & & &4.78E$-$02&0.865&1.409\\ 
 & & & & &  \\ \hline 
\end{tabular} 
 \caption{Energy loss for the thin g2p carbon dilution target. } 
\end{center} 
\end{table} 
 
\begin{table} 
\begin{center} 
\begin{tabular}{c|cccccc}\hline 
\multirow{4}{*}{\bf material} & \multirow{2}{*}{$X_0$} & \multirow{2}{*}{$\rho$} & \multirow{2}{*}{$\bar{X}_0$} & \multirow{2}{*}{$\ell$} & \multirow{2}{*}{$t$} & \multirow{2}{*}{\bf fraction} \\ 
& & & & & &  \\ 
& \multirow{2}{*}{$\mathrm{g/cm^2}$} & \multirow{2}{*}{$\mathrm{g/cm^3}$} & \multirow{2}{*}{$\mathrm{cm}$} & \multirow{2}{*}{$\mathrm{cm}$} & \multirow{2}{*}{$-$} & \multirow{2}{*}{(of total)} \\ 
& & & & & &  \\ \hline 
& & & & & &  \\ 
Beryllium Window  &65.190&1.85E+00&3.53E+01&3.81E$-$01&1.08E$-$03&0.136 \\ 
Helium Bag  &94.322&1.79E$-$04&5.28E+05&7.62E+00&1.44E$-$05&0.002 \\ 
Scat. Chamber  &24.011&2.70E+00&8.89E+00&1.78E$-$02&2.00E$-$03&0.252 \\ 
L$\mathrm{N_2}$ Shield  &24.011&2.70E+00&8.89E+00&3.80E$-$03&4.28E$-$04&0.054 \\ 
4K Shield  &24.011&2.70E+00&8.89E+00&1.30E$-$03&1.46E$-$04&0.018 \\ 
Target Nose  &24.011&2.70E+00&8.89E+00&1.27E$-$02&1.43E$-$03&0.180 \\ 
LHe in Nose &94.322&1.45E$-$01&6.50E+02&4.37E$-$01&6.72E$-$04&0.085 \\ 
LHe Target Cup  &94.322&1.45E$-$01&6.50E+02&1.42E+00&2.18E$-$03&0.274 \\ 
& & & & & &  \\ \hline 
& & & & & &  \\ 
\bf Total Before & \multicolumn{6}{c}{7.95E$-$03}  \\ 
& & & & & &  \\ \hline 
& & & & & &  \\ 
LHe Target Cup  &94.322&1.45E$-$01&6.50E+02&1.42E+00&2.18E$-$03&0.127 \\ 
LHe in Nose &94.322&1.45E$-$01&6.50E+02&4.37E$-$01&6.72E$-$04&0.039 \\ 
Target Nose  &24.011&2.70E+00&8.89E+00&1.27E$-$02&1.43E$-$03&0.083 \\ 
4K Shield  &24.011&2.70E+00&8.89E+00&1.30E$-$03&1.46E$-$04&0.009 \\ 
L$\mathrm{N_2}$ Shield  &24.011&2.70E+00&8.89E+00&3.80E$-$03&4.28E$-$04&0.025 \\ 
Scat. Chamber  &24.011&2.70E+00&8.89E+00&5.08E$-$02&5.71E$-$03&0.333 \\ 
Helium Bag  &94.322&1.79E$-$04&5.28E+05&1.71E+02&5.72E$-$03&0.333 \\ 
Kapton Window  &40.576&1.42E+00&2.86E+01&2.54E$-$02&8.89E$-$04&0.052 \\ 
& & & & & &  \\ \hline 
& & & & & &  \\ 
\bf Total After & \multicolumn{6}{c}{1.72E$-$02} \\  
& & & & & &  \\ \hline 
\end{tabular} 
 \caption{Radiation thicknesses for the g2p empty dilution target. } 
\end{center} 
\end{table} 
 
\begin{table} 
\begin{center} 
\begin{tabular}{c|ccccc}\hline 
\multirow{4}{*}{\bf material} & \multirow{2}{*}{$\rho$} & \multirow{2}{*}{$\ell$} & \multirow{2}{*}{$\xi$} & \multirow{2}{*}{mp} & \multirow{2}{*}{$dE$} \\ 
& & & & &  \\ 
 & \multirow{2}{*}{$\mathrm{g/cm^3}$} & \multirow{2}{*}{$\mathrm{cm}$} & \multirow{2}{*}{MeV} & \multirow{2}{*}{MeV} & \multirow{2}{*}{MeV} \\ 
& & & & &  \\ \hline 
& & & & &  \\ 
Beryllium Window &1.85E+00&3.81E$-$01&4.80E$-$03&0.077&0.136\\ 
Helium Bag &1.79E$-$04&7.62E+00&1.04E$-$04&0.002&0.004\\ 
Scat. Chamber &2.70E+00&1.78E$-$02&3.56E$-$03&0.054&0.099\\ 
L$\mathrm{N_2}$ Shield &2.70E+00&3.80E$-$03&7.61E$-$04&0.010&0.021\\ 
4K Shield  &2.70E+00&1.30E$-$03&2.60E$-$04&0.003&0.007\\ 
Target Nose  &2.70E+00&1.27E$-$02&2.54E$-$03&0.038&0.070 \\ 
LHe in Nose &1.45E$-$01&4.37E$-$01&4.86E$-$03&0.090&0.149 \\ 
LHe Target Cup  &1.45E$-$01&1.42E+00&1.57E$-$02&0.320&0.483 \\ 
& & & & &  \\ \hline 
& & & & &  \\  
\bf Total Before & & &3.26E$-$02&0.594&0.968\\ 
 & & & & &  \\ \hline 
& & & & &  \\ 
LHe Target Cup  &1.45E$-$01&1.42E+00&1.57E$-$02&0.320&0.483 \\ 
LHe in Nose &1.45E$-$01&4.37E$-$01&4.86E$-$03&0.090&0.149 \\ 
Target Nose  &2.70E+00&1.27E$-$02&2.54E$-$03&0.038&0.070 \\ 
4K Shield  &2.70E+00&1.30E$-$03&2.60E$-$04&0.003&0.007\\ 
L$\mathrm{N_2}$ Shield &2.70E+00&3.80E$-$03&7.61E$-$04&0.010&0.021\\ 
Scat. Chamber &2.70E+00&5.08E$-$02&1.01E$-$02&0.165&0.282\\ 
Helium Bag &1.79E$-$04&1.71E+02&1.60E$-$02&0.391&0.568\\ 
Kapton Window &1.42E+00&2.54E$-$02&2.84E$-$03&0.044&0.080\\ 
& & & & &  \\ \hline 
 & & & & &  \\ 
 \bf Total After & & &5.32E$-$02&1.061&1.660\\ 
 & & & & &  \\ \hline 
\end{tabular} 
 \caption{Energy loss for the g2p empty dilution target. } 
\end{center} 
\end{table} 
 
\begin{table} 
\begin{center} 
\begin{tabular}{c|cccccc}\hline 
\multirow{4}{*}{\bf material} & \multirow{2}{*}{$X_0$} & \multirow{2}{*}{$\rho$} & \multirow{2}{*}{$\bar{X}_0$} & \multirow{2}{*}{$\ell$} & \multirow{2}{*}{$t$} & \multirow{2}{*}{\bf fraction} \\ 
& & & & & &  \\ 
& \multirow{2}{*}{$\mathrm{g/cm^2}$} & \multirow{2}{*}{$\mathrm{g/cm^3}$} & \multirow{2}{*}{$\mathrm{cm}$} & \multirow{2}{*}{$\mathrm{cm}$} & \multirow{2}{*}{$-$} & \multirow{2}{*}{(of total)} \\ 
& & & & & &  \\ \hline 
& & & & & &  \\ 
Beryllium Window  &65.190&1.85E+00&3.53E+01&3.81E$-$01&1.08E$-$03&0.133 \\ 
Helium Bag  &94.322&1.79E$-$04&5.28E+05&7.62E+00&1.44E$-$05&0.002 \\ 
Scat. Chamber  &24.011&2.70E+00&8.89E+00&1.78E$-$02&2.00E$-$03&0.246 \\ 
L$\mathrm{N_2}$ Shield  &24.011&2.70E+00&8.89E+00&3.80E$-$03&4.28E$-$04&0.053 \\ 
4K Shield  &24.011&2.70E+00&8.89E+00&1.30E$-$03&1.46E$-$04&0.018 \\ 
Target Nose  &24.011&2.70E+00&8.89E+00&1.27E$-$02&1.43E$-$03&0.175 \\ 
LHe in Nose &94.322&1.45E$-$01&6.50E+02&4.37E$-$01&6.72E$-$04&0.082 \\ 
Target End Cap  &24.011&2.70E+00&8.89E+00&1.80E$-$03&2.00E$-$04&0.025 \\ 
LHe Target Cup  &94.322&1.45E$-$01&6.50E+02&1.41E+00&2.17E$-$03&0.267 \\ 
& & & & & &  \\ \hline 
& & & & & &  \\ 
\bf Total Before & \multicolumn{6}{c}{8.14E$-$03}  \\ 
& & & & & &  \\ \hline 
& & & & & &  \\ 
LHe Target Cup  &94.322&1.45E$-$01&6.50E+02&1.41E+00&2.17E$-$03&0.182 \\ 
Target End Cap  &24.011&2.70E+00&8.89E+00&1.80E$-$03&2.00E$-$04&0.017 \\ 
LHe in Nose &94.322&1.45E$-$01&6.50E+02&4.37E$-$01&6.72E$-$04&0.056 \\ 
Target Nose  &24.011&2.70E+00&8.89E+00&1.27E$-$02&1.43E$-$03&0.119 \\ 
4K Shield  &24.011&2.70E+00&8.89E+00&1.30E$-$03&1.46E$-$04&0.012 \\ 
L$\mathrm{N_2}$ Shield  &24.011&2.70E+00&8.89E+00&3.80E$-$03&4.28E$-$04&0.036 \\ 
Scat. Chamber  &24.011&2.70E+00&8.89E+00&5.08E$-$02&5.71E$-$03&0.477 \\ 
Helium Bag  &94.322&1.79E$-$04&5.28E+05&1.71E+02&3.24E$-$04&0.027 \\ 
Kapton Window  &40.576&1.42E+00&2.86E+01&2.54E$-$02&8.89E$-$04&0.074 \\ 
& & & & & &  \\ \hline 
& & & & & &  \\ 
\bf Total After & \multicolumn{6}{c}{1.20E$-$02} \\  
& & & & & &  \\ \hline 
\end{tabular} 
 \caption{Radiation thicknesses for the g2p dummy dilution target. } 
\end{center} 
\end{table} 
 
\begin{table} 
\begin{center} 
\begin{tabular}{c|ccccc}\hline 
\multirow{4}{*}{\bf material} & \multirow{2}{*}{$\rho$} & \multirow{2}{*}{$\ell$} & \multirow{2}{*}{$\xi$} & \multirow{2}{*}{mp} & \multirow{2}{*}{$dE$} \\ 
& & & & &  \\ 
 & \multirow{2}{*}{$\mathrm{g/cm^3}$} & \multirow{2}{*}{$\mathrm{cm}$} & \multirow{2}{*}{MeV} & \multirow{2}{*}{MeV} & \multirow{2}{*}{MeV} \\ 
& & & & &  \\ \hline 
& & & & &  \\ 
Beryllium Window &1.85E+00&3.81E$-$01&4.80E$-$03&0.077&0.136\\ 
Helium Bag &1.79E$-$04&7.62E+00&1.04E$-$04&0.002&0.004\\ 
Scat. Chamber &2.70E+00&1.78E$-$02&3.56E$-$03&0.054&0.099\\ 
L$\mathrm{N_2}$ Shield &2.70E+00&3.80E$-$03&7.61E$-$04&0.010&0.021\\ 
4K Shield  &2.70E+00&1.30E$-$03&2.60E$-$04&0.003&0.007\\ 
Target Nose  &2.70E+00&1.27E$-$02&2.54E$-$03&0.038&0.070 \\ 
LHe in Nose &1.45E$-$01&4.37E$-$01&4.86E$-$03&0.090&0.149 \\ 
Target End Cap  &2.70E+00&1.80E$-$03&3.55E$-$04&0.005&0.010 \\ 
LHe Target Cup  &1.45E$-$01&1.41E+00&1.57E$-$02&0.319&0.482 \\ 
& & & & &  \\ \hline 
& & & & &  \\  
\bf Total Before & & &3.30E$-$02&0.598&0.978\\ 
 & & & & &  \\ \hline 
& & & & &  \\ 
LHe Target Cup  &1.45E$-$01&1.41E+00&1.57E$-$02&0.319&0.482 \\ 
Target End Cap  &2.70E+00&1.80E$-$03&3.55E$-$04&0.005&0.010 \\ 
LHe in Nose &1.45E$-$01&4.37E$-$01&4.86E$-$03&0.090&0.149 \\ 
Target Nose  &2.70E+00&1.27E$-$02&2.54E$-$03&0.038&0.070 \\ 
4K Shield  &2.70E+00&1.30E$-$03&2.60E$-$04&0.003&0.007\\ 
L$\mathrm{N_2}$ Shield &2.70E+00&3.80E$-$03&7.61E$-$04&0.010&0.021\\ 
Scat. Chamber &2.70E+00&5.08E$-$02&1.01E$-$02&0.165&0.282\\ 
Helium Bag &1.79E$-$04&1.71E+02&2.34E$-$03&0.057&0.087\\ 
Kapton Window &1.42E+00&2.54E$-$02&2.84E$-$03&0.044&0.080\\ 
& & & & &  \\ \hline 
 & & & & &  \\ 
 \bf Total After & & &3.98E$-$02&0.731&1.189\\ 
 & & & & &  \\ \hline 
\end{tabular} 
 \caption{Energy loss for the g2p dummy dilution target. }
 \label{EDummy} 
\end{center} 
\end{table}

\chapter{\sc Preliminary Analysis of 2.5 T Settings}
\label{app:Appendix-I}

The 2.5 T kinematic settings ($E_0$ = 1.1, 1.7, and 2.2 GeV) represent the low $Q^2$ coverage of E08-027 (see Figure~\ref{kin}) for the transverse target field. These settings have not been previously analyzed in depth due to the dramatically lower target polarization of the 2.5 T target field\footnote{The statistical uncertainty approximately increases with the square of the overall decrease in the target polarization.}. This appendix serves as a first pass analysis of these settings and is not final. The goal of this preliminary analysis is to check the overall statistical quality of the data and identify systematic issues that should be addressed in future analyses. The appendix will follow the general procedure laid out in Chapter~\ref{ch:Results} and includes the combined statistics of both HRS.

\section{Asymmetry}
The following spectrometer acceptance cuts are placed on the data:
\begin{itemize}
\item $|$L.rec.dp$|$ $<$ 4\%
\item $|$L.tr.r\_y$|$ $<$ 80 mm
\item $|$L.tr.r\_x$|$ $<$ 800 mm
\item  $-$40 mrad $<$ L.tr.tg\_ph $<$ 40 mrad
\item  $-$60 mrad $<$ L.tr.tg\_th $<$ 80 mrad
\end{itemize}

\begin{figure}[htp]
\centering     
\subfigure[Run 3653, $P_0$ = 1.0 GeV]{\label{fig:LHRSAcc234234}\includegraphics[width=.70\textwidth]{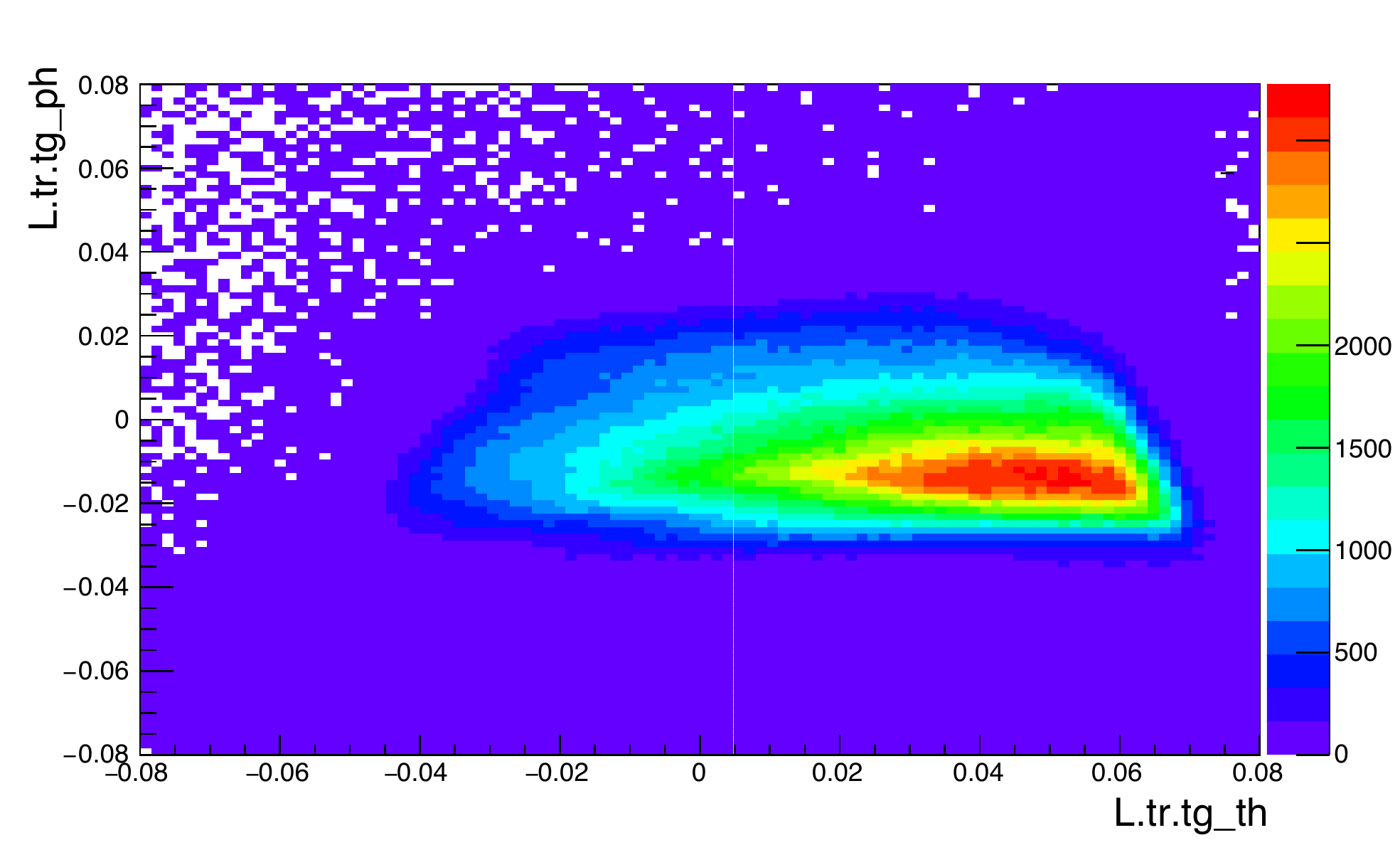}}
\qquad
\subfigure[Run 3554, $P_0$ = 1.44 GeV]{\label{fig:RHRSAcc23423423}\includegraphics[width=.70\textwidth]{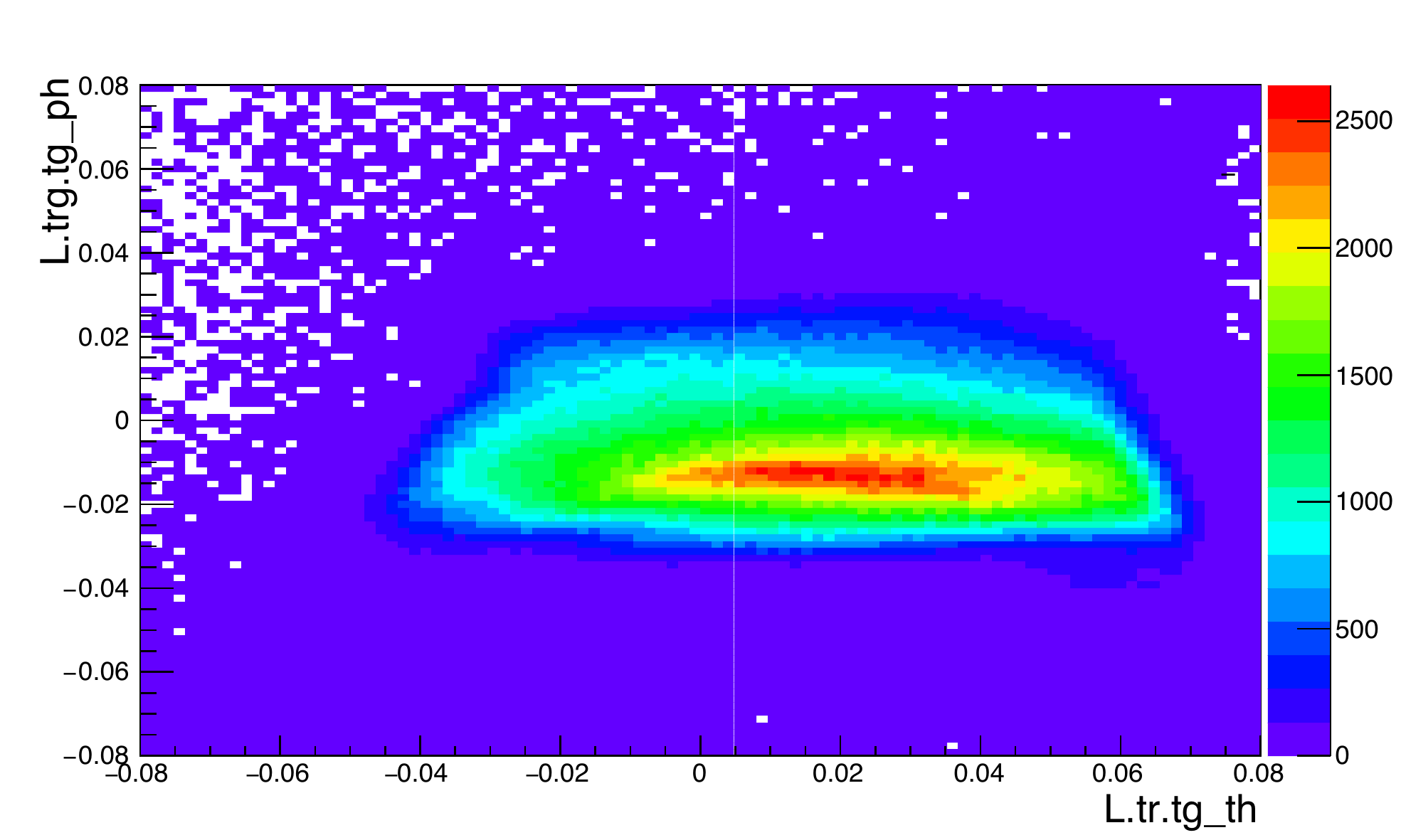}}
\qquad
\subfigure[Run 3459, $P_0$ = 2.07 GeV]{\label{fig:LHRSAccLong234234234}\includegraphics[width=.70\textwidth]{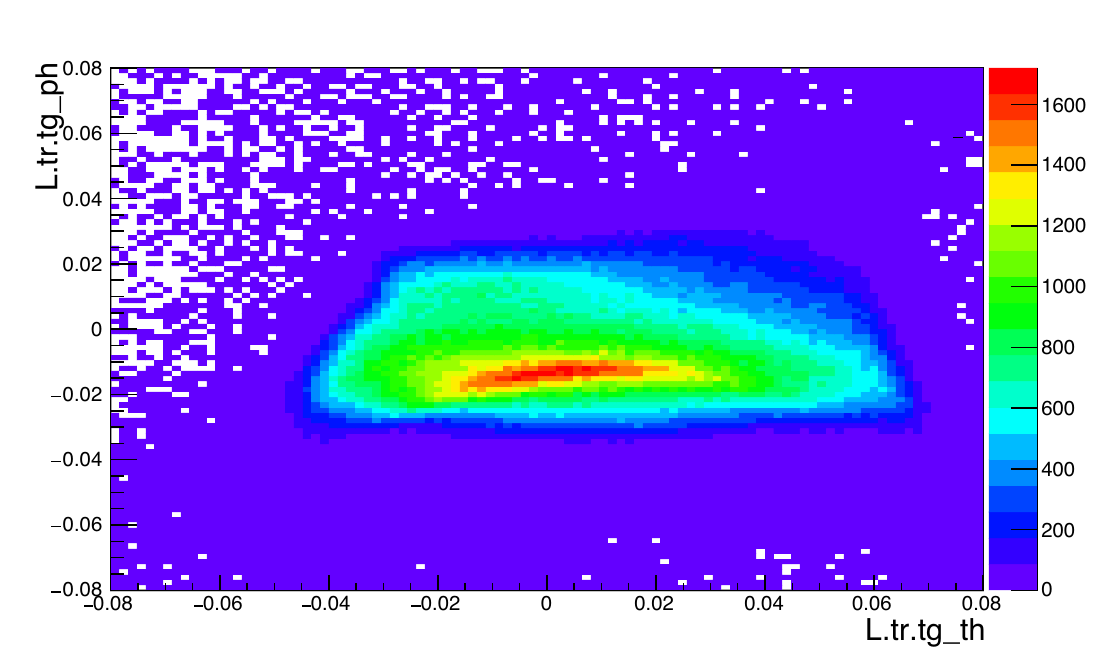}}
\caption{Acceptance cuts for $E_0$ = 2254 MeV 2.5 T on LHRS.}
\label{AccCuts25}
\end{figure}
These cuts are very similar to the 5 T cuts except that the theta cut is loosened from $-$40 mrad to $-$60 mrad. The reason for this is shown in Figure~\ref{AccCuts25}; the acceptance shape is not constant across scattered electron energies. The trend shown in Figure~\ref{AccCuts25} applies to all the 2.5 T data. The loose cut in this analysis is wide enough to account for the changing acceptance, but a varying cut could be used in the future to optimize the selection of good electrons.

\subsection{Scattering Angle Fit}
\begin{figure}[htp]
\centering     
\includegraphics[width=.80\textwidth]{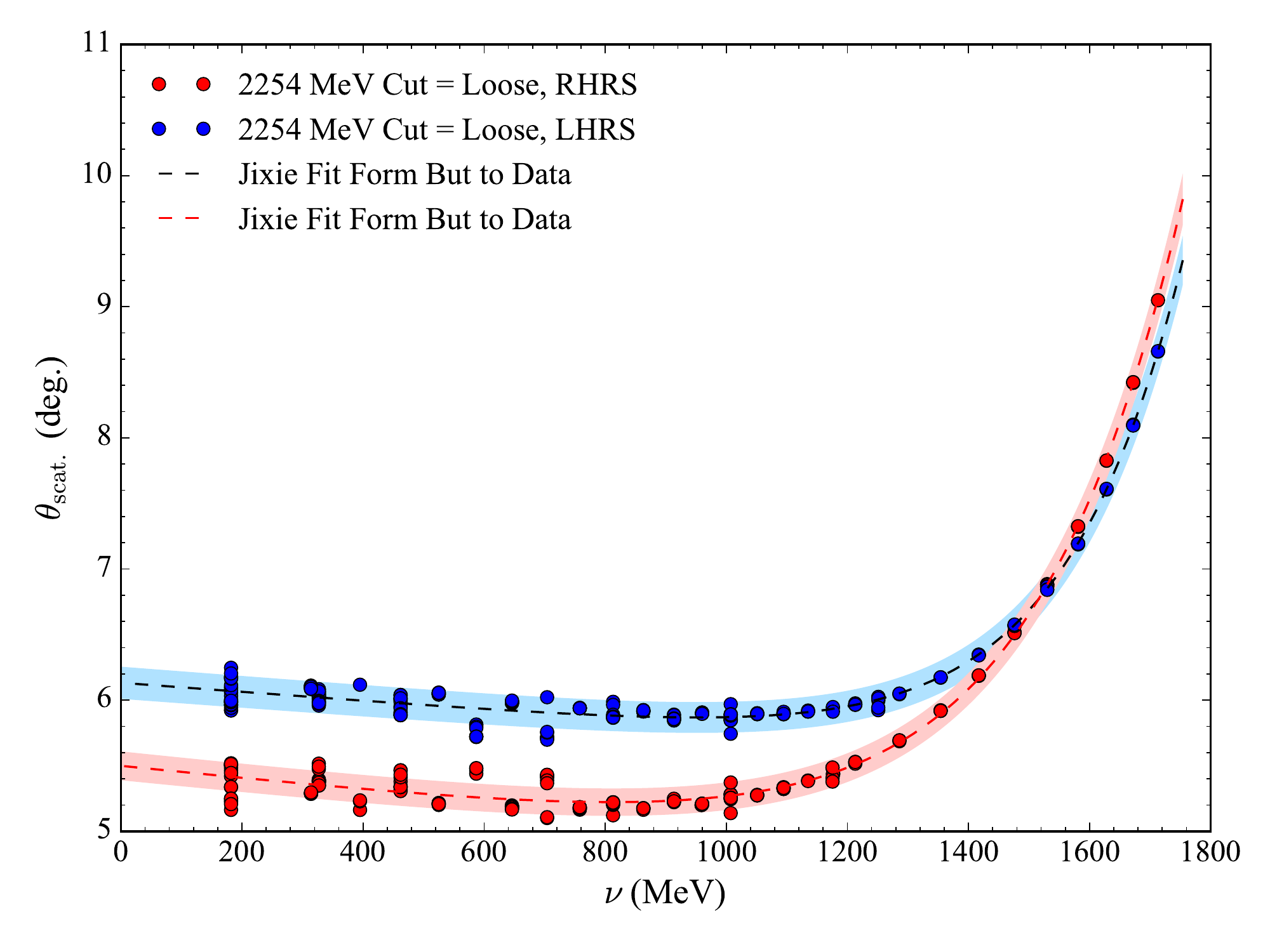}
\caption{Reconstructed scattering angle for the $E_0$ = 2254 MeV 2.5 T kinematic setting.}
\label{2254_25_Angle}
\end{figure}

 The symmetric hot-spot in Figure~\ref{AccCuts25} corresponds to relatively little variation in the scattering angle with $\nu$, and it is not until an electron energy of approximately 1 GeV that the scattering angle starts increasing rapidly. At the $P_0$ = 1 GeV point, the hot-spot is now concentrated in one part of the acceptance. The central scattering angle for the $E_0$ = 2254 MeV setting is shown in Figure~\ref{2254_25_Angle} and is generated using the same procedure described in Chapter~\ref{sec:Angfit}.

\begin{figure}[htp]
\centering     
\includegraphics[width=.80\textwidth]{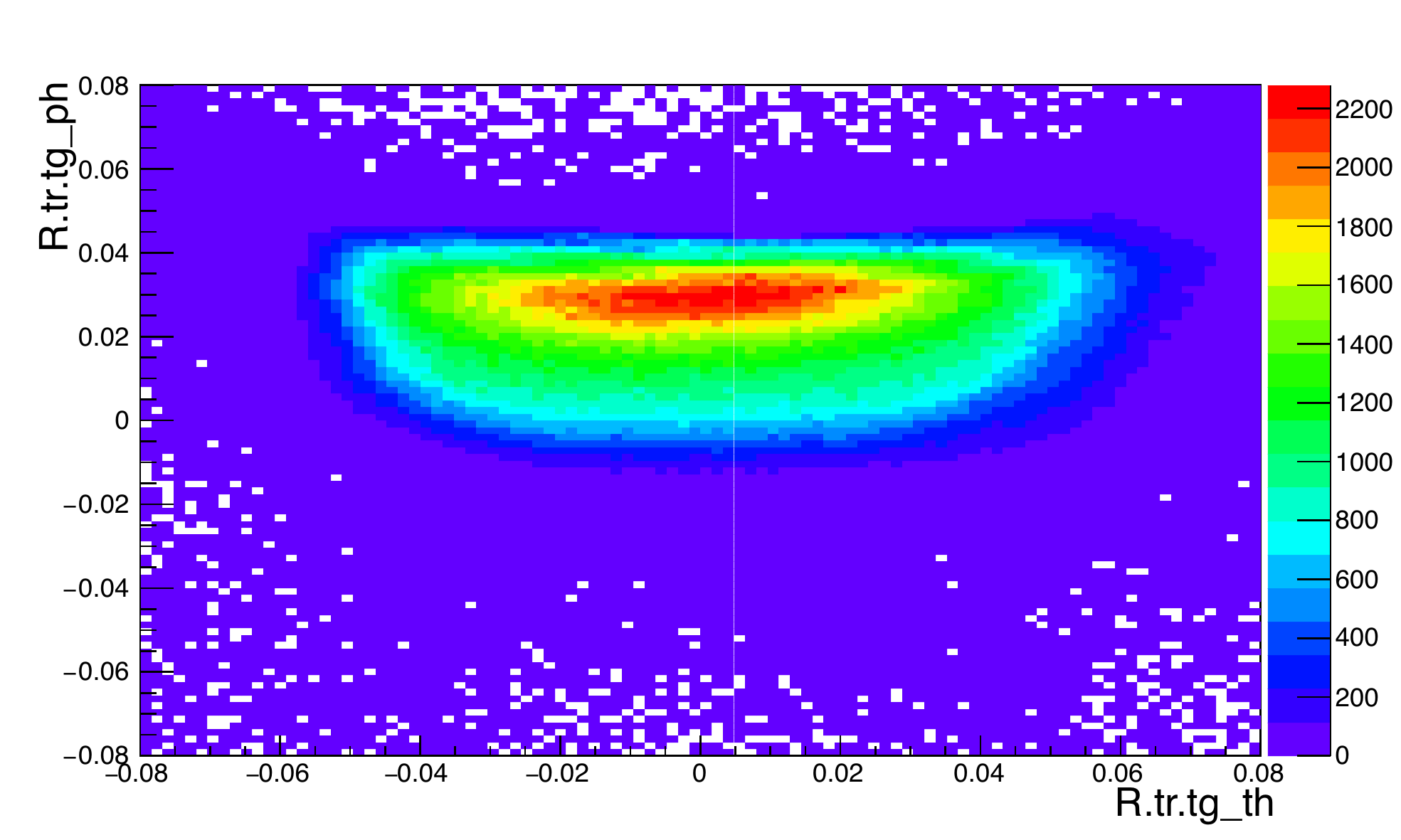}
\caption{Example acceptance for the RHRS at low $\nu$ for $E_0$ = 2254 MeV, 2.5 T.}
\label{22541_Acc}
\end{figure}

The difference between the LHRS and RHRS scattering angles is on the order of a degree at low $\nu$ in Figure~\ref{2254_25_Angle} and is also visible in the acceptance shape as a shift in phi target towards smaller phi values. This shift is seen in Figure~\ref{22541_Acc}. The large discrepancy between the two spectrometers will result in a systematic shift between the asymmetries generated on each arm. It is still an open question as to whether this systematic shift is dwarfed by the size of the statistical error bars, and it will need to be investigated further. 


\begin{figure}[htp]
\centering     
\subfigure[$E_0$ = 1711 MeV]{\label{fig:LHRSAcdfadsc234234}\includegraphics[width=.70\textwidth]{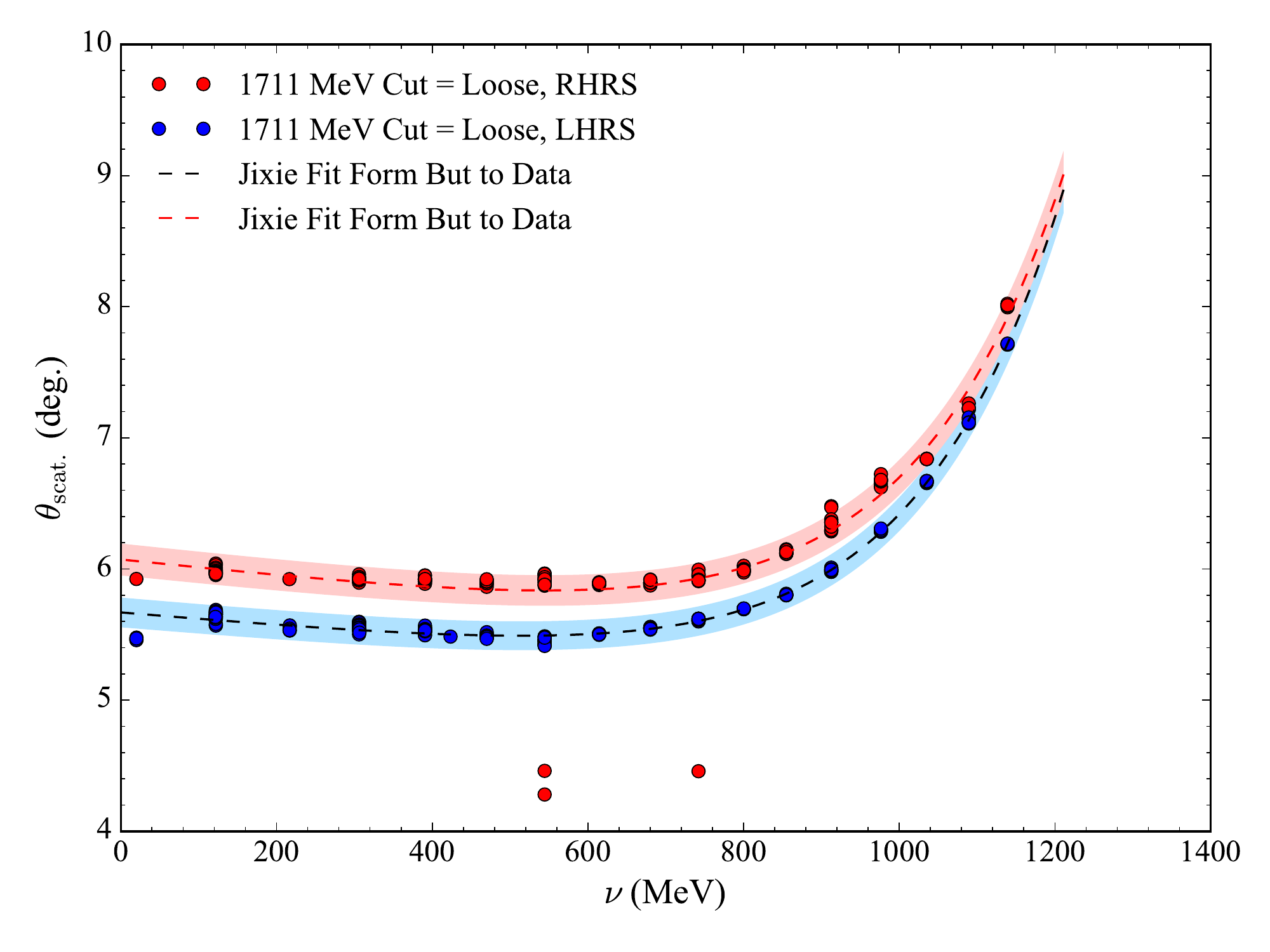}}
\qquad
\subfigure[$E_0$ = 1157 MeV]{\label{fig:RHRSAcadgagc23423423}\includegraphics[width=.70\textwidth]{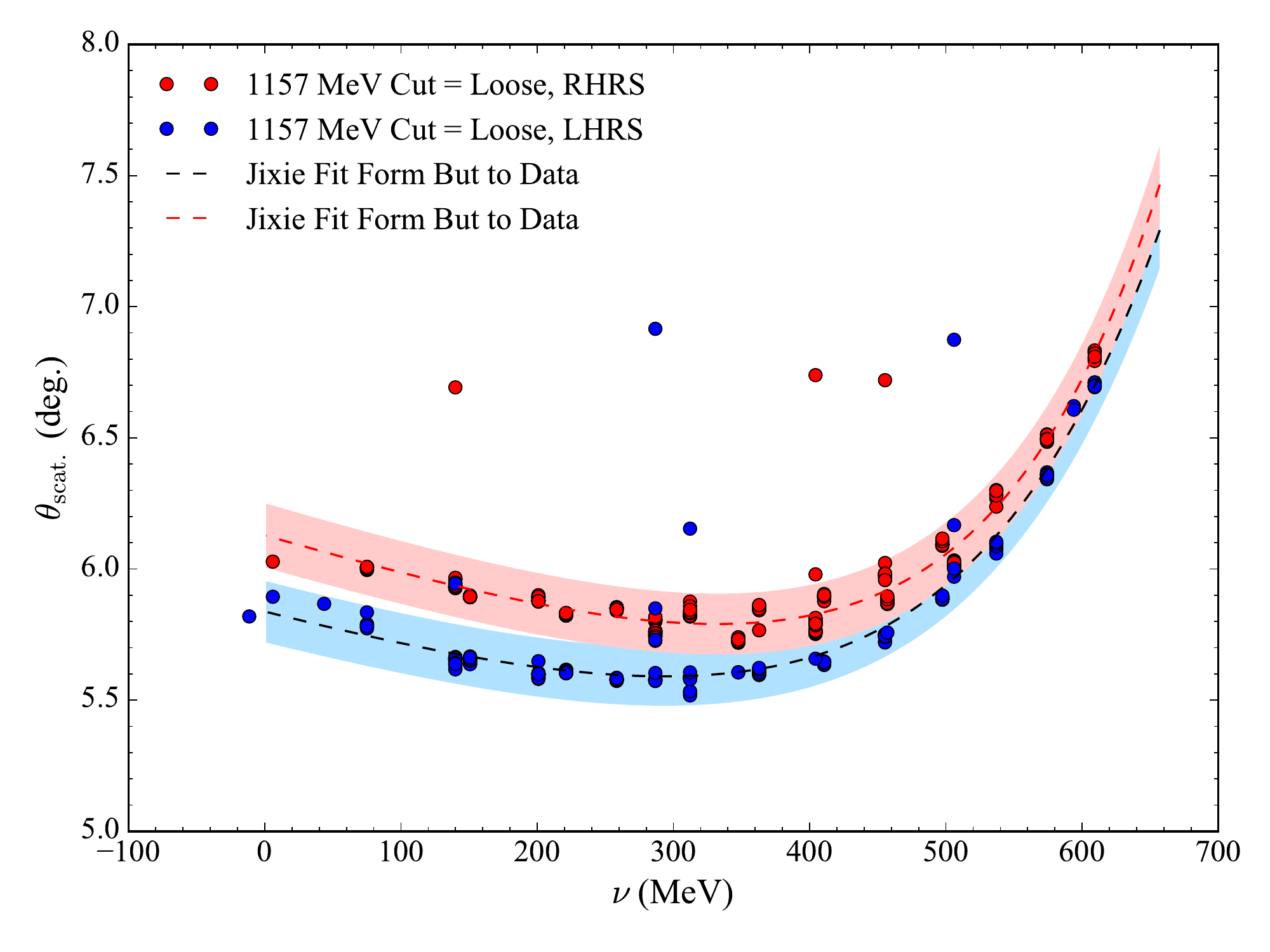}}
\caption{Reconstructed scattering angle for $E_0$ = 1711 MeV and $E_0$ = 1157 MeV.}
\label{Ang25}
\end{figure}

The remaining angle fits are shown in Figure~\ref{Ang25}. The angle difference between the LHRS and RHRS is much smaller at these settings and is consistent with the reconstruction results for the 5 T transverse settings. It is assumed that the systematic effect of this difference is negligible as it was for the 5 T settings. Ultimately, this assumption should be confirmed. The $E_0$ = 1157 MeV reconstruction exhibits odd behavior for a sizable portion of the runs and gives continuous bands of scattering angles at different momentum settings. These bands range in angle from approximately 6$^{\circ}$ to 20$^{\circ}$ and only exist for the LHRS. The effect on the fit in Figure~\ref{Ang25} is mitigated by cutting out any run with $\theta_{\mathrm{scat.}} > 7^{\circ}$ from the fit. This behavior should be checked for the final analysis. The preliminary fit parameters are in Table~\ref{table:fitparam25}.

\begin{table}[htp]
\begin{center}
\begin{tabular}{ l c c c r }\hline
 $E_0$ (MeV) & $p_0$ & $p_1$ & $p_2$ & $p_3$  \\[.2cm]  \hline 
 2254 & 3.839 & $-$4.989 & 5.341 & 0.351  \\[.2cm] 
 1711 & 3.965 & $-$5.222 & 4.755 & 0.530  \\[.2cm] 
 1157 & 4.161 & $-$6.497 & 4.041 & 1.523\\[.2cm]\hline
\end{tabular}
\caption{Fit parameters for scattering angle reconstruction at 2.5 T. }
\label{table:fitparam25}
\end{center}
\end{table}

\subsection{Preliminary Results}

\begin{figure}[htp]
\centering     
\subfigure[$E_0$ = 2254 MeV 2.5 T]{\label{fig:2254_25A}\includegraphics[width=.80\textwidth]{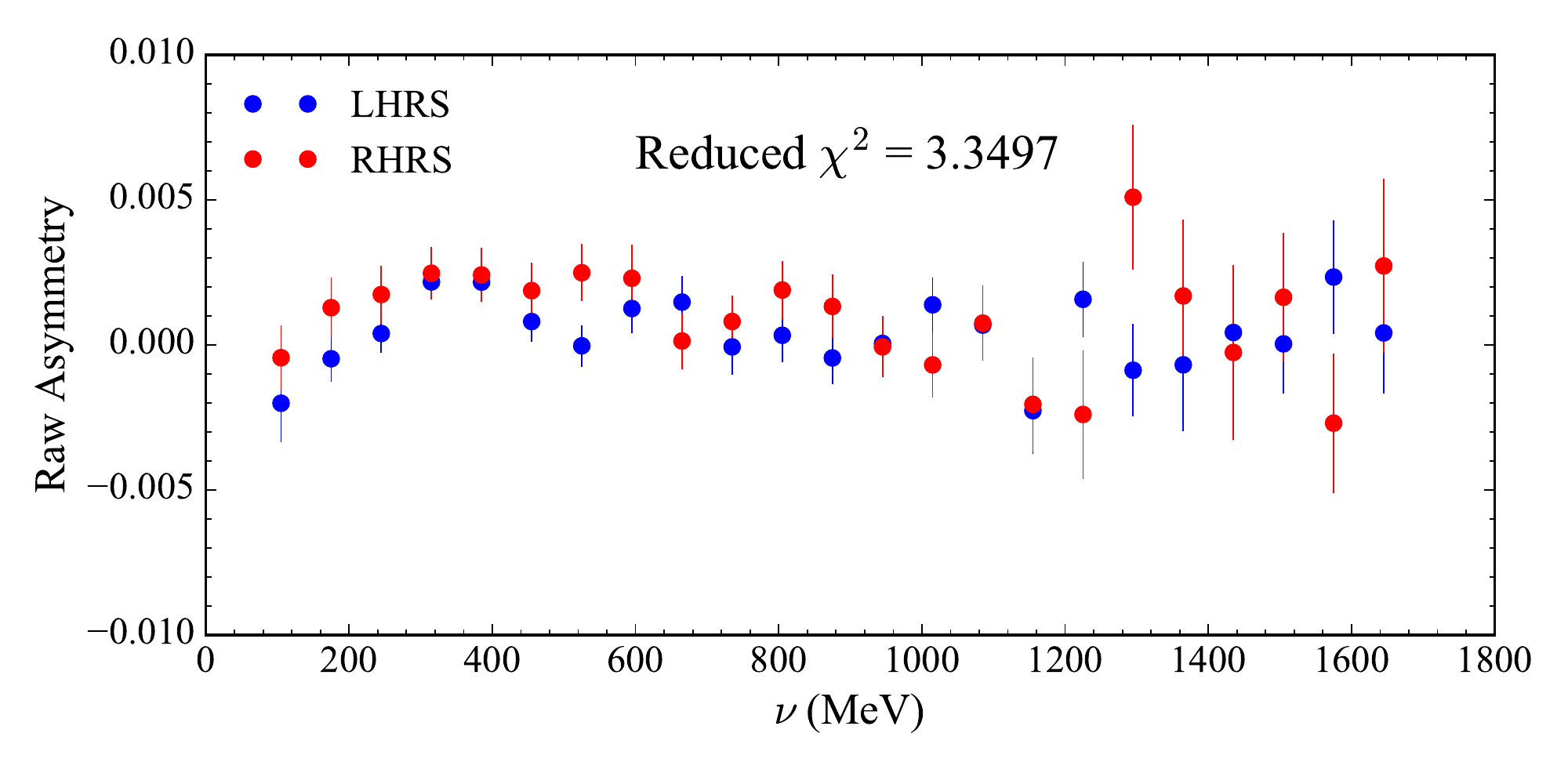}}
\qquad
\subfigure[$E_0$ = 1711 MeV 2.5 T]{\label{fig:1711_25A}\includegraphics[width=.80\textwidth]{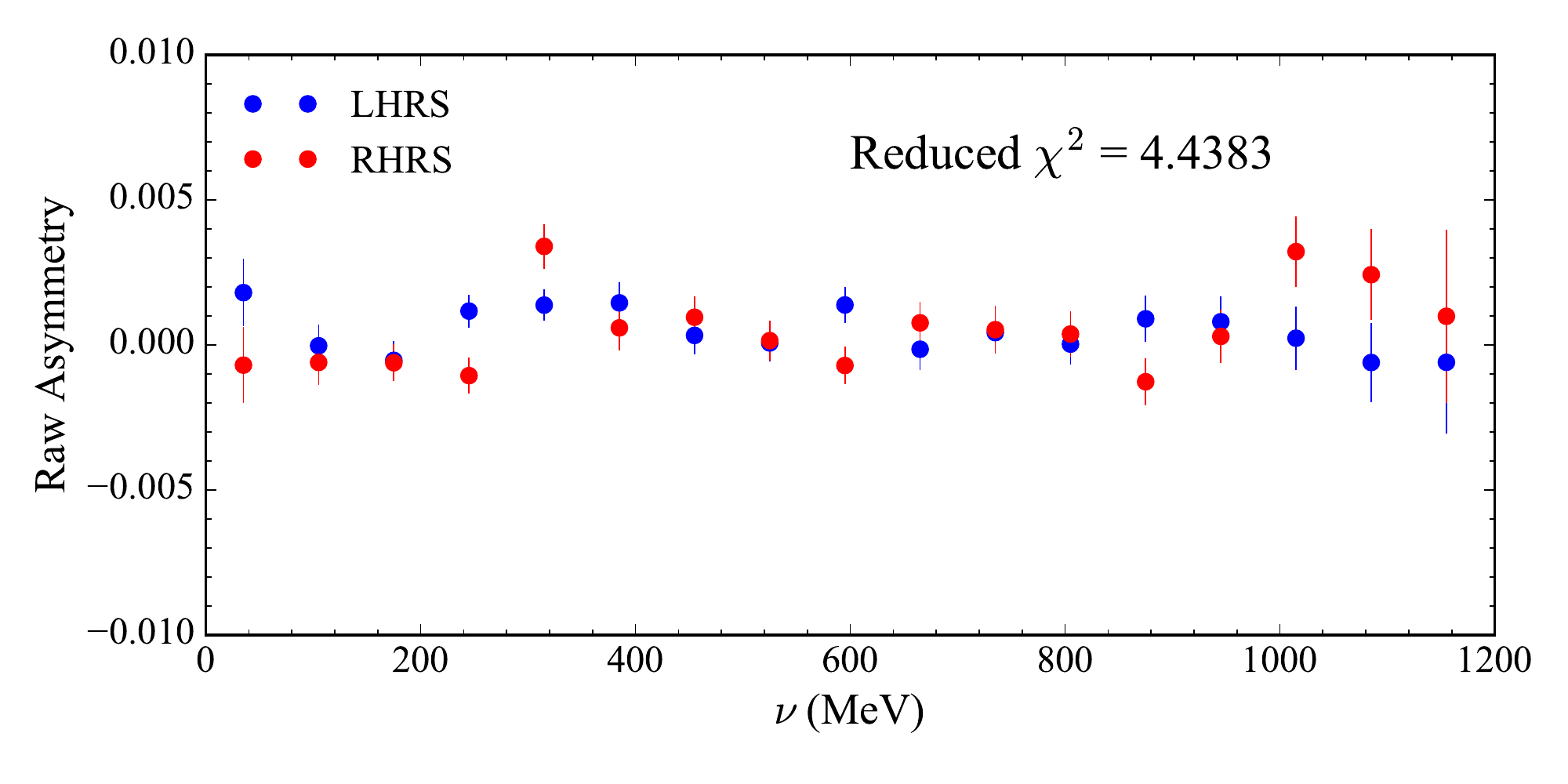}}
\qquad
\subfigure[$E_0$ = 1157 MeV 2.5 T]{\label{fig:1157_25A}\includegraphics[width=.80\textwidth]{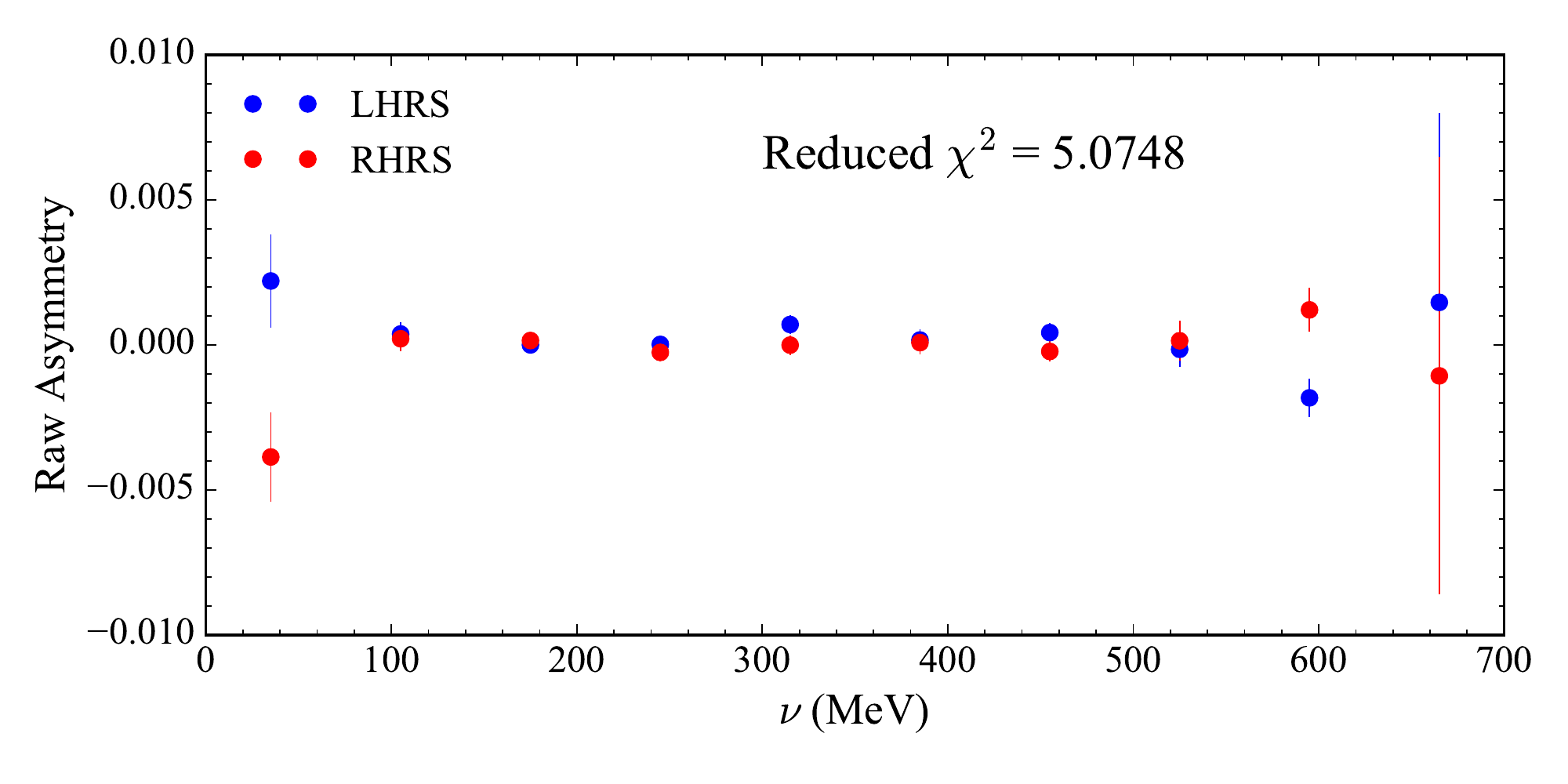}}
\caption{Raw asymmetry HRS comparison for the 2.5 T settings.}
\label{PrelimAsymComp}
\end{figure}

A preliminary comparison between the LHRS and RHRS asymmetries for the 2.5 T settings is shown in Figure~\ref{PrelimAsymComp}. The binning used is 70 MeV. A significant fraction of the runs at $P_0$ = 1.49 GeV for $E_0$
= 1711 MeV were taken with a mismatched septa and dipole current setting. This analysis includes these runs (RunQuality = -1 in the MySQL) but their effect on the asymmetry should be checked. The hope is that the ratio nature of the asymmetry cancels out any acceptance shifts from the mismatch.

The reduced $\chi^2$ of each comparison is greater than the previously discussed ideal value of $\chi^2_{\mathrm{red.}} \approx 1.3$. To resolve these discrepancies it is necessary to go through the asymmetries on a run-by-run basis to identify potential sources of the differences. Examples of sources from the 5 T analysis include runs with large livetime and charge asymmetries and edge effects from the acceptance. The angle difference at the $E_0$ = 2254 MeV setting is a likely source as is the septa and dipole mismatched runs at $E_0$ = 1711 MeV.   

The preliminary physics asymmetry results use a model dilution and packing fraction value of $pf$ = 0.60 because the analysis of both of those quantities is not complete. The dilution is produced from the model tune discussed in Appendix~\ref{app:Appendix-C} and a conservative 25\% systematic error is applied to the result. For this preliminary analysis, the RHRS asymmetries at $E_0$ = 2254 MeV are adjusted to match the LHRS scattering angle by a model scaling. In the scaling method, a model is run at each scattering angle fit and then the difference between the asymmetries is applied to the RHRS data. No estimate is provided by the residual systematic error after making this correction, but it needs to be determined in the final analysis.

\begin{figure}[htp]
\centering     
\subfigure[$E_0$ = 2254 MeV 2.5 T]{\label{fig:2254_25A}\includegraphics[width=.80\textwidth]{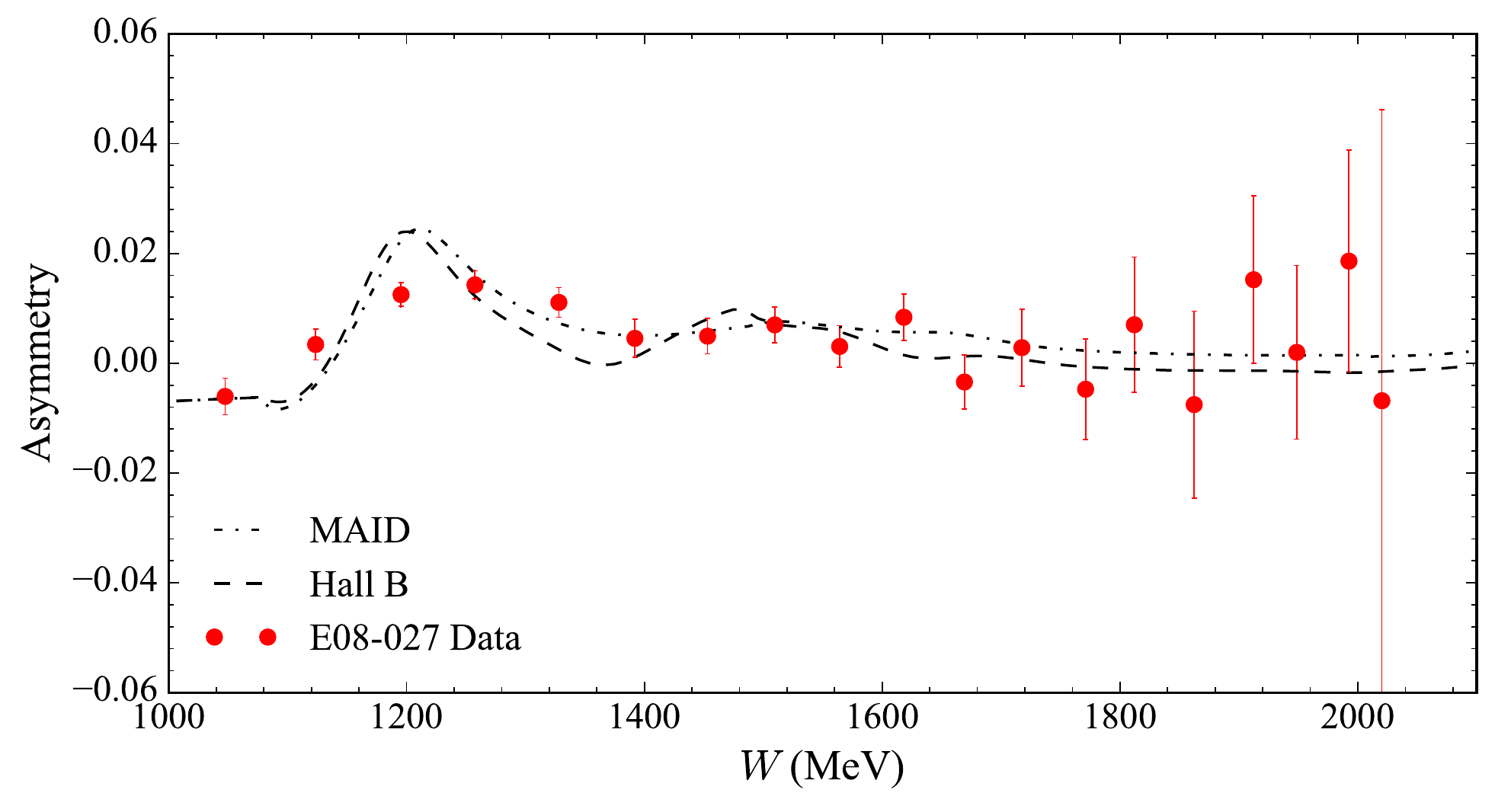}}
\qquad
\subfigure[$E_0$ = 1711 MeV 2.5 T]{\label{fig:1711_25A}\includegraphics[width=.80\textwidth]{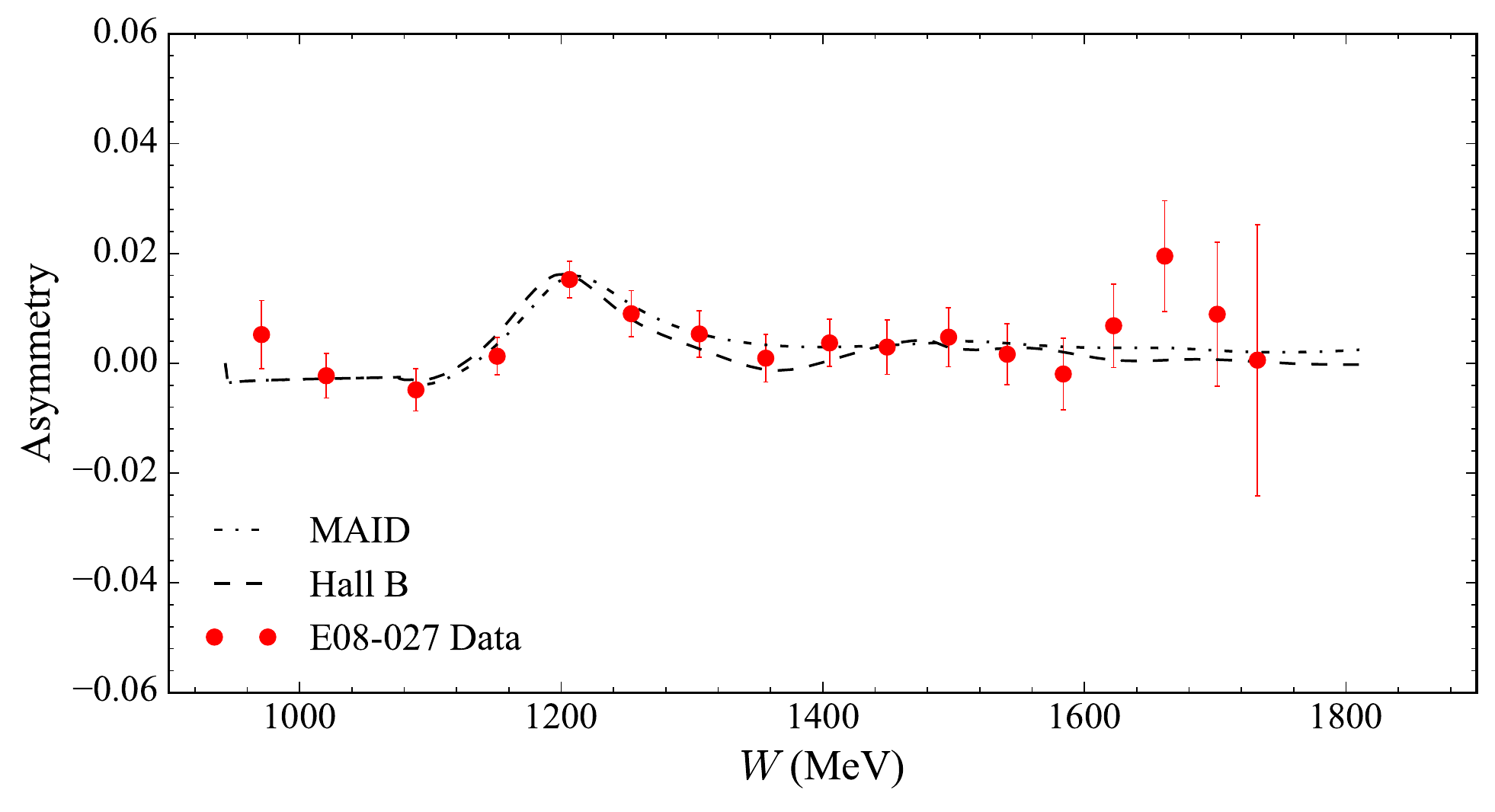}}
\qquad
\subfigure[$E_0$ = 1157 MeV 2.5 T]{\label{fig:1157_25A}\includegraphics[width=.80\textwidth]{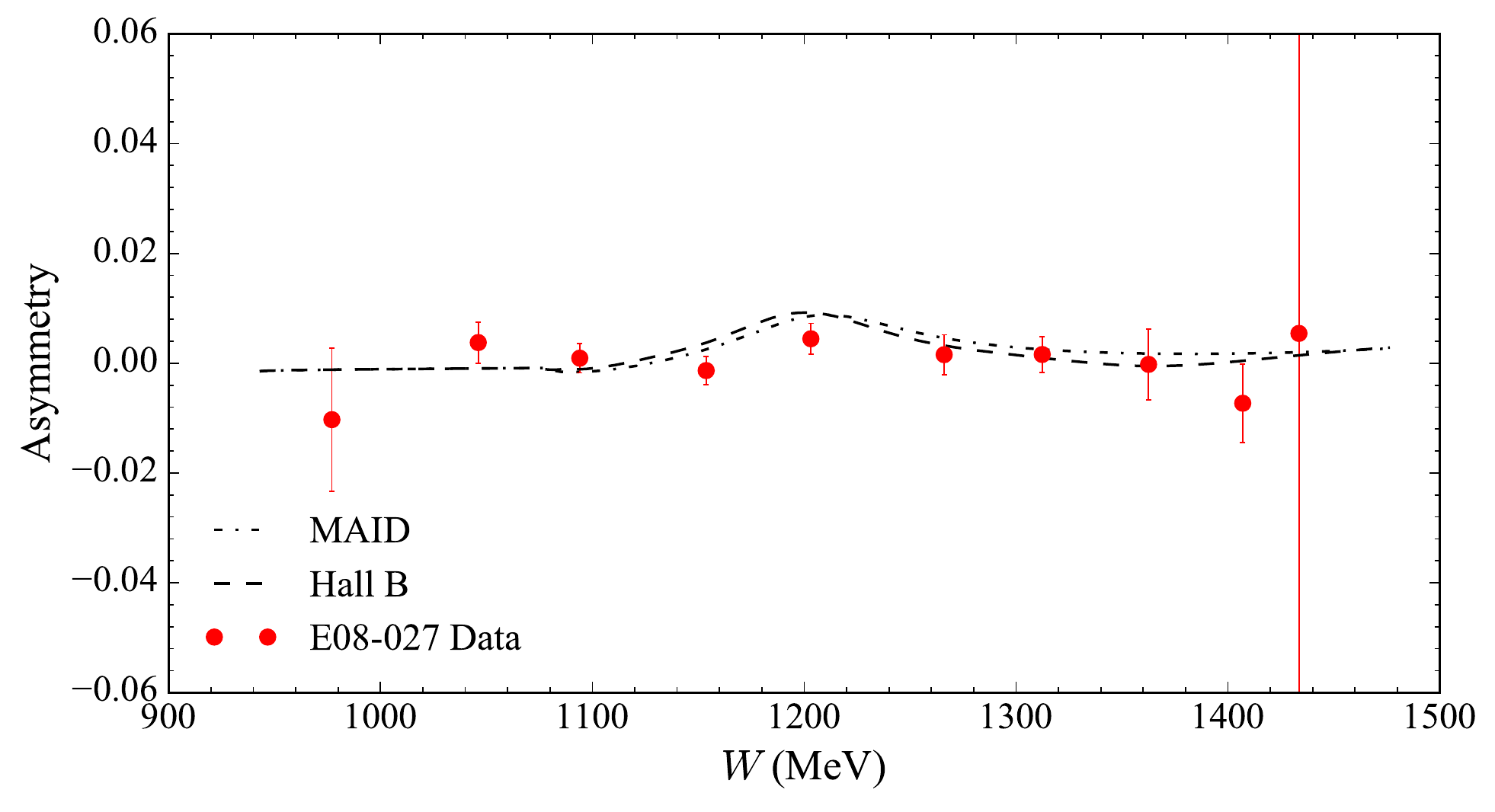}}
\caption{Preliminary asymmetries for the 2.5 T settings.}
\label{PrelimAsym}
\end{figure}

 The preliminary physics asymmetries are shown in Figure~\ref{PrelimAsym} and the inner error bars are statistical and the outer are the total error. The systematic error includes the 25\% contribution from the dilution mentioned above and the other sources discussed in Chapter~\ref{AsymSys}. The statistical error is the dominate source of error at each kinematic setting and data point. The $\Delta$ (1232) resonance  (and only this resonance) is visible at each setting. The $E_0$ = 1157 MeV setting suffers from poor statistics resulting from the smaller target polarization and also a smaller asymmetry. While the asymmetry has structure at this setting, it is still mostly consistent with zero.

\section{Polarized Cross Section Differences}

The unpolarized cross section contribution to the polarized cross section differences is provided by the same model used in the 5 T analysis. The $Q^2$ dependent systematic error is estimated from Figure~\ref{BostedModelComp} and is taken as 20\% for the 2.5 T settings. According to Figure~\ref{BostedModelComp} this is potentially an underestimate for the $E_0$ = 1157 MeV setting, but it will become apparent in the following discussions that this setting is severely statistics limited so the systematic errors have little bearing on the results.  
\begin{figure}[htp]
\centering     
\includegraphics[width=.80\textwidth]{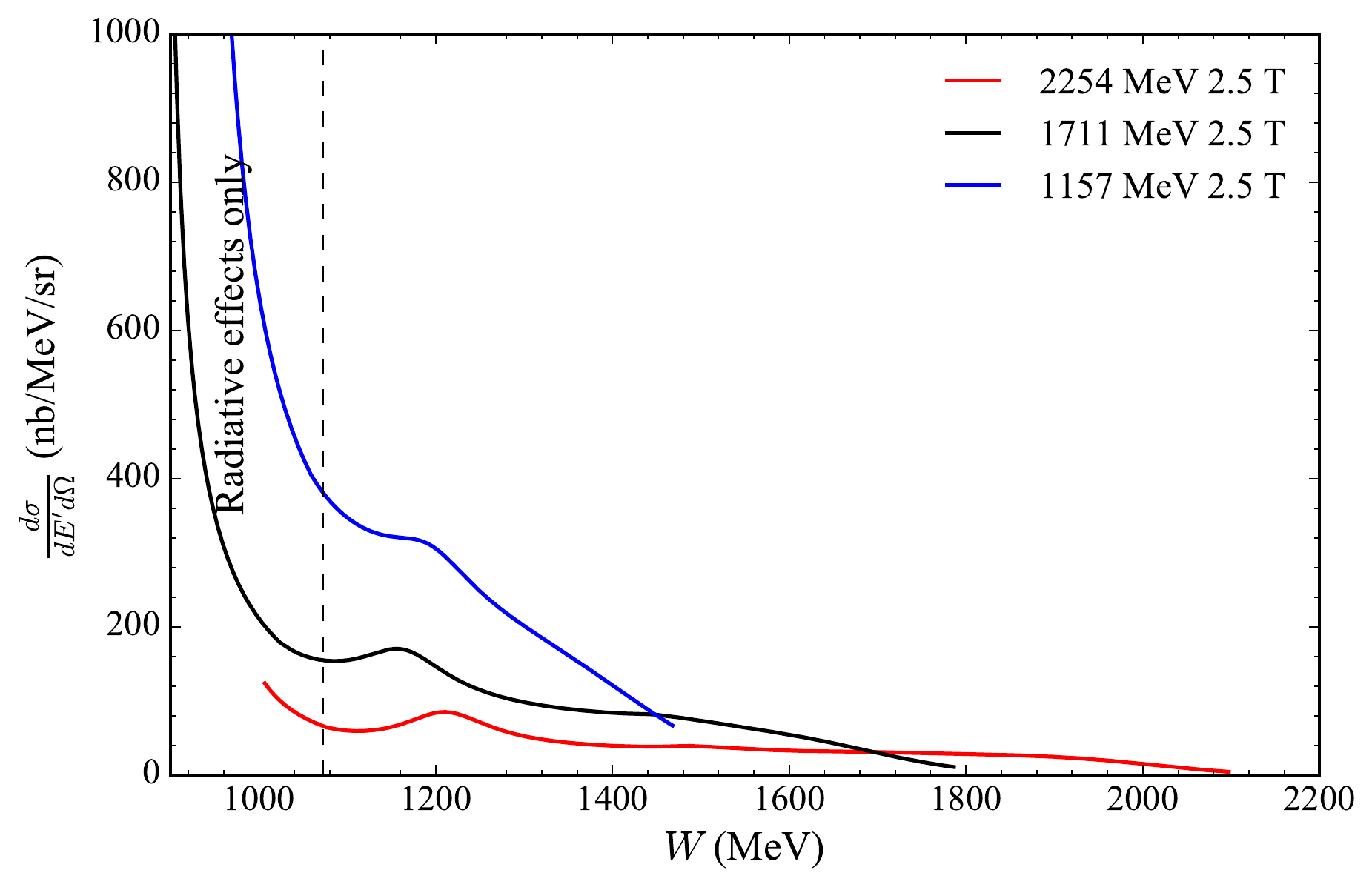}
\caption{Model generated radiated unpolarized cross sections for the 2.5 T settings.}
\label{Bosted_25T}
\end{figure}

The unpolarized model cross sections are shown in Figure~\ref{Bosted_25T}. The $E_0$ = 1157 MeV cross section is dominated by the unpolarized elastic tail. These radiative events will further increase any systematic effects during the error propagation after the tail subtraction. To mitigate this error, data was taken during the experiment on a shorter ammonia target cell. This had the effect of decreasing the radiation length. The amount of data taken (a few runs at each dipole setting) is not sufficient for an asymmetry calculation and the short-cell runs are not considered in this analysis, but they may be useful for the final analysis and radiative corrections.

\subsection{Radiative Corrections}
\begin{figure}[htp]
\centering     
\subfigure[$E_0$ = 2254 MeV 2.5 T]{\label{fig:2254_25DS}\includegraphics[width=.80\textwidth]{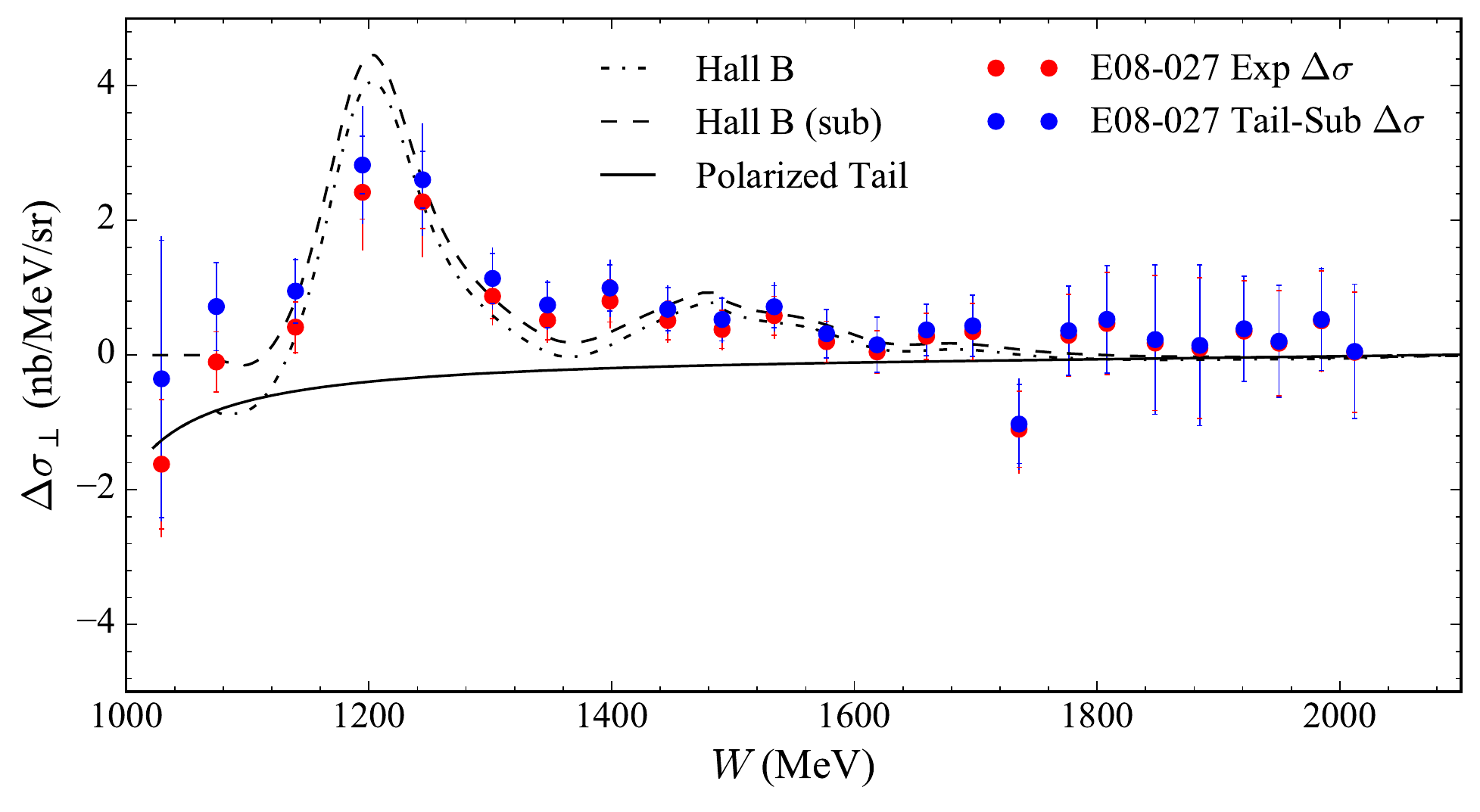}}
\qquad
\subfigure[$E_0$ = 1711 MeV 2.5 T]{\label{fig:1711_25DS}\includegraphics[width=.80\textwidth]{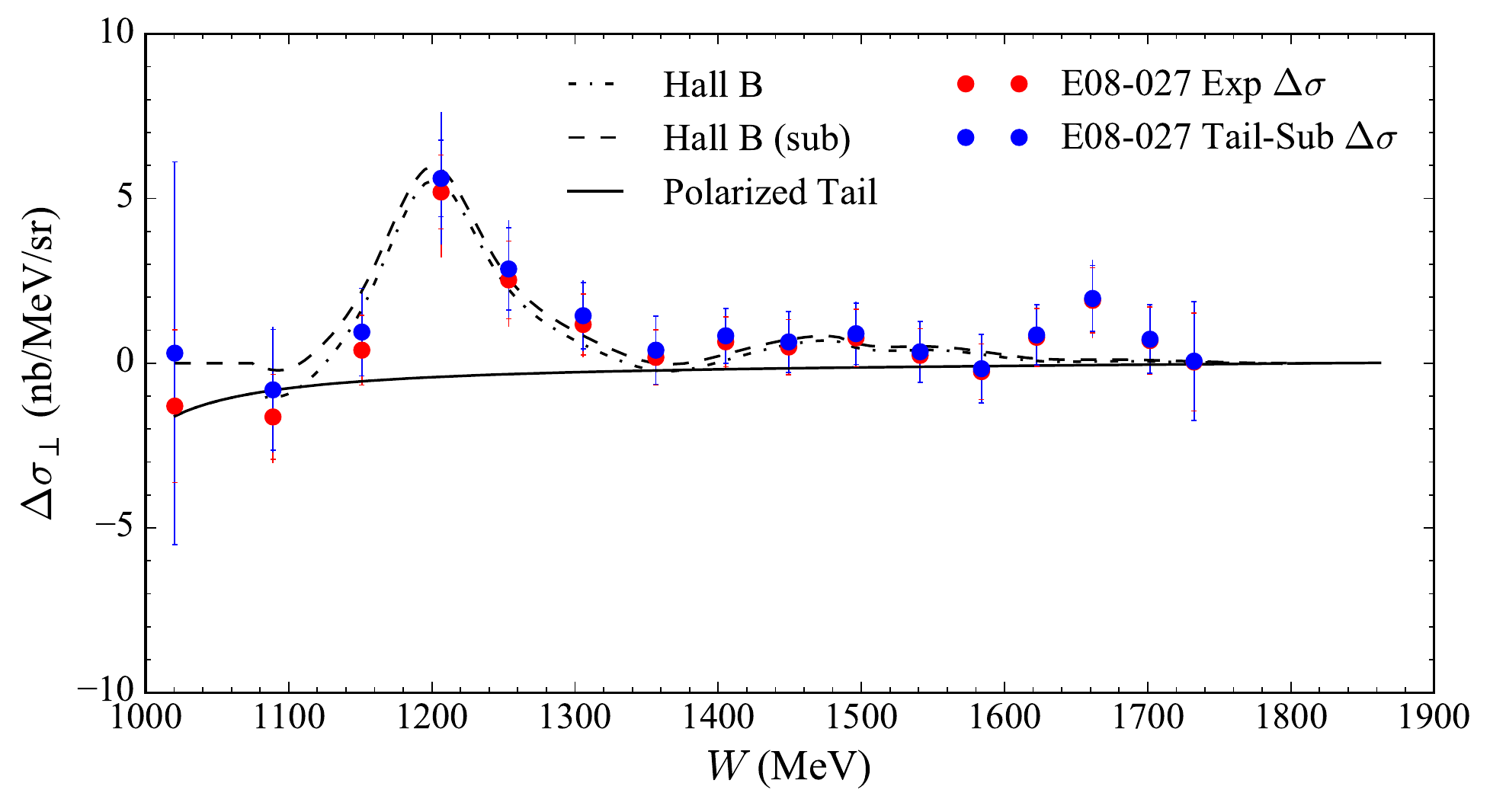}}
\qquad
\subfigure[$E_0$ = 1157 MeV 2.5 T]{\label{fig:1157_25DS}\includegraphics[width=.80\textwidth]{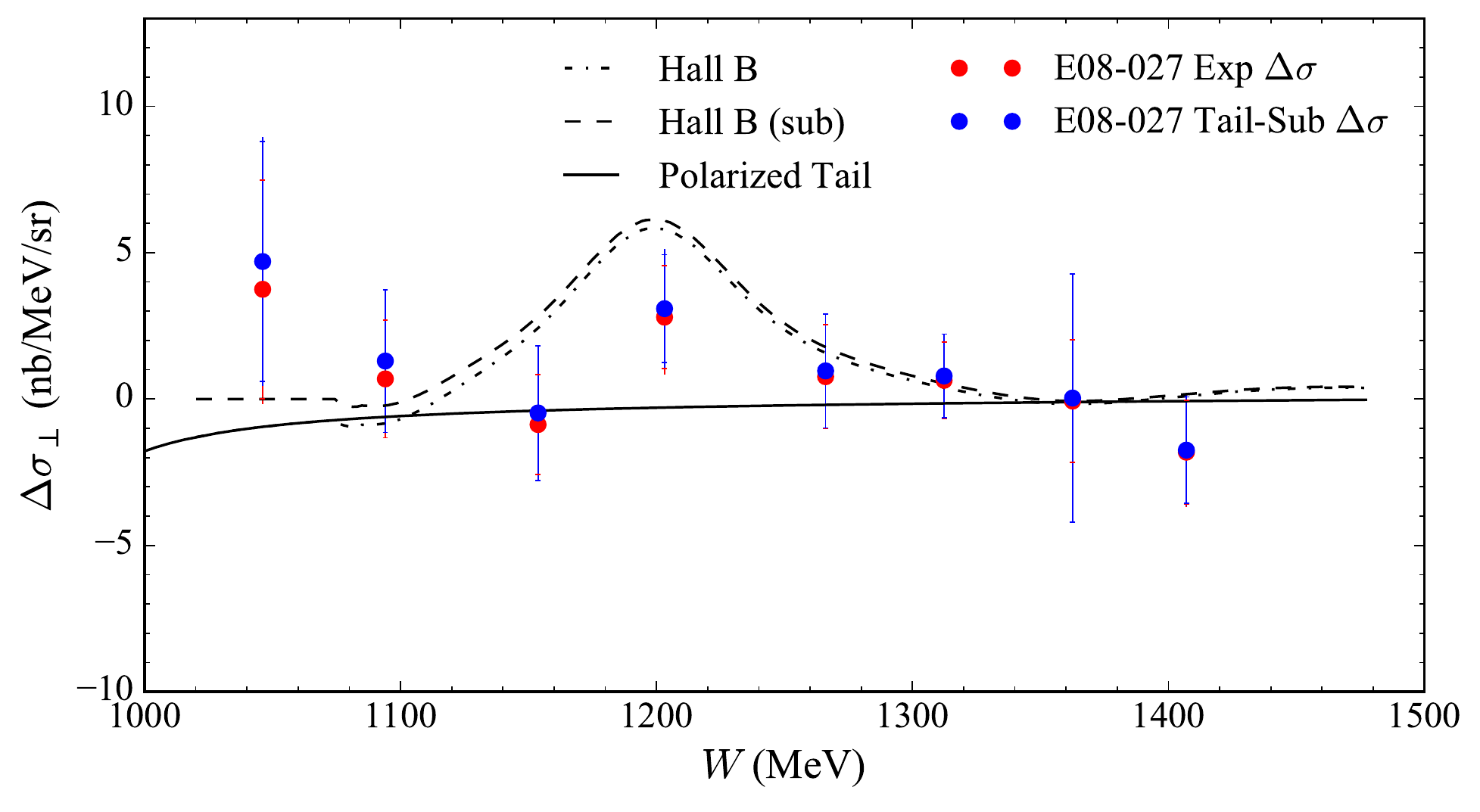}}
\caption{Preliminary polarized cross section differences for the 2.5 T settings.}
\label{PrelimDS}
\end{figure}

The polarized cross section differences before and after the polarized tail subtraction are shown in Figure~\ref{PrelimDS}. The inner error bars are statistical and the outer are the total error. The systematic error in the tail is determined following the procedures discussed in Chapter~\ref{result:RC} and the statistical error is modified after the tail subtraction according to equation~\eqref{stat_sub}. The systematic errors on the calculated tail are comparable with the 5 T settings and range mostly from 2-5\%. In some cases, where the elastic tail is very small, the error is larger (on the order of 20-30\%), but this is no cause for alarm because the errors are propagated absolutely and contribute little to the total error after subtraction. With the exception of $E_0$ = 1157 MeV, there is good agreement with the polarized tail and data below the pion production threshold. The resulting subtraction is consistent with zero. The sign of the radiative elastic tail is confirmed by comparison with elastic events.

\begin{figure}[htp]
\centering     
\subfigure[$E_0$ = 2254 MeV 2.5 T]{\label{fig:2254_25DSB}\includegraphics[width=.80\textwidth]{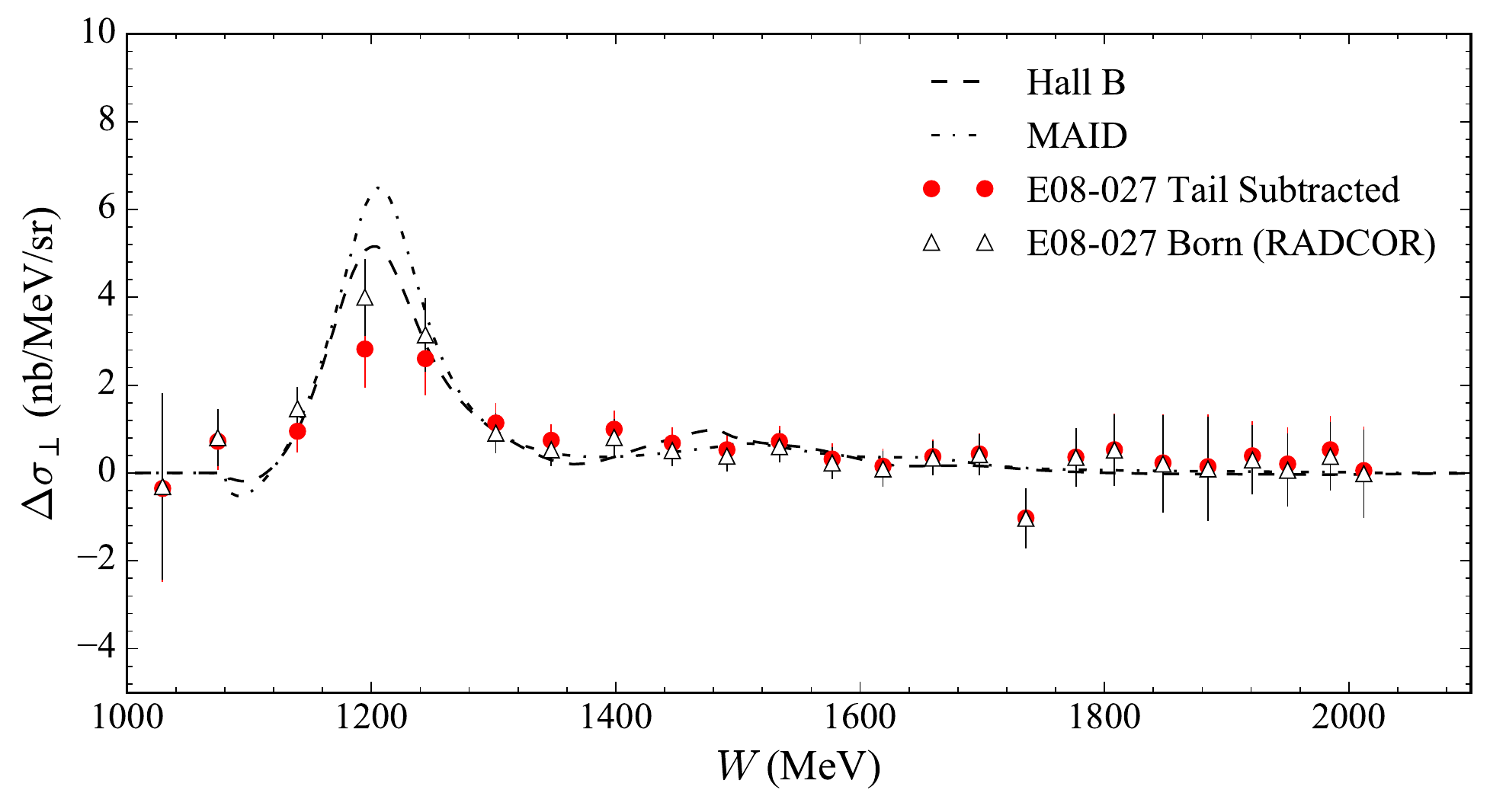}}
\qquad
\subfigure[$E_0$ = 1711 MeV 2.5 T]{\label{fig:1711_25DSB}\includegraphics[width=.80\textwidth]{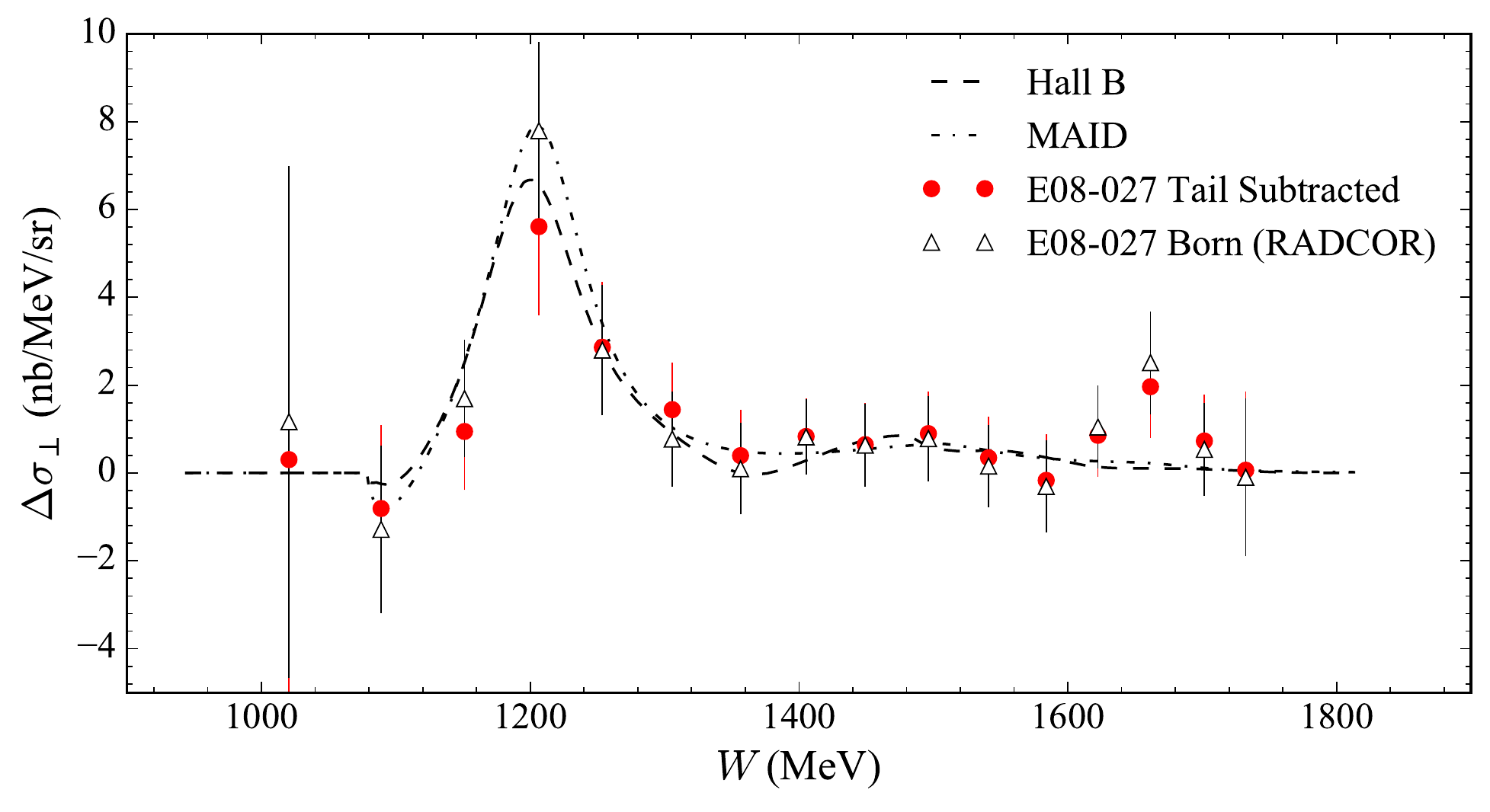}}
\qquad
\subfigure[$E_0$ = 1157 MeV 2.5 T]{\label{fig:1157_25DSB}\includegraphics[width=.80\textwidth]{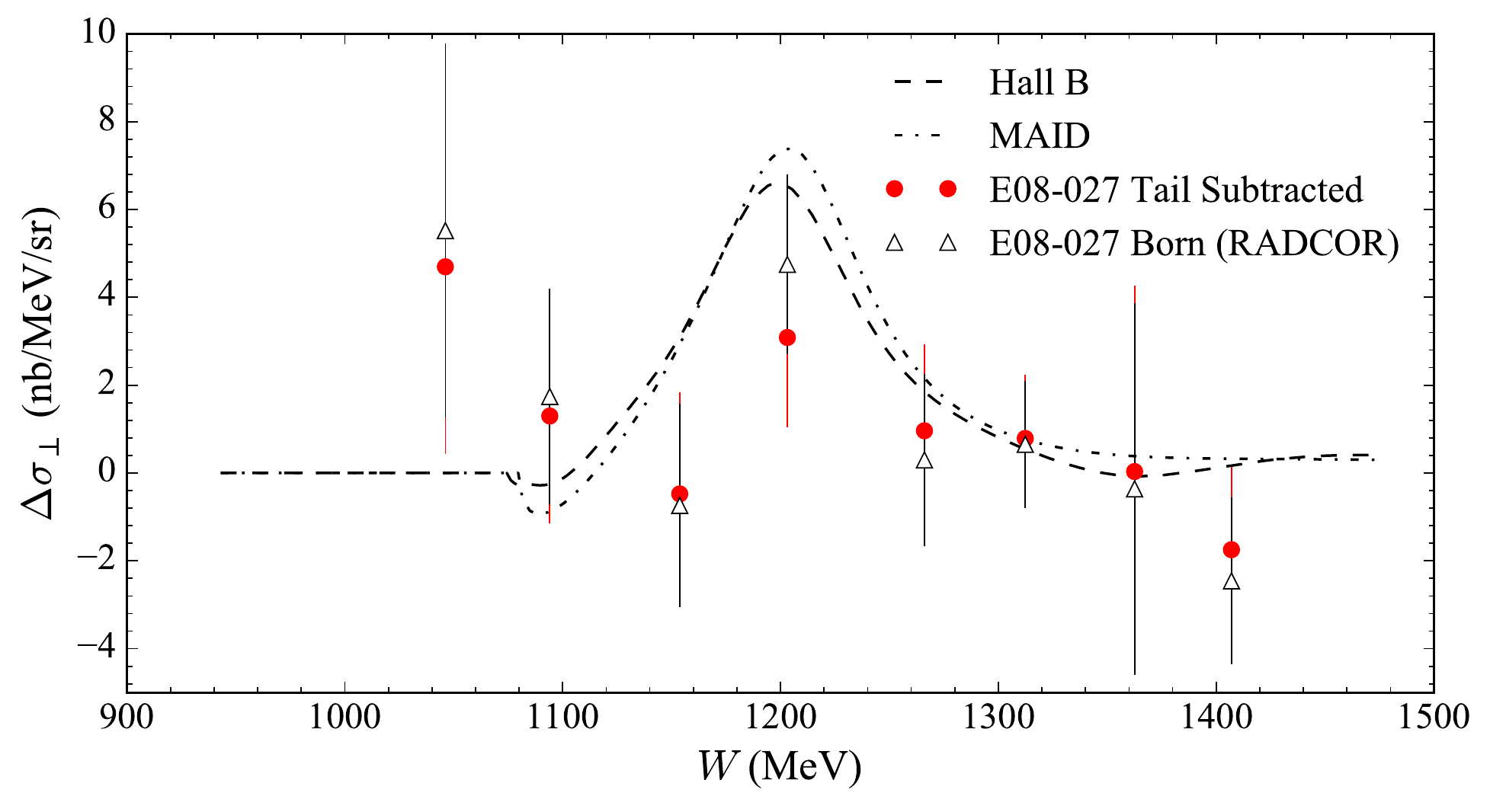}}
\caption{Preliminary Born polarized cross section differences for the 2.5 T settings.}
\label{PrelimBornDS}
\end{figure}

For this analysis the inelastic radiative corrections are carried out using the unpolarized formalism of RADCOR for the internal and external corrections. Referring back to Figure~\ref{BornDSResult}, the gained uncertainty from forgoing POLRAD is on the order of 5\% and tolerable for a preliminary analysis. In general, RADCOR overestimates the Born result when compared to the full treatment and should be kept in mind going forward. The preliminary Born polarized cross section differences for the 2.5 T setting are shown in Figure~\ref{PrelimBornDS}. The inner error bars are statistical and the outer are the total error. The systematic error for inelastic radiative correction is calculated using the methods from Chapter~\ref{result:RC} and the scale of the input modules used to aid the extrapolation are 0.95, 0.70 and 0.45. For the $E_0$ = 1711 MeV unfolding only one extrapolation spectrum is provided at $E_0$ = 1157 MeV and none is used to correct the $E_0$ = 1157 MeV data. Two spectra are used for $E_0$ = 2254 MeV. The use of unaided extrapolation at $E_0$ = 1157 MeV is justified in Figure~\ref{RCProc} and Figure~\ref{ErrorBarComp}, where data near the $\Delta$(1232) resonance is unaffected by extrapolation.

\section{Spin Structure Function Extraction and Moments}
The spin structure function, $g_2(x,Q^2)$, is extracted from the Born polarized cross section differences according to equation~\eqref{g2fromg1}. There is no currently published $g_1(x,Q^2)$ data in the kinematic region of the $E_0$ = 1711 MeV and $E_0$ = 1157 MeV datasets, so the estimated systematic error, taken as the absolute difference between the Hall B and MAID models should be confirmed for the final analysis. The kinematics of the $E_0$ = 2254 MeV 2.5 T setting match that of the $E_0$ = 2254 5 T longitudinal setting and those systematic errors carry over here.

\begin{figure}[htp]
\centering     
\subfigure[$E_0$ = 2254 MeV 2.5 T]{\label{fig:2254_25g2}\includegraphics[width=.80\textwidth]{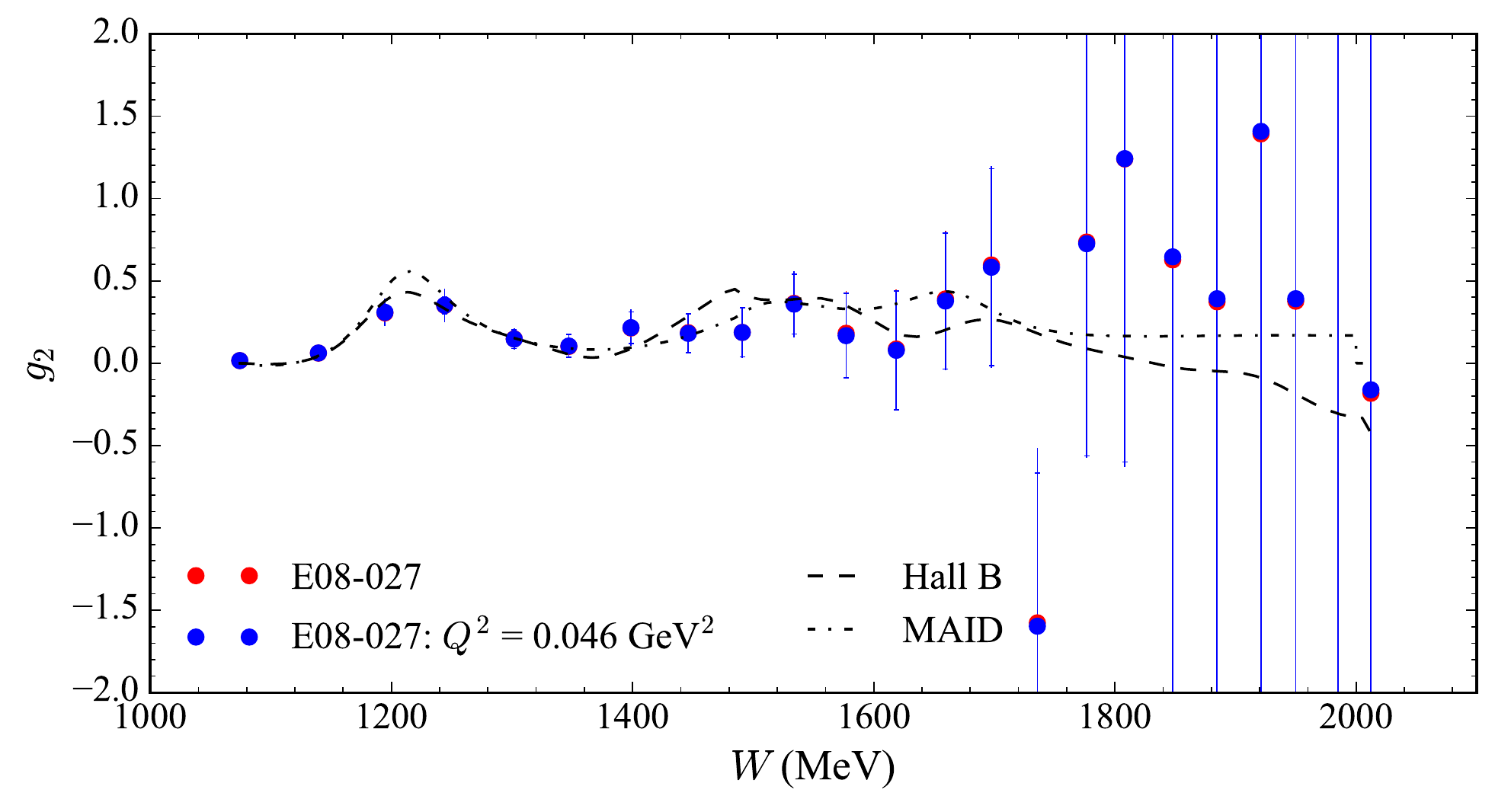}}
\qquad
\subfigure[$E_0$ = 1711 MeV 2.5 T]{\label{fig:1711_25g2}\includegraphics[width=.80\textwidth]{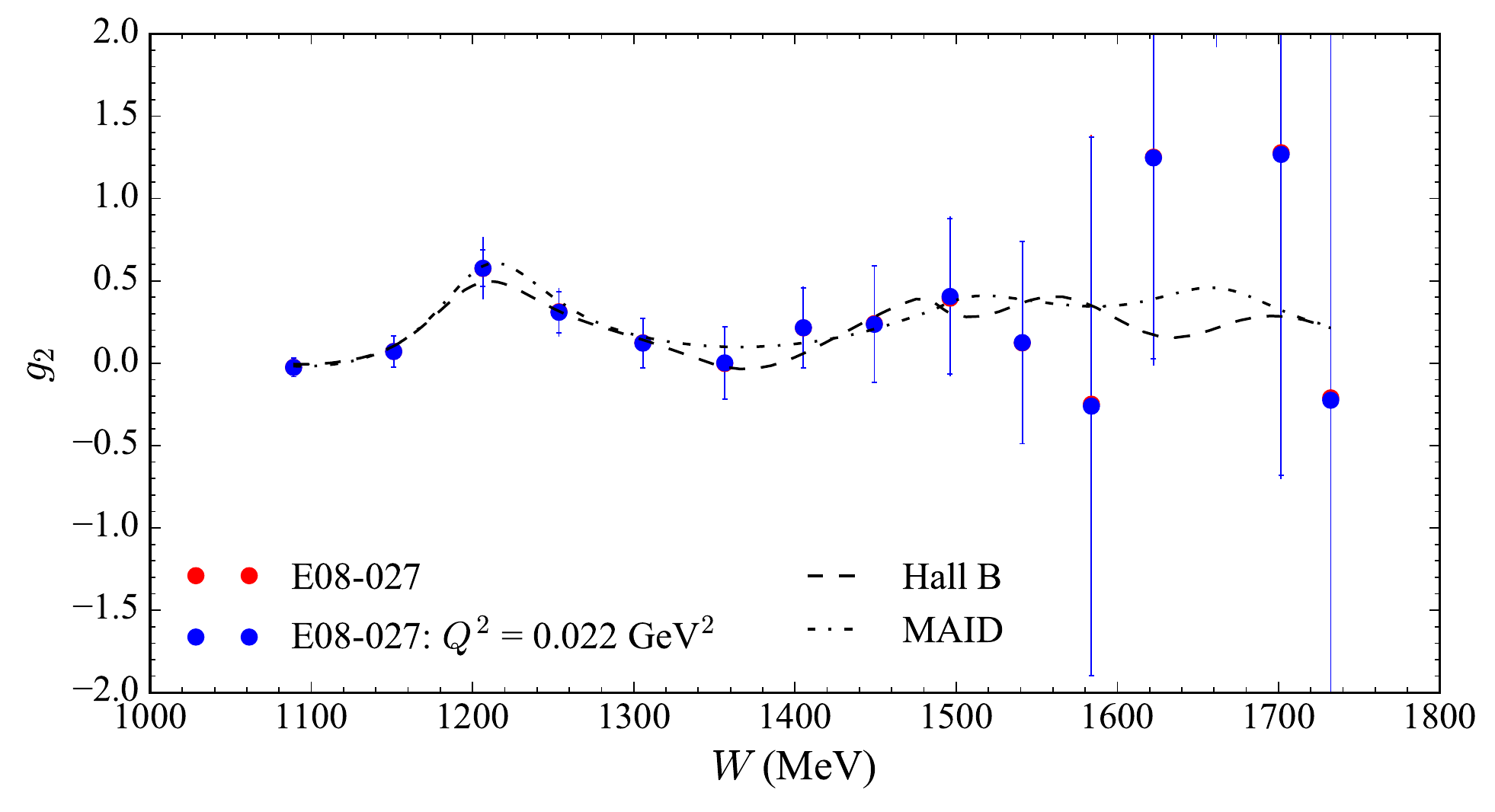}}
\qquad
\subfigure[$E_0$ = 1157 MeV 2.5 T]{\label{fig:1157_25g2}\includegraphics[width=.80\textwidth]{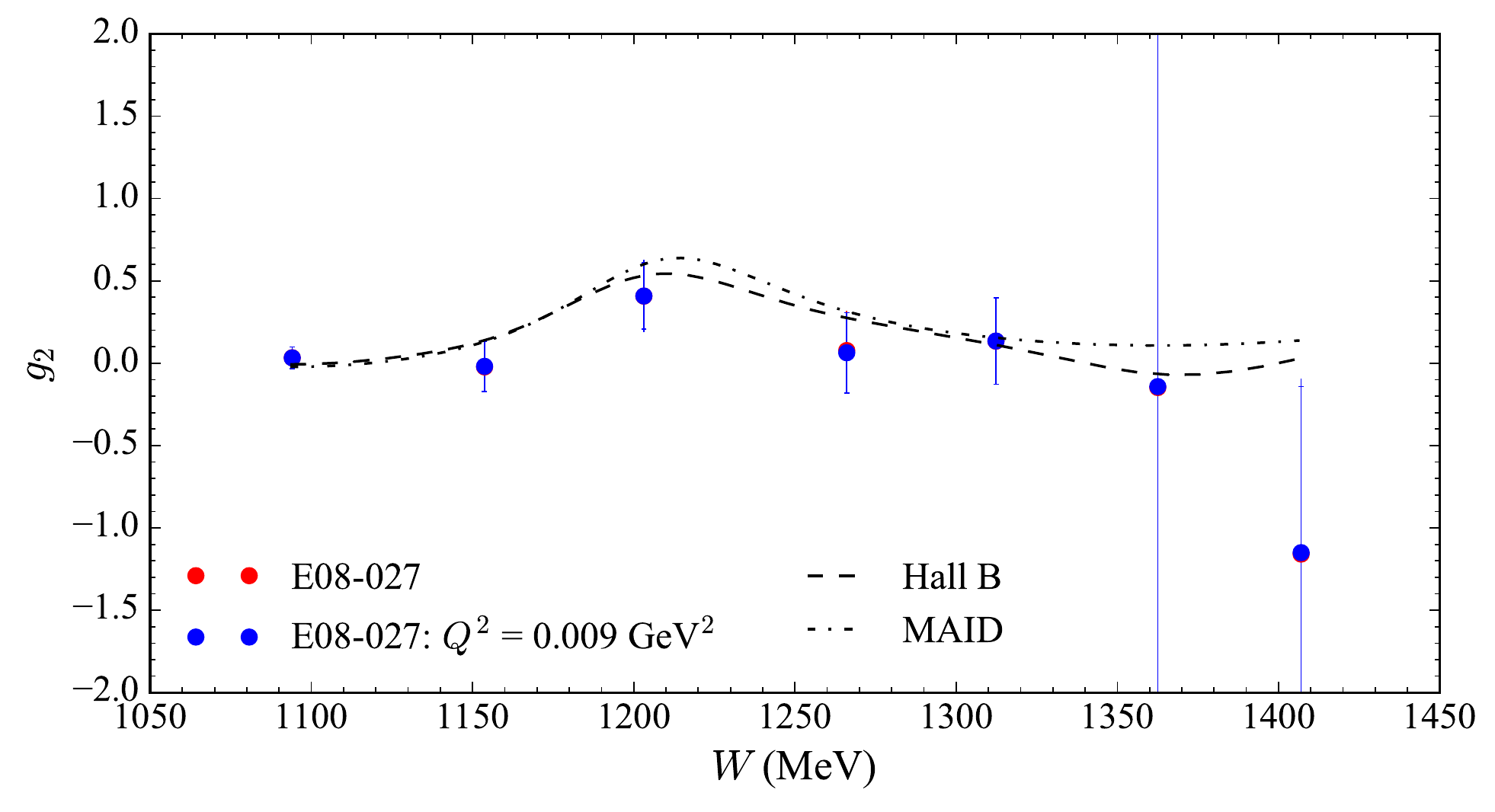}}
\caption{Preliminary $g_2(x,Q^2)$ for the 2.5 T settings.}
\label{Prelimg2}
\end{figure}

The results of the spin structure function extraction and adjustment to constant momentum transfer are shown in Figure~\ref{Prelimg2}. The inner error bars are statistical and the outer are the total uncertainty. The values of constant $Q^2$ are selected from the points nearest the $\Delta(1232)$ resonance in the data and correspond to 0.046 GeV$^2$, 0.022 GeV$^2$ and 0.009 GeV$^2$ for the $E_0$ = 2254 MeV, 1711 MeV and 1157 MeV settings respectively. The adjustment to constant $Q^2$ is done using the model method discussed in Chapter~\ref{EvolveMeth}. The additional systematic error from using this method is computed as the difference in the results from the Hall B and MAID models. For the final analysis, a study should be done, similar to that of the 5 T transverse settings, where one setting is evolved to another and the resulting difference is computed to gauge the systematic error. The nature of the model method (the $Q^2$ of the $\Delta(1232)$ resonance is held fixed) creates the most change in the structure functions at large $W$. For the 2.5 T setting this is where the statistical error bars are largest and the systematic error is less important. This high $W$ region also provides only a small contribution to the integrated moments as seen in Chapter~\ref{SumRulesMoments}. The moment results are most likely insensitive to anything but a gross error in the model evolution, but its error should be still be checked.

\subsection{First Moment of $g_2(x,Q^2)$}
\begin{figure}[htp]
\centering     
\includegraphics[width=.80\textwidth]{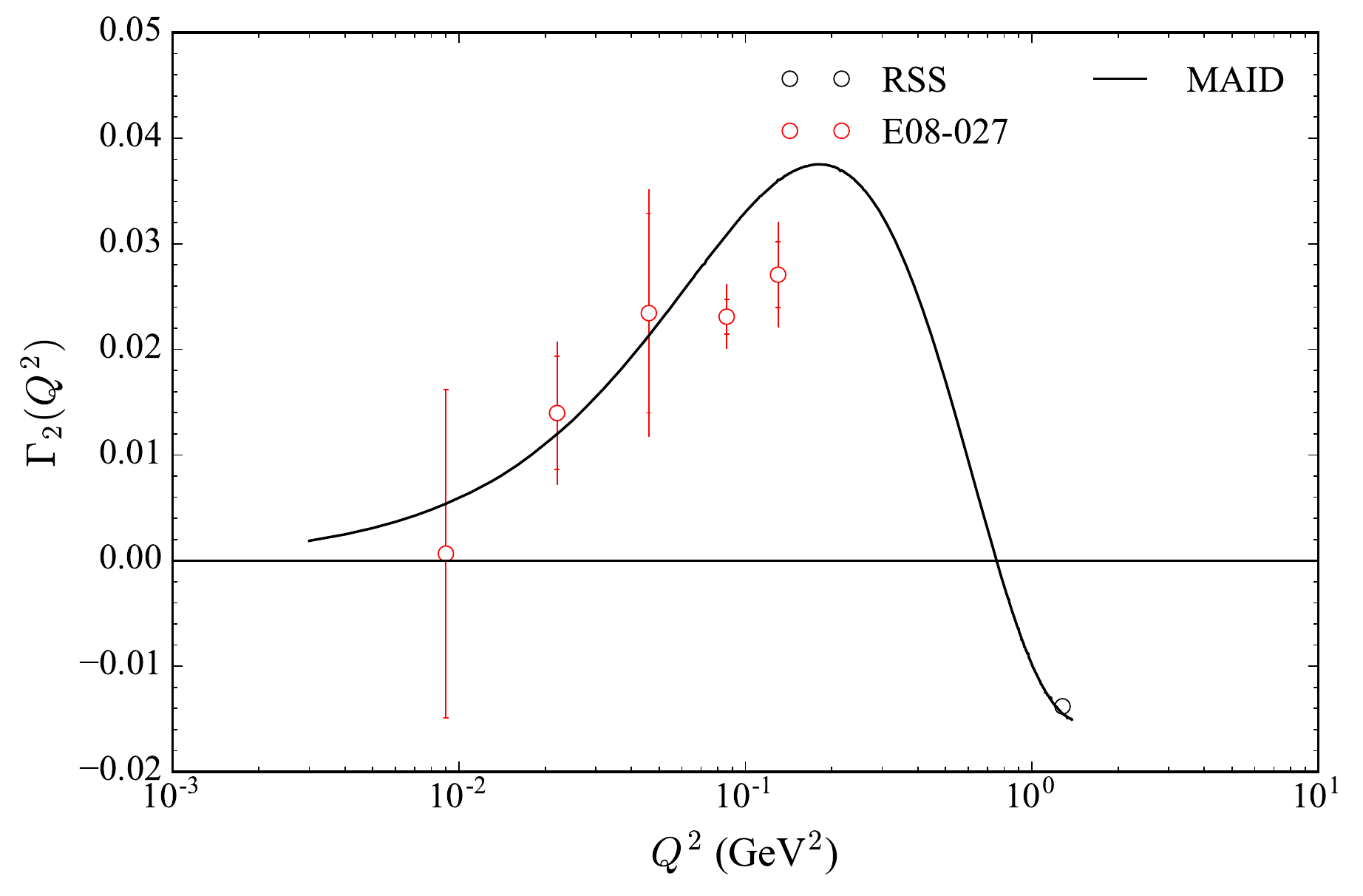}
\caption{E08-027 results for the first moment of $g_2(x,Q^2)$ at every kinematic setting.}
\label{Gamma2ALL}
\end{figure}

A good check on the overall quality of the data is to compute the first moment of $g_2(x,Q^2)$: 
\begin{equation}
\label{FirstG2}
\Gamma_2(Q^2) = \int_0^{x_{\mathrm{th}}} g_2(x,Q^2)dx\,,
\end{equation}
where the integration runs from zero to the pion production threshold. In reality, the data at a kinematic setting does not extend all the way to $x=0$, and the integrations here are truncated at the lowest $x$ available. This corresponds to the resonance region. For a full analysis (the B.C. Sum Rule for example) this low-$x$ region must be accounted for. The absence of any additional kinematic weighting in equation~\eqref{FirstG2} means that these results will have the best statistical uncertainty of any moment or sum rule. Put another way, if the errors on the first moment are consistent with zero then the higher order moments will be too. The results of the integration are shown in Figure~\ref{Gamma2ALL}. The MAID model is picked for comparison because it is also limited to the resonance region. The inner error bars are statistical and the outer are the total uncertainty. With the exception of $E_0$ = 1157 MeV, each setting is statistically different from zero.

\subsection{HyperFine Splitting Contribution}
\begin{figure}[htp]
\centering     
\includegraphics[width=.80\textwidth]{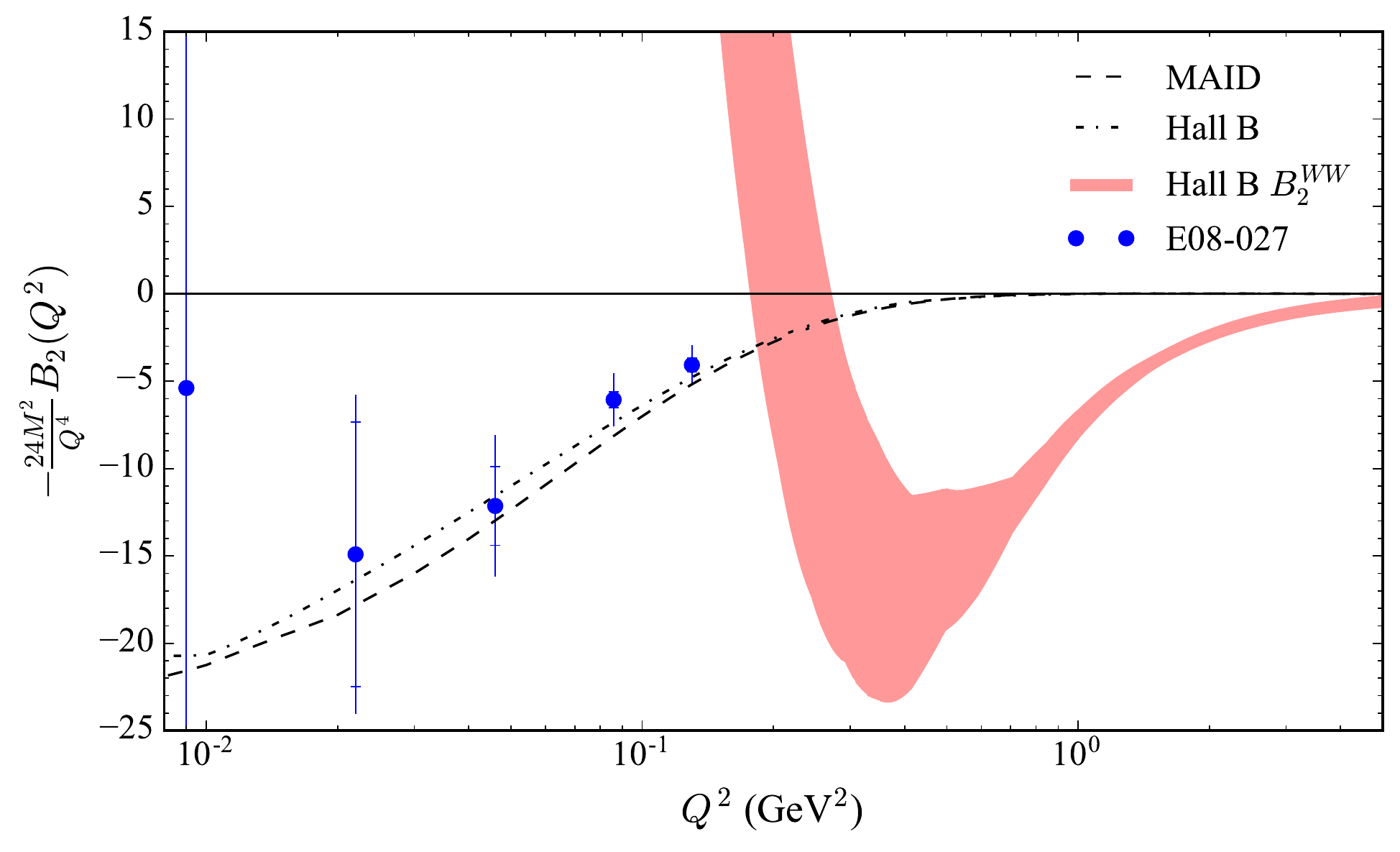}
\caption{E08-027 contribution to the $\Delta_2$ integrand at every kinematic setting.}
\label{B22ALL}
\end{figure}

The complete E08-027 measured contributions to the $\Delta_2$ integrand (see Chapter~\ref{HyperFineDATA} for a more complete description of the analysis) are shown in Figure~\ref{B22ALL}. The inner error bars are statistical and the outer are total error. Data from $E_0$ = 1157 MeV notwithstanding, the results again are statistically separated from zero. The data also push the $Q^2$ boundary lower into the unmeasured region and still find agreement (within uncertainties) with the two polarized models of MAID and Hall B.

The extended $Q^2$ range calculation from data of $\Delta_2$ is shown in Table~\ref{E08027Delta2_25}. At the level of statistical error the data is in agreement with both models. This preliminary analysis represents all of the E08-027 data and there is very little room for improvement on the statistical errors. It is relatively safe to conclude then that the E08-027 data cannot distinguish between the MAID and Hall B models. The 5 T data itself is also consistent with MAID and Hall B so nothing is lost by including all of the data. Both data sets (5 T only and all settings) still rule out the Hall B 2007 model used in all previous publications. 
\begin{table}[htp]
\begin{center}
\begin{tabular}{ l c   r  r r }\hline
  $\Delta_2$& $Q^2$ (GeV$^2$) & Result & Stat  & Sys    \\ \hline
  MAID &  (0.009,0.130)&   $-$1.32 & $-$  & $-$ \\
  Hall B           & (0.009,0.130)&   $-$1.20 & $-$ & $-$ \\
   E08-027   & (0.009,0.130)&   $-$1.15 & 0.52 & 0.31  \\ \hline

\end{tabular}
\caption{\label{E08027Delta2_25}Comparison of  the $\Delta_2$ contribution to the hydrogen hyperfine splitting over the entire E08-027 kinematic coverage.}
\end{center}
\end{table}

\section{Summary}
The lower target polarization of the 2.5 T settings left previously unanswered questions about the statistical quality of that data. This preliminary analysis addresses those concerns and shows that for two out of the three settings the data has enough statistics to warrant further and more detailed analysis. The bulk of this analysis will be addressing the possible sources of systematic errors that exist in the data, running a data quality check on the asymmetry to resolve HRS level differences, completing the dilution and packing fraction analysis and calculating unpolarized cross sections from data. If the simulated acceptance is not possible for these settings the unpolarized model discussed probably produces tolerable levels of additional uncertainty.  For moments involving $g_1(x,Q^2)$ the use of EG4~\cite{EG4} proton data will be necessary because the EG1~\cite{EG1b} data coverage ends with the $E_0$ = 2254 MeV 2.5 T setting.  Regardless, the lack of previous, current and future $g_2(x,Q^2)$ data make analysis of these settings an even more attractive and worthwhile endeavor.

\chapter{\sc Nitrogen Polarization in $^\mathrm{14}$NH$_{\mathrm{3}}$}
\label{app:Appendix-G}

\begin{figure}[htp]
\begin{center}
\includegraphics[scale=.90]{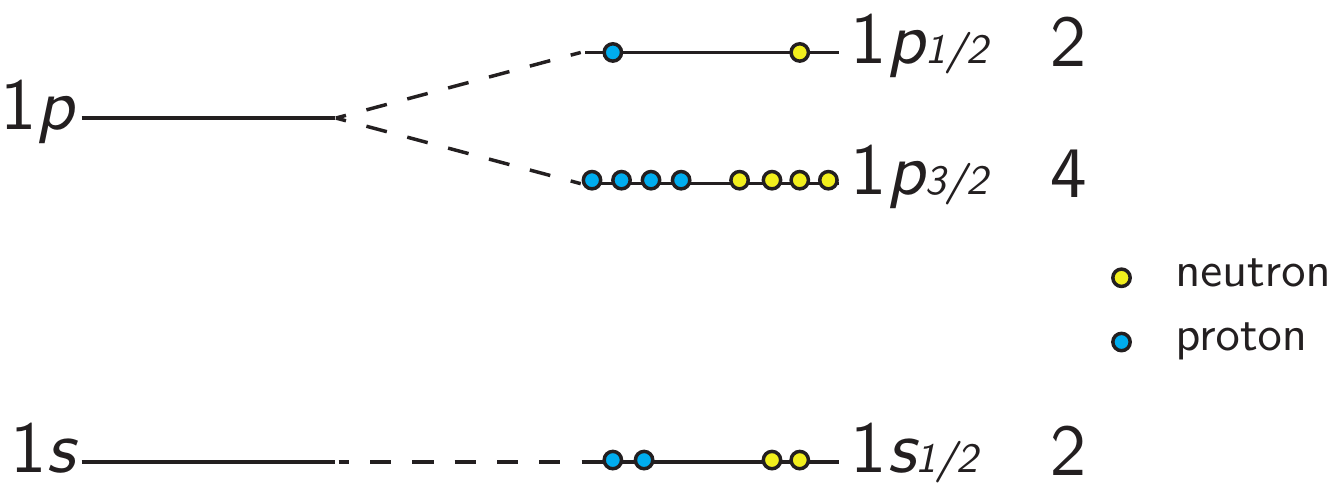}
\caption{\label{ShellNit}Shell model energy levels for the  $^\mathrm{14}$N nucleus. The number following the orbital notation is the multiplicity of that energy level.}
\end{center}
\end{figure}

In the nuclear shell model (see Figure~\ref{ShellNit}), the spin-one  $^\mathrm{14}$N nucleus has six sets of nucleons in the $1s$ and $1p_{3/2}$ shells and an additional unpaired proton and neutron in the  $1p_{1/2}$ energy shell~\cite{Shell}. These unpaired nucleons carry the spin of the nitrogen nucleus and are in an isospin triplet. Only the unpaired nucleons can be polarized,  so it is useful to look at the angular momentum decomposition of the $1p_{1/2}$ shell to determine the nucleon polarization with respect to the nitrogen nucleus. The free nucleons are separated into substates of the intrinsic spin ($s$ = 1/2) and the orbital angular momentum of the p-shell ($l=1$). 

Consider the $\Ket{I = 1,I_3 = 1}$ isospin triplet state. Both the proton and neutron must have $m_j = 1/2$ to account for this value of the nitrogen spin~\cite{Nitrogen}, and looking up the Clebsch-Gordan coefficients for $1\bigotimes\frac{1}{2}$ yields the following for one of the nucleons:
\begin{equation}
\label{eq:CG}
\begin{aligned}
\Ket{\tfrac{1}{2},\tfrac{1}{2}} &=  \sqrt{\tfrac{2}{3}}\Ket{1,1}\Ket{\tfrac{1}{2},-\tfrac{1}{2}}  -  \sqrt{\tfrac{1}{3}}\Ket{1,0}\Ket{\tfrac{1}{2},\tfrac{1}{2}}\,,
\end{aligned}
\end{equation}
where the decomposition is of the form $\Ket{J, m_j} = \mathrm{CG}\Ket{l,m_l}\Ket{s,m_s}$. By comparing values of $m_j$ and $m_s$ in equation~\eqref{eq:CG}, it is clear that the nucleon spin is aligned anti-parallel to the nitrogen spin a net 1/3 of the time. The result is the same for the $\Ket{I = 1,I_3 = -1}$ state, noting that both the proton and neutron must have $m_j = -1/2$. There is no contribution from the $I_3 = 0$ state because the proton and neutron are aligned/anti-aligned in equal measure. 

There are three free protons for every nitrogen nucleus in ammonia and the nitrogen only polarizes up to one sixth of the corresponding hydrogen polarization~\cite{PolHyper, Oscar}. The net contribution of the polarized protons in nitrogen to the measured target polarization (and spin asymmetry) is 1/3 $\times$ 1/3 $\times$ 1/6 $\approx$ 2\%. For example, a 90\% target polarization corresponds to approximately 88.2\% proton-ammonia polarization and 1.8\% proton-nitrogen polarization. The measured asymmetries must be corrected for both the dilution factor (see Chapter~\ref{EAsymm}) and also for the unwanted nitrogen polarization.

\chapter{\sc Parallel Asymmetry Analysis}
\label{app:Appendix-H}

\begin{figure}[htp]
\centering     
\subfigure[Raw]{\label{fig:Interp2254}\includegraphics[width=.80\textwidth]{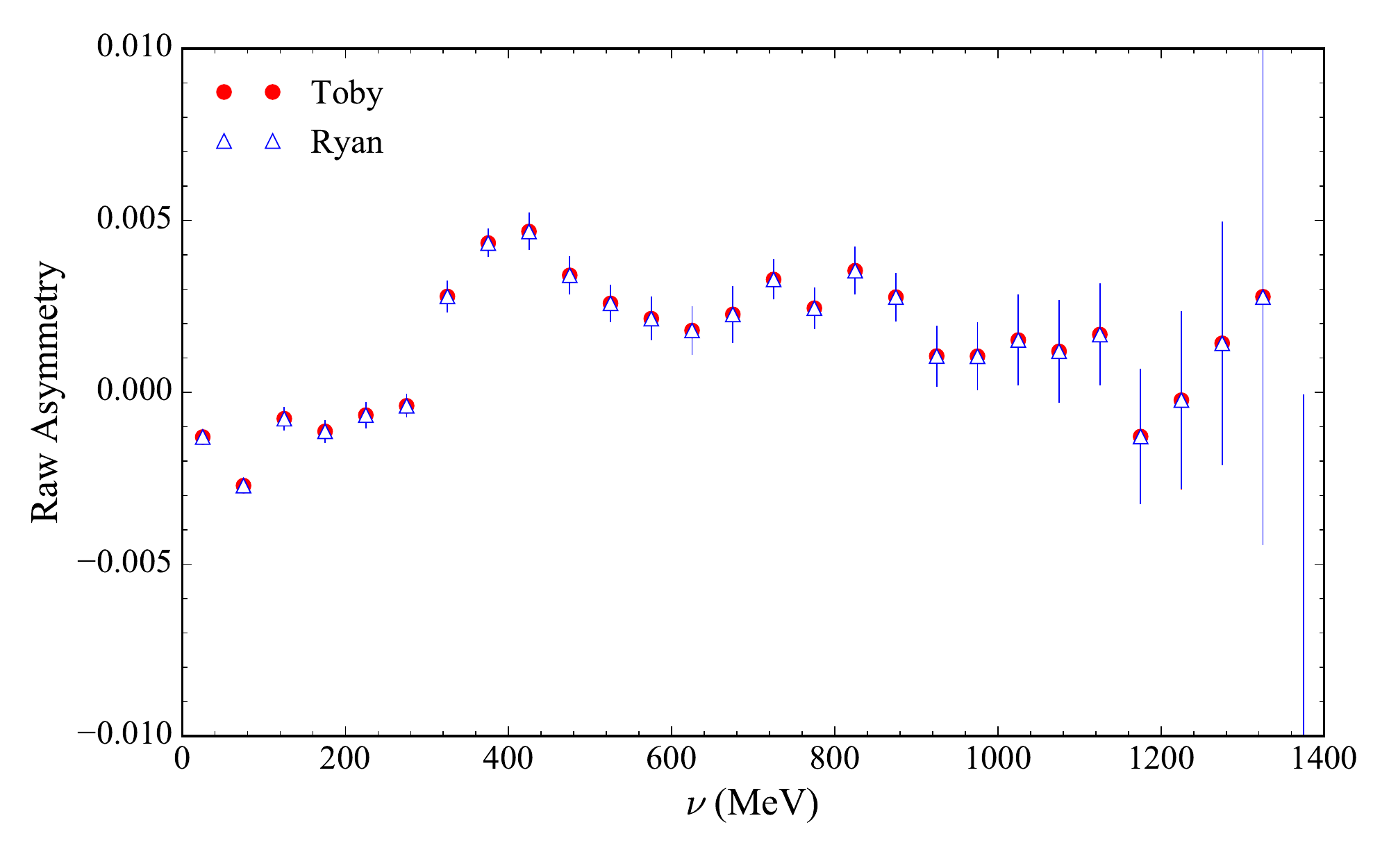}}
\qquad
\subfigure[Dilution applied]{\label{fig:Interp3350}\includegraphics[width=.80\textwidth]{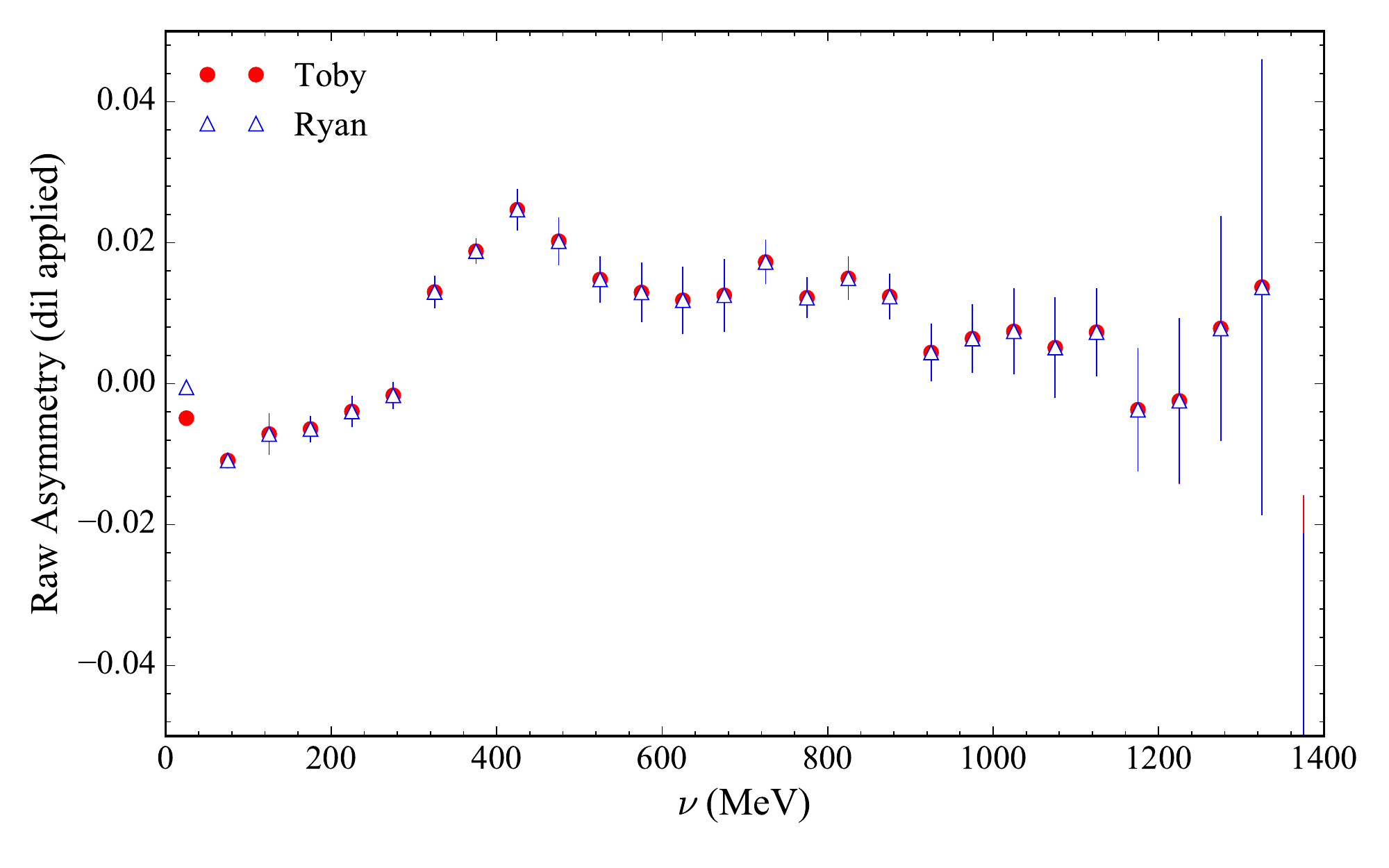}}
\caption{Parallel asymmetry analysis for $E_0$ = 2254 MeV 5T Transverse.}
\label{TComp2254trans}
\end{figure}

\begin{figure}[htp]
\centering     
\subfigure[Raw]{\label{fig:Interp2254}\includegraphics[width=.80\textwidth]{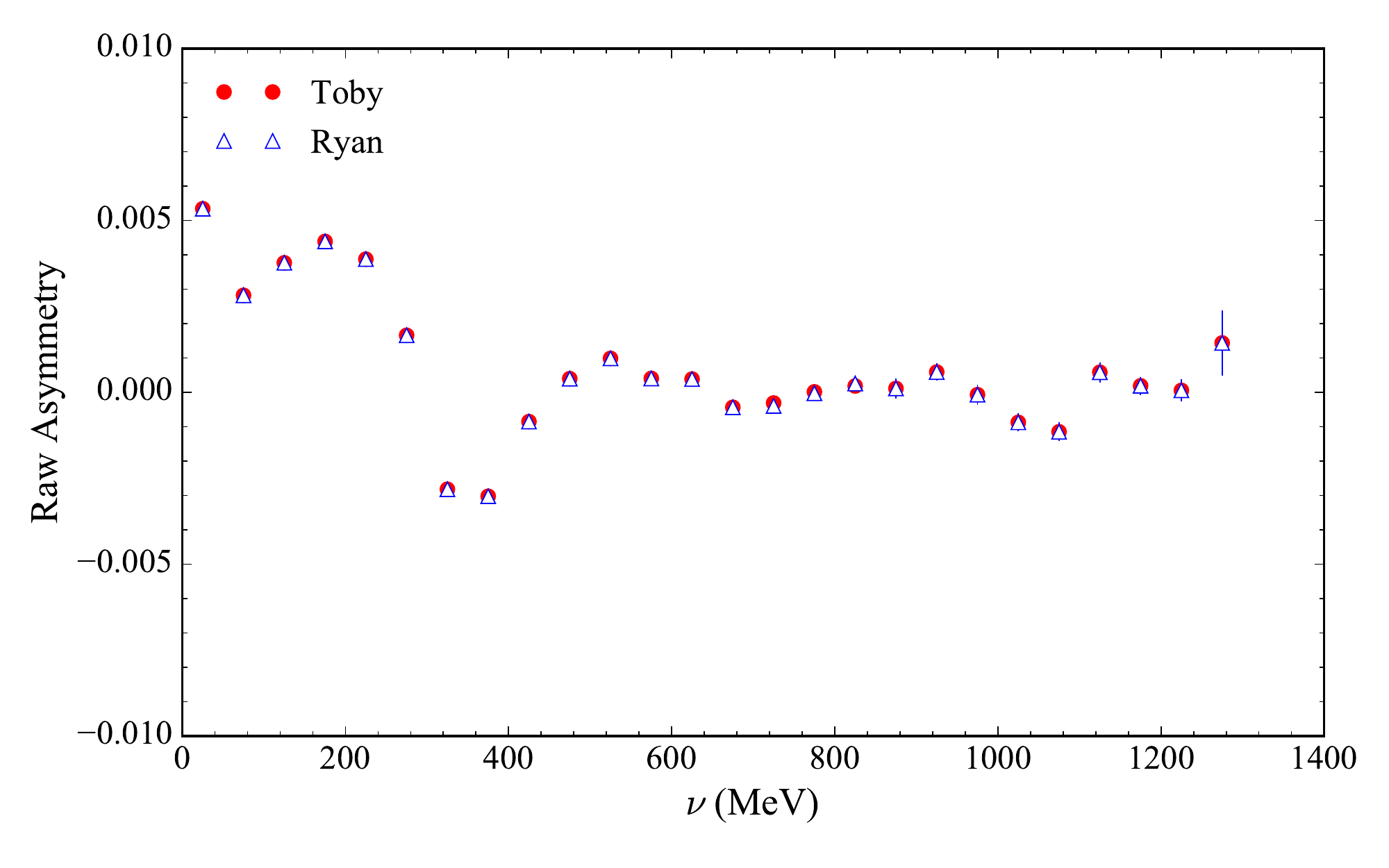}}
\qquad
\subfigure[Dilution applied]{\label{fig:Interp3350}\includegraphics[width=.80\textwidth]{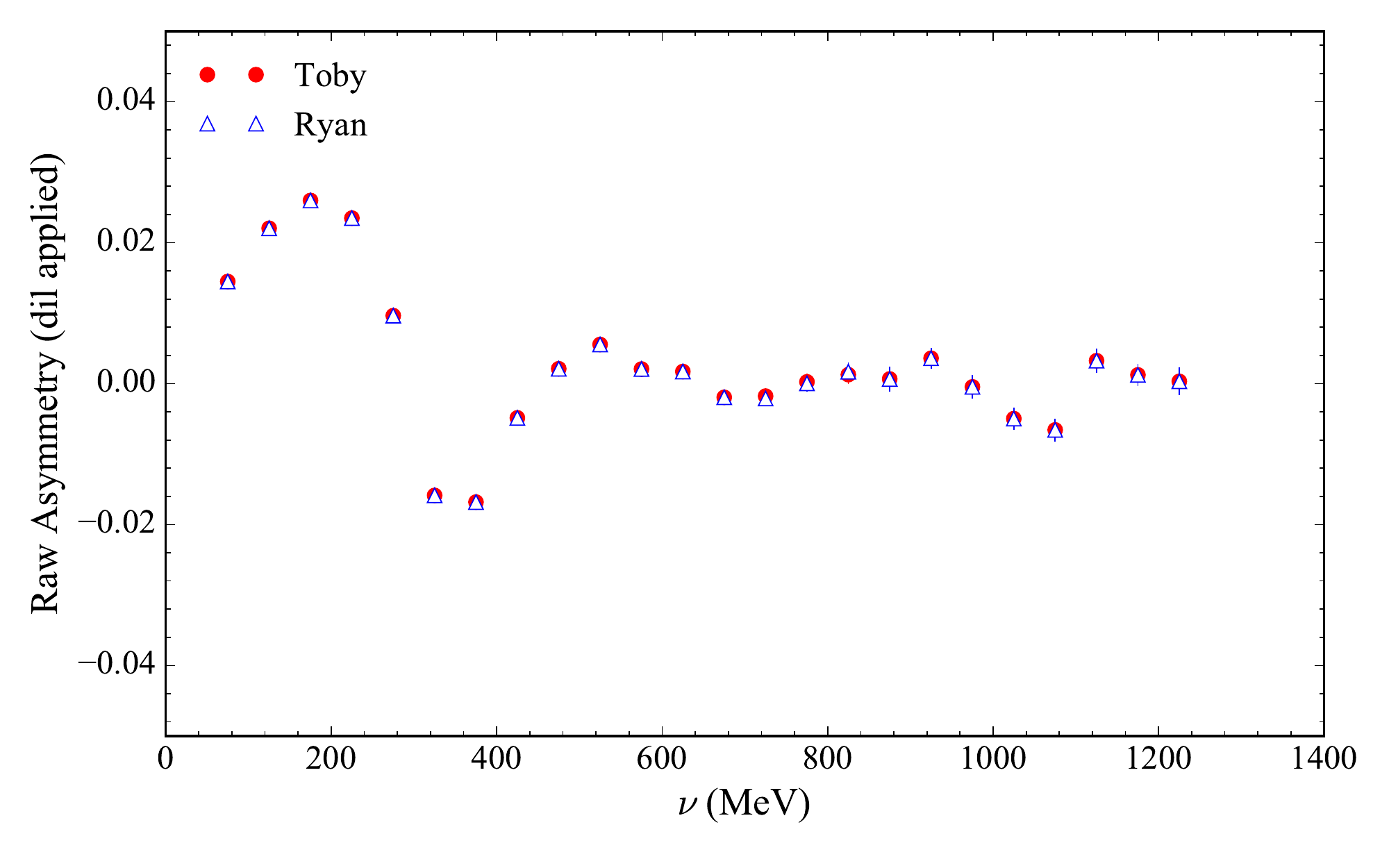}}
\caption{Parallel asymmetry analysis for $E_0$ = 2254 MeV 5T Longitudinal.}
\label{TComp2254long}
\end{figure}

\begin{figure}[htp]
\centering     
\subfigure[Raw]{\label{fig:Interp2254}\includegraphics[width=.80\textwidth]{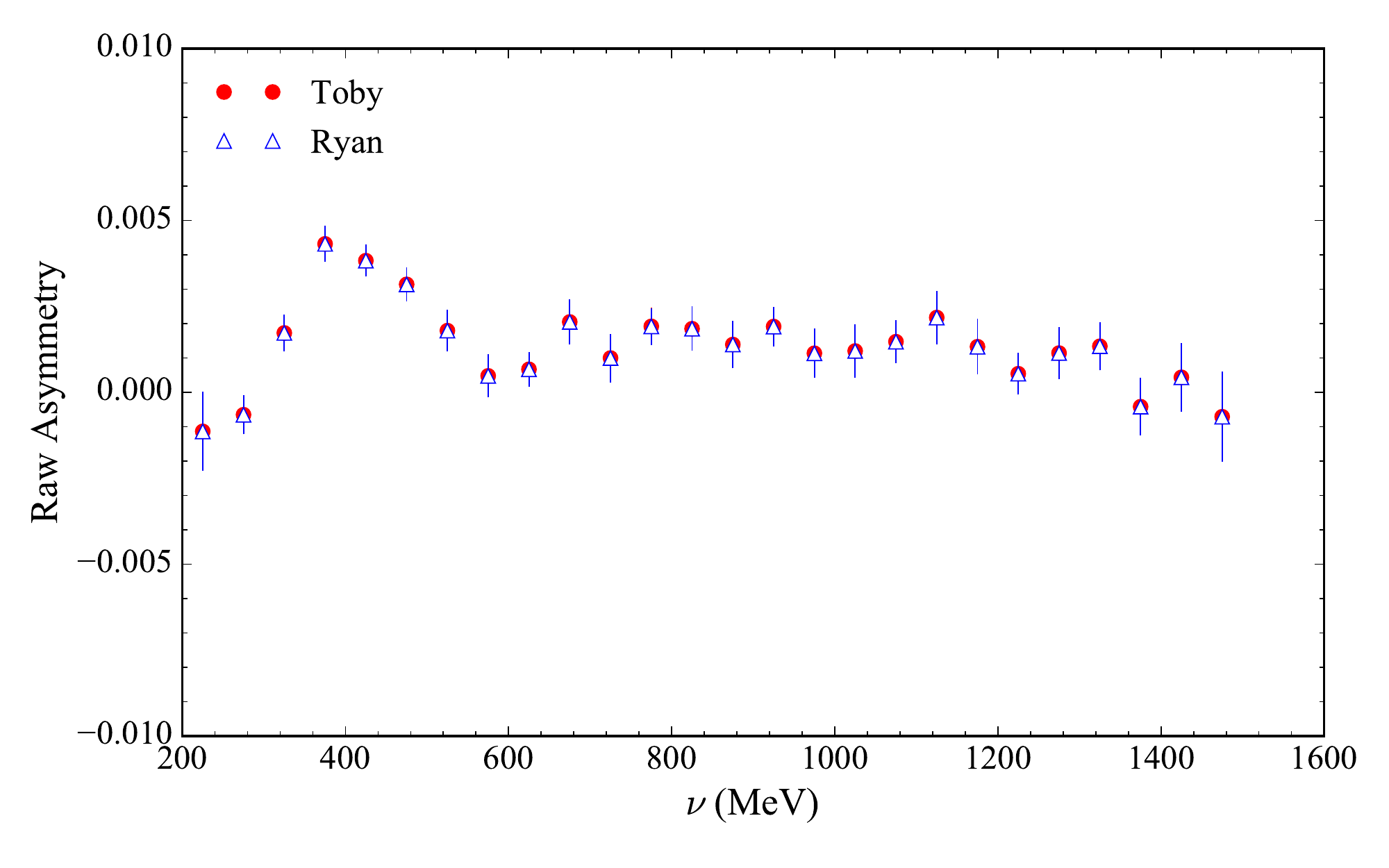}}
\qquad
\subfigure[Dilution applied]{\label{fig:Interp3350}\includegraphics[width=.80\textwidth]{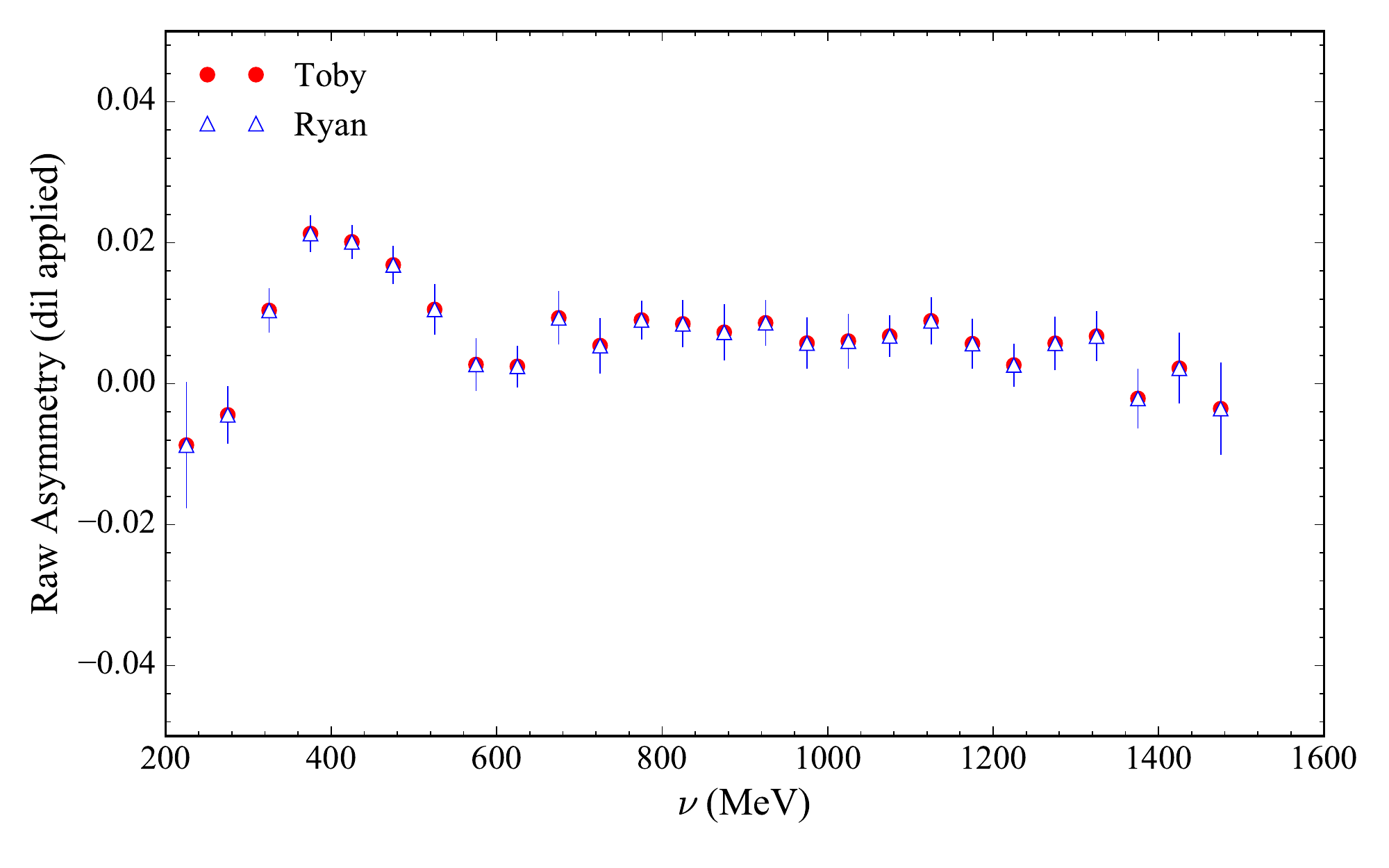}}
\caption{Parallel asymmetry analysis for $E_0$ = 3350 MeV 5T Transverse.}
\label{TComp3350trans}
\end{figure}

The results of a parallel analysis  for the asymmetry calculation are shown in Figure~\ref{TComp2254trans} through Figure~\ref{TComp3350trans}. The analysis of this author is represented by the blue triangles (Ryan), while the analysis from another collaborator is shown with the red circles (Toby)~\cite{TobyCom}. The purpose of this analysis is to provide a cross-check on the asymmetry generating procedure and is done for both the raw asymmetry (corrected for target polarization and beam polarization and sign only) and also with the dilution factor applied. The two analyses agree very well, provided they use the same exact run selection and acceptance cuts.

Initial comparisons showed some minor disagreement in Figure~\ref{TComp2254long}. The slight difference between the two asymmetries at the longitudinal setting was due to a slight difference in acceptance cuts applied to the data. For future analysis it is recommended to first confirm the choice of data cuts and run selection before doing a comparison. For reasonably close  data cuts and run choice two asymmetries will agree within statistical error bars, so this step is not critical. It is helpful though to root out the cause of small differences that might otherwise be resistant to changes in the actual analysis procedure. A tell-tale sign of differences in run selection is exact agreement in parts of the energy spectrum (momentum setting) and disagreement in other parts.

\chapter{\sc Hall B Model Parameters}
\label{app:Appendix-F}

The CLAS EG1b (Hall B) model described in Chapter~\ref{CLASHALLBMOD} has five input parameters that can be adjusted to change the output. The input parameters are described below with their default values:
\begin{itemize}
\item T = P, N, D : Choice of target type for the proton, neutron and deuteron respectively.
\item IPOL = 1 : Choice of $A_2(x,Q^2)$ DIS model. Default is a fit to existing world data.
\item IA1 = 4 : Choice of $A_1(x,Q^2)$ DIS model. Default is a fit to existing world data.
\item SFChoice = 11 : Choice of $F_1(x,Q^2)$ and $F_2(x,Q^2)$ model. Default is the Bosted-Mamyan-Christy fit.
\item AsymChoice = 11 : Choice of $A_1(x,Q^2)$ and $A_2(x,Q^2)$ resonance region model. Default is a fit to existing world data.
\end{itemize}

The authors of the model recommend  the following variations to estimate the systematic uncertainty of the model:

{\bf IPOL} : Run both options listed below and use the average deviation as the uncertainty.
\begin{itemize} 
\item IPOL = 2 : Add twist-3 term to $g_2(x,Q^2)$ in DIS.
\item IPOL = 5 : $A_2(x,Q^2)$ = 0 in DIS.
\end{itemize}

{\bf IA1} : Run both options listed below and use the average deviation as the uncertainty.
\begin{itemize} 
\item IA1 = 5 : Increase $A_1(x,Q^2)$ DIS model by its uncertainty.
\item IA1 = 6 : Decrease $A_1(x,Q^2)$ DIS model by its uncertainty.
\end{itemize}

{\bf AsymChoice} : Run both options listed below and add the deviations in quadrature.
\begin{itemize} 
\item AsymChoice = 12 : Small variation of the $A_1(x,Q^2)$ resonance region model.
\item AsymChoice = 13 : Previous version of the $A_1(x,Q^2)$ resonance region model.
\item AsymChoice = 14 : Replace $A_2(x,Q^2)$ resonance region model with MAID.
\item AsymChoice = 15 : Previous version of the $A_2(x,Q^2)$ resonance region model.
\end{itemize}

{\bf SFChoice} : Run both options listed below and linearly add the deviations.
\begin{itemize} 
\item SFChoice = 12 : Increase $F_1(x,Q^2)$ and $F_2(x,Q^2)$ by their uncertainty.
\item SFChoice = 13 : Change $R$ by its error and propagate to $F_1(x,Q^2)$, leaving $F_2(x,Q^2)$ unchanged.
\end{itemize}

There are a few additional input parameter choices that can be found by studying the comments in the source code, but they were not used for the systematic error analysis of this thesis.

\chapter{\sc Summary of Experimental Activities}
\label{app:Appendix-J}

My contributions to the E08-027 experiment are broken down into two categories: preparation for the experimental run-period and analysis of the collected data. I will discuss my major contributions to both categories below.

\noindent{\bf Experimental Preparation}\\
I was responsible for setting up and calibrating the spectrometer detector package and DAQ. This included generic cabling and detector check-out. The majority of my time was spent improving the DAQ rate for the Hall A HRS spectrometers. This was done through a combination of modeling the DAQ performance through Poisson probability theory and verifying the model with experimental data. This work resulted in a 50\% performance improvement in the efficiency of the Hall A DAQ, which was a record at the time.

\noindent{\bf Data Analysis}\\
My biggest contribution to the data analysis was in the understanding of the radiative corrections process. This work led to the most complete systematic error analysis (that I'm aware of) of the inclusive radiative corrections procedure. I was able to use these results to radiatively correct nitrogen data from the Small Angle GDH experiment. A publication of this data is currently in progress. My additional responsibilities included creating and managing the experimental MySQL database (including developing a web interface and multiple software libraries), trigger scintillator efficiency and DAQ deadtime calculations. My results are the most complete and closest to finalized when compared to any of the previous students.                                
\end{appendices}
\begin{singlespace}     
\addcontentsline{toc}{chapter}{\sc Bibliography} 
\bibliography{Thesis-UNH.bib}

\begin{thebibliography}{100}

\bibitem{Stern}
W.~{Gerlach} and O.~{Stern}.
\newblock {Das magnetische Moment des Silberatoms}.
\newblock {\em Zeitschrift fur Physik}, 9:353--355, 1922.

\bibitem{Dirac610}
P.~A.~M. Dirac.
\newblock {The Quantum Theory of the Electron}.
\newblock {\em Proceedings of the Royal Society of London A: Mathematical,
  Physical and Engineering Sciences}, 117(778):610--624, 1928.

\bibitem{Rutherford374}
E.~Rutherford.
\newblock {Bakerian Lecture. Nuclear Constitution of Atoms}.
\newblock {\em Proceedings of the Royal Society of London A: Mathematical,
  Physical and Engineering Sciences}, 97(686):374--400, 1920.

\bibitem{soddy1920name}
F.~Soddy.
\newblock {Name for the Positive Nucleus}.
\newblock {\em Nature}, 106(2668):502--503, 1920.

\bibitem{MagMom}
I.~Estermann, O.~C. Simpson, and O.~Stern.
\newblock {The Magnetic Moment of the Proton}.
\newblock {\em Phys. Rev.}, 52:535--545, 1937.

\bibitem{Feynman}
R.~P. Feynman.
\newblock {\em QED: The Strange Theory of Light and Matter}.
\newblock Alix G. Mautner Memorial Lectures. Princeton University Press,
  Princeton, NJ, 1988.

\bibitem{Thomas}
A.~W. Thomas and W.~Weise.
\newblock {\em {The Structure of the Nucleon}}.
\newblock Wiley, Berlin, 2001.

\bibitem{Greiner}
W.~Greiner, S.~Schramm, and E.~Stein.
\newblock {\em Quantum Chromodynamics}.
\newblock Springer, Berlin, 2002.

\bibitem{SpinCrisis}
B.~W. Filippone and X.-D. Ji.
\newblock {\em The Spin Structure of the Nucleon}, pages 1--88.
\newblock Springer US, Boston, MA, 2001.

\bibitem{SpinCrisis2}
C.~A. Aidala, S.~D. Bass, D.~Hasch, and K.~G. Mallot.
\newblock {The Spin Structure of the Nucleon}.
\newblock {\em Rev. Mod. Phys.}, 85:655--691, 2013.

\bibitem{Rutherford:1911}
E.~Rutherford.
\newblock {The Scattering of Alpha and Beta Particles by Matter and the
  Structure of the Atom}.
\newblock {\em Phil. Mag. Ser.6}, 21:669--688, 1911.

\bibitem{debroglie}
L.~De~Broglie.
\newblock {\em {Recherches sur la th{\'e}orie des Quanta}}.
\newblock Theses, {Migration - universit{\'e} en cours d'affectation}, 1924.

\bibitem{Bjorken:1968dy}
J.~D. Bjorken.
\newblock {Asymptotic Sum Rules at Infinite Momentum}.
\newblock {\em Phys. Rev.}, 179:1547--1553, 1969.

\bibitem{VinceT}
V.~Sulkosky.
\newblock {\em {The Spin Structure of $^3$He and the Neutron at Low $Q^2$: A
  Measurement of the Generalized GDH Integrand}}.
\newblock PhD thesis, The College of William and Mary, 2007.

\bibitem{Povh}
B.~Povh, K.~Rith, C.~Scholz, and F.~Zersche.
\newblock {\em {Particles and nuclei: An Introduction to the physical
  concepts}}.
\newblock 1995.

\bibitem{Quarks}
F.~Halzen and A.~D. Martin.
\newblock {\em {Quarks and Leptons: An Introductory Course in Modern Particle
  Physics}}.
\newblock Wiley, New York, NY, 1984.

\bibitem{Srednicki}
M.~Srednicki.
\newblock {\em Quantum Field Theory}.
\newblock Cambridge University Press, Cambridge, 2012.

\bibitem{Peskin}
M.~E. Peskin and D.~V. Schroeder.
\newblock {\em An Introduction to Quantum Field Theory}.
\newblock Westview Press, Boulder, CO, 1995.

\bibitem{Rosenbluth}
M.~N. Rosenbluth.
\newblock High energy elastic scattering of electrons on protons.
\newblock {\em Phys. Rev.}, 79:615--619, Aug 1950.

\bibitem{MOTT}
N.~F. Mott.
\newblock The scattering of fast electrons by atomic nuclei.
\newblock {\em Proceedings of the Royal Society of London. Series A, Containing
  Papers of a Mathematical and Physical Character}, 124(794):425--442, 1929.

\bibitem{Leader}
M.~Anselmino, A.~Efremov, and E.~Leader.
\newblock {The Theory and phenomenology of polarized deep inelastic
  scattering}.
\newblock {\em Phys.Rept.}, 261:1--124, 1995.

\bibitem{Weinberg:1995mt}
S.~Weinberg.
\newblock {\em {The Quantum theory of fields. Vol. 1: Foundations}}.
\newblock Cambridge University Press, 2005.

\bibitem{Hand}
L.~N. Hand.
\newblock {Experimental Investigation of Pion Electroproduction}.
\newblock {\em Phys. Rev.}, 129:1834--1846, 1963.

\bibitem{Gilman}
F.~J. Gilman.
\newblock {Kinematics and Saturation of the Sum Rules and Inequalities for
  Inelastic Electron-Nucleon Scattering}.
\newblock {\em Phys. Rev.}, 167:1365--1371, 1968.

\bibitem{PhotonPol}
D.~Drechsel, S.~S. Kamalov, and L.~Tiator.
\newblock {Gerasimov-Drell-Hearn Sum Rule and Related Integrals}.
\newblock {\em Phys. Rev. D}, 63:114010, 2001.

\bibitem{Feynman1988}
Richard~P. Feynman.
\newblock {\em The Behavior of Hadron Collisions at Extreme Energies}, pages
  289--304.
\newblock Springer Netherlands, Dordrecht, 1988.

\bibitem{PDG}
J.~Beringer et~al.
\newblock {Review of Particle Physics}.
\newblock {\em Phys. Rev. D}, 86:010001, 2012.

\bibitem{PhysRevLett.22.156}
C.~G. Callan and David~J. Gross.
\newblock High-energy electroproduction and the constitution of the electric
  current.
\newblock {\em Phys. Rev. Lett.}, 22:156--159, Jan 1969.

\bibitem{Ji}
X.-D. Ji.
\newblock {Physics of the $G_2$ Structure Function of the Nucleon}.
\newblock In {\em {Deep Inelastic Scattering and QCD. Proceedings, Workshop,
  Paris, France, April 24-28, 1995}}, pages 435--438, 1995.

\bibitem{Ashman}
J.~Ashman et~al.
\newblock {An Investigation of the Spin Structure of the Proton in Deep
  Inelastic Scattering of Polarised Muons on Polarised protons}.
\newblock {\em Nuclear Physics B}, 328(1):1 -- 35, 1989.

\bibitem{jacksonT}
J.~D. Jackson.
\newblock {\em Classical electrodynamics}.
\newblock Wiley, New York, {NY}, 3rd ed. edition, 1999.

\bibitem{Drechsel2}
D.~Drechsel, B.~Pasquini, and M.~Vanderhaeghen.
\newblock {Dispersion Relations in Real and Virtual Compton Scattering}.
\newblock {\em Phys. Rept.}, 378:99--205, 2003.

\bibitem{Drechsel}
D.~Drechsel.
\newblock {Spin Sum Rules and Polarizabilities}.
\newblock In {\em {Spin Structure at Long Distance. Proceedings,Workshop,
  Newport News, USA, March 12-13, 2009}}, volume 1155, pages 3--17, 2009.

\bibitem{Pantforder}
R.~Pantforder.
\newblock {\em {Investigations on the Foundation and Possible Modifications of
  the Gerasimov-Drell-Hearn Sum Rule}}.
\newblock PhD thesis, Bonn U., 1998.

\bibitem{Chen}
J.~P. Chen.
\newblock {Moments of Spin Structure Functions: Sum Rules and
  Polarizabilities}.
\newblock {\em Int. J. Mod. Phys.}, E19:1893--1921, 2010.

\bibitem{GDH}
S.~D. Drell and A.~C. Hearn.
\newblock {Exact Sum Rule for Nucleon Magnetic Moments}.
\newblock {\em Phys. Rev. Lett.}, 16:908--911, 1966.

\bibitem{BC}
R.~L. Jaffe and X.-D. Ji.
\newblock {Studies of the Transverse Spin-Dependent Structure Function
  ${g}_{2}(x, {Q}^{2})$}.
\newblock {\em Phys. Rev. D}, 43:724--732, 1991.

\bibitem{Bodo}
B.~Lampe and E.~Reya.
\newblock {Spin Physics and Polarized Structure Functions}.
\newblock {\em Physics Reports}, 332(1--3):1 -- 163, 2000.

\bibitem{Wilson}
K.~G. Wilson.
\newblock {Nonlagrangian Models of Current Algebra}.
\newblock {\em Phys. Rev.}, 179:1499--1512, 1969.

\bibitem{Jaffe}
R.~L. Jaffe.
\newblock {$g_{2}$--The Nucleon's Other Spin-Dependent Structure Function}.
\newblock {\em Comments Nucl. Part. Phys.}, 19(5):239--257, 1990.

\bibitem{Wandzura}
S.~Wandzura and F.~Wilczek.
\newblock {Sum Rules for Spin-Dependent Electroproduction--Test of Relativistic
  Constituent Quarks}.
\newblock {\em Physics Letters B}, 72(2):195 -- 198, 1977.

\bibitem{Proposal}
A.~Camsonne, J.P. Chen, D.~Crabb, K.~Slifer, et~al.
\newblock {A Measurement of $g_2^p$ and the Longitudinal-Transverse Spin
  Polarizability}.
\newblock \url{http://hallaweb.jlab.org/experiment/g2p/docs/PAC33/dlt.pdf}.

\bibitem{G1Twist}
H.~Kawamura, T.~Uematsu, and J.~Yasui, Y.and~Kodaira.
\newblock {Renormalization of Twist-Four Operators in QCD Bjorken and
  Ellis--Jaffe Sum Rules}.
\newblock {\em Modern Physics Letters A}, 12(02):135--143, 1997.

\bibitem{JaffeSum}
J.~Ellis and R.L. Jaffe.
\newblock {Sum Rule for Deep-Inelastic Electroproduction from Polarized
  Protons}.
\newblock {\em Phys. Rev. D}, 9:1444--1446, 1974.

\bibitem{Weinberg}
S.~Weinberg.
\newblock {Phenomenological Lagrangians}.
\newblock {\em Physica A: Statistical Mechanics and its Applications},
  96(1):327 -- 340, 1979.

\bibitem{Scherer}
S.~Scherer.
\newblock {Introduction to Chiral Perturbation Theory}.
\newblock {\em Adv.Nucl.Phys.}, 27:277, 2003.

\bibitem{Noether1918}
E.~Noether.
\newblock {Invariante Variationsprobleme}.
\newblock {\em {Nachrichten von der Gesellschaft der Wissenschaften zu
  G{\"o}ttingen, Mathematisch-Physikalische Klasse}}, 1918:235--257, 1918.

\bibitem{Goldstone}
Y.~Nambu and G.~Jona-Lasinio.
\newblock {Dynamical Model of Elementary Particles Based on an Analogy with
  Superconductivity}.
\newblock {\em Phys. Rev.}, 122:345--358, 1961.

\bibitem{Bernard}
V.~Bernard, N.~Kaiser, and U.-G. Meissner.
\newblock {Chiral Dynamics in Nucleons and Nuclei}.
\newblock {\em Int.J.Mod.Phys.}, E4:193--346, 1995.

\bibitem{ChiPT_1}
V.~Bernard, N.~Kaiser, and U.-G. Meissner.
\newblock {Small Momentum Evolution of the Extended Drell-Hearn-Gerasimov Sum
  rule}.
\newblock {\em Phys. Rev. D}, 48:3062--3069, 1993.

\bibitem{RelChiPT_1}
H.-B. Tang.
\newblock {A New Approach to Chiral Perturbation Theory for Matter Fields},
  1996.
\newblock arXiv:hep-ph/9607436.

\bibitem{HeavyChiPT_1}
E. and A.~V. Manohar.
\newblock {Baryon Chiral Perturbation Theory Using a Heavy Fermion Lagrangian}.
\newblock {\em Physics Letters B}, 255(4):558 -- 562, 1991.

\bibitem{HeavyChiPT_2}
E.~Jenkins and A.~V. Manohar.
\newblock {Chiral Corrections to the Baryon Axial Currents}.
\newblock {\em Physics Letters B}, 259(3):353 -- 358, 1991.

\bibitem{BaryonChiPT}
J.~Gasser, M.~E. Sainio, and A.~\v{S}varc.
\newblock {Nucleons with Chiral Loops}.
\newblock {\em Nuclear Physics B}, 307(4):779 -- 853, 1988.

\bibitem{Chi_Con}
P.~F. Bedaque.
\newblock {Chiral Perturbation Theory Analysis of Baryon Temperature Mass
  shift"}.
\newblock {\em Physics Letters B}, 387(1):1 -- 8, 1996.

\bibitem{Ji2}
X.-D. Ji, C.-W. Kao, and J.~Osborne.
\newblock {Generalized Drell-Hearn-Gerasimov Sum Rule at Order
  $\mathcal{O}$($p^4$) in Chiral Perturbation Theory}.
\newblock {\em Physics Letters B}, 472(1--2):1 -- 4, 2000.

\bibitem{Bernard2}
V.~Bernard, T.~R. Hemmert, and U.-G. Meissner.
\newblock {Spin Structure of the Nucleon at Low energies}.
\newblock {\em Phys. Rev. D}, 67:076008, 2003.

\bibitem{Bernard3}
V.~Bernard, T.~R. Hemmert, and U.-G. Meissner.
\newblock {Novel Analysis of Chiral Loop Effects in the Generalized
  Gerasimov-Drell-Hearn Sum rule}.
\newblock {\em Physics Letters B}, 545(1--2):105 -- 111, 2002.

\bibitem{Kao}
C.-W. Kao, T.~Spitzenberg, and M.~Vanderhaeghen.
\newblock {Burkhardt-Cottingham Sum Rule and Forward Spin Polarizabilities in
  Heavy Baryon Chiral Perturbation Theory}.
\newblock {\em Phys. Rev. D}, 67:016001, 2003.

\bibitem{BreitWigner}
G.~Breit and E.~Wigner.
\newblock {Capture of Slow Neutrons}.
\newblock {\em Phys. Rev.}, 49:519--531, 1936.

\bibitem{MAID2007}
D.~Drechsel, S.~S. Kamalov, and L.~Tiator.
\newblock {Unitary Isobar Model -- MAID2007}.
\newblock {\em The European Physical Journal A}, 34(1):69, 2007.

\bibitem{HallB}
N.~Guler.
\newblock {Measurement of Longitudinal Double Spin Asymmetries and Spin
  Structure Functions of the Deuteron in the CLAS EG1b Experiment}.
\newblock Technical report, CLAS EG1b Collaboration, 2012.
\newblock
  \url{https://userweb.jlab.org/~kuhn/EG1b/dPaper_old/eg1b-analysis.pdf}.

\bibitem{Bosted3}
M.~E. Christy and P.~E. Bosted.
\newblock Empirical fit to precision inclusive electron--proton cross sections
  in the resonance region.
\newblock {\em Phys. Rev. C}, 81:055213, 2010.

\bibitem{Bosted1}
P.~E. Bosted and M.~E. Christy.
\newblock {Empirical Fit to Inelastic Electron-Deuteron and Electron-Neutron
  Resonance Region Transverse Cross Sections}.
\newblock {\em Phys. Rev. C}, 77:065206, 2008.

\bibitem{Bosted2}
P.~E. Bosted and V.~Mamyan.
\newblock {Empirical Fit to Electron--Nucleus Scattering}, 2012.
\newblock arXiv:nucl-th/1203.2262.

\bibitem{Superscale}
C.~Maieron, T.~W. Donnelly, and I.~Sick.
\newblock {Extended Superscaling of Electron Scattering From Nuclei}.
\newblock {\em Phys. Rev. C}, 65:025502, 2002.

\bibitem{QFS}
J.S. O'Connell and Lightbody~J.W. Jr.
\newblock {Modeling Single Arm Electron Scattering and Nucleon Production From
  Nuclei by GeV Electrons}.
\newblock {\em Comput. Phys.; (United States)}, 2:3, 1988.

\bibitem{E143}
K.~Abe et~al.
\newblock {Measurements of the Proton and Deuteron Spin Structure Function
  ${\mathit{g}}_{2}$ and Asymmetry ${\mathit{A}}_{2}$}.
\newblock {\em Phys. Rev. Lett.}, 76:587--591, 1996.

\bibitem{E155}
P.L. Anthony et~al.
\newblock {Measurement of the Proton and Deuteron Spin Structure Functions
  $g_2$ and Asymmetry $A_2$}.
\newblock {\em Physics Letters B}, 458(4):529 -- 535, 1999.

\bibitem{SMC}
D.~Adams et~al.
\newblock {Spin Structure of the Proton from Polarized Inclusive Deep-Inelastic
  Muon-Proton Scattering}.
\newblock {\em Phys. Rev. D}, 56:5330--5358, 1997.

\bibitem{E155x}
P.L. Anthony et~al.
\newblock {Precision Measurement of the Proton and Deuteron Spin Structure
  Functions $g_2$ and Asymmetries $A_2$}.
\newblock {\em Physics Letters B}, 553(1--2):18 -- 24, 2003.

\bibitem{RSS}
F.~R. Wesselmann, K.~Slifer, S.~Tajima, et~al.
\newblock {Proton Spin Structure in the Resonance Region}.
\newblock {\em Phys. Rev. Lett.}, 98:132003, 2007.

\bibitem{Hermes}
A.~Airapetian et~al.
\newblock {Measurement of the Virtual-Photon Asymmetry $A_2$ and the
  Spin--Structure Function $g_2$ of the Proton}.
\newblock {\em The European Physical Journal C}, 72(3):1921, 2012.

\bibitem{SANE}
O.~A. Rondon.
\newblock {The RSS and SANE Experiments at Jefferson Lab}.
\newblock {\em AIP Conference Proceedings}, 1155(1):82--92, 2009.

\bibitem{Ashman2}
J.~Ashman et~al.
\newblock {A Measurement of the Spin Asymmetry and Determination of the
  Structure Function $g_1$ in Deep Inelastic Muon-Proton Scattering}.
\newblock {\em Physics Letters B}, 206(2):364 -- 370, 1988.

\bibitem{SMC2}
B.~Adeva et~al.
\newblock {Spin Asymmetries ${A}_{1}$ and Structure Functions ${g}_{1}$ of the
  Proton and the Deuteron From Polarized High Energy Muon Scattering}.
\newblock {\em Phys. Rev. D}, 58:112001, 1998.

\bibitem{SMC3}
B.~Adeva et~al.
\newblock {Spin Asymmetries ${A}_{1}$ of the Proton and the Deuteron in the Low
  $x$ and Low ${Q}^{2}$ Region From Polarized High Energy Muon Scattering}.
\newblock {\em Phys. Rev. D}, 60:072004, 1999.

\bibitem{COMPASS}
M.~G. Alekseev et~al.
\newblock {The Spin--Dependent Structure Function of the Proton and a Test of
  the Bjorken Sum Rule}.
\newblock {\em Physics Letters B}, 690(5):466 -- 472, 2010.

\bibitem{Hermesg1}
A.~Airapetian et~al.
\newblock {Precise Determination of the Spin Structure Function ${g}_{1}$ of
  the Proton, Deuteron, and Neutron}.
\newblock {\em Phys. Rev. D}, 75:012007, 2007.

\bibitem{E143_2}
K.~Abe et~al.
\newblock {Measurements of the Proton and Deuteron Spin Structure Functions
  ${g}_{1}$ and ${g}_{2}$}.
\newblock {\em Phys. Rev. D}, 58:112003, 1998.

\bibitem{E155g1}
P.L Anthony et~al.
\newblock {Measurements of the $Q^2$--Dependence of the Proton and Neutron Spin
  Structure Functions $g_1^p$ and $g_1^n$}.
\newblock {\em Physics Letters B}, 493(1Ð2):19 -- 28, 2000.

\bibitem{CLASg1}
K.V. Dharmawardane, S.E. Kuhn, P.~Bosted, Y.~Prok, et~al.
\newblock {Measurement of the $x$ and $Q^2$ Dependence of the Asymmetry $A_1$
  on the Nucleon}.
\newblock {\em Physics Letters B}, 641(1):11 -- 17, 2006.

\bibitem{EG1b}
Y.~Prok, P.~Bosted, V.D. Burkert, A.~Deur, K.V. Dharmawardane, G.E. Dodge, K.A.
  Griffioen, S.E. Kuhn, R.~Minehart, et~al.
\newblock {Moments of the spin structure functions and for $g_1^p$ and $g_1^d$
  for 0.05 $<$ $Q^2$ $<$ 3.0 GeV$^2$}.
\newblock {\em Physics Letters B}, 672(1):12 -- 16, 2009.

\bibitem{EG4}
M.~Battaglieri, R.~De Vita, A.~Deur, M.~Ripani, et~al.
\newblock {Jefferson Lab CLAS EG4 Run Group}.
\newblock \url{http://www.jlab.org/exp prog/proposals/03/PR03-006.pdf}.

\bibitem{g1global}
M.~Osipenko, S.~Simula, W.~Melnitchouk, P.~Bosted, V.~Burkert, E.~Christy,
  K.~Griffioen, C.~Keppel, S.~Kuhn, and G.~Ricco.
\newblock {Global Analysis of Data on the Proton Structure Function ${g}_{1}$
  and the Extraction of its Moments}.
\newblock {\em Phys. Rev. D}, 71:054007, 2005.

\bibitem{g1JP}
J.-P. Chen.
\newblock {Spin Sum Rules and Polarizabilities: Results From Jefferson Lab}.
\newblock In {\em {5th International Workshop on Chiral Dynamics, Theory and
  Experiment (CD06) Durham / Chapel Hill, North Carolina, September 18-22,
  2006}}, 2006.

\bibitem{E94010}
M.~Amarian et~al.
\newblock {Measurement of the Generalized Forward Spin Polarizabilities of the
  Neutron}.
\newblock {\em Phys. Rev. Lett.}, 93:152301, 2004.

\bibitem{ELSA}
H.~Dutz, K.~Helbing, J.~Krimmer, T.~Speckner, G.~Zeitler, et~al.
\newblock {First Measurement of the Gerasimov-Drell-Hearn Sum Rule for
  $^{1}\mathrm{H}$ from 0.7 to 1.8 GeV at ELSA}.
\newblock {\em Phys. Rev. Lett.}, 91:192001, 2003.

\bibitem{Krebs}
V.~Bernard, E.~Epelbaum, and U.-G. Krebs, H.and~Meissner.
\newblock {New Insights into the Spin Structure of the Nucleon}.
\newblock {\em Phys. Rev. D}, 87:054032, 2013.

\bibitem{Gold}
V.~Lensky, J.~M. Alarc\'on, and V.~Pascalutsa.
\newblock {Moments of Nucleon Structure Functions at Next--To--Leading Order in
  Baryon Chiral Perturbation Theory}.
\newblock {\em Phys. Rev. C}, 90:055202, Nov 2014.

\bibitem{BCElastic}
B.~Adeva et~al.
\newblock {Combined Analysis of World Data on Nucleon Spin Structure
  Functions}.
\newblock {\em Physics Letters B}, 320(3):400 -- 406, 1994.

\bibitem{ArringtonFit}
J.~Arrington.
\newblock {Implications of the Discrepancy Between Proton Form Factor
  Measurements}.
\newblock {\em Phys. Rev. C}, 69:022201, 2004.

\bibitem{KarlBC}
K.~Slifer, O.~A. Rond\'on, and others.
\newblock {Probing Quark-Gluon Interactions with Transverse Polarized
  Scattering}.
\newblock {\em Phys. Rev. Lett.}, 105:101601, 2010.

\bibitem{JPGamma1Plot}
J.-P. Chen.
\newblock {Moments of the Spin Structure Functions: Sum Rules and
  Polarizabilities}.
\newblock {\em International Journal of Modern Physics E}, 19(10):1893--1921,
  2010.

\bibitem{Burkert}
V.~D. Burkert and B.~L. Ioffe.
\newblock {On the $Q^2$ Variation of Spin-Dependent Deep-Inelastic
  Electron-Proton Scattering}.
\newblock {\em Physics Letters B}, 296(1):223 -- 226, 1992.

\bibitem{Soffer}
J.~Soffer and O.~Teryaev.
\newblock {QCD Radiative and Power Corrections and Generalized
  Gerasimov-Drell-Hearn Sum Rules}.
\newblock {\em Phys. Rev. D}, 70:116004, 2004.

\bibitem{hyper_data}
S~G Karshenboim.
\newblock What do we actually know about the proton radius?
\newblock {\em Canadian Journal of Physics}, 77(4):241--266, 1999.

\bibitem{Hyperfine}
A.~V. Volotka, V.~M. Shabaev, G.~Plunien, and G.~Soff.
\newblock {Zemach and Magnetic Radius of the Proton from the Hyperfine
  Splitting in Hydrogen}.
\newblock {\em The European Physical Journal D - Atomic, Molecular, Optical and
  Plasma Physics}, 33(1):23--27, 2005.

\bibitem{Fermi}
E.~Fermi.
\newblock {{\"U}ber die magnetischen Momente der Atomkerne}.
\newblock {\em Zeitschrift f{\"u}r Physik}, 60(5):320--333, 1930.

\bibitem{Hyperfine3}
V.~Nazaryan, C.~E. Carlson, and K.~Griffioen.
\newblock {Experimental Constraints on Polarizability Corrections to Hydrogen
  Hyperfine Structure}.
\newblock {\em Phys. Rev. Lett.}, 96:163001, 2006.

\bibitem{Zemach}
A.~C. Zemach.
\newblock {Proton Structure and the Hyperfine Shift in Hydrogen}.
\newblock {\em Phys. Rev.}, 104:1771--1781, 1956.

\bibitem{Hyperfine2}
C.~E. Carlson, V.~Nazaryan, and K.~Griffioen.
\newblock {Proton Structure Corrections to Electronic and Muonic Hydrogen
  Hyperfine Splitting}.
\newblock {\em Phys. Rev. A}, 78:022517, 2008.

\bibitem{CEBAF}
J.~Alcorn et~al.
\newblock {Basic Instrumentation for Hall A at Jefferson Lab}.
\newblock {\em Nucl. Instrum. Meth.}, A522:294--346, 2004.

\bibitem{Injector}
T.~B. Humensky et~al.
\newblock {SLAC's Polarized Electron Source Laser System and Minimization of
  Electron Beam Helicity correlations for the E-158 Parity Violation
  Experiment}.
\newblock {\em Nucl. Instrum. Meth.}, A521:261--298, 2004.

\bibitem{doped}
C.~Hernandez-Garcia, M.L. Stutzman, and P.G. O`Shea.
\newblock {Electron Sources for Accelerators}.
\newblock {\em Phys. Today}, 62:44, 2008.

\bibitem{Chao}
C.~Gu.
\newblock {Helicity Decoder for E08-027}.
\newblock Technical report, E08-027 Collaboration, 2014.
\newblock
  \url{http://hallaweb.jlab.org/experiment/g2p/technotes/E08027_TN2014_12.pdf}.

\bibitem{Tungsten}
{Tungsten Calorimeter Webpage}.
\newblock
  \url{http://hallaweb.jlab.org/experiment/LEDEX/camsonne/webpage_calorimeter/page_principale.html}.
\newblock Accessed: 2015-08-18.

\bibitem{Pengjia}
P.~Zhu.
\newblock {Beam Charge Measurement for $g_2^p$ Experiment}.
\newblock Technical report, E08-027 Collaboration, 2015.
\newblock
  \url{http://hallaweb.jlab.org/experiment/g2p/technotes/bcm_technote.pdf}.

\bibitem{MollerXS}
A.~Afanasev and A.~Glamazdin.
\newblock {Atomic Electron Motion for Moller Polarimetry in a Double Arm Mode}.
\newblock {\em Nucl. Instrum. Methods}, 1996.

\bibitem{MollerRaw}
{Results of Raw M{\o}ller Measurement}.
\newblock
  \url{http://hallaweb.jlab.org/equipment/moller/2012_raw_results_archive.html}.
\newblock Accessed: 2015-09-08.

\bibitem{Moller}
{E08-027 M{\o}ller Measurement}.
\newblock \url{http://hallaweb.jlab.org/equipment/moller/e08-027.html}.
\newblock Accessed: 2015-09-08.

\bibitem{ArcE}
{Arc Energy Measurement}.
\newblock \url{http://hallaweb.jlab.org/equipment/beam/energy/arc_web.html}.
\newblock Accessed: 2015-09-11.

\bibitem{Roblin}
Y.~Roblin.
\newblock {Beamline and Runorbit for g2p Readiness Review}.
\newblock Technical report, E08-027 Collaboration, 2012.
\newblock
  \url{https://hallaweb.jlab.org/wiki/index.php/G2p_Safety_Readiness_Review}.

\bibitem{Musson}
J.~Musson.
\newblock {Functional Description of Algorithms Used in Digital Receivers}.
\newblock Technical report, Jefferson Lab, 2014.
\newblock \url{https://wiki.jlab.org/ciswiki/images/2/23/JLAB-TN-14-028.pdf}.

\bibitem{Chao2}
C.~Gu.
\newblock {Target Field Mapping and Uncertainty Estimation}.
\newblock Technical report, E08-027 Collaboration, 2015.
\newblock
  \url{http://hallaweb.jlab.org/experiment/g2p/technotes/E08027_TN2015_13.pdf}.

\bibitem{Jie2}
J.~Liu.
\newblock {Magnetic Field Mapping on a Translation Table}.
\newblock Technical report, E08-027 Collaboration, 2013.
\newblock
  \url{http://hallaweb.jlab.org/experiment/g2p/collaborators/jie/2011_10_05_fieldmap_report/Target_Field_Map_Report.pdf}.

\bibitem{Zhu}
P.~Zhu et~al.
\newblock {Beam Position Reconstruction for the g2p Experiment in Hall A at
  Jefferson Lab}.
\newblock {\em Nucl. Instrum. Meth.}, A808:1 -- 10, 2016.

\bibitem{Pathria}
R.~K. Pathria and P.~D. Beale.
\newblock {\em {Statistical Mechanics}}.
\newblock Elsevier/Academic Press, Burlington, MA, 2011.

\bibitem{Averett}
T.~D. Averett, D~.G. Crabb, D.~B. Day, T.~J. Liu, J.~S. McCarthy, J.~Mitchell,
  O.~Rondon, D.~Zimmermann, I.~Sick, B.~Zihlmann, G.~Court, H.~Dutz, W.~Meyer,
  A.~Rijllart, S.~St. Lorant, J.~Button-Shafer, and J.~Johnson.
\newblock {A Solid Polarized Target for High-Luminosity Experiments }.
\newblock {\em Nucl. Instrum. Meth.}, A427(3):440 -- 454, 1999.

\bibitem{SE}
B.~Corzilius, A.~A. Smith, and R.~G. Griffin.
\newblock {Solid Effect in Magic Angle Spinning Dynamic Nuclear Polarization}.
\newblock {\em The Journal of Chemical Physics}, 137(5), 2012.

\bibitem{PolHyper}
D.~G. Crabb and W.~Meyer.
\newblock {Solid Polarized Targets For Nuclear and Particle Physics
  Experiments}.
\newblock {\em Annual Review of Nuclear and Particle Science}, 47(1):67--109,
  1997.

\bibitem{Tranny}
A.~Abragam and M.~Goldman.
\newblock {Principles of Dynamic Nuclear Polarisation}.
\newblock {\em Reports on Progress in Physics}, 41(3):395, 1978.

\bibitem{Levitt}
M.~H. Levitt.
\newblock {\em Spin Dynamics: Basics of Nuclear Magnetic Resonance, 2nd
  Edition}.
\newblock Wiley, Chicester, UK, 2008.

\bibitem{MRI}
D.~W. McRobbie.
\newblock {\em {MRI from Picture to Proton}}.
\newblock Cambridge University Press, Cambridge, 2007.

\bibitem{TargetPol}
J.~Pierce, J.~Maxwell, C.~Keith, et~al.
\newblock {Dynamically Polarized target for the g2p and gep Experiments at
  Jefferson Lab}.
\newblock {\em Physics of Particles and Nuclei}, 45(1):303--304, 2014.

\bibitem{Goldman}
M~Goldman.
\newblock {Measurement of Dipolar Energy in Nuclear Spin Systems}.
\newblock {\em Journal of Magnetic Resonance}, 17(3):393 -- 398, 1975.

\bibitem{Charpak}
G.~Charpak.
\newblock {Applications of Proportional Chambers and Drift Chambers in
  High-Energy Physics and Other Fields}.
\newblock {\em Nature}, 270:479--482, 1977.

\bibitem{Sauli}
F.~Sauli.
\newblock {Principles of Operation of Multiwire Proportional and Drift
  Chambers}, 1977.
\newblock {CERN-77-09}.

\bibitem{GASC}
M.~Iodice et~al.
\newblock {The CO2 Gas Cherenkov Detectors for the Jefferson Lab Hall-A
  Spectrometers}.
\newblock {\em Nucl. Instrum. Meth.}, A411(2):223 -- 237, 1998.

\bibitem{happex}
{HAPPEX Collaboration}.
\newblock \url{http://hallaweb.jlab.org/experiment/HAPPEX/}.
\newblock Accessed: 2015-08-18.

\bibitem{CODA}
{Jefferson Lab CODA Webpage}.
\newblock \url{https://coda.jlab.org/drupal/}.
\newblock Accessed: 2016-09-13.

\bibitem{EPICS}
Argonne~National Laboratory.
\newblock {Experimental Physics and Industrial Control System}.
\newblock \url{http://www.aps.anl.gov/epics/}.

\bibitem{Analyzer}
{Hall A Analyzer}.
\newblock \url{http://hallaweb.jlab.org/podd/}.
\newblock Accessed: 2017-01-07.

\bibitem{ROOT}
{ROOT Data Analysis Framework}.
\newblock \url{https://root.cern.ch/}.
\newblock Accessed: 2017-01-07.

\bibitem{HALOG}
{Hall A Electronic Log Book}.
\newblock \url{http://hallaweb.jlab.org/halog/log/html/logdir.html}.
\newblock Accessed: 2017-01-07.

\bibitem{g2pmysql}
{The g2p MySQL Web Database}.
\newblock \url{https://hallaweb.jlab.org/experiment/g2p/mysql/}.
\newblock Accessed: 2017-01-07.

\bibitem{MySQL}
{The g2p MySQL Wiki}.
\newblock \url{https://hallaweb.jlab.org/wiki/index.php/G2pmysql}.
\newblock Accessed: 2017-01-07.

\bibitem{Melissa}
M.~Cummings.
\newblock {Efficiency Studies and PID Cut Optimization}.
\newblock Technical report, E08-027 Collaboration, 2013.
\newblock
  \url{http://hallaweb.jlab.org/experiment/g2p/technotes/E08027_TN2013_04.pdf}.

\bibitem{Jie}
J.~Liu.
\newblock {VDC Multi-Track Efficiency Study}.
\newblock Technical report, E08-027 Collaboration, 2015.
\newblock
  \url{http://hallaweb.jlab.org/experiment/g2p/collaborators/jie/2013_09_24_multi_track_notes/vdc_eff_technotes.pdf}.

\bibitem{RyanTRIG}
R.~Zielinski.
\newblock {Trigger Scintillator Efficiency}.
\newblock Technical report, E08-027 Collaboration, 2014.
\newblock \url{https://userweb.jlab.org/~rbziel/TEff_Update.pdf}.

\bibitem{Vince}
V.~Sulkosky.
\newblock {Spectromer Optics Calibration for E97-110}.
\newblock Technical report, E97-110 Collaboration, 2005.
\newblock
  \url{http://hallaweb.jlab.org/experiment/E97-110/tech/optics_calib.ps.gz}.

\bibitem{Chao4}
C.~Gu.
\newblock {Spectromer Optics Study for E08-027}.
\newblock Technical report, E08-027 Collaboration, 2016.
\newblock
  \url{http://hallaweb.jlab.org/experiment/g2p/technotes/E08027_TN2016_18.pdf}.

\bibitem{Optics}
N.~Liyanage.
\newblock {Optics Calibration of the Hall A High Resolution Spectrometers using
  the New Optimizer}.
\newblock Technical report, Jefferson Laboratory, 2002.
\newblock
  \url{https://hallaweb.jlab.org/data_reduc/AnaWork2002/liyanage-ws2002.ps}.

\bibitem{Chao3}
C.~Gu.
\newblock {\em {The Spin Structure of the Proton at Low Q2: A Measurement of
  the Structure Function g2p}}.
\newblock PhD thesis, University of Virginia, 2016.

\bibitem{Min}
M.~Huang.
\newblock {Central Scattering Angle Measurement}.
\newblock Technical report, E08-027 Collaboration, 2014.
\newblock
  \url{http://hallaweb.jlab.org/experiment/g2p/collaborators/mhuang/technote/pointingSummary.pdf}.

\bibitem{Toby}
T.~Badman.
\newblock {g2p Target Polarization Analysis}.
\newblock Technical report, E08-027 Collaboration, 2016.
\newblock
  \url{http://hallaweb.jlab.org/experiment/g2p/collaborators/toby/technotes/target.pdf}.

\bibitem{N2Scale}
P.~E. Bosted, R.~Fersch, et~al.
\newblock {Ratios of $^{15}\mathrm{N}$/$^{12}\mathrm{C}$ and
  $^{4}\mathrm{He}$/$^{12}\mathrm{C}$ Inclusive Electroproduction Cross
  Sections in the Nucleon Resonance region}.
\newblock {\em Phys. Rev. C}, 78:015202, 2008.

\bibitem{TobyDil}
T.~Badman.
\newblock {g2p Dilution Analysis}.
\newblock Technical report, E08-027 Collaboration, 2016.
\newblock
  \url{http://hallaweb.jlab.org/experiment/g2p/collaborators/toby/technotes/dilution.pdf}.

\bibitem{Polrad}
I.~Akushevich, A.~Ilyichev, N.~Shumeiko, A.~Soroko, and A.~Tolkachev.
\newblock {POLARD 2.0 FORTRAN Code for the Radiative Corrections Calculation to
  Deep Inelastic Scattering of Polarized Particles}.
\newblock {\em Comput. Phys. Commun.}, 104:201--244, 1997.

\bibitem{Polrad2}
I.~V. Akushevich, A.~N. Ilyichev, and N.~M. Shumeiko.
\newblock {Radiative Electroweak Effects in Deep Inelastic Scattering of
  Polarized Leptons by Polarized Nucleons}.
\newblock {\em J. Phys.}, G24:1995--2007, 1998.

\bibitem{Polrad3}
A.~Afanasev, I.~Akushevich, and N.~Merenkov.
\newblock {Model Independent Radiative Corrections in Processes of Polarized
  Electron Nucleon Elastic Scattering}.
\newblock {\em Phys. Rev.}, D64:113009, 2001.

\bibitem{Polrad4}
A.~V. Afanasev, I.~Akushevich, A.~Ilyichev, and N.~P. Merenkov.
\newblock {QED Radiative Corrections to Asymmetries of Elastic $ep$ Scattering
  in Hadronic Variables}.
\newblock {\em Phys. Lett.}, B514:269--278, 2001.

\bibitem{TSAI}
Y.-S. Tsai.
\newblock {Radiative Corrections to Electron Scatterings}.
\newblock 1971.
\newblock
  {\url{http://www-public.slac.stanford.edu/sciDoc/docMeta.aspx?slacPubNumber=SLAC-PUB-0848}}.

\bibitem{MT}
L.~W. Mo and Y.~S. Tsai.
\newblock {Radiative Corrections to Elastic and Inelastic $\mathrm{ep}$ and
  $\mathrm{up}$ Scattering}.
\newblock {\em Rev. Mod. Phys.}, 41:205--235, 1969.

\bibitem{MillerT}
G.~Miller.
\newblock {\em {Inelastic Electron Scattering at Large Angles}}.
\newblock PhD thesis, SLAC, 1971.

\bibitem{Maximon}
L.~C. Maximon and D.~B. Isabelle.
\newblock {Radiative Tail in Elastic Electron Scattering}.
\newblock {\em Phys. Rev.}, 133:B1344--B1350, 1964.

\bibitem{Barreau}
P.~Barreau, M.~Bernheim, J.~Duclos, J.M. Finn, Z.~Meziani, J.~Morgenstern,
  J.~Mougey, D.~Royer, B.~Saghai, D.~Tarnowski, S.~Turck-Chieze, et~al.
\newblock {Deep-Inelastic Electron Scattering from Carbon}.
\newblock {\em Nuclear Physics A}, 402(3):515 -- 540, 1983.

\bibitem{Borie}
E.~Borie.
\newblock {Correction to the Formula for the Radiative Tail in Elastic Electron
  Scattering}.
\newblock {\em Lett. Nuovo Cim.}, 1S2:106--109, 1971.
\newblock [Lett. Nuovo Cim.1,106(1971)].

\bibitem{Loops}
J.~Schwinger.
\newblock {Quantum Electrodynamics. II. Vacuum Polarization and Self-Energy}.
\newblock {\em Phys. Rev.}, 75:651--679, 1949.

\bibitem{Miller}
G.~Miller et~al.
\newblock {Inelastic Electron-Proton Scattering at Large Momentum Transfers and
  the Inelastic Structure Functions of the Proton}.
\newblock {\em Phys. Rev. D}, 5:528--544, 1972.

\bibitem{Soft}
F.~Bloch and A.~Nordsieck.
\newblock {Note on the Radiation Field of the Electron}.
\newblock {\em Phys. Rev.}, 52:54--59, 1937.

\bibitem{Gramolin}
A.~V. Gramolin, V.~S. Fadin, A.~L. Feldman, R.~E. Gerasimov, D.~M. Nikolenko,
  I.~A. Rachek, and D.~K. Toporkov.
\newblock {A New Event Generator for the Elastic Scattering of Charged Leptons
  on Protons}.
\newblock {\em Journal of Physics G: Nuclear and Particle Physics},
  41(11):115001, 2014.

\bibitem{Bardin}
D.~{\relax Yu}. Bardin and N.~M. Shumeiko.
\newblock {An Exact Calculation of the Lowest Order Electromagnetic Correction
  to the Elastic Scattering}.
\newblock {\em Nucl. Phys.}, B127:242--258, 1977.

\bibitem{Akhundov}
A.~A. Akhundov, D.~{\relax Yu}. Bardin, and N.~M. Shumeiko.
\newblock {Electromagnetic Corrections to the Deep Inelastic mu p Scattering at
  High-Energies}.
\newblock {\em Sov. J. Nucl. Phys.}, 26:660, 1977.
\newblock [Yad. Fiz.26,1251(1977)].

\bibitem{Stein}
S.~Stein et~al.
\newblock {Electron Scattering at 4-Degrees with Energies of 4.5-GeV - 20-GeV}.
\newblock {\em Phys. Rev.}, D12:1884, 1975.

\bibitem{Williamson}
L.~C. Maximon and S.~E. Williamson.
\newblock {Piecewise Analytic Evaluation of the Radiative Tail from Elastic and
  Inelastic Electron Scattering}.
\newblock {\em Nucl. Instrum. Meth.}, A258:95--110, 1987.

\bibitem{Schwinger}
J.~Schwinger.
\newblock {Quantum Electrodynamics. III. The Electromagnetic Properties of the
  Electron: Radiative Corrections to Scattering}.
\newblock {\em Phys. Rev.}, 76:790--817, 1949.

\bibitem{Badalek}
B.~Badelek, D.~Bardin, K.~Kurek, and C.~Scholz.
\newblock {Radiative Correction Schemes in Deep Inelastic Muon Scattering}.
\newblock {\em Zeitschrift f{\"u}r Physik C Particles and Fields},
  66(4):591--599, 1995.

\bibitem{Dasu}
S.~Dasu et~al.
\newblock {Measurement of Kinematic and Nuclear Dependence of
  $R=\frac{{\ensuremath{\sigma}}_{L}}{{\ensuremath{\sigma}}_{T}}$ in Deep
  Inelastic Electron Scattering}.
\newblock {\em Phys. Rev. D}, 49:5641--5670, 1994.

\bibitem{RadLength}
J.~Singh and V.~Sulkosky.
\newblock {Radiation Thickness, Collisional Thickness, and Most Probable
  Collisional Energy Loss for E97-110}.
\newblock Technical report, E97-110 Collaboration, 2007.
\newblock
  \url{http://hallaweb.jlab.org/experiment/E97-110/tech/radlength_sagdhv130.pdf}.

\bibitem{Puckett}
A.~Puckett.
\newblock {\em {Recoil Polarization Measurements of the Proton Electromagnetic
  Form Factor Ratio to High Momentum Transfer}}.
\newblock PhD thesis, MIT, 2015.

\bibitem{JonesRSS}
M.~K. Jones et~al.
\newblock Proton ${G}_{E}/{G}_{M}$ from beam-target asymmetry.
\newblock {\em Phys. Rev. C}, 74:035201, 2006.

\bibitem{RTAIL}
R.~Altulmus and J.~Wise.
\newblock Rosetail.f.
\newblock FORTRAN Analysis Code.
\newblock Modified by K.~Slifer and S.~Choi.

\bibitem{RADCOR}
R.~R. Whitney.
\newblock Radcor.f.
\newblock FORTRAN Analysis Code.
\newblock Modified by K.~Slifer and J.~Singh.

\bibitem{POLCODE}
I.~Akushevich, A.~Ilyichev, N.~Shumeiko, A.~Soroko, and A.~Tolkachev.
\newblock Polsig.f.
\newblock FORTRAN Analysis Code.
\newblock Modified by K.~Slifer and S.~Choi.

\bibitem{Qiang}
Y.~Qiang.
\newblock {\em {Search for Pentaquark Partners $\Theta^{++}$, $\Sigma^0$ and
  $N^0$ in H (e,e'K $(\pi$)) X Reactions at Jefferson Lab Hall A}}.
\newblock PhD thesis, MIT, 2007.

\bibitem{JixieAngle}
J.~Zhang.
\newblock {Angle Fittings used by Karl in entry 109}.
\newblock \url{https://hallaweb.jlab.org/dvcslog/g2p/110}.

\bibitem{MelissaT}
M.~Cummings.
\newblock {\em {Investigating Proton Spin Structure: A Measurement of $g_2^p$
  at Low $Q^2$ }}.
\newblock PhD thesis, The College of William and Mary, 2016.

\bibitem{ChaoCom}
C.~Gu.
\newblock Private conversation.

\bibitem{HallCRes}
Hall~C Collaboration.
\newblock {Hall C Resonance Data Archive}.
\newblock \url{https://hallcweb.jlab.org/resdata/database/}.

\bibitem{ChristyCom}
E.~Christy.
\newblock Private conversation.

\bibitem{TobyCom}
T.~Badman.
\newblock Private conversation.

\bibitem{JAFit}
J.~Arrington.
\newblock {Implications of the Discrepancy Between Proton Form Factor
  Measurements}.
\newblock {\em Phys. Rev. C}, 69:022201, 2004.

\bibitem{BostedGE}
P.~E. Bosted.
\newblock {Empirical Fit to the Nucleon Electromagnetic Form Factors}.
\newblock {\em Phys. Rev. C}, 51:409--411, 1995.

\bibitem{HallBRes}
Hall~B Collaboration.
\newblock {CLAS Physics Database}.
\newblock \url{http://clas.sinp.msu.ru/cgi-bin/jlab/db.cgi/}.

\bibitem{KarlT}
K.~Slifer.
\newblock {\em {Spin Structure of $^3$He and the Neutron at Low $Q^2$; A
  Measurement of the Extended GDH Integral and the Burkhardt-Cottingham Sum
  Rule }}.
\newblock PhD thesis, Temple University, 2004.

\bibitem{PascaCom}
V.~Pascalutsa.
\newblock Private conversation.

\bibitem{BS}
S.~Choi.
\newblock Bs.f.
\newblock FORTRAN Analysis Code.
\newblock Improved Simpson method integration routine.

\bibitem{FerschT}
R.~Fersch.
\newblock {\em {Measurement of Inclusive Proton Double-Spin Asymmetries and
  Polarized Structure Functions }}.
\newblock PhD thesis, The College of William and Mary, 2008.

\bibitem{Yield}
E08-027 Collaboration.
\newblock {g2p Data Quality}.
\newblock \url{https://hallaweb.jlab.org/wiki/index.php/G2p_data_quality}.

\bibitem{JieBPM}
J.~Liu.
\newblock {Beam Positions and Yields Study}.
\newblock Technical report, E08-027 Collaboration, 2017.
\newblock
  \url{http://hallaweb.jlab.org/experiment/g2p/collaborators/jie/talks/bpm_notes.pdf}.

\bibitem{JieT}
J.~Liu.
\newblock {\em {The Proton Spin Structure Function $g_2$ at Low $Q^2$ and
  Applications of Polarized $^3$He }}.
\newblock PhD thesis, University of Virginia, 2017.

\bibitem{MinT}
M.~Huang.
\newblock {\em {A Measurement of the Proton Spin Structure Function $g_2^p$ at
  Low $Q^2$ }}.
\newblock PhD thesis, Duke University, 2016.

\bibitem{PengiaT}
P.~Zhu.
\newblock {\em {A Measurement of the Proton's Spin Structure Function $g_2$ at
  Low $Q^2$ }}.
\newblock PhD thesis, University of Science and Technology of China, 2015.

\bibitem{RDT}
R.~Michaels.
\newblock Private conversation.

\bibitem{Bob}
R.~Michaels.
\newblock {Deadtime FAQ}.
\newblock \url{http://hallaweb.jlab.org/equipment/daq/dtime_faq.html}.
\newblock Accessed: 2016-09-13.

\bibitem{TS}
E.~Jastrzembski, D.~J. Abbott, W.~G. Heyes, R.~W. MacLeod, C.~Timmer, and
  E.~Wolin.
\newblock {The Jefferson Lab Trigger Supervisor System}.
\newblock In {\em Real Time Conference, 1999. Santa Fe 1999. 11th IEEE NPSS},
  pages 538--542, 1999.

\bibitem{PUBMAYBE}
R.~Zielinski, V.~Sulkosky, K.~SLifer, and P.~Solvignon.
\newblock {Inclusive Unpolarized $^{14}$N and $^3$He Scattering at Jefferson
  Lab}.
\newblock Publication in process.

\bibitem{Leo}
W.~R. Leo.
\newblock {\em Techniques for Nuclear and Particle Physics Experiments: A
  How-to Approach. 2nd Revised Edition}.
\newblock Spring-Verlag, Berlin, 1994.

\bibitem{Fermi22}
E.~Fermi.
\newblock {The Ionization Loss of Energy in Gases and in Condensed Materials}.
\newblock {\em Phys. Rev.}, 57:485--493, 1940.

\bibitem{Density}
R.M. Sternheimer, M.J. Berger, and S.M. Seltzer.
\newblock {Density Effect for the Ionization Loss of Charged Particles in
  Various Substances}.
\newblock {\em Atomic Data and Nuclear Data Tables}, 30(2):261 -- 271, 1984.

\bibitem{Shell}
W.~W. True.
\newblock {Nitrogen-14 and the Shell Model}.
\newblock {\em Phys. Rev.}, 130:1530--1537, 1963.

\bibitem{Nitrogen}
B.~Adeva et~al.
\newblock {Measurement of Proton and Nitrogen Polarization in Ammonia and a
  Test of Equal Spin Temperature}.
\newblock {\em Nucl. Instrum. Meth.}, A419(1):60 -- 82, 1998.

\bibitem{Oscar}
O.~A. Rondon.
\newblock {Corrections to Nucleon Spin Structure Asymmetries Measured on
  Nuclear Polarized Targets}.
\newblock {\em Phys. Rev. C}, 60:035201, 1999.

\end{thebibliography}
\end{singlespace}
\end{document}